Frederik Ravn Klausen

# Random Problems in Mathematical Physics

PHD THESIS

DEPARTMENT OF MATHEMATICAL SCIENCES
UNIVERSITY OF COPENHAGEN

September 2023

Frederik Ravn Klausen
`frederik.ravn.klausen@gmail.com`
Department of Mathematical Sciences
University of Copenhagen
Universitetsparken 5
2100 Copenhagen
Denmark

| | |
|---|---|
| **Thesis title:** | Random Problems in Mathematical Physics |
| **Supervisor:** | Albert H. Werner<br>University of Copenhagen |
| **Assessment Committee:** | Professor Jan Philip Solovej<br>University of Copenhagen |
| | Professor Michael Aizenman<br>Princeton University |
| | Universitetslektor Jakob Björnberg<br>Chalmers and Göteborg University |
| **Date of Submission:** | July 3,<br>2023 |
| **Date of Defense:** | September 25,<br>2023 |
| **ISBN:** | 978-87-7125-217-0 |

*This thesis has been submitted to the PhD School of The Faculty of Science,
University of Copenhagen. It was supported by the Villum Foundation (Grant No. 10059 and
25452.)*

# Preface

I would like to thank everyone I had the pleasure to discuss mathematics with during my PhD. From my point of view, mathematics is best done in cooperation and over the course of the PhD I had the opportunity to discuss mathematics with a great many people. Thanks for these discussions, which have definitely helped me grow as a mathematician.

In particular, thanks to Albert for supervision, our outside meetings during the pandemic and ever-quick administrative support. Thanks to Aran and Peter for their enthusiasm and insights about mathematical physics both at ETH and in countless calls. Thanks to everyone at QMATH who was always up for discussing mathematics at the coffee machine. In fact, one of the papers of this thesis grew directly out of such a discussion. Thanks to Vincent for enjoying countless interruptions in our shared office. I would also like to thank Simone Warzel for hosting me in Munich several times in the second half of my PhD, and for plenty of discussions as well as countless lessons about research. Thanks to the anonymous referees of the papers, as their valuable contributions have greatly enhanced the quality of the papers. Ulrik also deserves credit for coming up with many of the ideas in this thesis through our many enlightening discussions.

I am grateful to the Villum Foundation for financial support, to Birgit and Nordisk Kollegium for physical support as well as the mental support of the world's best community at Stuen Syd.

Let me also thank Alex, Asbjørn, Boris, Helene, Jakob, Jens, Nikolaj, 2·Peter, Svend, Ulrik and Vincent for invaluable support, sharing the ups and downs of doing a PhD and comments on this thesis.

Finally, I would like to thank the people close to me for love and support, especially towards the completion of this work, where I appreciated it more than ever. I am forever grateful to Frederikke and Niels Kristian.

<div align="right">

Frederik Ravn Klausen
July, 2023

</div>

# Abstract

This PhD thesis deals with a number of different problems in mathematical physics with the common thread that they have probabilistic aspects. The problems all stem from mathematical studies of lattice systems in statistical and quantum physics; however beyond that, the selection of the concrete problems is to a certain extent arbitrary. This thesis consists of an introduction and seven papers.

In [**Mass**], we give a new proof of exponential decay of the truncated two-point correlation functions of the two-dimensional Ising model at the critical temperature in a magnetic field.

In [**MonCoup**], we provide counterexamples to monotonicity properties of the loop O(1) model and the (single, traced, sourceless) random current model. Additionally, we prove that the uniform even subgraph of the (traced, sourceless) double random current model has the law of the loop O(1) model.

In [**Kertész**], we prove strict monotonicity and continuity of the Kertész line for the random cluster model in the presence of a magnetic field implemented through a ghost vertex. Furthermore, we give new rigorous bounds that are asymptotically correct in the limit $h \to 0$.

In [**UEG**], we prove that the uniform even subgraph percolates in $\mathbb{Z}^d$ for $d \geq 2$, that the phase transition of the loop O(1) model on $\mathbb{Z}^d$ is non-trivial and we provide a polynomial lower bound on the correlation functions of both the loop O(1) model and single random current corresponding to a supercritical Ising model on $\mathbb{Z}^d$ whenever $d \geq 3$.

In [**MagQW**], we introduce a model for quantum walks on $\mathbb{Z}^2$ in a random magnetic field where the plaquette fields are i.i.d. random. We prove an a priori estimate and an exponential decay result of the expectations of fractional moments of the Green function.

In [**Spec**], we obtain a representation of generators of Markovian open quantum system with natural locality assumptions as a direct integral of finite range bi-infinite Laurent matrices with finite rank perturbations. We use the representation to calculate the spectrum of some infinite volume open quantum Lindbladians analytically and to prove gaplessness of the spectrum, absence of residual spectrum and a condition for convergence of finite volume spectra to their infinite volume counterparts.

In [**OpenLoc**], we consider a Markovian open quantum system where the terms in the generator are local. We prove that in the presence of any local dephasing in the system, then any steady state of the system will have exponentially decaying coherences. Furthermore, we prove for a general class of models that includes our motivating examples, that the results holds in expectation for large disorder, that is, a sufficiently strong random potential in the Hamiltonian. That result extends Anderson localization to open quantum systems.

# Sammenfatning

Denne Ph.D.-afhandling omhandler en række forskellige problemstillinger inden for matematisk fysik med det tilfælles at de alle har sandsynlighedsaspekter. Alle problemstillingerne stammer fra matematiske studier gittersystemer i statistisk fysik og kvantefysik, men derudover er udvælgelsen af de konkrete problemer i en vis udstrækning vilkårlig. Afhandlingen består af en introduktion og syv artikler.

I [Mass] giver vi et nyt bevis for eksponentielt henfald af de trunkerede topunktskorrelationsfunktioner af den todimensionelle Ising-model ved den kritiske temperatur i et magnetfelt.

I [MonCoup] giver vi modeksempler på monotonicitetsegenskaber for loop-O(1)-modellen og den (enkelte, sporede, divergensfri) tilfældige strøm. Vi beviser også, at den uniforme lige delgraf af den (sporede, divergensfri) dobbelte tilfældige strøm har samme lov som loop-O(1)-modellen.

I [Kertész] beviser vi streng monotonicitet og kontinuitet af Kertész-linjen for FK-perkolation i et magnetfelt implementeret gennem et spøgelsespunkt. Desuden giver vi nye bånd, der er asymptotisk korrekte i grænsen $h \to 0$.

I [UEG] beviser vi, at den uniforme lige delgraf perkolerer i $\mathbb{Z}^d$ for $d \geq 2$, at faseovergangen af loop-O(1)-modellen på $\mathbb{Z}^d$ er ikke-triviel, derudover giver vi et polynomielt nedre bånd for korrelationsfunktioner af både loop-O(1)-modellen og den enkelte tilfældige strøm svarende til en superkritisk Ising-model på $\mathbb{Z}^d$, med $d \geq 3$.

I [MagQW] introducerer vi en model for kvantegåture på $\mathbb{Z}^2$ i et tilfældigt magnetfelt, hvor magnetfelterne i hver plakette er uafhængigt identisk fordelte. Vi beviser et a priori-estimat og et eksponentielt henfaldsresultat af forventningerne til brøkmomenter af Greens-funktionen.

I [Spec] opnår vi en repræsentation af generatorer af markovske åbne kvantesystemer med naturlige lokalitetsantagelser som et direkte integral af biuendelige Laurentmatricer der har endelig rækkevidde med endelig rang pertubationer. Vi bruger repræsentationen til analytisk at beregne spektret af relevante eksempler Lindbladoperatorer i uendelig volumen. Derudover bruger vi repræsentationen til at bevise at spektrene aldrig har noget gab, at operatorerne aldrig har residualt spektrum og til at vise en betingelse for konvergens af spektrene i endelig volumen til deres modstykker i uendelige volumen.

I [OpenLoc], betragter vi et markovsk åbent kvantesystem, hvor alle led i generatoren er lokale. Vi beviser, at hvis der er lokal dephasing i systemet, så vil enhver stabil tilstand af systemet have eksponentielt henfaldende kohærenser. Ydermere beviser vi for en generel klasse af modeller, der inkluderer vores motiverende eksempler, at resultaterne holder i forventning, hvis der er et tilstrækkeligt stærkt tilfældigt potentiale i Hamiltonoperatoren. Derved udvider resultatet teorien om Anderson lokalisering til åbne kvantesystemer.

# Contributions and Structure

This thesis consists of two parts: An introduction and a collection of papers. First, we give an overall introduction to some probabilistic methods that we use throughout this thesis. The introduction deals mostly with proofs of exponential decay in lattice models using iterations, a theme which is sometimes known as the Hammersley paradigm. Therefore the introduction is a highly selective summary of methods which the later chapters build on and not in any way a representative review of the literature. Next, we give more technical introductions to each of the papers of this thesis.

The second part consists of the following papers.

During my PhD, I also was a (co-)author of the following projects that are not included in the this PhD-thesis.

# Contents

Contents

# Part I.

# Introduction

# 1. A Gentle Introduction: Probability in Mathematical Physics

In this chapter, we motivate and provide a simple introduction to some of the objects and techniques studied in this thesis using the Ising model and Bernoulli percolation as guiding examples. We invite the reader to look at pictures and ponder about probability as we go along. The only prerequisite is some degree of comfort with probabilities. Therefore, experts in the subject may choose to start directly in Section 2. For a much more in-depth introduction to classical statistical mechanics, see the inspiring introduction [FV17].

## 1.1. Intuition on Phase Transitions: A Look into the Ising Model

Phase transitions are a part of the everyday life of every human being. One of the most well-known phase transitions is the phase transition of water, which changes state from solid (ice) to liquid (water) and eventually to gas (vapor) as the temperature is increased.

The phase transition of water is so fundamental to us and to our understanding of temperature that we designed our temperature scales around it. At the same time, the phase transition of water is complex. Only one degree of temperature leads to an abrupt change of behavior. What is really going on when we make a cup of tea?

Phase transitions are a central object of study in statistical mechanics, where one approach is to find simple models that qualitatively exhibit some of the features of phase transitions. The prototypical example is the two-dimensional Ising model. Apart from its prototypical status in statistical mechanics, the purpose of introducing the Ising model here is two-fold. Firstly, it was in itself one of the main objects of study throughout this thesis (the papers [**Mass**], [**MonCoup**], [**Kertész**], [**UEG**]). In addition, some of the techniques that we use in [**MagQW**] and [**OpenLoc**] have direct simpler analogues in the case of the Ising model.

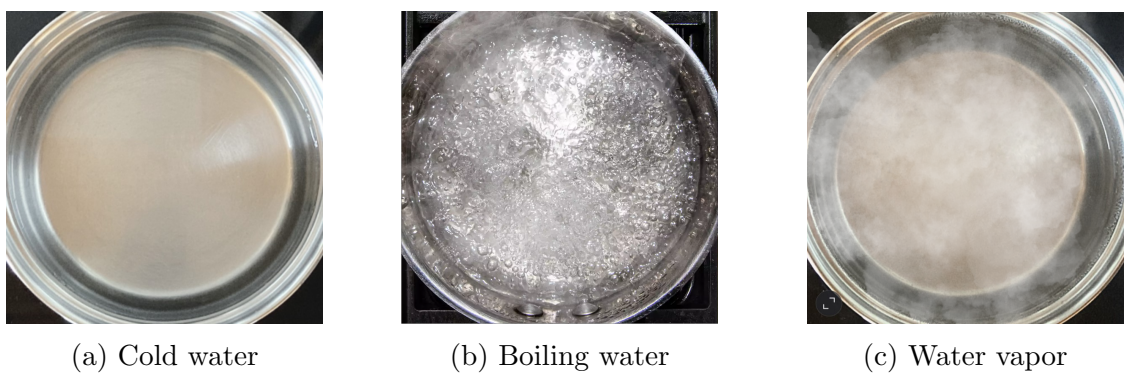

(a) Cold water       (b) Boiling water       (c) Water vapor

Figure 1.1.: Water at three different temperatures, arguably the most well-known phase transition. b) Image: Serious Eats / Amanda Suarez. Published with permission from seriouseats.com. All rights reserved.

### 1.1.1. The Energy of the Ising Model

The Ising model is a model of magnets. In school, it is often taught that a magnet is made up of mini-magnets. These mini-magnets, which point either up or down, we call spins, and we say the Ising model is a spin-model.

In the **ferromagnetic Ising model**, the energy of the system is lower whenever more mini-magnets point the same way as their neighbors. Thus, if we have a fixed network of mini-magnets, we can count the pairs where two neighbours point in opposite ways. More formally, a network can be specified by a graph $G$ with vertices $V$ and edges $E$, see Figure 1.2b) for an example.

Thus, we can say that the **energy**, which we will denote by the letter $H$, is given by

$H =$ number of pairs with mini-magnets pointing in opposite directions.

In the example in Figure 1.2, we see that $H = 3$ since there are three red edges between spins pointing in opposite directions.

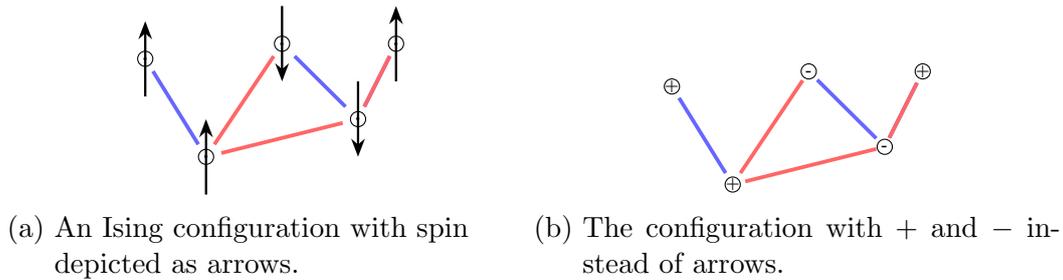

(a) An Ising configuration with spin depicted as arrows.

(b) The configuration with $+$ and $-$ instead of arrows.

Figure 1.2.: An Ising configuration with 5 spins (vertices $V$) and 5 designated pairs of edges (edges $E$) displayed with arrows and with $+$ and $-$ respectively. We have coloured the edges such that pairs of spins pointing the same way are blue and pairs pointing the opposite way are red.

For the ferromagnetic Ising model, it is easy to see that there are two states with the lowest energy $H = 0$, namely the one where all mini-magnets point up and the one where all mini-magnets point down, see Figure 1.3.

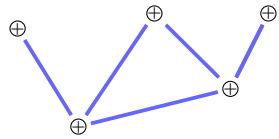

Figure 1.3.: The configuration with energy $H = 0$ where all spins point upwards.

Let us emphasize the **locality** of the model here. If we flip a mini-magnet then the only changes in energy come from edges adjacent to the vertex of a spin.

## 1.1.2. Relating Energy and Temperature

From the phase transitions of water we get an intuitive feeling that higher temperature means more movement in the system. Since any moving molecule has kinetic energy, it is natural to think that we can model higher temperatures of a system by higher energies. The central insight, that in some sense gave birth to the field of statistical mechanics, is that this perspective is particularly useful if we think of the system as **random**. In the case of the Ising model, it is most common to work with a fixed average energy, the so-called canonical ensemble.

The intuition described above is formalized through the Gibbs measure. The Gibbs measure is a particular probability measure. Probability measures, which play a central role in this thesis, constitute a systematic way of assigning probabilities to different events. The Gibbs measure for the Ising model we call $\mu_\beta$ and it assigns probabilities to different configurations of spins.

Under the Gibbs measure $\mu_\beta$ the probability of finding a given configuration $\sigma$ is proportional to the Boltzmann factors $e^{-\beta H(\sigma)}$, where $\beta > 0$ is a constant that we interpret as the inverse temperature. The constant of proportionality, or normalizing constant, $Z = \sum_\sigma e^{-\beta H(\sigma)}$ is also called the partition function. It is the sum of the Boltzmann factors for all configurations. It makes sure that the probabilities sum to 1. To summarize, the probability measure $\mu_\beta$ assigns a probability to a configuration $\sigma$ according to the formula

$$\mu_\beta[\sigma] = \frac{e^{-\beta H(\sigma)}}{Z}. \tag{1.1}$$

We motivate this specific assignment of probabilities in Section 1.3. Let us now give a simple example.

| Configuration: $\sigma$ | 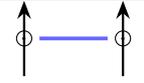 | 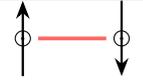 | 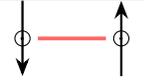 | 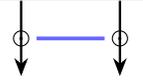 |
|---|---|---|---|---|
| Energy: $H(\sigma)$ | 0 | 1 | 1 | 0 |
| Boltzmann factor: $e^{-\beta H(\sigma)}$ | 1 | $e^{-\beta}$ | $e^{-\beta}$ | 1 |
| Probability: $\mu_\beta[\sigma]$ | $\frac{1}{2+2e^{-\beta}}$ | $\frac{e^{-\beta}}{2+2e^{-\beta}}$ | $\frac{e^{-\beta}}{2+2e^{-\beta}}$ | $\frac{1}{2+2e^{-\beta}}$ |

Table 1.1.: An example of a graph with two vertices and a single edge and the four possible confirgurations, their energies, Boltzmann factors and probabilities. Here the partition function is $Z = 2 + 2e^{-\beta}$.

Now notice that if the temperature is close to 0 then $\beta$ is very large and the probability that we are in a configuration with high energy is very small, which in turn means that the average energy will be small. In the simple example in Table 1.1, we see that the probability for the two configurations where the spins point the same way is close to $\frac{1}{2}$ and the remaining two have probabilities close to 0. Conversely, if $\beta$ is very small then all configurations will get approximately equal probability and the average energy will be large. In Table 1.1, we see that the probability of all four configurations becomes approximately $\frac{1}{4}$. More concretely, we can also calculate the average energy to be

$$1 \cdot \frac{e^{-\beta}}{2 + 2e^{-\beta}} + 1 \cdot \frac{e^{-\beta}}{2 + 2e^{-\beta}} = \frac{e^{-\beta}}{1 + e^{-\beta}}.$$

This expression tends to 0 when $\beta \to \infty$ and it tends to $\frac{1}{2}$ when $\beta \to 0$ and it thereby explains out interpretation of $\beta$ as the inverse temperature.

Now, the Ising model can be generalized to any graph, which by determining which minimagnets are neighbours. For example, Ising himself studied the one-dimensional model in [Isi24]. It turns out that a more interesting and complicated example is the **two-dimensional Ising model** to which we turn next.

### 1.1.3. The Two-Dimensional Ising Model

The graph of the two-dimensional Ising model is the square lattice. To illustrate it, we show an example of a configuration in Figure 1.4. The two-dimensional Ising model has a **phase transition** which has been subject of intense study [DC22]. In Figure 1.5, we see some different snapshots of random configurations from the two-dimensional Ising model for different values of $\beta$ showing a dramatic change of behavior as $\beta$ is increased.

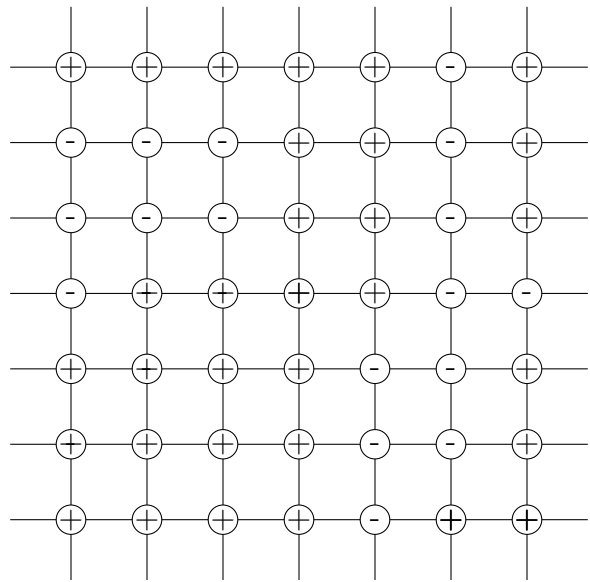

Figure 1.4.: An example of a part of the configuration of the 2D Ising model.

One way of mathematically study the phase transition exhibited in Figure 1.5 is through **correlation functions**. Correlation functions in different disguises play an important role throughout this thesis. In the simple case of the Ising model, we write the correlation function between a vertex $x$ and a vertex $y$ as

$$\langle \sigma_x \sigma_y \rangle_\beta = 2 \cdot \mu_\beta[\sigma_x = \sigma_y] - 1. \tag{1.2}$$

In other words, this quantity is two times the probability that the spins point the same way minus 1. Mathematically speaking, the correlation function is the expectation value of the product of two spins $\sigma_x \sigma_y$ under the probability measure $\mu_\beta$.

Notice that if $x$ and $y$ are completely independent of each other, then they point in the same way with probability $\frac{1}{2}$. In this case, we have $\langle \sigma_x \sigma_y \rangle_\beta = 0$. If they, on the other hand, always point the same way, then $\langle \sigma_x \sigma_y \rangle_\beta = 1$.

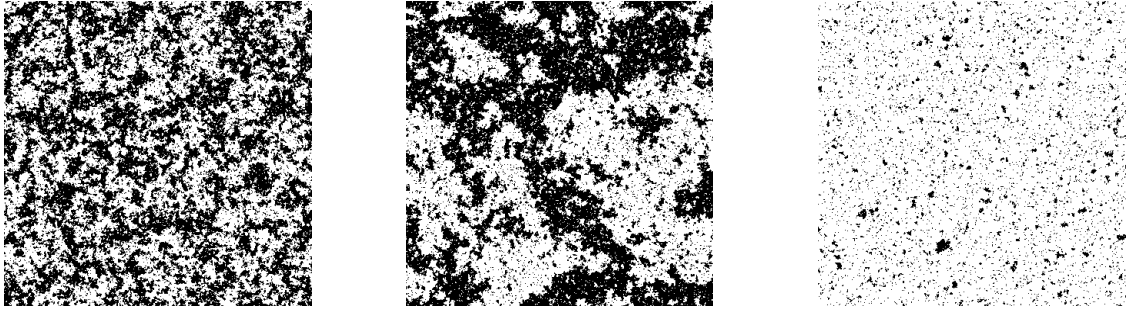

(a) A subcritical Ising config-uration. Here $\beta < \beta_c$.

(b) A critical Ising configura-tion. Here $\beta = \beta_c$.

(c) A supercritical Ising con-figuration. Here $\beta > \beta_c$.

Figure 1.5.: Three examples of a two-dimensional Ising model with varying inverse tem-perature $\beta$. Notice how the two-dimensional Ising model has a phase tran-sition. Here up-arrows are coloured black and down-arrows white (ⓒ Yvan Velenik).

For the Ising model, it is always the case that $\langle \sigma_x \sigma_y \rangle_\beta \geq 0$, which means that two spins are more likely to point the same way. This is known as the first Griffith's inequality after [Gri67]. Loosely speaking, the effect is strongest for neighbours. That means that if a spin at $x$ is pointing up then it is more likely that its neighbours are also pointing up.

In Figure 1.5 we see how spins closer to each other are more likely to point the same way. For all three figures, there are islands of + spins and − spins. In 1.5a) the islands are very small, in 1.5b) the islands are very large and in 1.5c) there is one big island of − spins with some lakes of + spins. This abrupt change signifies the phase transition. We say that the configuration in 1.5a) does not have long range order. This means that knowledge of the direction of one spin does not give knowledge of a spin far away, i.e. the correlation function $\langle \sigma_x \sigma_y \rangle_\beta$ tends to 0 when $x$ and $y$ become further and further apart. In contrast, the configuration in 1.5c) has long range order. There, a very large proportion of spins point the same way and so if one knows that a spin here is up, then it is more likely that a spin very far away is up, that is $\langle \sigma_x \sigma_y \rangle_\beta$ does not tend to 0. The situation in 1.5b) is in between, we say that it is **critical**. It is exactly at this point that the phase transition happen corresponding to the point where the water is boiling in Figure 1.1b).

Notice how in each picture, there is a sort of typical island-size or length scale of the system. This length scale, which we call the **correlation length**, must be related to the probability that two spins point the same way (which was captured by the correlation function). Below the critical temperature, it turns out [ABF87] that the correlation function is always **expo-nentially decaying**, that is, for some numbers $C$ and $\xi$

$$\langle \sigma_x \sigma_y \rangle_\beta \sim C e^{-\frac{|x-y|}{\xi}}. \tag{1.3}$$

This means if we pick $x$ and $y$ further and further apart then it becomes completely random whether they point the same way. This is the case in Figure 1.5a). The rate at which this randomness emerges is governed by the number $\xi$, which is the correlation length. For example, if $\xi$ is small (i.e. $\xi \approx 3$) then the clusters (islands) are very small, and if $\xi$ is large (i.e. $\xi \approx 3000$) then the clusters are rather large.

Exactly at the critical point $\beta = \beta_c$, the phase transition happens and $\xi = \infty$ and there is

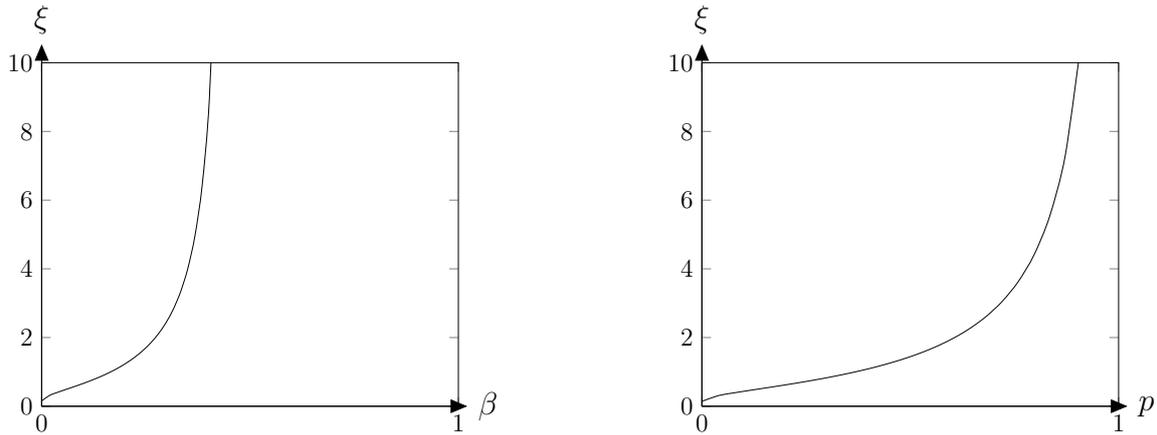

(a) The correlation length $\xi(\beta)$ of the two-dimensional Ising model.

(b) The correlation length $\xi(p)$ of one-dimensional Bernoulli percolation.

Figure 1.6.: The correlation length as a function of the parameter for a) the Ising model (schematically) and b) Bernoulli percolation. Notice how it diverges as $\beta \to \beta_c$ where $\beta_c = \frac{\ln\left(1+\sqrt{2}\right)}{2}$ (as proven by Onsager [Ons44]) and as $p \to p_c$ where $p_c = 1$.

no longer exponential decay. Instead, the correlation function decays polynomially. In Figure 1.6a the correlation length is shown as a function of $\beta$ and we see how it becomes infinite. Diverging correlation length is a hallmark of phase transitions.

Notice the analogy with boiling water (see Figure 1.1). Way below the phase transition there are a few water bubbles, and as we get closer to 100 degrees Celsius the bubbles become larger and larger. But once all the water is vaporized the bubbles are very small again.

### 1.1.4. A Brief Comment on Universality

The Ising model is arguably very crude and it does not resemble the microscopic properties of water or a real magnet. A physical motivation to nevertheless study the model is the concept of **universality**. The idea is that the behavior of the model does not depend much on the microscopic details, especially near the phase transition (see for example [DC22, Sec. 4.2.2] and references therein). According to [FV17, p.52], in the early days of statistical mechanics, the simple models were regarded as mostly interesting for mathematicians and lacking physical relevance. However, this perspective has changed. In the words of Friedli and Velenik [FV17, p.53]:

> One additional ingredient that played a key role in this change of perspective is the realization that, in the vicinity of a critical point, the behaviour of a system becomes essentially independent of its microscopic details, a phenomenon called universality. Therefore, in such a regime, choosing a simple model as the representative of the very large class of systems (including the more realistic ones) that share the same behaviour, allows one to obtain even a quantitative understanding of these real systems near the critical point.

As inspiring as we may find universality, we are also motivated by the beautiful mathematics that statistical mechanics has to offer. Beautiful mathematics that, as an additional perk, often arrives in the form of easy-to-state but difficult-to-solve problems that simultaneously can give rise to very rich mathematics (see e.g. [DC22] and references therein).

## 1.2. Mathematical Perspectives on Phase Transitions: Exploring Bernoulli Percolation

We now provide some perspectives on the mathematical study of phase transitions. In particular, we study the simpler model of Bernoulli percolation. The motivation for doing so is that many of the arguments that we use throughout this thesis have simpler analogous for Bernoulli percolation. Furthermore, Bernoulli percolation is arguably the simplest example of a percolation model, and percolation models play a substantial role in this thesis. One of the most studied percolation models is the random cluster model (or FK-representation), which is a percolation model that encodes the properties of the Ising model. In that way, it bridges between Bernoulli percolation and the Ising model. However, for simplicity and to focus on ideas rather than introducing all the models, we wait until Section 2 before introducing the random cluster model.

### 1.2.1. Definition of Bernoulli Percolation

In contrast to the Ising model, where spins could be up or down, Bernoulli percolation concerns edges that can be either open or closed. One can think of open edges as edges that are "switched on" and closed edges as edges that are "switched off". To mimic the example from Table 1.1 above, if the graph has two vertices and one edge $e$ then there are two configurations, one where $e$ is open and one where $e$ is closed. In Bernoulli percolation with parameter $p \in [0, 1]$ the probability that the edge is open is $p$ and that is closed is $1 - p$.

For a more complicated example consider the graph in Figure 1.7 with vertices that we label with the integers from $-4$ to 4 and edges between consecutive integers. On the sketch (Figure 1.7) the edges $(-4, -3)$, $(-3, -2)$, $(-2, -1)$, $(0, 1)$, $(2, 3)$, $(3, 4)$ are open and the dashed edges $(-1, 0)$, $(1, 2)$ are closed. The probability of that configuration would be $p^6(1-p)^2$ (since there are 6 open and 2 closed edges). Any way of opening some edges and closing others is called a configuration and we denote it with the letter $\omega$. We denote the set of all such configurations by $\Omega$. For a configuration $\omega \in \Omega$, we let $o(\omega)$ be the number of open edges and $c(\omega)$ denote the number of closed edges. In the example, this means that $o(\omega) = 6$ and $c(\omega) = 2$.

In a general graph $G = (V, E)$, Bernoulli percolation with parameter $p$ between 0 and 1 consists of opening each edge independently with probability $p$. Formally, this means that every configuration of open edges $\omega$ has the probability

$$\mathbb{P}_p[\omega] = p^{o(\omega)} \ (1-p)^{c(\omega)}.$$

For example, for $p = \frac{1}{2}$ this is the same as flipping a fair coin for every edge. If the coin is heads we open the edge and if it tails we close the edge. Without going into details, we mention that in [UEG] it is used extensively that $\mathbb{P}_{\frac{1}{2}}$ is the Haar measure on $\Omega$.

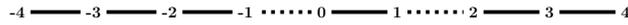

Figure 1.7.: A configuration of edges for Bernoulli percolation on $\mathbb{Z}$ (where we have only shown some of the points, the reader should think of the path as infinite in both directions). Here the full edges are open and the dashed edges are open.

In analogy with the correlation function for the Ising model introduced in (1.2) the correlation function of Bernoulli percolation we write as $\mathbb{P}_p[x \leftrightarrow y]$, which is the probability that there is a path of open edges connecting $x$ to $y$. In Chapter 2, we give more examples of correlation functions that are studied in this thesis.

The notion of phase transition is clearer for infinite system and in the case of Bernoulli percolation we aim to determine the probability of the existence of an infinite path of open edges. If there is such an infinite path we say that the model **percolates**. The term "percolation" should evoke the idea of water permeating a medium from one end to the other.

As the simplest examples, we consider Bernoulli percolation on $\mathbb{Z}$ and $\mathbb{Z}^2$. Here $\mathbb{Z}$ are the integers and thus one should think of the graph from Figure 1.7 extended to all the integers (with edges between consecutive integers). Similarly, the graph $\mathbb{Z}^2$ has vertices that consists of all pairs $(n, m)$ of integers $n$ and $m$ where there is an edge between two vertices $(n, m)$ and $(n', m')$ if and only if $|n - n'| + |m - m'| = 1$ (see Figure 1.4 or Figure 1.8). Furthermore, the (graph) distance between two vertices $x$ and $y$ is the minimal number of edges in a path between $x$ and $y$. The construction of the $d$-dimensional hypercubic lattice $\mathbb{Z}^d$, which is the main playground of this thesis, is similar.

On an infinite graph we can ask whether there is an infinite path of open edges starting from 0. We write the probability that such a path exists as $\mathbb{P}_p[0 \leftrightarrow \infty]$. It is often difficult to determine this probability as it involves taking infinitely many edges into account. There are however ways around this. As a first example of such an argument is Bernoulli percolation in one dimension, where the situation is simple.

## 1.2.2. One-Dimensional Bernoulli Percolation

Consider the one-dimensional example where some of the edges are shown in Figure 1.7. For any vertex $n$ in the graph we say that the event $\{0 \leftrightarrow n\}$ occurs if there is a path of open edges from 0 to $n$. Since there is only one possible path that could be open in this example it means that every edge from 0 to $n$ has to be open.

**Proposition 1.2.1.** *Consider $\mathbb{P}_{p,\mathbb{Z}}$, that is Bernoulli percolation on $\mathbb{Z}$ with parameter $p$. Then*

$$\mathbb{P}_{p,\mathbb{Z}}[0 \leftrightarrow n] = p^n.$$

*Proof.* Every edge $(i, i+1)$ between the vertices $i$ and $i+1$ has to be open for a path between 0 and $n$ to be open. The probability for each edge to be open is $p$. Since they are independent the probability for all $n$ edges to be open is $p^n$. $\qquad\square$

With this observation at hand we can consider the probability that $0 \leftrightarrow \infty$. If there is an infinite path starting from 0 then it must be infinite either in the positive or in the negative direction. Therefore, the probability that there is an infinite path from 0 must be smaller than the probability that there is a path from 0 to $n$ plus the probability that there is a path from 0 to $-n$. Thus,

$$\mathbb{P}_{p,\mathbb{Z}}[0 \leftrightarrow \infty] \leq \mathbb{P}_{p,\mathbb{Z}}[0 \leftrightarrow n] + \mathbb{P}_{p,\mathbb{Z}}[0 \leftrightarrow -n] = p^n + p^n = 2p^n.$$

When $n$ tends to $\infty$ then $2p^n$ tends to 0 if $p < 1$. On the other hand if, for $p = 1$ all edges are always open and therefore, there is always an infinite path starting at 0 (in other words, it happens with probability 1). These considerations prove the following proposition, which is our first proof of a phase transition!

**Proposition 1.2.2.** *Consider $\mathbb{P}_{p,\mathbb{Z}}$, that is Bernoulli percolation on $\mathbb{Z}$ with parameter $p$. Then,*

$$\mathbb{P}_{p,\mathbb{Z}}[0 \leftrightarrow \infty] = \begin{cases} 0 \text{ if } p < 1 \\ 1 \text{ if } p = 1 \end{cases}.$$

Since the $p$ that are allowed for in the model (and the only ones that make sense) are between 0 and 1 and the phase transition happens at $p = 1$ we say that the phase transition is **trivial**.

With a slight reformulation, allowing $p$ to be larger than 1, the model is perhaps more familiar.

**Example 1.2.3** (A too familiar example)**.** *Consider an epidemic with $I_0$ infected in the beginning. Let $R \in [0, \infty]$ be a parameter. Suppose that every infected individual in every time-step passes the disease onto $R$ new people. Thus, if there were $I_n$ infected individuals in the $n$'th timestep, then there are $RI_n$ infected individuals in the next time step. Therefore,*

$$I_{n+1} = RI_n.$$

*It follows that,*

$$I_n = R^n I_0.$$

*Thus, the situation mimics perfectly the situation in Proposition 1.2.1 (upon identifying $R = p$), but now values of $R > 1$ also make sense. The phase transition is still at $R = 1$. For $R < 1$ the model exhibits **exponential decay**, for $R = 1$ the number of newly infected individuals remains constant, and for $R > 1$ the number of new infections is **exponentially increasing**.*

Apart from highlighting that simple models are omnipresent the example teaches us a lesson on exponential decay. Namely, if we iteratively multiply a number less than 1 we obtain exponential decay.

As in (1.2) above we can also introduce a correlation function and a correlation length for Bernoulli percolation on $\mathbb{Z}$. The correlation function for two vertices $x, y \in \mathbb{Z}$ is given as

$$\mathbb{P}_p[x \leftrightarrow y] = p^{|x-y|} = e^{\log(p)|x-y|}. \tag{1.4}$$

Now, comparing (1.4) with the definition of the correlation length in (1.3) we see that the correlation length is given by $\xi = -\frac{1}{\log(p)}$. Again, we see that $\xi \to \infty$ as $p \to p_c = 1$ (see Figure

1.6b for a graphical illustration). Here $p_c$ is the critical point, which we can in general define by

$$p_c = \inf_{p \in [0,1]} \left\{ \mathbb{P}_p[0 \leftrightarrow \infty] > 0 \right\}.$$

We saw that Bernoulli percolation in one dimension $\mathbb{Z}$ is trivial since $p_c = 1$, but in two dimensions $\mathbb{Z}^2$, it holds that $0 < p_c < 1$ as we will now discuss.

### 1.2.3. Two-Dimensional Bernoulli Percolation

It turns out that the two-dimensional model is much more interesting and one can prove that the phase transition is non-trivial (that is $0 < p_c < 1$). In fact, a celebrated result by Kesten [Kes80] states that $p_c = \frac{1}{2}$. In Figure 1.8, we see how the subcritical, critical and supercritical phases looks.

Here (and throughout this thesis), $\Lambda_n$ will denote the box of size $2n \times 2n$ centered at 0: $\Lambda_n = \{x \in \mathbb{Z}^2 \mid |x| \leq n\}$ and the boundary $\partial \Lambda_n = \{x \in \mathbb{Z}^2 \mid |x| = n\}$. One way to go about proving the non-trivial phase transition is to establish for some $p_0 > 0$ that there exists a $c, \mu > 0$ such that

$$\mathbb{P}_{p_0, \mathbb{Z}^2}[0 \leftrightarrow \partial \Lambda_n] \leq ce^{-\mu n}. \tag{1.5}$$

Since any infinite path starting at 0 must pass through $\partial \Lambda_n$ for any $n \in \mathbb{N}$ it holds that for any $p \in [0,1]$

$$\mathbb{P}_p[0 \leftrightarrow \infty] \leq \mathbb{P}_p[0 \leftrightarrow \partial \Lambda_n].$$

Therefore, (1.5) implies that

$$\mathbb{P}_{p_0, \mathbb{Z}^2}[0 \leftrightarrow \infty] = 0.$$

One can say that if (1.5) holds then the probability that you go further and further away tends to 0. Therefore, the probability that you get infinitely far away must be 0. Hence, there is no infinite path starting from zero and so $p_0 < p_c$. Thus, the inequality (1.5) tells us that the model at $p_0$ is in the subcritical phase, which is illustrated in 1.8a).

This is a simple example of how proving exponential decay provides information of the phase of the system. As proving such exponential decay plays a central role in this thesis (e.g. for the Anderson model and random cluster model), so we delve into a slightly more complicated example, that highlights many points that will come up throughout this thesis.

### 1.2.4. Warm Up for Separating Surface Conditions: One-Step Bound

Denote the four points that are neighbours to 0 by $a_1, a_2, a_3, a_4$ and let $e_i$ be the edge between 0 and $a_i$ for each $i \in \{1, 2, 3, 4\}$, see also Figure 1.9b).

**Lemma 1.2.4.** *On the square lattice $\mathbb{Z}^2$ for every $p \in [0,1]$ it holds that*

$$\mathbb{P}_p[0 \leftrightarrow x] \leq p \sum_{i=1}^{4} \mathbb{P}_p[a_i \leftrightarrow x].$$

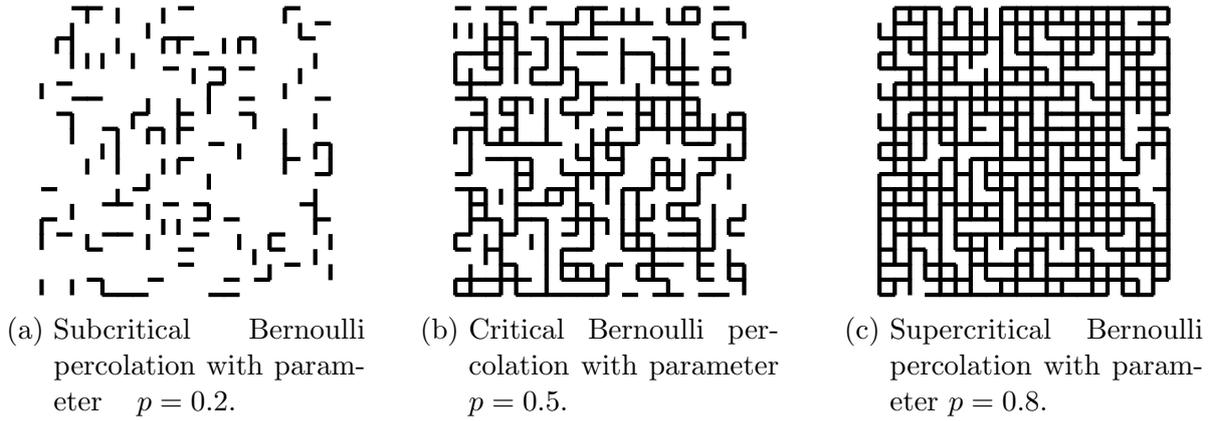

(a) Subcritical Bernoulli percolation with parameter $p = 0.2$.

(b) Critical Bernoulli percolation with parameter $p = 0.5$.

(c) Supercritical Bernoulli percolation with parameter $p = 0.8$.

Figure 1.8.: Three regimes for Bernoulli percolation on $\mathbb{Z}^2$, where the critical parameter is $p_c = 0.5$. Notice the striking similarities with Figures 1.1 and 1.5.

*Proof.* Notice first that $0 \leftrightarrow x$ if and only if for one $i \in \{1, 2, 3, 4\}$ the edge $e_i$ is open and there is an open path from $a_i$ to $x$ that does not use $e_i$.

Since the probability that one out of four events happens is smaller than the sum of the probabilities that each of them happens we can consider each of the four events $\{e_i \text{ is open}, a_i \overset{\{e_i\}^c}{\longleftrightarrow} x\}$ separately. Here $\{e_i\}^c$ means outside of $e_i$. This type of argument is know as a **union bound**.

Now, the event that there is a path from $a_i$ to $x$ not using $e_i$ does not in any way depend on whether $e_i$ is open or closed. I.e the two events are **independent** and the probabilities factorize, meaning that

$$\mathbb{P}_p[e_i \text{ is open }, a_i \overset{\{e_i\}^c}{\longleftrightarrow} x] = \mathbb{P}_p[e_i \text{ is open}] \ \mathbb{P}_p[a_i \overset{\{e_i\}^c}{\longleftrightarrow} x].$$

Furthermore, since the probability of having a path between $a_i$ and $x$ where it is allowed to use $e_i$ is larger than the probability of such a path where $e_i$ is not allowed then $\mathbb{P}_p[a_i \overset{\{e_i\}^c}{\longleftrightarrow} x] \leq \mathbb{P}_p[a_i \longleftrightarrow x]$. Hence, using $\mathbb{P}_p[e_i \text{ is open}] = p$ we obtain

$$\mathbb{P}_p[e_i \text{ is open}, a_i \overset{\{e_i\}^c}{\longleftrightarrow} x] \leq p \mathbb{P}_p[a_i \longleftrightarrow x].$$

Together with the union bound this establishes the lemma. $\qquad\square$

One-step bounds for other models play an independent role in this thesis. Here they mostly serve as a simple example of separating surface conditions that we discuss next. The many facets of separating surface conditions are the closest this thesis comes to a common theme.

## 1.2.5. A First Encounter with a Separating Surface Condition

A separating surface condition is a way to encapsulate the locality of the system. This way of thinking is used extensively throughout this thesis as we elaborate on in Section 2.2.3. If $x \in \mathbb{Z}^2$ is some vertex outside of $\Lambda_n$, then $\partial \Lambda_n$ is a separating surface, in the sense that any path from 0 to $x$ must cross $\partial \Lambda_n$.

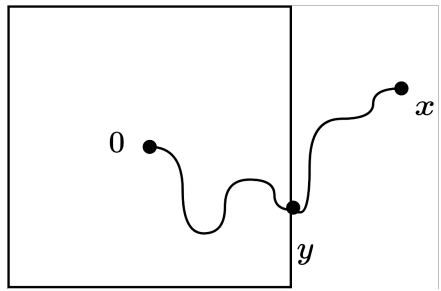

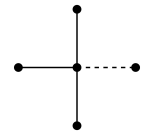

(a)  Illustration of the box $\Lambda_n$ acting as a separating surface between 0 and $x$. Any path from 0 to $x$ must visit a vertex $y \in \partial \Lambda_n$.

(b)  Example of the 4 edges in the single step bound. Independently, each of them is open with probability $p$.

Figure 1.9.: Two elements of the proof of exponential decay for two-dimensional Bernoulli percolation.

**Proposition 1.2.5.** *For every vertex $x \notin \Lambda_n$ it holds that*

$$\mathbb{P}_p[0 \leftrightarrow x] \leq \sum_{y \in \partial \Lambda_n} \mathbb{P}_p[0 \xleftrightarrow{\Lambda_n} y] \mathbb{P}_p[y \leftrightarrow x]. \tag{1.6}$$

Roughly speaking, Proposition 1.2.5 says that the probability that there is a path from 0 to $x$ is less than the probability that there is a path from 0 to the boundary of $\Lambda_n$ and then from the boundary to $x$ (see Figure 1.9a).

For an elementary proof we follow [DC18, Corollary 2.5]. Readers who are less experienced with probability may choose to skip the proof on their first reading.

*Proof.* Consider $\mathcal{C} = \{y \in \Lambda_n \mid y \xleftrightarrow{\Lambda_n} 0\}$ which is the connected component of 0 inside $\Lambda_n$. We will consider all different realizations of $\mathcal{C}$. So suppose that $C$ is a fixed set of edges in $\Lambda_n$ and $\mathcal{C} = C$. Then, if $0 \leftrightarrow x$ then any path from 0 to $x$ must exit $C$ the last time from some vertex $y \in C$, since $x \notin C$. This exit-vertex $y \in \partial \Lambda_n$, since if it was inside $\Lambda_n$ and not in the boundary, then the next vertex would be part of $C$. Thus, if $\{\mathcal{C} = C\}$ and $\{0 \leftrightarrow x\}$ then there is a vertex $y \in \partial \Lambda_n$ such that $\{0 \xleftrightarrow{C} y\}$ and $\{y \xleftrightarrow{C^c} x\}$. This argument means that

$$\{\mathcal{C} = C\} \cap \{0 \leftrightarrow x\} \subset \bigcup_{y \in \partial \Lambda_n} \{\mathcal{C} = C\} \cap \{0 \xleftrightarrow{C} y\} \cap \{y \xleftrightarrow{C^c} x\}.$$

Thus, by a union bound then

$$\mathbb{P}_p[\{\mathcal{C} = C\} \cap \{0 \leftrightarrow x\}] \leq \sum_{y \in \partial \Lambda_n} \mathbb{P}_p[\{\mathcal{C} = C\} \cap \{0 \xleftrightarrow{C} y\} \cap \{y \xleftrightarrow{C^c} x\}].$$

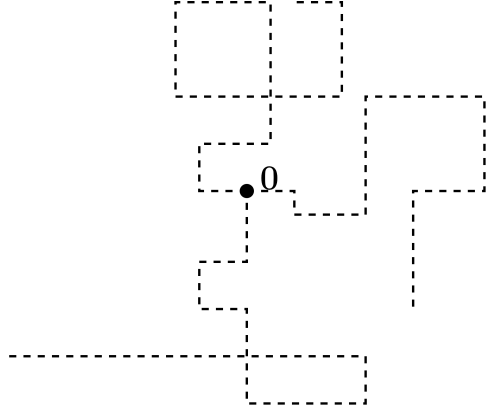

Figure 1.10.:  Example of some of the random walks some steps from the random walk expansion in (1.8).

As $\mathcal{C} \subset \Lambda_n$ and by considering all the different realizations of $\mathcal{C}$ we get that

$$
\begin{aligned}
\mathbb{P}_p[0 \leftrightarrow x] &= \sum_{C \subset \Lambda_n} \mathbb{P}_p[\{\mathcal{C} = C\} \cap \{0 \leftrightarrow x\}] \\
&\leq \sum_{C \subset \Lambda_n} \sum_{y \in \partial \Lambda_n} \mathbb{P}_p[\{\mathcal{C} = C\} \cap \{0 \overset{C}{\leftrightarrow} y\}] \mathbb{P}_p[y \overset{C^c}{\longleftrightarrow} x] \\
&\leq \sum_{y \in \partial \Lambda_n} \sum_{C \subset \Lambda_n} \mathbb{P}_p[\{\mathcal{C} = C\} \cap \{0 \overset{\Lambda_n}{\longleftrightarrow} y\}] \mathbb{P}_p[y \leftrightarrow x] \\
&= \sum_{y \in \partial \Lambda_n} \mathbb{P}_p[0 \overset{\Lambda_n}{\longleftrightarrow} y] \mathbb{P}_p[y \leftrightarrow x].
\end{aligned}
$$

$\square$

## 1.2.6. From Single-Step Bounds and Separating Surfaces to Exponential Decay

Using the arguments from the previous sections we can do an iteration argument that leads to the exponential decay in (1.5). We consider the particular case where the box $\Lambda_n$ consists of the five points in Figure 1.9b. For any vertex $v = (v_1, v_2) \in \mathbb{Z}^2$ let $\partial B_1(v)$ denote of distance 1 to $v$. Hence, by symmetry $\mathbb{P}_p[0 \overset{B_1(0)}{\longleftrightarrow} y] = p$ for every $y \in \partial B_1(0)$, see also Figure 1.9b.

Thereby, Lemma 1.2.4 can be stated as

$$
\mathbb{P}_p[0 \leftrightarrow x] \leq p \sum_{y \in \partial B_1(0)} \mathbb{P}_p[y \leftrightarrow x]. \tag{1.7}
$$

Notice also how a slight extension of Proposition 1.2.5 implies (1.7). Our objective now is to **iterate** the process and prove exponential decay by collecting the distance between 0 and $x$.

Using (1.7) again with $y$ instead of 0 yields

$$\mathbb{P}_p[0 \leftrightarrow x] \leq p^2 \sum_{y_1 \in \partial B_1(0)} \sum_{y_2 \in \partial B_1(y_1)} \mathbb{P}_p[y_2 \leftrightarrow x]. \tag{1.8}$$

One can interpret this bound as starting at 0, then walking to $y_1$, then from $y_1$ to $y_2$ and so on. In each step, we collect a factor of $p$ coming from $\mathbb{P}_p[y_1 \overset{B_1(y_1)}{\longleftrightarrow} y_2] = p$. Thus, if there are at least $|x|$ steps between 0 and $x$ we can iterate this $|x|$ times to obtain the following **random walk expansion** (see Figure 1.10)

$$\mathbb{P}_p[0 \leftrightarrow x] \leq p^{|x|} \sum_{y_1 \in \partial B_1(0)} \sum_{y_2 \in \partial B_1(y_1)} \cdots \sum_{y_{|x|} \in \partial B_1(y_{|x|-1})} \mathbb{P}_p[y_{|x|} \leftrightarrow x]. \tag{1.9}$$

Each of these sums has 4 terms and there are $|x|$ of them and since $\mathbb{P}_p[y_{|x|} \leftrightarrow x]$ is a probability then it is always less than 1. Therefore, all the sums are less than $4^{|x|}$ and we obtain the following lemma.

**Lemma 1.2.6.** *On $\mathbb{Z}^2$ it holds that*

$$\mathbb{P}_p[0 \leftrightarrow x] \leq (4p)^{|x|}. \tag{1.10}$$

*In particular, we see that if $p < \frac{1}{4}$ then $\mathbb{P}_p[0 \leftrightarrow x]$ has exponential decay.*

To obtain the estimate (1.5)

$$\mathbb{P}_{p_0, \mathbb{Z}^2}[0 \leftrightarrow \partial \Lambda_n] \leq c e^{-\mu n}$$

for $p_0 < \frac{1}{4}$ and all $n$ we use a union bound: If $0 \leftrightarrow \partial \Lambda_n$ then there must be at least one $x \in \partial \Lambda_n$ such that $0 \leftrightarrow x$ and thus, the probability of $0 \leftrightarrow \partial \Lambda_n$ is less than the sum of $\mathbb{P}_p[0 \leftrightarrow x]$ overall $x \in \partial \Lambda_n$. Thus, there exists $c, \mu > 0$ such that for any $n \in \mathbb{N}$,

$$\mathbb{P}_{p_0, \mathbb{Z}^2}[0 \leftrightarrow \partial \Lambda_n] \leq \sum_{x \in \partial \Lambda_n} \mathbb{P}_{p_0, \mathbb{Z}^2}[0 \leftrightarrow x] \leq 8n \cdot (4p_0)^n \leq c e^{-\mu n}$$

where we used that there are $8n$ points in $\partial \Lambda_n$ and each of them has distance at least $n$ to 0 so that we can use (1.10). The last inequality used that exponential decay dominates polynomial front factors, in the sense that for any polynomial $P$ and constants $c, \mu > 0$ it holds that $P(n)ce^{-\mu n} \leq c'e^{-\mu' n}$ for all positive integers $n$ for some constants $c', \mu' > 0$. So, it follows that for $p_0 < \frac{1}{4}$

$$\mathbb{P}_{p_0, \mathbb{Z}^2}[0 \leftrightarrow \infty] = 0.$$

In other words, $p_c \geq \frac{1}{4} > 0$.

Now, this is one of the two inequalities need to establish non-triviality of the phase transition of Bernoulli percolation on $\mathbb{Z}^2$. The other inequality, $p_c < 1$ is usually proven using a Peierls argument after [Pei36], since we do not use the argument in this thesis, we refer to [DC18, Theorem 1.1] for a nice exposition.

### 1.2.7. Generalized Iterations

We saw in the previous section that we proved exponential decay by keeping track of all the $n$-step walks from 0. Another way to keep track of the terms is generalized iterations. Define

$$f(n) = \sup_{x:|x| \geq n} \mathbb{P}_{p,\mathbb{Z}^2}[0 \leftrightarrow x].$$

Notice that $f(n+1) \leq f(n)$ for all positive integers $n$. From (1.7) and using translation invariance of $\mathbb{P}_{p,\mathbb{Z}^2}$ we obtain

$$\mathbb{P}_{p,\mathbb{Z}^2}[0 \leftrightarrow x] \leq 4p \sup_{y \in \partial B_1(0)} \mathbb{P}_{p,\mathbb{Z}^2}[y \leftrightarrow x] = 4p \sup_{y \in \partial B_1(0)} \mathbb{P}_{p,\mathbb{Z}^2}[0 \leftrightarrow (x-y)] \leq 4pf(|x|-1). \quad (1.11)$$

Taking the supremum, we obtain that

$$f(n) = \sup_{x:|x| \geq n} \mathbb{P}_{p,\mathbb{Z}^2}[0 \leftrightarrow x] \leq 4pf(n-1).$$

We can iterate this bound and since $f(0) \leq 1$, we obtain that

$$f(n) \leq (4p)^n.$$

We have thus obtained a new proof of Lemma 1.2.6 using generalized iterations. We will use the method of more generalized iterations repeatedly throughout this thesis. In the next section, we introduce the same method in a more general setting where it is known as the Hammersley paradigm.

## 1.3. The Origin of Probabilistic Models

Historically, modern probability theory was developed in parallel with statistical mechanics. As the title of this thesis indicates most of the problems studied in this thesis have probabilistic aspects. Therefore we briefly discuss how and why the notion of probability theory becomes relevant. Very roughly speaking, the probabilistic aspects arise in three ways in the problems in this thesis:

1. Lack of detailed knowledge of the system.

2. As a description of disorder in the system.

3. Inherently in the quantum system.

The three ways are interwoven in various ways, but let us briefly discuss each of them individually.

**1. Lack of detailed knowledge of the system.** From one point of view, the breakthrough of Boltzmann, Gibbs and Maxwell consisted of giving up on modelling the trajectories of individual properties of molecules in a gas and instead considering statistical behaviour. Broadly speaking, they thereby gave birth to statistical physics [Kle90]. In the case of gases, the microstates are no longer spins pointing up and down as we saw for the Ising model. Instead, the microstates are positions and velocities of particles, but the treatment is analogous. The central assumption is that if the total energy of the system is fixed, then each of the microstates with that total energy is equally likely. This is known as the microcanonical ensemble. The thermodynamic properties, such as the heat capacity, then derive statistically. The assumption can be argued (see [FV17, p. 489],[Jay57] and references therein for a discussion) to be an application of Laplace's principle of insufficient reason: *If there is no reason that any outcome is more likely than the others we assign them equal probabilities.* For example, if we have a standard die, there is no reason that any of the faces are preferred, so each face must have probability $\frac{1}{6}$.

It turns out, that in the **canonical ensemble** it is slightly easier to connect the microscopic phenomena to thermodynamics. In contrast to the microcanonical ensemble where the total energy is fixed, in the canonical ensemble one assumes that the average energy is fixed. Under that assumption, a calculation with Lagrange multipliers (see [FV17, Sec. 1.1.2]) leads us to the **Gibbs distribution** (cf. (1.1)), which serves as the basis of (modern) equilibrium statistical mechanics.

**2. As a description of disorder in the system.** The field of random operators, initiated by Anderson [And58], models disorder in quantum systems by introducing randomness. Think of a physical model for a perfect crystal. It may be that the model is very accurate, but real crystals are never perfect, so sometimes the perfect crystal is not what one should model. On the other hand, introducing imperfections into the model often makes the study of the model intractable. Furthermore, it may also be impossible to know the exact nature of the model since we lack knowledge of the state of the system. Again, we transfer the lack of knowledge of the state of the system into probabilistic aspects. So we resort to studying random models, where we can inquire about the average properties of the systems. One may also notice the similarity in spirit and time with the original paper on Bernoulli percolation by Broadbent and Hammersley [BH57].

**3. Inherently in the quantum system.** The introduction of randomness into quantum mechanics, as it is often taught (cf. the successful textbooks [SC95; GS18; NC02]), is slightly different. It is well explained by considering a simple quantum walk (without disorder). In the quantum walk, we model a particle that is spread out on the lattice, so that the particle is sort of everywhere at the same time. However, the **state of the system is not random**. But when we do a **measurement**, we will measure the quantum walk in a specific place with probabilities determined by the state of the system (see e.g. [NC02, sec. 2.2.3]).

We have seen how probabilities arise in several different ways in this thesis. We caution that this split-up may be somewhat artificial, as for example disorder in the system could also be thought of as lack of knowledge of the system.

## 1.4. Conclusion of the Gentle Introduction

We have seen that the Ising model and Bernoulli percolation provide rough models of phase transitions. The phase transition consisted of a clear demarcation between exponential decay and absence of exponential decay. We defined the correlation length as the inverse rate of exponential decay and saw that diverging correlation length was a hallmark of a phase transition. In the case of Bernoulli percolation in one dimension, we proved exponential decay and calculated the correlation length as a function of the parameter. Then we discussed the separating surface condition which could encapsulate the locality of our systems of interest (in our case the Ising model and Bernoulli percolation). We proved a separating surface condition for Bernoulli percolation and in the case of two-dimensional Bernoulli percolation we showed how it can be used to prove exponential decay (for $p < \frac{1}{4}$). Finally, we discussed the origin of randomness in the models studied in this thesis.

# 2. Classical and Quantum Lattice Models and Their Common Themes

In this chapter, we introduce the models that are the central objects of study in this thesis and discuss their common themes. We do not aim at giving a review of any of the models, but instead, we try to emphasize the similarities between the seemingly very different models. Therefore, this section will be substantially less self-contained than the previous sections, and we will refer the reader to some of the many excellent introductions to the topics: [Gri06; DC19; FV17] for the Ising model and its graphical representations, [AW15; Sto11; Kir07] for mathematical aspects of Anderson localization and [Man20] for open quantum systems. We also refer the reader to these introductions for many of the historical references.

## 2.1. Introduction to the Models and Their Generalized Correlation Functions

We start by introducing the models that are studied in this thesis. We already introduced Bernoulli percolation $\mathbb{P}_p$ and the Ising model $\mu_\beta$ and now we will also introduce its graphical representations: The loop O(1) model $\ell_x$, the random current model $\mathbf{P}_\beta$ and the random cluster model $\phi_{p,q,h}$. We consider the random cluster model in a magnetic field $h$ as well as its marginal on internal edges $\phi_{p,q,h}\mid_{\mathbb{Z}^d}$. In Table 2.1 we also give an overview of these models.

Afterwards, we introduce the Anderson model $H_\lambda$ and its Green function $G_\lambda$, as well as its unitary analogue $U$ and its Green function $G$ and a quantum walk in a random magnetic field $W$ and its Green function. Finally, we introduce open quantum systems briefly and consider the steady state of a local (disordered) open quantum system $\rho_\infty$.

The choice of exactly these models may to some extent be arbitrary, however, this also illustrates how general the overall ideas with locality and correlation lengths are.

Some examples of other models in this spirit that we do not discuss in this thesis are the XY-model, the Heisenberg model, O($n$) model, the clock model, self-avoiding walk, Gaussian Free Field and $\phi^4$-theory. For more information on these models see for example [PS19; WP20; Aiz82] and references therein.

All the models are defined on a graph $G = (V, E)$ finite or infinite. We are interested either in general graphs or we consider $G$ to be a subgraph of an infinite graph $\mathbb{G}$, where $\mathbb{G}$ is often the hypercubic lattice $\mathbb{Z}^d$.

In the spirit of the Hammersley paradigm (see [AW15, Chap. 9] and [Ham57]) each of the models have a generalized correlation function $\tau : G \times G \to \mathbb{R}_+$ which in the case of the percolation models is

$$\tau(x, y) = \mathbb{P}[x \leftrightarrow y], \tag{2.1}$$

for the spin models

$$\tau(x,y) = \mu[\sigma_x \sigma_y] \tag{2.2}$$

and for the random operators is derived from the Green function $G$

$$\tau(x,y) = \mathbb{E}[|G(x,y;z)|^s]. \tag{2.3}$$

Finally, in the case of steady states of open quantum systems, we always work in finite volume, and we set

$$\tau(x,y) = \mathbb{E}[|\rho_\infty(x,y)|], \tag{2.4}$$

for a steady state $\rho_\infty$.

### 2.1.1. Ising Model

Let us briefly summarize the introduction of the Ising model that we gave in Section 1.1.1 and simultaneously extend it to magnetic fields. The Ising model is a measure on the set of spin configurations $\sigma$. To define the Ising model on the graph $G = (V, E)$ we think of every vertex $v \in V$ having a spin $\sigma_v$ which is either $+1$ or $-1$. That is the configuration space is $\{-1, +1\}^V$.

The energy of a configuration $\sigma \in \{-1, +1\}^V$ for the Ising model in a magnetic field $h \in \mathbb{R}$, is given by

$$H(\sigma) = -\sum_{\substack{e \in E \\ e=(x,y)}} \sigma_x \sigma_y - h \sum_{v \in V} \sigma_v. \tag{2.5}$$

Notice that $\sigma_x$ and $\sigma_y$ always take values in $+1$ and $-1$ and therefore their product $\sigma_x \sigma_y$ is 1 if they are pointing the same way and it is $-1$ if they are pointing opposite ways.

Now, we go from energies to probabilities of configurations through the Gibbs measure, which assigns probabilities of configurations proportional to their Boltzmann factors $e^{-\beta H(\sigma)}$, where $\beta > 0$ is a parameter corresponding to the (inverse) temperature. Thus, we define the Ising probability measure $\mu_{\beta,h,G}$ by

$$\mu_{\beta,h,G}[\sigma] = \frac{e^{-\beta H(\sigma)}}{\sum_{\sigma'} e^{-\beta H(\sigma')}}. \tag{2.6}$$

Before we continue, let us introduce the Griffith's ghost vertex as it is central in both [Mass] and [Kertész]. For any graph $G = (V, E)$ we consider an extended graph $G_g = (V \cup \{g\}, E \cup E_g)$ where g is called the ghost vertex and $E_g = \cup_{v \in V} \{(v, g)\}$ are additional edges from any vertex $v$ in the graph to the ghost g. We will sometimes call the edges $E_g$ external edges. If we decide that $\sigma_g = 1$, that is the ghost is always spin up, then (2.5) reads,

$$H(\sigma) = -\sum_{\substack{e \in E \\ e=(x,y)}} \sigma_x \sigma_y - h \sum_{v \in V} \sigma_v \sigma_g = -\sum_{\substack{e \in E \cup E_g \\ e=(x,y)}} J_e \sigma_x \sigma_y. \tag{2.7}$$

with $J_e = h$ if $e$ is external and $J_e = 1$ otherwise.

### 2.1.2. Percolation Models

We saw that the Ising model is a probability measure $\{-1, +1\}^V$. In contrast, Bernoulli percolation considered configurations of open and closed edges. We define a function $\omega : E \to \{0, 1\}$ such that $\omega(e) = 1$ if $e$ is open and $\omega(e) = 0$ if $e$ is closed. Then, we can view $\omega$ as an element of $\{0, 1\}^E$. We will call $\Omega = \{0, 1\}^E$ the set of percolation configurations. The following recollection of terminology follows [**UEG**] closely.

There is a natural partial order $\preceq$ on $\Omega$ defined such that $\omega \preceq \omega'$ if for all $e \in E$ it holds that $\omega(e) \leq \omega'(e)$. Further, we say that an event $\mathcal{A} \subset \Omega$ is increasing if for all pairs $\omega, \omega' \in \Omega$ it holds that if $\omega \preceq \omega$ and $\omega \in \mathcal{A}$ then $\omega' \in \mathcal{A}$. For example, the event $\{x \leftrightarrow y\} \subset \Omega$ is increasing, since adding additional edges preserves connections

The notion of increasing events enables us to define a partial order on the probability measures on $\Omega$ (e.g. percolation measures). If $\nu_1, \nu_2$ are two percolation measures on $\Omega$ such that $\nu_1(\mathcal{A}) \leq \nu_2(\mathcal{A})$ for all increasing events $\mathcal{A}$, then we say that $\nu_1$ is stochastically dominated by $\nu_2$. Stochastic domination is also a partial order and we also denote it by $\preceq$.

Since stochastic domination and couplings play an important role in the papers [**Mass**], [**Kertész**] we explain some preliminary details that are not explained in the papers. One way to check stochastic domination is the existence of an increasing coupling: If $X$ and $Y$ are random variables on a background probability space with probability measure $\mathbb{P}$, $X \preceq Y$ almost surely and $X$ and $Y$ are distributed like $\nu_1$ and $\nu_2$ respectively then $\nu_1 \preceq \nu_2$. To see it, note that

$$\nu_1[\mathcal{A}] = \mathbb{P}[X \in \mathcal{A}] = \mathbb{P}[X \in \mathcal{A}, X \preceq Y] \leq \mathbb{P}[Y \in \mathcal{A}] = \nu_2[\mathcal{A}].$$

Interestingly, Strassen's theorem [Str65] tells us that in high generality the converse holds: If $\nu_1 \preceq \nu_2$, then there exists an increasing coupling. Finally, from the point of view taken in this thesis, the union of percolation measures is important and we define it as follows.

**Definition 2.1.1** (Union of two percolation measures). *For two percolation measures $\nu_1$ and $\nu_2$ we denote the measure sampled by taking the union of two independently sampled copies of $\nu_1$ and $\nu_2$ by $\nu_1 \cup \nu_2$. More formally, if $(X, Y) \sim \nu_1 \otimes \nu_2$ then we say that $\nu_1 \cup \nu_2$ is the law of $X \cup Y$, where an edge $e$ is open in $X \cup Y$ if it is open in either $X$ or $Y$. Notice that $\nu_1 \cup \nu_2$ stochastically dominates both $\nu_1$ and $\nu_2$.*

**Bernoulli percolation.** For Bernoulli percolation $\mathbb{P}_p$ each edge $e \in E$ is open with probability $p \in [0, 1]$ independently. This means that the probability of a configuration $\omega$ is

$$\mathbb{P}_p[\omega] = p^{o(\omega)}(1 - p)^{c(\omega)}.$$

Recall how we introduced Bernoulli percolation in more detail in Section 1.2. The simplest example of stochastic domination is that if $p > q$ then $\mathbb{P}_p \succeq \mathbb{P}_q$. To see that, we for edge $e$ let $U_e$ be independent $\mathsf{Unif}[0, 1]$ random variables. Define $X_p(e) = \mathbb{1}[U_e > p]$. Then $X_p \sim \mathbb{P}_p$ and $X_p \succeq X_q$ almost surely and so $\mathbb{P}_p \succeq \mathbb{P}_q$.

**The random cluster model.** The random cluster measure on a finite graph $G = (V, E)$ with a distinguished ghost vertex g, parameters $p \in (0, 1)$, $q > 1$, external field $h > 0$ and parameter $p_h = 1 - \exp\left(-\frac{q}{q-1}h\right)$, is the measure on $\{0, 1\}^{E_g}$ given by

$$\phi_{p,q,h,G}[\omega] = \frac{1}{Z_{p,q,h,G}} p^{o(\omega_{\mathrm{in}})}(1 - p)^{c(\omega_{\mathrm{in}})} p_h^{o(\omega_g)}(1 - p_h)^{o(\omega_g)} q^{\kappa(\omega)}, \qquad (2.8)$$

where $\omega_{\mathrm{g}}$ is the restriction of $\omega$ to the set of edges adjacent to g, $\omega_{\mathrm{in}}$ is the restriction of $\omega$ to the set of edges not adjacent to g, $o(\cdot)$ denotes the number of open edges and $\kappa(\omega)$ the number of components of $\omega$.

For integer $q \geq 2$ the random cluster model is a graphical representation of the Potts model in a magnetic field $h$. There, we have the Edwards-Sokal coupling [ES88], that we now explain for the case of the Ising model $q = 2$ with magnetic field $h = 0$. For any configuration $\omega$ sampled with respect to $\phi_{p,q=2,h,G}$ consider its connected components (also called clusters) $\mathcal{C}_1, \mathcal{C}_2, \ldots$. Then for each cluster, $\mathcal{C}_j$ we flip a fair coin. If it is heads, we give spin up (or $+1$) to all vertices in $\mathcal{C}_j$. If it is tails we give spins down (of -1) to all vertices. In that way we construct a spin configuration $\sigma \in \{-1, +1\}^V$. The theorem of Edwards and Sokal is then that the configuration has the distribution of the Ising model $\sigma \sim \mu_{\beta,G}$. From the Edwards-Sokal coupling one can quite fast deduce the following relation: (see e.g. [DC19, Cor. 1.4] for the details)

$$\mu_{q,h,\beta,G}[\sigma_x \sigma_y] = \phi_{p,q,h,G}[x \leftrightarrow y]. \tag{2.9}$$

We will not introduce the Potts model here, but only note that it is a spin model that generalizes the Ising model, in such a way that for $q = 2$ it is the Ising model. For an introduction to the Potts model, we refer the reader to the introduction of [Kertész]. Let us also note that in [Kertész] we dive deep into the stochastic domination relations that arise upon varying $p, q$ and $h$.

### 2.1.3. Graphical Representations of the Ising Model.

We now introduce the graphical representations of the Ising model in the sense of percolation models. We follow the introduction given in [MonCoup] which is rather non-standard, but this will ease the presentation here and illuminate the way the graphical representations are used throughout this thesis. In the introduction of [UEG] we give a more standard introduction to the models that the interested reader can use to cross-reference. For an overview over the models see Table 2.1.

**Loop $\mathrm{O}(1)$ model and uniform even graph.** An even subgraph of a finite graph $G = (V, E)$ as a subgraph $(V, F)$ such that $F \subset E$ where every vertex has even degree. The set of even subgraphs of a graph $G$ is denoted $\Omega_\emptyset(G)$. Notice that $(V, \emptyset)$ is always an even subgraph and so the set $\Omega_\emptyset(G)$ is always non-empty. The loop $\mathrm{O}(1)$ model $\ell_{x,G}$ which to every $\eta \in \Omega$ assigns the probability

$$\ell_x[\eta] = \frac{x^{o(\eta)}}{Z} \mathbb{1}[\eta \in \Omega_\emptyset(G)], \tag{2.10}$$

with $Z = \sum_{\eta \in \Omega_\emptyset(G)} x^{o(\eta)}$. The loop $\mathrm{O}(1)$ model $\ell_x$ is related to the Ising model with parameter $\beta$ whenever $x = \tanh(\beta)$. In particular, the value $\beta = \infty$ (zero temperature) corresponds to $x = 1$ and in that case, $\ell_x$ becomes the uniform even subgraph that we denote by UEG. The uniform even subgraph plays a major role in the paper [UEG] where we exhibit it as a Haar measure on the group of even graphs.

| **Model** | Ising | Bernoulli | Random Cluster | Random Current | loop O(1) |
|---|---|---|---|---|---|
| **Symbol** | $\mu$ | $\mathbb{P}$ | $\phi$ | $\mathbf{P}$ | $\ell$ |
| **Parameter** | $\beta$ | $p$ | $p$ | $\beta$ | $x$ |
| **Type** | Spin | Edge | Edge | (Multi)-Edge | Edge |
| **Weight** | $e^{-\beta H(\sigma)}$ | $\left(\frac{p}{1-p}\right)^{o(\omega)}$ | $2^{\kappa(\omega)}\left(\frac{p}{1-p}\right)^{o(\omega)}$ | $\prod_{e \in E}\frac{\beta^{\mathbf{n}(e)}}{\mathbf{n}(e)!}\mathbb{1}_{\partial \mathbf{n}=\emptyset}$ | $x^{o(\eta)}\mathbb{1}_{\partial \eta=\emptyset}$ |

Table 2.1.: Overview of the Ising model, its graphical representations and Bernoulli percolation. The parameters are related through $x = \tanh(\beta)$ and $p = 1 - e^{-2\beta}$. Notice that in the weight for the random current we have taken the conventional definition of the random current as a measure on multigraphs. For details see the introduction in [**UEG**].

**The FK-representation.** Setting $q = 2$ and $h = 0$ for the random cluster model in (2.8) is sometimes called the FK-representation (after [**FK72**]). It turns out that another way of viewing this model is by defining

$$\phi_x = \ell_x \cup \mathbb{P}_x, \tag{2.11}$$

where $\cup$ is the union of independent copies of the model as defined in Definition 2.1.1. Through the relation $x = \frac{p}{2-p}$ one can recover the definition in (2.8) above ([**MonCoup**,Theorem 8]).

**Random current model.** We can define the (traced, sourceless) *single random current* at inverse temperature $\beta$ as

$$\mathbf{P}_x = \ell_x \cup \mathbb{P}_{1-\sqrt{1-x^2}}. \tag{2.12}$$

The double random current model is particularly connected to the Ising model.

**Double random current model.** In a similar vein, we introduce the (sourceless, traced) double random current model as follows:

$$\mathbf{P}_x^{\otimes 2} = \mathbf{P}_x \cup \mathbf{P}_x. \tag{2.13}$$

The double random current is related directly to the Ising model by the relation

$$\mu_{\beta,G}[\sigma_x\sigma_y] = \mathbf{P}_x^{\otimes 2}[x \leftrightarrow y]^2, \tag{2.14}$$

see [**DC19**, (4.6)] for details. Traditionally, the random current expansion is introduced as a measure on multigraphs [**GHS70**; **Aiz82**] and the relation in (2.14) follows from that definition of the random current using the switching lemma. For this point of view see the introductions of [**Mass**] and [**UEG**].

Figure 2.1.: Overview of the couplings between the graphical representations of the Ising model. Each of the thick lines is either a union of the measure with Bernoulli percolation (horizontal) or with an independent copy of itself (vertical). The dashed lines indicate taking a uniform even subgraph.

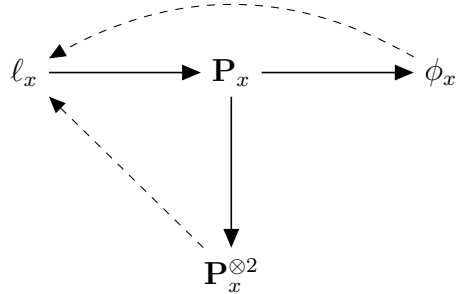

**Boundary conditions.** The models above were defined without introducing boundaries and boundary conditions - a point of view that we call free boundary conditions. The random cluster model and loop O(1) model with free boundary conditions are denoted $\phi_G^0 = \phi_G$ and $\ell_G^0 = \ell_G$ respectively. However, to use the locality of the graph boundary conditions are essential. If a graph $G = (V, E)$ has some boundary vertices $\partial G$, a boundary condition $\xi$ is a partition of the vertices $\partial G$, where vertices in the same element of the partition are identified giving rise to a new graph. The random cluster model and loop O(1) model with boundary conditions $\xi$ are denoted $\phi_G^\xi$ and $\ell_G^\xi$ respectively.

### 2.1.4. Relations Between Graphical Representations.

The graphical representations of the Ising model are related in various ways. The papers **[Mass]** and **[UEG]** rely heavily on these relations and one of the main results in **[MonCoup]** is an extension of the relations. The following theorem from **[UEG]** summarizes the relations. See the paper for details on parametrizations, but let us note that $x = \tanh(\beta)$ and $p = 1 - e^{-2\beta}$, so when we write $\ell_{\beta,G}^0$ we mean $\ell_x$ with $x = \tanh(\beta)$. We have sketched the relations in Figure 2.1 which is also from **[UEG]**.

**Theorem 2.1.2** (**[UEG**, Theorem 2.5]). *For any finite graph, $G = (V, E)$, the graphical representations of the Ising model are related in the following way.*

- $\ell_{\beta,G}^0 \cup \mathbb{P}_{1-\cosh(\beta)^{-1},G} = \mathbf{P}_{\beta,G}^\emptyset.$

- $\ell_{\beta,G}^0 \cup \mathbb{P}_{\tanh(\beta),G} = \phi_{\beta,G}^0.$

- $\mathbf{P}_{\beta,G}^{\otimes 2}[\mathrm{UEG}_\omega[\cdot]] = \ell_{\beta,G}^0[\cdot] = \phi_{\beta,G}^0[\mathrm{UEG}_\omega[\cdot]].$

The result that we prove in **[MonCoup]** is $\mathbf{P}_{\beta,G}^{\otimes 2}[\mathrm{UEG}_\omega[\cdot]] = \ell_{\beta,G}^0[\cdot]$ and references to the other results are **[GJ09**, Theorem 3.5], **[LW16]** **[Lis22**, Theorem 3.1], see also **[DC19**, Exercise 36] and the extension in **[Aiz+19**, Theorem 3.2].

## 2.1.5. Anderson Model

Having introduced the Ising model and its graphical representations we now turn to the Anderson model introduced by Anderson in [And58]. For the presentation, we follow [AW15].

The Anderson Hamiltonian $H$ defined in (2.15) below is a self-adjoint bounded operator on a (spin-less, single-particle) Hilbert space $\mathcal{H} = \ell^2(G)$ for some graph $G = (V, E)$. The operator $H_0$ is typically local and the prototypical example is the discrete Laplacian $H_0 = -\Delta$, defined by

$$\Delta = \sum_{e \in E, e=(x,y)} |x\rangle\langle y| + |y\rangle\langle x| - \sum_{v \in V} \deg(v) |v\rangle\langle v|,$$

where $\deg(v)$ is the degree of the vertex $v$. Here we also introduced Dirac notation where $|x\rangle$ denotes the standard basis vector $e_x$ of $\ell^2(G)$ for a vertex $x \in G$ and $\langle x|$ denotes the corresponding dual vector.

The operator $V$ is a random potential, that is, a diagonal operator that satisfies $V |x\rangle = V(x) |x\rangle$ for each position basis standard-vector $|x\rangle$ for $x \in V$. The values $V(x)$ are taken to be i.i.d. random with some distribution that is almost surely bounded and has density with respect to the Lebesgue measure.

Now, the Anderson Hamiltonian is then given by

$$H = H_0 + \lambda V, \tag{2.15}$$

where $\lambda > 0$ is the strength of the disorder. That is, the random potential models disorder in the system. The motivation for this point of view was introduced in Section 1.3.

The time evolution of a quantum particle starting at $y$ at time $t$ is given by $e^{-itH} |y\rangle$. Since $H$ is self-adjoint, $e^{-itH}$ is unitary and so the $\ell^2$-norm of $e^{-itH} |y\rangle$ is 1, which therefore allows an interpretation as the spread of probability mass. The surprising statement of Anderson localization is that (under suitable assumptions) the disordered system like the Anderson model is **localized**. That means that even in the limit $t \to \infty$ the probability mass does not spread out. More formally, there exists constants $A, \mu > 0$ such that for any $R > 0$

$$\sum_{y \in \mathbb{G}: d(x,y) \geq R} \mathbb{E}\left[\sup_{t \in \mathbb{R}} |\langle x, e^{-itH} y\rangle|^2\right] \leq A e^{-\mu R}. \tag{2.16}$$

The first-time reader should think that this is very surprising. A first explanation is that the statement is an effect of destructive interference of all the paths of the particles that escape the box.

The multiscale method has played a central role in the rigorous study of localization starting from Fröhlich and Spencer in [FS83]. In addition, Aizenman and Molchanov [AM93] invented the fractional moment method that can be used to approach (2.16), for example in the case of sufficiently large disorder $\lambda > 0$. In the papers [**OpenLoc**] and [**MagQW**], we rely heavily on fractional moment method.

In the analysis of Anderson localization the **Green function** $G(x, y; z)$ is fundamental. If $z \notin \sigma(H_\lambda)$ then the Green function is defined as

$$G(x, y; z) = \langle x, (H - z)^{-1} y\rangle.$$

Using the eigenfunction correlator (see [AW15, (7.6), (7.4)] upon taking expectations and using that $H$ is bounded), the Green function is related to dynamics since it holds for any $s \in (0, 1)$ that there exists $C_s > 0$ such that

$$\mathbb{E}\left[\sup_{t \in \mathbb{R}} \left| \langle x, e^{-itH} \, y \rangle \right| \right] \leq C_s \sup_{E \in \mathbb{R}} \liminf_{|\eta| \downarrow 0} \mathbb{E}\left[ |G(x, y; E + i\eta)|^s \right]. \tag{2.17}$$

Now, to obtain the localization result in (2.16) we set out to prove that for some $s \in (0, 1)$ there exist constants $C_s, \mu_s > 0$ such that for every $\eta > 0$, $E \in \mathbb{R}$ and

$$\mathbb{E}[|G(x, y; E + i\eta)|^s] \leq C_s e^{-\mu_s |x-y|}.$$

### 2.1.6. Unitary Anderson Model

More recently a unitary analogue of the Anderson model has been introduced [HJS09]. Since it was important for the paper [**MagQW**] we introduce it here. We follow the lecture notes by Stolz [Sto11], where we also refer to the reader for more information. The time-evolution of the Anderson model $e^{-itH}$ is unitary since $H$ is self-adjoint. The operator $e^{-itH}$ is in general not local. In contrast, the unitary Anderson model $U$ is a local operator defined by $U(\omega) = \mathbb{D}(\omega)S$, where $S$ is a band matrix (in other words $\langle x, Sy \rangle = 0$ whenever $|x - y| > r$ for some fixed number $r > 0$) and $\mathbb{D}(\omega)$ is a diagonal unitary operator with $e^{i\theta_k}$ as the $(k, k)$ matrix element. In the foundational paper on the unitary Anderson model [HJS09] the case where $\{\theta_k\}_{k \in \mathbb{Z}^d}$ are i.i.d. with bounded density with respect to the Lebesgue measure is considered.

It turns out that this model also exhibits localization. Analogously to (2.16), the system is dynamically localized if there exists constants $A, \mu > 0$ such that for any $R > 0$

$$\sum_{y \in \mathbb{G}: d(x,y) \geq R} \mathbb{E}\left[\sup_{n \in \mathbb{N}} |\langle x, U^n y \rangle|^2\right] \leq A e^{-\mu R}.$$

In some sense, if the self-adjoint model is a continuous time random walk then the unitary case is analogous to a discrete-time random walk.

There is no simple relation between the unitary and self-adjoint Anderson models, but the philosophy of the proofs is oftentimes the same:

The unitary Anderson model can be thought of as a quantum walk. Taking a step with the quantum walk corresponds to acting with the unitary matrix $U$. In this sense, $n$ steps of the walk are obtained by

$$U^n \ket{0} = \mathbb{D}(\omega)S\mathbb{D}(\omega)S \dots \mathbb{D}(\omega)S \ket{0}.$$

We see that in every step the walk evolves with $S$ and then gets a random phase with $\mathbb{D}(\omega)$.

### 2.1.7. Quantum Walk in a Magnetic Field

In the spirit of the unitary Anderson model, in [**MagQW**] we introduce a quantum walk on $\mathbb{Z}^2$ in a magnetic field (see Section 3.5 for details). In this setup, the particle has an internal degree of freedom corresponding to particle spin. The model is still of the form $W = \mathbb{D}(\omega)S$, where $\mathbb{D}(\omega)$ is a random diagonal matrix and $S$ is banded. However, instead of having the randomness directly in the phases, we imagine the walk taking place in a disordered magnetic field with i.i.d. fluxes through each plaquette in $\mathbb{Z}^2$ (where the distribution has bounded density with respect to the Lebesgue measure). The diagonal entries of $\mathbb{D}(\omega)$ are therefore no longer independent. Nevertheless, it turns out that the analysis of [HJS09] and [Joy12] can be amended to prove exponential decay of (the expectation of fractional moments of) the Green function.

### 2.1.8. Coherences in the Steady State of an Open Quantum System

The time evolution of a state $\rho$ in an open quantum system governed by the Lindblad master equation is given by

$$\frac{d\rho}{dt} = \mathcal{L}(\rho) = -i[H, \rho] + G \sum_k L_k \rho L_k^* - \frac{1}{2}(L_k^* L_k \rho + \rho L_k^* L_k). \tag{2.18}$$

Here $H$ is the Hamiltonian of the system, $L_k$ are the so-called Lindblad operators and $G > 0$ is a constant modelling the strength of dissipation. The generator $\mathcal{L}$ generates a completely positive, trace preserving (CPTP) map $e^{t\mathcal{L}}$ [Lin76; GKS76] which corresponds to the time-evolution of density matrices of quantum systems. In finite dimensions, it follows from Brouwers fixed point theorem (see [BN08] for a proof) that there always exists a steady state $\rho_\infty$, although it is not necessarily unique.

We are interested in (spin-less) single-particle systems on a lattice $\mathbb{Z}^d$. Thus, the corresponding Hilbert space is $\mathcal{H} = \ell^2(\mathbb{Z}^d)$, and since the system we are interested in is local we have a distinguished position basis that we denote $|x\rangle = e_x$ for $x \in \mathbb{Z}^d$. In the position basis, which is central to our setup both in [Spec] and [OpenLoc], we consider the cases where $L_k$ are local operators and $H = \sum_i h_i$ is a sum of local terms $h_i$.

The time evolution of an initial state $\rho_0$ is given by $e^{t\mathcal{L}}(\rho_0)$ and we are particularly concerned with the Abel average of the time evolution that is defined by

$$\rho_\varepsilon = \varepsilon \int_0^\infty e^{-t\varepsilon}\, e^{t\mathcal{L}_\Lambda}(\rho_0)dt = -\varepsilon(\mathcal{L}_\Lambda - \varepsilon)^{-1}(\rho_0). \tag{2.19}$$

It has an interpretation as the time evolution up to times $\frac{1}{\varepsilon}$.

Steady states $\rho_\infty$, that do not change under the evolution, satisfy the equation $\mathcal{L}(\rho_\infty) = 0$ and if the steady state is unique then $\rho_\varepsilon \to \rho_\infty$ as $\varepsilon \to 0$ for all initial states $\rho_0$. We are particularly interested in the matrix elements $\rho_\infty(x, y)$ of the steady state. In [OpenLoc] we prove (under suitable assumptions) that there exists $C, \mu > 0$ such that

$$|\rho_\infty(x, y)| \leq C e^{-\mu|x-y|}.$$

The off-diagonal matrix elements of the density matrix are often called the coherences of the system. Thus, the indicates exponential decay of coherences, a phenomenon that we will call exponential decoherence. Another way of phrasing it is that it proves that macroscopic superpositions in the position basis do not exist, which is an expression of classicality.

## 2.2. Common Themes: The Hammersley Paradigm

The purpose of the chapter is to introduce some intuition about some of the common techniques that will appear throughout this thesis.

Bernoulli percolation and the Ising model that we already discussed in the gentle intro serve as the guide for our intuition. The two-point functions tell us something about the **locality** and **correlations** of the system: How can one transfer knowledge from one part of the system to other parts of the system (e.g. in the case of the Ising model, if a spin is up, what does it tell about the probability that a spin is up very far away). This is captured in the notion of the correlation length.

### 2.2.1. Correlation Length

In all of these models, we explained the definition of the two-point functions $\tau$ above (cf. (2.1)- (2.4)). Suppose that $\tau$ has exponential decay. We define the **correlation length** by

$$\xi^{-1} = \limsup_{|x-y|\to\infty} \frac{-\log(\tau(x,y))}{|x-y|}. \tag{2.20}$$

With the techniques that we have at our disposal, we often only prove bounds of the form $\tau(x,y) \le Ce^{-\mu|x-y|}$ leading to an upper bound on the correlation length. An exception is in [**Mass**] we also have a matching lower bound (proven in [CJN20], with a more probabilistic proof in [CJ20]).

For the Anderson-type systems, the (inverse) rate of exponential decay of the Green function, which we call correlation length usually goes under the name of **localization length**. Now, following the discussion in Section 2.1.5, this means that the rate of exponential decay of the Green function corresponds to the localization length, which is the size of the approximative region where the particle will stay forever (cf. (2.16)). In the one-dimensional case, this is sometimes known as the Lyapunov exponent (cf. [AW15, Chap. 12]).

In [HJS09] it was proven how this picture carries over to the case of the unitary Anderson model. In [**MagQW**] we embark on this scheme and prove exponential decay of the Green function. However, we do not have a relation between the Green function and the eigenfunction correlator, so we cannot deduce any dynamical consequences. Efforts have been spent pursuing such a relation, for example, by generalizing the approach in [HJS09], however, this relation has not been obtained.

Finally, for steady states of open quantum systems, instead of correlation length, we will call the quantity $\xi$ defined in (2.20) the **coherence length**. Tt sets the length scale of coherences in the steady state $\rho_\infty$.

### 2.2.2. Local Mechanism

In our models of interest, the correlations of the system can often be investigated using a **local mechanism** of the system. The local mechanism is captured in the following Domain Markov Property (DMP) for the random cluster model. An analogous property holds for the Ising model $\mu_{\beta,G}$ (cf. [FV17, (3.26)]).

**Proposition 2.2.1** (Domain Markov Property, cf. Theorem 1.6 in [DC19])**.** *If* $G_1 = (V_1, E_1) \subseteq G_2 = (V_2, E_2)$ *are two finite graphs* $\omega_1 := \omega|_{E_1}$ *and* $\omega_2 := \omega\ |_{E_2\setminus E_1}$ *then for any boundary condition* $\xi$ *and any event* $A$ *depending on edges in* $G_1$*, it holds that*

$$\phi^\xi_{\beta,G_2}[\omega_1 \in A|\ \omega_2] = \phi^{\xi_{\omega_2}}_{\beta,G_1}[A]$$

*where v and w belong to the same element of* $\xi_{\omega_2}$ *if and only if they are connected by a path (that might have length 0) in* $((V_2 \setminus V_1), E_{\omega_2})/\sim_\xi$.

The content of the Domain Markov Property is that two regions can only influence each other through their boundaries. For the random current model and loop O(1) model, the situation is more complicated and here we do have a similar property, but it involves source constraints which means that complicate the picture. We discuss this in detail in [**UEG**, Remark 2.11].

Another instance of the local mechanism is the backbone exploration. If for the random current model and loop O(1) model we have two sources $x$ and $y$, that is, vertices that have fixed odd degrees then we know that there is necessarily a path between $x$ and $y$. For random current one must use the non-traced multigraph. The backbone exploration is an algorithmic way of exploring the path. The algorithm, that is essential in [**Mass**], is a way to keep track of the source constraints when exploring through the local mechanism.

A third instance of the local mechanism valid for the loop O(1) model and one of the main inventions in [**UEG**] is that for $x > x_c$ what happens outside some 'safety distance' does not affect the configuration much.

**Theorem 2.2.2** ([**UEG**, Theorem 1.3]). *For $x > x_c$, there exists $c > 0$ such that for any event $A$ which only depends on edges in $\Lambda_n$, we have*

$$|\ell_{x,\Lambda_k}^{\xi}[A] - \ell_{x,\mathbb{Z}^d}[A]| \leq \exp(-cn) \tag{2.21}$$

*for any boundary condition $\xi$ and any $k \geq 4n$. In particular, for $x > x_c$ and any sequence $\xi_k$ of boundary conditions, $\lim_{k \to \infty} \ell_{x,\Lambda_k}^{\xi_k} = \ell_{x,\mathbb{Z}^d}$ in the sense of weak convergence of probability measures.*

For the self-adjoint and unitary Anderson models, there are no direct analogues of the local mechanisms above. However, one can sometimes use finite range of the operators to transfer bounds on the (expectations of fractional moments of the) Green function in one volume to another (see [**AW15**, Chap. 11]). Furthermore, for the case of quantum walks, we employ a trick (see Lemma 3.1 in [**MagQW**]) where we in the boundary of a box interchange the local unitaries with unitaries corresponding to a fully localized walk. This then decouples (up to the Aharonov-Bohm effect) the Green function inside and outside the box. Common to these tricks is that they extensively use the resolvent equation (see (2.23) below).

For random open quantum systems, there is also a local mechanism at play. We use the finite range of the non-hermitian evolution as the deciding locality property. The local mechanism allows us to establish the separating surface condition, which is in some sense the core of the common themes of the papers of this thesis.

### 2.2.3. Separating Surface Condition

We already saw the example of a separating surface condition for Bernoulli percolation in Section 1.2.5. The reason for discussing this example already in the gentle introduction is that we use similar principles throughout this thesis. Let us give a more proper definition of what it means that our two-point function $\tau$ satisfies a separating surface condition. This is the essence of the Hammersley stratagem from [**AW15**, Chapter 9]. In the following, $S$ is a region that contains $x$, but not $y$. A separating surface condition for $\tau$ is a bound of the form

$$\tau[x,y] \leq \sum_{u \in \partial S, v \notin S} K(u,v)\tau_S[x,u]\tau[v,y]. \tag{2.22}$$

for all such vertices $x, y$. Here $K(u,v) \geq 0$ is non-negative and $\tau_S$ is the value of the correlation function in $S$.

For the Ising model and the FK-representation, the separating surface condition has the name of the Simon-Lieb inequality after [**Sim80**; **Lie04**]. For a nice proof using (auxiliary) random currents see [**Wil20**].

**Separating surface condition for Anderson type models.** For the (self-adjoint) Anderson model, the separating surface condition follows from the geometric resolvent equations [AW15, (11.12)] which we follow here. The approach was developed in [Aiz+01]. The approach turns out to be slightly more complicated than for the random cluster model and therefore, we need to introduce the one-step fattening $S^+$ of a set $S \subset \mathbb{Z}^d$, which we define as $S^+ = \{x \in \mathbb{Z}^d \mid \text{dist}(x, S) \leq 1\}$. Let further $\partial S^+ = S^+ \backslash S$. This notion also turns out to be important in [**MagQW**].

Here, we consider finite volume bounds and therefore, we also consider the Green function in finite volume $\Lambda$ which we denote by $G_\Lambda = (H - z)^{-1}$ for some $z \in \mathbb{C} \backslash \mathbb{R}$. We assume that $H = T + \lambda V$ where $V$ is onsite, $\lambda > 0$ is a constant, and $T$ has range 1. That is, $T(x, y) = 0$ whenever $|x - y| \geq 2$. We then split $T = T_\partial + T_0$ where $T_\partial$ is supported on edges going from $S$ to $\partial S^+$.

The resolvent equation is the following relation

$$(A - z)^{-1} - (B - z)^{-1} = (A - z)^{-1}(B - A)(B - z)^{-1}, \tag{2.23}$$

that holds whenever $A, B$ are bounded operators and $z \notin \sigma(A) \cup \sigma(B)$.

Using the resolvent equation twice (cf. [AW15, (11.10)]), we get that

$$G_\Lambda(x, y; z) = \sum_{\substack{(u, u') \in \partial S \\ (v, v') \in \partial S^+}} G_S(x, u; z) T(u, u') G_\Lambda(u', v'; z) T(v', v) G_{\Lambda \backslash S^+}(v, y; z).$$

One can interpret this as first walking inside $W$ from $x$ to $u$ with the smaller resolvent, then from $u'$ to $v'$ with the full resolvent and finally from $v$ to $y$ outside of $W$ again with a smaller resolvent.

Generally, the strategy is then to take fractional moments and expectations to obtain

$$\mathbb{E}[|G_\Lambda(x, y; z)|^s] \leq \sum_{\substack{(u, u') \in \partial S \\ (v, v') \in \partial S^+}} \mathbb{E}\left[|G_S(x, u; z)|^s |G_\Lambda(u', v'; z)|^s |G_{\Lambda \backslash S^+}(v, y; z)|^s\right]$$

and then find an excuse to get rid of the middle factor. This is usually done using a priori and decoupling estimates in some form (see the use of Corollary 8.4 in (11.15) in [AW15]). In the case of the unitary Anderson model and the proof of localization [HJS09] the same geometric resolvent equations were used and the approach thus fits into the Hammersley paradigm. Then the remaining Green functions are now Green functions only in $S$ and only in $\Lambda \backslash S^+$. Therefore, they are often independent (though in [**MagQW**] due to the Aharonov-Bohm effect, predicted in [AB59], they are not independent) and in the best of all worlds, we would obtain that

$$\mathbb{E}[|G_\Lambda(x, y; z)|^s] \leq C \sum_{(u, u') \in \partial S} \mathbb{E}[|G_S(x, u; z)|^s] \sum_{(v, v') \in \partial S^+} \mathbb{E}\left[|G_{\Lambda \backslash S}(v, y; z)|^s\right].$$

This is reminiscent of (2.22) and we could use it as a starting point for an iterative proof of exponential decay.

**Steady states of open quantum systems.** For the steady state of the open quantum systems we can also use a geometric resolvent equation to obtain exponential decay. For the Lindbladian $\mathcal{L}_\Lambda$ that we study, we consider it as a sum of two Lindbladians $\mathcal{L}_\Lambda = \mathcal{L}_\Lambda^0 + \mathcal{L}_\Lambda^\partial$ where $\mathcal{L}_\Lambda^\partial$ consists of all the terms that connect two given points $x$ and $y$ (see [**OpenLoc**] for details). Using the resolvent equation (2.23) then yields

$$(\mathcal{L}_\Lambda - \varepsilon)^{-1} = (\mathcal{L}_\Lambda^0 - \varepsilon)^{-1} + (\mathcal{L}_\Lambda^0 - \varepsilon)^{-1} \, \mathcal{L}_\Lambda^\partial (\mathcal{L}_\Lambda - \varepsilon)^{-1}.$$

Thus, by the definition of the Abel average from (2.19) we get that

$$\begin{aligned}
\rho_\varepsilon &= -\varepsilon (\mathcal{L} - \varepsilon)^{-1}(\rho_0) = -\varepsilon (\mathcal{L}_\Lambda^0 - \varepsilon)^{-1}(\rho_0) + (\mathcal{L}_\Lambda^0 - \varepsilon)^{-1} \, \mathcal{L}_\Lambda^\partial \left( (-\varepsilon)(\mathcal{L}_\Lambda - \varepsilon)^{-1}(\rho_0) \right) \\
&= \rho_\varepsilon^0 + (\mathcal{L}_\Lambda^0 - \varepsilon)^{-1} \, (\mathcal{L}_\Lambda^\partial(\rho_\varepsilon)),
\end{aligned}$$

where $\rho_\varepsilon^0$ is the Abel average corresponding to the evolution $\mathcal{L}_\Lambda^0$. Now, it turns out that the term $\rho_\varepsilon^0(x, y)$ vanishes as $\varepsilon \to 0$ and therefore we can collect exponential decay by analyzing the term $(\mathcal{L}_\Lambda^0 - \varepsilon)^{-1} \, (\mathcal{L}_\Lambda^\partial(\rho_\varepsilon))$ using the details of the decomposition $\mathcal{L}_\Lambda = \mathcal{L}_\Lambda^0 + \mathcal{L}_\Lambda^\partial$ (see [**OpenLoc**] for details).

### 2.2.4. From Separating Surface Conditions to Finite Size Criteria

An approach for obtaining an iterative proof of exponential decay is finding a **finite size criterion**. This is a statement checkable in finite volume that provides information about the infinite volume system. For example, we use a finite size criterion in Theorem 1.4 of [**Kertész**]. As shown in [**AW15**, Theorem 9.3], we may abstractly convert a separating surface condition on $\tau$ to a finite size criterion. Here, we generalize the iteration from the gentle introduction substantially, but further generalizations exist see [**AW15**, Theorem 9.3].

**Proposition 2.2.3** (Simplified finite volume criterion)**.** *Suppose that $\tau$ is uniformly bounded and translation invariant. Suppose* (2.22) *holds, that is for every finite set $S \subset \mathbb{Z}^d$, vertices $x \in S$ and $y \notin S$ then*

$$\tau[x, y] \le \sum_{u \in \partial S, v \notin S} K(u, v) \tau_S[x, u] \tau[v, y]. \tag{2.24}$$

*Assume in addition that $K$ satisfies $K(u, v) \le K$ and $K(u, v) = 0$ whenever $u \not\sim v$. If for some $S \subset \mathbb{Z}^d$, we have*

$$b(S) = \sum_{u \in \partial S} 2dK \tau_S(x, u) < 1, \tag{2.25}$$

*then there exists a $C, \xi > 0$ such that for all $x, y \in \mathbb{Z}^d$ it holds that*

$$\tau[x, y] \le C e^{-\frac{|x-y|}{\xi}}. \tag{2.26}$$

*Proof.* Consider the $S$ that satisfies (2.25) and let $\partial S^+$ be the set of all vertices outside $S$ that have an edge to a vertex in $\partial S$. Then,

$$\tau[x, y] \le \sum_{u \in \partial S, v \sim u} K(u, v) \tau_S[x, u] \tau[v, y] \le \left( \sum_{u \in \partial S} 2dK \tau_S(x, u) \right) \sup_{v \in \partial S^+} \tau[v, y].$$

Let $|S| = \mathrm{diam}(S) + 1$. Then, if $v \in \partial S^+$ and $x \in S$, it holds that $|v - y| \ge |x - y| - |S|$. Fixing $y$ and letting $f(n) = \sup_{x : |x-y| \ge n} \tau[x \leftrightarrow y]$ yields

$$f(n) = \sup_{x : |x-y| \ge n} \tau[x \leftrightarrow y] \le b(S) \sup_{x : |x-y| \ge n} \sup_{v \in \partial S^+} \tau[v, y] \le b(S) f(n - |S|).$$

Using the uniform bound $\tau[a, b] \le T$ and translation invariance we can **iterate** to obtain.

$$\tau[x \leftrightarrow y] \le f(|x - y|) \le Tb(S)^{\left\lfloor \frac{|x-y|}{|S|} \right\rfloor}.$$

By (2.25) this inequality finishes the proof. □

Similar reasoning as above was used in a new proof [DCT16] of sharpness of the Bernoulli and Ising phase transitions that was first proven by [ABF87; AB87]. The proof uses the finite volume and Lieb-Simon-type arguments extensively. For random currents in themselves, we do not know whether the phase transition is sharp, and it would be interesting if a finite size criterion could be used to prove sharpness of random currents on $\mathbb{Z}^d$. Proving that would settle some of our main conjectures that we work towards in [UEG].

The general point is that finite volume properties of the system can illuminate infinite systems and thereby provide information about phase transitions.

## 2.3. Concrete Strategies for Proving Exponential Decay

From one point of view of this thesis, there are two ways to prove exponential decay:

(relate)  Relate the model to another model where exponential decay is known.

(iterate)  Find an iteration (typically using the locality of the system, for example in the form of a separating surface condition).

In the papers of this thesis, we use both approaches.

(a) In [Mass], we prove exponential decay, by finding an iteration that uses the backbone exploration of the random current representation of the Ising model and whether it hits the ghost. Using a Domain Markov Property for random currents we can explore the backbone step by step. In each step, there is a probability that the backbone will be connected to the ghost vertex and by combining a conditioning argument with the Markov property, we can iterate to get exponential decay.

(b) In [UEG], we prove the absence of exponential decay using the coupling that relates the loop O(1) model to the random cluster model. This motivates the coupling that we prove for the double random current in [MonCoup]. The idea is that a wrap-around of the torus exists for the supercritical random cluster measure. That is a path that, informally speaking, goes all the way around the torus. Then, whenever we take a uniform even subgraph of the random cluster model, a wrap-around will still exist with probability $\frac{1}{2}$. The existence of such a wrap-around is not consistent with exponential decay on the torus. To transfer from the torus to $\mathbb{Z}^d$ we prove an exponential mixing result Theorem 2.2.2 that relates the periodic boundary conditions to the measure $\ell_{x,\mathbb{Z}^d}$.

(c) In [Kertész], we relate the Kertész percolation problem to random cluster percolation problem in $\mathbb{Z}^d$ without a magnetic field using stochastic domination. The bounds then restrict the regions where exponential decay can exist. We also show the existence of a finite size criterion, which gives us a lower bound on the Kertész line.

(d) In [**MagQW**], we tailor-suit the scheme from [HJS09; Joy12] using a geometric resolvent equation for finding an iteration to prove exponential decay of the Green function to the setting of quantum walks in random magnetic fields.

(e) In [**OpenLoc**], we use the geometric resolvent approach for steady-state localization to prove exponential decay by iteration. The resolvent equation, which has a physical interpretation as a relation between Abel averages, allows us to do a split-up into terms connecting two given points $x$ and $y$ and an evolution where the two points are separated. It turns out that the separate evolution is governed by the non-hermitian evolution and not the quantum jump terms. Using an iteration for the non-hermitian evolution gives us the exponential decay.

## 2.4. Conclusion of the Introduction

We have provided a brief introduction to the models studied in this thesis and (some of) their relations. We studied the correlations of the model and saw how many of the models studied have some notion of locality.

Furthermore, they satisfy a separating surface condition. By iterating the separating surface condition, we can, under some conditions, abstractly prove exponential decay of correlations, for example using a finite size criterion. Proving exponential decay determines the phase (that is, either subcritical or localized). If the model is exponentially decaying, we can define the correlation length, known as the localization length or coherence length in some of the concrete contexts.

# 3. Summaries of Papers

In this chapter we provide introductions to the papers included in this thesis.

## 3.1. [Mass] Mass scaling of the near-critical 2D Ising model using random currents

The paper [Mass] is co-authored with Aran Raoufi and it builds very heavily on work done during my master's thesis at ETH Zürich [Kla19]. However, the write-up and revisions of the paper and one of its central ideas (exploring the random current backbone always turning first) were done as a part of this PhD thesis. The following introduction builds on the introduction of [Mass] but is substantially different and dives more into the technicalities of the proof. For an online talk explaining the result and the proof see [KR21].

**Context.** The paper concerns the two-dimensional Ising model in a magnetic field $h$ exactly at the critical temperature $\beta_c$. We denote the corresponding correlation functions by $\langle \cdot \rangle_{\beta_c, h}$.

More precisely, the paper studies the near-critical regime, which is a way of rescaling the magnetic field and the lattice simultaneously so that one can obtain a continuum limit. Furthermore, a bound in the near-critical regime allows one to obtain a bound on the corresponding continuum field. However, since we can state the main result without mentioning the near-critical regime we do that for clarity. In the introduction of [Mass], rescaling in the near-critical regime is introduced.

The contribution of the paper is to provide a new proof of the following inequality:

**Theorem 3.1.1** ([Mass, Theorem 1.2]). *There exists* $B_0, C_0 \in (0, \infty)$ *such that for any* $0 < h < 1$ *and for all vertices* $x, y \in \mathbb{Z}^2$ *then*

$$\langle \sigma_y \sigma_x \rangle_{\beta_c, h} - \langle \sigma_y \rangle_{\beta_c, h} \langle \sigma_x \rangle_{\beta_c, h} \leq C_0 |x - y|^{-\frac{1}{4}} e^{-B_0 h^{\frac{8}{15}} |x - y|}.$$

The inequality was previously proven by Camia, Jiang and Newmann in [CJN20], using different methods that include the use of the conformal loop ensemble.

In [CJN20] a converse inequality is also proved using reflection positivity. A more probabilistic proof of the lower bound was given in [CJ20]. We note that this shows that the correlation length is finite, the mass gap exists and that the critical exponent of the correlation length equals $\frac{8}{15}$. Further, as it is explained in the introduction of the paper, the exponential decay proven in Theorem 1.1 directly translates into the scaling limit.

**Methods.** The proof of Theorem 3.1.1 uses the random current representation of the Ising model. More specifically, it uses the random current representation with a ghost vertex g, that we can use to express the truncated correlations $\langle \sigma_0; \sigma_x \rangle_{\beta_c,h} = \langle \sigma_0 \sigma_x \rangle_{\beta_c,h} - \langle \sigma_0 \rangle_{\beta_c,h} \langle \sigma_x \rangle_{\beta_c,h}$ in terms of the random current representation by

$$\langle \sigma_0; \sigma_x \rangle_{\beta_c,h} = \langle \sigma_0 \sigma_x \rangle_{\beta_c,h} \cdot \hat{\mathbf{P}}^{\{0,x\}}_{\beta_c,h} \otimes \hat{\mathbf{P}}^{\emptyset}_{\beta_c,h}[0 \nleftrightarrow \mathrm{g}] \leq \hat{\mathbf{P}}^{\{0,x\}}_{\beta_c,h}[0 \nleftrightarrow \mathrm{g}], \tag{3.1}$$

where the equality is obtained using the switching lemma and the inequality comes from stochastic domination.

Now, to study $\mathbf{P}^{\{0,x\}}_{\beta_c,h}[0 \nleftrightarrow \mathrm{g}]$ we know in $\mathbf{P}^{\{0,x\}}_{\beta_c,h}$ that $0$ is always connected to $x$ since these are the two only vertices with odd degree. The connection is potentially using the ghost and our job is to prove that the probability that happens is high. The way we do that is to **partially explore the backbone** of the random current $0$ to $x$. The backbone exploration is a way to explore the path from $0$ to $x$. If the explored backbone goes through the ghost g then we know that $0 \leftrightarrow \mathrm{g}$ in the random current.

To proceed, we divide the iteration into steps and we prove that in every step, no matter how the previous steps looked like, there is a positive probability that the backbone hits the ghost in the next step. Iterating yields the exponential decay of $\mathbf{P}^{\{0,x\}}_{\beta_c,h}[0 \nleftrightarrow \mathrm{g}]$ and hence of the truncated correlation function through (3.1).

The partial exploration is obtained using a Markov property for the random current in Theorem 2.4 and then the iteration is obtained in Proposition 3.2.

The exact details of the iteration are one of the main complications of the paper and it is also here that the idea of exploring the backbone in the mode of "trying to turn first" becomes important because trying to turn first means that no path can "cross" the explored edges. The details are given in the paper. It was noted by Vincent Tassion in [KR21] that the idea of this exploration extends to all planar graphs making the result here potentially more general than the original result proven in [CJN20].

To obtain that there is a positive probability in each step to hit the ghost we use a stochastic domination result [Aiz+19, Theorem 3.2] that relates the random current with sources $\mathbf{P}^{\{0,x\}}_{\beta_c,h}[\cdot]$ to the random cluster model $\phi_{\beta_c,h}[\cdot \mid 0 \leftrightarrow x]$.

It turns out that we are left with something that resembles the following question closely.

**Question 3.1.2.** *Suppose that $E$ is a set of edges in the left half of $\Lambda_n$. Let $L$ be the left boundary and $R$ be the right boundary of $\Lambda_n$. Does there exist a constant $C > 0$ such that for every $n$ and any boundary condition $\xi$ on $\partial \Lambda_n$ it holds that*

$$\phi^{\xi}_{\Lambda_n \setminus E}[0 \leftrightarrow R] \geq C \phi^{\xi}_{\Lambda_n \setminus E}[0 \leftrightarrow L] ?$$

With the local mechanism in mind, the statement is intuitive, but the author is not aware of any proof.

Finding a short proof of Question 3.1.2 would significantly shorten the complications that one would have to go through to obtain the Theorem 3.1.1.

The reason is that as soon as we can connect to some "free space" that is some parts where we did not yet explore anything, then it is not so complicated to prove that a region in free space has a positive probability to connect to the ghost. This result we also prove after a long detour using a recent near critical RSW-result [DCM22] and [CJN20, Lemma 2.4]. Underlying the approach is a repeated application of the usual critical RSW result [DCHN11].

## 3.2. [**MonCoup**] On monotonicity and couplings of random currents and the loop-O(1)-model

The paper [**MonCoup**] is a single-author paper and it builds to some extent on work done in my master's thesis at ETH Zürich [Kla19] supervised by Aran Raoufi. Roughly speaking, the paper consists of two parts. The first part consists of various counterexamples to monotonicity of the loop O(1) model and the random current representations, which I to some extent figured out during my master's thesis. The second part is a new coupling that states that sampling a uniform even subgraph as a subgraph of the double random current has the law of the loop O(1) model. The statement and the proof I figured out during my PhD.

**Context.** The paper considers graphical representations of the Ising model, namely the loop O(1) model and the random current representations (see Section 2.1.3). In contrast to the most standard graphical representation, the random cluster model, the loop O(1) model and the random current representations are not monotone. However, the double random current model does display some monotonicity, namely monotonicity of events of the type $\{a \leftrightarrow b\}$. It is still unclear, whether monotonicity holds in larger generality for the double current. In [GMM18] it was conjectured that $\ell_x$ is monotonic on even graphs (that is graphs where all vertices have even degrees). In this paper, we find a counter-example to monotonicity of both $\ell$ and $\mathbf{P}$ also for events of the form $\{a \leftrightarrow b\}$. We can state it as follows:

**Theorem 3.2.1** ([**MonCoup**, Sec. 2.2])**.** *There exists an even graph $G$ with vertices $a$ and $b$ such that the function $x \mapsto \ell_{x,G}[a \leftrightarrow b]$ is not monotone.*

In general, the total number of open edges of $\ell_x$ is monotone in $x$. Therefore, the monotonicity is a bit subtle. The example uses the existence of big loops that lead nowhere (see Figure 1 of the paper). The second main result of the paper is the new coupling between the double random current and the loop O(1) model.

**Theorem 3.2.2** ([**MonCoup**, Theorem 4])**.** *The law of the uniform even subgraph of the double random current measure $\mathbf{P}_\beta^{\otimes 2}$ has the law of the loop O(1) model $\ell_x$.*

It mimics a coupling from [GJ09] that shows the corresponding result is true also for the random cluster measure.

**Methods.** For the monotonicity result the main method is the explicit evaluation of polynomials. From one point of view, one of the contributions of the paper is highlighting how, for small graphs, probabilities of events of the type $\{a \leftrightarrow b\}$ can be calculated using polynomials.

For the coupling, the proof uses a result of Lis [Lis17, Theorem 3.2] that relates the double random current to the number of even subgraphs of a graph. It remains to find a good use case for the coupling. The only thing we prove is that the density of cyclic edges (that is edges that are parts of cycles) is the same for the random cluster model and the double random current.

## 3.3. [Kertész] Strict monotonicity, continuity and bounds on the Kertész line for the random-cluster model on $\mathbb{Z}^d$

The paper [Kertész] is joint work with Ulrik Thinggaard Hansen. While the problem is inspired by the work of my master's thesis which studied the Ising model in a magnetic field and how the corresponding random cluster model behaves, the present work was carried out during my PhD. The following introduction builds heavily on the introduction in [Kertész].

**Context.**    As we saw in (2.9) the random cluster model is a graphical representation of the Ising model,

$$\mu_{\beta,h,G}[\sigma_x \sigma_y] = \phi_{p,h,G}[x \leftrightarrow y]. \tag{3.2}$$

which as explained generalizes to the Potts model. In the random cluster picture, the magnetic field is implemented by adding a ghost vertex g, which is connected to all other vertices in the graph. Thereby we obtain the graph $G_g$ as explained in Section 2.1.1. Now, two vertices $x, y \in V$ can be connected using the ghost if there is a path of edges in $E \cup E_g$ from $x$ to $y$. But we could also consider whether they are connected without using the ghost, that is, if there is a path of edges in $E$ from $x$ to $y$. Whereas the thermodynamic phase transition coincides with a percolative phase transition with the ghost vertex included, instead the Kertész line separates two regions according to whether or not there is percolation without using the ghost vertex. Therefore, the Kertész line transition does not necessarily correspond to a thermodynamic phase transition (i.e. a point where the free energy is not analytic).

Whenever we fix two of the three parameters $p, q, h$ and vary the last, the model exhibits a (possibly trivial) percolation phase transition (without using the ghost) at points which we denote $p_c(q,h)$, $q_c(p,h)$ and $h_c(p,q)$ respectively.

Before continuing with the results of the paper we note what can be proven by stochastic domination in a straigtforward manner. Let us for clarity consider $q \in (1, \infty)$ fixed (although one of the main tricks in the paper is to vary $q$). In that case, the Kertész line is a line in the $(p,h)$-plane and the random cluster measure $\phi_{p,q,h}$ is increasing in both $p$ and $h$. We know that for $h \to \infty$ the internal marginal measures $\phi_{p,q,h}|_{\mathbb{Z}^d}$ converge to $\mathbb{P}_{p,\mathbb{Z}^d}$, that is Bernoulli percolation with parameter $p$. So we obtain a stochastic domination $\phi_{p,q,h}|_{\mathbb{Z}^d} \preceq \mathbb{P}_{p,\mathbb{Z}^d}$ for all $h \in [0, \infty)$. Thus, if $\mathbb{P}_{p,\mathbb{Z}^d}$ then $\phi_{p,q,h}|_{\mathbb{Z}^d}$ never percolates. Thus, $h_c(p,q) = \infty$ whenever $p < p_c(\mathbb{P}_{\mathbb{Z}^d})$. On the other hand, $\phi_{p,q,\mathbb{Z}^d} = \phi_{p,q,h=0}|_{\mathbb{Z}^d} \preceq \phi_{p,q,h}|_{\mathbb{Z}^d}$. That means that if $\phi_{p,q,\mathbb{Z}^d}$ percolates then $h_c(p,q) = 0$. That is the Kertész line is only non-trivial between the random cluster phase transition at $h = 0$ and the $p_c$ for Bernoulli percolation, so this is the region where we focus our attention (see for example Figure 3 of the paper for an illustration).

**Results.**    In the paper, we provide a unifying account of the problems on the Kertész line. The techniques are mostly inspired by techniques developed to study Bernoulli percolation.

**Strict monotonicity and continuity:** First, we use the techniques of [Gri95], which again build on the techniques from [AG91], to prove in the relevant regions the six maps of the form $q \mapsto p_c(q,h)$ are strictly monotone. This strict monotonicity implies that the Kertész line $h \mapsto p_c(q,h)$ is continuous. This proves in particular that $h_c(p) > 0$ for all $p \in (p_c(1,0), p_c(q,0))$ as was conjectured in [CJN18, Remark 4].

**Upper and lower bounds:** Second, we prove upper and lower bounds on the Kertész line complementing the bound given by Ruiz and Wouts in [RW08]. For simplicity, we state them in the simpler case of the Ising model $q = 2$ in dimension $d = 2$ the upper is given as

$$h_c(p) \leq \operatorname{arctanh}\left(\sqrt{\frac{2(1-p)^2}{p^2} - 1}\right).$$

The technique that we use to prove the upper bound is **stochastic domination**. The technical workhorse is a condition on $p, q$ and $h$ that allows us to know when $\phi_{p_1, q_1, h_1}$ stochastically dominates $\phi_{p_2, q_2, h_2}$. When we then use our knowledge of the phase transition for $h = 0$ the results can be used to infer stochastic domination.

For the lower bound, we prove the following **finite volume criterion**. Here $\mu = \frac{(2d+1)^{2d+1}}{(2d)^{2d}}$.

**Theorem 3.3.1** ([Kertész, Theorem 1.4]). *Suppose that $p < p_c(q, 0)$ and that $\delta = \mu^{-4^d}$ and let $k$ be the smallest natural number satisfying*

$$\phi^1_{p, q, 0, \Lambda_{3k}}[\Lambda_k \leftrightarrow \partial \Lambda_{3k}] < \frac{\delta}{2}.$$

*Then, there is no percolation at $(p, h)$ for*

$$p_h < 1 - \left(1 - \frac{\delta}{2}\right)^{1/|\Lambda_{3k}|}.$$

The theorem allows us to establish an, in principle explicit, lower bound on the Kertész line. Finally, we use a more standard cluster expansion for the Potts model to give bounds on when the pressure is analytic, that is the absence of a thermodynamic phase transition even in the presence of a magnetic field. One interesting observation about our bounds around $h = 0$ is that the upper bound has a vertical asymptote and the lower bound has a horizontal asymptote. An open problem for future research is the determination of the asymptote. In that regard, we conjecture the following

**Conjecture 3.3.2** ([Kertész, Conjecture 4.7]). *In the limit $p \to p_c$ it holds for some constant $c > 0$ that*

$$h_c(p) \sim c(p - p_c)^{\frac{15}{8}}.$$

# 3.4. [UEG] The Uniform Even Subgraph and Its Connection to Phase Transitions of Graphical Representations of the Ising Model

The paper [UEG] is joint work with Ulrik Thinggaard Hansen and Boris Kjær. It is also a part of the master's thesis by Boris [Kjæ23] which I co-supervised.

**Context.** The paper is concerned with the percolative properties of two graphical representations of the Ising model. In particular, it is inspired from [DC16, Question 1] where Duminil-Copin asked whether the single random current has a phase transition at the same point as the random-cluster model on $\mathbb{Z}^d$.

For $d = 2$ it follows rather easily from the result of [GMM18] using the coupling between the loop O(1) model and the random current model, since for $d = 2$ it turns out that already the loop O(1) model percolates. This motivates the investigation of how the percolative properties of both the loop O(1) model and random current model for $d \geq 3$. Aran Raoufi asked whether the uniform even subgraph of $\mathbb{Z}^d$ percolates as a toy problem towards [DC16, Question 1]. This is a toy problem because the uniform even subgraph corresponds to the loop O(1) model for $x = 1$.

**Results.** The first result of the [UEG] is to prove that it is indeed the case.

**Theorem 3.4.1** ([UEG, Theorem 1.1]). *For $d \geq 2$ the uniform even subgraph of $\mathbb{Z}^d$ percolates*

$$\mathrm{UEG}_{\mathbb{Z}^d}[0 \leftrightarrow \infty] > 0.$$

We can even strengthen the result to prove that the percolative phase transition of $\ell_x$ is non-trivial.

**Theorem 3.4.2** ([UEG, Theorem 1.2]). *Let $d \geq 2$. Then there exists an $x_0 < 1$ such that for all $x \in (x_0, 1]$ then*

$$\ell_{x,\mathbb{Z}^d}[0 \leftrightarrow \infty] > 0.$$

From increasing coupling between the loop O(1) model and the random cluster model we know that connection probabilities decay exponentially for $x < x_c$ where $x_c = \tanh(\beta_c)$. Thus, we are left with the question of determining whether $\ell_{x,\mathbb{Z}^d}$ percolates for $x \in [x_c, x_0]$. The main theorem of the paper [UEG] partially answers that question.

**Theorem 3.4.3** ([UEG, Theorem 1.5]). *Let $d \geq 2$ and $x > x_c$, then there exists a $C > 0$ such that for every $k$ and every $N \geq 3k$ and any boundary condition $\xi$, $\ell^{\xi}_{x,\Lambda_N}[0 \leftrightarrow \partial \Lambda_k] \geq \frac{C}{k}$. It follows that $\ell_{x,\mathbb{Z}^d}[|\mathcal{C}_0|] = \infty$.*

It follows that the same is true for the (sourceless, traced) single random current $\mathbf{P}_{\beta,\mathbb{Z}^d}$.

**Corollary 3.4.4** ([UEG, Corollary 1.6]). *For $\beta > \beta_c$, there exists a $C > 0$ such that for every $k$ and every $N \geq 3k$, then*

$$\mathbf{P}_{\beta,\Lambda_N}[0 \leftrightarrow \partial \Lambda_k] \geq \frac{C}{k}.$$

*Moreover, the expected cluster size of the cluster of 0 in $\mathbf{P}_{\beta,\mathbb{Z}^d}$ is infinite.*

We note that the same results hold on the hexagonal lattice and there we know there is no percolation for $\ell_x$ for all $x \in [0, 1]$, so to improve the results one would need to use the structure of $\mathbb{Z}^d$.

**Methods.** The techniques in play in the paper are very diverse. The first results of the paper we prove by constructing a condition that ensures that the marginal of the uniform even subgraph is distributed as Bernoulli percolation with parameter $\frac{1}{2}$. In the case of finite graphs, it was known that the edge in the complement of a spanning tree of the graph will have marginal $\mathbb{P}_{\frac{1}{2}}$. To deal with the same problem for infinite graphs nicely we develop an algebraic approach where we exhibit the uniform even subgraph as the Haar measure on the group of even graphs with a symmetric difference as the group operation. Then we can give a criterion for when the marginal of the Haar measure UEG becomes the $\mathbb{P}_{\frac{1}{2}}$, which is the Haar measure on the group of all graphs with symmetric difference as group operation. The proof of Theorem 3.4.2 follows the same philosophy and uses in addition [LSS97, Theorem 0.0].

For the main theorem of the paper, the proof consists of two parts. In the first part, we prove the following mixing result of the loop O(1) model. The construction uses combinatorial insight into the uniform even graph combined with the existence of a very dense cluster in the supercritical random cluster model coming from Pisztora's construction [Pis96] for random-cluster models in dimension $d \geq 3$.

**Theorem 3.4.5** ([UEG, Theorem 1.3])**.** *For $x > x_c$, there exists $c > 0$ such that for any event $A$ which only depends on edges in $\Lambda_n$, we have*

$$|\ell^{\xi}_{x,\Lambda_k}[A] - \ell_{x,\mathbb{Z}^d}[A]| \leq \exp(-cn) \tag{3.3}$$

*for any boundary condition $\xi$ and any $k \geq 4n$. In particular, for $x > x_c$, the loop O(1) model on $\mathbb{Z}^d$ admits a unique infinite volume measure.*

The second part is orthogonal in the sense that it uses the torus very specifically. To get a sense of the argument consider any percolation configuration $\omega$ on the torus and a uniform even subgraph $\eta$ of $\omega$. Then, if $\omega$ has a loop $\gamma$ wrapping around the torus, then the symmetric difference $\eta \triangle \gamma$ also has the law of the uniform even subgraph of $\omega$. We further know that either $\eta$ or $\eta \triangle \gamma$ has a loop wrapping around the torus (by the combinatorics of the problem that is equivalent to the ground state of the toric code is 4-fold degenerate). Since there is a long loop, it must pass through at least one of the vertices on a given hyperplane and thus by translation invariance we obtain $\ell^{\text{per}}_{x,\Lambda_n}[0 \leftrightarrow \partial \Lambda_n] \geq \frac{c}{n^{d-1}}$. A technical construction extends the result to the following bound that ensures the infinite expected cluster sizes.

**Theorem 3.4.6** ([UEG, Theorem 1.4])**.** *Let $x > x_c$. Then, there exists $c > 0$ such that $\ell^{\text{per}}_{x,\Lambda_n}[0 \leftrightarrow \partial \Lambda_n] \geq \frac{c}{n}$ for all $n$.*

Combining these two theorems yields Theorem 3.4.3.

# 3.5. [MagQW] Quantum Walks in Random Magnetic Fields

The manuscript [MagQW] is joint work with Christopher Cedzich and Albert H. Werner. In the manuscript, we introduce a model for quantum walks in a random magnetic field. The introduction here follows the first sections of [MagQW].

**Context.** In [HJS09] Hamza, Joye and Stolz introduced the unitary Anderson model and a framework for proving localization using the fractional moment method in the unitary case. This approach was also used by Joye in [Joy12] to prove localization of a quantum walk. The approach entails proving first a priori estimate on the expectation of fractional moments of the Green function, then proving exponential decay (expectations of the fractional moments) of the Green function and finally proving that dynamical localization of the walk follows from exponential decay of the fractional moments.

**Model.** We consider the Hilbert space $\mathcal{H} = \ell^2(\mathbb{Z}^2) \otimes \mathbb{C}^2$ corresponding to a particle on the lattice $\mathbb{Z}^2$ with an internal degree of freedom. We define the unitary so-called coin operators $C_1, C_2$ by

$$C_i = \mathbb{1}_{\ell^2(\mathbb{Z}^2)} \otimes \begin{pmatrix} c_{11}^i & c_{12}^i \\ c_{21}^i & c_{22}^i \end{pmatrix}, \tag{3.4}$$

which is the same local unitary that acts on the internal degree of freedom on all sites simultaneously. Further, we consider shift operators $S_\alpha$ for $\alpha \in \{1, 2\}$ defined by

$$S_\alpha \left| x, \pm \right\rangle = \left| x \pm e_\alpha, \pm \right\rangle. \tag{3.5}$$

Now, the deterministic walk-operator $W_0$ is given by

$$W_0 = S_1 C_1 S_2 C_2. \tag{3.6}$$

Finally, the random quantum walk operator that is our object of interest is given by

$$W = D(\omega) W_0, \tag{3.7}$$

where $D(\omega)$ is a diagonal unitary operator, satisfying $D(\omega) \left| x, \pm \right\rangle = e^{-i\theta^\pm(x)} \left| x, \pm \right\rangle$. The phases $\theta^\pm$ correspond to the phase that a particle acquired traversing the edges that it just traversed with the previous action of the operator $W_0$, see the paper for details. In particular, note that the existence of the phases is due to a magnetic field being non-zero. Indeed, the magnetic field $F(x)$ is distributed such that the flux through each plaquette is i.i.d. random with density with respect to the Lebesgue measure bounded from above and below.

A special set of coins corresponds to walks with bound orbits. They are given by

$$\mathcal{C}_r = \left\{ (C_1, C_2) \in U(2) \times U(2) : c_{11}^i = c_{22}^i = 0, \left| c_{11}^j \right| = \left| c_{22}^j \right| = 1, \{i, j\} = \{1, 2\} \right\}. \tag{3.8}$$

and we will call them reflecting coins. The reader should think of them as corresponding to the infinite disorder in the self-adjoint case. In further analogy with the self-adjoint case, we will try to prove localization close to the reflecting coins, which corresponds to large disorder in the analogy.

**Result.** In **[MagQW]** we embark on generalizing the framework to include also the case of the quantum walk in the magnetic field. We first obtain an a priori estimate on the expectation of fractional moments of the Green function by generalizing the corresponding argument in [HJS09] from rank-2 perturbations to rank-4 perturbations. Interestingly, the proof involves a slight detour into studying the pseudo-spectrum of dissipative operators, in the spirit of the results from [Aiz+06]. After obtaining the a priori estimate we turn to exponential decay of the Green function, which is the main theorem of the paper. Here $W(C_1, C_2)$ is the unitary of the walk stemming from the coins $C_1$ and $C_2$ and $W(C_1^r, C_2^r)$ is the unitary stemming from a reflecting coin.

**Theorem 3.5.1** (**[MagQW]**, Theorem 2.3])**.** *There exists $\varepsilon > 0$ such that if*
$$\|W(C_1, C_2) - W(C_1^r, C_2^r)\| < \varepsilon \text{ for some } (C_1^r, C_2^r) \in \mathcal{C}_r \text{ then there are constants } \mu, C > 0$$
*such that for all $s \in (0, \frac{1}{3})$ and all $x, y \in \mathbb{Z}^2 \times \{-1, 1\}$ it holds for all $z \in (\frac{1}{2}, 2)$ that*

$$\mathbb{E}\left[\left|\langle x, (W - z)^{-1} y \rangle\right|^s\right] \leq C e^{-\mu|x-y|}.$$

**Methods.** In the proof by [HJS09], independence of the phases is essential. The proof uses the resolvent equation to can obtain a geometric decoupling between the Green function inside the box $\Lambda_L$, which we denote by $G^L$ and the Green function outside the box $\Lambda_{L+3}$ denoted by $G^{L+3}$ (the argument is substantially more complicated, for details see the paper).

However, since we are working with independent fluxes and not independent $A$-fields there is not quite independence between the $A$-fields inside a box and the $A$-fields outside the box. This is due to the Aharonov-Bohm effect where a particle moving the two different ways around a box would experience a phase change corresponding to the total flux through the box [AB59]. Therefore, we would only expect the Green function inside and outside a box to be conditionally independent given the total flux through the box. By conditional independence, we can factorize the (fractional moments of the) Green function inside and outside the box.

**Lemma 3.5.2** (**[MagQW]**, Theorem 7.6], Factorization using Aharonov-Bohm effect)**.** *Let $f_L$ be the density of the random variable $F_L = \sum_{x \in \Lambda_L} F(x)$ representing the total flux through $\Lambda_L$. For $|y| \geq L + 2$ and $u \in \Lambda_L, v \in \Lambda_{L+3}^c$ we have that*

$$\mathbb{E}\left[\left|\langle 0, G^L u \rangle\right|^s \left|\langle v, G^{L+3} y \rangle\right|^s\right] \leq \left\|\frac{1}{f_L}\right\|_\infty \mathbb{E}\left[\left|\langle 0, G^L u \rangle\right|^s\right] \mathbb{E}\left[\left|\langle v, G^{L+3} y \rangle\right|^s\right]$$

*for all $0 < s < 1$.*

Then, we embark on the resampling strategy from [HJS09], again the lack of independence yields additional complications and it becomes important that the phases (or rather pairs of phases) have bounded conditional distribution given all the other phases of the system. Finally, by using an iteration strategy also employed in [Joy12] we obtain the exponential decay of the fractional moments of the Green function.

The last step would then entail going from fractional moments estimate to dynamical localization of the walk. Despite many efforts, it turns out that the proof of [HJS09] does not easily generalize to the case where the phases are equal in pairs (which means that the conditional distribution is not bounded with respect to the Lebesgue measure).

## 3.6. [Spec] Spectra of generators of Markovian evolution in the thermodynamic limit: From non-Hermitian to full evolution via tridiagonal Laurent matrices

The paper [Spec] is joint work with Albert H. Werner.

**Context.**   The paper concerns the spectra of single-particle translation-invariant generators of Lindblad semigroups in infinite volume. In finite volume, the spectra of Lindblad generators (henceforth Lindbladians) and in particular the spectral gap yields information on the speed of relaxation of a Lindblad semigroup towards the steady state subspace. The corresponding problems in infinite volume is arguably an understudied area of mathematical physics.

In this paper, we study the spectra directly in infinite volume. For simplicity, we work with the Hilbert space $\mathcal{H} = \ell^2(\mathbb{Z})$. The infinite volume Lindbladian can be defined from the Lindblad form a priori as an operator on for example the trace class operators $\mathrm{TC}(\mathcal{H})$, the Hilbert-Schmidt operators $\mathrm{HS}(\mathcal{H})$ or the compact or bounded operators. Before coming to the main theorem we make some comments on the different spaces and the notion of spectral independence of such operators. In the rest of the paper, the Lindbladian is mainly considered as an operator on $\mathrm{HS}(\mathcal{H})$.

**Results.**   In the main theorem, we find an isometric isomorphism of Hilbert spaces $\mathcal{J}: \mathrm{HS}(\ell^2(\mathbb{Z})) \to \int_{[0,2\pi]}^{\oplus} \ell^2(\mathbb{Z})_q dq$ (see Section 3.2 of the paper for an introduction to direct integrals) such that upon conjugation with this isometric isomorphism the Lindbladian takes a particularly nice form.

**Theorem 3.6.1** ([Spec, Theorem 3.8])**.** *Suppose that $\mathcal{L}$ is of the form* (2.18) *with Lindblad operators $L_k$ satisfying locality and translation invariance assumptions. For an isometric isomorphism $\mathcal{J}: \mathrm{HS}(\ell^2(\mathbb{Z})) \to \int_{[0,2\pi]}^{\oplus} \ell^2(\mathbb{Z})_q dq$ then*

$$\mathcal{J}\mathcal{L}\mathcal{J}^* = \int_{[0,2\pi]}^{\oplus} T(q) + F(q) dq,$$

*with $T(q)$ a bi-infinite r-diagonal Laurent operator and $F(q)$ a finite rank operator with finite range for each $q \in [0, 2\pi]$.*

Using the main theorem, we can prove both abstract consequences and compute the spectrum in concrete cases.

An important technical result that we believe is of independent interest is the following theorem. Using the notion of pseudospectrum it generalizes a related theorem proven in the self-adjoint case in [RS78, XIII.85]. The pseudospectrum $\sigma_\varepsilon(A)$ of an operator $A$ is defined by

$$\sigma_\varepsilon(A) = \{z \in \mathbb{C} \mid \|A - z\| \geq \varepsilon^{-1}\},$$

and it is essential for the study for non-normal operators [TE05].

**Theorem 3.6.2** ([**Spec**, Theorem 3.12])**.** *Let* $I \subset \mathbb{R}$ *be an interval and* $\mathcal{H} = \int_I^{\oplus} \mathcal{H}_q dq$ *for some family of separable Hilbert spaces* $\{\mathcal{H}_q\}_{q \in I}$. *Suppose that* $\{A(q)\}_{q \in I}$ *is a measurable family of bounded operators, such that* $A(q)$ *acts on* $\mathcal{H}_q$ *and* $A = \int_I^{\oplus} A(q) dq \in \mathcal{B}(\mathcal{H})$. *Then for all* $\varepsilon > 0$ *it holds that*

$$\sigma(A) \subset \bigcup_{q \in I}^{\text{ess}} \sigma_\varepsilon(A(q)) \ \text{and} \ \sigma(A) = \bigcap_{\varepsilon > 0} \left( \bigcup_{q \in I}^{\text{ess}} \sigma_\varepsilon(A(q)) \right).$$

Combining the two theorems we get the following corollary that we can use to compute spectra explicitly.

**Corollary 3.6.3** ([**Spec**, Corollary 3.14])**.** *Let* $\mathcal{L} \in \mathcal{B}(\text{HS}(\mathcal{H}))$ *be a Lindbladian of the form* (2.18) *satisfying assumption* $\mathcal{A}_2 a)$ *and* $\mathcal{A}_2 b)$ *and let* $T(q)$ *and* $F(q)$ *be as in Theorem 3.6.1. Then*

$$\sigma(\mathcal{L}) = \bigcup_{q \in [0, 2\pi]} \sigma(T(q) + F(q)). \tag{3.9}$$

*Furthermore, for the non-Hermitian evolution* $\mathcal{T} = \int_{[0,2\pi]}^{\oplus} T(q) dq$ *it holds that*

$$\sigma(\mathcal{T}) = \bigcup_{q \in [0, 2\pi]} \sigma(T(q)), \ \text{and} \ \sigma(\mathcal{T}) \subset \sigma(\mathcal{L}).$$

Abstractly, we then prove that the residual spectrum of the Lindbladians we study is always empty. Using tools from the complex analysis, we prove that the Lindbladians are either gapless or have an infinite dimensional kernel. Finally, we give a condition for convergence of the finite volume spectra with periodic boundary conditions to their infinite volume counterparts.

We go on to use Corollary 3.6.3 to compute the infinite volume spectra of some operators that have been studied in the physics literature with periodic boundary conditions [Zni15; EG05a; EG05b]. In particular, we put the observations into a more general light. We continue by studying certain systems with non-normal dissipators that were recently associated with localization in open quantum systems [Yus+17]. We prove that the spectrum of the Non-Hermitian evolution in itself is gapless (that is without considering the quantum jump terms). Finally, we prove some bounds for the spectra in random potentials. In particular, we consider an analogue of the Kunz-Soulliard theorem for open quantum systems.

## 3.7. [**OpenLoc**] Exponential decay of coherences in steady states of open quantum systems with large disorder

The paper [**OpenLoc**] is joint work with Simone Warzel.

**Context.** The framework of the paper is a single-particle open quantum system with Markovian evolution. The form of the generator of the evolution (see cf. (2.18) in Section 2.1.8) was found in [Lin76; GKS76]. We assume that the terms in the Lindbladian have uniformly bounded range. Thereby, the system aquires a local mechanism so that in spirit our intuition from Section 2.2.2 applies.

As we saw in Section 2.1.5 for the Anderson model, disorder can be modelled through a random potential with strength $\lambda > 0$. The Hamiltonian is then $H = H_0 + \lambda V$, where $V$ is a random potential defined through $V |x\rangle = V(x) |x\rangle$ for each basis vector $|x\rangle$ of the position basis and $\lambda > 0$ is a parameter describing the strength of the disorder. Here $V(x)$ are i.i.d. random variables with bounded and compactly supported densities.

In [FS16] it is shown that a random potential slows the evolution of the open quantum system from ballistic to diffusive (in a certain sense). More recently, localization in open quantum systems was investigated numerically in [Yus+17] for a special example of Lindblad generators that create coherences. This example was also investigated in [DROZ11] in the context of dissipative engineering [VWC09]. Another example, is the Anderson model with local dephasing, that we studied from a spectral point of view in [**Spec**].

**Results.** For any $\varepsilon > 0$ and state $\rho_0$ we define the Abel average $\rho_\varepsilon$ by

$$\rho_\varepsilon = \varepsilon \int_0^\infty e^{-t\varepsilon} \, e^{t\mathcal{L}}(\rho_0) dt = -\varepsilon(\mathcal{L} - \varepsilon)^{-1}(\rho_0). \tag{3.10}$$

Since $e^{t\mathcal{L}}(\rho_0)$ is the time evolution of the state $\rho$ until time $t$. We get that $\rho_\varepsilon$ can be interpreted as a time average up to timescales of $\frac{1}{\varepsilon}$.

We prove the result on a finite $\Lambda$, but we emphasize that the constants are uniform in $\Lambda$. In addition to the locality assumption, we have a weak gap assumption on the non-hermitian evolution that entails that the gap with a Dirichlet boundary condition in at least one point closes at most polynomially fast in the thermodynamic limit. We give more details in on the assumption in the paper and we show that our motivating examples stemming from dissipative engineering satisfy the assumption.

**Theorem 3.7.1** ([**OpenLoc**, Theorem 3.1]). *Let $\mathcal{L}$ satisfy the assumptions outlined above. For sufficiently large disorder $\lambda > 0$, there exist constants $C, \mu > 0$ such that for any connected set $\Lambda \subset \mathbb{Z}^d$ and any $x, y \in \Lambda$, $\varepsilon \in (0, 1)$, initial state $\rho_0$, and any $s \in (0, 1)$ there exists a $C_s > 0$ such that*

$$\mathbb{E}|\rho_\varepsilon(x, y)| \leq Ce^{-\mu|x-y|} + \varepsilon^{2s-1} \, C_s. \tag{3.11}$$

*Furthermore, for any measurable choice of steady state $\omega \mapsto \rho_\infty(\omega)$ of $\mathcal{L}_\Lambda$ it holds that*

$$\mathbb{E}|\rho_\infty(x, y)| \leq Ce^{-\mu|x-y|}. \tag{3.12}$$

In the case of local dephasing with rate $\gamma > 0$, the gap of the non-hermitian evolution is constant and we obtain the following deterministic strengthening of the theorem. Again, we give more details in the paper.

**Theorem 3.7.2** ([**OpenLoc**, Theorem 3.2])**.** *Let $\mathcal{L}$ satisfy the assumptions outlined above. Then there exist $C, \mu > 0$ such that for any $\Lambda \subset \mathbb{Z}^d$, any $x, y \in \Lambda$, initial state $\rho_0$, $\varepsilon \in (0, 1)$, and any $s \in (0, 1)$ there exists a $C_s > 0$ such that*

$$|\rho_\varepsilon(x, y)| \leq Ce^{-\mu|x-y|} + \varepsilon^{2s-1} \, C_s. \tag{3.13}$$

*In particular, for any steady state $\rho_\infty$ of $\mathcal{L}_\Lambda$ then*

$$|\rho_\infty(x, y)| \leq Ce^{-\mu|x-y|}. \tag{3.14}$$

**Methods.** The methods of the paper are inspired by the fractional moment approach to Anderson localization pioneered in [AM93] and many technical ideas stem from [AW09]. We do a split up of the Lindbladian $\mathcal{L}$ that allows us to reduce bound the Abel averaged time evolution only in terms of the effective non-Hermitian evolution so that we do not have to take quantum jump terms into account. Then we check that the proof of fractional moments of the Green function at large disorder generalizes from the Hamiltonian case to the the non-Hermitian evolution. After that, we use the locality of $\mathcal{L}$ through the use of the resolvent equation (as outlined in Section 2.2.3 above) to transfer the exponential decay of the non-hermitian evolution to the Abel averaged time evolution.

# Part II.

# Papers

# 1. Mass Scaling of the Near-Critical 2D Ising Model using Random Currents

# Mass scaling of the near-critical 2D Ising model using random currents

### Frederik Ravn Klausen, Aran Raoufi

#### Abstract

We examine the Ising model at its critical temperature with an external magnetic field $ha^{\frac{15}{8}}$ on $a\mathbb{Z}^2$ for $a, h > 0$. A new proof of exponential decay of the truncated two-point correlation functions is presented. It is proven that the mass (inverse correlation length) is of the order of $h^{\frac{8}{15}}$ in the limit $h \to 0$. This was previously proven with CLE-methods in [1]. Our new proof uses instead the random current representation of the Ising model and its backbone exploration. The method further relies on recent couplings to the random cluster model [2] as well as a near-critical RSW-result for the random cluster model [3].

## 1 Introduction

The square lattice Ising model [4] suggested by Lenz [5] is the archetypal statistical physics model undergoing an order/disorder phase transition. It has been subject of intense study in the past century [6, 7], starting with Periels' proof of the existence of a phase transition [8] and Onsager's calculation of the free energy [9]. The rigorous understanding of the critical two-dimensional Ising model has advanced tremendously in the past decade starting with the breakthroughs [10, 11] and with the subsequent works (see, for example, [12]).

One of the questions that remained unsolved until recently is obtaining the speed of the decay of the truncated correlations in the near-critical two-dimensional Ising model. For $a \in (0, 1]$ and $h > 0$ the near critical regime is defined to be the Ising measure on the lattice $a\mathbb{Z}^2$ with the parameter $\beta = \beta_c(\mathbb{Z}^2)$ and external field $a^{15/8} h$. We denote the corresponding correlation functions with $\langle . \rangle_{a,h}$. The following theorem is proved in [1] using the scaling limit of the FK-Ising model which was proved to exist in [13] and its connections to the conformal loop ensemble [14]. See also the review [15].

**Theorem 1.1** *There exists $B_0, C_0 \in (0, \infty)$ such that for any $a \in (0, 1]$ and $h > 0$ with $ha^{\frac{15}{8}} \le 1$,*

$$0 \le \langle \sigma_x \sigma_y \rangle_{a,h} - \langle \sigma_x \rangle_{a,h} \langle \sigma_y \rangle_{a,h} \le C_0 a^{\frac{1}{4}} |x - y|^{-\frac{1}{4}} e^{-B_0 h^{\frac{8}{15}} |x-y|}.$$

*Accordingly, for $a = 1$ the result on $\mathbb{Z}^2$ is that for any $h \in [0, 1)$,*

$$\langle \sigma_y \sigma_x \rangle_{1,h} - \langle \sigma_y \rangle_{1,h} \langle \sigma_x \rangle_{1,h} \le C_0 |x - y|^{-\frac{1}{4}} e^{-B_0 h^{\frac{8}{15}} |x-y|}.$$

In this paper we prove Theorem 1.2 from which we can deduce Theorem 1.1.

**Theorem 1.2** *For any $h > 0$ and $a \le 1$ there are functions $C(h) > 0$ and $m(h) > 0$ independent of $a > 0$ such that for any $x, y \in a\mathbb{Z}^2$ it holds that*

$$\langle \sigma_x \sigma_y \rangle_{a,h} - \langle \sigma_x \rangle_{a,h} \langle \sigma_y \rangle_{a,h} \le C(h) a^{\frac{1}{4}} |x - y|^{-\frac{1}{4}} e^{-m(h)|x-y|}.$$

The proof uses first a partial exploration of the backbone of random currents, then a recent coupling between the random current measure with sources and the random cluster model [2]. The proof utilises a new result that extends a result on crossing probabilities for the critical random cluster model to the near critical regime [3].

Before diving into the details, we briefly show how Theorem 1.1 follows (as explained in [1]) from Theorem 1.2.

*Proof of Theorem 1.1.* Let $H$ be such that $a = H^{\frac{8}{15}}$ and $h = 1$. Let $\langle \sigma_0; \sigma_x \rangle = \langle \sigma_0 \sigma_x \rangle_{a,h} - \langle \sigma_0 \rangle_{a,h} \langle \sigma_x \rangle_{a,h}$. Then from Theorem 1.2

$$\langle \sigma_0; \sigma_x \rangle_{H^{\frac{8}{15}},1} \leq C(1) H^{\frac{2}{15}} |x|^{-\frac{1}{4}} e^{-m(1)|x|}$$

for $x \in H^{\frac{8}{15}} \mathbb{Z}^2$. Using the relation $\langle \sigma_0; \sigma_x \rangle_{H^{\frac{8}{15}},1} = \langle \sigma_0; \sigma_{x'} \rangle_{1,H}$ whenever $x' = \frac{x}{H^{\frac{8}{15}}}$ we obtain

$$\langle \sigma_0; \sigma_{x'} \rangle_{1,H} \leq C(1) |x'|^{-\frac{1}{4}} e^{-mH^{\frac{8}{15}}|x'|}$$

for $x' \in \mathbb{Z}^2$. Rescaling back to $a\mathbb{Z}^2$ yields the result. $\qquad\square$

In [1] a converse inequality is also proved using reflection positivity. A more probabilistic proof of the lower bound was given in [16]. We note that this shows that the correlation length is finite, the mass gap exists and that critical exponent of the correlation length equals $\frac{8}{15}$. Further, as it is explained in [1] the exponential decay proven in Theorem 1.1 directly translates into the scaling limit.

Indeed, as in [1] if $\Phi^{a,h}$ is the near critical magnetization field given by

$$\Phi^{a,h} = a^{\frac{15}{8}} \sum_{x \in a\mathbb{Z}^2} \sigma_x \delta_x$$

with $\{\sigma_x\}_{x \in a\mathbb{Z}^2} \in \{0,1\}^{a\mathbb{Z}^2}$, it was proven in Theorem 1.4 of [13] that $\Phi^{a,h}$ converges in law to a continuum (generalized) random field $\Phi^h$. Let $C_0^\infty(\mathbb{R}^2)$ denote the set of smooth functions with compact support and let $\Phi^h(f)$ be $\Phi^h$ paired against $f \in C_0^\infty(\mathbb{R}^2)$. Then as in [1] it holds that

**Corollary 1.3** *Let $f, g \in C_0^\infty(\mathbb{R}^2)$, then there are $B_0, C_0 \in (0, \infty)$ such that*

$$\left| Cov(\Phi^h(f), \Phi^h(g)) \right| \leq C_0 \int \int_{\mathbb{R}^2 \times \mathbb{R}^2} |f(x)| |g(x)| |x-y|^{-\frac{1}{4}} e^{-B_0 h^{\frac{8}{15}}|x-y|} dx dy.$$

Starting with [17], there has in the physics community been much interest in the masses of the Ising model [18] including possible connections to the exceptional Lie Algebra $E_8$ [19] which has been investigated also experimentally [20, 21, 22] and numerically [23]. On the mathematical side, exponential decay was first rigorously proven in [24] and in [25] a linear upper bound for the mass was proven. Proving the correct scaling exponent is a further step towards rigorous results in this direction. For further rigorous developments see also [26].

## 2 Preliminaries

We start by briefly introducing the Ising model and its random cluster and random current representations that we will use to prove the result. Let $G = (V, E)$ be a finite graph. Then for each spin configuration $\sigma \in \{\pm 1\}^V$ and $h \geq 0$ define the energy

$$H(\sigma) = -\sum_{xy \in E} \sigma_x \sigma_y - h \sum_{x \in V} \sigma_x,$$

where $h$ describes the effect of an external magnetic field. For each $A \subset V$ we let $\sigma_A = \prod_{x \in A} \sigma_x$ and define the correlation function as

$$\langle \sigma_A \rangle = \frac{\sum_{\sigma \in \{\pm 1\}^V} \sigma_A \exp(-\beta H(\sigma))}{Z}$$

where $Z = \sum_{\sigma \in \{\pm 1\}^V} \exp(-\beta H(\sigma))$ is the partition function. In what follows, we will be concerned with the Ising model on the graph $a\mathbb{Z}^2$ which is obtained by taking the thermodynamic limit of finite graphs. For discussions about the thermodynamic limit we refer the reader to [27].

In both representations we implement the magnetic field using Griffiths' ghost vertex g. This means that we consider the graph $G_{\text{ghost}} = (V \cup \{g\}, E \cup E_g)$ where $E_g = \cup_{v \in V} \{e_{vg}\}$ are additional edges from every original vertex $v$ to the ghost vertex g (see for example [7]). We will refer to the edges $E$ as *internal* edges and to the edges $E_g$ as *ghost* edges.

**The random current representation**

Let us now introduce the random current representation which is a very effective tool in the study of the Ising model [28, 29, 30, 2, 31, 32, 33, 34]. Further information can be found in [7] and [35]. The central building blocks in the random current representation of the Ising model on a graph $G_{\text{ghost}} = (V \cup \{g\}, E \cup E_g)$ are the currents $\mathbf{n} \in \mathbb{N}_0^{E \cup E_g}$. For each current $\mathbf{n}$ we can define its sources $\partial \mathbf{n}$ as the $v \in V$ where $\sum_{xv \in E \cup E_g} \mathbf{n}_{xv}$ is odd. Let further the *weight* of each current $\mathbf{n}$ be given by

$$w(\mathbf{n}) = \prod_{xy \in E} \frac{\beta^{\mathbf{n}_{xy}}}{\mathbf{n}_{xy}!} \prod_{xg \in E_g} \frac{(\beta h)^{\mathbf{n}_{xg}}}{\mathbf{n}_{xg}!}.$$

A simple identity which connects the random currents to the Ising model is given as (2.4a) in [31]

$$\langle \sigma_0 \sigma_x \rangle = \frac{\sum_{\partial \mathbf{n} = \{0, x\}} w(\mathbf{n})}{\sum_{\partial \mathbf{n} = \emptyset} w(\mathbf{n})}. \tag{1}$$

Further, given a current $\mathbf{n}$ define the traced current $\hat{\mathbf{n}} \in \{0, 1\}^{E \cup E_g}$ by $\hat{\mathbf{n}}(e) = 0$ if $\mathbf{n}(e) = 0$ and $\hat{\mathbf{n}}(e) = 1$ if $\mathbf{n}(e) > 0$. Then $\mathbb{P}_G^A(\mathbf{n})$, the random current measure with sources $A \subset V$, is the probability measure that satisfies $\mathbb{P}_G^A(\mathbf{n}) \propto w(\mathbf{n}) 1_{\{\partial \mathbf{n} = A\}}$. If $A$ and $B$ are either vertices in or subsets of $V \cup \{g\}$ we denote the event that they are connected in a configuration $\omega \in \{0, 1\}^{E \cup E_g}$ by $A \leftrightarrow B$, meaning that one vertex of $A$ is connected to one vertex of $B$.

The traced random current measure $\hat{\mathbb{P}}_G^A$ gives each $\omega \in \{0,1\}^{E \cup E_g}$ the probability

$$\hat{\mathbb{P}}_G^A(\omega) = \sum_{\partial \mathbf{n} = A} \mathbb{P}_G^A(\mathbf{n}) 1_{\{\hat{\mathbf{n}} = \omega\}}.$$

To ease the notation in what follows define $\hat{\mathbb{P}}_G^{\{0,x\}} \otimes \hat{\mathbb{P}}_G^{\emptyset}$ to be the probability measure which assigns each $\omega \in \{0,1\}^{E \cup E_g}$ the probability

$$\hat{\mathbb{P}}_G^{\{0,x\}} \otimes \hat{\mathbb{P}}_G^{\emptyset}(\omega) = \frac{1}{Z_{\emptyset} Z_{\{0,x\}}} \sum_{\partial \mathbf{n} = \{0,x\}, \partial \mathbf{m} = \emptyset} w(\mathbf{n}) w(\mathbf{m}) 1[\widehat{\mathbf{n} + \mathbf{m}} = \omega].$$

On the square lattice $a\mathbb{Z}^2$ in a magnetic field $a^{15/8} h$ we denote the non traced and traced single current measures by $\mathbb{P}_{a,h}^A$ and $\hat{\mathbb{P}}_{a,h}^A$ respectively. The main part of what follows proves exponential decay of truncated correlations, but first we obtain the correct front factor $a^{\frac{1}{4}}$. We do a similar trick as in [1] where we set the magnetic field $h$ to 0 in the boxes of radius 1 around 0 and $x$ and call that magnetic field $\vec{h}$.

**Proposition 2.1** *We have*

$$\langle \sigma_0; \sigma_x \rangle_{a,h} \leq \langle \sigma_0; \sigma_x \rangle_{a,\vec{h}} = \langle \sigma_0 \sigma_x \rangle_{a,\vec{h}} \cdot \hat{\mathbb{P}}_{a,\vec{h}}^{\{0,x\}} \otimes \hat{\mathbb{P}}_{a,\vec{h}}^{\emptyset}(0 \nleftrightarrow \mathrm{g}) \leq C a^{\frac{1}{4}} \hat{\mathbb{P}}_{a,\vec{h}}^{\{0,x\}}(0 \nleftrightarrow \mathrm{g}).$$

*Proof.* The first inequality follows from the GHS inequality [36]. The second step used the switching lemma [36] and (1). Since the event $\{0 \nleftrightarrow \mathrm{g}\}$ is decreasing the probability increases when $\hat{\mathbb{P}}_{a,\vec{h}}^{\emptyset}$ is removed. Then the last inequality is a standard application of equation (1-arm) below (see also [37]). □

**The random cluster model**

Each configuration $\omega \in \{0,1\}^{E \cup E_g}$ corresponds to a (spanning) subgraph of $G_{ghost}$. For each $e \in E \cup E_g$ if $w_e = 1$ we say that $e$ is *open* and if $w_e = 0$ we say that $e$ is *closed*. There is a natural partial order $\preceq$ on the configurations where $\omega \preceq \omega'$ if $\omega'$ can be obtained from $\omega$ by opening edges. An event $\mathcal{A}$ is *increasing* if for any $\omega \preceq \omega'$ it holds that $\omega \preceq \omega'$ implies $\omega' \in \mathcal{A}$. Let further $k(\omega)$ be the number of clusters of vertices of the configuration $\omega$.

The random cluster model with *free boundary conditions* $\phi_G^0$ is a percolation measure on the finite graph $G_{ghost} = (V \cup \{\mathrm{g}\}, E \cup E_g)$ such that for every $\omega \in \{0,1\}^{E \cup E_g}$

$$\phi_G^0(\omega) \propto 2^{k(\omega)} \prod_{e \in E \cup E_g} \frac{p_e}{1 - p_e}$$

where $p_e = 1_{\{e \text{ is internal and open}\}} (1 - \exp(-2\beta)) + 1_{\{e \text{ is ghost edge and open}\}} (1 - \exp(-2\beta h)) + \frac{1}{2} 1_{\{e \text{ closed}\}}$. In what follows, we will consider the *free* random cluster model on some finite subsets $\Lambda$ of $a\mathbb{Z}^2$ and we will denote that measure by $\phi_\Lambda^{0,a}$ at the same time fixing $\beta = \beta_c = \frac{\log(1 + \sqrt{2})}{2}$. Let $\Lambda_k(x)$ denote the box with side length $k$ around some point $x \in a\mathbb{Z}^2$ and let $\Lambda_k = \Lambda_k(0)$. Notice that $\Lambda_k$ only depends on the distance in $\mathbb{R}^2$ which is not affected when $a$ changes. Further, let $A_{n,m}(x) = \Lambda_m(x)/\Lambda_n(x)$ be the $(n,m)$ annulus around $x$ and $A_{n,m} = A_{n,m}(0)$. The random cluster model has many nice properties that we will use in what follows. Since the boundary conditions are free the random cluster model has stochastic domination in terms of the domain. This means that if $\Lambda_1 \subset \Lambda_2$ then for any increasing event $\mathcal{A}$,

$$\phi_{\Lambda_1}^{0,a}(\mathcal{A}) \leq \phi_{\Lambda_2}^{0,a}(\mathcal{A}). \tag{MON}$$

Further, the (FKG)-inequality [7, Theorem 1.6] states that for increasing events $\mathcal{A}, \mathcal{B}$ then

$$\phi_\Lambda^{0,a}(\mathcal{A} \cap \mathcal{B}) \geq \phi_\Lambda^{0,a}(\mathcal{A})\phi_\Lambda^{0,a}(\mathcal{B}). \tag{FKG}$$

We note that the 1-arm exponent for the random cluster model [37, Lemma 5.4] is given by

$$C_1 a^{\frac{1}{8}} \leq \phi_{\Lambda_1}^{0,a}(0 \leftrightarrow \partial\Lambda_1) \leq C_2 a^{\frac{1}{8}}. \tag{1-arm}$$

The following result was proven in [1] and it will also prove useful for us.

**Lemma 2.2** *([1], Proposition 1) Suppose that configuration of internal edges $\omega$ has clusters $C_1, \ldots, C_n$. Then*

$$\phi_{\Lambda_3}^{0,a}(C_i \leftrightarrow \mathrm{g}|\omega) = \tanh(ha^{\frac{15}{8}}|C_i|)$$

*and the events $\{C_i \leftrightarrow \mathrm{g}\}$ given $\omega$ are mutually independent.*

Finally, we state a connection between the random currents with sources and the random cluster model.

**Theorem 2.3** *([2], Theorem 3.2) Let $\{X(e)\}_{e \in E}$ be independent Bernoulli percolation with parameter $(1 - \exp(-\beta_e))$ with $\beta_e = \beta$ for $e \in E$ and $\beta_e = \beta h$ for $e \in E_\mathrm{g}$. Then define for each $e \in E \cup E_\mathrm{g}$ the configuration*

$$\omega(e) = \max\{\hat{\mathbf{n}}(e), X(e)\}.$$

*where $\hat{\mathbf{n}}$ has the law of $\hat{\mathbb{P}}^{\{x,y\}}$ the traced random current with sources $\partial\mathbf{n} = \{x,y\}$. Then $\omega$ has the law of $\phi_G^0(\cdot \mid x \leftrightarrow y)$ which is the random cluster measure conditioned on $\{x \leftrightarrow y\}$. Hence, if $\mathcal{A}$ is a decreasing event then*

$$\hat{\mathbb{P}}^{\{x,y\}}(\mathcal{A}) \geq \phi_G^0(\mathcal{A} \mid x \leftrightarrow y).$$

A key part in our result is the backbone exploration which we turn to next.

**Backbone exploration**

Let us first define the partially explored backbone. Suppose that $\mathbf{n}$ is some current with sources $\partial\mathbf{n} = \{x,y\}$. Then there is a path between $x$ and $y$ with $\mathbf{n}(e)$ odd for all edges $e$ along the path. The backbone is an algorithmic way of step-by-step constructing such a path starting from $x$ until it hits some set of vertices $A \supset \{\mathrm{g}, y\}$ [1]. To do that, we define the sets of (explored) edges $\emptyset = S_0 \subset S_1 \ldots$ inductively. For each $i \geq 0$ the set $S_i$ is defined in such a way that $\mathbf{n}$ restricted to $S_i$ has sources $\{x\} \triangle \{u_i\}$ for some vertex $u_i$. We will say that the backbones path up to step $i$ is $x = u_0, u_1, \ldots u_i$. If $u_i \in A$, let $S_{i+1} = S_i$ (and hence $S_k = S_i$ for all $k \geq i$).

---

[1]In the constructions in the litterature the set corresponding to $A$ usually does not necessarily contain the ghost, but for our purpose in this paper we include it.

Figure 1: The situation in the backbone exploration with the incoming edge $e_{u_{i-1},u_i}$ coloured blue. The vertices $v_R, v_L, v_S$ are respectively to the right, left and straight of $u_i$ with respect to the incoming edge. As usual g denotes the ghost.

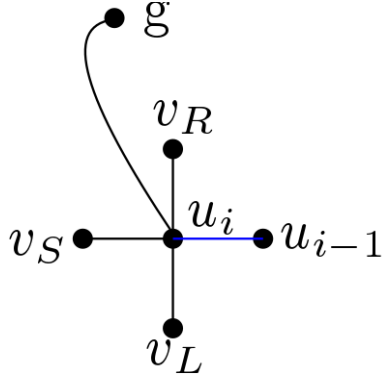

If $u_i \notin A$ we continue as follows. If $i = 0$ we consider the five edges incident to $u_0 = x$. Order them as $e_0, e_1, \ldots, e_4$ with $e_0 = e_{xg}$ and the other edges in arbitrary order. Since $x$ is a source of $\mathbf{n}$ there is at least one $i$ such that $\mathbf{n}(e_i)$ is odd. Let $k$ be the least such $i$ and let $S_1 = \{e_0, \ldots e_k\}$. Then $u_1$ is such that $e_k = e_{u_0 u_1}$. In words, the backbone explored the edges $\{e_0, \ldots e_k\}$ and walked to the vertex $u_1$.

For $i \geq 1$ we call the edge $e_{u_{i-1},u_i}$ the *incoming* edge to the vertex $u_i$. We can define an order on the remaining edges such that $(e_0, e_1, e_2, e_3) = (e_{u_i g}, e_{u_i v_R}, e_{u_i v_L}, e_{u_i v_S})$ where $e_{u_i v_R}, e_{u_i v_L}, e_{u_i v_S}$ denote the edges that are right, left and straight with respect to the incoming edge. See also Figure 1.

Now, let $k$ be the least $i$ such that $\mathbf{n}(e_i)$ is odd and $e_i \notin S_i$. Notice that since $u_i \notin A$ and $\partial \mathbf{n} = \{x, y\}$ there is always at least one such $i$. Define $S_{i+1} = S_i \cup \{e_0, \ldots e_k\}$. Then $u_{i+1}$ is such that $e_k = e_{u_i, u_{i+1}}$. In words, the backbone walks on edges $e$ of odd $\mathbf{n}(e)$ exploring in each step first the edge to the ghost and after that the edges right, left and straight with respect to the incoming edge in that order. The backbone path is the path of explored vertices $x = u_0, u_1, \ldots$ and the explored backbone in step $i$ is $S_i$.

This sequence $\{S_i\}_{i \in \mathbb{N}}$ stabilizes after a finite number of steps and we call the terminating set the *backbone starting from $x$ explored up to $A$* and denote it by $\bar{\gamma}_{x,A}(\mathbf{n})$. There is a path from $x$ to $A$ along the vertices $x = u_0, u_1, \ldots, u_{\text{end}}$ with $u_{\text{end}} \in A$ such that every edge $e$ in the path obeys that $e \in \bar{\gamma}_{x,A}(\mathbf{n})$ and has $\mathbf{n}(e)$ odd. We call this path $\gamma_{x,A}(\mathbf{n})^2$. The vertex $u_{\text{end}}$ we call $\gamma_{x,A}^{\text{end}}(\mathbf{n})$. If $u_{\text{end}}$ is the ghost g we say that *the backbone hits the ghost.*

In what follows, we will work with events of the form $\mathcal{Q} = \{F = \bar{\gamma}_{0,\Gamma}(\mathbf{n})\}$ where $\Gamma \supset \{g, x\}$ is a set of vertices, and $F$ is a set of edges. Notice that by construction we can tell whether the explored backbone is $F$ by looking only at the edges in $F$ which means that $1_{\mathcal{Q}}(\mathbf{n}) = 1_{\mathcal{Q}}(\mathbf{n}_F)$ where $\mathbf{n}_F$ is the current restricted to the set $F$.

The partial backbone exploration is useful because of the following Markov property.

---

[2] Note that if $h = 0$ then $\bar{\gamma}_{x,\{g,y\}}(\mathbf{n})$ explores some edges in and around the path $\gamma_{x,\{g,y\}}(\mathbf{n})$ from $x$ to $y$ where $\mathbf{n}(e)$ is odd for all traversed edges.

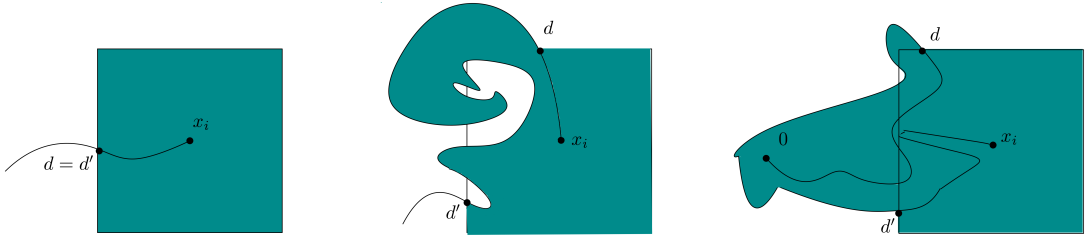

Figure 2: The explored backbone sketched in black together with the points $d$ and $d'$ in three different cases. The domain $D_i$ coloured green. Note that in the first case $d = d'$ and that in the last picture the position of 0 is not to scale.

**Theorem 2.4** *Let* $\Gamma \supset \{g, x\}$ *be a set of vertices,* $F$ *be a set of edges,* $\mathcal{Q} = \{F = \bar{\gamma}_{0,\Gamma}(\mathbf{n})\}$ *and on the event* $\mathcal{Q}$ *let* $\tilde{x} = \gamma_{0,\Gamma}^{end}(\mathbf{n})$ *be the unique vertex in* $\Gamma$ *connected to* $x$ *in* $\gamma_{0,\Gamma}(\mathbf{n})$*. Let* $\mathcal{A}$ *be any event such that* $1_{\mathcal{A}}(\mathbf{n}_{\Lambda})1_{\mathcal{Q}}(\mathbf{n}_{\Lambda}) = 1_{\mathcal{A}}(\mathbf{n}_{\Lambda/F})1_{\mathcal{Q}}(\mathbf{n}_F)$*. Then whenever* $\mathbb{P}_{\Lambda}^{\{0,x\}}(\mathcal{Q}) > 0$ *it holds that*

$$\mathbb{P}_{\Lambda}^{\{0,x\}}(\mathcal{A} \mid \mathcal{Q}) = \mathbb{P}_{\Lambda \setminus F}^{\{\tilde{x},x\}}(\mathcal{A}).$$

*Proof.* That $\mathbf{n} \in \mathcal{Q}$ means that the explored backbone of $\mathbf{n}$ up to $\Gamma$ is $F$. Thus, $F$ is the terminating set of the sequence $\{S_i\}_{i \in \mathbb{N}}$. Thus, on the event $\mathcal{Q}$ the current $\mathbf{n}$ restricted to $F$ must have sources $\partial \mathbf{n}_F = \{0\} \triangle \{u_{\text{end}}\} = \{0, \tilde{x}\}$. Since $\mathbf{n} = \mathbf{n}_{\Lambda \setminus F} + \mathbf{n}_F$ it holds that

$$\{0, x\} = \partial \mathbf{n} = \partial \mathbf{n}_{\Lambda \setminus F} \triangle \partial \mathbf{n}_F = \partial \mathbf{n}_{\Lambda \setminus F} \triangle \{0, \tilde{x}\}.$$

So for $\mathbf{n} \in \mathcal{Q}$ then $\partial \mathbf{n}_{\Lambda \setminus F} = \{\tilde{x}, x\}$. The map $\mathbf{n} \mapsto (\mathbf{n}_F, \mathbf{n}_{\Lambda \setminus F})$ is a bijection from $\{\mathbf{n} \in \mathcal{Q} \mid \partial \mathbf{n} = \{0, x\}\}$ to $\{(\mathbf{n}_F, \mathbf{n}_{\Lambda \setminus F}) \mid \mathbf{n}_F + \mathbf{n}_{\Lambda \setminus F} \in \mathcal{Q}, \partial \mathbf{n}_F = \{0, \tilde{x}\}, \partial \mathbf{n}_{\Lambda \setminus F} = \{\tilde{x}, x\}\}$ with inverse $(\mathbf{n}_F, \mathbf{n}_{\Lambda \setminus F}) \mapsto \mathbf{n}_F + \mathbf{n}_{\Lambda \setminus F}$. Thus, for any function $f : \mathbb{N}_0^{E \cup E_g} \to \mathbb{R}$ it holds that

$$\sum_{\partial \mathbf{n} = \{0, x\}} f(\mathbf{n}) 1_{\mathcal{Q}}(\mathbf{n}) = \sum_{\substack{\partial \mathbf{n}_F = \{0, \tilde{x}\} \\ \partial \mathbf{n}_{\Lambda \setminus F} = \{\tilde{x}, x\}}} f(\mathbf{n}) 1_{\mathcal{Q}}(\mathbf{n}).$$

Since $w(\mathbf{n}_F + \mathbf{n}_{\Lambda \setminus F}) = w(\mathbf{n}_F) \cdot w(\mathbf{n}_{\Lambda \setminus F})$ and the fact that $\mathcal{Q}$ only depends on edges in $F$,

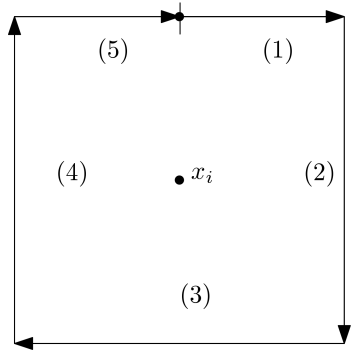

Figure 3: Sketch the order $\preceq$ on $\partial\Lambda_1(x_i)$ used when defining $D_i$. Two points $a, b \in \partial\Lambda_1(x_i)$ have $a \preceq b$ if the number of the segment of $a$ is smaller than the number of the segment of $b$ or if $a$ and $b$ are inside the same segment and $a$ is earlier than $b$ with respect to the arrow on that segment.

the double sum below factorizes and

$$
\begin{aligned}
\mathbb{P}_\Lambda^{\{0,x\}}(\mathcal{A} \mid \mathcal{Q}) &= \frac{\sum_{\partial \mathbf{n} = \{0,x\}} w(\mathbf{n}) 1_{\mathcal{A}}(\mathbf{n}) 1_{\mathcal{Q}}(\mathbf{n})}{\sum_{\partial \mathbf{n} = \{0,x\}} w(\mathbf{n}) 1_{\mathcal{Q}}(\mathbf{n})} \\
&= \frac{\sum_{\partial \mathbf{n}_F = \{0,\tilde{x}\}, \partial \mathbf{n}_{\Lambda \setminus F} = \{\tilde{x},x\}} w(\mathbf{n}_F + \mathbf{n}_{\Lambda \setminus F}) 1_{\mathcal{A}}(\mathbf{n}) 1_{\mathcal{Q}}(\mathbf{n})}{\sum_{\partial \mathbf{n}_F = \{0,\tilde{x}\}, \partial \mathbf{n}_{\Lambda \setminus F} = \{\tilde{x},x\}} w(\mathbf{n}_F + \mathbf{n}_{\Lambda \setminus F}) 1_{\mathcal{Q}}(\mathbf{n})} \\
&= \frac{\sum_{\partial \mathbf{n}_F = \{0,\tilde{x}\}} w(\mathbf{n}_F) 1_{\mathcal{Q}}(\mathbf{n}_F) \sum_{\partial \mathbf{n}_{\Lambda \setminus F} = \{\tilde{x},x\}} w(\mathbf{n}_{\Lambda \setminus F}) 1_{\mathcal{A}}(\mathbf{n}_{\Lambda / F})}{\sum_{\partial \mathbf{n}_F = \{0,\tilde{x}\}} w(\mathbf{n}_F) 1_{\mathcal{Q}}(\mathbf{n}_F) \sum_{\partial \mathbf{n}_{\Lambda \setminus F} = \{\tilde{x},x\}} w(\mathbf{n}_{\Lambda \setminus F})} \\
&= \frac{\sum_{\partial \mathbf{n}_{\Lambda \setminus F} = \{\tilde{x},x\}} w(\mathbf{n}_{\Lambda \setminus F}) 1_{\mathcal{A}}(\mathbf{n}_{\Lambda \setminus F})}{\sum_{\partial \mathbf{n}_{\Lambda \setminus F} = \{\tilde{x},x\}} w(\mathbf{n}_{\Lambda \setminus F})} = \mathbb{P}_{\Lambda \setminus F}^{\{\tilde{x},x\}}(\mathcal{A}).
\end{aligned}
$$

$\square$

## 3 Main result

We now prove Theorem 1.2 given a result that we then prove later.

Before starting the proof, we go through some notation that we will use throughout the main section. Let $n$ be such that $x \in A_{9n,9(n+1)}$. In what follows, we will consider the case $|x| \geq 36$ which means that $n \geq 4$. In the proof of Theorem 1.2 we tie it together with the case $|x| < 36$. We also let $\Lambda$ be any box which contains $\Lambda_{9(n+1)}$. Everything we prove will be independent of this $\Lambda$. Later we let $\Lambda \Uparrow \mathbb{Z}^2$.

We will explore the backbone partially in steps up to the annuli $A_{9i,9(i+1)}$. Suppose that in this exploration the backbone does not hit the ghost g, which we can assume in our application. Then define $x_i = \gamma_{0,\Lambda_{9i}^c}^{\text{end}}(\mathbf{n}) \in A_{9i,9(i+1)}$. Thus, $x_i$ is random variable corresponding to the first vertex the backbone hits in the $i$-th annulus of the form $A_{9i,9(i+1)}$, see also Figure 5.

Further, to ease notation we let $\bar{\gamma}_i = \bar{\gamma}_{0,\Lambda_{9i}^c}(\mathbf{n})$. Thus, $\bar{\gamma}_i$ is the set of edges explored until the backbone hits the $i$-th annulus. Similarly, to ease notation let $\gamma_i = \gamma_{0,\Lambda_{9i}^c}(\mathbf{n})$.

Thus, $\gamma_i$ is the path $u_0, \ldots, u_{\mathrm{end}} = x_i = \gamma_{0,\Lambda_{9i}^c}^{\mathrm{end}}(\mathbf{n})$ explored until the backbone hits the $i$-th annulus. Let $\mathcal{Q}_{\bar{\gamma}_i}$ be the event that the backbone explored is $\bar{\gamma}_i$.

A technical detail is that to account for removing the magnetic field in the box of size 1 around $x$ we explore the backbone partially also from $x$ until it leaves the box of size 2. Denote the explored backbone $\bar{\gamma}_{x,\Lambda_2(x)^c}(\mathbf{n})$ by $\bar{\gamma}_x$, name the first point hit outside $\Lambda_2(x)$ by $\tilde{x} = \gamma_{x,\Lambda_2(x)^c}^{\mathrm{end}}(\mathbf{n})$ and let the event that the explored backbone is $\bar{\gamma}_x$ be denoted $\mathcal{Q}_{\bar{\gamma}_x}$.

The set $\bar{\gamma}_i$ contains the path $\gamma_i$ from 0 to $x_i$. Thus, $\bar{\gamma}_i$ will intersect $\partial \Lambda_1(x_i)$ one or more times. Since $x_i$ is the first vertex of the annulus $A_{9i,9(i+1)}$ to be intersected the set $\bar{\gamma}_i \cap \partial\Lambda_1(x_i)$ is contained within one half of $\partial\Lambda_1(x_i)$. Let $d, d'$ denote the points in $\bar{\gamma}_i \cap \partial\Lambda_1(x_i)$ that are most clockwise and anticlockwise with respect to some way of walking around $\partial\Lambda_1(x_i)$, see also Figure 2.

More formally, we consider an order $\preceq$ of the points in $\partial\Lambda_1(x_i)$ and then define $d, d'$ to be the minimal and maximal element of $\bar{\gamma}_i \cap \partial\Lambda_1(x_i)$ with respect to this ordering. Let us define the order in the case where $x_i$ is in the right side of the annulus (which is the case $x_1, x_2, x_3$ on Figure 5). Generalising to the other cases is straightforward. For the defininition, we split $\partial\Lambda_1(x_i)$ and define the order with respect to the segments and arrows as shown on Figure 3.

Now, define $\Omega_i = \{v \in \partial\Lambda_1(x_i) \mid d' \preceq v \preceq d\}$ and let $\tilde{\Omega}_i = \partial\Lambda_1(x_i) \backslash \Omega_i$. Let $\Sigma_i$ be the graph obtained by removing $\bar{\gamma}_i \cup \bar{\gamma}_x$ from $\Lambda$. Then we can define the domain $D_i$ to be the connected component of $x_i$ in the graph induced by the vertices of $\Sigma_i$ without $\tilde{\Omega}_i$. See also Figure 2. In the following claim we show how our order of exploration with respect to the incoming edge implies that $D_i$ is contained in $\Lambda_{9(i+1)}$.

**Claim 3.1** *The set $D_i \subset \Lambda_{9(i+1)}$ and it only depends on the current $\mathbf{n}_{\Lambda_{9i}}$.*

*Proof.* Since the vertex $d$ is explored by the backbone it is either on the backbone path $\gamma_i$ in which case we let $v = d$. Otherwise, there is an edge $e_{dv}$ from $d$ to a vertex on the backbone path that we call $v$. Similarly, we can define a vertex $v'$ taking $d'$ as the starting point. Let $P_i$ be the subpath of $\gamma_i$ which goes either from $v$ to $v'$ or from $v'$ to $v$ and extended by the edges $\{e_{dv}\}$ and/or $\{e'_{d'v'}\}$ if $d, d'$ are not on the backbone path. By construction the path $P_i$ is edge self-avoiding, but we do not know that $P_i$ is vertex self-avoiding and hence non self-intersecting. However, due to the way we explore the edges of the backbone with respect to the incoming edge if there is a vertex which is hit by the backbone path twice (i.e. $u_i = u_j$ for some $i \neq j$) then the backbone path must turn $90°$ twice at that vertex. This means that we can deform the path slightly to be non-intersecting (see Figure 4).

Since $P_i$ is a path between $d$ and $d'$ and further $\tilde{\Omega}_i$ is also a path between $d$ and $d'$ along $\Lambda_1(x_i)$, by definition of $d$ and $d'$, $P_i$ and $\tilde{\Omega}_i$ do not intersect. Thus, if we glue them together then $P_i \cup \tilde{\Omega}_i$ is a closed non-intersecting path, which therefore encloses a domain $Q_i \subset \Lambda_{9(i+1)}$. Now, assume for contradiction that $\delta : x_i \to Q_i^c$ is a path. Since we have removed all the vertices in $\tilde{\Omega}_i$ including $d$ and $d'$ it is impossible for $\delta$ to exit $Q_i$ through a vertex in $\tilde{\Omega}_i$. Since in the backbone exploration we remove the explored edges on the backbone path all remaining vertices have degree at most 2 in $\Sigma_i \backslash \tilde{\Omega}_i$. It is only possible to have a path exiting $Q_i$ if it crosses the path $P_i$ through a vertex of degree 2. The way we explore the backbone implies that if two edges remain they must have an angle of $90°$. Therefore, the remaining edges do not cross $P_i$ and hence there is no such path $\delta$. Since $D_i$ is the connected component of $x_i$ it holds that $D_i \subset Q_i$. □

Figure 4: Sketch of the case where the backbone explores the same vertex twice as in the proof of Claim 3.1. Suppose that the backbone entered and left the vertex along edges of the same colour (orange or blue). Since we in the backbone exploration always explore turning to the right with respect to the incoming edge first the backbone path can never cross itself. Thus, a non-crossing deformation as shown is always possible.

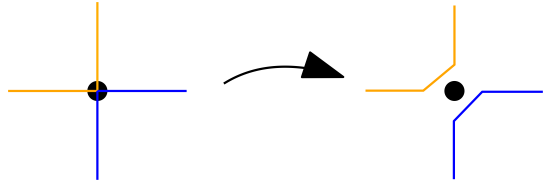

To finish the setup we finally define

$$\mathcal{R}_i^* = \{\mathbf{n} \mid \partial\Lambda_2(x_i) \overset{A_{2,4}(x_i) \ \backslash (D_i \cup \bar{\gamma}_i)}{\longleftrightarrow\!\!\!\!/} \partial\Lambda_4(x_i)\}.$$

Notice that if $\hat{\mathbf{n}}_1 = \hat{\mathbf{n}}_2$ for two currents $\mathbf{n}_1, \mathbf{n}_2$ and $\mathbf{n}_1 \in \mathcal{R}_i^*$ then $\mathbf{n}_2 \in \mathcal{R}_i^*$. Thus, $\mathcal{R}_i^*$ only depends on the traced and not on the full current. Define the corresponding connection event either for the traced current or for the random cluster measure by

$$\hat{\mathcal{R}}_i^* = \{\partial\Lambda_2(x_i) \overset{A_{2,4}(x_i) \ \backslash (D_i \cup \bar{\gamma}_i)}{\longleftrightarrow\!\!\!\!/} \partial\Lambda_4(x_i)\} \subset \{0,1\}^{A_{2,4}(x_i) \ \backslash (D_i \cup \bar{\gamma}_i)}.$$

Notice that $\hat{\mathcal{R}}_i^* = \{\hat{\mathbf{n}} \mid \mathbf{n} \in \mathcal{R}_i^*\}$. Hence, it holds that

$$\mathbb{P}_{\Lambda \backslash (\bar{\gamma}_i \cup \bar{\gamma}_x), \vec{h}}^{\{x_i, \tilde{x}\}} (\mathcal{R}_i^*) = \hat{\mathbb{P}}_{\Lambda \backslash (\bar{\gamma}_i \cup \bar{\gamma}_x), \vec{h}}^{\{x_i, \tilde{x}\}} \left( \hat{\mathcal{R}}_i^* \right). \tag{2}$$

The following proposition will yield the main result given Proposition 3.3 below. The idea of the proof is first to use the backbone exploration and then show that for every macroscopic step, with a strictly positive probability there is a connection to the ghost. To get the correct front factor $a^{\frac{1}{4}}$, we also partially explore the backbone also from the end around $x$ until $\tilde{x} = \gamma_{x, \Lambda_2(x)^c}^{\text{end}}(\mathbf{n})$.

**Proposition 3.2** *Suppose that, for all $\tilde{x} \in \partial\Lambda_2(x)$, all realisations of the backbone $\bar{\gamma}_x$ from $x$ to $\tilde{x}$, all $1 \le i \le n$ and realisations of the backbone from $0$ to $x_i$ denoted $\bar{\gamma}_i$ and $a \le 1$,*

$$\mathbb{P}_{\Lambda \backslash (\bar{\gamma}_i \cup \bar{\gamma}_x), \vec{h}}^{\{x_i, \tilde{x}\}} (\mathcal{R}_i^*) \ge c$$

*uniformly in any (x-dependent) $\Lambda$ sufficiently large. Then*

$$\langle \sigma_0 ; \sigma_x \rangle_{a,h} \le C a^{\frac{1}{4}} \exp\left( -M(h) \, |x| \right)$$

*for $|x| \ge 36$ and where $M(h)$ does not depend on $a$.*

*Proof.* Let $\mathcal{H}$ be the event that the backbone explored from $0$ hits $x$ ( i.e. does not hit the ghost). Notice that $\{\tilde{x} = \text{g}\} \cap \mathcal{Q}_{\bar{\gamma}_x} \cap \mathcal{H} = \emptyset$ so when we condition $\mathcal{H}$ on all possible events $\mathcal{Q}_{\bar{\gamma}_x}$ we can omit those where $\tilde{x} = \text{g}$. In other words, if $\tilde{x} = \text{g}$ then the backbone explored from 0 would necessarily hit the ghost since after the partial backbone exploration then g and 0 would be the only two vertices with odd degree.

Further, $1_{\mathcal{H}}(\mathbf{n}_\Lambda)1_{\mathcal{Q}_{\bar{\gamma}_x}}(\mathbf{n}_\Lambda) = 1_{\mathcal{H}}(\mathbf{n}_{\Lambda\setminus\bar{\gamma}_x})1_{\mathcal{Q}_{\bar{\gamma}_x}}(\mathbf{n}_{\bar{\gamma}_x})$ so by the backbone exploration Theorem 2.4

$$\mathbb{P}^{\{0,x\}}_{\Lambda,\vec{h}}(\mathcal{H}) = \sum_{\bar{\gamma}_x} \mathbb{P}^{\{0,x\}}_{\Lambda,\vec{h}}(\mathcal{H} \mid \mathcal{Q}_{\bar{\gamma}_x})\mathbb{P}^{\{0,x\}}_{\Lambda,\vec{h}}(\mathcal{Q}_{\bar{\gamma}_x}) = \sum_{\bar{\gamma}_x}\mathbb{P}^{\{0,\tilde{x}\}}_{\Lambda\setminus\bar{\gamma}_x,\vec{h}}(\mathcal{H})\mathbb{P}^{\{0,x\}}_{\Lambda,\vec{h}}(\mathcal{Q}_{\bar{\gamma}_x}) \tag{3}$$

where we from now on assume that $\tilde{x} \neq$ g which means that $\tilde{x} \in \partial\Lambda_2(x)$.

For each $1 \leq i \leq n$ let $\mathcal{G}_i$ be the event that the backbone hits the annulus $A_{9i,9(i+1)}$ before hitting the ghost. Given a current configuration in $\mathcal{G}_i$ we know that the vertices $x_j$ exist for $1 \leq j \leq i$. Further, if we define $\mathcal{G}_0 = \{\mathbf{n} \mid \partial\mathbf{n} = \{0,x\}\}$ then $\mathcal{G}_{i+1} \subset \mathcal{G}_i$ for each $0 \leq i \leq n-1$ as well as $\{0 \not\leftrightarrow \text{g}\} \subset \mathcal{H} \subset \mathcal{G}_n$. Since $\mathbb{P}^{\{0,\tilde{x}\}}_{\Lambda\setminus\bar{\gamma}_x,\vec{h}}(\mathcal{H}) \leq \mathbb{P}^{\{0,\tilde{x}\}}_{\Lambda\setminus\bar{\gamma}_x,\vec{h}}(\mathcal{G}_n)$ and bounding $\mathbb{P}^{\{0,\tilde{x}\}}_{\Lambda\setminus\bar{\gamma}_x,\vec{h}}(\mathcal{H})$ uniformly in $\bar{\gamma}_x$ bounds $\mathbb{P}^{\{0,x\}}_{\Lambda,\vec{h}}(\mathcal{H})$ through (3), it means that to bound $\mathbb{P}^{\{0,x\}}_{\Lambda,\vec{h}}(0 \not\leftrightarrow \text{g})$ it suffices to show exponential decay of $\mathbb{P}^{\{0,\tilde{x}\}}_{\Lambda\setminus\bar{\gamma}_x,\vec{h}}(\mathcal{G}_n)$ uniformly in $\bar{\gamma}_x$. Notice that given $\mathcal{Q}_{\bar{\gamma}_{i-1}}$ then $\mathcal{G}_i$ depends only on edges in $\Lambda\setminus(\bar{\gamma}_{i-1} \cup \bar{\gamma}_x)$ so

$$1_{\mathcal{G}_i}(\mathbf{n}_{\Lambda\setminus\bar{\gamma}_x})1_{\mathcal{Q}_{\bar{\gamma}_{i-1}}}(\mathbf{n}_{\Lambda\setminus\bar{\gamma}_x}) = 1_{\mathcal{G}_i}(\mathbf{n}_{\Lambda\setminus(\bar{\gamma}_{i-1}\cup\bar{\gamma}_x)})1_{\mathcal{Q}_{\bar{\gamma}_{i-1}}}(\mathbf{n}_{(\bar{\gamma}_{i-1}\cup\bar{\gamma}_x)})$$

and by the backbone exploration Theorem 2.4

$$\mathbb{P}^{\{0,\tilde{x}\}}_{\Lambda\setminus\bar{\gamma}_x,\vec{h}}(\mathcal{G}_n) = \left[\prod_{i=1}^n \mathbb{P}^{\{0,\tilde{x}\}}_{\Lambda\setminus\bar{\gamma}_x,\vec{h}}(\mathcal{G}_i \mid \mathcal{G}_{i-1})\right]\mathbb{P}^{\{0,\tilde{x}\}}_{\Lambda\setminus\bar{\gamma}_x,\vec{h}}(\mathcal{G}_0)$$

$$= \prod_{i=1}^n \sum_{\mathcal{Q}_{\bar{\gamma}_{i-1}}\in\mathcal{G}_{i-1}} \mathbb{P}^{\{0,\tilde{x}\}}_{\Lambda\setminus\bar{\gamma}_x,\vec{h}}(\mathcal{G}_i \mid \mathcal{Q}_{\bar{\gamma}_{i-1}})\mathbb{P}^{\{0,\tilde{x}\}}_{\Lambda\setminus\bar{\gamma}_x,\vec{h}}(\mathcal{Q}_{\bar{\gamma}_{i-1}} \mid \mathcal{G}_{i-1})$$

$$\leq \prod_{i=1}^n \sum_{\mathcal{Q}_{\bar{\gamma}_{i-1}}\in\mathcal{G}_{i-1}} \mathbb{P}^{\{x_{i-1},\tilde{x}\}}_{\Lambda\setminus(\bar{\gamma}_{i-1}\cup\bar{\gamma}_x),\vec{h}}(\mathcal{G}_i)\mathbb{P}^{\{0,\tilde{x}\}}_{\Lambda\setminus\bar{\gamma}_x,\vec{h}}(\mathcal{Q}_{\bar{\gamma}_{i-1}} \mid \mathcal{G}_{i-1}) \leq (1-c)^{n-2}$$

where in the last step we used $\mathbb{P}^{\{x_{i-1},\tilde{x}\}}_{\Lambda\setminus(\bar{\gamma}_{i-1}\cup\bar{\gamma}_x),\vec{h}}(\mathcal{G}_i) \leq \mathbb{P}^{\{x_{i-1},\tilde{x}\}}_{\Lambda\setminus(\bar{\gamma}_{i-1}\cup\bar{\gamma}_x),\vec{h}}(\mathcal{R}^{*c}_{i-1}) \leq 1-c$. Here, the first of these inequalities follow since on the event $\mathcal{G}_i \subset \mathcal{R}^{*c}_{i-1}$ which again follows by contraposition since on the event $\mathcal{R}_{i-1}$ the backbone must hit the ghost before hitting the annulus $A_{9i,9(i+1)}$. Now, by the remarks in the beginning of the proof it follows from Proposition 2.1 that

$$\langle\sigma_0;\sigma_x\rangle_{\Lambda,h,a} \leq Ca^{\frac{1}{4}}\mathbb{P}^{\{0,x\}}_{\Lambda,\vec{h}}(0 \not\leftrightarrow \text{g}) \leq Ca^{\frac{1}{4}}(1-c)^{n-2} = Ca^{\frac{1}{4}}\exp(-M(h)\,|x|).$$

The inequality passes to the infinite volume limit since the constants are independent of $\Lambda$. $\qquad\square$

Next, we will move from the random current event $\mathcal{R}^*_i$ to the traced current or random cluster event $\hat{\mathcal{R}}^*_i$. In the remaining section we will prove the following proposition which only concerns the random cluster model. Here and in all of the following by $\geq c$ we mean larger than a (possibly different each time) strictly positive constant which is uniform in $a \leq 1$, $h$ and the explored backbones $\bar{\gamma}_i$ and $\bar{\gamma}_x$ .

Figure 5: Sketch of the overall approach. Either the backbone goes through the ghost or we can define the $x_i$ as shown. Later, we show that it is probable that $x_i$ is connected to the ghost within each of the black squares.

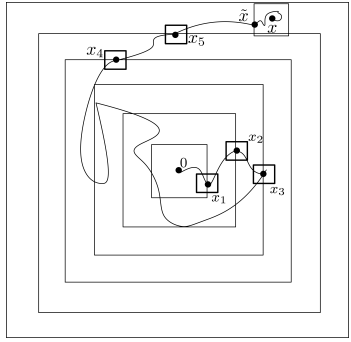

**Proposition 3.3** *There is a $h_0 > 0$ such that for all realisations of the backbone $\bar{\gamma}_i$ up to $x_i$, all $\bar{\gamma}_x$ explored until $\tilde{x} \in \partial \Lambda_2(x)$ and all $a \leq 1, h < h_0$ then*

$$\phi^{0,a}_{\Lambda \setminus (\bar{\gamma}_i \cup \bar{\gamma}_x), \tilde{h}} \left( \hat{\mathcal{R}}_i^* \mid x_i \leftrightarrow \tilde{x} \right) \geq c$$

*uniformly in any $\Lambda$ sufficiently large around $x$.*

Now, collecting the results we can prove the main theorem assuming Proposition 3.3 which is in the language of the random cluster model where many more tools are available than for random currents most notably the RSW.

*Proof of Theorem 1.2.* By (2), the monotone coupling from Theorem 2.3 along with the fact that $\hat{\mathcal{R}}_i^*$ is a decreasing event and Proposition 3.3 we get that

$$\mathbb{P}^{\{x_i, \tilde{x}\}}_{\Lambda \setminus (\bar{\gamma}_i \cup \bar{\gamma}_x), \tilde{h}} \left( \mathcal{R}_i^* \right) = \hat{\mathbb{P}}^{\{x_i, \tilde{x}\}}_{\Lambda \setminus (\bar{\gamma}_i \cup \bar{\gamma}_x), \tilde{h}} (\hat{\mathcal{R}}_i^*) \geq \phi^{0,a}_{\Lambda \setminus (\bar{\gamma}_i \cup \bar{\gamma}_x), \tilde{h}} \left( \hat{\mathcal{R}}_i^* \mid x_i \leftrightarrow \tilde{x} \right) \geq c.$$

Thus, we can apply Proposition 3.2. In Proposition 3.3 we only have the result for sufficiently small $h$, but this suffices by the GHS-inequality [36]. To account for the constraint $|x| \geq 36$ in Proposition 3.2 and get the correct front factor notice that from the GHS inequality and Proposition 5.5 in [37], for some $B > 0$, it holds for all $x \in a\mathbb{Z}^2$ that

$$\langle \sigma_0; \sigma_x \rangle_{a,h} \leq \langle \sigma_0; \sigma_x \rangle_{a,0} \leq B \left( \frac{a}{|x|} \right)^{\frac{1}{4}}.$$

Using that for $|x| \geq K(h)$ it holds that

$$\langle \sigma_0; \sigma_x \rangle_{a,h} \leq C a^{\frac{1}{4}} \exp \left( -\frac{M(h)}{2} |x| \right) \exp \left( -\frac{M(h)}{2} |x| \right) \leq C \left( \frac{a}{|x|} \right)^{\frac{1}{4}} \exp \left( -\frac{M(h)}{2} |x| \right).$$

Now, by putting $C(h) = \max\{B, C\} \exp \left( \frac{M(h)}{2} K(h) \right)$ and $m(h) = \frac{M(h)}{2}$ our main result Theorem 1.2 follows.

$\square$

# 4    Proof of Proposition 3.3

Recall that $\Sigma_i = \Lambda \backslash (\bar{\gamma}_i \cup \bar{\gamma}_x)$. To ease the notation here and in what follows we define $\phi_{\Sigma_i} = \phi_{\Sigma_i, \bar{h}}^{0,a} = \phi_{\Lambda \backslash (\bar{\gamma}_i \cup \bar{\gamma}_x), \bar{h}}^{0,a}$. Further, for a set $\Gamma$ let $x \overset{\Gamma^+}{\leftrightarrow} y$ denote the event that $x$ and $y$ are connected in $\Gamma \cup \{g\}$ and similarly by $x \overset{\Gamma}{\leftrightarrow} y$ that $x$ and $y$ are connected in $\Gamma$ *not using the ghost*. Define the domain $T_i$ to be all points in $D_i$ as well as all points in $\Lambda_2(x_i)$ that can be reached from $x_i$ without using edges in $\bar{\gamma}_i$ or $\partial \Lambda_2(x_i)$. Further, define $\{x_i \overset{T_i^+}{\leftrightarrow} g\}$ to be the event that $x_i$ is connected within the domain $T_i$ to some vertex $v$ where the edge from $v$ to g is open. Define $\tilde{D}$ and $\tilde{T}$ similarly to $D_i$ and $T_i$ with $\tilde{x}$ instead of $x_i$. Define also $A_i = \partial \Lambda_2(x_i) \cap T_i^c$ and $\tilde{A} = \partial \Lambda_2(\tilde{x}) \cap \tilde{T}^c$.

**Proposition 4.1** *Suppose for some $1 < i < n$ that $\phi_{\Sigma_i} \left( x_i \overset{T_i^+}{\leftrightarrow} g \mid x_i \leftrightarrow A_i \right) \geq c$ and $\phi_{\Sigma_i} \left( \tilde{x} \overset{\tilde{T}^+}{\leftrightarrow} g \mid \tilde{x} \leftrightarrow \tilde{A} \right) \geq c$. Then for all $h \leq h_0$ it holds that $\phi_{\Sigma_i} \left( \hat{\mathcal{R}}_i^* \mid x_i \leftrightarrow \tilde{x} \right) \geq c$.*

We first state the recent result that the random cluster model still has the RSW property at scales up to the correlation length. Here we need the wired boundary condition $\phi^1$ which is introduced in for example [7].

**Lemma 4.2** *([3], Lemma 8.5) For any sufficiently large $C > 0$, there is an $\varepsilon > 0$ such that if $n \geq 1$ and $H \geq 0$ are such that $Hn^2 \phi_{\Lambda_n, H=0}^{0,a=1}(0 \leftrightarrow \partial \Lambda_n) \leq \varepsilon$ then*

$$\phi_{A_{n,2n}, H=0}^{1,a=1}(\Lambda_n \not\leftrightarrow \partial \Lambda_{2n}) \leq C \phi_{A_{n,2n}, H}^{1,a=1}(\Lambda_n \not\leftrightarrow \partial \Lambda_{2n}).$$

Translating the lemma into our setting yields the following lemma.

**Lemma 4.3** *There exists a $h_0 > 0$ such that for $a \leq 1$ and $h \leq h_0$ it holds that*

$$\phi_{\Sigma_i}(\hat{\mathcal{R}}_i^*) \geq c.$$

*where $c > 0$ is (as always) independent of $0 < a \leq 1, \Sigma_i$ and $0 \leq h < h_0$. Further, for any event $E$ depending only on edges in $D_i$ it holds that*

$$\phi_{\Sigma_i}(\hat{\mathcal{R}}_i^* \mid E) \geq c.$$

*Proof.* If $n = \frac{2}{a}$ and $H = a^{\frac{15}{8}} h$ then using equation (1-arm) for some $c > 0$ it holds that

$$Hn^2 \phi_{\Lambda_n, H=0}^{1,a=1}(0 \leftrightarrow \partial \Lambda_n) \leq c(an)^{\frac{15}{8}} h = c 2^{\frac{15}{8}} h \leq \varepsilon$$

which can be satisfied by choosing $h$ sufficiently small (independent of $a$). Therefore

$$\phi_{A_{2,4}, h}^{1,a}(\Lambda_2 \not\leftrightarrow \partial \Lambda_4) = \phi_{A_{\frac{2}{a}, \frac{4}{a}}, ha^{\frac{15}{8}}}^{1,1}(A_{\frac{2}{a}} \not\leftrightarrow \partial A_{\frac{4}{a}}) \geq C \phi_{A_{\frac{2}{a}, \frac{4}{a}}, h=0}^{1,1}(A_{\frac{2}{a}} \not\leftrightarrow \partial A_{\frac{4}{a}}) = C \phi_{A_{2,4}, h=0}^{1,a}(\Lambda_2 \not\leftrightarrow \partial \Lambda_4).$$

Since $\hat{\mathcal{R}}_i^*$ is decreasing it follows by (MON) and the RSW for usual rectangles [37] that

$$\phi_{\Sigma_i, \bar{h}}^{0,a}(\hat{\mathcal{R}}_i^*) \geq \phi_{\Sigma_i, h}^{0,a}(\hat{\mathcal{R}}_i^*) \geq \phi_{\Lambda, h}^{0,a}(\hat{\mathcal{R}}_i^*) \geq \phi_{\Lambda, h}^{0,a}(\Lambda_2(x_i) \not\leftrightarrow \partial \Lambda_4(x_i)) \geq \phi_{\Lambda, h}^{1,a}(\Lambda_2(x_i) \not\leftrightarrow \partial \Lambda_4(x_i))$$
$$\geq \phi_{A_{2,4}, h}^{1,a}(\Lambda_2 \not\leftrightarrow \partial \Lambda_4) \geq c \phi_{A_{2,4}, h=0}^{1,a}(\Lambda_2 \not\leftrightarrow \partial \Lambda_4) \geq c.$$

Using comparison between boundary conditions, and that because of the argument in Claim 3.1 removing all explored edges of the backbone acts as a free boundary condition it holds that

$$\phi_{\Sigma_i, \bar{h}}^{0,a}(\hat{\mathcal{R}}_i^* \mid E) \geq \phi_{\Sigma_i \setminus D_i, h}^{1 \text{ on } \partial \Lambda_1(x_i), 0 \text{ else}}(\hat{\mathcal{R}}_i^*) \geq \phi_{A_{2,4},h}^{1,a}(\Lambda_2 \not\leftrightarrow \partial \Lambda_4) \geq c.$$

$\square$

Next, we need the general result that we can do mixing also with a magnetic field at scales up to the correlation length and that we, up to constants, can decorrelate events in $T_i$ from events in $(T_i \cup \Lambda_4(x_i))^c$. Define $J_i = (T_i \cup \Lambda_4(x_i)) \setminus \bar{\gamma}_i$ and $\tilde{J} = (\tilde{T} \cup \Lambda_4(\tilde{x})) \setminus \bar{\gamma}_x$. Define also $\hat{\mathcal{R}}_\sim^*$ similarly to $\hat{\mathcal{R}}_i^*$ as an event in the vicinity of $\tilde{x}$ instead of $x_i$.

**Lemma 4.4** *(Mixing) Let $E_1, E_2$ be increasing events that only depend on edges in the boxes $T_i \cup \Lambda_2(x_i)$ and $\tilde{T} \cup \Lambda_2(\tilde{x})$ respectively, then for $1 \leq i \leq n - 2$ it holds that*

$$\phi_{\Sigma_i}(E_1 \cap E_2) \asymp \phi_{J_i, \bar{h}}^{0,a}(E_1) \phi_{\tilde{J}, \bar{h}}^{0,a}(E_2).$$

*Similarly, if $l \leq 1$, $z_1, z_2$ are such that $\Lambda_{2l}(z_1) \cap \Lambda_{2l}(z_2) = \emptyset$ and $E_1, E_2$ are increasing events that only depend on edges in the boxes $\Lambda_l(z_1), \Lambda_l(z_2)$ respectively, then*

$$\phi_{\Sigma_i}(E_1 \cap E_2) \asymp \phi_{\Lambda_{2l}(z_1) \setminus (\bar{\gamma}_i \cup \bar{\gamma}_x), \bar{h}}^{0,a}(E_1) \phi_{\Lambda_{2l}(z_2) \setminus (\bar{\gamma}_i \cup \bar{\gamma}_x), \bar{h}}^{0,a}(E_2).$$

*Proof.* We prove the first statement first. Define $E = E_1 \cap E_2$. It follows from Lemma 4.3 that

$$c\phi_{\Sigma_i}(E) \leq \phi_{\Sigma_i}(E \mid \hat{\mathcal{R}}_i^*)$$

and similarly we can condition on $\hat{\mathcal{R}}_\sim^*$. Using that closed dual paths inside the annulus give rise to monotonicity properties as free boundary conditions (which is for example proven in Lemma 11 in [1]) we obtain that

$$c\phi_{\Sigma_i}(E) \leq \phi_{\Sigma_i}(E_1, E_2 \mid \hat{\mathcal{R}}_i^*, \hat{\mathcal{R}}_\sim^*) \leq \phi_{J_i \cup \tilde{J}, \bar{h}}^{0,a}(E_1, E_2) = \phi_{J_i, \bar{h}}^{0,a}(E_1) \phi_{\tilde{J}, \bar{h}}^{0,a}(E_2).$$

Since the reverse inequality is (FKG) and (MON) the first result follows. The second assertion follows mutatis mutandis using that the estimates from the proof of Lemma 4.3 which by Lemma 4.2 also work on smaller scales and using the event $\{\partial \Lambda_l(z_i) \not\leftrightarrow \partial \Lambda_{2l}(z_i)\}$ for $i = 1, 2$ instead of $\hat{\mathcal{R}}_i^*$ and $\hat{\mathcal{R}}_\sim^*$. $\square$

With the lemmas proven we continue to the proof of the Proposition 4.1. In Figure 6 we sketch the situation we would like to prove is the case. The idea being that if we condition on $x_i \leftrightarrow \tilde{x}$ possibly using the ghost, then we can only prove that the event $\hat{\mathcal{R}}_i^*$ happens with constant probability if we can make sure that the connection indeed happens using the ghost. The connection to the ghost we get from the assumption if $x_i$ and or $\tilde{x}$ connects to a point at a distance of at least 2.

*Proof of Proposition 4.1.* First, notice that

$$\phi_{\Sigma_i}\left(\hat{\mathcal{R}}_i^* \mid x_i \overset{\Sigma_i^+}{\leftrightarrow} \tilde{x}\right) \geq \phi_{\Sigma_i}\left(\hat{\mathcal{R}}_i^* \mid x_i \overset{T_i}{\leftrightarrow} g, \tilde{x} \overset{\tilde{T}}{\leftrightarrow} g\right) \phi_{\Sigma_i}\left(x_i \overset{T_i}{\leftrightarrow} g, \tilde{x} \overset{\tilde{T}}{\leftrightarrow} g \mid x_i \overset{\Sigma_i^+}{\leftrightarrow} \tilde{x}\right).$$

Figure 6: Sketch of the idea behind the proof of Proposition 4.1. In bold we sketch the clusters of $x_i$ and $\tilde{x}$ and the explored backbones so far with thin lines. The connections to the ghost are indicated with straight edges and the event $\hat{\mathcal{R}}_i^*$ is indicated with the dotted ring around the cluster of $x_i$.

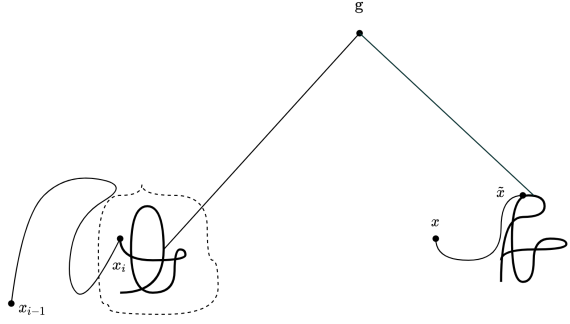

From the mixing argument in Lemma 4.3 it follows that $\phi_{\Sigma_i}\left(\hat{\mathcal{R}}_i^* \mid x_i \overset{T_1}{\leftrightarrow} g, \tilde{x} \overset{\tilde{T}}{\leftrightarrow} g\right) \geq c$.

Thus, we just need to prove that $\phi_{\Sigma_i}\left(x_i \overset{T_1}{\leftrightarrow} g, \tilde{x} \overset{\tilde{T}}{\leftrightarrow} g\right) \geq c\phi_{\Sigma_i}\left(x_i \overset{\Sigma^+}{\leftrightarrow} \tilde{x}\right)$. Notice that

$$\{x_i \overset{\Sigma^+}{\leftrightarrow} \tilde{x}\} \subset \left(\{x_i \overset{T_1}{\leftrightarrow} g\} \cup \{x_i \leftrightarrow A_i\}\right) \cap \left(\{\tilde{x} \overset{\tilde{T}}{\leftrightarrow} g\} \cup \{\tilde{x} \leftrightarrow \tilde{A}\}\right).$$

Now, by first a union bound, then (Mixing) and the assumption and finally (MON) and (FKG)

$$\begin{aligned}
\phi_{\Sigma_i}\left(x_i \overset{\Sigma^+}{\leftrightarrow} \tilde{x}\right) &\leq \phi_{\Sigma_i}\left(x_i \overset{T_1}{\leftrightarrow} g, \tilde{x} \overset{\tilde{T}}{\leftrightarrow} g\right) + \phi_{\Sigma_i}\left(x_i \overset{T_1}{\leftrightarrow} g, \tilde{x} \leftrightarrow \tilde{A}\right) \\
&\quad + \phi_{\Sigma_i}\left(x_i \leftrightarrow A_i, \tilde{x} \overset{\tilde{T}}{\leftrightarrow} g\right) + \phi_{\Sigma_i}\left(x_i \leftrightarrow A_i, \tilde{x} \leftrightarrow \tilde{A}\right) \\
&\leq 4C\phi_{J_i,\tilde{h}}^{0,a}\left(x_i \overset{T_1}{\leftrightarrow} g\right) \phi_{\tilde{J},\tilde{h}}^{0,a}\left(\tilde{x} \overset{\tilde{T}}{\leftrightarrow} g\right) \leq c\phi_{\Sigma_i}\left(x_i \overset{T_1}{\leftrightarrow} g, \tilde{x} \overset{\tilde{T}}{\leftrightarrow} g\right).
\end{aligned}$$

$\square$

We now turn to the main technical part for proving Proposition 3.3. From now on, we will assume, for notational reasons, without loss of generality that $x_i$ is on the right side of the inner boundary of the annulus $A_{9i,9(i+1)}$, i.e. that $x_i = (\frac{9i}{2}, y)$ for some $y$ such that $-\frac{9i}{2} \leq y \leq \frac{9i}{2}$. Then define $L = [1,2] \times [0, \frac{1}{10}] + x_i$ and $\tilde{L}$ similarly to be a rectangle in the vicinity of $\tilde{x}$.

**Lemma 4.5** *For each $1 < i < n$ it holds that*

$$\phi_{\Sigma_i}(x_i \leftrightarrow L \mid x_i \leftrightarrow A_i) \geq c \text{ as well as } \phi_{\Sigma_i}(\tilde{x} \leftrightarrow L \mid \tilde{x} \leftrightarrow \tilde{A}) \geq c.$$

Let $C(x_i)$ be number of points in $\Lambda_2(x_i)$ connected to $x_i$ without using edges to the ghost and $N_{M,i} = \{C(x_i) \geq M\}$ for each $M \in \mathbb{N}$.

**Lemma 4.6** *Let $k > 0$. Then,*

$$\phi_{\Sigma_i}\left(N_{ka^{-\frac{15}{8}},i} \mid x_i \leftrightarrow L, x_i \leftrightarrow A_i\right) \geq c.$$

Figure 7: Sketch of the topological rectangle $T$ with marked points $a, b, c$ and $d$ from Lemma 4.7. The points $e$ and $f$ are on the segment $(ad)$ and the path $\gamma$ is inside $T$.

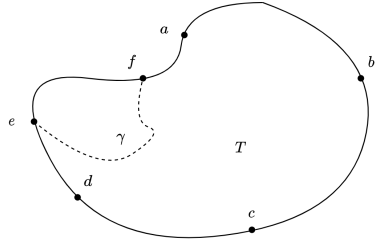

Then let us start proving the lemmas. To do that we need to show that crossings of topological rectangles exist with constant probability. This is done in ([38], Theorem 1.1) if the discrete extremal length $l_\Omega[(ab), (cd)]$ is bounded. From ([38],(3.7)) (see also [39]) we have the following characterisation of the discrete extremal length

$$l_\Omega((ab), (cd)) = \sup_{g: E(\Omega) \to \mathbb{R}_+ \cup \{0\}} \frac{\left(\inf_{\gamma:(ab) \leftrightarrow (cd)} \sum_{e \in \gamma} g_e\right)^2}{\sum_e g_e^2}$$

where the supremum is over all non-negative, not identically zero functions on the edges. Using this representation we obtain the following lemma (which is equivalent to Rayleigh's monotonicity law).

**Lemma 4.7** *Let $T$ be a topological rectangles with marked points $(abcd)$. Let $e, f$ be points on $(ad)$ and $\gamma$ a path from $e$ to $f$ inside $T$ (see Figure 7). Let $\tilde{T}$ be the points in $T$ reachable from $b$ in $T \backslash \gamma$. Note $\tilde{T}$ is a new topological rectangle with marked points $(abcd)$. Then,*

$$l_T((ad), (bc)) \geq l_{\tilde{T}}((ad), (bc)) \qquad \text{as well as} \qquad l_T((ab), (cd)) \geq l_{\tilde{T}}((ab), (cd)).$$

*Proof.* We prove the first inequality, the second follows similarly. Since the graph $\tilde{T}$ is finite the supremum and infimum are attained and we get some maximizing function $\tilde{g}$ for $\tilde{T}$. Now, define the function $g$ by extending $\tilde{g}$ with $g(e) = 0$ whenever $e \notin T \backslash \tilde{T}$. Then,

$$l_{\tilde{T}}((ad) \leftrightarrow (bc)) = \frac{\left(\inf_{\gamma:(ad) \leftrightarrow (bc) \text{ in } \tilde{T}} \sum_{e \in \gamma} \tilde{g}_e\right)^2}{\sum_e \tilde{g}_e^2} = \frac{\left(\inf_{\gamma:(ad) \leftrightarrow (bc) \text{ in } T} \sum_{e \in \gamma} g_e\right)^2}{\sum_e g_e^2} \leq l_T((ad) \leftrightarrow (bc)).$$

The second equality follows since any path $\gamma : (ad) \leftrightarrow (bc)$ in $T$ has a subpath $\tilde{\gamma} : (ad) \leftrightarrow (bc)$ in $\tilde{T}$. The inequality follows since the function $g$ is just one element in the supremum defining the discrete extremal length. $\qquad \square$

Using Lemma 4.7 we can now prove Lemma 4.5

*Proof of Lemma 4.5.* Recall the definition of $d, d'$ from the beginning of section 3. Define the explored vertices $\mathcal{V}$ of the backbone to be all vertices with at least one incident explored edge. Then define $U$ to be the set of vertices in $V(\Lambda) \backslash (\mathcal{V} \cup D_i)$ with at least one edge to $\mathcal{V}$. Since $\gamma_i \cap \Lambda_{2,R}(x_i) = \emptyset$ there exists at least one $*$-path (i.e. a path that can also jump diagonally) $P_i^*$ in $U$ from $d$ to a vertex in $\partial \Lambda_2^c(x_i)$. From such a $*$-path $P_i^*$ we can construct a usual path $P_i$ just going around the plaquette every time $P_i^*$ jumps diagonally. Let the first vertex that $P_i$ hits in $\Lambda_2(x_i)^c$ be $d_1$ and denote henceforth the path $P_i$ by $(dd_1)$. Define $d'_1$ similarly following the outside of the backbone from $d'$.

Now, let $\Lambda_1(x_i)_R$ denote the right half of the box $\Lambda_1(x_i)$. Define $a_i = x_i + (1,-1)$, $b_i = x_i + (2,-1)$, $a'_i = x_i + (1,1)$ and $b'_i = x_i + (2,1)$ see Figure 8. Define $T_{i,1} = \Lambda_2(x_i) \backslash D_i$. Then let $\mathcal{S}_i \in \{0,1\}^{E(T_{i,1})}$ be the event defined by

$$\mathcal{S}_i = \left\{ (a_ib_i) \overset{T_{i,1}}{\leftrightarrow} (dd_1) \right\} \bigcap \left\{ (a'_ib'_i) \overset{T_{i,1}}{\leftrightarrow} (d'd'_1) \right\} \bigcap \left\{ (a_ia'_i) \overset{L}{\leftrightarrow} (b_ib'_i) \right\}.$$

I.e. $\mathcal{S}_i$ ensures that any path from $x_i$ to $\Lambda_{9i+1}^c$ will intersect a cluster of open edges that in particular hits $L$. We claim that

**Claim 4.8** $\phi_{\Sigma_i}(\mathcal{S}_i) \geq c$.

*Proof.* We prove that each of the three events defining $\mathcal{S}_i$ has a positive probability. That

$$\phi_{\Sigma_i} \left( (a_ia'_i) \overset{L}{\leftrightarrow} (b_ib'_i) \right) \geq c$$

follows from RSW for usual rectangles [37]. Thus, by symmetry it suffices to prove

$$\phi_{\Sigma_i}((a_ib_i) \overset{T_{i,1}}{\leftrightarrow} (dd_1)) \geq c.$$

Notice that the path $(dd_1)$ does not leave the left half of the box $\Lambda_2(x_i)$ since the backbone is only in the left half. If we consider a new topological rectangle $T_{i,2}$ to be $\Lambda_2(x_i)_L$ union the top-right quarter of $A_{1,2}(x_i)$ with the four marked points $a_i, b_i, d, d_i$ and where we use the part of $\partial \Lambda_1(x_i)$ from $x_i + (0,1)$ until $d$ as the boundary twice as shown on Figure 8, then the path $(dd_1)$ has the form of $\gamma$ in Lemma 4.7 so we conclude that

$$l_{T_{i,1}}\left( (a_ib_i), (dd_1) \right) \leq l_{T_{i,2}}\left( (a_ib_i), (dd_1) \right).$$

Define $c_i = x_i + (0,-1)$ and $d_i = x_i + (0,-2)$. Then $c_i, d_i$ are on the segment $(dd_1)$ and thus

$$l_{T_{i,2}}\left( (a_ib_i), (dd_1) \right) \leq l_{T_{i,2}}\left( (a_ib_i), (c_id_i) \right) \leq l_{T_{i,3}}\ \left( (a_ib_i), (c_id_i) \right) \leq c$$

where we in the last step considered a new topological rectangle $T_{i,3}$ where we used the part of $A_{1,2}(x_i)$ enclosed by $(a_ib_i)$ and $(c_id_i)$ as shown on Figure 8 which has bounded discrete extremal length. Therefore $l_{T_{i,1}}\left( (a_ib_i), (dd_1) \right) \leq c$ which means by ([38], Theorem 1.1) if that

$$\phi_{\Sigma_i} \left( (a_ib_i) \overset{T_{i,1}}{\leftrightarrow} (dd_1) \right) \geq \phi_{T_{i,1},\tilde{h}}^{0,a} \left( (a_ib_i) \overset{T_{i,1}}{\leftrightarrow} (dd_1) \right) \geq c.$$

That $\phi_{\Sigma_i}(\mathcal{S}_i) \geq c$ then follows from (FKG). $\qquad \square$

Now, to finish the proof of the lemma note that since $\{x_i \leftrightarrow A_i\} \cap \mathcal{S}_i \subset \{x_i \leftrightarrow L\}$ then by (FKG)

$$\phi_{\Sigma_i}\left( x_i \leftrightarrow L \mid x_i \leftrightarrow A_i \right) \geq \phi_{\Sigma_i}\left( \mathcal{S}_i \mid x_i \leftrightarrow A_i \right) \geq \phi_{\Sigma_i}\left( \mathcal{S}_i \right) \geq c.$$

$\qquad \square$

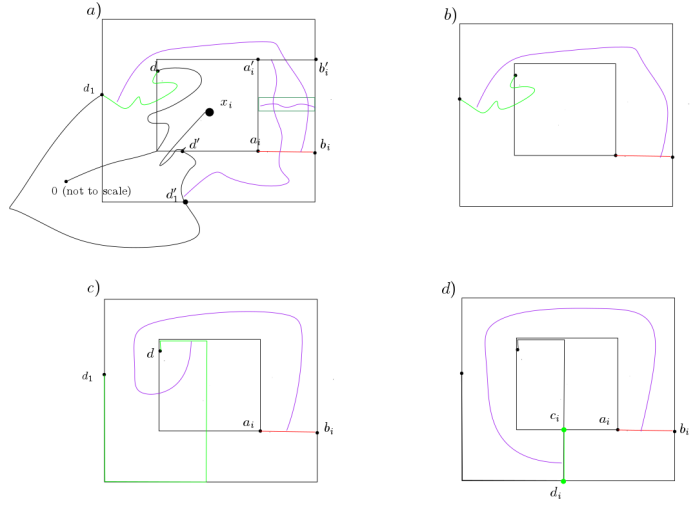

Figure 8: a) The backbone and the paths constructed in the steps in the proof of Lemma 4.5. b) The path $(a_i b_i) \leftrightarrow (d d_1)$ is (strictly) inside the domain $T_{i,1}$. c) The domain $T_{i,2}$ and the path $(a_i b_i) \leftrightarrow (d d_1)$. d) The domain $T_{i,3}$ and an example of the path $(a_i b_i) \leftrightarrow (c_i d_i)$

We end by proving Lemma 4.6.

*Proof of Lemma 4.6.* We use some ideas from Lemma 3.1 in [40]. Consider a square $B = \Lambda_{\frac{1}{10}}(\frac{3}{2}, \frac{1}{20}) + x_i$ corresponding with the previous lemma such that $L$ passes through $B$. Define the event $\mathcal{S}$ to be $\mathcal{S}_i$ where there is also a crossing of each of the four (overlapping) rectangles that make up the annulus $A_{\frac{1}{10}, \frac{2}{10}}(\frac{3}{2}, \frac{1}{20})$ around $B$ as shown on Figure 9. By RSW for usual rectangles and (FKG) we know that $\phi_{\Sigma_i}(\mathcal{S}) \geq c$. By the definition of $\mathcal{S}_i$ from Lemma 4.5 $\{x_i \leftrightarrow L\} \cap \mathcal{S} = \{x_i \leftrightarrow B\} \cap \mathcal{S}$ and so by (FKG) and Lemma 4.5 we get that

$$\phi_{\Sigma_i}\left(x_i \leftrightarrow B \mid x_i \overset{T_i}{\leftrightarrow} A_i, \mathcal{S}\right) \geq \phi_{\Sigma_i}\left(x_i \leftrightarrow L \mid x_i \overset{T_i}{\leftrightarrow} A_i, \mathcal{S}\right) \geq c.$$

Now, since

$$\phi_{\Sigma_i}\left(N_{ka^{-\frac{15}{8}},i} \mid x_i \leftrightarrow L, x_i \overset{T_i}{\leftrightarrow} A_i\right) \geq \phi_{\Sigma_i}\left(N_{ka^{-\frac{15}{8}},i} \mid x_i \leftrightarrow B, x_i \overset{T_i}{\leftrightarrow} A_i, \mathcal{S}\right) \phi_{\Sigma_i}\left(x_i \leftrightarrow B \mid x_i \overset{T_i}{\leftrightarrow} A_i, \mathcal{S}\right)$$

it suffices to prove that

$$\phi_{\Sigma_i}\left(N_{ka^{-\frac{15}{8}},i} \mid x_i \leftrightarrow B, x_i \overset{T_i}{\leftrightarrow} A_i, \mathcal{S}\right) \geq c.$$

This follows from a second moment estimate. So let $N_B = \sum_{z \in B} 1_{z \leftrightarrow x_i}$. First using the fact that $\mathcal{S} \cap \{x_i \leftrightarrow L\} \cap \{z \leftrightarrow \partial \Lambda_{\frac{3}{10}}(x)\} \subset \{x_i \leftrightarrow z\}$ and then by (FKG), (MON) and

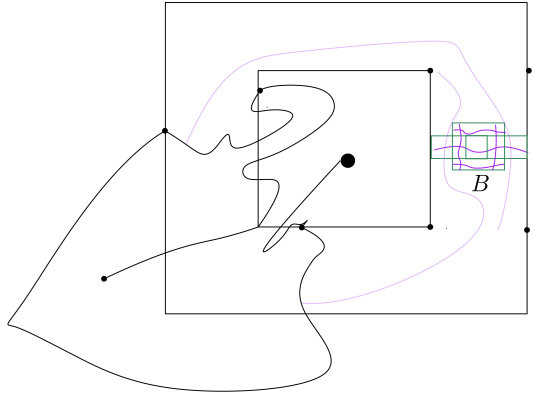

Figure 9: To the right we sketch the situation in Lemma 4.6. We keep the paths that we have seen exist in Lemma 4.5 with constant probability. Then we use RSW to construct some more paths encircling the box $B$ as shown. Later in the proof we split the box $B$ up into dyadic annuli.

(1-arm) we get

$$
\begin{aligned}
\mathbb{E}(N_B \mid x_i \leftrightarrow B, x_i \overset{T_i}{\nleftrightarrow} A_i, \mathcal{S}) &= \sum_{z \in B} \phi_{\Sigma_i} \left( x_i \leftrightarrow z \mid x_i \leftrightarrow B, x_i \overset{T_i}{\nleftrightarrow} A_i, \mathcal{S} \right) \\
&\geq \sum_{z \in B} \phi_{\Sigma_i} \left( z \leftrightarrow \partial \Lambda_{\frac{3}{10}}(z) \mid x_i \leftrightarrow B, x_i \overset{T_i}{\nleftrightarrow} A_i, \mathcal{S} \right) \\
&\geq \sum_{z \in B} \phi^0_{\Lambda_{\frac{3}{10}}(z)} \left( z \leftrightarrow \partial \Lambda_{\frac{3}{10}}(z) \right) = \frac{1}{100a^2} C a^{\frac{1}{8}} = c a^{-\frac{15}{8}}.
\end{aligned}
$$

Let us then consider the second moment. First we use that $\mathcal{S} \cap \{x_i \leftrightarrow B\} \cap \{x_i \overset{T_i}{\nleftrightarrow} A_i\} = \mathcal{S} \cap \{x_i \overset{T_i}{\nleftrightarrow} A_i\}$ and $\phi_{\Sigma_i}(\mathcal{S}) \geq c$ and FKG in the form

$$
\phi_{\Sigma_i}(\mathcal{S}, x_i \overset{T_i}{\nleftrightarrow} A_i) \geq \phi_{\Sigma_i}(\mathcal{S}) \phi_{\Sigma_i}(x_i \overset{T_i}{\nleftrightarrow} A_i) \geq c \phi_{\Sigma_i}(x_i \overset{T_i}{\nleftrightarrow} A_i)
$$

and then we do a dyadic summation for each $z$ partitioning $B$ into annuli $A_{2^{k-1}, 2^k}(z)$ for $k$ such that $-m \leq k \leq 0$ where $m \in \mathbb{N}$ is chosen such that $A_{2^{k-1}, 2^k}(x) = \emptyset$ for $k < -m$.

$$
\mathbb{E}(N_B^2 \mid x_i \leftrightarrow B, x_i \overset{T_i}{\nleftrightarrow} A_i, \mathcal{S}) = \sum_{z, y \in B} \phi_{\Sigma_i} \left( x_i \leftrightarrow z, x_i \leftrightarrow y \mid x_i \leftrightarrow B, x_i \overset{T_i}{\nleftrightarrow} A_i, \mathcal{S} \right)
$$

$$
\leq \frac{c}{\phi_{\Sigma_i}(x_i \overset{T_i}{\nleftrightarrow} A_i)} \sum_{z, y \in B} \phi_{\Sigma_i} \left( x_i \leftrightarrow z, x_i \leftrightarrow y, x_i \overset{T_i}{\nleftrightarrow} A_i \right)
$$

$$
\leq \frac{c}{\phi_{\Sigma_i}(x_i \overset{T_i}{\nleftrightarrow} A_i)} \sum_{z \in B} \sum_{k=-m}^{0} \sum_{y \in A_{2^{k-1}, 2^k}(z)} \phi_{\Sigma_i} \left( z \leftrightarrow \partial \Lambda_{2^{k-2}}(z), y \leftrightarrow \partial \Lambda_{2^{k-2}}(y), x_i \leftrightarrow \partial \Lambda_{\frac{1}{20}, \frac{1}{20}}(x_i) \right)
$$

$$
\leq \frac{c \phi_{\Sigma_i}(x_i \leftrightarrow \partial \Lambda_{\frac{1}{20}, \frac{1}{20}}(x_i))}{\phi_{\Sigma_i}(x_i \overset{T_i}{\nleftrightarrow} A_i)} \sum_{z \in B} \sum_{k=-m}^{0} \sum_{y \in A_{2^{k-1}, 2^k}(z)} \phi_{\Sigma_i} \left( z \leftrightarrow \partial \Lambda_{2^{k-2}}(z) \right) \phi_{\Sigma_i} \left( y \leftrightarrow \partial \Lambda_{2^{k-2}}(y) \right)
$$

$$
\leq \sum_{z \in B} \sum_{k=-m}^{0} c_1 \left( \frac{2^k}{a} \right)^2 \left( \frac{a}{2^{k-2}} \right)^{\frac{2}{8}} = c_2 a^{-2} a^{-2} a^{\frac{1}{4}} \sum_{k=-m}^{0} 2^{2k - \frac{k}{4}} = c_3 a^{-\frac{15}{4}}
$$

where we also used (1-arm) several times and that (Mixing) holds at all scales smaller than our fixed macroscopic scale. The conclusion follows from the Paley-Zygmund inequality

$$\phi_{\Sigma_i}\left(N_{ka^{-\frac{15}{8}},i} \mid x_i \leftrightarrow B, x_i \overset{T_i}{\leftrightarrow} A_i, \mathcal{S}\right) \geq ca^{\frac{15}{4}-\frac{15}{4}} = c.$$

$\square$

Finally, let us prove Proposition 3.3.

*Proof of Proposition 3.3.* By Proposition 4.1 it suffices to prove $\phi_{\Sigma_i}\left(x_i \overset{T_i^+}{\leftrightarrow} \mathrm{g} \mid x_i \leftrightarrow A_i\right) \geq$ $c$ and the similar inequality for $\tilde{T}$ and $\tilde{x}$ which will follow in the same way. First, notice that by Lemma 4.5 and Lemma 4.6

$$\phi_{\Sigma_i}\left(N_{ka^{-\frac{15}{8}},i} \mid x_i \leftrightarrow A_i\right) \geq \phi_{\Sigma_i}\left(N_{ka^{-\frac{15}{8}},i} \mid x_i \leftrightarrow L, x_i \leftrightarrow A_i\right)\phi_{\Sigma_i}\left(x_i \leftrightarrow L \mid x_i \leftrightarrow A_i\right) \geq c \cdot c.$$

Now, we can make the following observation using Lemma 2.2 to find

$$\begin{aligned}
\phi_{\Sigma_i}(x_i \overset{T_i^+}{\leftrightarrow} \mathrm{g} \mid x_i \leftrightarrow A_i) &\geq \phi_{\Sigma_i}\left(N_{ka^{-\frac{15}{8}},i}, x_i \overset{T_i^+}{\leftrightarrow} \mathrm{g} \mid x_i \leftrightarrow A_i\right) \\
&= \phi_{\Sigma_i}\left(x_i \overset{T_i^+}{\leftrightarrow} \mathrm{g} \mid N_{ka^{-\frac{15}{8}},i}, x_i \leftrightarrow A_i\right)\phi_{\Sigma_i}\left(N_{ka^{-\frac{15}{8}},i} \mid x_i \leftrightarrow A_i\right) \\
&\geq \tanh\left(ka^{-\frac{15}{8}}ha^{\frac{15}{8}}\right)c \geq c(k,h). \qquad \square
\end{aligned}$$

# Acknowledgments

The authors would like to thank Wendelin Werner for establishing collaboration and support through the SNF Grant 175505. The first author would like to thank the Swiss European Mobility Exchange program as well as the Villum Foundation for support through the QMATH center of Excellence(Grant No. 10059) and the Villum Young Investigator (Grant No. 25452) programs. Further, thanks to Hugo Duminil-Copin and Ioan Manolescu for discussions and to the anonymous referees for many excellent comments.

# 2. On monotonicity and couplings of random currents and the loop-$\mathrm{O}(1)$-model

# On monotonicity and couplings of random currents and the loop-O(1)-model

Frederik Ravn Klausen

**Abstract**

Using recent couplings, we provide counterexamples to monotonicity properties of percolation models related to graphical representations of the Ising model. We further prove a new coupling of the double random current model to the loop-O(1)-model.

## 1 Introduction

The field of graphical representations of the Ising model has been developing rapidly in the last 50 years. Examples of well-studied models include the random cluster model [1], [2], the random current representation [3], [4], [5], [6] as well as the loop-O(1)-model [7]. The motivation for studying these models is that one can link percolative and connectivity properties to phase transitions and correlation functions of the Ising model. Usually the models are introduced in their own right, but for the purpose here we define the models directly as percolation models via their couplings to the loop-O(1)-model [8, Exercise 36], [9, Theorem 3.5], [10] [11]. The framework provided by these recent couplings greatly simplifies the proofs and counterexamples given here.

For the definitions, we need the notion of an even subgraph. Given a finite graph $G = (V, E)$ an *even subgraph of* (V,F) is a spanning subgraph where each $v \in V$ is incident to an even number of edges in $F$. The set of even subgraphs of a graph $G$ is denoted $\mathcal{E}_\emptyset(G)$. A natural probability measure on $\mathcal{E}_\emptyset(G)$ is the *loop-O(1)-model* $l_x$ which is given by

$$l_x(g) = \frac{1}{Z} x^{|g|}, \text{ for each } g \in \mathcal{E}_\emptyset(G) \tag{1}$$

with $Z = \sum_{g \in \mathcal{E}_\emptyset(G)} x^{|g|}$. Here $|g|$ denotes the number of edges in $g$ and $x = \tanh(\beta)$ as in [11].

For two configurations $\omega_1, \omega_2$ of edges in the graph $G$ we let $\omega_1 \cup \omega_2$ be the configuration with the union of the edges in $\omega_1$ and $\omega_2$ and for two probability measures $\mu_1, \mu_2$ on the configurations of edges let $\mu_1 \cup \mu_2$ be the measure corresponding to sampling independent configurations with the law of $\mu_1, \mu_2$ and taking the union of the two configurations. We define $\mu^{\otimes 2} = \mu \cup \mu$ and let $\mathbb{P}_x$ be Bernoulli edge percolation with parameter $x \in (0, 1)$. Then we can define the (traced, sourceless) *single random current* at inverse temperature $\beta$ as

$$P_x = l_x \cup \mathbb{P}_{1-\sqrt{1-x^2}} \tag{2}$$

and the (FK-Ising, $q = 2$) *random cluster model* with $p = 1 - \exp(-2\beta)$ by

$$\phi_\beta = \phi_x = l_x \cup \mathbb{P}_x. \tag{3}$$

In Appendix A3 we describe how the previous definitions are related to the standard definitions in the literature. Using that $\mathbb{P}_p \cup \mathbb{P}_q = \mathbb{P}_{p+q-pq}$ we obtain as in (traced, sourceless) *double random current* which is the union of two independent single random currents is given by

$$P_x^{\otimes 2} = l_x^{\otimes 2} \cup \mathbb{P}_{x^2} \tag{4}$$

and similarly the double random cluster model is $\phi_x^{\otimes 2} = l_x^{\otimes 2} \cup \mathbb{P}_{x(2-x)}$.

These models are motivated by being graphical representations of the Ising model. In particular, we have the following relation between correlation functions [8, (1.5), (4.6)]

$$P_x^{\otimes 2}(x \leftrightarrow y) = \langle \sigma_x \sigma_y \rangle^2 = \phi_{\frac{\beta}{2}}(x \leftrightarrow y)^2 \quad x, y \in V. \tag{5}$$

where $\sigma_x, \sigma_y$ are Ising spins and $\langle \cdot \rangle$ is the expectation of the Ising measure.

Many tools are available for the random cluster model in particular monotonicity and the FKG-inequality [8, Theorem 1.6]. In the following, we investigate the monotonicity properties of the other models some of which turn out to be less well behaved than the random cluster model.

The first question is motivated by the fact that, by monotonicity of Ising correlations or of the random cluster model, it follows that $x \mapsto P_x^{\otimes 2}(x \leftrightarrow y) = \langle \sigma_x \sigma_y \rangle^2 = \phi_{\frac{\beta}{2}}(x \leftrightarrow y)^2$ is increasing. Here and in the following, we define $\{A \leftrightarrow B\}$ to be the event that at least one vertex in $A$ is connected to a vertex in $B$.

**Question 1.** *[12, Question 2] Is the map $x \mapsto P_x^{\otimes 2}(A \leftrightarrow B)$ increasing for any subsets $A, B \subset V$.*

In general, we are also interested in whether the model in itself is monotonic. The next question is related and with Theorem 3 in mind it can be seen as an easier example of the question from before.

**Conjecture 2.** *[13, Conjecture 5.1] Let $G$ be an even graph then $l_x$ is monotonic.*

In the following, we give a counterexample to Conjecture 2 and we give partial results and counterexamples towards Question 1. Most notably we show that the single random current measure is *not* monotonic. In general, for each of the above mentioned percolation models, we try to understand whether the FKG inequality, monotonicity, monotonicity of connection events $\{A \leftrightarrow B\}$ and monotonicity of singleton connection events $\{a \leftrightarrow b\}$, hold. We denote these properties by FKG, MON, CON and SING respectively.

The double current is more well behaved as it satisfies (SING), we further prove a new coupling for the double current and discuss its monotonicity properties. Some of the information in the theorems and counterexamples given is summed up in Table 1.

## 2 Counterexamples

In this section we give counterexamples to the FKG and monotonicity for the models $l_x, P_x$, and $l_x^{\otimes 2}$ .

Table 1: Overview of monotonicity properties.

| Case | $l_x$ | $P_x$ | $\phi_x$ | $l_x^{\otimes 2}$ | $P_x^{\otimes 2}$ | $\phi_x^{\otimes 2}$ |
|------|-------|-------|----------|-------------------|-------------------|----------------------|
| $p(x)$ | 0 | $1 - \sqrt{1-x^2}$ | $x$ | 0 | $x^2$ | $x(2-x)$ |
| FKG | × | × | ✓ | × | ? | ✓ |
| MON | × | × | ✓ | × | ? | ✓ |
| CON | × | × | ✓ | × | ? | ✓ |
| SING | × | × | ✓ | × | ✓ | ✓ |

Figure 1: The graph used to construct counterexamples to FKG. The three segments consist of $n, m$ and $l$ edges respectively. In some examples we let $n = l$.

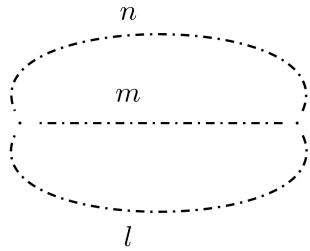

$n$

$m$

$l$

**2.1 Counterexamples to FKG.** Consider the graph in Figure 1 with $n = l$. The partition function corresponding to this graph is

$$Z = \sum_{g \in \mathcal{E}_\emptyset(G)} x^{|g|} = 1 + x^{n+m} + x^{n+m} + x^{2n}.$$

Let $X_1$ be the event that all edges in the upper $n + m$ loop are open and let $X_2$ be the event that all edges in the lower $n + m$ loop are open.

**2.1.1 Loop-**O(1)**-model:** Notice that $l_x(X_1 \cap X_2) = 0$, whereas $l_x(X_1) > 0, l_x(X_2) > 0$ and there is no positive association. A counterexample on an even graph is given in [13].

**2.1.2 Single random currents:** For the traced single random current we can use the same example and we get with $p(x) = 1 - \sqrt{1-x^2}$ that

$$Z P_x(X_2) = Z P_x(X_1) = p^{n+m} + x^{n+m} + x^{n+m} p^n + x^{2n} p^m$$

and

$$Z P_x(X_2 \cap X_1) = p^{2n+m} + 2 \cdot x^{n+m} p^n + x^{2n} p^m.$$

Now, since $Z \to 1$ in the limit $x \to 0$ the function

$$x \mapsto P_x(X_2 \cap X_1) - P_x(X_2) P_x(X_1) = \frac{Z P_x(X_2 \cap X_1) - Z P_x(X_2) Z P_x(X_1)}{Z^2} + \frac{(Z-1) P_x(X_2 \cap X_1)}{Z^2}$$

Figure 2: The graph in question shown as the rightmost graph along with its eight even subgraphs. We let the outer paths be $n$ edges long and the inner paths be $m$ edges long. The nodes $a$ and $b$ are marked with dots. We list number of edges of each subgraph, the corresponding weights and whether $a$ and $b$ are connected in the subgraph.

| Subgraph | · ·  | · · | ◇ | · | · | · | · | $n$ $m$ $m$ $n$ |
|---|---|---|---|---|---|---|---|---|
| Edges | 0 | $2n$ | $2m$ | $n+m$ | $n+m$ | $n+m$ | $n+m$ | $2m+2n$ |
| Weight | 1 | $x^{2n}$ | $x^{2m}$ | $x^{n+m}$ | $x^{n+m}$ | $x^{n+m}$ | $x^{n+m}$ | $x^{2m+2n}$ |
| $\{a \leftrightarrow b\}$ | × | × | ✓ | × | × | × | × | ✓ |

becomes negative for sufficiently small $x$. Thus, FKG is not satisfied. Using the same example, but a slightly more complicated analysis the same counterexample works for $l_x^{\otimes 2}$ we give the details in the Appendix.

For the *double random current* this example cannot be used to find a counterexample since indeed the two events $X_1$ and $X_2$ are positively associated for all $n$ and $m$.

**2.2 Counterexamples to monotonicity.** We now give a counterexample to Conjecture 2. Consider the graph shown rightmost on Figure 2 with $n$ edges along the outer paths and $m$ edges along the inner paths for some even $m$. Let further $a$ and $b$ be the two points shown. Then there are 8 possible even subgraphs as shown on Figure 2 where we have also listed their corresponding weigths. We see that $Z = 1 + x^{2n} + x^{2m} + 4x^{n+m} + x^{2n+2m}$ and as $a$ and $b$ are connected in the 3rd and the last graph only and therefore

$$l_x(a \leftrightarrow b) = \frac{x^{2m} + x^{2m+2n}}{Z}.$$

Now, numerical inspection shows that this function is not monotonic for $m = 2$ and $n \geq 8$. This provides a counterexample to Conjecture 2. The intuition behind the counterexample is that when $x$ increases we get an interval where it is more likely to sample one of the graphs with $n$ vertices where $a$ and $b$ are not connected. Similarly, we can use the coupling to the loop-O(1)-model to calculate the probability that $a$ is connected to $b$ for the single random current and the double loop-O(1)-model. We do the calculation in the appendix and show the plot of the functions in Figure 3 where we see that they are not monotone. Remark that $P_x^{\otimes 2}(a \leftrightarrow b)$ is always monotone since it satisfies SING.

Figure 3: Plot of the non-monotonous function for the event $\{a \leftrightarrow b\}$ and $l_x$ and $(n, m) = (18, 2)$ (left), $P_x$ and $(n, m) = (2000, 300)$ (middle) as well as $l_x^{\otimes 2}$ and $(n, m) = (38, 2)$ (right).

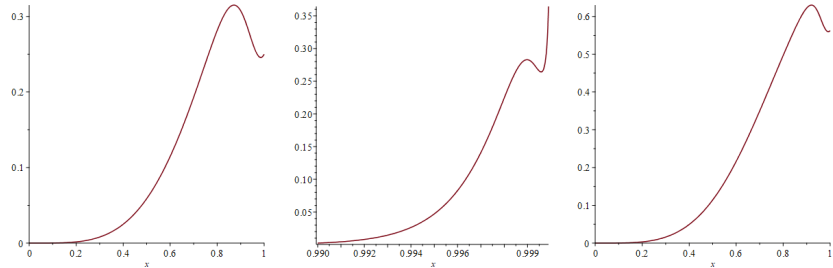

Figure 4: Overview of the couplings and correspondingly the implication diagram for monotonicity and FKG in view of Theorem 3. Each of the thick lines is either a union of the measure with Bernoulli-percolation (horizontal) or with an independent copy of itself (vertical). Since we have monotonicity and FKG for Bernoulli percolation it follows from Theorem 3 that we obtain the properties also along the thick lines (for some suitable graphs and parameters where we have monotonicity or FKG already). We have indicated the couplings from Theorem 4 and [9, Corollary 3.6] obtained by picking an even subgraph uniformly at random with dashed lines.

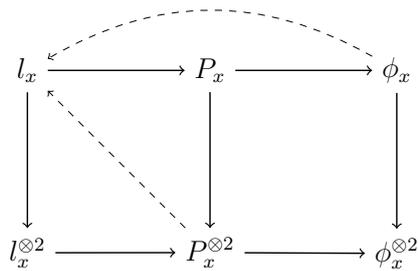

## 3 Monotonicity and FKG are stable under unions

One may observe that the counterexample in Figure 3 grow larger in size when going from the $l_x$ to $P_x$. The intuition being that the additional independent Bernoulli percolation, relating these models, 'enhances' properties of monotonicity and positive association. This intuition is in line with the following theorem which we will use to fill out Table 1. Further, since we know MON and FKG for percolation the implications along the arrows in Figure 4 follow.

**Theorem 3.** *Suppose that $\mu_x$ and $\nu_x$ are percolation measures monotonic in $x$. Let $\mu_x \cup \nu_x$ have the law of the union of two sets independently sampled. If $\mu_x$ and $\nu_x$ both satisfy FKG then $\mu_x \cup \nu_x$ satisfies FKG.*

*Proof.* If $x_1 < x_2$, then $\mu_{x_2}$ stochastically dominates $\mu_{x_1}$. We then use Strassen's characterization of stochastic domination. Let $P_\mu(\eta, \omega)$ be a coupling of the two. Where $\eta \sim \mu_{x_1}$, $\omega \sim \mu_{x_2}$ and $P_\mu(\{\eta \leq \omega\}) = 1$. Similarly, let $P_\nu(\tilde{\eta}, \tilde{\omega})$ be a coupling of $\nu_{x_1}$ and $\nu_{x_2}$. Then $\mu_{x_1} \cup \nu_{x_1}$ has the law of $\eta \cup \tilde{\eta}$ whereas $\mu_{x_2} \cup \nu_{x_2}$ has the law of $\omega \cup \tilde{\omega}$. So the measure distributed as $(\eta \cup \tilde{\eta}, \omega \cup \tilde{\omega})$ is a coupling with the correct marginals and such that $\eta \cup \tilde{\eta} \leq \omega \cup \tilde{\omega}$ almost surely. This proves the correct stochastic domination. The proof of the FKG-part is given in [14]. □

## 4 A new double current coupling

The counterexamples presented above do not seem to work for the double current. This vaguely suggest that the double current is similar in qualitative behaviour to the random cluster model, whereas the single current is more reminiscent of the loop-O(1)-model. The double current also has the same phase transition as the random cluster model, but it is not known to be the case for the single current [12, Question 1]. This intuition is further supported by the following new coupling for the double current which is exactly the same coupling as for the random cluster model [9, Corollary 3.6].

**Theorem 4.** *Sample a uniform even subgraph from the traced double current. That has the law of the loop-O(1)-model.*

*Proof.* We use the characterization from Theorem 3.2 in [11] which gives that

$$P_x^{\otimes 2}(\omega) = \frac{1}{Z^2} |\mathcal{E}_\emptyset(\omega)| \sum_{\omega_1 \subset \omega, \omega_1 \in \mathcal{E}_\emptyset(G)} x^{|\omega_1|} x^{2|\omega_2|} (1-x^2)^{|E|-|\omega|}.$$

Here $\omega_2 = \omega/\omega_1$ and compared to Theorem 3.2 in [11] there is an additional sum since a slightly different measure is considered there. The probability of sampling an even subgraph $\eta$ is given as

$$P_{coup}(\eta) = \sum_{\omega \supset \eta} \frac{1}{|\mathcal{E}_\emptyset(\omega)|} P_x^{\otimes 2}(\omega)$$

$$= \frac{1}{Z^2} \sum_\omega \sum_{\omega_1 \in \mathcal{E}_\emptyset(G)} 1_{\eta \subset \omega} 1_{\omega_1 \subset \omega} x^{|\omega_1|} x^{2|\omega_2|} (1-x^2)^{|E|-|\omega|}.$$

Now, we can interchange the two sums and then sum over all edges from $\omega_1 \cup \eta$ and upwards. To do that, we let $k$ denote the number of open edges in $\omega$ addition to $\omega_1 \cup \eta$. Then

$$P_{coup}(\eta) = \frac{1}{Z^2} \sum_{\omega_1 \in \mathcal{E}_\emptyset(G)} \sum_{k=0}^{|E| - |\eta \cup \omega_1|} \binom{|E| - |\eta \cup \omega_1|}{k} x^{|\omega_1|} x^{2(|\eta \cup \omega_1| + k - |\omega_1|)} (1 - x^2)^{|E| - |\eta \cup \omega_1| - k}$$

$$= \frac{1}{Z^2} \sum_{\omega_1 \in \mathcal{E}_\emptyset(G)} x^{-|\omega_1|} x^{2|\eta \cup \omega_1|} = \frac{1}{Z^2} \, x^{|\eta|} \sum_{\omega_1 \in \mathcal{E}_\emptyset(G)} x^{|\eta \triangle \omega_1|} = \frac{1}{Z^2} Z x^{|\eta|} = l_x(\eta)$$

where we used the bionomial theorem and the fact that for a fixed even subgraph $\eta$ the map $\omega_1 \mapsto \eta \triangle \omega_1$ is a bijection on the set of even subgraphs.

$\square$

**4.1 Monotonicity of 1-edge events.** We prove that the event $\{e$ is open$\}$ is monotonic in the loop-O(1)-model. We say that $e$ is *cyclic* if $e$ is part of a loop. Hence by an analogous argument as in [9, Corollary 3.6] we obtain the following Corollary.

**Corollary 5.** *It holds that*

$$\frac{1}{2} P_x^{\otimes 2}(e \text{ open, cyclic}) = l_x(e \text{ open}) = \frac{1}{2} \phi_x(e \text{ open, cyclic})$$

*and thus by monotonicity of $\phi_x$ one edge events are monotone.*

*Proof.* For any configuration $\omega$ if $e$ is open and cyclic in $\omega$ it means that it is part of a loop $K$. The map $\eta \mapsto \eta \triangle K$ is a bijection between the set of even subgraphs of $\omega$ that contain $e$ and the even subgraphs of $\omega$ that do not contain $e$. Thus, $e$ is part of exactly half of the even subgraphs of $\omega$ and the probability that $e$ is still open when we pick an even subgraph uniformly at random is exactly $\frac{1}{2}$. The first equality then follows from Theorem 4 and the second from the similar coupling from the random cluster model in [9, Theorem 3.1] as in the proof of [9, Corollary 3.6].

$\square$

We now show how monotonicity of the 1-edge event for the other models follows.

**Corollary 6.** *Suppose that $p \colon [0,1] \to [0,1], x \mapsto p(x)$ is non-decreasing and differentiable. Then*

$$x \mapsto (l_x \cup \mathbb{P}_{p(x)})(e \text{ open}) \text{ as well as } x \mapsto (l_x \cup l_x \cup \mathbb{P}_{p(x)})(e \text{ open})$$

*are increasing.*

*Proof.* We have that

$$\left(l_x \cup \mathbb{P}_{p(x)}\right)(e \text{ open}) = \frac{1}{\sum_{g \in \mathcal{E}_\emptyset(G)} x^{|g|}} \sum_{g \in \mathcal{E}_\emptyset(G)} x^{|g|} \left(1_{e \in g} + p(x)(1_{e \notin g})\right)$$

$$= l_x(e \text{ open}) + p(x)(1 - l_x(e \text{ open})).$$

Hence it follows that $(l_x \cup \mathbb{P}_{p(x)})(e \text{ open})$ has positive derivative by Corollary 5 since

$$l_x(e \text{ open})'(1 - p(x)) + p(x)'(1 - l_x(e \text{ open})) \geq 0.$$

To see the second part notice that $l_x^{\otimes 2}(e \text{ open}) = l_x(e \text{ open})\,(2 - l_x(e \text{ open}))$ and hence $l_x^{\otimes 2}(e \text{ open})' \geq 0$. $\qquad\square$

## 5   Discussion

On trees $P_x^{\otimes 2} \sim \mathbb{P}_{x^2}$ and $\phi_x \sim \mathbb{P}_x$ so the models are not the same, but we might ask if the law of the cyclic edges is the same for the two models now that the probabilities that an edge is part of a loop is the same. To see that this is not the case consider again the example in Figure 1. First, we split up a configuration $\omega = \omega_c \dot\cup \omega_s$ where $\omega_c$ are all the cyclic edges in $\omega$ and $\omega_s$ are the rest. Then look at

$$P_x^{\otimes 2}(|\omega_c| = l + m) = \frac{1}{Z^2}\left(x^{2(l+m)} + 2x^{l+m} + x^{2(l+m)}\right) \cdot (1 - x^{2n})$$

whereas

$$\phi_x(|\omega_c| = l + m) = \frac{1}{Z}(x^{l+m} + x^{l+m})(1 - x^n)$$

and hence

$$\frac{\phi_x(|\omega_c| = l + m)}{P_x^{\otimes 2}(|\omega_c| = l + m)} = \underbrace{(1 + x^{n+l} + x^{n+m} + x^{l+m})}_{Z}\frac{2x^{l+m}}{(1 + x^n)2x^{l+m}(1 + x^{l+m})} \neq 1.$$

We conclude that the distribution of the loops are not quite the same for the double current and the random cluster model.

With the results at hand, we can now summarize our findings. In conclusion from the basic results in the introduction, the counterexamples and Theorem 3 used as shown on Figure 4 we can establish the properties in Table 1. With this overview at hand, it is natural to ask add the following questions in addition to Question 1.

**Question 7.** *Is the double random current $P_x^{\otimes 2}$ monotonic, does it satisfy FKG?*

Notice that our counterexamples do not work for $P_x^{\otimes 2}$. Further, monotonicity of the random cluster model can be proven using Holley's criterion for example in the form of Lemma 1.5 in [8] . However, this fails for the double random current model for example on the graph considered in Figure 1. It is further noted that the FKG-lattice condition is not satisfied [15].

Let us mention that apart from the coupling shown above and the fact that we know that SING is satisfied then it also holds that $x^{o(g_1)+o(g_2)}p(x)^{|E|-o(g_1 \cup g_2)} = x^{2|E|}x^{-o(g_1 \triangle g_2)}$ whenever $p(x) = x^2$. Hence the double random current plays very well together with the structure on the space of even subgraphs. From the coupling we see that the loops of the double current are related to the loops of the random cluster model and we hope that the couplings described here could help resolving Question 7. Further, it might inspire couplings also for the double-current measure with sources as the one in [16, Thm 3.2].

# Acknowledgments

The author would like to thank the Swiss European Mobility Exchange program as well as the Villum Foundation for support through the QMATH center of Excellence(Grant No. 10059) and the Villum Young Investigator (Grant No. 25452) programs. Further, thanks to Peter Wildemann for discussions and proofreading and to the anonymous referee for many helpful comments.

# Appendix

## A1: Counterexample to FKG for $l_x^{\otimes 2}$

*Double loop-O(1)-model:* We use the same sets as in the counterexample for $l_x$. In Tables 2 and 3 all the possible configurations and whether they belong to $X_1$ and $X_2$ are listed. Thus, the corresponding probabilties are

$$l_x^{\otimes 2}(X_1) = \frac{2x^{n+m} + 3x^{2(n+m)} + 4x^{3n+m}}{Z^2} = l_x^{\otimes 2}(X_2)$$

whereas

$$l_x^{\otimes 2}(X_1 \cap X_2) = \frac{2x^{2(n+m)} + 4x^{3n+m}}{Z^2}.$$

Hence when we compare and use that $Z = 1 + O(x)$ we see that

$$l_x^{\otimes 2}(X_1)l_x^{\otimes 2}(X_2) - l_x^{\otimes 2}(X_1 \cap X_2)Z^2 = 4x^{2n+2m} + o(x^{2n+2m}) - 2x^{2(n+m)} + o(x^{2n+2m})$$
$$= 2x^{2n+2m} + o(x^{2n+2m})$$

which is positive for $x$ sufficiently small. Again since $Z \geq 1$ this means that FKG is not satisfied.

## A2: Explicit formulas for $l_x^{\otimes 2}(a \leftrightarrow b)$ and $P_x(a \leftrightarrow b)$

*Single random currents:* To see that for single random currents SING is not generally satisfied we continue the example from the loop-O(1)-model which is shown in Figure 2 with $n = l$. and in addition sample percolation with probability $p(x) = 1 - \sqrt{1 - x^2}$ on each edge. Let also $m' = \frac{m}{2}$ then we can calculate the probabilities using the coupling.

The main difficulty is when we get the empty subgraph. In that case we split up and count after how often we open all the edges in the segments of length $m'$ (from $a, b$ to the one of the vertices with valence four). If we open three of more $m'$-segments $a$ and $b$ will be connected - this happens with probability $4p^{3m'}(1-p^{m'})+p^{4m'}$. If we open at most one $m'$-segment $a$ and $b$ are not connected. Further, there are 4 ways of opening exactly two $m'$ segments each with probability $p^{2m'}(1-p^{m'})^2$ in two of the combinations $a$ and $b$ are connected and in the other two they are connected with probability $2p^n - p^{2n}$. So if we in the loop-O(1)-model sample the empty subgraph the probability that $a$ and $b$ are connected after sampling percolation is $f(p) = \left(4p^{3m'}(1-p^{m'})+p^{4m'}+p^{2m'}(1-p^{m'})^2\left(2+2(2p^n-p^{2n})\right)\right)$.

Table 2: Double Loop-O(1)-model: Overview of when the event $X_1$ is satisfied. The rows and columns represent the even subgraph sampled in each of the two independent copies of $l_x$.

|           | $\emptyset$ | n+m upper | n+m lower | 2n |
|-----------|:-----------:|:---------:|:---------:|:--:|
| $\emptyset$ |           | ✓         |           |    |
| n+m upper | ✓           | ✓         | ✓         | ✓  |
| n+m lower |             | ✓         |           | ✓  |
| 2n        |             | ✓         | ✓         |    |

Table 3: Double Loop-O(1)-model: Overview of when $X_1 \cap X_2$ is satisfied.

|           | $\emptyset$ | n+m upper | n+m lower | 2n |
|-----------|:-----------:|:---------:|:---------:|:--:|
| $\emptyset$ |           |           |           |    |
| n+m upper |             |           | ✓         | ✓  |
| n+m lower |             | ✓         |           | ✓  |
| 2n        |             | ✓         | ✓         |    |

That means the total probability becomes

$$P_x(a \leftrightarrow b) = \frac{1 \cdot f(p) + x^{2n} \cdot (2p^{m'} - p^{2m'})^2 + x^{2m} + 4 \cdot x^{n+m} \cdot (2p^{m'} - p^{2m'}) + x^{2n+2m}}{Z}.$$

Here we have plotted $P_x(a \leftrightarrow b)$ in Figure 3 for $(n,m) = (2000, 300)$ where we see that the function is not monotone.

*Double loop*-O(1)-*model:* For $l_x^{\otimes 2}$ we have to written out the $8 \times 8$ table in Table 4 where we plot all pairs of even subgraphs. Using the Table we obtain the following function.

$$l_x^{\otimes 2}(\{a \leftrightarrow b\}) = \frac{1}{Z^2}\left(2x^{2m}Z - x^{4m} + 2x^{2n+2m}Z - x^{4n+4m} - 2 \cdot x^{2n+4m} + 8x^{2n+2m}\right).$$

We have plotted the function for $(n,m) = (2, 18)$ in Figure 3 showing that monotonicity and in particular SING also fails for $l_x^{\otimes 2}$.

## A3: Couplings

In this appendix, we show how the definitions above correspond to the usual definitions of the (traced, sourceless) single random current and the random cluster model. Thereby giving a slightly different and self-contained proof of the relations from [8, Exercise 36] see also [9, Theorem 3.5], [10] [11, Theorem 3.1]. Using $x = \tanh(\beta)$ recovers our definitions.

Table 4: Double Loop-O(1)-model and all pairs of even subgraphs. We list whether $\{a \leftrightarrow b\}$. With the arrows we indicate if the upper or low path of length $n$ or $m$ is open.

| | $\emptyset$ | $2m$ | $n\uparrow+m\uparrow$ | $n\uparrow+m\downarrow$ | $n\downarrow+m\uparrow$ | $n\downarrow+m\downarrow$ | $2n$ | $2n+2m$ |
|---|---|---|---|---|---|---|---|---|
| $\emptyset$ | | ✓ | | | | | | ✓ |
| $2m$ | ✓ | ✓ | ✓ | ✓ | ✓ | ✓ | ✓ | ✓ |
| $n\uparrow+m\uparrow$ | | ✓ | | ✓ | | ✓ | | ✓ |
| $n\uparrow+m\downarrow$ | | ✓ | ✓ | | ✓ | | | ✓ |
| $n\downarrow+m\uparrow$ | | ✓ | | ✓ | | ✓ | | ✓ |
| $n\downarrow+m\downarrow$ | | ✓ | ✓ | | ✓ | | | ✓ |
| $2n$ | | ✓ | | | | | | ✓ |
| $2n+2m$ | ✓ | ✓ | ✓ | ✓ | ✓ | ✓ | ✓ | ✓ |

**Theorem 8.** *Defining the (traced, sourceless) single random current and the random cluster model as in [8] it holds that*

- $l_{\tanh(\beta)} \cup \mathbb{P}_{1-\cosh(\beta)^{-1}} = P_{\tanh(\beta)}$

- $l_{\tanh(\beta)} \cup \mathbb{P}_{\tanh(\beta)} = \phi_\beta.$

*Proof.* To prove the first relation, consider all sourceless currents $m$ such that $\hat{m} = n$. For each sourceless current $m$ let $u(m)$ be the set of edges with an uneven number. Since the current is sourceless $u(m)$ has to be an even spanning subgraph and conversely for every even spanning subgraph $\gamma$ gives rise to some $m$ with $u(m) = \gamma$. Summing the weight of all such $m$ we find that the relative weight of $\gamma$ is

$$\sum_{m|u(m)=\gamma} w(m) = \left(\prod_{e\in\gamma}\sum_{n_e\geq 0}\frac{\beta^{2n_e+1}}{(2n_e+1)!}\right)\left(\prod_{e\notin\gamma}\sum_{n_e\geq 0}\frac{\beta^{2n_e}}{(2n_e)!}\right) = \sinh(\beta)^{o(\gamma)}\cosh(\beta)^{|E|-o(\gamma)} \propto \tanh(\beta)^{o(\gamma)}.$$

Notice that this is exactly the relative weight of $\gamma$ in the loop-O(1)-model. Now, given that the uneven edges are $\gamma$ to compute the relative weight of $n$ we need a positive number for all remaining (even) edges of $n$ and the current has to be 0 at the edges that are closed in $n$. This has weights $\frac{\cosh(\beta)-1}{\cosh(\beta)}$ and $\frac{1}{\cosh(\beta)}$ for each edge respectively. Since the edges are independent given their parity this is exactly Bernoulli percolation with parameter $1 - \cosh(\beta)^{-1}$.

For the second relation, notice that $\frac{x}{1-x} = \frac{\tanh(\beta)}{1-\tanh(\beta)} = \frac{p}{2(1-p)}$ with $p = 1 - \exp(-2\beta)$. Let $\eta$ be an even spanning subgraph of $\omega$. There are $E - o(\eta)$ closed edges left and we have to open

$o(\omega) - o(\eta)$ of them. Hence $E - o(\omega)$ have to stay closed. Thus,

$$\sum_{\partial\eta=\emptyset} \mathbb{P}(\omega \mid \eta) l_\beta(\eta) = \sum_{\partial\eta=\emptyset, \eta \leq \omega} x^{o(\omega)-o(\eta)}(1-x)^{E-o(\omega)} x^{o(\eta)}$$

$$\propto \left(\frac{x}{1-x}\right)^{o(\omega)} \sum_{\partial\eta=\emptyset, \eta\leq\omega} 1 \propto \left(\frac{p}{2(1-p)}\right)^{o(\omega)} 2^{k(\omega)+o(\omega)} \propto p^{o(\omega)}(1-p)^{c(\omega)} 2^{k(\omega)} \propto \phi_p(\omega)$$

where we used that a graph with $k$ connected components, $|E|$ edges and $|V|$ vertices has $2^{k+|E|-|V|}$ even spanning subgraphs.

$\square$

It now follows that

$$P_x^{\otimes 2} = P_x \cup P_x = l_x \cup \mathbb{P}_{1-\sqrt{1-x^2}} \cup l_x \cup \mathbb{P}_{1-\sqrt{1-x^2}} = l_x^{\otimes 2} \cup \mathbb{P}_{2(1-\sqrt{1-x^2})-(1-\sqrt{1-x^2})^2} = l_x^{\otimes 2} \cup \mathbb{P}_{x^2}$$

where we used that $\mathbb{P}_p \cup \mathbb{P}_q = \mathbb{P}_{p+q-pq}$ as stated two lines below (3). That $\phi_x^{\otimes 2} = l_x^{\otimes 2} \cup \mathbb{P}_{x(2-x)}$ follows by a similar and simpler computation.

# 3. Strict monotonicity, continuity and bounds on the Kertész line for the random-cluster model on $\mathbb{Z}^d$

# Strict monotonicity, continuity and bounds on the Kertész line for the random-cluster model on $\mathbb{Z}^d$

### Ulrik Thinggaard Hansen and Frederik Ravn Klausen

#### Abstract

Ising and Potts models can be studied using the Fortuin–Kasteleyn representation through the Edwards-Sokal coupling. This adapts to the setting where the models are exposed to an external field of strength $h > 0$. In this representation, which is also known as the random-cluster model, the Kertész line is the curve which separates two regions of the parameter space defined according to the existence of an infinite cluster in $\mathbb{Z}^d$. This signifies a geometric phase transition between the ordered and disordered phases even in cases where a thermodynamic phase transition does not occur. In this article, we prove strict monotonicity and continuity of the Kertész line. Furthermore, we give new rigorous bounds that are asymptotically correct in the limit $h \to 0$ complementing the bounds from [J. Ruiz and M. Wouts. On the Kertész line: Some rigorous bounds. Journal of Mathematical Physics, 49:053303, May 2008, [1]] , which were asymptotically correct for $h \to \infty$. Finally, using a cluster expansion, we investigate the continuity of the Kertész line phase transition.

## 1 Introduction

The random-cluster model [2] has been under intense investigation the last 30 years. The model generalises the Fortuin-Kasteleyn graphical representation of the Ising model [3] to a graphical representation of all Potts models.

Highlights in the investigations in two dimensions are the calculation of the critical point for all $1 \le q < \infty$ in [4], sharpness of the phase transition [5], the scaling relations of critical exponents [6] as well as the rigorous determination of the domain of parameters where the phase transition is continuous [7, 8]. In higher dimensions, or in the presence of a magnetic field, results are scarcer with most recent efforts focusing on the near-critical planar regime [9, 10, 11, 12].

A magnetic field is implemented in the random-cluster model using Griffith's ghost vertex, which is an additional vertex connected to all other vertices in the graph. The Kertész line then separates two regions which are defined according to whether or not there is percolation without using the ghost vertex. The Kertész line transition need not correspond to a thermodynamic phase transition (i.e. a point where the free energy is not analytic). For example, by the Lee-Yang Theorem, the free energy of the Ising model is analytic for all $h \ne 0$ [13, 14]. Thus, passing through the domain where $h > 0$, one never encounters a thermodynamic phase transition. Nevertheless, in the random-cluster model, the percolative behaviour may change and thereby signify a geometric phase transition which separates an ordered phase from a disordered phase even at $h > 0$. This phase transition defines the Kertész line.

The Kertész line was first studied by Kertész in [15] and further discussed in [1, 16, 17, 18]. Kertész noted that there is no contradiction between the analyticity of the free energy and the geometric phase transition. It was proven in [16] that for large $q$ and small $h > 0$, where the phase transition is first order, that the Kertész line and the thermodynamic phase transition line coincide.

An interesting feature of the problem is that it involves three variables $(p, q, h)$ and its study has to delve into the trade-off between the decorrelating effect of $h$ on the edges in $\mathbb{Z}^d$, the order-enhancing effect of the inverse temperature governed by the parameter $p$ and the monotonically decreasing behaviour in $q$.

## Organisation of the paper and main results

The random-cluster measure on a finite graph $G = (V, E)$ with a distinguished vertex $\mathfrak{g}$, parameters $p \in (0, 1)$ and $q > 1$ and external field $h > 0$ is the measure on $\{0, 1\}^E$ given by

$$\phi_{p,q,h,G}[\{\omega\}] = \frac{1}{Z_{p,q,h,G}} p^{o(\omega_{\mathrm{in}})} (1-p)^{c(\omega_{\mathrm{in}})} p_h^{o(\omega_{\mathfrak{g}})} (1-p_h)^{o(\omega_{\mathfrak{g}})} q^{\kappa(\omega)},$$

where $\omega_{\mathfrak{g}}$ is the restriction of $\omega$ to the set of edges adjacent to $\mathfrak{g}$, $\omega_{in}$ is the restriction of $\omega$ to the set of edges not adjacent to $\mathfrak{g}$, $o(\cdot)$ denotes the number of open edges and $\kappa(\cdot)$ the number of components. Finally, $p_h = 1 - \exp\left(-\frac{q}{q-1}h\right)$.

In our paper, $G$ will be a subgraph of $\mathbb{Z}^d$ with a single external vertex $\mathfrak{g}$ (called the ghost) added, which is connected to every other vertex. As one fixes two of the three parameters $p, q, h$ and varies the third, the model exhibits a (possibly trivial) percolation phase transition at a point which we denote $p_c(q, h)$, $q_c(p, h)$ and $h_c(p, q)$ respectively. The Kertész line is exactly the set of such points of phase transition. A more precise definition will be given later. We shall often omit one of the variables from notation. In such a case, we consider the omitted variable fixed.

After catching the reader up on some preliminaries, our first order of business in Section 3 is to use the techniques from [19] to prove the following:

**Theorem 1.1.** *Let $d \geq 2$. Then, the maps $q \mapsto p_c(q, h)$, $p \mapsto q_c(p, h)$ and $h \mapsto q_c(p, h)$ are strictly increasing and the map $h \mapsto p_c(q, h)$ is strictly decreasing. Furthermore, $q \mapsto h_c(p, q)$ is strictly increasing on $(q_c(p, 0), \infty)$ and $p \mapsto h_c(p, q)$ is strictly decreasing on $(p_c(1, 0), p_c(q, 0))$.*

Continuity follows from the strict monotonicity and thus, that the Kertész line is aptly named - it is, indeed, a curve. In particular, this proves that $h_c(p) > 0$ for all $p \in (p_c(1, 0), p_c(q, 0))$ as was conjectured in [17, Remark 4].

**Corollary 1.2.** *The Kertész line $h \mapsto p_c(q, h)$ (and $p \mapsto h_c(p, q)$) is continuous.*

*Proof.* Suppose that $h \mapsto p_c(q, h)$ were strictly decreasing but discontinuous at $h_0$. Denote by $p_c^+ = \sup_{h > h_0} p_c(q, h)$ and by $p_c^- = \inf_{h < h_0} p_c(q, h)$. Then, we see that for $p \in (p_c^+, p_c^-)$, there is no percolation at $(p, h)$ for $h < h_0$, but there *is* percolation for $h > h_0$. Accordingly, $p \mapsto h_c(p, q)$ would be constant on $(p_c^-, p_c^+)$. Thus, the corollary follows by contraposition. $\square$

Then, in Section 4, we obtain new upper and lower bounds on the Kertész line improving those of [1]. In particular, we obtain bounds that asymptotically tend to the correct value in the limit $h \to 0$ complementing the bounds from [1], which asymptotically matched the correct value for $h \to \infty$.

**Theorem 1.3.** *There exists an explicit function $\operatorname{arctanh}_q$ such that, for any $q \in [1, \infty)$ and any $d \geq 2$, it holds that*

$$h_c(p) \leq \operatorname{arctanh}_q \left( \sqrt{\frac{1}{q-1} \left( \frac{q}{q_c(p,0)} - 1 \right)} \right).$$

*In particular, in the special case of the planar Ising model $q = 2$, this reduces to*

$$h_c(p) \leq \operatorname{arctanh} \left( \sqrt{\frac{2}{q_c(p,0)} - 1} \right) = \operatorname{arctanh} \left( \sqrt{\frac{2(1-p)^2}{p^2} - 1} \right).$$

Our methods include a generalisation of a recent lemma from [10] concerning the probability of clusters in $\omega_{\text{in}}$ connecting to the ghost which allows us to obtain a stochastic domination result. This allows to bound the model at $(p, q, h)$ from below by the model at $(p, q'(h), 0)$ for some explicit value of $q'(h)$. The game is then to make $h$ large enough that $q'(h) < q_c(p)$, so that we achieve percolation. We note that our results here transfer to improve results on the random field Ising model, which was analysed using the Kertész line in [17].

In order to get a lower bound for the Kertész line, we employ techniques from [20] to obtain the following: Set $\mu := \frac{(2d+1)^{2d+1}}{(2d)^{2d}}$ and $\Lambda_k = [-k, k]^d \cap \mathbb{Z}^d$.

**Theorem 1.4.** *Suppose that $p < p_c(q, 0)$ and that $\delta = \mu^{-4^d}$ and let $k$ be the smallest $k$ such that*

$$\phi^1_{p,q,0,\Lambda_{3k}}[\Lambda_k \leftrightarrow \partial \Lambda_{3k}] < \frac{\delta}{2}.$$

*Then, there is no percolation at $(p, h)$ for*

$$p_h < 1 - \left( 1 - \frac{\delta}{2} \right)^{1/|\Lambda_{3k}|}.$$

In Section 5 , we adapt the cluster expansion to this particular setup and use it to prove that the phase transition across the Kertész line is continuous when $h$ is sufficiently large. This complements the results of [16], where the Pirogov-Sinai theory was used to prove that the phase transition is discontinuous for sufficiently large $q$ and sufficiently small $h > 0$. More precisely, let $\mathfrak{Z} = \lim_{n \to \infty} \frac{\log(Z_{\Lambda_n})}{|\Lambda_n|}$ denote the pressure. We then prove the following:

**Theorem 1.5.** *There is a function $h_0(q, d)$ such that for $h > h_0(q, d)$, $\mathfrak{Z}$ is an analytic function of $p$. Explicitly,*

$$h_0(q) = \begin{cases} \left( 1 + \frac{1}{q-1} \right)^{-1} (\log(2) + \log(q-1) + 2d + (2d+1)\log(2d+1) - 2d\log(2d)) & q \geq 2 \\ \left( 1 + \frac{1}{q-1} \right)^{-1} (\log(q) + 2d + (2d+1)\log(2d+1) - 2d\log(2d)) & q \in (1, 2). \end{cases}$$

We have plotted the function $h_0$ in Figure 5.

Finally, in Section 6, we provide an outlook on further questions on the Kertész line. We discuss continuity and to what extent the Kertész line always coinsides with the line of maximal correlation length. Finally, we briefly other models, namely the Loop-$O(1)$ and random current models, for which there is a natural notion of a Kertész line, but which lack the sort of monotonicity which is crucial to the study of the random-cluster model.

The aim of this paper is two-fold. On the one hand, we aim to answer questions about the Kertész line of the Ising model. On the other, we try to provide a unifying account of the problem, a display of the many different (at times, rather standard) techniques from around the field of percolation that may aid in attacking the problem.

## 2  Preliminaries

We first introduce the Potts model, of which the random-cluster model is a graphical representation. We follow the notation of [21] which we also refer to for further information.

### Potts and random-cluster models

For some finite graph $G = (V, E)$ the Ising model is a probability distributions on the configuration space $\Omega = \{-1, +1\}^V$. The $q$-state Potts model is a generalisation on the configuration space $(\mathbb{T}^q)^V$, where $\mathbb{T}^q$ is defined as follows:

**Definition 2.1.** *We define $\mathbb{T}^q$ to be the vertices of a $q$-simplex in $\mathbb{R}^{q-1}$ containing $\mathbf{1} := (1, 0, 0, 0, ..., 0)$ as a vertex and such that*

$$\langle x, y \rangle = \begin{cases} 1 & x = y \\ -\frac{1}{q-1} & else \end{cases}$$

*for all vertices $x$ and $y$ of the simplex.*

Note that one recovers the spin space $\{-1, +1\}$ of the Ising model for $q = 2$. Throughout, for a subgraph $G = (V, E)$ of some graph $\mathbb{G} = (\mathbb{V}, \mathbb{E})$, we shall denote by $\partial_e G$ the edge boundary of $\mathbb{G}$, i.e. the set of edges in $\mathbb{E}$ with one end-point in $V$ and one end-point outside it. Similarly, the vertex boundary $\partial_v G$ of $G$ is the set of vertices in $V$ incident to edges in $\partial_e G$. The Potts Hamiltonian with boundary condition $b$, where $b = 0$ corresponds to free boundary conditions and $b = 1$ corresponds to boundary condition of all spins pointing in the same direction, is defined as follows.

**Definition 2.2.** *For a finite subgraph $G = (V, E)$ of an infinite graph $\mathbb{G}$, $b \in \{0, 1\}$, and $\sigma \in \mathbb{T}_q^V$, we define the Hamiltonian*

$$\mathcal{H}^b(\sigma) = -\left( \sum_{e=(i,j) \in E} \langle \sigma_i, \sigma_j \rangle + \mathbb{1}_{\{b=1\}} \sum_{\substack{e=(i,j) \in \partial_e G \\ i \in V}} \langle \sigma_i, \mathbf{1} \rangle \right)$$

*We define the q-state Potts Model partition function with boundary condition $b \in \{0, 1\}$, inverse temperature $\beta$, and external field $h$ as*

$$Z^{b,q}_{\beta,h}(G) = \sum_{\sigma \in \mathbb{T}^V_q} e^{-\beta \mathcal{H}^b(\sigma) + h \sum_i \langle \sigma_i, \mathbf{1} \rangle}$$

*The probability of a configuration $\sigma \in \mathbb{T}^V_q$ is then given by*

$$\mu^{b,q}_{\beta,h}[\{\sigma\}] = \frac{1}{Z^{b,q}_{\beta,h}(G)} e^{-\beta \mathcal{H}^b(\sigma) + h \sum_i \langle \sigma_i, \mathbf{1} \rangle}.$$

The random-cluster model is a graphical representation of the Potts model which is of independent interest. In particular, its correlation functions define an interpolation between those of the Potts model for non-integer $q$.

**Definition 2.3.** *For a finite subgraph $G = (V, E)$ of an infinite graph $\mathbb{G}$, $q \geq 1$, vector $\boldsymbol{p} = \{p_e\}_{e \in E} \in [0, 1]^E$ and partition $\xi$ of $\partial_v G$, the random-cluster measure on $G$ with boundary condition $\xi$ is the probability measure $\phi^\xi_{\boldsymbol{p},q,G}$ on $\{0, 1\}^E$ which, to each $\omega \in \{0, 1\}^E$, assigns the probability*

$$\phi^\xi_{\boldsymbol{p},q,G}[\{\omega\}] = \frac{1}{Z^\xi_{\boldsymbol{p},q,G}} q^{\kappa^\xi(\omega)} \prod_{e \in E} p_e^{\omega(e)} (1 - p_e)^{1 - \omega(e)},$$

*where $\kappa^\xi$ is the number of connected components in the graph*

$$G^\xi_\omega = (V, E_\omega) / \sim_\xi,$$

*$E_\omega := \{e \in E \mid \omega(e) = 1\}$, $\sim_\xi$ is the equivalence relation with equivalence classes given by $\xi$ and $Z^\xi_{\boldsymbol{p},q,G}$ is a normalising constant called the partition function.*
*$E_\omega$ is called the set of **open edges** and dually, $E \setminus E_\omega$ is called the set of **closed edges**.*

**Remark 2.4.** *Two boundary conditions are of special interest. These are the wired respectively free boundary conditions denoted as $\xi = 1$ and $\xi = 0$ respectively. $\xi = 1$ corresponds to the trivial partition where all boundary vertices belong to the same class and $\xi = 0$ corresponds to the trivial partition where all boundary vertices belong to distinct classes.*

**Proposition 2.5** (Domain Markov Property)**.** *If $G_1 = (V_1, E_1) \subseteq G_2 = (V_2, E_2)$ are two finite subgraphs of an infinite graph $\mathbb{G}$, we write $\omega_1 := \omega|_{E_1}$ and $\omega_2 := \omega \mid_{E_2 \setminus E_1}$. Then,*

$$\phi^\xi_{\boldsymbol{p},q,G_2}[\omega_1 \in A \mid \omega_2] = \phi^{\xi_{\omega_2}}_{\boldsymbol{p},q,G_1}[A]$$

*where $v$ and $w$ belong to the same element of $\xi_{\omega_2}$ if and only if they are connected by a path (that might possibly have length 0) in $((V_2 \setminus V_1), E_{\omega_2}) / \sim_\xi$.*

**Definition 2.6.** *For an infinite graph $\mathbb{G} = (\mathbb{V}, \mathbb{E})$, we say that a probability measure $\phi_{\boldsymbol{p},q}$ on $\{0, 1\}^{\mathbb{E}}$ is an **infinite-volume random-cluster measure** on $\mathbb{G}$ if, for any finite subgraph $G = (V, E)$ of $\mathbb{G}$, we write $\omega_1 = \omega|_E$ and $\omega_2 = \omega|_{E^c}$ and have*

$$\phi_{\boldsymbol{p},q}[\omega_1 \in A \mid \omega_2] = \phi^{\xi_{\omega_2}}_{\boldsymbol{p},q,G}[A],$$

*where $v$ and $w$ belong to the same element of $\xi_{\omega_2}$ if and only if they are connected by a path in $(\mathbb{V} \setminus V, E_{\omega_2})$.*

**Remark 2.7.** *Two natural infinite volume measures occur as monotonic limits of finite volume ones. For an increasing sequence $G_n = (V_n, E_n)$ with $\mathbb{G} = \cup_{n=1}^{\infty} G_n$, we define*

$$\phi^1_{\boldsymbol{p},q,\mathbb{G}}[A] = \lim_{n \to \infty} \phi^1_{\boldsymbol{p},q,G_n}[A]$$

$$\phi^0_{\boldsymbol{p},q,\mathbb{G}}[A] = \lim_{n \to \infty} \phi^0_{\boldsymbol{p},q,G_n}[A]$$

*for all increasing[1] events A depending only on finitely many edges (so that the probabilities on the right-hand side are well-defined eventually). It is easy to check that the limit does not depend on the choice of sequence $G_n$. Since such events are intersection-stable and generate the product $\sigma$-algebra of $\{0,1\}^{\mathbb{E}}$, this determines the two (possibly equal) infinite volume measures uniquely. That these limits define probability measures is a standard consequence of Banach-Alaoglu, since $\{0,1\}^{\mathbb{E}}$ is compact.*

For our purposes, the infinite graph $\mathbb{G} = (\mathbb{V}, \mathbb{E})$ will always be an augmented version of $\mathbb{Z}^d$ where we add an extra vertex $\mathfrak{g}$, called the *ghost vertex*, such that the set of vertices for each $d \geq 2$ becomes $\mathbb{V} = \mathbb{Z}^d \cup \{\mathfrak{g}\}$ and the set of edges is

$$\mathbb{E} = \{(x,y) \in \mathbb{Z}^d| \; \|x - y\|_\infty = 1\} \cup \{(x, \mathfrak{g})| \; x \in \mathbb{Z}^d\},$$

with the former set denoting the usual nearest neighbour edge set $\mathbb{E}_d$ of $\mathbb{Z}^d$, which we shall refer to as the *inner edges*, and the latter denoting the so-called *ghost edges* $\mathbb{E}_{\mathfrak{g}}$ after Griffiths (see, for example, [21]).

Similarly, we will work with the implicit assumption that a finite subgraph $G = (V, E)$ of $\mathbb{G}$ always has edge sets of the form

$$E = \{(x,y) \in \mathbb{E}| \; x, y \in V\},$$

i.e. $G$ is the subgraph induced by $V$.

This also gives a natural partition of $E = E_{\text{in}} \cup E_{\mathfrak{g}}$ similar to that of $\mathbb{E}$.

One class of finite subgraphs of special interest is the class of boxes

$$\Lambda_k(v) = \{w \in \mathbb{Z}^d| \; \|v - w\|_\infty \leq k\} \cup \{\mathfrak{g}\},$$

for a fixed vertex $v \in \mathbb{V}$.

The connection of the random-cluster model to the Potts Model goes through the Edwards-Sokal coupling (see, for instance, [21]). For any finite subgraph $G$ of $\mathbb{G}$, we have

$$Z^b_{\boldsymbol{p},q,G} = e^{-\beta|E| - h|V|} Z^{b,q}_{\beta,h,G}$$

for the specific choice of edge parameters

$$p_e = \begin{cases} p := 1 - \exp\left(-\frac{q}{q-1}\beta\right) & e \in \mathbb{E}_d \\ p_h := 1 - \exp\left(-\frac{q}{q-1}h\right) & e \in \mathbb{E}_{\mathfrak{g}}. \end{cases} \tag{1}$$

---

[1] In the interest of the flow of the article, we have postponed the formal definition to Definition 2.9 below.

In keeping with this, we will simply write measure

$$\phi_{p,q,h,G}^{\xi}[\{\omega\}] = \frac{1}{Z_{p,q,h,G}^{\xi}} p^{o(\omega_{in})}(1-p)^{c(\omega_{in})} p_h^{o(\omega_{\mathfrak{g}})}(1-p_h)^{c(\omega_{\mathfrak{g}})},$$

with $o(\omega_{in})$, respectively $c(\omega_{in})$, denoting the number of open, respectively closed edges, in $E_{\text{in}}$ - and similarly for the ghost edges.

**Remark 2.8.** *The case $q = 1$, henceforth called Bernoulli percolation, is somewhat particular. Here, the state of the edges becomes a product measure and (1) no longer gives a translation between an external field strength $h$ and an edge parameter $p_h$. As such, for Bernoulli percolation, we shall instead directly write*

$$\mathbb{P}_{p,p_h,G} := \phi_{\boldsymbol{p},1,G},$$

*where $p_e = p$ for $e \in E_{\text{in}}$ and $p_e = p_h$ for $e \in E_{\mathfrak{g}}$.*
*Note that, by independence, $\mathbb{P}_{p,p_h,G}$ is the same for any boundary condition and thus, we drop it from the notation.*

Throughout, we shall think of $h$ as an enhancing parameter boosting the number of interior edges rather than being an edge parameter. This is because the percolation properties of the ghost vertex $\mathfrak{g}$ itself are trivial. In other words, when studying the random-cluster model, if $\omega \sim \phi_{p,q,h,\mathbb{G}}$, we are interested in the percolation phase transition of the marginal $\omega_{in} := \omega|_{\mathbb{E}_d}$, the distribution of which we will denote simply by $\phi_{p,q,h,\mathbb{Z}^d}$. More precisely, we define the critical parameter $p_c = p_c(q,h,d)$ as

$$p_c := \inf\{p \mid \theta(p,q,h) > 0\},$$

where $\theta(p,q,h) = \phi_{p,q,h,\mathbb{Z}^d}^1[0 \leftrightarrow \infty]$ is the probability that 0 is part of an infinite connected component of inner edges (the cluster of the ghost vertex $\mathfrak{g}$ is trivially infinite almost surely for $h > 0$). When $\theta(p,q,h) > 0$ (and correspondingly, an infinite cluster exists almost surely), we say that the model *percolates*.

The random-cluster model has many nice properties. First of all, it is a graphical representation of the Potts model in the sense that for all $x, y \in V$ of some subgraph $G$ of $\mathbb{G}$, whether finite or infinite, (see [21, (1.5)]),

$$\phi_{p,q,h,G}^b[x \leftrightarrow y] = \mu_{\beta,h,G}^{b,q}[\sigma_x \cdot \sigma_y],$$

where $x \leftrightarrow y$ denotes the event that there is an open path connecting $x$ and $y$ in $(V, E_\omega)$ and $b \in \{0, 1\}$. For any graph $G = (V, E)$ (finite or infinite), there is a natural partial order $\preceq$ on the space of percolation configurations $\{0, 1\}^E$. We say that $\omega \preceq \omega'$ if $\omega(e) \leq \omega'(e)$ for all $e \in E \cup E_{\mathfrak{g}}$. This furthermore gives us a notion of events which respect the partial order.

**Definition 2.9.** *We call an event $A$ increasing if, for any $\omega \in A$, it holds that $\omega \preceq \omega'$ implies $\omega' \in A$.*
*For two percolation measures $\phi_1, \phi_2$ we say that $\phi_1$ is stochastically dominated by $\phi_2$ if $\phi_1[A] \leq \phi_2[A]$ for all increasing events $A$. We will also denote this order relation as $\phi_1 \preceq \phi_2$.*

The study of increasing events turns out to be natural due to the fact that they are all positively correlated (see [21]).

**Proposition 2.10** (FKG). *For any two increasing events $A, B$, any $(p, q, h)$, any boundary condition $\xi$ and graph $G$, we have*

$$\phi_{p,q,h,G}^{\xi}[A]\phi_{p,q,h,G}^{\xi}[B] \leq \phi_{p,q,h,G}^{\xi}[A \cap B].$$

*In particular, $\phi_{p,q,h,G}^{\xi} \preceq \phi_{p,q,h,G}^{\xi}(\cdot|A)$.*

Furthermore, the random-cluster model carries many other natural monotonicity properties in the sense of stochastic domination (see [21]), making it a natural dependent percolation processes to study.

**Theorem 2.11.** *For the random-cluster model on some subgraph $G$ of $\mathbb{G}$ (finite or infinite), the following relations hold:*

i) $\phi_{p_0,q_0,h_0,G}^{\xi}$ *is monotonic in $\xi$ in the sense that if $\xi'$ is a finer partition than $\xi$, we have*

$$\phi_{p,q,h,G}^{\xi'} \preceq \phi_{p,q,h,G}^{\xi}$$

*for any parameters $(p, q, h)$.*

ii) $\phi$ *is increasing in $p$, $h$ and decreasing in $q$, i.e. if $p' \geq p$ and $h' \geq h$ and $q' \leq q$, we have*

$$\phi_{p,q,h,G}^{\xi} \preceq \phi_{p',q',h',G}^{\xi}$$

*for any boundary condition $\xi$.*

iii) *The random-cluster model is comparable to Bernoulli percolation in the following sense: for $p_e$ still given by (1), we have*

$$\mathbb{P}_{\tilde{p},\tilde{p}_h,G} \preceq \phi_{p,q,h,G}^{\xi} \preceq \mathbb{P}_{p,p_h,G}$$

*for any boundary condition $\xi$, where $\tilde{p} = \frac{p}{p+q(1-p)}$ and $\tilde{p}_h = \frac{p_h}{p_h+q(1-p_h)}$.*

It follows from *iii)* that for $d \geq 2, q \geq 1, h \geq 0$, the random-cluster model always has a non-trivial phase transition, meaning that $p_c \in (0, 1)$.

Furthermore, any of these stochastic domination relations can be realised as a so-called *increasing coupling*. That is, for any two of the above measures $\mu$ and $\nu$, if $\mu \preceq \nu$, there exists a measure $\mathbb{P}$ on $\{(\omega_1, \omega_2) | \omega_1, \omega_2 \in \{0,1\}^E\}$ with the property that $\mathbb{P}[\omega_1 \preceq \omega_2] = 1$ and for any event $A$,

$$\mathbb{P}[\omega_1 \in A] = \mu[A]$$
$$\mathbb{P}[\omega_2 \in A] = \nu[A]$$

For an explicit construction, see [21, Lemma 1.5].

## Definition of the Kertész line

In the following we introduce the Kertész line following [17], see also [1, 15].

**Definition 2.12.** *Suppose that $q$ and $d$ are fixed. Then the Kertész line is defined by*

$$h_c(p) = h_c(p, q, d) = \sup\{h \geq 0 \mid \theta(p, q, h) = 0\},$$

*i.e. $h_c(p)$ is the largest $h$ such that $p \leq p_c(q, h, d)$.*

The facts that $\{0 \leftrightarrow \infty\}$ is an increasing event and $\phi^1_{p,q,h,\mathbb{Z}^d}$ is monotone in $p$ imply that the Kertész line is monotonically decreasing in $p$. Furthermore, translation invariance as well as monotonicity of $\phi^1_{p,q,h,\mathbb{Z}^d}$ in $h$ implies that the Kertész line separates $\{(p, h) \mid p \geq 0, h \geq 0\}$ into two regions with and without an infinite cluster of inner edges. If $p > p_c := p_c(q, 0, d)$, then monotonicity in $h$ implies that $h_c(p) = 0$. Stochastic domination of the random-cluster model by Bernoulli bond percolation (see Theorem 2.11) implies that if $p < p_B := p_c(1, 0, d)$, then $h_c(p) = \infty$ (see [17], Section 3).

In [17, Theorem 7-8], it is proven that if $p \in (p_B, p_c)$, then $0 < h_c(p) < \infty$ except for the case where $p$ is close to $p_c$ where strict positivity of the Kertész line is not proven. This will be proven in the next section, thereby settling the question of whether the Kertész line for the random-cluster model on $\mathbb{Z}^d$ is always non-trivial.

## Probability of connecting to the ghost given configuration of inner edges

In the following, we generalise [10, Lemma 2.4], which is stated only in the case of FK-Ising (i.e. $q = 2$), to the setting of the random-cluster model for $q \in [1, \infty)$. This generalisation, which is straight forward, was first noted in the appendix of [22]. The lemma and some of the bounds that follow from it are more easily stated if we define the function $\tanh_q : [0, \infty) \to [0, 1)$ by

$$\tanh_q(x) = \frac{1 - e^{-2x}}{(q-1)e^{-2x} + 1}.$$

Notice that $\tanh_2 = \tanh$, that $\tanh_q$ is strictly increasing, satisfies $\tanh_q([0, \infty)) = [0, 1)$ and has an inverse which we call $\operatorname{arctanh}_q : [0, 1) \to [0, \infty)$ given by

$$\operatorname{arctanh}_q(x) = \frac{1}{2} \log \left( \frac{(q-1)x + 1}{1 - x} \right).$$

Then, we can state the following lemma.

**Lemma 2.13.** *Suppose that $G$ is a finite graph and that a configuration of inner edges $\omega_{in}$ has clusters $C_1, \ldots, C_n$. Then, for each $1 \leq i \leq n$,*

$$\phi^0_{p,q,h,G}[C_i \leftrightarrow \mathfrak{g} \mid \omega_{\text{in}}] = \tanh_q(h|C_i|)$$

*and the events $\{C_i \leftrightarrow \mathfrak{g}\}$ are mutually independent given $\omega_{\text{in}}$.*

*Proof.* First, we look at inner edges and compute

$$Z^0_{p,q,h,G}\phi^0_{p,q,h,G}[\{\omega_{\text{in}}\}] = p^{o(\omega_{\text{in}})}(1-p)^{c(\omega_{\text{in}})}q\prod_{j=1}^{n}\left(\sum_{f\in\{0,1\}^{C_j}}p_h^{o(f)}(1-p_h)^{c(f)}(q\,\mathbb{1}_{\{C_j\nleftrightarrow\mathfrak{g}\}}+\mathbb{1}_{\{C_j\leftrightarrow\mathfrak{g}\}})\right)$$

with $o(f)$ respectively $c(f)$ denoting the number of open respectively closed ghost edges in $C_i$. From this decomposition, conditional independence follows. Now, to obtain the desired form, note that we get a factor of $q$ from the cluster $C_i$ if and only if all the edges are closed and thus, we can write the sum as

$$\sum_{f\in\{0,1\}^{C_i}}p_h^{o(C_i(f))}(1-p_h)^{c(C_i(f))}(q\,\mathbb{1}_{\{C_i\nleftrightarrow\mathfrak{g}\}})+\mathbb{1}_{\{C_i\leftrightarrow\mathfrak{g}\}})$$

$$= q(1-p_h)^{|C_i|}+\sum_{f\in\{0,1\}^{C_i}}p_h^{o(C_i(f))}(1-p_h)^{c(C_i(f))}-(1-p_h)^{|C_i|}$$

$$= (q-1)(1-p_h)^{|C_i|}+1 = (q-1)e^{-2h|C_i|}+1.$$

This then means that

$$Z^0_{p,q,h,G}\phi^0_{p,q,h,G}[\{\omega_{\text{in}}\}] = p^{o(\omega_{\text{in}})}(1-p)^{c(\omega_{\text{in}})}q\prod_{j=1}^{n}\left((q-1)e^{-2h|C_j|}+1\right).$$

Then, we are ready to compute

$$\phi^0_{p,q,h,G}[C_i\leftrightarrow\mathfrak{g}\mid\omega_{\text{in}}] = \frac{\phi^0_{p,q,h,G}[\{C_i\leftrightarrow\mathfrak{g}\}\cap\{\omega_{\text{in}}\}]}{\phi^0_{p,q,h,G}[\{\omega_{\text{in}}\}]} = \frac{1-e^{-2h|C_i|}}{(q-1)e^{-2h|C_i|}+1} = \tanh_q(h|C_i|),$$

which proves the formula. □

Since the lemma is true uniformly in finite volume it transfers to the infinite volume limit.

## 3 Strict monotonicity

We prove that the Kertész line is strictly monotone in all of its parameters. All the following results are uniform in the boundary conditions, so we shall suppress them for ease of notation. The proofs all follow the strategy from [19] rather closely.

In doing so, we will need the following elementary lemma:

**Lemma 3.1.** *Let $G=(V,E)$ be a finite graph, $F$ some finite, totally ordered set, $\nu$ a probability measure on $\{\eta:E\to F\}$ and $(U_j)_{1\leq j\leq|E|}$ a family of i.i.d. uniform random variables. For any enumeration $(e_j)_{1\leq j\leq|E|}$ of $E$, if we recursively define*

$$X(e_1) = \min\left\{f\in F\,\Big|\,U_1\leq\sum_{g\leq f}\nu[\eta(e_1)=g]\right\}$$

$$X(e_{j+1}) = \min\left\{f\in F\,\Big|\,U_{j+1}\leq\sum_{g\leq f}\nu[\eta(e_{j+1})=g|\,\eta(e_i)=X(e_i)\,\forall i\leq j]\right\},$$

*then* $X \sim \nu$.

*Proof.* This follows from the Law of Total Probability as soon as we establish that $X(e)$ has the correct marginal for every $e$. For $e = e_1$, we simply see that

$$\mathbb{P}[X(e_1) = f] = \mathbb{P}\left[\sum_{g < f} \nu[\eta(e_1) = g] < U_1 \leq \sum_{g \leq f} \nu[\eta(e_1) = g]\right] = \nu[\eta(e_1) = f].$$

The same calculation shows that $X(e_{j+1})$ has the correct conditional law given $(X(e_i))_{1 \leq i \leq j}$ and thus, the proposition follows by finite induction. $\qquad\square$

Next, in order to prove Theorem 1.1, we prove the following technical proposition, which is in the spirit of [19, Theorem 2.3] (where a similar comparison result between $p-$ and $q-$derivatives at $h = 0$ is proved).

**Proposition 3.2.** *There exist strictly positive smooth functions* $\alpha, \gamma : (0,1) \times (0,\infty) \times (1,\infty) \to \mathbb{R}$ *such that, for any finite subgraph $G$ of $\mathbb{G}$ containing $\mathfrak{g}$ and any increasing event $A$ depending only on the edges of $G$, we have*

$$\alpha(p,q,h) \frac{\partial}{\partial h} \phi_{p,q,h,G}[A] \leq \frac{\partial}{\partial p} \phi_{p,q,h,G}[A] \leq \gamma(p,q,h) \frac{\partial}{\partial h} \phi_{p,q,h,G}[A].$$

*Proof.* We are going to give the explicit construction of $\gamma$. The construction of $\alpha$ is analogous. A direct calculation shows that

$$\frac{\partial}{\partial p} \phi_{p,q,h,G}[A] = \frac{1}{p(1-p)} \operatorname{Cov}[o(\omega_{\text{in}}), \mathbb{1}_A]$$

$$\frac{\partial}{\partial h} \phi_{p,q,h,G}[A] = \frac{1}{p_h(1-p_h)} \operatorname{Cov}[o(\omega_{\mathfrak{g}}) \mathbb{1}_A].$$

If $(\omega_1, \omega_2)$ denotes an increasing coupling with marginals $\omega_1 \sim \phi_{p,q,h,G}$ and $\omega_2 \sim \phi_{p,q,h,G}(\cdot|A)$, we get that

$$\operatorname{Cov}[o(\omega_{\text{in}}), \mathbb{1}_A] = \phi_{p,q,h,G}[A]\mathbb{E}[o(\omega_{2,\text{in}}) - o(\omega_{1,\text{in}})]$$

$$\operatorname{Cov}[o(\omega_{\mathfrak{g}}), \mathbb{1}_A] = \phi_{p,q,h,G}[A]\mathbb{E}[o(\omega_{2,\mathfrak{g}}) - o(\omega_{1,\mathfrak{g}})].$$

Thus, we are finished if we can establish that $\mathbb{E}[o(\omega_{2,\text{in}}) - o(\omega_{1,\text{in}})]$ and $\mathbb{E}[o(\omega_{2,\mathfrak{g}}) - (\omega_{1,\mathfrak{g}})]$ are comparable.

This, however, turns out to be easy, because

$$\mathbb{E}[o(\omega_{2,\text{in}}) - o(\omega_{1,\text{in}})] = \sum_{e \in E_{\text{in}}} \mathbb{P}[\omega_2(e) = 1, \omega_1(e) = 0] \qquad (2)$$

$$\mathbb{E}[o(\omega_{2,\mathfrak{g}}) - o(\omega_{1,\mathfrak{g}})] = \sum_{e \in E_{\mathfrak{g}}} \mathbb{P}[\omega_2(e) = 1, \omega_1(e) = 0]. \qquad (3)$$

Throughout, for $e \in E$, we let $B_e$ denote the event $[\omega_2(e) = 1, \omega_1(e) = 0]$. For $v \in V \setminus \{\mathfrak{g}\}$, let $B_v$ denote the event that all neighbouring edges of $v$ in $\mathbb{Z}^d$ are closed in $\omega_1$ and $\mathfrak{G}_v$ denote the event that the ghost edge $(v, \mathfrak{g})$ is open in $\omega_1$.

If $v$ and $w$ are the two end-points of $e$, we claim that

$$\mathbb{P}[B_v \cap \mathfrak{G}_w | B_e] \geq (1 - p)^{2d-1} \frac{p_h}{p_h + q(1 - p_h)}.$$

To see this, we apply Lemma 3.1 with $F = \{(0,0), (1,0), (1,1)\}$ (with the obvious ordering) and any enumeration such that $e_1 = e$, $e_2 = (w, \mathfrak{g})$ and $e_3, ..., e_{2d+1}$ are some enumeration of the other neighbours of $v$. Allowing ourselves the slight abuse of notation of keeping our letters, we get that

$$\mathbb{P}[B_v \cap \mathfrak{G}_w | B_e] \geq \mathbb{P}\left[\left\{U_2 \geq \frac{p_h}{p_h + q(1 - p_h)}\right\} \cap \bigcap_{j=3}^{2d+1} \{U_j \leq (1 - p)\}\,\middle|\, B_e\right],$$

where we have used $iii$) in Theorem 2.11 to bound the conditional distribution of $\omega_1$ from above and from below by Bernoulli percolation. To establish the above claim, we simply note that $B_e$ is measurable with respect to the $\sigma$-algebra generated by $U_1$.

Now, on the event $B_v \cap \mathfrak{G}_w \cap B_e$, the end-points of $(v, \mathfrak{g})$ are disconnected in $\omega_1$ and so, we get

$$\mathbb{P}[\omega_1((v, \mathfrak{g})) = 1 | B_e \cap B_v \cap \mathfrak{G}_w] = \phi^0_{p,q,h,\{v,\mathfrak{g}\}}[(v, \mathfrak{g})] = \frac{p_h}{p_h + q(1 - p_h)},$$

by applying Lemma 3.1 in a similar fashion.

Meanwhile, the end-points are connected in $\omega_2$, so that

$$\mathbb{P}[\omega_2((v, \mathfrak{g}) = 1) | B_e \cap B_v \cap \mathfrak{G}_w] \geq \phi^1_{p,q,h,\{v,\mathfrak{g}\}}[(v, \mathfrak{g})] = p_h,$$

where we have used the FKG-inequality (Proposition 2.10). All in all, we get that

$$\mathbb{P}[B_{(v,\mathfrak{g})} | B_e] \geq \left(p_h - \frac{p_h}{p_h + q(1 - p_h)}\right) \frac{p_h}{p_h + q(1 - p_h)} (1 - p)^{2d-1} := \gamma'$$

Thus, for every inner edge $e$, pick an end-point $v(e)$. Since any given vertex $v$ belongs to at most $2d$ different inner edges, we get that

$$\gamma' \mathbb{E}[o(\omega_{2,\text{in}}) - o(\omega_{1,\text{in}})] \leq \sum_{e \in E_{\text{in}}} \mathbb{P}[B_{(v(e),\mathfrak{g})} | B_e] \mathbb{P}[B_e] \leq 2d \sum_{v \in V} \mathbb{P}[B_{(v,\mathfrak{g})}] \leq 2d \sum_{e \in E_{\mathfrak{g}}} \mathbb{P}[B_e].$$

Defining $\gamma := 2d \frac{p_h(1 - p_h)}{\gamma' p(1 - p)}$ completes the construction. $\qquad\square$

Meanwhile, the following is a direct analogue of [19, Theorem 2.3] for the case with $h > 0$.

**Proposition 3.3.** *There exist strictly positive smooth functions* $\alpha, \gamma : (0, 1) \times (0, \infty) \times (1, \infty) \to \mathbb{R}$ *such that, for any finite subgraph $G$ of $\mathbb{G}$ containing $\mathfrak{g}$ and any increasing event $A$ depending only on the edges of $G$, we have*

$$-\alpha(p, q, h)\frac{\partial}{\partial q}\phi_{p,q,h,G}[A] \leq \frac{\partial}{\partial p}\phi_{p,q,h,G}[A] + \frac{\partial}{\partial h}\phi_{p,q,h,G}[A] \leq -\gamma(p, q, h)\frac{\partial}{\partial q}\phi_{p,q,h,G}[A].$$

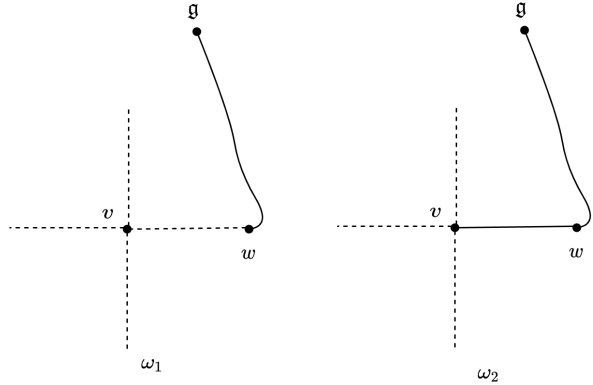

Figure 1: Sketch of the event $B_v \cap \mathfrak{G}_w \cap B_e$ from the proof of Proposition 3.2. With dotted lines we denote closed edges and with solid lines open edges. The edge in question $(v, \mathfrak{g})$ is not depicted. The probability that it is open differs in the two cases $\omega_1$ and $\omega_2$.

*Proof.* Similarly to the previous result, one obtains via a direct calculation that

$$\frac{\partial}{\partial q} \phi_{p,q,h,G}[A] = \frac{1}{q} \text{Cov}[\mathbb{1}_A, \kappa],$$

where $\kappa$ denotes the number of components of a configuration (counted according to the appropriate boundary conditions). Once again, letting $(\omega_1, \omega_2)$ denote some increasing coupling of $\omega_1 \sim \phi_{p,q,h,G}$ and $\omega_2 \sim \phi_{p,q,h,G}[\cdot|A]$, we get

$$-\text{Cov}[\mathbb{1}_A, \kappa] = \phi_{p,q,h,G}[A]\mathbb{E}[\kappa(\omega_1) - \kappa(\omega_2)].$$

It is immediate that

$$\kappa(\omega_1) - \kappa(\omega_2) \leq o(\omega_{2,\text{in}}) - o(\omega_{1,\text{in}}) + o(\omega_{2,\mathfrak{g}}) - o(\omega_{1,\mathfrak{g}}),$$

which allows one to construct $\alpha$ using (2) from the proof of the previous proposition.

For the other inequality, let $N_I(\omega_1, \omega_2)$ denote the number of vertices which are isolated in $\omega_1$ but not in $\omega_2$. Then, again, since $\omega_1 \preceq \omega_2$, it holds that

$$\kappa(\omega_1) - \kappa(\omega_2) \geq \frac{1}{2} N_I(\omega_1, \omega_2) = \frac{1}{2} \sum_{v \in V} \mathbb{1}_{I_v},$$

where $I_v$ is the event that $v$ is isolated in $\omega_1$ but not in $\omega_2$. Just like before, for every edge $e$ incident to $v$, one can argue the existence of some smooth function $\gamma'$ such that

$$\mathbb{P}[I_v|B_e] \geq \gamma'.$$

Summing over them all, we get that

$$\mathbb{P}[I_v] \geq \frac{\gamma'}{2d} \sum_{e \sim v} \mathbb{P}[B_e],$$

Using that any edge is incident to exactly two vertices, we have

$$\mathbb{E}[\kappa(\omega_1) - \kappa(\omega_2)] \geq \frac{2\gamma'}{4d} \sum_{e \in E} \mathbb{P}[B_e] \geq \frac{\gamma'}{2d} \left( \mathbb{E}[o(\omega_{2,\mathrm{in}}) - o(\omega_{1,\mathrm{in}})] + \mathbb{E}[o(\omega_{2,\mathfrak{g}}) - o(\omega_{1,\mathfrak{g}})] \right),$$

from which $\gamma$ may be constructed easily. $\qquad\square$

Finally, we are in position to prove the main result.

*Proof of Theorem 1.1 :* We are going to prove that $h \mapsto p_c(q, h)$ is strictly decreasing. The other statements are proven completely analogously. Consider $q$ fixed and let $\alpha(p, h)$ be the function from Proposition 3.2. For $(p, h)$, let $\ell_{p,h} = (\ell_{p,h}^1, \ell_{p,h}^2) : (T_{p,h}^-, T_{p,h}^+) \to (0, 1) \times (0, \infty)$ be the integral curve of the vector field $(-1, \alpha)$ started at $(p, h)$, i.e. a maximal solution to

$$\frac{d}{dt} \ell_{p,h}(t) = (-1, \alpha(\ell_{p,h}(t)))$$
$$\ell_{p,h}(0) = (p, h).$$

The existence of such a solution is completely standard ODE fare. For instance, one may appeal to the Picard-Lindelöf Theorem.

Now, for any finite subgraph $G$ of $\mathbb{G}$ containing $\mathfrak{g}$ and any increasing event $A$, we find that

$$\frac{d}{dt} \phi_{\ell_{p,h}^1(t), q, \ell_{p,h}^2(t), G}[A] = \frac{d}{dt} \ell_{p,h}(t) \cdot \nabla \phi_{\ell_{p,h}^1(t), q, \ell_{p,h}^2(t), G}[A] \leq 0,$$

by construction of $\alpha$. Since $A$ was arbitrary, we conclude that $\phi_{\ell_{p,h}^1(t_1), q, \ell_{p,h}^2(t_1), G} \preceq \phi_{\ell_{p,h}^1(t_2), q, \ell_{p,h}^2(t_2), G}$ for $t_1 < t_2$. Furthermore, since $G$ was arbitrary, this extends to any infinite volume limits.

Let $h < h'$ be such that $\ell_{p_c(h), h}(t)$ intersects $\mathbb{R} \times \{h'\}$ and denote by $(\tilde{p}, h')$ the point of intersection. Since $\alpha > 0$, the second coordinate of $\ell_{p,h}(t)$ is strictly increasing in $t$ and the first coordinate strictly decreasing. Thus, $\tilde{p} < p_c(h)$.

For any $p > \tilde{p}$, let $(\hat{p}, h)$ denote the intersection of $\ell_{p,h'}(t)$ with $\mathbb{R} \times \{h\}$. Then, since the paths of the different integral curves are either identical or non-intersecting, we must have that $\hat{p} > p_c(h)$. Thus, we have that $\phi_{\hat{p}, q, h}$ almost surely percolates.

However, $\phi_{\hat{p}, q, h} \preceq \phi_{p, q, h'}$, so we conclude that $\phi_{p, q, h'}$ almost surely percolates. We conclude that

$$p_c(h') \leq \tilde{p} < p_c(h),$$

which is what we wanted. $\qquad\square$

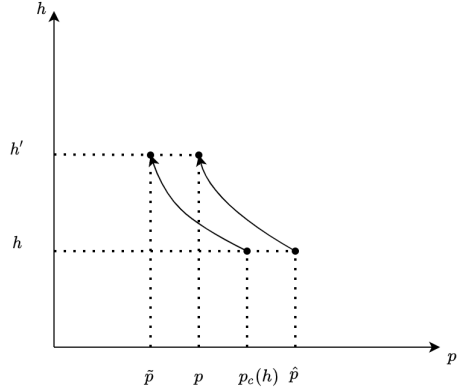

Figure 2: The points $(p_c(h), h)$, $(\tilde{p}, h')$, $(p, h')$, $(\hat{p}, h)$ as well as the integral curves from the proof of Theorem 1.1.

# 4 Stochastic Domination and bounds

In this section, we investigate conditions for stochastic domination. Since the event $\{0 \leftrightarrow \infty\}$ is increasing, stochastic domination results directly transfer to bounds on the Kertész line. Here and in the following, let $r_i = \frac{p_i}{1-p_i}$ and $H_i = \frac{p_{h_i}}{1-p_{h_i}}$ for $i \in \{1, 2\}$.

**Theorem 4.1.** *Suppose that $0 \leq p_2 \leq p_1 \leq 1$, $q_i \in [1, \infty)$, $h_i \geq 0$ for $i \in \{1, 2\}$ and that for all positive integers $n, m$ it holds that*

$$\frac{r_2}{q_2} \left( (q_2 - 1) \tanh_{q_2}(nh_2) \tanh_{q_2}(mh_2) + 1 \right) \leq \frac{r_1}{q_1} \left( (q_1 - 1) \tanh_{q_1}(nh_1) \tanh_{q_1}(mh_1) + 1 \right).$$

*Then, for any increasing event $A$ that only depends on the inner edges, we have*

$$\phi_{p_2, q_2, h_2}[A] \leq \phi_{p_1, q_1, h_1}[A].$$

Letting $p_1 = p_2$ and $h_2 = 0$ and using that $\tanh_{q_1}$ is increasing, we obtain the following results from which our bounds will follow.

**Corollary 4.2.** *Suppose that $q_1, q_2 \in [1, \infty)$ and $h_1 \geq 0$ are such that*

$$\frac{1}{q_2} \leq \frac{1}{q_1} \left( (q_1 - 1) \tanh_{q_1}(h_1)^2 + 1 \right).$$

*Then, for any increasing event $A$ that only depends on the inner edges,*

$$\phi_{p, q_2, 0}[A] \leq \phi_{p, q_1, h_1}[A].$$

In order to obtain the result, we modify an argument from [2, Theorem 3.21].

*Proof of Theorem 4.1.* Again, we prove the result uniformly over all finite graphs $G$. This then transfers to any infinite-volume limits.

Let $X$ be an increasing random variable which only depends on inner edges. Then,

$$\phi_{p_j,q_j,h_j,G}[X] = \frac{1}{Z_j} \sum_{\omega_{\mathrm{in}} \in \{0,1\}^{E_{\mathrm{in}}}} r_j^{o(\omega_{\mathrm{in}})} X(\omega_{\mathrm{in}}) \sum_{\omega_\partial \in \{0,1\}^{E_\partial}} H_j^{o(\omega_\partial)} q_j^{\kappa(\omega)},$$

where we have abbreviated $Z_j := (1 - p_j)^{|E|} (1 - p_{h_j})^{|V|} Z_{p_j,q_j,h_j,G}^\xi$.

Now, define the random variable $Y$ depending on inner edges $\omega_{\mathrm{in}}$ by

$$Y(\omega_{\mathrm{in}}) = \frac{r_1^{o(\omega_{\mathrm{in}})} \sum_{\omega_\partial} H_1^{o(\omega_\partial)} q_1^{\kappa(\omega)}}{r_2^{o(\omega_{\mathrm{in}})} \sum_{\omega_\partial} H_2^{o(\omega_\partial)} q_2^{\kappa(\omega)}}.$$

This lets us rewrite

$$\phi_{p_1,q_1,h_1,G}[X] = \frac{Z_2}{Z_1} \frac{1}{Z_2} \sum_{\omega_{\mathrm{in}}} X[\omega_{\mathrm{in}}] Y(\omega_{\mathrm{in}}) r_2^{o(\omega_{\mathrm{in}})} \sum_{\omega_\partial} H_2^{o(\omega_\partial)} q_2^{\kappa(\omega)} = \frac{Z_2}{Z_1} \phi_{p_2,q_2,h_2,G}[XY].$$

Letting $X = 1$, one gets

$$1 = \phi_{p_1,q_1,h_1,G}[1] = \frac{Z_2}{Z_1} \sum_{\omega_{\mathrm{in}}} Y(\omega_{\mathrm{in}}) r_2^{o(\omega_{\mathrm{in}})} \sum_{\omega_\partial} H_2^{o(\omega_\partial)} q_w^{\kappa(\omega)},$$

from which we can conclude

$$\phi_{p_1,q_1,h_1,G}[X] = \frac{\phi_{p_2,q_2,h_2,G}[XY]}{\phi_{p_2,q_2,h_2,G}(Y)}.$$

Now, if we can prove that $Y$ is increasing, then by FKG, we get that $\phi_{p_2,q_2,h_2,G}[X] \leq \phi_{p_1,q_1,h_1,G}[X]$, since $X$ is increasing. Since $X$ was arbitrary, this would prove the stochastic domination between the marginals on $E_{\mathrm{in}}$.

Therefore, in the following, we focus on $Y$ and whether it is increasing. Let an edge $e = (x,y)$ be given. Then, the condition for $Y$ to be increasing is $Y(\omega_{\mathrm{in}}) \leq Y(\omega_{\mathrm{in}}^e)$ where $\omega^e$ denotes the configuration $\omega$ where we have opened the edge $e$ if it was closed in $\omega$. If we let $\{x \leftrightarrow y\}$ denote the event that the end-points $x, y$ of $e$ are connected (possibly using the ghost), then

$$\kappa(\omega^e) = \kappa(\omega) \, \mathbb{1}_{\{x \leftrightarrow y\}}(\omega) + (\kappa(\omega) - 1) \, \mathbb{1}_{\{x \nleftrightarrow y\}}(\omega).$$

Thus, we get that $Y$ is increasing if and only if

$$Y(\omega_{\mathrm{in}}) \leq Y(\omega_{\mathrm{in}}^e) = \frac{r_1^{o(\omega_{\mathrm{in}}^e)} \sum_{\omega_\partial} H_1^{o(\omega_\partial)} \left( q_1^{\kappa(\omega)} \, \mathbb{1}_{\{x \leftrightarrow y\}}(\omega) + \frac{1}{q_1} q_1^{\kappa(\omega)} \, \mathbb{1}_{\{x \nleftrightarrow y\}}(\omega) \right)}{r_2^{o(\omega_{\mathrm{in}}^e)} \sum_{\omega_\partial} H_2^{o(\omega_\partial)} \left( q_2^{\kappa(\omega)} \, \mathbb{1}_{\{x \leftrightarrow y\}}(\omega) + \frac{1}{q_2} q_2^{\kappa(\omega)} \, \mathbb{1}_{\{x \nleftrightarrow y\}}(\omega) \right)}.$$

Notice that for $i \in \{1, 2\}$ any event $A$ and configuration of internal edges $\omega_{\mathrm{in}}$, then

$$\phi_{p_i, q_i, h_i, G}[A \mid \omega_{\mathrm{in}}] = \frac{\phi_{p_i, q_i, h_i, G}[A \cap \{\omega_{\mathrm{in}}\}]}{\phi_{p_i, q_i, h_i, G}[\{\omega_{\mathrm{in}}\}]} = \frac{r_i^{o(\omega_{\mathrm{in}})} \sum_{\omega_{\mathfrak{d}}} H_i^{o(\omega_{\mathfrak{d}})} q_i^{\kappa(\omega)} \mathbb{1}_A}{r_i^{o(\omega_{\mathrm{in}})} \sum_{\omega_{\mathfrak{d}}} H_i^{o(\omega_{\mathfrak{d}})} q_i^{\kappa(\omega)}} = \frac{\sum_{\omega_{\mathfrak{d}}} H_i^{o(\omega_{\mathfrak{d}})} q_i^{\kappa(\omega)} \mathbb{1}_A}{\sum_{\omega_{\mathfrak{d}}} H_i^{o(\omega_{\mathfrak{d}})} q_i^{\kappa(\omega)}},$$

which makes the inequality above equivalent to

$$r_2 \left( \left(1 - \frac{1}{q_2}\right) \phi_{p_2, q_2, h_2, G}[x \leftrightarrow y \mid \omega_{\mathrm{in}}] + \frac{1}{q_2} \right) \leq r_1 \left( \left(1 - \frac{1}{q_1}\right) \phi_{p_1, q_1, h_1, G}[x \leftrightarrow y \mid \omega_{\mathrm{in}}] + \frac{1}{q_1} \right). \tag{4}$$

If $x$ and $y$ are connected with the inner edges then the condition is just $r_2 \leq r_1$ which again is equivalent to $p_1 \geq p_2$. If we let $C_x$ be the cluster of $x$ in $\omega_{\mathrm{in}}$, then we can rewrite the previous inequality using Lemma 2.13 to obtain

$$\frac{r_2}{q_2} \left( (q_2 - 1) \tanh_{q_2}(|C_x|h_2) \tanh_{q_2}(|C_y|h_2) + 1 \right) \leq \frac{r_1}{q_1} \left( (q_1 - 1) \tanh_{q_1}(|C_x|h_1) \tanh_{q_1}(|C_y|h_1) + 1 \right).$$

The main statement now follows since $|C_x|$ and $|C_y|$ are positive integers. $\qquad\square$

## 4.1  Upper bound on the Kertész line

For any dimension $d \geq 2$ the random-cluster model has a phase transition for $q \in [1, \infty)$ at some $p_c(q)$. For example it is proven [4] for $d = 2$ that $p_c(q, 0) = \frac{\sqrt{q}}{1 + \sqrt{q}}$. It is proven in [19] that $p_c(q)$ is strictly monotone and Lipschitz continuous in $q$ as a function from $[1, \infty) \to [p_B, 1)$ for all $d \geq 2$.[2] Therefore, it has an inverse function $q_c : [p_B, 1) \to [1, \infty)$ such that $(p, q_c(p))$ is critical. When $d = 2$, inverting the relation before yields $q_c(p, 0) = \left(\frac{p}{1-p}\right)^2$.

*Proof of Theorem 1.3.* We know that for every $\varepsilon > 0$ then there is percolation at $(p + \varepsilon, q_c(p), 0)$. Then, by Theorem 4.1 we get that there is also percolation for $(p + \varepsilon, q, h)$ as long as

$$\frac{1}{q_c(p, 0)} \leq \frac{1}{q} \left( (q - 1) \tanh_q(h)^2 + 1 \right).$$

Then, for fixed $q$, since $\tanh_q$ is increasing, the minimal $h$ where the inequality is satisfied is the $h$ that satisfies the equality $\sqrt{\frac{1}{q-1} \left(\frac{q}{q_c(p,0)} - 1\right)} = \tanh_q(h)$. Using again that $\tanh_q$ is strictly monotone yields the theorem. $\qquad\square$

Analytically, we can see that the bound has the correct behaviour in the limits. The bound becomes infinite whenever $\frac{1}{q-1} \left(\frac{q}{q_c(p,0)} - 1\right) = 1$, i.e. when $q_c(p, 0) = 1$, which again means that $p = p_B$. Similarly, the bound tends to 0 whenever $\frac{1}{q-1}\left(\frac{q}{q_c(p,0)} - 1\right) = 0$, i.e. $q = q_c(p, 0)$, which means that $p = p_c(q, 0)$. In Figure 3 we have plotted this upper bound. Notice that, in contrast to the bounds from [1], it gives the correct limit when $h \to 0$.

---

[2] One may see that $p_c(q)$ is surjective as follows: For any $p$, the probability of crossing the annulus $\Lambda_3 \setminus \Lambda_1$ can be made arbitrarily small by increasing $q$. The methods from Section 4.3 may then be employed to see that there is no percolation at $(p, q)$.

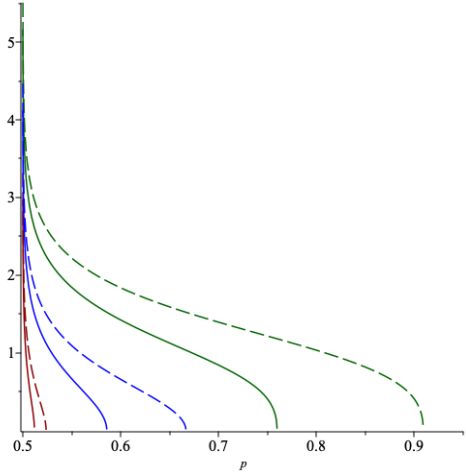

Figure 3: The upper bound for the Kertész line in $d = 2$ proven in Theorem 1.3 (solid) as well as Theorem 4.4 (dashed) for $q = 1.1, 2, 10$ in red, blue and green respectively. The solid upper bound tends to $p_c(q, 0)$ in the limit $h \to 0$. All the bounds have the correct limit as $h \to \infty$.

**Remark 4.3.** *One may note that our techniques allow one to use something about the random-cluster model for $q \neq 2$ to gain information about the Ising model where $q = 2$.*

### Upper bound using the Bernoulli percolation threshold

Before, we used knowledge about the phase transition at $h = 0$ to infer knowledge about the Kertész line at all $h \geq 0$. In the following, we use a similar trick where we use that if $p = p_B + \varepsilon$ where $p_B(d)$ is the critical $p$ of Bernoulli percolation, then there is percolation without using the ghost for $h$ sufficiently large. In $d = 2$ we can then use Kesten's celebrated result that $p_B(2) = \frac{1}{2}$ (see [23]). For $d = 2$ the method does not produce a better bound than the bound in Theorem 1.3, but if one only has knowledge about the phase transition of Bernoulli percolation in higher dimensions and not the random-cluster model, it can produce a better bound than Theorem 1.3.

**Theorem 4.4.** *Let $q \in [1, \infty)$ and $r_B = \frac{p_B}{1 - p_B}$ where $p_B(d)$ is the critical parameter of Bernoulli*

percolation in $d$ dimensions. Then, the following upper bound on the Kertész line holds:

$$h_c(p) \leq \operatorname{arctanh}_q \left( \sqrt{\frac{1}{q-1} \left( q r_B \frac{1-p}{p} - 1 \right)} \right).$$

In particular, if $d = 2$ and $q = 2$ corresponding to the planar Ising model we obtain

$$h_c(p) \leq \operatorname{arctanh} \left( \sqrt{2 \frac{1-p}{p} - 1} \right).$$

*Proof.* In Theorem 4.1 we let $(p_1, q_1, h_1) = (p_1, q, h_1)$ for some $p_1 \geq p_B$ as well as $(p_2, q_2, h_2) = (p_B + \varepsilon, q, N)$ for some small $\varepsilon > 0$ and arbitrarily large $N$. Then the condition for stochastic domination is that

$$r_2 \left( (q-1) \tanh_q(h_2 n) \tanh_q(h_2 m) + 1 \right) \leq r_1 \left( (q-1) \tanh_q(h_1 n) \tanh_q(h_1 m) + 1 \right) \qquad (5)$$

for all positive integers $n, m$. Notice that we have $\tanh_q(h_2 n) \leq 1$ and therefore

$$r_2 \left( (q-1) \tanh_q(h_2 n) \tanh_q(h_2 m) + 1 \right) \leq q r_2.$$

Since $n, m \geq 1$ then $\tanh_q(h_1 n) \geq \tanh_q(h_1)$ and thus

$$r_1 \left( (q-1) \tanh_q(h_1)^2 + 1 \right) \leq r_1 \left( (q-1) \tanh_q(h_1 n) \tanh_q(h_1 m) + 1 \right).$$

Thus, it is sufficient for (5) and therefore stochastic domination that

$$q r_2 \leq r_1 \left( (q-1) \tanh_q(h_1)^2 + 1 \right).$$

Picking $r_2 = r_B = \frac{p_B}{1 - p_B}$ as well as recalling that $r_1 = \frac{p}{1-p}$ we obtain that it is sufficient that

$$\operatorname{arctanh}_q \left( \sqrt{\frac{1}{q-1} \left( q r_B \frac{1-p}{p} - 1 \right)} \right) \leq h_1.$$

To finish the proof, we note that there is percolation at $(p_2, q_2, h_2)$ so long as $N$ is chosen large enough. □

## 4.2 Lower bound on the Kertész line

In this section, we give a lower bound on the Kertész line following the strategy in [20] for proving exponential decay of cluster sizes (which, in particular, implies a lack of percolation). The arguments rely on nothing but sharpness of the subcritical phase (as is known from [5]).

**Lemma 4.5.** *Let $S$ be a finite subset of $\mathbb{Z}^d$. Then, there exists a subset $T_S \subseteq S$ such that $|T_S| \geq \frac{|S|}{4^d}$ and for every $v, w \in T_S$, we have that $\|v - w\|_{L^1} \geq 4$.*

**Remark 4.6.** *By considering $S = \Lambda_k$ and letting $k \to \infty$, we get that the bound is sharp.*

*Proof.* Note that

$$S = \bigcup_{\tau \in [0,3]^d} S \cap (\tau + 4\mathbb{Z}^d),$$

implying that there exists $\tau_0 \in [0,3]^d$ such that $|S \cap (\tau_0 + 4\mathbb{Z}^d)| \geq \frac{|S|}{4^d}$. It is easily seen that $T_S := S \cap (\tau_0 + 4\mathbb{Z}^d)$ also has the desired separation property. $\square$

In the following, we let $A_n$ denote the set of connected subsets of $\mathbb{Z}^d$ containing 0 and exactly $n$ vertices. It is a classic result that $|A_n|$ has an exponential growth rate. For instance, Lemma 5.1 in [24] shows that $|A_n| \leq \left(\frac{(2d+1)^{2d+1}}{(2d)^{2d}}\right)^n := \mu^n$.

*Proof of Theorem 1.4.* For each $v \in 2k\mathbb{Z}^d$, define $X(v) = \mathbb{1}[\Lambda_k(v) \leftrightarrow \partial \Lambda_{3k}(v)]$, which is a site percolation process on $2k\mathbb{Z}^d$.

Given a $v \in 2k\mathbb{Z}^d$, we let $C_\omega(v)$ be the cluster of $v$ in $\omega_{in}$, $C_X(v)$ be the set of $w \in 2k\mathbb{Z}^d$ such that $d(w, C_\omega(v)) \leq k$. Note that if $|C_\omega(v)| > 3^d |\Lambda_k|$, then every $w \in C_X(v)$ is open in $X$. Furthermore, if $|C_\omega(v)| \geq N$ then $|C_X(v)| \geq \left\lfloor \frac{N}{|\Lambda_k|} \right\rfloor := n$.

To account for the magnetic field, we will need to control the density of ghost edges. For each $v \in 2k\mathbb{Z}^d$, we say that $v$ is *good* if every ghost edge in $\Lambda_{3k}(v)$ is closed. Otherwise, we say that $v$ is *bad*. Interchangeably, we will say that the box itself is good respectively bad. We denote the process of bad boxes by $\mathcal{B}$, i.e. $\mathcal{B}(v) = \mathbb{1}[v \text{ is bad}]$.

We will now let $N > 3^d |\Lambda_k|$ and bound the quantity $\phi_{p,q,h}(|C_\omega(v)| \geq N) \leq \phi_{p,q,h}(|C_X(v)| \geq n)$. If $|C_X(v)| \geq n$, then we know that there is an open connected $S$ in $X$ containing $v$ and exactly $n$ vertices. Hence, by a union bound,

$$\phi_{p,q,h,\mathbb{G}}[|C_X(v)| \geq n] \leq \sum_{\substack{S \subseteq 2k\mathbb{Z}^d, \\ S/(2k) \in A_n}} \phi_{p,q,h}[S \text{ open in } X].$$

Now, by Lemma 4.5, we can pick a thinned set $T_S$ of at least $\frac{|S|}{4^d}$ vertices, such that for $w, w' \in T_S$, we have $\Lambda_{3k}(w) \cap \Lambda_{3k}(w') = \{\mathfrak{g}\}$ [3]. For a given $w \in T_S$, we have, by the Domain Markov Property, that if $E_k^c(w)$ denotes the complement of the edges between vertices in $\Lambda_{3k}(w)$,

$$\phi_{p,q,h,\mathbb{G}}[X(w) = 1|\ \omega|_{E_k^c(w)}] \leq \phi_{p,q,h,\mathbb{G}}[\mathcal{B}(w) = 1|\ \omega|_{E_k^c(w)}] + \phi_{p,q,0,\Lambda_{3k}(w)}^1[X(w) = 1].$$

Furthermore, by Theorem 2.11, we have

$$\phi_{p,q,h,\mathbb{G}}[\mathcal{B}(w) = 1|\ \omega|_{E_k^c(w)}] \leq 1 - (1 - p_h)^{|\Lambda_{3k}|}.$$

Consequently,

$$\phi_{p,q,h,\mathbb{G}}\left[\bigcap_{w \in T_S} \{X(w) = 1\}\right] \leq \prod_{w \in T_S} \left(\phi_{p,q,0,\Lambda_{3k}(w)}^1[X(w) = 1] + 1 - (1 - p_h)^{|\Lambda_{3k}|}\right).$$

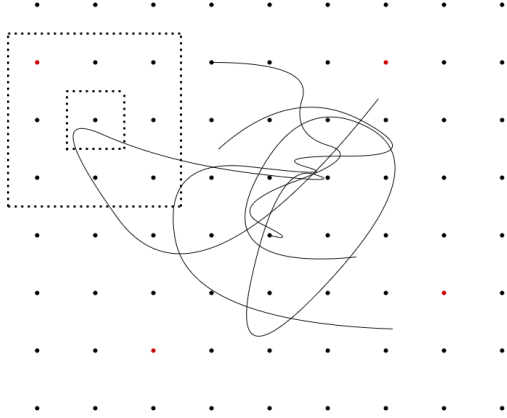

Figure 4: A representation of the coarse-graining scheme in the proof of Theorem 1.4. The dots illustrate the grid $2k\mathbb{Z}^2$. With the dotted lines we indicate the boxes $\Lambda_k(v)$ and $\Lambda_{3k}(v)$ around a point $v$. Notice that since the cluster $S$ enters the box $\Lambda_k(v)$ this means that $X(v) = 1$. With the red dots we indicate bad sites in $2k\mathbb{Z}^2$. In the proof, we use that, for sufficiently small $h$, these bad sites are very uncommon.

Adding all of this together yields

$$
\begin{aligned}
\phi_{p,q,h,\mathbb{G}}\big[|C_X(v)| \geq n\big] &\leq \sum_{\substack{S \subseteq 2k\mathbb{Z}^d, \\ S/(2k) \in A_n}} \phi_{p,q,h,\mathbb{G}}[S \text{ open in } X] \\
&\leq \sum_{\substack{S \subseteq 2k\mathbb{Z}^d, \\ S/(2k) \in A_n}} \phi_{p,q,h,\mathbb{G}}[\cap_{w \in T_S}\{X(w) = 1\}] \\
&\leq \sum_{\substack{S \subseteq 2k\mathbb{Z}^d, \\ S/(2k) \in A_n}} \prod_{w \in T_S} \big(\phi^1_{p,0,\Lambda_{3k}(w)}[X(w) = 1] + 1 - (1 - p_h)^{|\Lambda_{3k}|}\big) \\
&\leq \sum_{\substack{S \subseteq 2k\mathbb{Z}^d, \\ S/(2k) \in A_n}} \big(\phi^1_{p,q,0,\Lambda_{3k}(w)}[X(w) = 1] + 1 - (1 - p_h)^{|\Lambda_{3k}|}\big)^{|S|/4^d} \\
&\leq \Big(\mu^{4^d}\big(\phi^1_{p,q,0,\Lambda_{3k}(0)}[X(0) = 1] + 1 - (1 - p_h)^{|\Lambda_{3k}|}\big)\Big)^{n/4^d},
\end{aligned}
$$

---

[3]Notice that, since $\Lambda_{3k}(w)$ and $\Lambda_{3k}(w')$ only have a single vertex in common, the wired measure on the union is a product measure of the wired measure on each box.

which decays to 0 if $1 - (1-p_h)^{|\Lambda_{3k}|} < \frac{\delta}{2}$. This is attained for

$$p_h < 1 - \left(1 - \frac{1}{2\mu^{4^d}}\right)^{1/|\Lambda_{3k}|} = 1 - \left(1 - \frac{\delta}{2}\right)^{1/|\Lambda_{3k}|}.$$

□

We conclude this section with a discussion of the bounds and the tangent at $(p_c, q, 0)$.

## 4.3 Discussion of the tangent at $(p_c, q, 0)$ for $d = 2$

We now use the correlation length to get a slightly more explicit lower bound under the conjectural assumption of the existence of critical exponents. In the following we consider fixed $q \in [1, 4]$. As in [6, (1.5)] we define the correlation length $\xi$ for $h = 0$ for $p < p_c$ by

$$\xi = - \lim_{n\to\infty} \frac{n}{\log\left(\phi^1_{p,q,0}\left(0 \leftrightarrow \partial\Lambda_n\right)\right)}.$$

By a union bound, we have

$$\phi^1_{p,q,0}\left[0 \leftrightarrow \partial\Lambda_{3n}\right] \leq \phi^1_{p,q,0,\Lambda_{3n}}\left[\Lambda_n \leftrightarrow \partial\Lambda_{3n}\right] \leq \phi^1_{p,q,0,\Lambda_{3n}}\left[\cup_{x\in\partial_v\Lambda_n}(x \leftrightarrow \partial\Lambda_{3n})\right] \leq 8n\phi^1_{p,q,0,\Lambda_{3n}}[0 \leftrightarrow \partial\Lambda_{2n}]$$

implying that

$$e^{-\frac{3n}{\xi}+o(n)} \leq \phi^1_{p,q,0,\Lambda_{3n}}\left(\Lambda_n \leftrightarrow \partial\Lambda_{3n}\right) \leq e^{-\frac{2n}{\xi}+o(n)}.$$

It is further expected that there exist constants $C_1, C_2, C > 0$ such that

$$C_1 e^{-C\frac{n}{\xi}} \leq \phi^1_{p,q,0,\Lambda_{3n}}\left(\Lambda_n \leftrightarrow \partial\Lambda_{3n}\right) \leq C_2 e^{-C\frac{n}{\xi}}. \tag{6}$$

uniformly in $p$ for fixed $q$. In the following, we will assume (6). Then,

$$C_2 e^{-C\frac{n}{\xi}} < \frac{\mu^{-16}}{2}$$

is satisfied for

$$n > \frac{16}{C}\xi\log(\mu) + C_3.$$

So choose $k = \frac{16}{C}\xi\log(\mu) + C_3 + 1$. Now, by Bernoulli's inequality, we see that

$$\frac{1}{18\mu^{16}k^2} < 1 - \left(1 - \frac{1}{2\mu^{16}}\right)^{1/|\Lambda_{3k}|}.$$

Thus, it is sufficient for the assumption in Theorem 1.4 that

$$p_h < \frac{1}{18\mu^{16}k^3} = \frac{1}{18\mu^{16}(\frac{16}{C}\xi\log(\mu) + C_3 + 1)^2} = \frac{1}{C_4\xi^2 + O(\xi)}.$$

in the limit $\xi \to \infty$. Now, in the planar case $d = 2$, the correlation length $\xi(p)$ is conjectured [6] to have the form

$$\xi(p) \sim C|p - p_c|^{-\nu}$$

for some critical $q$ dependent exponent $\nu$ which has the form

$$\nu(q) = \frac{2\arccos\left(-\frac{\sqrt{q}}{2}\right)}{6\arccos\left(-\frac{\sqrt{q}}{2}\right) - 3\pi} \in \left[\frac{2}{3}, \frac{4}{3}\right]$$

for $q \in [1, 4]$. In the particular case of the Ising model, the conjecture is that $\nu(2) = 1$. Thus, under that conjecture using $\mu = \frac{5^5}{4^4}$, our condition becomes

$$p_h < \frac{1}{18\mu^{16}k^2} = C|p - p_c|^{2\nu}$$

for some constant $C > 0$. Using the fact that $p_h(h) = 1 - e^{-2h}$ is approximately linear in $h$ for small $h$, we see that the tangent of the bound is asymptotically flat as $p \to p_c$. The bound has horizontal tangent and this holds for all $1 \leq q \leq 4$ since it is conjectured that $2\nu(q) > 1$ for all $q \in [1, 4]$.

In conclusion, we notice that both our upper and lower bounds tend to the correct value $(p_c, q, 0)$ as $h \to 0$. Thereby, the bounds complement those of [1], that are best for large $h$, i.e. around the Bernoulli percolation threshold. However, in our bounds the asymptote for the lower bound is horizontal and for the upper bound it is vertical as shown in Figure 3. Since the lower bound is asymptotically horizontal, this leaves open the natural question of what the inclination of the tangent is at the point $(p_c, q, 0)$.

Numerical evidence for the planar Ising case [25] observed $\frac{\beta}{\beta_c} - 1 \sim ch^{\kappa}$ in the limit $h \to 0$ for a $\kappa = 0.534(3)$, where it is noticed that $\frac{1}{\beta\delta} = \frac{8}{15} \approx 0.533$ (for the definition of the critical exponents $\beta, \delta$ see [25]). Now, using that $\beta - \beta_c$ and $p - p_c$ have a linear relationship when both are small yields that $p - p_c \sim ch^{\kappa}$ or equivalently that

**Conjecture 4.7.** *In the limit $p \to p_c$ it holds for some constant $c > 0$ that*

$$h_c(p) \sim c(p - p_c)^{\frac{15}{8}}.$$

This indicates that the lower bound is almost optimal and one would have to improve the upper bound. In particular this leaves plenty of further work concerning the asymptotic behaviour of the Kertész line for $h \to 0$. A potential starting point for this program could be [6, Lemma 8.5] which was also used in [12] to gain knowledge about the correlation length in a non-zero magnetic field.

# 5 Continuity of the Kertész line phase transition

There are several ways to define continuity of a phase transition: One relates to whether or not $\theta(p_c) = 0$ (or, equivalently, $\mu^1_{\beta_c}(\sigma_0 \cdot \mathbf{1}) = 0$) and another to whether the infinite-volume measures $\phi^1_{p_c}$ and $\phi^0_{p_c}$ coincide or not (or whether $\mu^1_{\beta_c} = \mu^0_{\beta_c}$). A third perspective pertains to the regularity of the pressure

$$\mathfrak{Z} = \lim_{n\to\infty} \frac{\log(Z^1(\Lambda_n))}{|\Lambda_n|},$$

where $Z^1$ denotes the partition function of either the random-cluster model or the Potts model. Indeed, one can show that $\mathfrak{Z}$ is convex as a function of $p$ (or $\beta$) and hence, admits left and right derivatives everywhere (see [21, Exercise 21]). These take the form

$$\frac{\partial}{\partial p^+}\mathfrak{Z} = \phi^1_{p,q,h}[\omega_e] \qquad \frac{\partial}{\partial p^-}\mathfrak{Z} = \phi^0_{p,q,h}[\omega_e],$$

where $\phi^b_{p,q,h}[\omega_e]$ denotes the probability that a given inner edge $e$ is open [4].
Furthermore, by Proposition 4.6 in [2], $\phi^1_{p,q,h,\mathbb{Z}^d} = \phi^0_{p,q,h,\mathbb{Z}^d}$ if and only if these probabilities agree. Accordingly, these two measures are different at $p_c$ if and only if $\mathfrak{Z}$ fails to be $C^1$. Furthermore, non-uniqueness of the infinite volume measure would imply that $\theta(p_c) > 0$. To see this last implication, one might simply note that for any increasing event $A$ depending only on finitely many edges,

$$\phi^1_{p,q,h,\mathbb{Z}^d}[A \cap (0 \not\leftrightarrow \infty)] = \sup_k \phi^1_{p,q,h,\mathbb{Z}^d}[A \cap (0 \not\leftrightarrow \Lambda_k)] \leq \sup_k \phi^0_{p,q,h,\Lambda_k}[A] = \phi^0_{p,q,h,\mathbb{Z}^d}[A]$$

and hence, $\phi^1_{p,q,h} = \phi^0_{p,q,h}$ if $\phi^1_{p,q,h}[0 \leftrightarrow \infty] = 0$. In particular, this means that a discontinuous geometric phase transition implies a (first order) thermodynamic one. By contraposition, this means that the analyticity of the pressure, which we prove below, rules out only thermodynamic phase transitions.

The converse direction, that $\theta(p_c) > 0$ if and only $\phi^1_{p_c} \neq \phi^0_{p_c}$, is subtle, since a general proof would imply that $\theta(p_c, 1, 0) = 0$, which remains perhaps the single largest open question in all of percolation theory. Indeed, in [26], an example is given of several random-cluster models which have unique infinite volume measures at $p_c$, but such that $\theta(p_c) > 0$, meaning that the converse being true would have to rely upon the specific structure of $\mathbb{Z}^d$.
Thus, we shall spend this section focusing on the characterisation of phase transition via regularity of the pressure $\mathfrak{Z}$. Following [14], we extract a cluster expansion for the random-cluster model in order to prove Theorem 1.5.

For the proof, we start with the case of the Potts model.

**Lemma 5.1.** *For any finite subgraph $G$ of an infinite graph $\mathbb{G}$ and $\beta, h > 0$, we have*

$$Z^{1,q}_{\beta,h}(G) = e^{\beta|E|}e^{h|V|} \sum_{G'=(V',E')\subseteq G} e^{-\beta|E'|}\tilde{Z}^{0,q-1}_{\beta,0}(G')e^{-(1+\frac{1}{q-1})(h|V'|+\beta|\partial_e G'|)}$$

$$:= e^{\beta|E|+h|V|}\Xi^q_{\beta,h}(G),$$

---

[4]By translation invariance of the infinite-volume measure, this probability does not depend on the choice of inner edge.

where $\tilde{Z}^{0,q-1}$ is constructed by taking spins amongst the $q-1$ spins in $\mathbb{T}_q$ different from $\mathbf{1}$.

*Proof.* For $\sigma \in \mathbb{T}_q^V$, let $G'(\sigma)$ denote the subgraph of $G$ with vertex-set $\{i \in V \mid \sigma_i \neq \mathbf{1}\}$. Then, clearly

$$Z_{\beta,h}^{1,q}(G) = \sum_{G'=(V',E')\subseteq G} \sum_{\substack{\sigma \in \mathbb{T}_q \\ G'(\sigma)=G'}} e^{-\beta\mathcal{H}^1(\sigma)+h\sum_{i\in V}\langle\sigma_i,\mathbf{1}\rangle}$$

$$= \sum_{G'=(V',E')\subseteq G} e^{\beta|E\setminus(E'\cup\partial_e G')|} e^{h|V\setminus V'|} e^{-\frac{\beta}{q-1}|\partial_e G'|} e^{-\frac{h}{q-1}|V'|} \sum_{\sigma\in(\mathbb{T}_q\setminus\{\mathbf{1}\})^{V'}} e^{-\beta H^0(\sigma)}$$

$$= e^{\beta|E|} e^{h|V|} \sum_{G'=(V',E')\subseteq G} e^{-h(1+\frac{1}{q-1})|V'|} e^{-\beta(1+\frac{1}{q-1})|\partial_e G'|} e^{-|E'|} \tilde{Z}_{\beta,0}^{0,q-1},$$

which is what we wanted. $\qquad\square$

If $H$ and $H'$ are two finite, disjoint subgraphs of a larger graph $G$, then clearly, $\tilde{Z}_{\beta,h}^{b,q}(H \cup H') = \tilde{Z}_{\beta,h}^{b,q}(H)\tilde{Z}_{\beta,h}^{b,q}(H')$. Hence, by decomposing $G'$ into a tuple of connected components $\mathcal{S} = (S_1, S_2, ..., S_m)$, we get that

$$\Xi_{\beta,h}^q(G) = \sum_{\mathcal{S}\in\Gamma} \prod_{S\in\mathcal{S}} w_{h,\beta}^q(S) \prod_{S\neq S'\in\mathcal{S}} \delta(S,S'),$$

where $\Gamma$ is the set of connected subgraphs of $G$ and for $S = (V_S, E_S)$,

$$w_{\beta,h}^q(S) = e^{-\beta|E_S|}(S)e^{-(1+\frac{1}{q-1})(h|V_S|+\beta|\partial_e S|)}\tilde{Z}_{\beta,0}^{0,q-1}$$

and

$$\delta(S,S') = \mathbb{1}_{\{d(S,S')>1\}},$$

where $d(S,S') = \inf_{x\in S, y\in S'} \|x-y\|_{\ell^1}$. Thus, we have written $\Xi_{\beta,h}^q(G)$ as the partition function of a polymer model with hardcore interactions.

It is worth noting that, since $q \geq 2$,

$$\tilde{Z}_{\beta,0}^{0,q-1}(S) \leq (q-1)^{|V_S|} e^{\beta|E_S|},$$

so that, in fact,

$$w_{\beta,h}^q(S) \leq (q-1)^{|V_S|} e^{-(1+\frac{1}{q-1})(h|V_S|+\beta|\partial_e S|)}.$$

**Lemma 5.2.** *Assume that $G \subseteq \mathbb{Z}^d$, let $B(S)$ denote the set of vertices within $\ell^1$ distance 1 of $S$ and $a(S) = |B(s)|$. Then, if $\Gamma^{\infty,d}$ is the set of all connected subgraphs of $\mathbb{Z}^d$ and $S'$ denotes some fixed element of $\Gamma$, we get that*

$$\sum_{S\in\Gamma} w_{\beta,h}^q(S)e^{a(S)}(1-\delta(S,S')) \leq a(S') \max_{j\in B(S')} \sum_{j\in S\in\Gamma^{\infty,d}} w_{\beta,h}^q(S)e^{a(S)}.$$

*Proof.* Note that the only positive contributions to the sum on the left-hand side come from $S$ such that $d(S, S') \geq 1$. Thus,

$$\sum_{S \in \Gamma} w_{\beta,h}^q(S) e^{a(S)}(1 - \delta(S, S')) \leq \sum_{j \in B(S')} \sum_{j \in S \in \Gamma} w_{\beta,h}^q(S) e^{a(S)},$$

from which the lemma follows immediately. $\qquad\square$

Recall the function $h_0$ from Theorem 1.5.

**Lemma 5.3.** *For any $h > h_0(q, d)$, we have*

$$\sum_{0 \in S \in \Gamma^{\infty,d}} w_{\beta,h}^q(S) e^{a(S)} < 1.$$

**Remark 5.4.** *By translation invariance, it suffices to consider $j = 0$.*

*Proof.* We see that

$$\sum_{0 \in S \in \Gamma^{\infty,d}} w_{\beta,h}^q(S) e^{a(S)} = \sum_{k=1}^{\infty} e^{-(1 + \frac{1}{q-1})hk} \sum_{\substack{0 \in S \in \Gamma^{\infty,d} \\ |V_S| = k}} e^{(a(S))} e^{-\beta|E_S|} \tilde{Z}_{\beta,0}^{0,q-1}(S) e^{-(1 + \frac{1}{q-1})\beta|\partial_e S|}$$

$$\leq \sum_{k=1}^{\infty} (q-1)^k e^{-(1 + \frac{1}{q-1})hk} e^{2dk} |\{S \in \Gamma^{\infty,d} | \ |V_S| = k, \ 0 \in V_S\}|$$

$$\leq \sum_{k=1}^{\infty} (q-1)^k e^{-(1 + \frac{1}{q-1})hk} e^{2dk} \left( \frac{(2d+1)^{2d+1}}{(2d)^{2d}} \right)^k,$$

where, in the last line, we have once again used Lemma 5.1 in [24]. The right hand side is, of course, a geometric sum which is convergent with a sum less than 1 for $h > h_0(q, d)$. Thus, this case of Theorem 1.5 follows from [14, Theorem 5.4]. $\qquad\square$

**Corollary 5.5.** *The above results carry over to the partition function of the random-cluster model with wired boundary conditions for non-integer $q$.*

*Proof.* For integer $q$, note that, by the Edwards-Sokal Coupling, we have

$$\Xi_{\beta,h}^q(G) = e^{-\beta|E|} e^{-h|V|} Z_{\beta,h}^{1,q}(G) = \sum_{\omega_{in} \in \{0,1\}^E} \sum_{\omega_\mathfrak{d} \in \{0,1\}^V} p^{o(\omega_{in})} (1-p)^{c(\omega_{in})} p_h^{o(\omega_\mathfrak{d})} (1-p_h)^{c(\omega_\mathfrak{d})} q^{\kappa^1(\omega, \omega_\mathfrak{d})}$$

for $\beta = -\frac{q-1}{q} \log(1-p)$ and $h = -\frac{q-1}{q} \log(1-p_h)$.

Note that only the pairing of $\beta$ and $p$ depends on the inner product in $\mathbb{T}_q$, whereas the role of the factor of $q^{\kappa(\omega)}$ is to cancel with the uniformly random colouring in the Edwards-Sokal coupling.

Therefore, when considering fewer colours, we still have

$$e^{-\beta|E_S|} \tilde{Z}_{\beta,0}^{0,q-1}(S) = \sum_{\omega \in \{0,1\}^{E_s}} p^{o(\omega)} (1-p)^{c(\omega)} (q-1)^{\kappa(\omega)},$$

Hence, redefining

$$w_{p,h}^q(S) = \sum_{\omega \in \{0,1\}^{E_s}} p^{o(\omega)} (1-p)^{c(\omega)} (q-1)^{\kappa(\omega)} e^{-(1+\frac{1}{q-1})(h|V_S| + \beta(p)|\partial_e S|)}$$

represents the random-cluster model partition function with wired boundary conditions as a polymer model for arbitrary $q \geq 1$. Here, we have inverted the formula (1) to get $\beta(p) = \frac{q-1}{q} \log(1-p)$.

For $q \geq 2$, since any vertex belongs to at most one component, it remains true that

$$\sum_{\omega \in \{0,1\}^{E_s}} p^{o(\omega)}(1-p)^{c(\omega)}(q-1)^{\kappa(\omega)} \leq \sum_{\omega \in \{0,1\}^{E_s}} p^{o(\omega)}(1-p)^{c(\omega)}(q-1)^{|V_S|} = (q-1)^{|V_S|},$$

and so, we retain the convergent cluster expansion above with the same bounds.

For $q \in (1,2)$, we instead use that $(q-1)^{\kappa(\omega)} \leq (q-1)$ to get

$$\sum_{\omega \in \{0,1\}^{E_s}} p^{o(\omega)}(1-p)^{c(\omega)}(q-1)^{\kappa(\omega)} \leq q-1.$$

Thus, in this case,

$$\sum_{0 \in S \in \Gamma^{\infty,d}} w_{p,h}^q(S) e^{a(S)} \leq (q-1) \sum_{k=1}^{\infty} e^{-(1+\frac{1}{q-1})hk} e^{2dk} \left( \frac{(2d+1)^{2d+1}}{(2d)^{2d}} \right),$$

and we once again get convergence for $h > h_0(q,d)$. $\qquad \square$

In [16], discontinuity was proven using the Pirogov-Sinai theory. There, it was proven for $d \geq 2, q \geq 1, h \geq 0$ that if

$$c_d(1 + (q-1)e^{-h})^{-\frac{1}{2d}} < 1$$

for some inexplicit constant $c_d$, which only depends on the dimension, then the phase transition on the Kertész line is discontinuous - i.e. $\mathfrak{Z}$ fails to be $C^1$. See also analogous results for the Potts model from [27]. To plot this in the $(q,h)$ plane, we can rearrange it into the condition that $h < \log\left( \frac{q-1}{c_d^{2d}-1} \right)$.

# 6 Outlook

We end by giving an outlook introducing further problems on Kertész line which lie in natural continuation of our work.

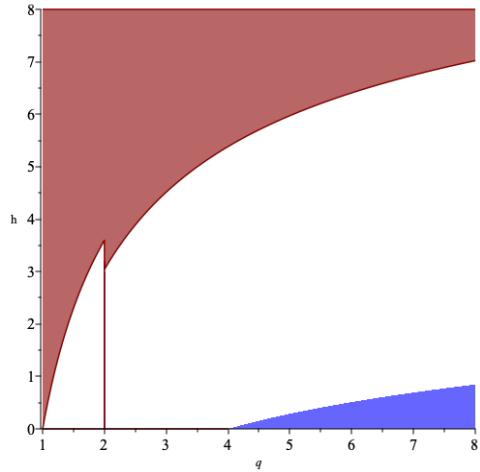

Figure 5: Plot of regions of discontinuity and continuity in the $(q, h)$ plane. For any $q \geq 1$ and $h \geq 0$ there is a unique $p(q, h)$ such that $(p(q, h), q, h)$ is critical. We colour the plane red or blue according to whether the phase transition at that point is a continuous or discontinuous. In Theorem 1.5 above we prove continuity of the phase transition in the contiguous region coloured red. Continuity on the line $q = 2$ follows from the Lee-Yang theorem [13] and for $h = 0$ and $q \in [1, 4]$ from the explicit proof [7]. The blue region is not proven to be discontinuous since the constant $c_d$ is not explicit. Instead, the blue region is obtained from using the best possible constant in the bound from [16] (this constant is $c_d = \sqrt{2}$ to match the change between continuous and discontinuous phase transition at $h = 0$ for $q = 4$).

**Continuity of discontinuity**

For $h = 0$, the quantity $\phi^1_{q,p_c(q)}(0 \leftrightarrow \infty) \searrow 0$ as $q \searrow 4$. We conjecture that this is also the case for $h > 0$:

**Conjecture 6.1.** *The set of $(q, h)$ such that the function $p \mapsto \mathfrak{Z}$ is not $C^1$ is open.*

One could choose to view this as continuity of the gap

$$\phi^1_{p_c(q,h),q,h,\mathbb{Z}^d}[0 \leftrightarrow \infty] - \phi^0_{p_c(q,h),q,h,\mathbb{Z}^d}[0 \leftrightarrow \infty].$$

Continuity of the gap for $h = 0$ was proven in [7, 8] by explicit computation of the critical magnetization which was continuous around $q = 4$. One might note that in the special case of $d = 2$, the techniques of [20] do apply more or less verbatim to the case where $h > 0$, implying that the regime of the exponential decay of the truncated wired measure is open in the $(p, h)$-plane. Seeing as this is a purely infinite volume phenomenon, however, makes attacking it quite delicate.

Furthermore, we conjecture the following.

**Conjecture 6.2.** *For each $q \in [1, \infty)$ the map $h \mapsto \phi^1_{p_c(q,h),q,h,\mathbb{Z}^d}[0 \leftrightarrow \infty]$ is decreasing.*

Under this conjecture, we would see that the discontinuity in the $(q, h)$ plane (as plotted on Figure 5) would be a contiguous region. This would further establish the existence of a tri-critical point defined by

$$h_c(q) = \inf\{h \geq 0 \mid \phi^1_{p_c(q,h),q,h,\mathbb{Z}^d}[0 \leftrightarrow \infty] = 0\}.$$

In the physics literature [28], there are some numerical studies which give rise to predictions of the universality class of this phase transition which at this point in time seems out of reach. One simple point that we can make along these lines is that if $1 \leq q \leq 4$ and $h \to 0$ then, by easy stochastic domination arguments, it holds that $\phi^1_{(p_c(q,h),q,h)}[0 \leftrightarrow \infty] \to 0$. The conjecture would imply that $\phi^1_{p_c(q,h),q,h,\mathbb{Z}^d}[0 \leftrightarrow \infty]$ is identically 0.

**Pseudo-critical line**

Since the Kertész line characterizes a geometric rather than a thermodynamic phase transition it is a priori not clear whether any signs of the phase transition transfer from the random-cluster model to the Potts model. Of course, in the regime of discontinuous phase transition (large $q$ and small $h$), the discontinuity implies that the free energy is not $C^1$ and therefore, the existence of a thermodynamic phase transition.

However, in the case where the Kertész line does not necessarily correspond to a thermodynamic phase transition we can ask whether any feature of the Kertész line can be observed in the Potts model. For general $q \in (1, \infty)$, one might, as in [25], define a *pseudocritical line* as the line where the susceptibility is maximal in $p$. Since divergent susceptibility is an indicator of criticality, it is a natural question to ask whether the psedocritical line and the Kertész line coincide, i.e. whether the susceptibility is maximal exactly on the Kertész line or not. If it is not the case, as it is conjectured in [25], it would be interesting to investigate whether, at $q > 4$, the susceptibility peaks at a different value than the critical line and why. Similarly, one might consider the line of the maximal correlation length and ask to what extent that line coincides with the Kertész line.

Finally, one may ask about the critical exponents of the Kertész line. For example, it is claimed in [25] that in the planar Ising case $q = d = 2$, the critical exponents correspond to Bernoulli percolation rather than those of the FK-Ising.

## Kertész line for the random current and loop O(1) models

It is also natural to consider the Kertész line problem for the random current and loop O(1) models. Let $P_h^{\emptyset}$ denote the random current measure and let $P_h^{\otimes 2}$ denote the sourceless double random current (see [21] for definitions of the measures). Just as we have done in this article, a magnetic field can be implemented with a ghost vertex, allowing us to once again consider the percolation phase transition. However, these models lack monotonicity [29], making the problem much more intricate. However, the monotonicity required to establish the existence of the Kertész line is much weaker than the overall monotonicity where the counterexamples of [29] apply. This leads us to the following conjectures that would establish the existence of the Kertész line for random currents.

**Conjecture 6.3.** *For any $d \geq 2$ the functions $h \mapsto P_h^{\emptyset}\left[0 \overset{\mathbb{Z}^d}{\leftrightarrow} \infty\right]$ and $h \mapsto P_h^{\otimes 2}\left[0 \overset{\mathbb{Z}^d}{\leftrightarrow} \infty\right]$ are increasing.*

For the loop O(1) model (see for example [30] for a definition in a magnetic field) the problem may be more intricate since the lack of monotonicity appears stronger [29]. We note that in the limit $h \to \infty$, the ghost edges are almost always open. In the limit, this changes the parity constraint of the marginal on the inner edges from percolation where all vertices are conditioned to have even degree into percolation where all vertices are conditioned to have odd degree.

If we denote the critical $p$ for odd and even percolation by $p_{c,\text{odd}}$ and $p_{c,\text{even}}$ respectively, then if it is the case that $p_{c,\text{odd}} > p_{c,\text{even}}$, this would be an illustration of how non-monotone the loop O(1) model is. We note that in the planar case $d = 2$ and $h = 0$, it is proven [31] that $p_{c,even} = 1 - p_c(2,0)$, but to our knowledge, nothing is known about odd percolation.

From this result it follows from the couplings ([21, Exercise 36]) that the single and double random currents also have the same phase transition for $d = 2$. Using these increasing couplings, one can infer bounds on the Kertész line between the models. Thus, bounding the Kertész line for random currents or the loop O(1) models may be a way to obtain better bounds on the Kertész line for the random-cluster model than presented here and in [1]. Studying the Kertész line for random currents may also shed some light upon the problem of whether the single current and double current have the same phase transition ([32, Question 1]).

## Acknowledgements

The first author acknowledges funding from Swiss SNF. The second author thanks the Villum Foundation for support through the QMATH center of Excellence (Grant No.10059) and the Villum Young Investigator (Grant No.25452) programs. The authors would like to thank Ioan Manolescu for discussions and Bergfinnur Durhuus for comments on an early draft of the paper. Furthermore, the authors are very grateful to an anonymous referee for detailed comments.

# 4. The Uniform Even Subgraph and Its Connection to Phase Transitions of Graphical Representations of the Ising Model

# The Uniform Even Subgraph and Its Connection to Phase Transitions of Graphical Representations of the Ising Model

Ulrik Thinggaard Hansen, Boris Kjær, Frederik Ravn Klausen

### Abstract

The uniform even subgraph is intimately related to the Ising model, the random-cluster model, the random current model and the loop O(1) model. In this paper, we first prove that the uniform even subgraph of $\mathbb{Z}^d$ percolates for $d \geq 2$ using its characterisation as the Haar measure on the group of even graphs. We then tighten the result by showing that the loop O(1) model on $\mathbb{Z}^d$ percolates for $d \geq 2$ on some interval $(1 - \varepsilon, 1]$. Finally, our main theorem is that the loop O(1) model and random current models corresponding to a supercritical Ising model are always at least critical, in the sense that their two-point correlation functions decay at most polynomially and the expected cluster sizes are infinite.

## 1 Introduction

The Ising model has been extensively studied for the past 100 years. A central tool in the recent study of the model has been its graphical representations, the random-cluster model [28], the random current model [1, 32] and the high-temperature expansion [57], the latter also known as the loop O(1) model.

Even more recently, it has turned out that the uniform even subgraph of a graph is intimately related to the graphical representations of the Ising model. The uniform even subgraph of a finite graph $G$ is the uniform measure on the set of (spanning) subgraphs of $G$ with even degree at every vertex. It holds that the loop O(1) model can be sampled both as a uniform even subgraph of the random-cluster model [33] and as a uniform even subgraph of the (traced) double random current model [41].

The couplings may serve as one motivation for the study of the uniform even subgraph and we will see that this perspective does, in fact, give new information about the percolative properties of the graphical representations of the Ising model.

Furthermore, we will see that the uniform even subgraph is very natural. Indeed, we may consider the group of subgraphs of a given graph with symmetric difference of sets of edges as the group operation. The uniform even subgraph is then nothing but the Haar measure on the subgroup of even graphs.

In between the loop O(1) model and the random-cluster model, we find the (traced single) random current model, a graphical representation which has been central in the latest developments of the Ising model both in the planar case [42], in higher dimensions [20, 4] and in even higher generality [54, 2, 26]. As with the random-cluster model, the random current model has recently become an object of inherent interest [21, 23, 22].

In [17, Question 1] it was asked whether the single random current has a phase transition at the same point as the random-cluster model on $\mathbb{Z}^d$. From one point of view, since it is known that the random-cluster model can be obtained from the single random current by adding edges independently at random, a positive answer would imply that the added edges do not shift the phase transition. This, in turn, would run counter to common heuristics in statistical mechanics.

On the other hand, we will exploit a combinatorial fact about even subgraphs of the torus to prove that the correlations of the loop O(1) model, and hence the single random current, decay at most polynomially fast below the critical temperature of the random-cluster model. In models with positive association, this often signifies criticality. In particular, if the single random current were known to have a sharp phase transition, then our results would imply that the phase transition of the random current would coincide with that of the random-cluster model (and therefore, also the Ising model and the double random current). As such, this may be taken as evidence towards a positive answer to [17, Question 1].

## 1.1 Overview of the results and sketch of proofs

In this paper, we let $\ell_{x,G}^{\xi}$, $\mathbf{P}_{\beta,G}$ and $\phi_{p,G}^{\xi}$ denote the loop O(1), the random current model and the random-cluster model respectively. These are all models of random graphs, formal definitions of which, as well as details on their respective parametrisations, will be provided in Section 2. For the time being, all we need to know about them is the following:

For a finite graph $G = (V, E)$, the uniform even subgraph of $G$ (henceforth $\mathrm{UEG}_G$) is the uniform probability measure on the set of spanning even subgraphs of $G$. If $\omega \sim \phi_{p,G}^{\xi}$ and $\eta$ is a uniform even subgraph of $\omega$, then there exists $x = x(p)$ such that $\eta \sim \ell_{x,G}^{\xi}$. It is a classical result that the random-cluster model on subgraphs of $\mathbb{Z}^d$, for $d \geq 2$ undergoes a phase transition at a parameter $p_c \in (0, 1)$, separating a regime of all clusters being small from one where there exists an infinite cluster [51]. In the latter case, we say that the model *percolates*. We denote $x_c = x(p_c)$.

Our first contribution is to give an abstract characterisation of the uniform even subgraph of infinite graphs, an application of which is the following theorem. To fix notation, for a vertex $v$ of a random graph, we write $\mathcal{C}_v$ for the connected component of $v$, write $v \leftrightarrow w$ for the event $\mathcal{C}_v = \mathcal{C}_w$ and write $v \leftrightarrow \infty$ for the event that $|\mathcal{C}_v| = \infty$.

**Theorem 1.1.** *For $d \geq 2$,*

$$\mathrm{UEG}_{\mathbb{Z}^d}[0 \leftrightarrow \infty] > 0.$$

Previously, this was only known in $d = 2$ and proving percolation of the uniform even subgraph is a toy problem, the solution to which might shed some light upon the question about the phase transitions of random currents[1]. We solve the toy problem by showing a criterion for marginals of the uniform even subgraph to be Bernoulli distributed at parameter $\frac{1}{2}$ (see Lemma 3.5). The proof technique extends to showing that the phase transition of the loop O(1) model is non-trivial for $\mathbb{Z}^d$, $d \geq 3$.

[1] UTH and FRK are grateful to Franco Severo and Aran Raoufi for posing the toy problem.

**Theorem 1.2.** *Suppose that $d \geq 2$. Consider the loop O(1) model $\ell_{x,\mathbb{Z}^d}$ with parameter $x \in [0,1]$. Then there exists an $x_0 < 1$ such that*

$$\ell_{x,\mathbb{Z}^d} \, [0 \leftrightarrow \infty] > 0,$$

*for all $x \in (x_0, 1)$.*

Towards proving our main theorem, the main technical contribution is the insight that the loop O(1) model is insensitive to boundary conditions. In the following, $\Lambda_n := [-n,n]^d \cap \mathbb{Z}^d$ denotes the box of size $n$ around 0 and $\mathbb{T}_n^d = \Lambda_n/(2n\mathbb{Z}^d)$ the associated torus. We say that a supergraph $G$ extends $\Lambda_n$ if the induced graph of $V(\Lambda_n)$ in $G$ is $\Lambda_n$ and $\partial_V^G(\Lambda_n) = \partial_V^{\mathbb{Z}^d}(\Lambda_n)$.

**Theorem 1.3.** *For $x > x_c$, there exists $c > 0$ such that for any $n \in \mathbb{N}$ and any event $A$ which only depends on edges in $\Lambda_n$ and any $G$ which is a supergraph extending $\Lambda_{4n}$, then*

$$|\ell_{x,G}^\xi[A] - \ell_{x,\mathbb{Z}^d}[A]| \leq \exp(-cn),$$

*for any boundary condition $\xi$. In particular, for $x > x_c$ and any sequence $\xi_k$ of boundary conditions, $\lim_{k \to \infty} \ell_{x,\Lambda_k}^{\xi_k} = \ell_{x,\mathbb{Z}^d}$ in the sense of weak convergence of probability measures.*

Due the different behaviour of boundary conditions (cf. Section 2.1.1) in the loop O(1) model, mixing does not follow immediately. However, to complete the picture, mixing is proven in Theorem 4.11.

In the second step towards the main theorem we exploit the topology of the torus to produce an essential lower bound.

**Theorem 1.4.** *Let $x > x_c$. Then, there exists $c > 0$ such that $\ell_{x,\mathbb{T}_n^d}[0 \leftrightarrow \partial\Lambda_n] \geq \frac{c}{n}$ for all $n$.*

Combining Theorem 1.3 and Theorem 1.4 proves our main theorem:

**Theorem 1.5.** *Let $d \geq 2$ and $x > x_c$. Then, there exists a $C > 0$ such that for every $k$ and every $N \geq 4k$ and any boundary condition $\xi$, $\ell_{x,\Lambda_N}^\xi[0 \leftrightarrow \partial\Lambda_k] \geq \frac{C}{k}$. It follows that, $\ell_{x,\mathbb{Z}^d}[|\mathcal{C}_0|] = \infty$.*

From the couplings that we state in Theorem 2.5, the same result follows for the (sourceless, traced) single random current $\mathbf{P}_\beta$ for any $\beta > \beta_c$, where $\beta_c$ is the Ising critical inverse temperature.

**Corollary 1.6.** *Let $d \geq 2$ and $\beta > \beta_c$. Then, there exists a $C > 0$ such that $\mathbf{P}_{\beta,\Lambda_N}[0 \leftrightarrow \partial\Lambda_k] \geq \frac{C}{k}$, for every $k$ and every $N \geq 4k$. It follows that $\mathbf{P}_{\beta,\mathbb{Z}^d}[|\mathcal{C}_0|] = \infty$.*

Together with previous results (sharpness, couplings, $d = 2$), the main theorem establishes the almost complete phase diagram for the loop O(1) and (single) random current model on $\mathbb{Z}^d$ (see Figure 1). We also argue the case of the hexagonal lattice $\mathbb{H}$ in Section 5.2.

The overall strategy of our approach is simple. We consider the random-cluster model $\phi_p$ for $p > p_c$ on the $d$-dimensional torus of size $n$. Since $p$ is supercritical, it is fairly easy to prove the existence of a simple path wrapping around the torus once - henceforth called a wrap-around. Whenever a wrap-around $\gamma$ exists in a random-cluster configuration $\omega$, if $\eta$ is a uniform even subgraph of $\omega$, then $\gamma \triangle \eta$ also has the law of the uniform even subgraph of $\omega$. However, using the topology of the torus, we can prove that the number of wrap-arounds of the torus modulo 2 of the two configurations $\eta, \gamma \triangle \eta$ are different. Therefore, with probability at least $\frac{1}{2}$, there is at least one wrap-around in the uniform even subgraph (and hence, in the loop O(1) model and therefore also the random current model). Since $\eta$ has the distribution of the loop O(1) model, this lets us provide lower bounds for connection probabilities for the loop O(1) model on the torus.

The main technical part of the paper then lies in proving Theorem 1.3, which allows us to transfer the connections of the torus into the space $\mathbb{Z}^d$. We, in turn, prove this theorem by a) proving that the uniform even subgraph is generically not very sensitive to boundary conditions and b) recalling some classical literature on the connectivity of the supercritical random-cluster model.

This trick of exhibiting large clusters through the topology of the torus was previously employed to prove a polynomial lower bound for the escape probability of the Lorenz mirror model in [44]. One may view the XOR-trick employed in [16] as another instance. However, the trick comes in many disguises. For example, it rears its head as the four-fold degeneracy in the ground state of the toric code [40], which is an important candidate for the implementation of quantum error correction.

As a final aside, one may wonder just how sensitive the loop O(1) model is to the topology it is placed in. Our application of the above trick is highly sensitive to topology and does not readily generalise to a direct proof on $\mathbb{Z}^d$. On the other hand, Theorem 1.3 seems to tell us that the topology largely does not matter. In Section 6.4, we investigate this topological sensitivity by proving that removing certain edges in the hexagonal lattice can change the loop O(1) model from not percolating to percolating.

# 2 Preliminaries: Graphical representations of the Ising model and their connections

In this section, we introduce the classical ferromagnetic Ising model $\mathbf{I}$ and its graphical representations: the random-cluster model $\phi$, the (sourceless, traced) random current model $\mathbf{P}$ and the loop O(1) model $\ell$.

Furthermore, we will be concerned with the uniform even subgraph UEG, Bernoulli percolation $\mathbb{P}$ and the double random current $\mathbf{P}^{\otimes 2}$, the latter of which is obtained as the union of two independent copies of the single random current $\mathbf{P}$.

The models are defined, first on finite graphs, then suitably extended to models on infinite graphs. For any graph $G = (V, E)$, we denote the space of *percolation configurations* $\Omega(G) = \{0, 1\}^E$, which is identified with $\mathcal{P}(E)$ under the map $\omega \mapsto E_\omega := \omega^{-1}(\{1\})$. For $\omega \in \Omega(G)$, the associated spanning subgraph is $(V, E_\omega)$. A measure on $\Omega = \Omega(G)$ will be called a percolation measure. The Ising model $\mathbf{I}$ is a measure on $\{-1, 1\}^V$, also called a spin model, while its graphical representations $\phi, \mathbf{P}, \ell$, as well as $\mathbb{P}$, are percolation measures. The models are related

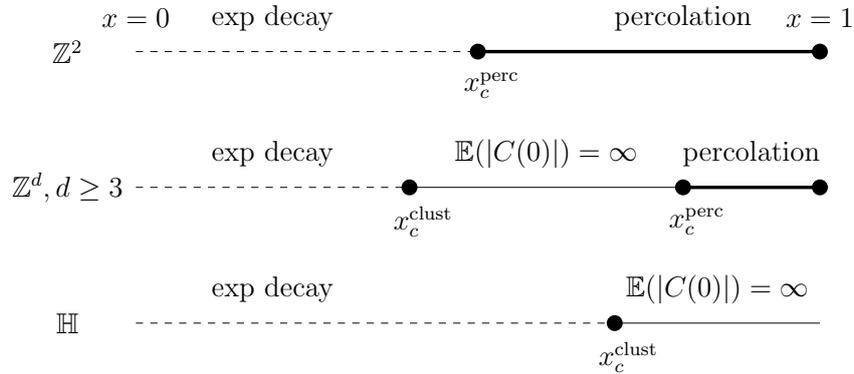

Figure 1: The phase diagram of the loop O(1) model and the single random current on $\mathbb{Z}^2$ and $\mathbb{Z}^d, d \geq 3$ as well as the loop O(1) model on the hexagonal lattice $\mathbb{H}$, all three in the ferromagnetic regime $x \in [0, 1]$. What remains to be proven in the phase diagram is whether $x_c^{\mathrm{perc}} = x_c^{\mathrm{clust}}$ for $\mathbb{Z}^d, d \geq 3$ (as conjecture in Conjecture 6.6). Note also that the case $x = 1$ corresponds to the uniform even subgraph UEG, which percolates for $\mathbb{Z}^d, d \geq 2$, but not for $\mathbb{H}$.

in a myriad ways, the most prominent of which are:[2]

- The random current model $\mathbf{P}$ and the random-cluster model $\phi$ are obtained from the loop O(1) model $\ell$ by adding additional edges independently at random (that is, according to $\mathbb{P}$).

- The loop O(1) model $\ell$ is obtained as a uniform even subgraph of either the random-cluster model $\phi$ or the double random current model $\mathbf{P}^{\otimes 2}$.

- For any $v, w \in V$,
  $$\mathbf{P}^{\otimes 2}[v \leftrightarrow w] = \langle \sigma_v \sigma_w \rangle^2 = \phi[v \leftrightarrow w]^2.$$
  These equalities mean that $\mathbf{P}^{\otimes 2}, \mathbf{I}$ and $\phi$ have a common phase transition at the critical inverse temperature $\beta_c$ of the Ising model.

- In two dimensions, the loop O(1) model $\ell$ is the law of the interfaces of the Ising model.

We summarise the couplings in Figure 2 (partially borrowed from [41]).

### 2.0.1 The Ising model

The much celebrated Ising model, introduced by Lenz [46], is a paradigmatic example of a model which undergoes a phase transition. The Ising model on a finite graph $G = (V, E)$ is a probability measure on the configuration space $\{-1, +1\}^V$. The energy of a configuration $\sigma \in \{-1, +1\}^V$ is

$$H(\sigma) = -\sum_{(v,w) \in E} \sigma_v \sigma_w.$$

---

[2] Details including parametrisations are given in Theorem 2.5 and Table 1.

With parameter $\beta \in [0, \infty]$, called the inverse temperature, the probability of a configuration is

$$\mathbf{I}_{\beta,G}[\sigma] = \frac{\exp(-\beta H(\sigma))}{Z_{\beta,G}},$$

where $Z_{\beta,G} = \sum_\sigma \exp(-\beta H(\sigma))$ is a normalisation constant called the partition function. This extends to the case $\beta = \infty$ by weak continuity.

In keeping with the literature, we write $\langle \sigma_v \sigma_w \rangle_{\beta,G}$ for the correlation function $\mathbf{I}_{\beta,G}[\sigma_v \sigma_w]$, i.e. the expectation of the random variable $\sigma_v \sigma_w$ under the measure $\mathbf{I}_{\beta,G}$.

## 2.1 The graphical representations

We start off by fixing some terminology. For a percolation configuration $\omega \in \Omega$, we say that $e$ is *open* in $\omega$ if $\omega(e) = 1$ and $e$ is *closed* if $\omega(e) = 0$. There is a canonical partial order on $\Omega$ given by pointwise comparison, that is, $\omega \preceq \omega'$ if $\omega(e) \leq \omega'(e)$ for all $e \in E$. We say, an event $\mathcal{A} \subset \Omega$ is *increasing* if for all pairs $\omega, \omega' \in \Omega$, it holds that if $\omega \in \mathcal{A}$ and $\omega \preceq \omega$, then $\omega' \in \mathcal{A}$. The notion of increasing events enables us to define a partial order on percolation measures on $\Omega$ by declaring $\nu_1 \preceq \nu_2$, if $\nu_1(\mathcal{A}) \leq \nu_2(\mathcal{A})$ for all increasing events $\mathcal{A}$. In this case, we say that $\nu_2$ *stochastically dominates* $\nu_1$.

Further, for two percolation measures $\nu_1$ and $\nu_2$, we let $\nu_1 \cup \nu_2$ denote the measure sampled as the union of two independently sampled copies of $\nu_1$ and $\nu_2$. That is, if $(\omega_1, \omega_2) \sim \nu_1 \otimes \nu_2$, then $\nu_1 \cup \nu_2$ is the law of $\omega_1 \cup \omega_2$ which is defined by $E_{\omega_1 \cup \omega_2} = E_{\omega_1} \cup E_{\omega_2}$. Note that $\nu_1 \cup \nu_2 \succeq \mu_1, \nu_2$.

### 2.1.1 Boundary conditions

If $G = (V, E) \subset (\mathbb{V}, \mathbb{E})$ is a subgraph, the edge boundary is $\partial_e G = \{(v, w) \in \mathbb{E} | v \in V, w \notin V\}$. Similarly, the vertex boundary is $\partial_v G = \{v \in V | \exists w \ (v, w) \in \partial_e G\}$.

For a graph with boundary $\partial_v G \subset V$, a (topological) boundary condition is a partition of $\partial_v G$. In physics parlance, the boundary vertices belonging to the same class according to the partition are wired together.

There is a partial order on boundary conditions given by fineness. If $\xi$ is a finer partition than $\xi'$, we write $\xi \preceq \xi'$. There is an initial and a terminal boundary condition with respect to this order, namely the *free* boundary condition where all vertices of $\partial_v G$ belong to distinct classes, and the *wired* boundary condition consisting of only one class. We denote these by 0 and 1 so that $0 \preceq \xi \preceq 1$ for any boundary condition $\xi$. Given a boundary condition $\xi$ we may define the quotient multigraph $G/\sim_\xi$ by identifying vertices according to the partition, and similarly for any subgraph of $G$. So for $\omega \in \Omega(G)$, we obtain $\omega^\xi \in \Omega(G/\sim_\xi)$. Define $\kappa^\xi(\omega)$ to be the number of connected components of $\omega^\xi$. The connected components are often referred to as *clusters*.

In the presence of a percolation model $\nu_G$, where $G$ ranges over subgraphs of $\mathbb{G}$, we consider the types of boundary conditions arising from the inclusions $G \subset G' \subset \mathbb{G}$. Fixing the state $\omega|_{G' \setminus G} = \psi$ of the configurations on the complement of $G$, we say the measure $\nu_{G'}[\cdot|\psi]$, as a percolation measure on $G$, has *exploratory boundary conditions*. Proposition 2.3 below relates these to topological boundary conditions on $G$ for the random-cluster model, but for loop O(1) and random current models, there is no such connection. We also consider the marginal measure $\nu_{G'}|_G$ which is the average over all exploratory boundary conditions. We say that this, as a percolation measure on $G$, has *marginal boundary conditions*.

### 2.1.2 The random-cluster model

The most well-studied graphical representation of the Ising model is the random-cluster model. For $p \in [0, 1]$, finite graph $G$, and boundary condition $\xi$, it is the probability measure given by

$$\phi_{p,G}^{\xi}[\omega] \propto 2^{\kappa^{\xi}(\omega)} \left( \frac{p}{1-p} \right)^{o(\omega)},$$

where $o(\omega) = |E_{\omega}|$ denotes the number of open edges in $\omega$. Whenever the boundary condition is omitted, we assume the free boundary condition, i.e. $\phi_{p,G} = \phi_{p,G}^0$.

The random-cluster model is related to the Ising model through the Edwards-Sokal coupling [27]: Suppose $\omega \sim \phi_{p,G}$ and $\sigma$ is obtained from $\omega$ by independently assigning $+$ and $-$ spins to the clusters of $\omega$. Then, $\sigma \sim \mathbf{I}_{\beta,G}$. Conversely, $\omega$ can be sampled from $\sigma$ by taking each edge $e = (v, w)$ such that $\sigma_v = \sigma_w$ and opening it with probability $p$. From the Edwards-Sokal coupling, it follows that

$$\langle \sigma_v \sigma_w \rangle_{\beta,G} = \phi_{p,G}[v \leftrightarrow w], \tag{1}$$

for $p = 1 - e^{-2\beta}$ (cf. [19, Corollary 1.4]). In particular, (1) implies that there is long-range order for the Ising model if and only if there are large clusters in the random-cluster model. Thus, the physical properties of the Ising model may be studied via the graphical properties of the random-cluster model. Towards this end, the random-cluster model has several monotonicity properties which makes it amenable to analysis. While this paper is not focused on proving something new about the random-cluster model, these properties will nonetheless play a crucial role as we explore the other graphical representations through the couplings. Here, we repeat a tailored version of the more general statement in [36]. For these and more results about the random-cluster model, we refer the reader to [19] and [34].

**Proposition 2.1.** *For the random-cluster model on a finite subgraph $G$ of a graph $\mathbb{G}$ (finite or infinite), the following relations hold:*

i) *The measure $\phi_{p,G}^{\xi}$ is monotone in $\xi$ in the sense that if $\xi \preceq \xi'$, then*

$$\phi_{p,G}^{\xi} \preceq \phi_{p,G}^{\xi'},$$

*for any parameter $p \in [0, 1]$.*

*ii) The measure $\phi_{p,G}^\xi$ is increasing in $p$ i.e. if $p \leq p'$, then*

$$\phi_{p,G}^\xi \preceq \phi_{p',G}^\xi,$$

*for any boundary condition $\xi$.*

*iii) The random-cluster model is comparable to Bernoulli percolation in the following sense:*

$$\mathbb{P}_{\tilde{p},G} \preceq \phi_{p,G}^\xi \preceq \mathbb{P}_{p,G},$$

*for any boundary condition $\xi$, where $\tilde{p} = \frac{p}{2-p}$.*

**Proposition 2.2** (FKG inequality)**.** *Let $G$ be a finite graph, and $A$ and $B$ increasing events, then*

$$\phi_{p,G}^\xi[A \cap B] \geq \phi_{p,G}^\xi[A]\phi_{p,G}^\xi[B]$$

*for any boundary condition $\xi$ and $p \in [0,1]$.*

**Proposition 2.3** (Domain Markov Property)**.** *If $G_1 = (V_1, E_1) \subseteq G_2 = (V_2, E_2)$ are two finite subgraphs of an infinite graph $\mathbb{G}$, we write $\omega_1 := \omega|_{E_1}$ and $\omega_2 := \omega|_{E_2 \setminus E_1}$. Then, for any boundary condition $\xi$ and any event $A$ depending on edges in $G_1$, it holds that*

$$\phi_{p,G_2}^\xi[\omega_1 \in A | \, \omega_2] = \phi_{p,G_1}^{\xi_{\omega_2}}[A],$$

*where $v, w \in V_1$ belong to the same element of $\xi_{\omega_2}$, if and only if they are connected (or possibly equal) in $(V_2, E_{\omega_2})/ \sim_\xi$.*

### 2.1.3 Construction of infinite volume measures

The monotonicity in boundary conditions combined with the Domain Markov Property allows us to define infinite volume measures in the following way. For a sequence of finite graphs $G_n \Uparrow \mathbb{G} = (\mathbb{V}, \mathbb{E})$ for some infinite graph $\mathbb{G}$ it holds that the marginals of $\phi_{G_n,p}^1$ on a fixed finite subset $\Lambda$ are monotonically decreasing. In other words, if $\mathcal{A}$ is an increasing event that depends only on edges in $\Lambda$, then $\{\phi_{G_n,p}^1(\mathcal{A})\}_{n \in \mathbb{N}}$ is monotonically decreasing. Similarly, the sequence $\{\phi_{G_n,p}^0(\mathcal{A})\}_{n \in \mathbb{N}}$ is monotonically increasing. Since they are bounded, they have limits $\phi_{p,\mathbb{G}}^1(\mathcal{A})$ and $\phi_{p,\mathbb{G}}^0(\mathcal{A})$ respectively. One can check that the limit does not depend on the choice of sequence $G_n$. Since the set of increasing events is intersection-stable and generates the product $\sigma$-algebra of $\{0,1\}^{\mathbb{E}}$, the two (possibly equal) infinite volume measures are uniquely determined. These limits define probability measures follows as the space of probability measures is compact, which is a standard consequence of Banach-Alaoglu.

On the other hand, any measure $\nu$ on $\{0,1\}^{\mathbb{E}}$ which almost surely has the Domain Markov Property (cf. Proposition 2.3) could be called an infinite volume random-cluster measure. However, such a measure would necessarily satisfy $\phi^0 \preceq \nu \preceq \phi^1$ and by [54, Corollary 3], $\phi_{p,\mathbb{Z}^d}^0 = \phi_{p,\mathbb{Z}^d}^1$ for all $p$. Accordingly, there is a unique infinite volume measure and, we shall drop the boundary conditions from our notation in the infinite volume case and merely write $\phi_{p,\mathbb{Z}^d}$. A similar construction defines the infinite volume Ising measures $\mathbf{I}_{\beta,\mathbb{Z}^d}^0, \mathbf{I}_{\beta,\mathbb{Z}^d}^+, \mathbf{I}_{\beta,\mathbb{Z}^d}^-$. With the infinite volume measures at hand, the following is obtained from (1) in a more or less straightforward manner.

**Proposition 2.4.** *Let* $\mathbb{G} = \mathbb{Z}^d$ *and let* $\beta \geq 0$ *be given. Then there is long-range order in the Ising model, i.e. there exists a* $c > 0$ *such that for all* $v, w \in \mathbb{V}$,

$$\langle \sigma_v \sigma_w \rangle_{\beta, \mathbb{Z}^d} \geq c,$$

*if and only if there is an infinite cluster almost surely, or equivalently,*

$$\phi_{\beta, \mathbb{Z}^d}[0 \leftrightarrow \infty] > 0. \qquad (2)$$

On the hypercubic lattice $\mathbb{Z}^d$, for $d \geq 2$, there exists a unique sharp phase transition [51, 2]. This means that there exists a $\beta_c$ such that $\langle \sigma_v \sigma_w \rangle$ decays exponentially in the distance between $v$ and $w$ for all $\beta < \beta_c$ and that there is long-range order for all $\beta > \beta_c$.

### 2.1.4 Even subgraphs

Both the random current model and loop O(1) model are defined in terms of even subgraphs of a graph. A graph is said to be *even* if every vertex degree is even (and in particular finite). We denote the set of all percolation configurations corresponding to even graphs by $\Omega_\emptyset$.

A related notion often used in the context of multigraphs in the random current literature is that of sources. For any (multi-)graph $H = (V_H, E_H)$ we say that the *sources* of $H$, denoted $\partial H$, is the set of vertices of odd degree. An even graph is then a graph such that $\partial H = \emptyset$. Just like we identified spanning subgraphs of a given graph $G = (V, E)$ with the space of percolation configurations $\Omega$, so we identify a configuration $\mathbf{n} \in \mathbb{N}_0^E$ with a multigraph, and such a multigraph is called a *current*. If $\partial \mathbf{n} = \emptyset$, we say that the current is *sourceless*.

### 2.1.5 The random current model

We now briefly introduce the random current model. For a more complete exposition, see [17] or [19]. We first define the model on a finite graph $G = (V, E)$. To introduce the random current model, we define the weight

$$w_\beta(\mathbf{n}) = \prod_{e \in E} \frac{\beta^{\mathbf{n}_e}}{\mathbf{n}_e!}.$$

Then the random current with source set $A$, denoted $\mathbf{P}_{\beta, G}^A$, is the probability measure on $\mathbb{N}_0^E$, given by

$$\mathbf{P}_{\beta, G}^A[\mathbf{n}] \propto w_\beta(\mathbf{n}) \mathbb{1}_{\{\partial \mathbf{n} = A\}}.$$

Since we can view any deterministic boundary condition as a free boundary condition on an appropriate graph, we will mostly work with free random currents $\mathbf{P}_{\beta, G}^A = \mathbf{P}_{\beta, G}^{A, 0}$. The following relation (cf. [19, (4.5)]) provides the first relation of the random current model to the Ising model

$$\langle \sigma_v \sigma_w \rangle_{\beta, G} = \frac{\sum_{\mathbf{n} | \partial \mathbf{n} = \{v, w\}} w_\beta(\mathbf{n})}{\sum_{\mathbf{n} | \partial \mathbf{n} = \emptyset} w_\beta(\mathbf{n})}.$$

From the multigraph valued random current, one derives a percolation measure called the traced random current. For any current $\mathbf{n} \in \mathbb{N}_0^E$, the corresponding traced current $\hat{\mathbf{n}}$ is defined by $\hat{\mathbf{n}}(e) = \mathbb{1}_{\mathbf{n}(e)>0}$ for each $e \in E$. Of prime import to the trace operation is the fact that it does not change connectivities, i.e. $v \leftrightarrow w$ in $\mathbf{n}$ if and only if $v \leftrightarrow w$ in $\hat{\mathbf{n}}$. In the following, we will only discuss the traced random current without sources, which we will denote $\mathbf{P}_{\beta,G}$.

Finally, a key player in the random-current literature is the *(sourceless) double random current*, defined by

$$\mathbf{P}_{\beta,G}^{\otimes 2} := \mathbf{P}_{\beta,G} \cup \mathbf{P}_{\beta,G}.$$

As a celebrated consequence of the switching lemma (see [19, Lemma 4.3]), one can prove that

$$\langle \sigma_v \sigma_w \rangle_{\beta,G}^2 = \mathbf{P}_{\beta,G}^{\otimes 2}[v \leftrightarrow w]. \tag{3}$$

Thus, the double random current, just like the random-cluster model, has the onset of large clusters at the critical point of the Ising model. One of the main motivations for this paper is to investigate for which infinite graphs $\mathbb{G}$ this is also true of the single random current $\mathbf{P}_{\beta,\mathbb{G}}$. A positive answer would be implied by a positive answer for the loop O(1) model, which we shall now introduce.

### 2.1.6 The loop O(1) model

For a finite graph $G = (V, E)$, we now define the loop O(1) model $\ell_{x,G}^\xi$ for parameter $x \in [0,1]$ as a measure on $\Omega(G)$. For $x \in [0,1]$, the loop O(1) model $\ell_{x,G}^\xi[\eta]$ is defined by

$$\ell_{x,G}^\xi[\eta] \propto x^{o(\eta)} \mathbb{1}_{\partial \eta^\xi = \emptyset}.$$

In particular, for $x = 1$, $\ell_{1,G}^0$ is the uniform measure on $\Omega_\emptyset(G)$, which we denote by $\mathrm{UEG}_G$. We note that $\ell_{x,G}^0$ is the high temperature expansion of the Ising model for $x = \tanh(\beta)$.

Notice how we may choose to view the sourceless random current $\mathbf{P}_{\beta,G}$ and the loop O(1) model $\ell_{x,G}$ as respectively independent Poisson and a Bernoulli variable on each edge conditioned on the sum over the valences on edges adjacent to a given vertex being even. This point of view was used in e.g. [56].

For planar graphs, the loop O(1) model is the law of the interfaces of a corresponding Ising model on the faces of the graph. This is discussed further in Section 5.1.

### 2.1.7 Bernoulli percolation

Although it is not a graphical representation of the Ising model, we consider also Bernoulli percolation with parameter $p \in [0,1]$, which we will denote by $\mathbb{P}_p$. This is the percolation measure where every edge $e \in E$ is open with probability $p$ independently. The model was introduced in [12] and has since been subject to intense study [18]. Here it will mainly play an auxiliary role.

### 2.1.8 Coupling the graphical representations

To state the couplings, let us define the uniform even subgraph of a probability measure.

For a probability measure $\mu$ on $\Omega$, consider the measure obtained by first sampling $\mu$ and then sampling the UEG of the first sample. That is, for every even graph $\eta \in \Omega_\emptyset$, consider $\mathrm{UEG}_\omega[\eta]$ as a function of $\omega \in \Omega$, so

$$\mu[\mathrm{UEG}_\omega[\eta]] = \sum_{\omega \in \Omega} \mathrm{UEG}_\omega[\eta]\mu[\omega] = \sum_{\omega \in \Omega} \frac{\mathbb{1}_{\eta \preceq \omega}}{|\Omega_\emptyset[\omega]|}\mu[\omega].$$

Now, we state the couplings from [19, Exercise 36]. The original references are [33, Theorem 3.5], [49] [48, Theorem 3.1], [41, Theorem 4.1]. We refer to Table 1 for the translations between the various parametrisations.

**Theorem 2.5.** *For any finite graph $G = (V, E)$, the graphical representations of the Ising model are related as follows:*

- $\ell_{x,G}^0 \cup \mathbb{P}_{1-\cosh(\beta)^{-1}, G} = \mathbf{P}_{\beta, G}$

- $\ell_{x,G}^0 \cup \mathbb{P}_{\tanh(\beta), G} = \phi_{p,G}^0$

- $\mathbf{P}_{\beta, G}^{\otimes 2}[\mathrm{UEG}_\omega[\cdot]] = \ell_{x,G}^0[\cdot] = \phi_{p,G}^0 \ [\mathrm{UEG}_\omega[\cdot]]$

**Remark 2.6.** *We note that even though the couplings here are only formulated with free boundary conditions, the measures with boundary conditions correspond to the free measure on the quotient of the original graph. As such, the couplings in the theorem also hold for e.g. wired boundary conditions.*

*Proof.* For a proof of the first two points, see [41, Theorem A.1]. The first equality in the last point is [41, Theorem 4.1] and the other equality is the statement of [33, Theorem 3.5]. We give a proof in our notation. For any even graph $\eta$, it holds that

$$\phi_{p,G}^0[\mathrm{UEG}_\omega[\eta]] \propto \sum_{\omega \in \Omega} \frac{2^{\kappa(\omega)}\mathbb{1}_{\eta \preceq \omega}}{|\Omega_\emptyset(\omega)|}\left(\frac{p}{1-p}\right)^{o(\omega)} \propto \sum_{\omega \in \Omega}\left(\frac{p}{2(1-p)}\right)^{o(\omega)}\mathbb{1}_{\eta \preceq \omega} \propto \mathbb{P}_{x,G}[\eta \preceq \omega] \propto x^{o(\eta)} = \ell_{x,G}^0[\eta],$$

where $\mathbb{P}_{x,G}$ denotes the expectation under Bernoulli percolation with parameter $x = \tanh(\beta) = \frac{p}{2-p}$ and we used the fact that $|\Omega_\emptyset(\omega)| = 2^{\kappa(\omega)+|o(\omega)|-|V|}$. $\square$

The relation $|\Omega_\emptyset(\omega)| = 2^{\kappa(\omega)+|o(\omega)|-|V|}$ is classical and one of its proofs goes as follows: if $e$ is part of a loop $l$, then $\eta \mapsto \eta \triangle l$ is a bijection between the set of even subgraphs with $e$ open and the set of even subgraphs with $e$ closed. Since the latter can be identified with $\Omega_\emptyset((V, E \setminus \{e\}))$, it follows that adding an extra edge $e$ to a connected graph doubles the number of even subgraphs. This bijection will play a central role in the rest of the paper.

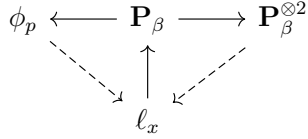

Figure 2: Overview of the couplings between the graphical representations of the Ising model. Dashed arrows point towards the distribution obtained from taking a uniform even subgraph. Full arrows indicate a union with another percolation measure.

One might also take the coupling in Theorem 2.5 as definition of the loop O(1) model with boundary condition, so

$$\ell_{x,G}^{\xi}[\cdot] \ = \phi_{p,G}^{\xi} \ [\mathrm{UEG}_\omega[\cdot]]. \tag{4}$$

Notice that this definition is consistent with our previous definition of $\ell_{x,G}^{\xi}$.

We can also take this approach to define the loop O(1) model in infinite volume, as was noted in [8, Remark 3.16], where we define the uniform even subgraph on an infinite graph $\mathbb{G}$ as the Haar measure on the group of even graphs. This allows us to define the loop O(1) model on an infinite graph as

$$\ell_{x,\mathbb{G}}[\cdot] \ = \phi_{p,\mathbb{G}} \ [\mathrm{UEG}_\omega[\cdot]]. \tag{5}$$

The equivalence of this definition to those given in [33, 8] is discussed in Section 3.2. There, we will also see that $\mathrm{UEG}_\mathbb{G}$ is in a certain sense unique when $\mathbb{G}$ is one-ended. This is case for the infinite cluster of $\omega \sim \phi_{p,\mathbb{Z}^d}$ when $d \geq 2$ by the Burton-Keane theorem [13, Theorem 2].

The literature [5, 38] also has yet another construction of the infinite volume loop O(1) model which uses a relation of the loop O(1) model to the gradient Ising model. Fortunately, the uniqueness of the infinite volume measure proven in Theorem 1.3 implies that the constructions agree. We remark that (5) also gives an independent construction of the infinite volume random current measure as

$$\mathbf{P}_{\beta,\mathbb{Z}^d} = \ell_{x,\mathbb{Z}^d} \cup \mathbb{P}_{1-\cosh(\beta)^{-1},\mathbb{Z}^d}.$$

## 2.2 The phase transitions of the graphical representations

Having introduced the various graphical representations, we now turn our attention to their phase transitions and how they are connected. Suppose that $\mathbb{G}$ is an infinite graph embedded in $\mathbb{R}^d$ on which $\mathbb{Z}^d$ acts by translation by a $\mathbb{Z}$-linearly independent family of vectors $(v_j)_{1\leq j \leq d}$. We will call such a graph $d$-periodic (or bi-periodic for $d = 2$). We will mainly be concerned with $\mathbb{G} = \mathbb{Z}^d$ for $d \geq 2$ or $\mathbb{G} = \mathbb{H}$ where $\mathbb{H}$ is the hexagonal lattice in two dimensions.

We consider one of the parametrised families of translation invariant infinite volume measures $\nu \in \{\ell_{x,\mathbb{G}}^0, \mathbf{P}_{\beta,\mathbb{G}}, \mathbf{P}_{\beta,\mathbb{G}}^{\otimes 2}, \phi_{p,\mathbb{G}}^0\}$. Further, whenever we parametrise a measure by a parameter that is not its natural parametrisation, e.g. using $\ell_\beta$ instead of $\ell_x$, we implicitly use the bijections between the parametrisations that are summarised in Table 1. For each of the models, there is a *percolative phase transition* defined by

$$\beta_c^{\mathrm{perc}}(\nu) = \inf\{\beta \geq 0 \mid \nu_\beta[0 \leftrightarrow \infty] > 0\},$$

with the convention $\inf \emptyset = \infty$. Recall that if $\nu_\beta[0 \leftrightarrow \infty] > 0$, we say that the model *percolates* at $\beta$.

|   | $\beta$ | $p$ | $x$ |
|---|---|---|---|
| $\beta$ |  | $\frac{1}{2}\log(1-p)$ | $\operatorname{arctanh}(x)$ |
| $p$ | $1-e^{-2\beta}$ |  | $\frac{2x}{2x+1}$ |
| $x$ | $\tanh(\beta)$ | $\frac{p}{2(1-p)}$ |  |

Table 1: The parameters $\beta, p, x$ in standard parametrisations of $\mathbf{I}_\beta, \mathbf{P}_\beta, \phi_p, \ell_x$ are always assumed to relate according to the table whenever they occur in relation to one another. This over-determination is convenient for the individual parametrisations and follows the literature standard.

One may also consider related but a priori different phase transitions corresponding to expected cluster sizes and regimes of exponential decay. These are defined by

$$\beta_c^{\mathrm{clust}}(\nu) = \inf\{\beta \geq 0 \mid \nu_\beta[|C(0)|] = \infty\},$$

as well as

$$\beta_c^{\mathrm{exp}}(\nu) = \sup\{\beta \geq 0 \mid \exists c, C > 0, \forall v \in \mathbb{G} : \nu_\beta[0 \leftrightarrow v] \leq Ce^{-c|v|}\}.$$

For translation invariant measures $\nu$, it is straightforward to see that

$$\beta_c^{\mathrm{exp}}(\nu) \leq \beta_c^{\mathrm{clust}}(\nu) \leq \beta_c^{\mathrm{perc}}(\nu).$$

Furthermore, one says that a phase transition is *sharp* if $\beta_c^{\mathrm{exp}}(\nu) = \beta_c^{\mathrm{perc}}(\nu)$. In general, the phase transition of the random-cluster model on a lattice is sharp (see [24]), but examples exist of lattice models that are not sharp, with the famous example of the intermediate phase described in the Berezinski-Kosterlitz-Thouless phase transition predicted in [10, 43] and rigorously proven in [30].

The relations (2), (3) together with sharpness of the Ising correlation function, proven in [2], imply that

$$\beta_c^{\mathrm{perc}}(\phi) = \beta_c^{\mathrm{clust}}(\phi) = \beta_c^{\mathrm{exp}}(\phi) = \beta_c = \beta_c^{\mathrm{perc}}(\mathbf{P}^{\otimes 2}) = \beta_c^{\mathrm{clust}}(\mathbf{P}^{\otimes 2}) = \beta_c^{\mathrm{exp}}(\mathbf{P}^{\otimes 2}), \qquad (6)$$

where $\beta_c$ is the critical inverse temperature of the Ising model.

The main concern of this paper is the phase transition of the two remaining graphical representations, $\ell_x$ and $\mathbf{P}_\beta$. The couplings from Theorem 2.5 immediately imply stochastic domination

$$\ell_\beta \preceq \mathbf{P}_\beta \preceq \phi_\beta. \qquad (7)$$

Thus, for $\# \in \{\mathrm{perc}, \mathrm{clust}, \mathrm{exp}\}$, it holds that

$$\beta_c^\#(\ell) \geq \beta_c^\#(\mathbf{P}) \geq \beta_c^\#(\phi) = \beta_c.$$

For the special case of the graph $\mathbb{Z}^d$, it was asked in [17, Question 1] whether the percolative phase transition for the random current is the same as for the Ising model.

**Question 2.7.** *For $\mathbb{G} = \mathbb{Z}^d$ does it hold that $\beta_c^{\mathrm{perc}}(\mathbf{P}) = \beta_c$?*

As one of the findings of this paper, we note that percolative loop O(1) phase transition depends a lot on the lattice. Therefore, in the investigation of the random current, one might allow oneself the freedom of also considering the phase transitions $\beta_c^{\mathrm{clust}}$ and $\beta_c^{\mathrm{exp}}$. Our main result (cf. Theorem 1.5) implies that for $\mathbb{G} = \mathbb{Z}^d$ and $d \geq 2$ or $\mathbb{G} = \mathbb{H}$, then

$$\beta_c^{\mathrm{clust}}(\ell) = \beta_c.$$

In particular,

$$\beta_c^{\mathrm{clust}}(\ell) = \beta_c^{\mathrm{exp}}(\ell) = \beta_c^{\mathrm{exp}}(\mathbf{P}) = \beta_c^{\mathrm{clust}}(\mathbf{P}) = \beta_c.$$

**Remark 2.8** (On generality)**.** *The techniques we employ mostly rely on the fact that $\mathbb{Z}^d$ descends to a graph on the torus, which allows us to utilise the non-trivial topology of the torus to control the existence of large clusters. As such, the result should carry over to locally finite $d$-periodic graphs embedded in $\mathbb{R}^d$.*

*The reason for not reflecting this in our statement is a) making it more accessible and b) the fact that a lot of the literature on the Ising model that we use is stated specifically for $\mathbb{Z}^d$ - for instance, [20, 11, 52]. We expect no particular difficulty in extending those results to lattices with suitable symmetries but prefer not to get bogged down on the possible generality of the results on which we rely.*

*One may note that none of the above papers are necessary for treating the planar case where planar duality (introduced in Section 5.1) and sharpness of the phase transition does the necessary work for us. In particular, our results apply to the hexagonal lattice, which we shall discuss further in Section 5.2.*

*The assumption of having some structure is necessary for the result, as there are examples of graphs where the phase transitions for the loop O(1) model $\ell$ and the random current model $\mathbf{P}$ are non-unique [35].*

## 2.3 All phase transitions coincide for $\mathbb{Z}^2$

On the square lattice, the overall picture is well-understood. For the Ising model, the existence of a phase transition was proven by Peierls [51] and the exact value of $\beta_c = \frac{1}{2} \log\left(1 + \sqrt{2}\right)$ was proven by Onsager [50].

The uniqueness of the phase transition at the point $\beta_c$ for the loop O(1) model was noted in [31]. The corresponding result for the single random current follows directly by the coupling in Theorem 2.5.

First, we recall the notion of ∗-connectivity for planar graphs. Two vertices $v, w$ of $\mathbb{Z}^2$ are said to be ∗-adjacent if they are adjacent to the same face. Accordingly, we have a notion of ∗-paths. A set of vertices $W$ is said to be ∗-connected if any pair of points $v, w \in W$ can be joined by a ∗-path contained in $W$. For an Ising configuration $\sigma \in \{-1, +1\}^{V(\mathbb{Z}^2)}$, a ∗-connected +-cluster is a maximal ∗-connected set of vertices that are coloured +.

**Theorem 2.9** ([31, Theorem 1.3]). *It holds that $\beta_c^{\mathrm{perc}}(\ell_{\mathbb{Z}^2}) = \beta_c$. It follows that $\beta_c^{\#}(\nu) = \beta_c$ for all $\nu \in \{\ell_{\mathbb{Z}^2}, \mathbf{P}_{\mathbb{Z}^2}, \mathbf{P}_{\mathbb{Z}^2}^{\otimes 2}, \phi_{\mathbb{Z}^2}\}$ and $\# \in \{\mathrm{perc}, \mathrm{clust}, \exp\}$.*

*Proof.* In [37], Higuchi proved that for the Ising model on $\mathbb{Z}^2$, if $\beta < \beta_c$, then there exists an infinite ∗-connected +-cluster and no infinite ordinary + cluster almost surely. Since the loop O(1) model $\ell_{x, \mathbb{Z}^2}$ is the law of the interfaces (see Section 5.1 for details on planar duality) of the Ising model, having an infinite ∗-connected +-cluster which is not an ordinary infinite cluster means that $\ell_{x, \mathbb{Z}^2}$ percolates. The second statement follows from (7) and (6). □

Since the uniform even subgraph is the loop O(1) model $\ell_x$ for $x = 1$, the results settle percolation of $\mathrm{UEG}_{\mathbb{Z}^2}$.

**Corollary 2.10.** $\mathrm{UEG}_{\mathbb{Z}^2}$ *percolates.*

For completeness, we give an independent proof of Higuchi's statement about the co-existence of infinite ∗-components. The proof is inspired by [14, Proposition 2]. Let $\mathcal{C}_\infty^{+,*}$ be the event that there is an infinite ∗-connected +-cluster and let $\mathcal{C}_\infty^{-,*}$ be the event that there is an infinite ∗-connected −-cluster.

Finally, we remind the reader that a boundary condition for the Ising model on a finite subgraph $G$ of $\mathbb{Z}^2$ is given by fixing a configuration on the vertices adjacent to $G$. Thus, we get an Ising model on $G$ conditioned on these extra boundary spins. Just as in the case of boundary conditions for percolation configurations, these boundary conditions are partially ordered (see e.g. [29, Section 3.6.2]).

**Proposition 2.11.** *Let $\beta < \beta_c$ then $\mathbf{I}_{\beta, \mathbb{Z}^2}[\mathcal{C}_\infty^{-,*}] = \mathbf{I}_{\beta, \mathbb{Z}^2}[\mathcal{C}_\infty^{+,*}] = 1$.*

*Proof.* We let $A(n) = \Lambda_{2n} \backslash \Lambda_n$ be an annulus. If there is a ∗-circuit in $A(n)$ of − spins encircling the inner boundary, then we say that the annulus $A(n)$ is *good*. Note that $A(n)$ is not good if and only if there is a (usual) path of +-spins connecting the inner boundary to the outer boundary.

Let us first prove that

$$\mathbf{I}_{\beta, A(n)}^+[A(n) \text{ is good}] \to 1. \tag{8}$$

Let $v \in \partial \Lambda_n$ and let $\mathcal{C}_v^+$ be the +-cluster of $v$. Then, since $\beta < \beta_c$, by sharpness [37, Theorem 3], there exists a $c > 0$ such that $\mathbf{I}_{\beta, A(n)}^+[|\mathcal{C}_v^+| \geq n] \leq e^{-cn}$. Thus, by a union bound we obtain (8) as follows

$$\mathbf{I}_{\beta, A(n)}^+[A(n) \text{ is not good}] \leq \sum_{v \in \partial \Lambda_n} \mathbf{I}_{\beta, A(n)}^+[|\mathcal{C}_v^+| \geq n]$$
$$\leq 8n e^{-cn} \to 0.$$

By monotonicity in boundary conditions, the same convergence holds for arbitrary boundary conditions.

Now, we look at all $(n, 2n)$-annuli with centers in $n\mathbb{Z}^2$, i.e. $\{A(n) + nk\}_{k \in \mathbb{Z}^2}$. For every $k \in \mathbb{Z}^2$ define $X(k) = \mathbb{1}[A(n) + nk$ is good]. Then, we consider $\{X(k)\}_{k \in \mathbb{Z}^2}$ as a spin model and show that it percolates for sufficiently large $n$. It follows from (8) and DMP for the Ising model that

$$\mathbf{I}_{\beta, \mathbb{Z}^2}\left[X(k) = 1 \mid \{X(j)\}_{j \in \mathbb{Z}^2 : |j-k| \geq 4}\right] \geq \mathbf{I}^+_{\beta, A(n)}\left[A(n) \text{ is good}\right] \to 1,$$

so we can use [47, Theorem 0.0] to dominate the process $\{X(k)\}_{k \in \mathbb{Z}^2}$ from below by independent Bernoulli random variables with some parameter $p_n$ where $p_n \to 1$ as $n \to \infty$. Hence, we can choose $n$ such that $p_n$ is above the threshold for site percolation (which is strictly smaller than 1 [51]). By planarity, for every edge $(v, w)$ of $\mathbb{Z}^2$, if $A(n) + nv$ and $A(n) + nw$ are both good, then the corresponding circuits of $-$'s must intersect. In particular, percolation of the good annuli implies an infinite *-cluster of $-$ spins. Thus, $\mathbf{I}_{\beta, \mathbb{Z}^2}[\mathcal{C}^{-, *}_\infty] = 1$ and by spin flip symmetry, it follows that $\mathbf{I}_{\beta, \mathbb{Z}^2}[\mathcal{C}^{+, *}_\infty] = 1$. □

# 3 The uniform even subgraph

This section starts by collecting general properties of the uniform even subgraph. Afterwards, we apply the theory to prove that the uniform even subgraph of $\mathbb{Z}^d$ percolates for $d \geq 3$, and give further results on even percolation in the last part of the section. Section 3.2 gives a detailed comparison between different constructions of uniform even subgraphs on infinite graphs.

In the treatment of the uniform even subgraph, we take a generalist point of view and regard "uniform" as synonymous with "invariant with respect to a group action". For a given graph $G = (V, E)$, the space $\Omega(G)$ is a group under point-wise addition modulo 2, indeed a $\mathbb{Z}_2$-vector space. This addition corresponds to symmetric difference on sets of edges, denoted $\triangle$, and we will use these notions interchangeably. The symmetric difference of two even graphs is again even, and so is the empty subgraph, so the set of even subgraphs $\Omega_\emptyset(G)$ is a closed $\mathbb{Z}_2$-linear subspace of $\Omega(G)$. We define the uniform even subgraph to be the (normalised) Haar measure on this group of even subgraphs.

The measure thus constructed is known in the literature as the free uniform even subgraph, when the graph is finite, and when the graph is infinite, it is known as the wired uniform even subgraph. While previously studied constructions of limit measures coincide with the Haar measure (see section Section 3.2) this property has to the best of our knowledge not been emphasised before, e.g. in [33, 8]. We demonstrate its merits in Theorem 1.1 where we prove that the uniform even subgraph of $\mathbb{Z}^d$ percolates for $d \geq 3$.

## 3.1 Marginals of the UEG

We will repeatedly let $G = (\mathbb{V}, E)$ be a spanning subgraph of an infinite graph $\mathbb{G} = (\mathbb{V}, \mathbb{E})$. A standard approach to infinite volume measures in statistical mechanics is to consider conditional distributions (e.g. the conditional distribution of a configuration $\omega$ in $G$ given a configuration $\omega_0$ in $G^c = (\mathbb{V}, \mathbb{E} \setminus E))$ rather than marginals [29, p.270]. This approach has merit in the treatment of the random-cluster model with Proposition 2.3 as a cornerstone of the theory. However, for the uniform even subgraph, and in turn the loop O(1) model, conditional probabilities are less useful. Specifically, $\ell_{x,G}[\cdot \mid \omega_1^c]$ is not, in general, a loop O(1) model with (topological) boundary conditions as defined in (4) since the sources on the boundary may force edges within $G$. Instead, it is more tractable to consider the marginal of the measure $\mathrm{UEG}_{\mathbb{G}}$ on $G$ - that is, $\mathrm{UEG}_{\mathbb{G}}|_G$. As an indication, the uniform even subgraph of $G$ with wired boundary conditions can (under suitable conditions) be realised as $\mathrm{UEG}_{\mathbb{G}}|_G$ whereas wired boundary conditions are extremal for the random-cluster model (see Section 2.1.1).

The $\mathbb{Z}_2$-vector space $\Omega$ is compact in the product topology[3] so it admits a unique Haar measure normalised to probability. Notice that this Haar probability measure is $\mathbb{P}_{1/2}$ (the relation to construction via Kolmogorov's theorem is discussed in Remark 3.12).

Every closed subgroup of $\Omega$ is also Abelian and compact, so it has a corresponding Haar probability measure. If $G$ is locally finite, we can define the source map $\partial : \{0,1\}^E \to \{0,1\}^V \cong \mathcal{P}(V)$ by mapping edges to incident vertices and extending linearly and continuously. All graphs appearing in this work are assumed to be locally finite. The source map $\partial$ is a continuous homomorphism so the group of even subgraphs, given by

$$\Omega_\emptyset(G) = \{\omega \in \Omega(G)\mid \partial\omega = \emptyset\} = \ker \partial,$$

is a closed subgroup (note that this agrees with our previous definition of $\Omega_\emptyset$). More generally, for $K \subset \{0,1\}^{\mathbb{V}}$, we write

$$\Omega_K(G) = \{\omega \in \Omega(G)\mid \partial\omega \in K\}.$$

The set $K$ can be thought of as a set of admissible source configurations and $\Omega_K(G)$ is a group when $K$ is a group. For example, the source set $\{\omega \in \Omega\mid \partial\omega = A\}$, where a specific source configuration is fixed, is a group if and only if $A = \emptyset$. For convenience, we carry forth the notation $\Omega_\emptyset$ when $K$ is the trivial subgroup $\{\emptyset\}$. The uniform measure on this group, i.e. the Haar probability measure, is denoted $\mathrm{UEG}_G$. Due to Proposition 3.2 below, when $K$ is a group, we refer to $\Omega_K$ as the even subgraphs with *marginal* boundary conditions $K$.

For a fixed subgraph $G = (\mathbb{V}, E) \subset \mathbb{G} = (\mathbb{V}, \mathbb{E})$ the inclusion $\iota_E : E \to \mathbb{E}$ induces the projection $\pi_G : \{0,1\}^{\mathbb{E}} \to \{0,1\}^E$ also known as the restriction to $E$. By construction of the product topology, the projection is continuous and the marginal on $G$ of any Borel measure $\mu$ on $\{0,1\}^E$ is defined to be the pushforward along $\pi_G$, denoted $\mu|_G$ [4].

Crucially, $\pi_G$ is a homomorphism, which, along with the following general fact[5], explains the functorial structure of the marginals of uniform measures.

**Lemma 3.1.** *Let $\Gamma, \Gamma'$ be Abelian compact topological groups and $f : \Gamma \to \Gamma'$ a continuous homomorphism. Let $\mu$ be a Haar probability measure on $\Gamma$. If $f$ is surjective, then $f_*\mu$ is a Haar probability measure on $\Gamma'$. For a general continuous homomorphism, we obtain the Haar measure on the image $f(\Gamma)$.*

---

[3] equivalently, the topology of pointwise convergence or the topology generated by cylinder events.   [4] This is not be confused with the plain restriction of a measure to a subspace, viz. $\Omega(G) \subset \Omega(\mathbb{G})$. That is, $\mu(A) \neq \mu|_G(A) = \mu(\pi_G^{-1}(A))$ for general measurable $A \subset \Omega(G)$.   [5] Commutativity in this context is only used to ensure uniqueness of invariant measures. A more general statement is true for invariant measures on non-Abelian groups but this is not relevant here.

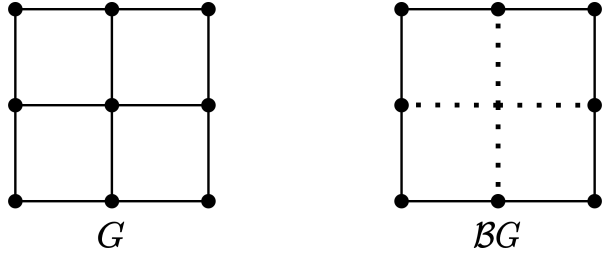

Figure 3: An example of the graph $\mathcal{B}G$. Here, $\partial_v G$ should be inferred from viewing $G$ as a subgraph of the square lattice $\mathbb{Z}^2$.

*Proof.* The homomorphism property of $f$ implies translation invariance of $f_*\mu$ on its support. The rest of the claim follows from the uniqueness of the Haar measure. □

We now turn our attention to extending [33, Theorem 2.6] by including infinite subgraphs and adding a description of the marginal:

**Proposition 3.2.** *Let $G = (\mathbb{V}, E) \subset \mathbb{G}$ and set $\mathcal{M} = \partial \circ \pi_G(\Omega_\emptyset)$. Then,*

$$\pi_G(\Omega_\emptyset(\mathbb{G})) = \Omega_\mathcal{M}(G).$$

*Moreover, the marginal, $\mathrm{UEG}_\mathbb{G}|_G$, is given by the uniform (Haar) measure on $\Omega_\mathcal{M}(G)$.*

*Proof.* Denote $\Gamma = \pi_G(\Omega_\emptyset)$. Since these groups are vector spaces, by rank-nullity,

$$\Gamma \cong \Omega_\emptyset(G) \oplus \Gamma/\Omega_\emptyset(G).$$

Recall that $\Omega_\emptyset(G) = \ker \partial|_{\Omega(G)}$, so the cosets of $\Gamma/\Omega_\emptyset(G)$ are characterised by their boundary. By the isomorphism theorem, we identify $\Gamma/\Omega_\emptyset(G) \cong \mathcal{M} \subset \{0,1\}^\mathbb{V}$. Then,

$$\Gamma = \{\omega : \Omega(G) | \ \partial \omega \in \mathcal{M}\}.$$

The statement now follows from Lemma 3.1. □

For any $E \subset \mathbb{E}$, the induced graph $G(E) = (V, E) \subset \mathbb{G}$ has vertices

$$V = \{v \in \mathbb{V} | \ v \text{ is not isolated in } (\mathbb{V}, E)\}.$$

We define the boundary graph as $\mathcal{B}G = (\partial_v G, \delta_e G) \subset G(E)$ where $\delta_e G \subset E$ are the edges in $E$ between two vertices of $\partial_v G$ (see Figure 3). The set of wired even subgraphs is

$$\Omega_{\partial_v}(G) = \{\omega \in \Omega(G) | \ \partial \omega \subset \partial_v G\},$$

corresponding to $\Omega_\mathcal{M}$ with $\mathcal{M} = \{0,1\}^{\partial_v G}$, and $\mathcal{M}$ is considered the wired marginal boundary condition. We write $\mathrm{UEG}_G^1$ for the uniform probability measure on $\Omega_{\partial_v}(G)$. For finite graphs with boundary, the trivial boundary condition $\mathcal{M} = 0$ corresponds to $\Omega_{\mathcal{M}} = \Omega_{\emptyset}$ which is called the set of free even subgraphs. Generalising this notion to free even subgraphs of infinite graphs requires some care, and is dealt with in Section 3.2. Finally, it is instructive to note that $\mathcal{M} = \partial \circ \pi_G(\Omega_{\emptyset}) \subset \{0,1\}^{\mathbb{V}}$ need not be generated by vectors of the form $\mathbb{1}_v + \mathbb{1}_w$ for $v, w \in \mathbb{V}$; take for example $\mathbb{G}$ to be a cycle graph and $G$ given by two edges apart from each other.

**Remark 3.3.** *We have not given a definition of topological boundary conditions for uniform even subgraphs, cf. Section 2.1.1. Conversely, the description of marginal boundary conditions, in terms of the groups $\mathcal{M}$, is valid for uniform even subgraphs, but not for the three graphical representations of the Ising model we consider.*

**Definition 3.4.** *Let $G = (\mathbb{V}, E) \subset \mathbb{G}$ and $S \subset \mathcal{P}(\mathbb{E})$. We say $S$ separates edges of $G$ if for each $e \in E$ there is $s \in S$ such that $s \cap E = \{e\}$.*
*If $E$ is finite, we say that $G$ is essentially free if $\mathcal{B}G = (\partial_v G, \delta_e G)$ is connected.*

**Lemma 3.5.** *Let $G = (\mathbb{V}, E) \subseteq \mathbb{G}$ and assume $\Omega_{\emptyset}(\mathbb{G})$ separates edges of $\delta_e G$. If $G$ is finite and essentially free, then*

$$\pi_G(\Omega_{\emptyset}(\mathbb{G})) = \Omega_{\partial_v}(G) \tag{9}$$

*and hence, $\mathrm{UEG}_{\mathbb{G}}|_G = \mathrm{UEG}_G^1$.*
*If $G$ is finite or infinite and $\Omega_{\emptyset}(\mathbb{G})$ separates edges of $G$, then*

$$\pi_G(\Omega_{\emptyset}) = \Omega(G)$$

*and hence, $\mathrm{UEG}_{\mathbb{G}}|_G = \mathbb{P}_{\frac{1}{2}, G}$.*

*Proof.* Let $G \subset \mathbb{G}$ possibly infinite. For $E' \subset E$, $\Omega((\mathbb{V}, E'))$ is generated by subgraphs with a single edge, so if $\Omega_{\emptyset}(\mathbb{G})$ separates edges of $E'$, then

$$\Omega((\mathbb{V}, E')) \subset \pi_G(\Omega_{\emptyset}(\mathbb{G})). \tag{10}$$

This is because $\pi_G$ restricted to $\Omega_{\emptyset}(\mathbb{G})$ is a homomorphism of groups and the generators of $\Omega((\mathbb{V}, E'))$ are in the range.

Assume now $G$ is finite and essentially free. Observe that

$$\Omega_{\partial_v}(G) = \Omega_{\emptyset}(G) + \Omega((\mathbb{V}, \delta_e G)),$$

since if $\omega \in \Omega_{\partial_v}(G)$ there are $\tilde{\omega} \in \Omega_{\emptyset}(G)$ and $\eta \in \Omega(\mathcal{B}G)$ such that $\omega = \tilde{\omega} \triangle \eta$. Indeed, since $\partial \omega \subset \partial_v G$ is finite, it partitions into pairs of vertices $(v_i, w_i)$. Since $\mathcal{B}(G)$ is connected, it contains a path $\gamma_i$ from $v_i$ to $w_i$ for each $i$ and taking $\eta := \Delta_i \gamma_i$, we see that $\tilde{\omega} := \omega \triangle \eta \in \Omega_{\emptyset}(G)$. This proves (9) when $\Omega_{\emptyset}(\mathbb{G})$ separates edges of $\partial_e G$, and the marginal measure is thus the wired uniform even measure by Proposition 3.2.

Allowing $G$ to be infinite, if $\Omega_{\emptyset}(\mathbb{G})$ separates edges of $G$, the claim again follows from (10) and Proposition 3.2. $\qquad\square$

Our first result follows as an application of Lemma 3.5.

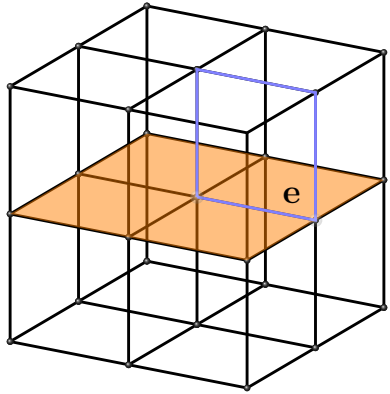

Figure 4: A subset of the lattice $\mathbb{Z}^3$, with a $\mathbb{Z}^2$ sheet highlighted with orange. For an edge $e$ in the sheet, we show the even graph (loop in blue) that we use to prove that $\Omega_\emptyset(\mathbb{Z}^3)$ separates edges of $\mathbb{Z}^2$.

**Theorem 1.1.** *For $d \geq 2$,*

$$\mathrm{UEG}_{\mathbb{Z}^d}[0 \leftrightarrow \infty] > 0.$$

*Proof.* We show that $\mathbb{Z}^d$ contains a subgraph $H$ such that $\Omega_\emptyset(\mathbb{Z}^d)$ separates edges of $H$ and such that $\mathbb{P}_{\frac{1}{2}, H}$ percolates. Then, by Lemma 3.5, the marginal $\mathrm{UEG}_{\mathbb{Z}^d}|_H$ percolates and hence, so does $\mathrm{UEG}_{\mathbb{Z}^d}$.

In the case $d \geq 4$, $\mathbb{P}_{\frac{1}{2}, \mathbb{Z}^{d-1}}$ percolates (this for example follows from Proposition 3.14 , but was originally proven in [15]) and therefore, we can choose $H$ to be a hyperplane of co-dimension 1. Every edge in the hyperplane $H$ is part of a plaquette[6] which is not contained in the hyperplane, so $\Omega_\emptyset(\mathbb{Z}^d)$ separates edges of $H$, see Figure 4.

When $d = 3$, a hyperplane will not suffice since $\mathbb{P}_{\frac{1}{2}, \mathbb{Z}^2}$ does not percolate [39]. However, if we take all possible edges among $\mathbb{Z}^2 \times \{0, 1\} \subset \mathbb{Z}^3$ to define the spanning subgraph $H$ (see Figure 5), then $\mathbb{P}_{\frac{1}{2}, H}$ percolates. We defer the proof of this fact to Proposition 3.14

We can separate edges on each sheet of $\mathbb{Z}^2 \times \{0, 1\}$ as before (with the loop around the plaquette sticking out upwards on the upper plane and with the loop around the plaquette sticking out downwards for each edge on the lower plane). Meanwhile, the connecting edges between the two sheets can be separated by doubly infinite cycles orthogonal to the slabs as sketched in Figure 5. $\qquad \square$

To apply the argument of the preceding proof to arbitrary graphs $G \subset \mathbb{G}$, criteria are needed for when $\Omega_\emptyset(\mathbb{G})$ separates edges of $G$. A set $\mathcal{F} \subset \Omega(\mathbb{G})$ is finitary if any given edge is contained in finitely many $f \in \mathcal{F}$. In these terms, the criterion is that there exists a finitary and linearly independent set $\mathcal{F} \subset \Omega_\emptyset(\mathbb{G})$ and an injective map $E \to \mathcal{F}$ mapping every edge to a cycle containing the edge itself.

If $\mathbb{G}$ were finite and connected, a necessary and sufficient condition for edge separation would be for $\mathbb{G}$ to have a spanning tree contained in $(\mathbb{V}, \mathbb{E} \setminus E)$ as demonstrated in [33]. Indeed, a basis for the cycles of a connected graph $\mathbb{G}$ may be constructed from the edges in the complement of a spanning tree as follows: The end-points of a given edge in the complement of a spanning tree are connected through a unique path in the spanning tree which together with the edge defines a cycle. Since each edge outside the spanning tree appears in exactly one cycle, the set thus defined is linearly independent. One may manually check that it is spanning.

---

[6] Recall that a plaquette of $\mathbb{Z}^d$ is a cycle of length 4.

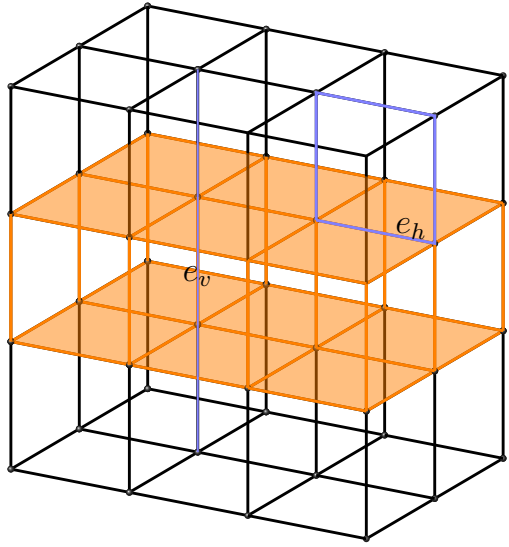

Figure 5: The situation in the proof in the case $d = 3$ in Theorem 1.1 with the edges in the subgraph $\mathbb{Z}^2 \times \{0, 1\}$ coloured orange (except for the two edges $e_h$ and $e_v$). The horizontal edge $e_h$ is separated by the light blue cycle, whereas the edge $e_v$ is separated by the bi-infinite path, the first part of which is also coloured light blue.

One observes that this is equivalent to the condition that $(\mathbb{V}, \mathbb{E} \setminus E)$ is connected. This extends to the case where $\mathbb{G}$ is infinite as follows:

**Proposition 3.6.** *Suppose that $G \subset \mathbb{G}$ such that every connected component of $\mathbb{G} \setminus G$ is infinite. Then,*

$$\Omega_{\partial_v}(\mathbb{G})|_G = \Omega_{\partial_v}(G)$$

*and it follows that*

$$\mathrm{UEG}_{\mathbb{G}}|_G = \mathrm{UEG}_G^1.$$

*Proof.* One inclusion is clear, so let $\omega \in \Omega_{\partial_v} G$ so that $\partial \omega \subset \partial_v G$. Since $G$ is finite, so is $\partial_v G$ and any $v \in \partial_v G$ has a neighbour in $\mathbb{G} \setminus G$ by definition. Therefore, to each $v \in \partial \omega$, we may choose an infinite path $\gamma_v$ in $\mathbb{G} \setminus G$ with $\partial \gamma_v = v$. Define the configuration $\omega' = \omega \triangle_{v \in \partial \omega} \gamma_v$. Now, $\partial \omega' = \emptyset$ and $\omega'|_G = \omega$. The last claim follows from Proposition 3.2. $\qquad \square$

Another application of Lemma 3.5 is the following:

**Corollary 3.7.** *For all $d \geq 2$ and any $k$, $\mathrm{UEG}_{\Lambda_{k+1}}^0 |_{\Lambda_k} = \mathrm{UEG}_{\Lambda_{k+1}}^1 |_{\Lambda_k} = \mathrm{UEG}_{\Lambda_k}^1.$*

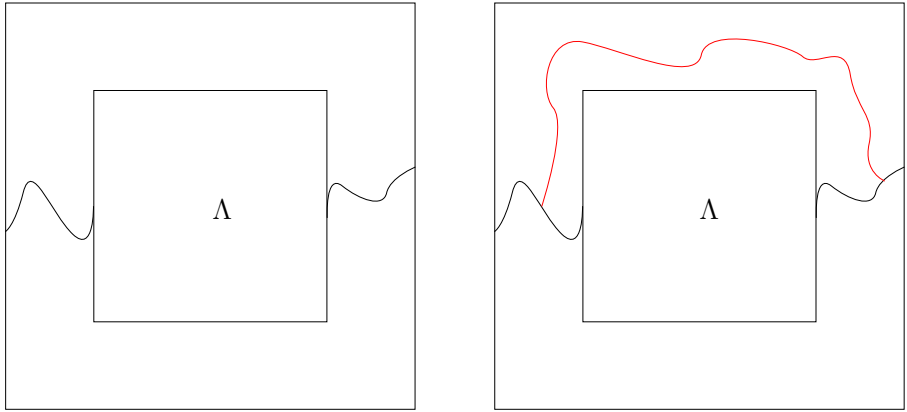

Figure 6: An example of two supergraphs of the finite graph $\Lambda$. On the left, the marginal of $\text{UEG}^1$ on $\Lambda$ is different from the marginal of UEG due to the lack of a *connected* separating surface between the inner and outer boundary. On the right, the two marginals are the same due to the presence of the red path.

*Proof.* Note that $\mathcal{B}\Lambda_k$ is connected. Furthermore, $\Omega_\emptyset(\Lambda_{k+1}) \subset \Omega_{\partial_v}(\Lambda_{k+1})$ separates edges on $\delta_e\Lambda_k$ since every edge is singled out by the plaquette directly above it in the orthogonal direction to the hyperplane in which it lies. Thus, $\Lambda_k \subset \Lambda_{k+1}$ satisfies Lemma 3.5 in the finite setting. $\square$

This result, while simple in itself, runs counter to common intuition in statistical mechanics because it shows that the uniform even subgraph is extremely insensitive to boundary conditions despite being a highly dependent model.

We continue by showing a general condition under which this insensitivity to boundary conditions occurs. On a graph $\mathbb{G} = (\mathbb{V}, \mathbb{E})$, we say that a path $\eta \subset E$ connects $G_1 = (\mathbb{V}, E_1)$ and $G_2 = (\mathbb{V}, E_2)$ if its end-points are non-isolated vertices in $G_1, G_2$ respectively. Recall that $W \subset \mathbb{V}$ is called a separating surface between disjoint edge sets $E_1, E_2$ if any path connecting $(\mathbb{V}, E_1)$ and $(\mathbb{V}, E_2)$ visits $W$.

**Proposition 3.8.** *Assume $E_1, E_2, E_3 \subset E$ are pairwise disjoint and that $E_1, E_2$ are finite. Let $G_j = (V, \bigcup_{i \leq j} E_i)$. If the induced graph $W$ of $E_2$ is connected and a separating surface $W$ between $E_1$ and $E_3$, then $\pi_{G_1}(\Omega_\emptyset(G_2)) = \pi_{G_1}(\Omega_\emptyset(G_3))$.*

*Hence, $\text{UEG}_{G_2}|_{G_1} = \text{UEG}_{G_3}|_{G_1}$.*

*Proof.* It suffices to check $\pi_{G_1}\Omega_\emptyset(G_3) \subset \pi_{G_1}\Omega_\emptyset(G_2)$. Without loss of generality, assume $\gamma \in \Omega_\emptyset(G_3)$ is connected. If $\pi_{G_1}(\gamma) = \gamma|_{E_1} = \gamma$, there is nothing to prove.

Parametrising $\gamma$ as an edge self-avoiding closed path, the first edge of which lies in $E_1$ gives rise to a partition, $S$, of $\gamma|_{E_2 \cup E_3}$ consisting of maximal subpaths contained in $E_1 \cup E_2$. We claim that for each $s \in S$, there is $h_s \in \Omega_\emptyset((V, E_2 \cup E_3))$ such that $s \triangle h \subset E_2$.

Indeed, if $s \subset E_2$, $h_s = \emptyset$. On the other hand, if $s$ intersects $E_3$, by the fact that $W$ is a separating surface, $s_{|E_2}$ splits into maximal subpaths with respect to the parametrisation whereof the first and last segment visit $W$.

Since $W$ is connected, these two segments are connected by a path $c \subset E_2$. Furthermore, we may take $c$ such that it only intersects the given two segments of $s$ in its end-points $\gamma(t_0), \gamma(t_1)$ in such a way that $\gamma([t_0, t_1]) \subset s$. Then, $h_s = \gamma([t_0, t_1]) \triangle c$ is even and satisfies $s \triangle h_s \subset E_2$. Let $\tilde{h} = \triangle_s h_s \in \Omega_\emptyset(G_2)$ and $\tilde{\gamma} = \gamma \triangle \tilde{h}$. Then, $\tilde{\gamma} \in \Omega_\emptyset(G_2)$ and $\tilde{\gamma}_{|E_1} = \gamma_{|E_1}$ which completes the proof. $\qquad\square$

This tells us that, in the presence of a uniform even subgraph of $G_3$, observing the edges of $G_1$ gives no information about the even subgraph on the edges of $G_3$. This holds in the strongest possible sense:

**Corollary 3.9.** *Assume $E_1, E_2, E_3 \subset E$ are pairwise disjoint and that $E_1, E_2$ are finite. Let $G_j = (V, \bigcup_{i \leq j} E_i)$. If the induced graph $W$ of $E_2$ is connected and a separating surface $W$ between $E_1$ and $E_3$, and $\eta \sim \mathrm{UEG}_{G_3}$, then $\eta|_{E_1} \perp \eta|_{E_3}$.*

*Proof.* For $j \in \{1, 3\}$ fix $\eta_j \in \pi_{(V, E_j)}(\Omega_\emptyset(G_3))$ and define the set $\Omega_\emptyset^{\eta_1, \eta_3}$ as the subset of $\Omega_\emptyset(G_3)$ which agrees with $\eta_j$ on $E_j$. Since $E_1$ and $E_2$ are finite, one may note that

$$\mathrm{UEG}_{G_3}[\mathbb{1}_{\eta|_{E_1} = \eta_1} \mid \eta|_{E_3}] = \frac{|\Omega_\emptyset^{\eta_1, \eta|_{E_3}}|}{\sum_{\eta_1' \in \pi_{(V, E_1)}(\Omega_\emptyset(G_3))} |\Omega_\emptyset^{\eta_1', \eta|_{E_3}}|}.$$

Thus, the claim follows if we can prove that $\Omega_\emptyset^{\eta_1, \eta_3}$ is a translate of $\Omega_\emptyset^{0, \eta_3}$, where $0$ denotes the empty graph. By Proposition 3.8, there exists $\eta \in \Omega_\emptyset^{\eta_1, 0}$. The translation map, $\eta' \mapsto \eta' \triangle \eta$ on $\Omega_\emptyset(G_3)$, is injective, and maps $\Omega_\emptyset^{\eta_1, \eta_3}$ into $\Omega_\emptyset^{0, \eta_3}$ and $\Omega_\emptyset^{0, \eta_3}$ into $\Omega_\emptyset^{\eta_1, \eta_3}$. $\qquad\square$

Proposition 3.8 applies to $\Lambda_k$ and $\mathbb{Z}^d \setminus \Lambda_{2k}$ with the annulus in between as the separating surface, illustrated in Figure 6. To extend this application to a supercritical random-cluster configuration, our first order of business in Section 4 will be to prove that the presence of a connected separating surface in an annulus is exponentially likely in the supercritical random-cluster model. This is instrumental to proving Theorem 1.3.

## 3.2 Constructions of uniform even subgraphs in infinite volume

Defining UEG as a Haar measure immediately provides a construction of the uniform even subgraph of an infinite graph. However, it is standard to regard percolation measures as limits of finitely supported measures [29, Chap. 6] and this is, indeed, the approach of [33, 8] for the uniform even subgraph. For the remainder of this section, $\mathbb{G}$ is an infinite graph. In this section, we show how both the limit of the free and wired measures may be realised as limits of Haar measures. The infinite group of free even subgraphs is a limit of an increasing sequence of subgroups while the infinite group of wired even subgraphs is determined by a decreasing sequence of quotients.

To begin with, let us briefly review a standard approach. It was observed in [33] that any locally finite connected graph admits a finitary generating set $\mathcal{C} \subset \Omega_\emptyset(\mathbb{G})$. Since $\mathcal{C}$ is finitary, the sums

$$\sum_{C \in \mathcal{A}} C = \sum_{C \in \mathcal{C}} \mathbb{1}_{\mathcal{A}}(C) C \tag{11}$$

are pointwise convergent for any $\mathcal{A} \subset \mathcal{C}$. The second assertion is that $\Omega_\emptyset(\mathbb{G}) = \overline{\mathrm{span}(\mathcal{C})}$. The set $\mathcal{C}$ can be chosen as a Schauder basis of $\Omega_\emptyset$ and properties of such sets are studied in [8]. The observation is used for sampling a uniform even graph by (uniformly) sampling the subsets $\mathcal{A} \subset \mathcal{C}$. This is done by replacing the coefficients $\mathbb{1}_{\mathcal{A}}(C)$ in the sum (11) with independent random Bernoulli-1/2 variables $\epsilon_C$. Observe, that the construction of this measure amounts to defining a surjective homomorphism $\Phi^\mathcal{C} : \mathbb{Z}_2^\mathbb{N} \to \Omega_\emptyset(\mathbb{G})$ (implicitly enumerating $\mathcal{C}$) and applying Lemma 3.1 to obtain the Haar on $\Omega_\emptyset(\mathbb{G})$ measure as $\Phi_*^\mathcal{C}\mu$. One speaks of sampling the coefficients $\epsilon_C$ sequentially, implying that one is really taking the weak limit of the measures $(\Phi^\mathcal{C}|_{\mathbb{N}_{\leq N}})_*\mu$ for increasing $N$. These measures on $\Omega(\mathbb{G})$ are finitely supported and approximate $\Phi_*^\mathcal{C}\mu$. Indeed, since $\mathcal{C}$ is finitary, the expectation of any local event (or so-called cylinder event on a finite cylinder) is eventually constant as $N \to \infty$.

With UEG defined a priori as the Haar measure (and not as a limit or as $\Phi_*^\mathcal{C}\mu$ for specific $\mathcal{C}$), we are led to consider approximations of UEG by finitely supported measures. In addition, we may ask whether the approximations are *local* in the following sense: a measure is local if it agrees with the pushforward of a measure $\mu$ on $\Omega(G)$ along the natural inclusion $\Omega(G) \subset \Omega(\mathbb{G})$ for a finite set $E$ and $G = (\mathbb{V}, E)$. If the finitary basis $\mathcal{C}$ consists of finite cycles, we may consider $\Phi_*^\mathcal{C}\mu$ to be a limit of local measures. In [8], it is recognised that when $\mathcal{C}$ is a subset of the finite subgraphs, it can at most generate the set of free uniform graphs $\Omega_\emptyset^0 \subset \Omega_\emptyset(\mathbb{G})$ (see eq. (12)), which is a proper subset of $\Omega_\emptyset(\mathbb{G})$ if $\mathbb{G}$ has more than one end (a fact which will be discussed at the end of this section).

A natural approach to locally approximating UEG is to consider $\mathrm{UEG}_G$ for all $G = (\mathbb{V}, E) \subset \mathbb{G}$ where $E$ is finite. The simple observation that $\Omega_\emptyset(G) \subset \Omega_\emptyset(G')$ whenever $G \subset G'$ shows that these groups form an increasing net ordered by inclusion of finite subgraphs[7]. Consequently, the Haar measures $\mathrm{UEG}_G$ converge due to a general result.

**Theorem 3.10.** *Let $H$ be a compact Abelian group and let $(\Gamma_\alpha)_{\alpha \in \mathcal{I}}$ be an increasing net of closed sub-groups of $H$. The Haar measures $\mu_\alpha$, on $\Gamma_\alpha$ respectively, converge weakly to the Haar measure on $\Gamma = \overline{\cup_{\alpha \in \mathcal{I}} \Gamma_\alpha}$.*

*Proof.* The measures $\mu_\alpha$ extend to measures on $\Gamma$ by push-forward under the inclusion $\Gamma_\alpha \subset \Gamma$. By compactness of the space of probability measures on $H$, it suffices to establish that all accumulation points of $(\mu_\alpha)$ agree. Thus, let $\nu$ be an accumulation point of $\mu_\alpha$. Since $\Gamma_\alpha \preceq \Gamma_\beta$ for all $\alpha \preceq \beta$, $\nu$ is invariant under translation by any element of $\cup_{\alpha \in \mathcal{I}} \Gamma_\alpha$.

Now let $(g_j)_{j \in \mathcal{J}} \subset \cup_{\alpha \in \mathcal{I}} \Gamma_\alpha$ be a net converging to $g \in \Gamma$. Let $f$ be a continuous function on $\Gamma$, and note that the uniform continuity of $f$ implies that $f(x - g_j) \to f(x - g)$ uniformly.

---

[7] The reader may choose to think of a sequence of finite subgraphs $G_1 \subset G_2 \subset \ldots$ such that $\mathbb{G} = \bigcup_{n \in \mathbb{N}} G_n$ rather than the net of all finite subgraphs.

Therefore,

$$\lim_{j \in \mathcal{J}} \nu[f(x - g_j)] = \nu[f(x - g)].$$

Since $\nu[f(x - g_j)] = \nu[f]$ for all $j$, we conclude that $\nu$ is a probability measure with support on $\Gamma$ which is invariant under translation by elements of $\Gamma$. Therefore, $\nu$ is the unique Haar measure on $\Gamma$. $\qquad\square$

Define the set of free even subgraphs of $\mathbb{G}$

$$\Omega^0_\emptyset(\mathbb{G}) = \overline{\bigcup_{G \subset \mathbb{G} \text{ finite}} \Omega_\emptyset(G)} = \overline{\bigcup_{n \in \mathbb{N}} \Omega_\emptyset(G_n)}, \tag{12}$$

where $(G_n)_{n \in \mathbb{N}}$ is any sequence of finite subgraphs, $G_n \Uparrow \mathbb{G}$. The Haar measure on $\Omega^0_\emptyset(\mathbb{G})$, denoted $\mathrm{UEG}^0_\mathbb{G}$, is called the free uniform even subgraph. Theorem 3.10 shows that $\mathrm{UEG}^0_\mathbb{G} = \lim \mathrm{UEG}_{G_n}$ weakly. Furthermore, the existence of a finitary basis implies that this approximation is eventually constant on local events.

On the other hand, we refer to $\Omega_\emptyset(\mathbb{G})$ as the set of *wired* even subgraphs [8]. We shall now see how $\Omega_\emptyset(\mathbb{G})$ is approximated by local wired even subgraphs giving some justification to this name. Recall that the set of wired even subgraphs $\Omega_{\partial_v}(G)$ for a subgraph $G \subset \mathbb{G}$ is defined in terms of the boundary $\partial_v G$ with respect to $\mathbb{G}$. If $G \subset G' \subset \mathbb{G}$, typically $\Omega_{\partial_v}(G) \nsubseteq \Omega_{\partial_v}(G')$ in contrast to the situation for $\Omega_\emptyset$. Instead, the arrows are reversed, that is $\Omega_{\partial_v}(G)$ is (or at least contains) a quotient of $\Omega_{\partial_v}(G')$. Recall that by the isomorphism theorem, finite quotients correspond to homomorphisms with finite range such as the projections $\pi_G|_{\Omega_\emptyset(\mathbb{G})}$ onto $\pi_G(\Omega_\emptyset(\mathbb{G}))$ in our setting. Indeed, $\Omega_\emptyset(\mathbb{G})$ is profinite which is a way of saying that it is determined (up to isomorphism) by its (category of) finite quotients. In particular, write

$$\Omega_\emptyset(\mathbb{G}) = \varprojlim \pi_G(\Omega_\emptyset(\mathbb{G})).$$

for the cofiltered projective limit over the net of finite subgraphs ordered by inclusion. We do not unravel the definition here but remark that it is the smallest object admitting suitable projections and refer the reader to [53] for a presentation within the category of measure spaces. It was noted in [33, Theorem 2.6] that the projections $\pi_G|_{\Omega_\emptyset(\mathbb{G})}$ with corresponding marginals $\mathrm{UEG}_\mathbb{G}|_G$ for all finite $G \subset \mathbb{G}$ determine $\Phi^{\mathbb{c}}_*\mu$ by Kolmogorov's extension theorem. Again, this is a consequence of a general result for Haar measures.

**Theorem 3.11.** *Let* $\Gamma = \varprojlim \Gamma_\alpha$ *be a profinite group. Then, the projective limit of the normalised Haar measures on* $\Gamma_\alpha$ *exists and identifies with the consequently unique normalised Haar measure on* $\Gamma$.

*Proof.* Let $\mu_\alpha$ denote the Haar probability measure on $\Gamma_\alpha$, and $\pi_\alpha : \Gamma \to \Gamma_\alpha$ the projection for all $\alpha$. The conditions of [53, Theorem 3.4] are trivially satisfied for finite sets $\Gamma_\alpha$ so the existence of a unique regular Borel probability measure $\mu$ on $\Gamma$ is granted, such that $\mu_\alpha = (\pi_\alpha)_*\mu$ for all $\alpha$ and satisfying inner regularity with respect to cylinder sets, which amounts to

$$\mu[U] = \lim_\alpha \mu_\alpha[\pi_\alpha(U)], \tag{13}$$

for $U \subset \Gamma$ open. The left and right invariance of $\mu$ follows from that of $\mu_\alpha$ and (13) since $\pi_\alpha$ is a homomorphism for each $\alpha$, so $\mu_\alpha(\pi_\alpha(gUh)) = \mu_\alpha(\pi_\alpha(g)\pi_\alpha(U)\pi_\alpha(h)) = \mu_\alpha(U)$ for $g, h \in \Gamma$. Thus, $\mu$ is a Haar measure on $\Gamma$. Any Haar measure on $\Gamma$ shares the properties of $\mu$, so the uniqueness of $\mu$ according to [53, Theorem 3.4] implies the uniqueness of the Haar measure. □

**Remark 3.12.** *There are several approaches to constructing Haar measures, and Theorem 3.11 with the groundwork in [53, Theorem 3.4] is among them. This result is not new but included for the benefit of the reader. Note also, that [53, Theorem 3.4] generalises Kolmogorov's theorem. In particular, Kolmogorov's theorem provides a construction of $\mathbb{P}_{\frac{1}{2}}$ which coincides with the Haar measure on $\Omega(\mathbb{G}) = \varprojlim \Omega(G)$ by Theorem 3.11.*

Returning to the setting $G \subset \mathbb{G}$, observe that since $\pi_G(\Omega_\emptyset(\mathbb{G})) \not\subset \Omega_\emptyset(\mathbb{G})$, the uniform measure on $\pi_G(\Omega_\emptyset(\mathbb{G}))$ does not push forward to a measure on $\Omega_\emptyset(\mathbb{G})$, but it does push forward to $\Omega(\mathbb{G})$ along the inclusion $\pi_G(\Omega_\emptyset(\mathbb{G})) \subset \Omega(\mathbb{G})$. With this in mind, the convergence in (13) can be realised as weak convergence of measures on $\Omega(\mathbb{G})$. Therefore, we obtain a local approximation of UEG$_\mathbb{G}$ by uniform measures on the groups $\pi_G(\Omega_\emptyset(\mathbb{G}))$. As it turns out, on a connected graph, we may take approximating measures to be wired measures, UEG$_G^1$. Indeed, if $G_n \Uparrow \mathbb{G}$, such that each $G_n$ is finite and every component of $(\mathbb{V}, \mathbb{E} \setminus E_n)$ is infinite, then it follows from Proposition 3.6 that UEG$_G$ is approximated by a sequence of local wired uniform even subgraphs. We summarise the discussion in a theorem.

**Theorem 3.13.** *Let $\mathbb{G}$ be a locally finite, infinite connected graph. The Haar measures on $\Omega_\emptyset(\mathbb{G})$ and $\Omega_\emptyset^0(\mathbb{G})$ as probability measures on $\Omega(\mathbb{G})$ are weak limits of local wired and free uniform even subgraphs respectively.*

Finally to compare the free and the wired even subgraphs, let $G_n$ be as before and consider the commuting diagram consisting of inclusion and restriction maps

$$
\begin{array}{ccccccccc}
\ldots & \longrightarrow & \Omega_\emptyset(G_n) & \longrightarrow & \Omega_\emptyset(G_{n+1}) & \longrightarrow & \ldots & \longrightarrow & \Omega_\emptyset^0(\mathbb{G}) \\
& & \downarrow & & \downarrow & & & & \downarrow \\
\ldots & \longleftarrow & \Omega_{\partial_v}(G_n) & \longleftarrow & \Omega_{\partial_v}(G_{n+1}) & \longleftarrow & \ldots & \longleftarrow & \Omega_\emptyset(\mathbb{G})
\end{array}
$$

This diagram characterises the limiting groups. Recall that an *end* is an equivalence class of *rays*, that is, vertex self-avoiding paths with a single end-point where two rays are equivalent if there exists another self-avoiding path intersecting both rays infinitely often. Equivalently, the set of ends is the limit

$$\mathfrak{e}(\mathbb{G}) = \varprojlim \mathcal{C}^\infty(\mathbb{G} \setminus G),$$

taken over all finite $G \subset \mathbb{G}$, where $\mathcal{C}^\infty(\mathbb{G} \setminus G)$ denotes the set of infinite components of the graph $\mathbb{G} \setminus G$, and $\mathcal{C}_x(\mathbb{G} \setminus G')$ is identified with $\mathcal{C}_x(\mathbb{G} \setminus G)$ for all $x \in \mathbb{V}$ when $G \subset G'$.

If $\mathfrak{e}(\mathbb{G})$ is finite, one can check that $n$ can be chosen large enough such that there is an $M > n$ such that for all $m > M$,

$$\Omega_{\partial_v}(G_n)/\pi_{G_n}(\Omega_\emptyset(G_m)) \cong \{f \in \{0,1\}^{\mathfrak{e}(\mathbb{G})} | \sum_{e \in \mathfrak{e}(\mathbb{G})} f(e) = 0\} \tag{14}$$

Therefore, (14) is also satisfied for $\Omega_\emptyset(\mathbb{G})/\Omega_\emptyset^0(\mathbb{G})$ extending the results of [8].

### 3.3   Percolation on $\mathbb{Z}^2$

The proof of percolation of $\mathrm{UEG}_{\mathbb{Z}^3}$ relies on the following result on Bernoulli percolation:

**Proposition 3.14.** *The critical parameter for Bernoulli percolation on $\mathbb{Z}^2 \times \{0,1\}$ is strictly smaller than $\frac{1}{2}$. In particular, $\mathbb{P}_{\frac{1}{2},\mathbb{Z}^2 \times \{0,1\}}$ percolates.*

*Proof.*   We shall see that this is a relatively straightforward corollary of the fact that $p_c(\mathbb{P}_{\mathbb{Z}^2}) = \frac{1}{2}$ [39]. To see this, we define a map $T : \{0,1\}^{E(\mathbb{Z}^2 \times \{0,1\})} \to \{0,1\}^{E(\mathbb{Z}^2)}$ as follows: For $\omega \in \{0,1\}^{E(\mathbb{Z}^2 \times \{0,1\})}$, and an edge $e = (v,w) \in E(\mathbb{Z}^2)$, we set $e$ to be open in $T\omega$ if either $e^0 := ((v,0),(w,0))$ is open in $\omega$ or each of the three edges

$$\{v^\uparrow, e^1, w^\uparrow\}$$

is open in $\omega$. Here, $e^1 := ((v,1),(w,1))$ and $v^\uparrow := ((v,0),(v,1))$.

Now, for $\omega \sim \mathbb{P}_{p,\mathbb{Z}^2 \times \{0,1\}}$, the distribution of $T\omega$ is not quite Bernoulli percolation on $\mathbb{Z}^2$, since each vertical edge appears in multiple plaquettes. But the upshot is that any two vertices $v,w$ of $\mathbb{Z}^2$, $v$ and $w$ are connected in $T\omega$ only if $(v,0)$ and $(w,0)$ are connected in $\omega$.

In order to get rid of this dependence, let $G$ be the multigraph obtained from $\mathbb{Z}^2 \times \{0,1\}$ by replacing every vertical edge $v^\uparrow$ by four parallel edges $(v)^{\uparrow,e}$ indexed by the four non-vertical edges $e$ adjacent to $(v,1)$ (see Figure 7). We consider the measure $\mathbb{P}_G^{\boldsymbol{p}}$ on $\{0,1\}^{E(G)}$ where each edge is open independently with probability $p$ for all non-vertical edges and probability $\frac{p}{4}$ for the new vertical edges. For each $v \in \mathbb{Z}^2$, by a union bound, the probability that some edge above $(v,0)$ is open is at most $p$. Therefore, for every pair of vertices $v,w \in V(G) = V(\mathbb{Z}^2 \times \{0,1\})$, we have

$$\mathbb{P}_{p,\mathbb{Z}^2 \times \{0,1\}}[v \leftrightarrow w] \geq \mathbb{P}_G^{\boldsymbol{p}}[v \leftrightarrow w]. \tag{15}$$

Using this, we shall define a similar map $\tilde{T} : \{0,1\}^{E(G)} \to \{0,1\}^{E(\mathbb{Z}^2)}$ by declaring an edge $e = (v,w)$ open in $\tilde{T}\omega$ if either $e^0$ is open in $\omega$ or each of the three edges

$$\{v^{\uparrow,e^1}, e^1, w^{\uparrow,e^1}\}$$

is open in $\omega$. Now, because of the splitting of the vertical edges, we get that $\tilde{T}\omega \sim \mathbb{P}_{\tilde{p},\mathbb{Z}^2}$ for $\omega \sim \mathbb{P}_G^{\boldsymbol{p}}$. Inclusion-exclusion yields that

$$\tilde{p} = p + p\left(\frac{p}{4}\right)^2 - p^2\left(\frac{p}{4}\right)^2.$$

For $p = \frac{1}{2}$ we find $\tilde{p} = \frac{1}{2} + \frac{1}{256} > \frac{1}{2}$, and by continuity, $\tilde{p} > \frac{1}{2}$ in a neighbourhood of $p = \frac{1}{2}$. Accordingly, $\tilde{T}\omega$ percolates for all $p$ in this neighbourhood. As before, connections in $\tilde{T}\omega$ imply connections in $\omega$, which in turn, implies that $\omega$ percolates. By (15), so does $\mathbb{P}_{p,\mathbb{Z}^2 \times \{0,1\}}$, which is what we wanted. $\qquad\square$

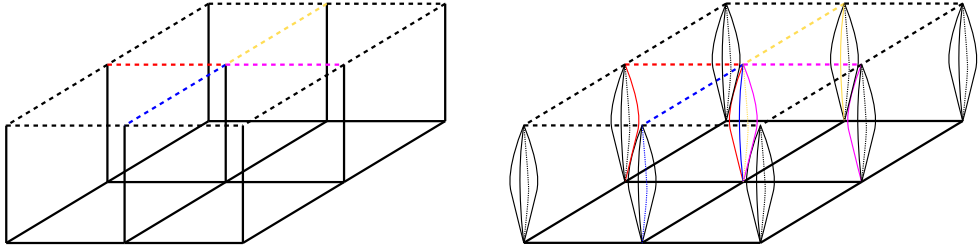

Figure 7: In order to get independence of $T\omega$, we replace every vertical edge of $\mathbb{Z}^2 \times \{0, 1\}$ by a parallel family of four vertical edges with lower edge weights. In order to construct $\tilde{T}\omega$, we then colour the new vertical edges according to the horizontal ones.

### 3.3.1 Even and odd percolation

The proofs that $\text{UEG}_{\mathbb{Z}^2}$ percolates (Corollary 2.10) and that $\text{UEG}_{\mathbb{Z}^d}$ percolates for $d \geq 3$ (Theorem 1.1) are rather different, and neither of the proofs extends to cover the other case. Let us take a moment to ponder the differences between the two methods.

First notice that one can sample the uniform even subgraph of $\mathbb{Z}^2$ by placing a fair coin on each plaquette, flipping them, and taking the symmetric difference of all the plaquettes where the coin landed heads up. As $\mathbb{Z}^2$ is self-dual, the distribution of the coins is site percolation on $\mathbb{Z}^2$. Since site percolation on $\mathbb{Z}^2$ does not percolate at parameter $p = \frac{1}{2}$, the clusters of coins showing heads are all finite. Therefore, the infinite component of the UEG arises as a union of finite clusters of heads meeting at plaquettes that share only a vertex and not an edge. This is intimately related to the existence of vertices of degree 4. Attempting to investigate the importance of vertices of degree 4, we prove that the uniform even subgraph of any bi-periodic trivalent planar graph does not percolate in Proposition 5.3. However, in Proposition 6.1 we construct a (non-amenable, non-planar) trivalent graph $\mathbb{J}$ such that $\text{UEG}_{\mathbb{J}}$ percolates.

In parallel to the case of the uniform even subgraph, whenever a finite graph $G$ allows a dimerisation, that is, a perfect matching, we may consider the uniform odd subgraph. Since $\mathbb{Z}^d$ allows a dimerisation, one way to define the uniform odd subgraph of $\mathbb{Z}^d$ is by taking the symmetric difference of the UEG and a fixed dimerisation. A more general characterisation of the uniform odd subgraph is that it is the unique probability measure on the co-set of odd subgraphs (supposing this is non-empty) which is invariant under the action of the even subgraphs.

Since Bernoulli-$\frac{1}{2}$ percolation is the Haar measure on the space of percolation configurations, it is invariant under taking the symmetric difference with any deterministic set. Therefore, the symmetric difference of a dimerisation with the edges in a hyperplane as in the proof of Theorem 1.1 is still $\mathbb{P}_{\frac{1}{2}}$ distributed. Thus, the proof of Theorem 1.1 generalises and the uniform

| | even | odd |
|:---:|:---:|:---:|
| $\mathbb{Z}^2$ | ✓ | ? |
| $\mathbb{Z}^d, d \geq 3$ | ✓ | ✓ |
| $\mathbb{H}$ | × | × |

Table 2: Overview of percolation of even and odd percolation on the hypercubic and hexagonal lattices

odd subgraph percolates on $\mathbb{Z}^d$ for $d \geq 3$.

On the contrary, the proof of Corollary 2.10 does not generalise to the uniform odd subgraph of $\mathbb{Z}^2$ and to our knowledge, it is still open whether odd percolation has the same phase transition as even percolation in $\mathbb{Z}^2$, the consequences of which are discussed in [36, Section 6]. Furthermore, we note in Proposition 5.5 and Proposition 5.7 that neither the uniform even nor the uniform odd subgraph of the hexagonal lattice percolates. We summarise our knowledge of even and odd percolation in Table 2.

Finally, recall from Theorem 2.9 that the uniform even subgraph of $\mathbb{Z}^2$ percolates. However, if we consider the inclusion $\mathbb{Z}^2 \subset \mathbb{Z}^3$, then the marginal of $\mathrm{UEG}_{\mathbb{Z}^3}$ on $\mathbb{Z}^2$ is $\mathbb{P}_{\frac{1}{2}, \mathbb{Z}^2}$ by Lemma 3.5. Since $\mathbb{P}_{\frac{1}{2}, \mathbb{Z}^2}$ does not percolate, there is a certain non-monotonicity of (percolation of) the uniform even subgraph. In [35], we show an even stronger non-monotonicity statement of UEG and $\ell_x$ on general graphs.

## 3.4 The loop O(1) model percolates for large $x \in [0, 1]$

For monotone measures, given a single point of percolation, there is an interval of parameters for which there is percolation. This does not apply to the loop O(1) model, which has negative association, but we can still bootstrap the strategy from Theorem 1.1 to get an interval of percolation points.

**Theorem 1.2.** *Suppose that $d \geq 2$. Consider the loop O(1) model $\ell_{x, \mathbb{Z}^d}$ with parameter $x \in [0, 1]$. Then there exists an $x_0 < 1$ such that*

$$\ell_{x, \mathbb{Z}^d} [0 \leftrightarrow \infty] > 0,$$

*for all $x \in (x_0, 1)$.*

**Remark 3.15.** *A similar statement for the single random current $\mathbf{P}_\beta$ follows from the fact that $\mathbf{P}_\beta$ stochastically dominates $\ell_x$ cf. Theorem 2.5. However, this also follows without Theorem 1.2 since $\mathbf{P}$ dominates a Bernoulli percolation.*

*Proof.* For $d = 2$, the statement follows from the stronger result in Theorem 2.9.

The rest of the argument follows the strategy of Theorem 1.1. We are going to show that the marginal of $\ell_{x, \mathbb{Z}^d}$ on a suitably chosen set is bounded from below by a supercritical Bernoulli percolation. Again, we divide into cases according to whether $d = 3$ or $d \geq 4$.

For $d \geq 4$, consider a hyperplane $\mathbb{Z}^{d-1} \subset \mathbb{Z}^d$. Let $p \in (0,1)$ and consider the random-cluster model $\phi_{p,\mathbb{Z}^d}$. Then, by Proposition 2.1 the measure $\phi_{p,\mathbb{Z}^d}$ dominates Bernoulli percolation with parameter $\tilde{p} = \frac{p}{2-p}$, i.e. $\mathbb{P}_{\tilde{p},\mathbb{Z}^d} \preceq \phi_{p,\mathbb{Z}^d}$. Equivalently, there exists a coupling $(\omega, \tilde{\omega})$ such that $\omega \sim \phi_{p,\mathbb{Z}^d}$, $\tilde{\omega} \sim \mathbb{P}_{\tilde{p},\mathbb{Z}^d}$ and $\omega \preceq \tilde{\omega}$ almost surely.

Now, for any edge $e \in \mathbb{Z}^{d-1}$, we say that $e$ is *good* if $e$ is open, the loop around the plaquette containing $e$ just above $e$ in $\mathbb{Z}^d$ is open, see Figure 4. The probability that this loop is open in $\tilde{\omega} \sim \mathbb{P}_{\tilde{p}}$ is $\tilde{p}^4$. Define the process $Z_e\left((\omega, \tilde{\omega})\right) = \mathbb{1}_{\{e \text{ is good}\}}(\tilde{\omega})$ for each $e \in E$. Then, $Z_e \perp Z_{e'}$ if $|e - e'| \geq 2$. Therefore, by [47, Theorem 0.0], the process $Z$ stochastically dominates some $W \sim \mathbb{P}_{q,\mathbb{Z}^d}$ where $q \to 1$ if $p \to 1$. Defining now $Q_e\left((\omega, \tilde{\omega})\right) = \mathbb{1}_{\{e \text{ is good}\}}(\omega)$, then $Q_e \geq Z_e \geq W_e$ almost surely and therefore $Q \succeq W$.

Next, we apply the relation (5) between $\phi_{p,\mathbb{Z}^d}$ and $\ell_{x,\mathbb{Z}^d}$ which consists of taking a uniform even subgraph. Under this coupling, conditionally on $\phi_{p,\mathbb{Z}^d}$, edges that are separated by cycles become independent by Lemma 3.5. Therefore, we must have that $\ell_{x,\mathbb{Z}^d}|_{\mathbb{Z}^{d-1} \times \{0\}} \succeq \mathbb{P}_{\frac{q}{2}, \mathbb{Z}^{d-1} \times \{0\}}$.

Thus, for any fixed $a \in (0, \frac{1}{2})$ if we pick $p$ close enough to 1 then $\frac{q}{2} > a$. Since $d - 1 \geq 3$, we can once again use the fact that the edge percolation threshold for Bernoulli percolation on $\mathbb{Z}^3$ is strictly less than $\frac{1}{2}$ and so, $\ell_{x,\mathbb{Z}^d}|_{\mathbb{Z}^{d-1} \times \{0\}}$, and therefore also $\ell_{x,\mathbb{Z}^d}$, percolates.

For $d = 3$, we apply the same argument as in the proof of the $d = 3$ in Theorem 1.1, where we replace the hyperplane by $\mathbb{Z}^2 \times \{0, 1\}$ and use Proposition 3.14. $\qquad \square$

# 4 Proof of Theorem 1.5

Before we get our hands dirty with the details of the proof of our main theorem, we begin this section by giving a brief overview of the arguments that go into the proof. We do so to emphasise the underlying topological ideas, for the benefit of the impatient reader, as well as preparing the road ahead. Then, we review some technical aspects of the Ising literature, and set up the machinery that we shall need. Together with the previous results on the marginals of UEG, this enables the proof of Theorem 1.3 as well as Theorem 1.4. The latter requires an adaptation of the multi-valued mapping principle, to be discussed.

## 4.1 Road map and torus basics

We saw in Corollary 3.7 that for $\mathbb{Z}^d$, the boundary conditions under which we take the uniform even subgraph do not matter as soon as we take a single step away from the boundary due to the richness of loops in the graph. For the supercritical random-cluster model, we can hardly expect a result quite as powerful, but it is known that we have a local uniqueness property of the model ensuring that any suitably nice finite piece of the lattice is going to have a single large cluster. Thus, with the caveat that we might have to take a bit more than a single step away from the boundary, we should also expect the boundary conditions we are working with not to matter. In the rest of the section, we fix $d \geq 3$.

At the same time, even graphs have some nice interplay with the topology of the torus. For a hyperplane $H$ in $\mathbb{T}_n^d$ orthogonal to the $e_1$ direction and $v \in H$, we shall call the edge $(v, v + e_1)$ outgoing and the edge $(v, v - e_1)$ incoming. One observes that for any even subgraph $G = (V, E)$

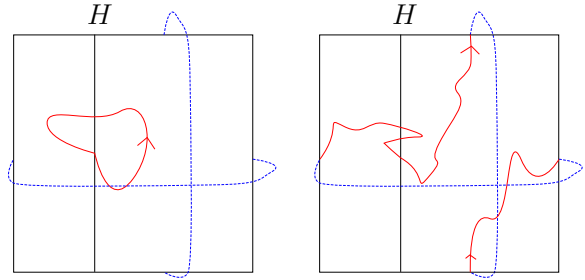

Figure 8: A simple, short, topologically trivial path pictured left and a long path wrapping all the way around the torus on the right. A path wraps all the way around the torus an odd number of times in the $e_1$ direction if and only if it contains an odd number of the outgoing edges from $H$.

of $\mathbb{T}_n^d$, any outgoing edge $e$ in $E$ must be connected in $\mathbb{T}_n^d \setminus H$ either to another outgoing edge or to one of the incoming edges. In the latter case, $e$ is part of a cluster winding all the way around the torus in one direction, and such a cluster has size at least[8] $n$. This must be the case for *some* outgoing edge $e$ if the number of outgoing edges in $E$ is odd, see Figure 8. This discussion is completely analogous to the observation for the mirror model made in [44] and motivates the following definition:

**Definition 4.1.** *We say that a loop is **simple** if it is a path from a vertex to itself such that every other vertex in the path is only visited once. A simple loop $\gamma$ in $\mathbb{T}_n^d$ is a **wrap-around** if it contains an odd number of outgoing edges of a hyperplane $H$ orthogonal to the $e_1$-direction.*

*We say that an even subgraph $G = (V, E)$ of $\mathbb{T}_n^d$ is **non-trivial** if $E$ contains a wrap-around. Otherwise, say that $G$ is **trivial**.*

**Remark 4.2.** *Following the above discussion, if $\gamma$ contains an odd number of outgoing edges of the hyperplane $H$, then it must also do so for any hyperplane $H'$ parallel to $H$. As such, the above definition does not depend on the choice of hyperplane. Of course, a loop might wrap around the torus in several directions, and we could define $j$-wrap-arounds for every cardinal direction in $\mathbb{Z}^d$, but since we do not use the different choices of direction in our proofs at all, we omit them from the definition.*

To illustrate our use of wrap-arounds, we warm up with the following lemma:

**Lemma 4.3.** *Let $G = (V, E)$ be a trivial even subgraph of $\mathbb{T}_n^d$ and let $\gamma$ be a wrap-around. Then, $G\Delta\gamma := (V, E\Delta\gamma)$ is non-trivial.*

*Proof.* Fix a hyperplane $H$ and note that, since the number of outgoing edges in $\gamma$ is odd, the parity of outgoing edges from $H$ is different in $G$ than it is in $G\Delta\gamma$. This immediately implies the statement. □

---

[8] The astute reader might wonder why we do not simply make this argument on the sphere (corresponding to the wired random-cluster measure). It is exactly this lower bound on the size of topologically non-trivial clusters which fails for the wired measure.

One immediate consequence hereof is the following:

**Corollary 4.4.** *Let $G$ be a fixed, not necessarily even, subgraph of $\mathbb{T}_n^d$ which contains a wrap-around and let* $\mathtt{NT}$ *denote the event on $\Omega_\emptyset(G)$ that a percolation configuration is non-trivial. Then,*

$$\mathrm{UEG}_G[\mathtt{NT}] \geq \frac{1}{2}.$$

*Proof.* Fix a wrap-around $\gamma$ in $G$ and let $\eta \sim \mathrm{UEG}_G$. By the Haar measure property, we have that $\eta \overset{d}{=} \eta \Delta \gamma$. However, by Lemma 4.3, at least one of $\eta$ and $\eta \Delta \gamma$ is non-trivial. Accordingly, by a union bound,

$$1 = \mathrm{UEG}_G[(\eta \in \mathtt{NT}) \cup (\eta \Delta \gamma \in \mathtt{NT})] \leq 2\,\mathrm{UEG}_G[\mathtt{NT}].$$

$\square$

Thus, if we can exhibit wrap-arounds in the random-cluster model, we get long clusters in the loop O(1) model with positive probability. By using translation invariance of the random-cluster model on the torus, we get a lower bound on the probability of having a long cluster passing through exactly the vertex 0. This is the main idea of our proof, although we shall be slightly more clever in our application of Lemma 4.3 to improve the bound we get.

**Definition 4.5.** *For an even subgraph $G$ of $\mathbb{T}_n$, we denote by $\mathcal{C}_{\mathtt{NT}}$ the union of the non-trivial connected components of $G$.*

By translation invariance of the random-cluster model on the torus,

$$\ell_{x,\mathbb{T}_n^d}[0 \in \mathcal{C}_{\mathtt{NT}}] = \ell_{x,\mathbb{T}_n^d}\left[\frac{|\mathcal{C}_{\mathtt{NT}}|}{|\mathbb{T}_n^d|}\right].$$

Hence, our goal in Section 4.5 shall be to lower bound this quantity.

## 4.2 Local Uniqueness

In order to exploit the uniform even subgraph to say something intelligent about the loop O(1) model, we have to build up some technical machinery for the random-cluster model. This section is dedicated to doing just that. As such, the work herein mostly consists in massaging results from the literature into a form more amenable to our needs. The arguments are slightly technical and a reader who is more eager to get to the proofs of our main theorems might choose to skip it on a first reading. Apart from the proof of Theorem 1.3, the results that we shall be needing later on are Lemmata 4.6, 4.8 and 4.14. As a first ingredient, one may note that the construction in [20, Theorem 1.3] implies equally well the following lemma:

**Lemma 4.6.** *For any $p > p_c$, there exists $c > 0$ such that for any $n \in \mathbb{N}$ and any event $A$ depending only on edges of $\Lambda_n$,*

$$\left|\phi_{p,\Lambda_{2n}}^\xi[A] - \phi_{p,\mathbb{Z}^d}[A]\right| < \exp(-cn)$$

*for any boundary condition $\xi$.*

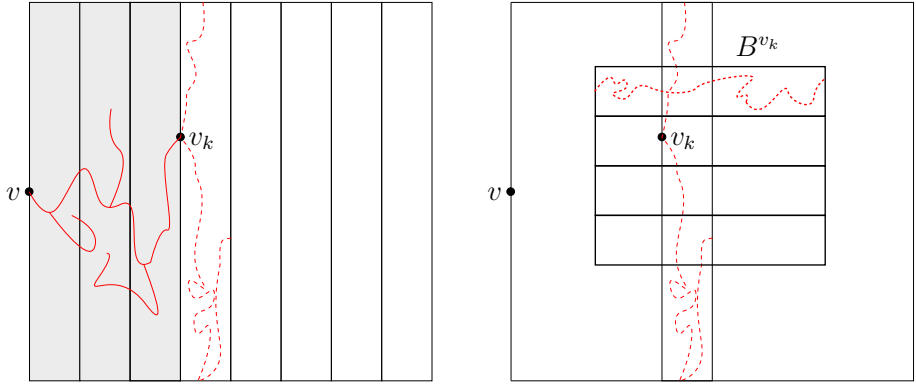

Figure 9: On the left: A sketch of the event $E_v$ from the side. When exploring a cluster traversing the annulus from the inside to a given face, we have a probability to connect to all other faces of a given box every time we enter a new strip (corresponding to a translate of $\tilde{S}_h$). This probability is uniform in the past configuration. On the right: How to apply the same argument in the last direction. Using the fact that a crossing from top to bottom of $B^{v_k}$ must intersect many transversal slabs, it is unlikely that this happens without the cluster of $v_k$ also connecting to the left and right sides of the box.

Our second input comes from Pisztora's construction of a Wulff theory for random-cluster models in arbitrary dimension [52]. Let $\mathfrak{C}_{n,L,\varepsilon,\theta} \subset \{0,1\}^{E(\Lambda_n)}$ be the event that there exists a cluster $\mathcal{C}_{\max}$ in $\Lambda_n$ such that

- $\mathcal{C}_{\max}$ is the unique cluster in $\Lambda_n$ touching all faces of $\partial \Lambda_n$.

- $|\mathcal{C}_{\max}| \geq (\theta - \varepsilon)n^d$.

- There are at most $\varepsilon n^d$ vertices in $\Lambda_n$ that do not lie on $\mathcal{C}_{\max}$ and that lie on clusters larger than $L$.

Combining [52, Theorem 1.2] and [11, Theorem 2.1] yields the following:

**Proposition 4.7.** *For all $p > p_c$, if $\theta := \phi_{p,\mathbb{Z}^d}[0 \leftrightarrow \infty]$, for all $0 < \varepsilon < \theta/2$, there exist $L$ and $c > 0$ such that*

$$\phi_{p,\Lambda_n}^\xi[\mathfrak{C}_{n,L,\varepsilon,\theta}] \geq 1 - \exp\left(-cn^{d-1}\right)$$

*for all $n \in \mathbb{N}$ and every boundary condition $\xi$.*

A first consequence of this is a result on annular domains, which is essential for showing that the loop $O(1)$ model is not very sensitive to boundary conditions.

**Lemma 4.8** (Local Uniqueness). *For $d \geq 3$ and every $p > p_c$, there exists $c > 0$ with the following property: If $\mathtt{UC}_n$ is the event that $\omega|_{\Lambda_{2n} \setminus \Lambda_n}$ has a unique cluster crossing from $\Lambda_n$ to $\partial \Lambda_{2n}$, then*

$$\phi_{p,\Lambda_{2n}}^{\xi}[\mathtt{UC}_n] > 1 - \exp(-cn)$$

*for all $n \in \mathbb{N}$ and every boundary condition $\xi$.*

**Remark 4.9.** *Note that the above statement is also true for $d = 2$ by sharpness and duality arguments (see Section 5.1) and for $d = 1$ since here, there is no $p > p_c$. However, the proof we give below very specifically uses the fact that $d \geq 3$.*

**Remark 4.10.** *In the following, we are juggling several constants. Our convention here and throughout will be to remark upon the changing of the value of a constant $c$ by denoting the new one $c'$ the first time it appears. Afterwards, to prevent notational bloat, we shall revert to simply writing $c$.*

*Proof.* The strategy for the proof comes in two steps: First, we show that with probability exponentially close to 1, there is a unique large cluster in $\Lambda_{2n} \setminus \Lambda_n$ of large volume and then we show that, again with high probability, any crossing must be part of this one big cluster.

For any finite set $\mathcal{B}$ of translates $B_j$ of $\Lambda_{n/2}$, observe the auxiliary graph $G_{\mathcal{B}}$ with vertices $j \in \{1, ..., |\mathcal{B}|\}$ and an edge $(j,l)$ if $B_j \cap B_l$ contains a translate of $\Lambda_{n/4}$. For each edge $(j,l)$, let $B_{j,l}$ denote a choice of such a translate. For our purposes, $\mathcal{B}$ will be the cover of $\Lambda_{2n} \setminus \Lambda_n$ consisting of all translates of $\Lambda_{n/2}$ contained in $\Lambda_{2n} \setminus \Lambda_{n-1}$. Since the centre of every such translate lies on $\partial \Lambda_{3n/2}$, we get that

$$|\mathcal{B}| \leq C n^{d-1}. \tag{16}$$

For each $j$, let $\mathfrak{C}^j$ denote the event that the corresponding translate of the event $\mathfrak{C}_{n/2,L,\theta/4^d,\theta}$ from Proposition 4.7 occurs in $B_j$. Similarly, for an edge $(j,l)$ of $G_{\mathcal{B}}$, let $\mathfrak{C}^{j,l}$ denote the event that the corresponding translate of $\mathfrak{C}_{n/4,L,\theta/4^d,\theta}$ occurs in $B_{j,l}$.

By a union bound, Proposition 4.7 and (16), we see that

$$\phi_{p,\Lambda_{2n}}^{\xi}\left[\left(\cap_j \mathfrak{C}^j\right) \cap \left(\cap_{(j,l) \in E(G_{\mathcal{B}})} \mathfrak{C}^{j,l}\right)\right] \geq 1 - Cn^{d-1}\exp\left(-c(n/2)^{d-1}\right) - C^2 n^{2(d-1)}\exp\left(-c(n/4)^{d-1}\right),$$

which is at least $1 - \exp\left(-c'n^{d-1}\right)$ for an adjusted value $c'$.

For future reference, we shall abbreviate $\mathfrak{C}^{\max} = \left(\cap_j \mathfrak{C}^j\right) \cap \left(\cap_{(j,k) \in E(G_{\mathcal{B}})} \mathfrak{C}^{j,k}\right)$. In conclusion,

$$\phi_{p,\Lambda_{2n}}^{\xi}[\mathfrak{C}^{\max}] \geq 1 - \exp(-cn). \tag{17}$$

Now, for $(j,k) \in E(G_{\mathcal{B}})$, on the event $\mathfrak{C}^j \cap \mathfrak{C}^k$, there is a unique large cluster $\mathcal{C}^j$ contained in $B_j$ of size at least $\theta/4^d(n/2)^d$ and likewise for $k$. However, on $\mathfrak{C}^{j,k}$, there is a cluster $\mathcal{C}^{j,k}$ in $B^{j,k}$ of size $\frac{4^d - 1}{4^d}\theta\left(\frac{n}{4}\right)^d > \theta/4^d\left(\frac{n}{2}\right)^d$. Accordingly, $\mathcal{C}^{j,k} \subseteq \mathcal{C}^j \cap \mathcal{C}^k$. Since $G_{\mathcal{B}}$ is connected, we get that, on $\mathfrak{C}^{\max}$, all the $\mathcal{C}^j$ are part of one big cluster.

Now, for the second part of the argument, let $v \in \partial \Lambda_n$ be given and let $\mathcal{C}_v$ denote the cluster of $v$ in $\omega|_{\Lambda_{2n} \setminus \Lambda_n}$. Let $\mathcal{E}_v$ denote the event that $v$ is connected inside of $\Lambda_{2n} \setminus \Lambda_n$ to $\partial \Lambda_{2n}$ and for any $j$, let $\mathfrak{A}_j^v$ be the event that $\mathcal{C}_v \cap B_j$ contains a cluster which touches every face of $\partial B_j$. We denote by $\mathfrak{A}^v$ the event that $\mathfrak{A}_j^v$ occurs for some $j$. We wish to show that if $\mathcal{C}_v$ crosses $\Lambda_{2n} \setminus \Lambda_n$, then with high probability, it must touch all faces of some $B_j$, that is

$$\phi_{p,\Lambda_{2n}}^\xi [\cup_{v \in \partial \Lambda_n} \mathcal{E}_v \setminus \mathfrak{A}^v] \le e^{-cn}. \tag{18}$$

Let us first see how (18) finishes the proof. Notice that

$$\mathfrak{C}^{\max} \setminus (\cup_{v \in \partial \Lambda_n} \mathcal{E}_v \setminus \mathfrak{A}^v) \subset \mathtt{UC}_n,$$

since on the former event, any cluster which crosses $\Lambda_{2n} \setminus \Lambda_n$ touches all faces of $B_j$ for some $j$, any such cluster must be the same as $\mathcal{C}^j$, and all the $\mathcal{C}^j$ are part of the same cluster. The lemma then follows by combining (18) and (17).

Thus, let us establish (18). First, by [11, Theorem 2.1], there exists a constant $h$ such that $\phi_{p,S_h}^0$ percolates[9], where $S_h$ denotes the slab $\{0, ..., h\}^{d-2} \times \mathbb{Z}^2$. In particular, if $\tilde{S}_h := \{0, ..., h\} \times \mathbb{Z}^{d-1}$, we get that $\phi_{p,\tilde{S}_h}^0$ also percolates. For the rest of the proof, we shall assume, without loss of generality, that $n > h$. Combining [52, Lemma 3.3] with the FKG inequality (Proposition 2.2), there exists a $c > 0$, depending only on $p$ such that, for any hyper-rectangle $B$ in $\mathbb{Z}^d$ and any vertex $w$, if $\mathfrak{A}_B^h(w)$ denotes the event that the cluster of $w$ in $\tilde{S}_h + w$ touches all faces of $B \cap (\tilde{S}_h + w)$, we have $\phi_{p,\tilde{S}_h+w}^0[\mathfrak{A}_B^h(w)] \ge c$.

Now, suppose $v$ is connected to $\{\langle w, e_1 \rangle = 2n\}$, and call this event $\mathcal{E}_v^\uparrow$. On $\mathcal{E}_v^\uparrow$, we must have that $\mathcal{C}_v$ crosses $\frac{n}{h}$ disjoint translates of $\tilde{S}_h$. To use this, we explore the cluster of $v$ from $\Lambda_k$ to $\{\langle w, e_1 \rangle = 2n\}$ one translate of $\tilde{S}_h$ at a time and denote by $v_k$ the first vertex of $\mathcal{C}_v$ we encounter in $v + khe_1 + \tilde{S}_h$. See Figure 9.

Let $B^{v_k}$ denote some $B_j$ such that $v_k, v_k + he_1 \in B^{v_k}$. Similarly, let $\mathrm{Past}(v_k)$ denote the state of all previously discovered edges (open or closed) and let $E(\mathrm{Past}(v_k))$ denote the set of discovered edges. By the Domain Markov Property (Proposition 2.3), we have that

$$\phi_{p,\mathbb{Z}^d}[\cdot | \ \mathrm{Past}(v_k)] = \phi_{p,\mathbb{Z}^d \setminus E(\mathrm{Past}(v_k))}^{\xi(\mathrm{Past}(v_k))}[\cdot],$$

where $\xi(\mathrm{Past}(v_k))$ are the boundary conditions which are wired on the component of $v$ in $\mathrm{Past}(v_k)$ and free otherwise. Since for any probability measure $\nu$ and events $U, V$ with $\nu(U \cap V) > 0$, we have $\nu[\cdot | \ U \cap V] = \nu_V[\cdot | \ U]$, where $\nu_V[\cdot] = \nu[\cdot | \ V]$, we conclude that

$$\phi_{p,\mathbb{Z}^d}[\mathfrak{A}_{B^{v_k}}^h(v_k) | \ \mathrm{Past}(v_k), \mathcal{E}_v^\uparrow] = \phi_{p,\mathbb{Z}^d \setminus E(\mathrm{Past}(v_k))}^{\xi(\mathrm{Past}(v_k))}[\mathfrak{A}_{B^{v_k}}^h(v_k) | \tilde{\mathcal{E}}_v^\uparrow],$$

where $\omega \in \tilde{\mathcal{E}}_v^\uparrow$ if $\omega \cup \mathrm{Past}(v_k) \in \mathcal{E}_v^\uparrow$ (and $\mathrm{Past}(v_k)$ is identified with its open edges). Since $\tilde{\mathcal{E}}_v^\uparrow$ is increasing, we can apply the FKG inequality to get that

$$\phi_{p,\mathbb{Z}^d \setminus E(\mathrm{Past}(v_k))}^{\xi(\mathrm{Past}(v_k))}[\mathfrak{A}_{B^{v_k}}^h(v_k) | \ \tilde{\mathcal{E}}_v^\uparrow] \ge \phi_{p,\mathbb{Z}^d \setminus E(\mathrm{Past}(v_k))}^{\xi(\mathrm{Past}(v_k))}[\mathfrak{A}_{B^{v_k}}^h(v_k)].$$

---

[9] Due to the finite size of the graph in all but two directions, the boundary conditions under which we take the infinite volume limit on $S_h$ matter, unlike in the case of $\mathbb{Z}^d$.

Applying the comparison between boundary conditions (cf. Proposition 2.1$ii$), we see that

$$\phi_{p,\mathbb{Z}^d \setminus E(\text{Past}(v_k))}^{\xi(\text{Past}(v_k))}[\mathfrak{A}_{B^{v_k}}^h(v_k)] \geq \phi_{p,\tilde{S}_h + v_k}^0[\mathfrak{A}_{B^{v_k}}^h(v_k)] \geq c$$

Iterating on the above, we see that, conditionally on $\mathcal{E}_v^\uparrow$, $\mathfrak{A}_{B^{v_k}}^h(v_k)$ occurs for some $k$ with probability at least $1 - (1-c)^{n/h} > 1 - e^{-c'n}$ for some appropriate choice of $c'$. On the event $\mathfrak{A}_{B^{v_k}}^h(v_k)$, we have that the cluster of $v_k$ in $\mathcal{C}_v \cap B^{v_k}$ touches all faces of $B^{v_k}$ except possibly those orthogonal to $e_1$. However, we may apply a similar exploration argument to get that

$$\phi_{p,\mathbb{Z}^d}[\mathfrak{A}^{v_k} \mid \mathfrak{A}_{B^{v_k}}^h(v_k)] > 1 - e^{-cn},$$

see Figure 9. Note that if $\kappa$ denotes the first $k$ such that $\mathfrak{A}_{B^{v_k}}^h(v_k)$ occurs, then

$$\mathcal{E}_v^\uparrow \cap (\kappa < \infty) \cap (\mathfrak{A}^{v_\kappa}) \subseteq \mathfrak{A}^v$$

and therefore, on $\mathcal{E}_v^\uparrow \setminus \mathfrak{A}^v$, either $\kappa = \infty$ or there is some $k$ and a $w$ on the boundary of $v + khe_1 + S_h$ such that $\mathfrak{A}_{B^w}^h(w) \setminus \mathfrak{A}^w$ occurs. Therefore, a union bound shows that

$$\phi_{p,\mathbb{Z}^d}[\mathcal{E}_v^\uparrow \setminus \mathfrak{A}^v] \leq \phi_{p,\mathbb{Z}^d}[\kappa = \infty \mid \mathcal{E}_v^\uparrow] + \sum_{k=1}^{n/h} \sum_{w \in \partial(v + khe_1 + S_h)} \phi_{p,\mathbb{Z}^d}[\mathfrak{A}_{B^w}^h(w) \setminus \mathfrak{A}^w] \leq (1 + Cn^d)e^{-cn} \leq e^{-c'n}.$$

The argument in the case where $v$ is connected to another face is similar. Thus, summing over the $2d$ faces of $\Lambda_{2n}$, we get that

$$\phi_{p,\mathbb{Z}^d}[\mathcal{E}_v \setminus \mathfrak{A}^v] \leq 2de^{-c'n} \leq e^{c''n}.$$

By yet another union bound,

$$\phi_{p,\mathbb{Z}^d}[\cup_{v \in \partial \Lambda_n} \mathcal{E}_v \setminus \mathfrak{A}^v] \leq Cn^{d-1}e^{-c'n} \leq e^{-c''n}$$

for an adjusted constant $c''$. By Lemma 4.6, we get (18). □

## 4.3 Insensitivity to boundary conditions and mixing of the loop O(1) model

Lemma 4.8 is the random-cluster analogue of the connectivity property that we used for the uniform even graph of $\mathbb{Z}^d$ in Corollary 3.7. Combined with Lemma 4.6, we get that the loop O(1) model is insensitive to boundary conditions:

**Theorem 1.3.** *For $x > x_c$, there exists $c > 0$ such that for any $n \in \mathbb{N}$ and any event $A$ which only depends on edges in $\Lambda_n$ and any $G$ which is a supergraph extending $\Lambda_{4n}$, then*

$$|\ell_{x,G}^\xi[A] - \ell_{x,\mathbb{Z}^d}[A]| \leq \exp(-cn),$$

*for any boundary condition $\xi$. In particular, for $x > x_c$ and any sequence $\xi_k$ of boundary conditions, $\lim_{k \to \infty} \ell_{x,\Lambda_k}^{\xi_k} = \ell_{x,\mathbb{Z}^d}$ in the sense of weak convergence of probability measures.*

*Proof.* For a percolation configuration $\omega$, note that the event $\mathtt{UC}_n$ from Lemma 4.8 is equal to the event that there is a cluster in $\omega|_{\Lambda_{2n}\setminus\Lambda_n}$ containing a separating surface between $\omega|_{\Lambda_n}$ and $\omega|_{\Lambda_{4n}\setminus\Lambda_{2n}}$. In other words, Proposition 3.8 applies, so that whenever $\omega \in \mathtt{UC}_n$, $\mathrm{UEG}_\omega[A] = \mathrm{UEG}_{\omega|_{\Lambda_{2n}}}[A]$. It follows that $\mathbb{1}_{\mathtt{UC}_n}(\omega)\mathrm{UEG}_\omega[A]$ is a random variable which is measurable with respect to the state of $\omega$ on edges in $\Lambda_{2n}$. Furthermore, it is positive and bounded from above by 1. Denoting by $d_{\mathrm{TV}}$ the total variation distance between probability measures, we can conclude:

$$
\begin{aligned}
|\ell_{x,G}^\xi[A] - \ell_{x,\mathbb{Z}^d}[A]| &= \left| \phi_{x,G}^\xi[\mathrm{UEG}_\omega[A]] - \phi_{x,\mathbb{Z}^d}[\mathrm{UEG}_\omega[A]] \right| \\
&\leq \left| \phi_{x,G}^\xi[\mathbb{1}_{\mathtt{UC}_n(\omega)}\mathrm{UEG}_\omega[A]] - \phi_{x,\mathbb{Z}^d}[\mathbb{1}_{\mathtt{UC}_n}(\omega)\mathrm{UEG}_\omega[A]] \right| + 2 - \phi_{x,G}^\xi[\mathtt{UC}_n] - \phi_{x,\mathbb{Z}^d}[\mathtt{UC}_n] \\
&\leq d_{\mathrm{TV}}(\phi_G^\xi|_{\Lambda_{2n}}, \phi_{\mathbb{Z}^d}|_{\Lambda_{2n}}) + \exp(-cn) \\
&\leq \exp(-c'n) + \exp(-cn) \\
&\leq \exp(-c''n),
\end{aligned}
$$

where in the second inequality, we used Lemma 4.8 and in the third, we used Lemma 4.6. $\square$

By a similar argument, one may show an actual mixing result on $\mathbb{Z}^d$.

**Theorem 4.11.** *For $x > x_c$, there exists $c > 0$ such that for any $n \in \mathbb{N}$ and any events $A$ and $B$ such that $A$ depends only on the edges in some box $v_A + \Lambda_n$ and $B$ depends only on the edges in some box $v_B + \Lambda_n$, for two vertices $v_A$ and $v_B$ such that $|v_A - v_B| \geq 6n$, then*

$$
|\ell_{x,\mathbb{Z}^d}[A \cap B] - \ell_{x,\mathbb{Z}^d}[A]\ell_{x,\mathbb{Z}^d}[B]| < \exp(-cn).
$$

*Proof.* Denote by $\mathtt{UC}_n^A$ the event that $\omega|_{v_A + \Lambda_{2n}\setminus\Lambda_n}$ has a unique cluster crossing from inner to outer radius and define $\mathtt{UC}_n^B$ similarly. By Lemma 4.8, we have that

$$
\ell_{x,\mathbb{Z}^d}[A \cap B] = \phi_{x,\mathbb{Z}^d}[\mathrm{UEG}_\omega[A \cap B]] = \phi_{x,\mathbb{Z}^d}[\mathbb{1}_{\mathtt{UC}_n^A}(\omega)\mathbb{1}_{\mathtt{UC}_n^B}(\omega)\mathrm{UEG}_\omega[A \cap B]] + O(\exp(-cn)). \tag{19}
$$

Denoting by $\omega_A = \omega|_{v_A + \Lambda_{2n}}$ and $\omega_B = \omega|_{v_B + \Lambda_{2n}}$, we can apply Corollary 3.9 to get that

$$
\begin{aligned}
\mathbb{1}_{\mathtt{UC}_n^A}(\omega)\mathbb{1}_{\mathtt{UC}_n^B}(\omega)\mathrm{UEG}_\omega[A \cap B] &= \mathbb{1}_{\mathtt{UC}_n^A}(\omega)\mathbb{1}_{\mathtt{UC}_n^B}(\omega)\mathrm{UEG}_\omega[A]\mathrm{UEG}_\omega[B] \\
&= \mathbb{1}_{\mathtt{UC}_n^A}(\omega_A)\mathbb{1}_{\mathtt{UC}_k^B}(\omega_B)\mathrm{UEG}_\omega[A]\mathrm{UEG}_\omega[B]. \tag{20}
\end{aligned}
$$

As $\mathbb{1}_{\mathtt{UC}_n^A}(\omega_A)\mathrm{UEG}_\omega[A]$ is a positive, measurable function of $\omega_A$ bounded from above by 1, so we shall once again attempt to bound a total variation distance. Let $(\tilde{\omega}_A, \tilde{\omega}_B)$ a coupling of two percolation configurations with the same marginals as $(\omega_A, \omega_B)$ such that $\tilde{\omega}_A$ and $\tilde{\omega}_B$ are independent. Then, by [20, Corollary 1.4], we have that there exists a coupling $P$ of the two such that $\tilde{\omega}_B = \omega_B$ almost surely and

$$
P[\omega_A \neq \tilde{\omega}_A \mid \omega_B] < \exp(-cn).
$$

Equivalently,

$$
d_{\mathrm{TV}}((\omega_A, \omega_B), (\tilde{\omega}_A, \tilde{\omega}_B)) < \exp(-cn),
$$

and accordingly,

$$\phi_{x,\mathbb{Z}^d}[\mathbb{1}_{\mathtt{UC}_n^A}(\omega_A)\mathbb{1}_{\mathtt{UC}_n^B}(\omega_B)\mathrm{UEG}_\omega[A]\mathrm{UEG}_\omega[B]]$$
$$=\phi_{x,\mathbb{Z}^d}[\mathbb{1}_{\mathtt{UC}_n^A}(\omega_A)\mathrm{UEG}_\omega[A]] \times \phi_{x,\mathbb{Z}^d}[\mathbb{1}_{\mathtt{UC}_n^B}(\omega_B)\mathrm{UEG}_\omega[A]\mathrm{UEG}_\omega[B]] + O(\exp(-cn)). \qquad (21)$$

Now, to finish, we note that

$$\ell_{x,\mathbb{Z}^d}[A]\ell_{x,\mathbb{Z}^d}[B] = \phi_{x,\mathbb{Z}^d}[\mathrm{UEG}_\omega[A]] \times \phi_{x,\mathbb{Z}^d}[\mathrm{UEG}_\omega[B]]$$
$$= \phi_{x,\mathbb{Z}^d}[\mathbb{1}_{\omega \in \mathtt{UC}_n^A(\omega_A)}\mathrm{UEG}_\omega[A]] \times \phi_{x,\mathbb{Z}^d}[\mathbb{1}_{\omega \in \mathtt{UC}_n^B(\omega_B)}\mathrm{UEG}_\omega[B]] + O(\exp(-cn)). \qquad (22)$$

Combining (19), (20), (21) and (22) yields the desired. $\qquad \square$

## 4.4 Existence of many wrap-arounds

Theorem 1.3 enables us to apply our observations from Section 4.1 by first arguing directly on the torus and then saying that the model on $\mathbb{Z}^d$ does not look too different. First, we prepare for proving the existence of sufficiently many wrap-arounds.

**Lemma 4.12.** *There exists a continuous function $f : (0,1)^2 \to (0,1)$ with the following property:*

*Let $G = (V, E)$ be a finite graph and $\xi$ a boundary condition. For $p_1 < p_2$, there exists an increasing coupling $P$ between $\omega_1 \sim \phi_{p_1,G}^\xi$ and $\omega_2 \sim \phi_{p_2,G}^\xi$ such that for any random finite set of edges $F \subseteq \omega_2$ measurable with respect to $\omega_2$, we have*

$$P[\omega_1(e) = 0 \ \forall e \in F(\omega_2) \mid \omega_2] \geq P[f(p_1, p_2)^{|F(\omega_2)|} \mid \omega_2].$$

*Proof.* Pick an ordering $(e_j)_{1 \leq j \leq |E|}$ of the edges and let $U_j$ be an i.i.d. family of uniforms on $[0,1]$. For $i \in \{1, 2\}$, define the target $\mathtt{t}_{e_j,i} : \{0,1\}^{\{e_1,\dots,e_{j-1}\}} \to (0,1)$ as the conditional probability under $\phi_{p_i}$ that the edge $e_j$ is open given the state of the previous edges, i.e.

$$\mathtt{t}_{e_j,i}(\mathfrak{w}) = \phi_{p_i,G}^\xi[\omega_{e_j} \mid \omega_{e_l} = \mathfrak{w}_{e_l} \ \forall l \leq j-1].$$

Then, recursively setting

$$\omega_i(e_j) = \mathbb{1}_{U_j \leq \mathtt{t}_{e_j,i}(\omega_i|_{\{e_1,\dots,e_{j-1}\}})}$$

yields an increasing coupling between the two random graphs $\omega_i \sim \phi_{p_i,G}^\xi$.

Let us first remark that if we can prove that

$$\mathtt{t}_{e_j,2}(\mathfrak{w}) - \mathtt{t}_{e_j,1}(\mathfrak{w}') \geq \min\left\{p_2 - p_1, \frac{p_2}{2-p_2} - \frac{p_1}{2-p_1}\right\} \qquad (23)$$

deterministically for any $\mathfrak{w} \succeq \mathfrak{w}'$, then we are done.

To see this, note that the event that $\omega_1(e) = 0$ for every $e \in F$ conditional on $F$ is the event that $\mathbf{t}_{e,1} < U_e$ conditional on $U_e \leq \mathbf{t}_{e,2}$ for all $e \in F$, conditional on $F$. Since $\mathbf{t}_{e,2} \leq p_2$ (cf. Proposition 2.1$iii$)), we can set

$$f(p_1, p_2) := \frac{1}{p_2} \min \left\{ p_2 - p_1, \frac{p_2}{2 - p_2} - \frac{p_1}{2 - p_1} \right\},$$

and the rest is merely computation.

Accordingly, let us establish (23). Since, $\mathbf{t}$ is increasing in $\mathfrak{w}$, we can assume without loss of generality that $\mathfrak{w} = \mathfrak{w}'$. Now, if $A_e$ denotes the event that the end-points of $e$ are connected in $\omega \setminus \{e\}$, we have for $i \in \{1, 2\}$ that

$$\mathbf{t}_{e_j,i}(\mathfrak{w}) = p_i \phi_{p_i,G}^\xi[A_{e_j} \mid \omega_{e_l} = \mathfrak{w}_{e_l} \; \forall l \leq j - 1] + \frac{p_i}{2 - p_i} \left( 1 - \phi_{p_i,G}^\xi[A_{e_j} \mid \omega_{e_l} = \mathfrak{w}_{e_l} \; \forall l \leq j - 1] \right),$$

whence, since $p_2 > p_1$ and $A_e$ is increasing,

$$
\begin{aligned}
\mathbf{t}_{e_j,2}(\mathfrak{w}) - \mathbf{t}_{e_j,1}(\mathfrak{w}) &= (p_2 - p_1) \, \phi_{p_1,G}^\xi[A_{e_j} \mid \omega_{e_l} = \mathfrak{w}_{e_l} \; \forall l \leq j - 1] \\
&\quad + \left( \frac{p_2}{2 - p_2} - \frac{p_1}{2 - p_1} \right) \left( 1 - \phi_{p_2,G}^\xi[A_{e_j} \mid \omega_{e_l} = \mathfrak{w}_{e_l} \; \forall l \leq j - 1] \right) \\
&\quad + \left( p_2 - \frac{p_1}{2 - p_1} \right) \left( \phi_{p_2,G}^\xi[A_{e_j} \mid \omega_{e_l} = \mathfrak{w}_{e_l} \; \forall l \leq j - 1] - \phi_{p_1,G}^\xi[A_{e_j} \mid \omega_{e_l} = \mathfrak{w}_{e_l} \; \forall l \leq j - 1] \right) \\
&\geq \min \left\{ p_2 - p_1, \frac{p_2}{2 - p_2} - \frac{p_1}{2 - p_1} \right\}.
\end{aligned}
$$

$\square$

This enables us to reprove the classical result [3, Lemma 4.2] in Bernoulli percolation for the random-cluster model, allowing for the control of the number of crossings in a box.

For an event $A$ and $r > 0$, let $I_r(A)$ denote the set of $\omega$ such that the Hamming distance from $\omega$ to $\Omega \setminus A$ is at least $r$, i.e. changing the state of any $r - 1$ edges of $\omega$ cannot produce a configuration outside of $A$. For $A$ the event that $\Lambda_n$ contains a crossing between two opposite faces, we remark that $I_r(A)$ is exactly the event that $\omega$ contains at least $r$ edge-disjoint crossings.

**Lemma 4.13.** *There exists a continuous function $f : (0,1)^2 \to (0,1)$ such that for any increasing event $A$, finite graph $G$, boundary condition $\xi$, $r \in \mathbb{N}$ and $p_1 < p_2$, then*

$$f(p_1, p_2)^r \left( 1 - \phi_{p_2,G}^\xi[I_r(A)] \right) \leq 1 - \phi_{p_1,G}^\xi[A].$$

*Proof.* Let $\omega_2 \sim \phi_{p_2,G}^\xi$ and note that, on the event $\omega_2 \notin I_r(A)$, there exists a (possibly empty) set of edges $F$ such that $|F| \leq r$, every edge in $F$ is open and $\omega_2 \setminus F \notin A$. This set is not necessarily unique, but we may simply posit some rule for resolving ambiguities. Under any such choice, we see that $F$ becomes measurable with respect to $\omega_2$.

Hence, letting $P$ denote the coupling from Lemma 4.12, we see that

$$1 - \phi_{p_1,G}^\xi[A] \geq P[\omega_2 \notin I_r(A), \omega_1 \notin A]$$
$$\geq P[\omega_1(e) = 0 \ \forall e \in F \,|\, \omega_2 \notin I_r(A)] \left(1 - \phi_{p_2,G}^\xi[I_r(A)]\right)$$
$$\geq f(p_1, p_2)^r \left(1 - \phi_{p_2,G}^\xi[I_r(A)]\right),$$

which is what we wanted. $\qquad\square$

**Lemma 4.14.** *For a percolation configuration $\omega$ on $\mathbb{T}_n^d$, let $\boldsymbol{N}$ denote the maximal number of edge disjoint wrap-arounds in $\omega$. Then, for every $p > p_c$, there exist $\alpha, c > 0$ such that it holds that*

$$\phi_{p,\mathbb{T}_n^d}[\boldsymbol{N} \leq \alpha n^{d-1}] \leq \exp\left(-cn^{d-1}\right)$$

*for every $n$.*

*Proof.* The proof essentially follows in two steps, which we outline heuristically here: Pick $\delta > 0$ small enough and show that under $\phi_{p-\delta,\mathbb{T}_n^d}$, there is a crossing winding around the torus once with high probability. Then, we will use the previous lemma to argue that, under $\phi_{p,\mathbb{T}_n^d}$, there is a large number of such crossings with high probability.

Let $\delta < p - p_c$ and, for $j \in \{0, 1\}$, observe the two boxes $B_j = (\Lambda_{n-1} + jne_1)/2n\mathbb{Z}^d$ and let $\mathfrak{C}^j$ denote the event that the relevant translate of the event $\mathfrak{C}_{n-1,L,\theta,\theta/4^d}$ from Proposition 4.7 occurs in $B_j$. Furthermore, let $\tilde{\mathfrak{C}}^j$ denote the event that the relevant translate of $\mathfrak{C}_{n-1/2,L,\theta,\theta/4^d}$ occurs in $\tilde{B}_j := \Lambda_{n-1/2} + (-1)^j \frac{n}{2} e_1$. Note that $\tilde{B}_j \subseteq B_0 \cap B_1$.

By a union bound and Proposition 4.7, we have

$$\phi_{p-\delta,\mathbb{T}_n^d}\left[\mathfrak{C}^0 \cap \mathfrak{C}^1 \cap \tilde{\mathfrak{C}}^0 \cap \tilde{\mathfrak{C}}^1\right] \geq 1 - 2\exp\left(-cn^{d-1}\right) - 2\exp\left(-c(n/2)^{d-1}\right) \geq 1 - \exp\left(-c'n^{d-1}\right) \tag{24}$$

for a slightly smaller constant $c' > 0$. Let $\mathtt{SL}$ be the event that there is a simple loop wrapping around the torus once in the sense of Definition 4.1. Let us show that

$$\mathfrak{C}^0 \cap \mathfrak{C}^1 \cap \tilde{\mathfrak{C}}^0 \cap \tilde{\mathfrak{C}}^1 \subset \mathtt{SL}. \tag{25}$$

On the event $\mathfrak{C}^0 \cap \mathfrak{C}^1$, for each $j \in \{0, 1\}$, there is a unique large cluster $\mathcal{C}^j$ contained in each $B_j$, and it contains a path from the left side of $B_j$ to its right side. Furthermore, they are the unique clusters of size at least $\frac{\theta}{4^d}(n-1)^d$ in $B_0$ and $B_1$ respectively. Finally, on $\tilde{\mathfrak{C}}^j$, there is a cluster $\tilde{\mathcal{C}}^j$ in $\tilde{B}_j$ of size $\frac{4^d-1}{4^d}\theta\left(\frac{n-1}{2}\right)^d > \frac{\theta}{4^d}(n-1)^d$. All in all, on $\mathfrak{C}^0 \cap \mathfrak{C}^1 \cap \tilde{\mathfrak{C}}^0 \cap \tilde{\mathfrak{C}}^1$, then $\tilde{\mathcal{C}}^j \subseteq \mathcal{C}^0 \cap \mathcal{C}^1$. That is, all the large clusters intersect.

Let us now construct a simple loop of open edges wrapping around the torus once. In $\mathcal{C}^j$, there is a path $\gamma^j$ from the left to the right face of $B^j$, which is connected by a path to both $\tilde{\mathcal{C}}^0$ and $\tilde{\mathcal{C}}^1$. By using the latter paths, we may glue $\gamma^0$ and $\gamma^1$ together, which yields a cluster containing a simple loop wrapping around the torus once and (25) follows. See Figure 10.

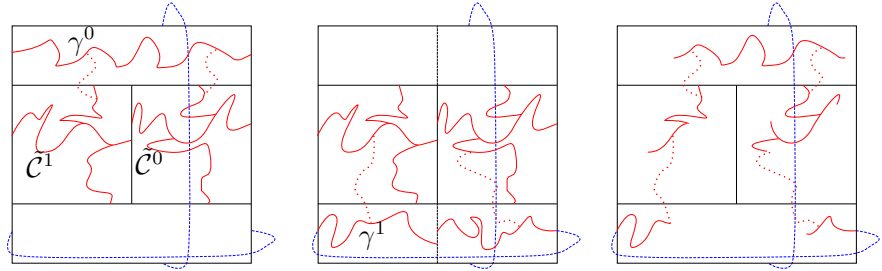

Figure 10: The construction of a simple loop wrapping around the torus once. On the left, on $\mathcal{C}^0 \cap \tilde{\mathcal{C}}^0 \cap \tilde{\mathcal{C}}^1$, the path $\gamma^0$ is connected to both of the clusters $\tilde{\mathcal{C}}^0$ and $\tilde{\mathcal{C}}^1$ via the dotted paths. In the middle panel, on $\mathcal{C}^1 \cap \tilde{\mathcal{C}}^0 \cap \tilde{\mathcal{C}}^1$, the same is true of the path $\gamma^1$. From this, a wrap-around may be extracted as displayed in the last picture.

Since a loop cannot vanish by adding additional edges, $\mathtt{SL}$ is increasing. Thus, by Lemma 4.13, we have a continuous function $f : (0,1)^2 \to (0,1)$ such that, for every $\alpha > 0$,

$$\exp\left(-c'n^{d-1}\right) \geq 1 - \phi_{p-\delta, \mathbb{T}_n^d}[\mathtt{SL}] \geq f(p-\delta, p)^{\alpha n^{d-1}} \left(1 - \phi_{p, \mathbb{T}_n^d}[I_{\alpha n^{d-1}}(\mathtt{SL})]\right),$$

where we also used (24) and (25). Accordingly,

$$\phi_{p, \mathbb{T}_n^d}[I_{\alpha n^{d-1}}(\mathtt{SL})] \geq 1 - \exp\left(\left(\alpha \log(1/f(p-\delta, p)) - c'\right) n^{d-1}\right),$$

yielding that

$$\phi_{p, \mathbb{T}_n^d}[I_{\alpha n^{d-1}}(\mathtt{SL})] \geq 1 - \exp\left(c''n^{d-1}\right)$$

for an adjusted value $c'' > 0$ and suitably small $\alpha$.

All that is left to do is to notice that $I_{\alpha n^{d-1}}(\mathtt{SL})$ is the event that there exist at least $\alpha n^{d-1}$ edge-disjoint simple loops wrapping around the torus exactly once, which is a sub-event of $(\boldsymbol{N} \geq \alpha n^{d-1})$. $\qquad\square$

## 4.5 Infinite expected cluster sizes from torus wrap-arounds

Finally, we are in position to prove our main result following the outline given in Section 4.1: We exploit Lemma 4.3 and Lemma 4.14 to lower bound the number of large clusters in the loop O(1) model on the torus. Then, we use Lemma 4.6 and Lemma 4.8 to compare the loop O(1) model on the torus to the one on $\mathbb{Z}^d$ with arbitrary boundary conditions.

First of all, we employ a modification of the multi-valued mapping principle to make more clever use of Lemma 4.3. The multi-valued mapping principle is a very general piece of combinatorial technology which allows for a rough sort of counting.

In its essence, the principle generalises the idea that if there exists a $k$-to-1 map $f$ from a set $A$ to a set $B$, then $|A| = k|B|$. A way of envisioning this is as a bipartite graph with the elements of $A$ and $B$ as vertices and an edge between $a$ and $b$ if $f(a) = b$. In general, if you have a bipartite graph with bipartition $(A, B)$ such that the degree of any vertex in $A$ is at least $n$ and the degree of any vertex in $B$ is at most $k$, then $n|A| \leq k|B|$. The following argument essentially generalises this fact to the setting where $A$ might also have internal edges. Recall that $\mathcal{C}_{\mathtt{NT}}$ is the union of the clusters containing a wrap-around.

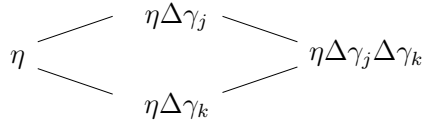

Figure 11: A small subgraph of the auxiliary graph $\mathfrak{G}_{\gamma,\eta_0}$. An initial even subgraph is exposed to the symmetric difference with all possible combinations of our initial wrap-arounds. It turns out that most of such combinations yield a high number of vertices on non-trivial clusters.

**Lemma 4.15.** *For any $c > 0$ and any $\varepsilon \in (0, \frac{c}{2d})$, there exists $\delta > 0$ such that the following holds:*

*For any $n$, and any subgraph $G$ of $\mathbb{T}_n$ with $cn^{d-1}$ edge-disjoint wrap-arounds $(\gamma_j)_{1 \leq j \leq cn^{d-1}}$, we have*

$$\mathrm{UEG}_G[|\mathcal{C}_{\mathtt{NT}}| \geq \varepsilon n^{d-1}] \geq \delta.$$

*Proof.* Let $G$ be a subgraph of $\mathbb{T}_n$ with $cn^{d-1}$ edge-disjoint wrap-arounds $(\gamma_j)_{1 \leq j \leq cn^{d-1}}$. For a subset $A \subset \{1, 2, \ldots, cn^{d-1}\}$, let $\gamma_A := \Delta_{a \in A} \gamma_A$. As such, $\eta_0 \in \Omega_{\emptyset}(\mathbb{T}_n)$, $\eta_0 \Delta \gamma_A$ is the symmetric difference of $\eta_0$ with $\gamma_j$ for each $j \in A$. Consider the following auxiliary graph $\mathfrak{G}_{\gamma,\eta_0}$ isomorphic to the Cayley graph with $(\gamma_j)$ as generators: The vertices of $\mathfrak{G}_{\gamma,\eta_0}$ are $\eta_0 \Delta \gamma_A$, where $A$ ranges over the power-set of $\{1, .., cn^{d-1}\}$ and $\eta_1$ is adjacent to $\eta_2$ if $\eta_1 \Delta \eta_2$ is equal to $\gamma_j$ for some $j$. See Figure 11.

Note that if $\eta_0 \sim \mathrm{UEG}_G$, $\eta_0 \overset{d}{=} \eta_0 \Delta \gamma_A$ for every $A$. Accordingly if $\tilde{\eta}$ is a uniform vertex of $\mathfrak{G}_{\gamma,\eta_0}(\eta_0)$, $\eta_0 \overset{d}{=} \tilde{\eta}$. The upshot of this is that it suffices to argue that, deterministically, a high proportion of the vertices in $\mathfrak{G}_{\gamma,\eta_0}$ have the property that $|\mathcal{C}_{\mathtt{NT}}| \geq \varepsilon n^{d-1}$.

Let $\mathcal{S}_\varepsilon$ denote the set of vertices in $\mathfrak{G}_{\gamma,\eta_0}$ such that $|E(\mathcal{C}_{\mathtt{NT}})| < d\varepsilon n^{d-1}$. Note that any $\eta \in \mathcal{S}_\varepsilon$ has $E(\mathcal{C}_{\mathtt{NT}}(\eta))$ intersect $\gamma_j$ for at most $d\varepsilon n^{d-1}$ different $j$. For any $j$ such that $\gamma_j$ does not intersect $E(\mathcal{C}_{\mathtt{NT}}(\eta))$, we can apply Lemma 4.3 to the subgraph $(V(\mathbb{T}_n), \eta \setminus E(\mathcal{C}_{\mathtt{NT}}(\eta)))$ to get that there is at least one wrap-around in $\eta \Delta \gamma_j$ which uses no edges from $\mathcal{C}_{\mathtt{NT}}(\eta)$ and no wrap-around in $\eta$ which is not in $\eta \Delta \gamma_j$. All in all, we get that

$$|E(\mathcal{C}_{\mathtt{NT}}(\eta \Delta \gamma_j))| \geq |E(\mathcal{C}_{\mathtt{NT}}(\eta))| + n.$$

Accordingly, $\eta$ has at least $(c - d\varepsilon)n^{d-1}$ neighbours $\eta'$ such that $|E(\mathcal{C}_{\mathtt{NT}}(\eta'))| - |E(\mathcal{C}_{\mathtt{NT}}(\eta))| \geq n$. Let $\alpha$ denote the number of such edges, that is,

$$\alpha = |\{\eta \in \mathcal{S}_\varepsilon, \, j \leq cn^{d-1} | \, |E(\mathcal{C}_{\mathtt{NT}}(\eta \Delta \gamma_j))| \geq |E(\mathcal{C}_{\mathtt{NT}}(\eta))| + n\}|$$

Define $\alpha_{\mathrm{ext}}$ to be the number of edges with one end-point in $\mathcal{S}_\varepsilon$ and the other in $\mathfrak{G}_{\gamma,\eta_0} \setminus \mathcal{S}_\varepsilon$ and let $\alpha_{\mathrm{int}}$ denote the number of such edges with both end-points in $\mathcal{S}_\varepsilon$. Since each $\eta \in \mathfrak{G}_{\gamma,\eta_0} \setminus \mathcal{S}_\varepsilon$ has at most $cn^{d-1}$ neighbours in $\mathcal{S}_\varepsilon$, we have that

$$\alpha_{\text{ext}} \leq cn^{d-1} |\mathfrak{G}_{\gamma,\eta_0} \setminus \mathcal{S}_\varepsilon|.$$

On the other hand, any $\eta \in \mathcal{S}_\varepsilon$ has at most $d\varepsilon n^{d-1}$ neighbours with smaller $E(\mathcal{C}_{\text{NT}})$. Accordingly,

$$\alpha_{\text{int}} \leq d\varepsilon n^{d-1} |\mathcal{S}_\varepsilon|.$$

Adding this up, we get that

$$(c - d\varepsilon)n^{d-1} |\mathcal{S}_\varepsilon| \leq \alpha = \alpha_{\text{ext}} + \alpha_{\text{int}} \leq cn^{d-1} |\mathfrak{G}_{\gamma,\eta_0} \setminus \mathcal{S}_\varepsilon| + d\varepsilon n^{d-1} |\mathcal{S}_\varepsilon|,$$

implying that

$$\frac{|\mathfrak{G}_{\gamma,\eta_0} \setminus \mathcal{S}_\varepsilon|}{|\mathcal{S}_\varepsilon|} \geq \frac{c - 2d\varepsilon}{c},$$

or, equivalently, that

$$\frac{|\mathfrak{G}_{\gamma,\eta_0} \setminus \mathcal{S}_\varepsilon|}{|\mathfrak{G}_{\gamma,\eta_0}|} \geq 1 - \frac{1}{2}\left(\frac{c}{c - d\varepsilon}\right) := \delta.$$

Now, for $\eta \in \mathfrak{G}_{\gamma,\eta_0} \setminus \mathcal{S}_\varepsilon$, we see that

$$2d\varepsilon n^{d-1} \leq 2|E(\mathcal{C}_{\text{NT}}(\eta)| = \sum_{v \in V(\mathcal{C}_{\text{NT}}(\eta))} \deg(v) \leq 2d|V(\mathcal{C}_{\text{NT}}(\eta))|,$$

implying that $|V(\mathcal{C}_{\text{NT}}(\eta))| \geq \varepsilon n^{d-1}$. To finish, observe that since $\tilde{\eta}$ was a uniform vertex of $\mathfrak{G}_{\gamma,\eta_0}$,

$$\text{UEG}_G[|\mathcal{C}_{\text{NT}}| \geq \varepsilon n^{d-1}] = P[\tilde{\eta} \in \mathfrak{G}_{\gamma,\eta_0} \setminus \mathcal{S}_\varepsilon] \geq \delta.$$

$\square$

We can now conclude the polynomial lower bound for the loop O(1) model on the torus.

**Theorem 1.4.** *Let $x > x_c$. Then, there exists $c > 0$ such that $\ell_{x,\mathbb{T}_n^d}[0 \leftrightarrow \partial\Lambda_n] \geq \frac{c}{n}$ for all $n$.*

*Proof.* By translation invariance,

$$\ell_{x,\mathbb{T}_n^d}[0 \leftrightarrow \partial\Lambda_n] \geq \ell_{x,\mathbb{T}_n^d}[0 \in \mathcal{C}_{\text{NT}}] = \ell_{x,\mathbb{T}_n^d}\left[\frac{|\mathcal{C}_{\text{NT}}|}{|\mathbb{T}_n^d|}\right].$$

Let $\boldsymbol{N}$ denote the maximal number of disjoint wrap-arounds on the torus. For any given $c > 0$, choose $\varepsilon \in (0, c/2d)$ and let $\delta$ be chosen as in as in Lemma 4.15. Then, by Theorem 2.5,

$$
\begin{aligned}
\ell_{x,\mathbb{T}_n^d}\left[\frac{|\mathcal{C}_{\text{NT}}|}{|\mathbb{T}_n^d|}\right] &= \phi_{x,\mathbb{T}_n^d}\left[\text{UEG}_\omega\left[\frac{|\mathcal{C}_{\text{NT}}|}{|\mathbb{T}_n^d|}\right]\right] \\
&\geq \phi_{x,\mathbb{T}_n^d}\left[\mathbb{1}_{\boldsymbol{N} \geq cn^{d-1}}(\omega)\text{UEG}_\omega\left[\frac{|\mathcal{C}_{\text{NT}}|}{|\mathbb{T}_n^d|}\right]\right] \\
&\geq \phi_{x,\mathbb{T}_n^d}\left[\mathbb{1}_{\boldsymbol{N} \geq cn^{d-1}}(\omega)\frac{\varepsilon n^{d-1}}{cn^d}\delta\right] \\
&= \phi_{x,\mathbb{T}_n^d}[\boldsymbol{N} \geq cn^{d-1}] \cdot \frac{\varepsilon\delta}{cn},
\end{aligned}
$$

where, in the second inequality, we used Lemma 4.15.

By Lemma 4.14, we have that $\phi_{x,\mathbb{T}_n^d}[\boldsymbol{N} \geq cn^{d-1}] \geq 1 - \exp\left(-cn^{d-1}\right)$ for some value of $c > 0$, which finishes the proof. $\qquad\square$

Combining the insensitivity to external topology we proved for the loop O(1) model in Theorem 1.3, this lower bound readily transfers from the torus to Euclidean space and we obtain Theorem 1.5.

The diameter of a cluster $\mathcal{C}$ in $\mathbb{Z}^d$ is given by $\mathrm{diam}(\mathcal{C}) = \sup_{x,y \in \mathcal{C}} |x - y|$. Recall that $\mathcal{C}_0$ is the cluster of 0.

**Corollary 4.16.** *For $x > x_c$ then*

$$\ell_{x,\mathbb{Z}^d}[\mathrm{diam}(\mathcal{C}_0)] = \infty.$$

*Proof.* We have that

$$\ell_{x,\mathbb{Z}^d}[\mathrm{diam}(\mathcal{C}_0)] \geq \sum_{k=1}^{\infty} \ell_{x,\mathbb{Z}^d}[\mathrm{diam}(\mathcal{C}_0) \geq k] = \sum_{k=1}^{\infty} \ell_{x,\mathbb{Z}^d}[0 \leftrightarrow \partial\Lambda_k],$$

and the right-hand side diverges by Theorem 1.5. $\qquad\square$

# 5 The loop O(1) model on the hexagonal lattice $\mathbb{H}$ and other bi-periodic planar graphs

In this section, we first prove a no-go theorem for percolation of the loop O(1) model on a class of planar graphs. Then, we focus on the hexagonal lattice and discuss how to adapt the arguments of Section 4 apply to it, despite the fact (as we shall see) that the model does not percolate. This confirms that the arguments of Section 4 alone are not strong enough to prove percolation, e.g. on $\mathbb{Z}^d$.

## 5.1 Characterisation of percolation in the planar case.

Using the more complete theory of planar percolation models, we can answer the question of whether $\beta_c^{\mathrm{clust}}(\ell) = \beta_c^{\mathrm{perc}}(\ell)$ for planar graphs with suitable symmetries with the aid of the arguments of [31] - up to generalising some well-established results in the literature. Unfortunately, the answer varies with the graph.

First off, recall the notion of planar duality. To an embedded planar graph $\mathbb{G} = (\mathbb{V}, \mathbb{E})$, we associate its dual graph $\mathbb{G}^* = (\mathbb{V}^*, \mathbb{E}^*)$, where $\mathbb{V}^*$ is the set of faces of $\mathbb{G}$ and every edge $e \in \mathbb{E}$ is associated to a dual edge $e^* \in \mathbb{E}^*$ between the two faces adjacent to $e$. One can check that $\mathbb{G}^*$ can be embedded into the plane by identifying a given face with a prescribed point in its interior.

For spin models on $\{-1, +1\}^{\mathbb{V}}$ as well as percolation measures on $\{0, 1\}^{\mathbb{E}}$, we get dual models on $\{-1, +1\}^{\mathbb{V}}$ and $\{0, 1\}^{\mathbb{E}^*}$ respectively by flipping spins and considering *-connections instead ordinary connections in the former case and by setting $\omega^*(e^*) = 1 - \omega(e)$ in the latter.

In this picture, the dual model of $\phi_{p,\mathbb{G}}$ is $\phi_{p^*,\mathbb{G}^*}$, where $p$ and $p^*$ satisfy the duality relation (cf. [19, Proposition 2.17])

$$\frac{pp^*}{(1-p)(1-p^*)} = 2. \tag{26}$$

The spirit of the relationship between the loop O(1) model and the planar Ising model goes back to Kramers and Wannier [45]. Below, $x^*$ is the $x$-parameter obtained from Table 1 by plugging in $p^*$:

**Proposition 5.1.** *Suppose that $\mathbb{G}$ is a planar graph. Then, for every $\beta$, $\ell_{x^*,\mathbb{G}}$ is the law of the interfaces of an Ising model with $+$ boundary conditions on $\mathbb{G}^*$ at inverse temperature $\beta$.*

More precisely, in this picture, the planar loop O(1) model on $\mathbb{G}$ can be coupled with the Ising model on $\mathbb{G}^*$ as the pair $(\eta, \sigma)$, where, for a given dual edge $e^* = (f, g)$, we set $\eta(e) = 1$ if and only if $\sigma_f \sigma_g = -1$. This extends to $\beta = 0$, where $\eta \sim \mathrm{UEG}_{\mathbb{G}}$ and $\sigma$ assigns $+$ and $-$ spins independently with probability $\frac{1}{2}$.

Now, we turn our attention to the class of bi-periodic embedded planar graphs, for which we have the following non-coexistence result:

**Theorem 5.2** (Non-co-existence). *For any bi-periodic planar graph $\mathbb{G} = (\mathbb{V}, \mathbb{E})$, there exists no translation-invariant measure $\mu$ on either $\{-1, +1\}^{\mathbb{V}}$ or $\{0, 1\}^{\mathbb{E}}$ such that*

*a) $\mu$ has FKG.*

*b) $\mu$ has a unique infinite component and an infinite dual component almost surely.*

This theorem was originally proven in [55, Theorem 9.3.1], and we refer the reader to [24, Theorem 1.5] for a short, independent proof. Since any such $\mathbb{G}$ must be amenable, the Burton-Keane argument [13] applies to all models below to show that if there is an infinite (primal or dual) component, it must be unique.

First, let us note that there is a robust subclass of bi-periodic planar graphs for which percolation of the loop O(1) model is impossible.

**Proposition 5.3.** *The loop O(1) model on any trivalent bi-periodic graph $\mathbb{G}$ does not percolate, i.e. for every vertex $v \in \mathbb{V}$ and all $x \in [0, 1]$, then*

$$\ell_{x,\mathbb{G}}[v \leftrightarrow \infty] = 0.$$

*Proof.* Percolation of the loop O(1) model implies the existence of an infinite cluster of open edges, which on a trivalent graph $\mathbb{G}$, is an infinite simple path. In the dual picture, this would imply the co-existence of infinite components of $+$ and $-$ for the Ising model on $\mathbb{G}^*$. Now, $\mathbb{G}^*$ is a bi-periodic planar graph, so by Theorem 5.2 there can be no co-existence of such infinite components in the Ising model (at finite or infinite temperature). $\square$

On a bi-periodic planar graph $\mathbb{G}$ with vertices of higher degree, we would expect exponential decay of the size of $+$ clusters on $\mathbb{G}^*$, similarly to [37]. Heuristically, the subcritical Ising model on $\mathbb{G}^*$ should behave roughly like Bernoulli site percolation at parameter $\frac{1}{2}$ due to the exponential decay of correlations. In general, $*$-connections are easier than ordinary connections, so by essential enhancement (see [7, 9]), the critical parameter for site percolation, henceforth $p_c^{\mathrm{site}}$, on $\mathbb{G}$ should be strictly greater than $\frac{1}{2}$. [10]

Accordingly, in this range of parameters, the loop O(1) model on $\mathbb{G}^*$ should percolate by the same arguments as those given in [31]. On the other hand, for a supercritical Ising model on $\mathbb{G}^*$, we can use the Edwards-Sokal coupling to get that there is an infinite component of either $+$'s or $-$'s which dominates the infinite component of the random-cluster model. Since $\mathbb{G}^*$ is bi-periodic, this excludes the co-existence necessary for the loop O(1) model to percolate.

We believe that following the program outlined above one could obtain a proof of the following conjecture:

**Conjecture 5.4.** *Suppose that $\mathbb{G}$ is a bi-periodic planar graph such that $p_c^{\mathrm{site}}(\mathbb{G}^*) > \frac{1}{2}$. Then,*

$$\beta_c^{\mathrm{exp}}(\ell_{\mathbb{G}}) = \beta_c^{\mathrm{per}}(\ell_{\mathbb{G}}) = \beta_c(\phi_{\mathbb{G}}).$$

## 5.2 Phase transition on the hexagonal lattice $\mathbb{H}$

In order to illustrate the robustness of our arguments on the torus, we are going to apply them to the hexagonal lattice, where we know percolation of the loop O(1) model is impossible due to Proposition 5.3.

For normalisation purposes, we embed the triangular lattice as the lattice generated by the edges $1, e^{i\frac{\pi}{3}}$ and $e^{i\frac{2\pi}{3}}$ and consider the hexagonal lattice as its dual. Since the triangular lattice is invariant under translation by $1$ and $e^{i\frac{\pi}{3}}$, then so is the hexagonal lattice. Hence, consider the linear map $T$ which fixes $1$ and maps $i$ to $e^{i\frac{\pi}{3}}$ and the tilted box $\Lambda_k^{\mathbb{H}} = T\Lambda_k$. On this box, we can observe the quotient where $v \sim w$ if and only if $v - w \in 2k\left(\mathbb{Z} + e^{i\frac{\pi}{3}}\mathbb{Z}\right)$, which corresponds to a graph embedded on the torus on which the random-cluster model is automorphism-invariant. Since the hexagonal lattice also has a reflection symmetry in the line $\{\mathrm{Im}\, z = 0\}$, we see that the toric graph thus defined is vertex-transitive and thus, $\ell_{x,\Lambda_k^{\mathbb{H}}/\sim}[0 \in \mathcal{C}_{\mathtt{NT}}] = \ell_{x,\Lambda_k^{\mathbb{H}}/\sim}[v \in \mathcal{C}_{\mathtt{NT}}]$ for any $v \in \Lambda_k^{\mathbb{H}}$.

As such, all of our arguments from above carry through so long as we can justify a version of Lemma 4.8 and Lemma 4.14 for the hexagonal lattice. This, however, is downstream from sharpness of the random-cluster phase transition (see [24]) on the triangular lattice, as we shall now sketch. All the arguments are rather standard and slightly orthogonal to the themes of this paper. As such, we shall assume rough familiarity with them and only provide cursory details. We direct the interested reader to [19].

Indeed, to get a version of Lemma 4.8, note that, for $p > p_c$, the probability of having a dual crossing from $\Lambda_n^{\mathbb{H}}$ to $\partial\Lambda_{2n}^{\mathbb{H}}$ on $\mathbb{H}$ is exponentially unlikely. However, the existence of two disjoint clusters crossing the tilted annulus would imply the existence of such a dual crossing.

---

[10] This is not necessarily true e.g. if $\mathbb{G}^*$ is obtained by periodically attaching finite graphs to the vertices of a triangulation, but $p_c^{\mathrm{site}}(\mathbb{G}^*) > \frac{1}{2}$ should be the generic case.

Similarly, the probability of having a dual top-to-bottom crossing in $\Lambda_k^{\mathbb{H}}$ is exponentially unlikely, implying that the probability of having a left-to-right crossing in $\Lambda_k^{\mathbb{H}}$ is exponentially close to 1, which allows us to apply Lemmas 4.12 and 4.13 to get a version of Lemma 4.14.

Once these are off the ground, the rest of the arguments of our paper carry through without issue, yielding $\beta_c^{\mathrm{clust}}(\ell_{\mathbb{H}}) = \beta_c^{\mathrm{exp}}(\ell_{\mathbb{H}}) = \beta_c(\phi)$ for the loop O(1) model on $\mathbb{H}$. Combined with Proposition 5.3, we obtain the full phase diagram that we summarise below. We have also plotted the situation in Figure 1.

**Proposition 5.5.** *For the loop* O*(1) model on* $\mathbb{H}$,

$$\beta_c^{\mathrm{clust}}(\ell_{\mathbb{H}}) = \beta_c^{\mathrm{exp}}(\ell_{\mathbb{H}}) = \beta_c(\phi_{\mathbb{H}}),$$

*while*

$$\beta_c^{\mathrm{perc}}(\ell_{\mathbb{H}}) = \infty.$$

Thus, for the loop O(1) model, we know exactly what happens on the hexagonal lattice $\mathbb{H}$. However, for the random current measure, the story is quite different. Due to stochastic domination from below by Bernoulli percolation (see Theorem 2.5), there is a percolative phase transition at some $\beta_c^{\mathrm{perc}}(\mathbf{P}_{\mathbb{H}}) < \infty$.

Indeed, all connected even subgraphs of a trivalent graph are simple cycles, which constrains $\ell_{x,\mathbb{H}}^0$ rather heavily, but does not affect $\mathbf{P}_{x,\mathbb{H}}$ a priori. As such, it is natural to suspect that while the behaviour of the loop O(1) model is sensitive to the graph, the behaviour of the random current measure is generic.

**Conjecture 5.6.** *All phase transitions of the random current and random-cluster model on the hexagonal lattice coincide:*

$$\beta_c^{\mathrm{perc}}(\mathbf{P}_{\mathbb{H}}) = \beta_c(\phi_{\mathbb{H}}).$$

As a final aside, we add a note on odd percolation on the hexagonal lattice. One might recall from Section 3.3 that the uniform odd subgraph can be obtained by fixing a deterministic dimerisation and then taking the symmetric difference with a uniform even subgraph, but on an odd graph, one checks that the uniform odd graph also arises as the complement of a uniform even graph.

**Proposition 5.7.** *For any trivalent, bi-periodic graph* $\mathbb{G}$*, the uniform odd subgraph of* $\mathbb{G}$ *does not percolate. In particular, the uniform odd subgraph of* $\mathbb{H}$ *does not percolate.*

*Proof.* Since the uniform odd graph of $\mathbb{G}$ is the complement of a uniform even one, we have a coupling $(\eta, \sigma)$ of a uniform odd subgraph $\eta$ of $\mathbb{G}$ and $\mathbf{I}_{0,\mathbb{G}^*}$ by setting $\eta(e) = 1$ for $e^* = (f, g)$ if and only if $\sigma_f \sigma_g = 1$. Since the monochromatic clusters are surrounded by simple loops in the complement of $\eta$ and $\mathbb{G}$ is trivalent, we get that $\eta$ percolates if and only if the monochromatic clusters of $\sigma$ percolate. However, by spin-symmetry, this would imply the co-existence of an infinite cluster of +'s and an infinite cluster of −'s. Since this does not happen, we conclude that $\eta$ does not percolate. □

# 6 Perspective

We conclude by discussing some more general properties percolation of the UEG and some open problems.

## 6.1 General characterisations of percolation of the UEG

As with any model without positive association, getting a grip of the general behaviour of the UEG is a priori a daunting task, but one might hope for some semi-robust arguments that allow one to handle large classes of graphs, as for the planar case or the case where the graph $G$ in question has a subgraph $H$ where $\Omega_\emptyset$ separates edges and $p_c(\mathbb{P}_{p,H}) < \frac{1}{2}$.

We saw in Section 5.1 that trivalence prevents the UEG on bi-periodic planar graphs from percolating. However, as we now show, it is not true that trivalence is an obstruction to percolation in general. Consider the trivalent supergraph $\mathbb{J}$ of $\mathbb{N}$, adding the edges $(1, 10)$, $(1, 100)$ and $(n, 10^{n+1})$ for all $n$ that are not powers of 10. One may note that $\mathbb{J}$ is non-amenable and non-planar.

**Proposition 6.1.** $\mathrm{UEG}_{\mathbb{J}}$ *percolates. In particular, there exists an infinite trivalent graph for which the uniform even subgraph percolates.*

*Proof.* We follow the construction of the uniform even subgraph from [8] and construct a basis for $\Omega_\emptyset(\mathbb{J})$ from a spanning tree. Pick the spanning tree in $\mathbb{J}$ which is simply the original graph $\mathbb{N}$ and for every other edge $e$ (henceforth called 'external'), let $C_e$ be the simple loop in $\mathbb{J}$ given by joining $e$ with the unique path in $\mathbb{N}$ connecting the end-points of $e$. One may check that the $C_e$ form a locally finite basis of $\Omega_\emptyset(\mathbb{J})$. One may note, in particular, that every external edge belongs to a unique such loop.

Thus, similar to Equation (11), for every external edge $e$, we can let $\epsilon_e$ be an i.i.d. family of Bernoulli-$\frac{1}{2}$ variables and sample the UEG of $\mathbb{J}$ as

$$\sum_e \epsilon_e C_e.$$

Due to the trivalence of $\mathbb{J}$, this gives the cluster of 1 the following random walk type representation on $\mathbb{N}$: First, we set $x_0 := 1$ and reveal the states of $\epsilon_{(1,10)}$ and $\epsilon_{(1,100)}$. If both are 0, then the cluster of 1 is trivial and the process ends. Otherwise, we set $x_1$ equal to the largest number $j$ such that $\epsilon_{(1,j)} = 1$ and set $e_1 = (1, j)$.

Now, recursively, given that the process arrived at $x_j$ through the edge $e_j$, since the cluster of 1 is a simple loop in $\mathbb{J}$, exactly one of the other edges adjacent to $x_j$ must be open. We proceed by cases:

    *i)* We have already revealed the state of one of the neighbouring edges. In this case, we know the unique open edge among the two and set $e_{j+1}$ equal to this edge and $x_{j+1}$ equal to the other end-point of $e_{j+1}$. Either $x_{j+1} = 1$, and the process terminates, or it is not, in which case the process continues.

ii) $e_j$ is an external edge and we have not revealed the state of $(x_j - 1, x_j)$. In this case, we check the sum of $\epsilon_e$ for every external edge $e = (n, m)$ such that $n < x_j < m$ (these are exactly the external edges such that $(x_j - 1, x_j) \in C_e$). If the sum is even, then $(x_j - 1, x_j)$ is closed, so $(x_j, x_j + 1)$ must be open and vice versa. Since we have not yet revealed the state of $(x_j - 1, x_j)$, the conditional parity of the sum is a Bernoulli random variable. In conclusion, with probability $\frac{1}{2}$, $(x_j - 1, x_j)$ is open, and we set $e_{j+1} = (x_j - 1, x_j)$ and $x_{j+1} = x_{j-1}$. Otherwise, we set $e_{j+1} = (x_j, x_j + 1)$ and $x_{j+1} = x_j + 1$. The upshot is that the process goes left with probability $\frac{1}{2}$ and right otherwise.

iii) $e_j$ is not an external edge and the state of the unique external edge $e^{x_j}$ going through $x_j$ has not yet been revealed. In this case, if $\epsilon_{e^{x_j}} = 1$, we set $e_{j+1} = e^{x_j}$ and $x_{j+1}$ equal to the other end-point of $e^{x_j}$. Otherwise, the edge $e_{j+1} := (x_j, x_j + (x_j - x_{j-1}))$ is open, and we set $x_{j+1} = x_j + (x_j - x_{j-1})$. The upshot is that the process takes the external edge it just arrived at with probability $\frac{1}{2}$, and otherwise it continues along $\mathbb{N}$ in the same direction as it has been travelling thus far.

The upshot of the upshots is that the walk arrives somewhere via an external edge, turns left or right with probability $\frac{1}{2}$ and keeps walking for a geometric number of steps until it encounters an open external edge. The probability that 1 is connected to $\infty$ is then at least the probability that $x_j$ never has the form $10^n$ at times $j$ where $e_j$ is not an external edge (note that 1, 10 and 100 do have this form). This is achieved if $B_n$ happens for every $n$, where $B_n$ denotes the event that there is an open external edge in $[10^n + 1, 10^{n+1} - 1]$.

Since the marginal of the open external edges is Bernoulli $\frac{1}{2}$ percolation, we get that

$$\mathrm{UEG}_{\mathbb{J}}[B_n] \geq 1 - 2^{10^{n+1} - 10^n - 2} \geq 1 - 2^{8 \cdot 10^n}$$

A union bound now yields that

$$\mathrm{UEG}_{\mathbb{J}}[1 \leftrightarrow \infty] \geq \mathrm{UEG}_{\mathbb{J}}[\cap_n A_n] \geq 1 - \sum_{n=0}^{\infty} 2^{-8 \cdot 10^n} > 0.$$

<div align="right">□</div>

Secondly, one might wonder about the links between the behaviour of Bernoulli percolation and that of the UEG. For instance, one might have a suspicion that if a graph $G$ is easily disconnected in the sense that the percolation threshold $p_c(\mathbb{P}_{p,G})$ is very close to 1, this lack of connectivity might also impact the UEG. This, too, turns out to be false. In order to prove this, we start with a result stating that the infinite cluster of the slightly supercritical random-cluster model is easily disconnected. Recall, how we defined the dual parameter in (26).

**Proposition 6.2.** *For every $\delta \in (0, 1)$, there exists $r > 0$ such that $\phi_{p,\mathbb{Z}^2} \cup \mathbb{P}_{\delta,\mathbb{Z}^2}$ percolates for every $p > p_c(\phi_{p,\mathbb{Z}^2}) - r$. In particular, if $p < (p_c - r)^*$ and $\omega \sim \phi_{p,\mathbb{Z}^2}$, then $q_c(\mathbb{P}_q(\omega)) > 1 - \delta$ almost surely.*

*Proof.* For fixed $p, \delta \in (0, 1)$ and a subgraph $H$ of $\mathbb{Z}^2$, note that $\nu_{p,\delta,H}^{\xi} := \phi_{p,H}^{\xi} \cup \mathbb{P}_{\delta,H}$ is a monotonic percolation measure on $H$ and hence, so is its dual measure. Therefore, we can apply the general sharpness arguments from [24] to see that, for fixed $p$, there exists a $\delta_c(p) \in [0, 1]$ such that for $\delta < \delta_c$, $\nu_{p,\delta,H}^1$ has exponential decay and for $\delta > \delta_c$, $(\nu_{p,\delta,H}^0)^*$ has exponential decay.

Since the phase transition of $\phi_{\mathbb{Z}^2}$ on $\mathbb{Z}^2$ is continuous [58, 25], $\phi_{p_c,\mathbb{Z}^2}^0$ does not have exponential decay, and we get that $\delta_c(p_c(\phi_{p,\mathbb{Z}^2})) = 0$. In particular, for any $\delta$ and any $c \in (0, 1)$, there exists an $n$ such that

$$(\nu_{p_c,\delta,\Lambda_{2n}}^0)^*[A(n) \text{ is good}] > 1 - c,$$

where $A(n) = \Lambda_{2n} \setminus \Lambda_n$ and $A(n)$ is said to be good if it does not contain a dual open crossing from $\Lambda_n$ to $\partial\Lambda_{2n}$. Just like in the Theorem 2.9, this inequality is sufficient to extract an infinite cluster for $c$ sufficiently small.

However, for fixed $\delta > 0$, $\nu_{p,\delta,\Lambda_{2n}}^0[A(n) \text{ is good}]$ is a continuous function of $p$ and hence, there exists some $r > 0$ such that $\nu_{p,\delta,\Lambda_{2n}}^0[A(n) \text{ is good}] > 1 - c$ remains true for $p > p_c - r$. In particular, $\nu_{p,\delta,\mathbb{Z}^2}$ percolates for all such $p$ and its dual has exponential decay. This gives the first conclusion.

To get the second conclusion, we move to the dual picture and see that $\nu_{p,\delta,\mathbb{Z}^2}^*$ is the distribution of $\omega^* \setminus \zeta$, where $\zeta \sim \mathbb{P}_{\delta,(\mathbb{Z}^2)^*}$ and independent of $\omega$. This, however, is exactly Bernoulli-$(1 - \delta)$ percolation on $\omega^*$. Since $\nu_{p,\delta,\mathbb{Z}^2}^*$ has exponential decay, we conclude that $\mathbb{P}_{1-\delta,\omega}$ does not percolate almost surely, which is what we wanted. $\square$

We note that the proposition also holds for the random-cluster model for $1 \le q \le 4$, but we do not know whether it holds in $\mathbb{Z}^d$ for $d \ge 3$. Finally, from the extended version of the Theorem from [31] we obtain the following Corollary.

**Corollary 6.3.** *For every $\varepsilon > 0$, there exists an infinite graph $G$ with $q_c(\mathbb{P}_{q,G}) > 1 - \varepsilon$ such that the uniform even subgraph of $G$ percolates.*

*Proof.* For given $\varepsilon$, we may apply Proposition 6.2 to pick $\delta$ such that $\mathbb{P}_{1-\varepsilon,\omega}$ does not percolate almost surely for $\omega \sim \phi_{p_c+\delta,\mathbb{Z}^2}$. However, by [31, Theorem 1.3], $\ell_{p_c+\delta,\mathbb{Z}^2}$, which is the UEG of $\omega$, percolates almost surely.

In particular, there must exist at least one realisation $G$ of $\omega$ which has $q_c(\mathbb{P}_{q,G}) > 1 - \varepsilon$ and the property that UEG$_G$ percolates almost surely. $\square$

## 6.2 The situation for $p = p_c$

Our proof of infinite expectation of cluster sizes of $\ell_{p,\mathbb{Z}^d}$ required that $p > p_c$. For $p < p_c$ there is exponential decay by stochastic domination by $\phi_p$. Here, we briefly discuss the situation for $p = p_c$.

For $d \ge 3$ and $p = p_c$, we do have a polynomial lower bound for connection probabilities, since by [19, Theorem 4.8], there exists $c, C > 0$ such that

$$\frac{c}{|v|^{d-1}} \le \langle \sigma_0 \sigma_v \rangle_{\beta_c, \mathbb{Z}^d} \le \frac{C}{|v|^{d-2}},$$

where we note that a more general, and for $d \geq 5$ tighter, bound is proven in [6]. Therefore, $\phi_{p_c, \mathbb{Z}^d}[|\mathcal{C}_0|] = \infty$. On the other hand, it does not percolate by continuity of the Ising phase transition [25]. Heuristically, a model with infinite expected cluster sizes should percolate as soon as any independent density of edges is added to it. Therefore, with the coupling from Theorem 2.5 in mind, we do not expect that $\ell_{x_c, \mathbb{Z}^d}[|\mathcal{C}_0|] = \infty$.

However, for any $d \geq 2$, we still believe that connection probabilities of $\ell_{x_c, \mathbb{T}_n^d}$ satisfy polynomial lower bounds in the volume of the torus. It remains, however, a difficulty to transfer the result to $\mathbb{Z}^d$, since we cannot use Pizstora's construction to exhibit separating surfaces in $\phi_{p_c, \mathbb{Z}^d}$. For $d = 2$, polynomial bounds on the existence can be achieved with ordinary RSW theory, which suffices for establishing a polynomial lower bound on cluster sizes in $\ell_{x_c, \mathbb{Z}^2}$. In other dimensions, though, no similar tool exists in the literature to our knowledge. We summarise our expectations in the following conjecture:

**Conjecture 6.4.** Let $d \geq 3$. Then we expect that $\ell_{x_c, \mathbb{Z}^d}[|\mathcal{C}_0|] < \infty$ and that there exists some $a, b > 0$ such that
$$\frac{a}{|v|^b} < \ell_{x_c, \mathbb{Z}^d}[0 \leftrightarrow v].$$

## 6.3 Remaining questions for the phase diagram of $\ell$ and P on $\mathbb{Z}^d$

In Theorem 1.5 and Proposition 5.5 we managed to prove a condition akin to criticality all the way down to the random-cluster phase transition for the loop O(1) model on both $\mathbb{Z}^d$ and $\mathbb{H}$. However, on $\mathbb{Z}^2$, the transition from exponential decay to percolation happens at one point, whereas it never happens for $\mathbb{H}$. This motivates the following question that asks whether a proper intermediate regime can exist:

**Question 6.5.** Does there exist a lattice $\mathbb{L}$ such that the loop O(1) model $\ell_{x, \mathbb{L}}$ for $x \in [0, 1]$ has a non-trivial intermediate regime, i.e. an interval $(a, b) \subset [0, 1]$ and points $0 < x_0 < a$ and $b < x_1 < 1$ such that $\ell_{x_0, \mathbb{L}}$ has exponential decay, $\ell_{x_1, \mathbb{L}}$ percolates, and $\ell_{x, \mathbb{L}}$ neither has exponential decay nor percolates for $x \in (a, b)$?

Even with our main Theorem in mind, our motivating problem of interest [17, Question 1] is still left open for $\mathbb{Z}^d$ with $d \geq 3$. Since the random current model stochastically dominates the loop O(1) model, a positive answer would follow from sharpness for $\ell_x$. One reason to suspect this on $\mathbb{Z}^d$ goes as follows:

The estimate in Theorem 1.4 is particularly crude. We essentially only use that one may increase the size of the non-trivial clusters by taking the symmetric difference with a wrap-around which intersects only trivial clusters. However, due to the existence of vertices of degree at least 4, there are plenty of scenarios where acting by a wrap-around increases the size of the non-trivial cluster even when the two intersect. As such, we conjecture the following, which was noted in [31] (cf. Theorem 2.9):

**Conjecture 6.6.** For $\mathbb{Z}^d$, $d \geq 3$ it holds that $\beta_c^{\mathrm{perc}}(\ell) = \beta_c^{\mathrm{exp}}(\ell)$.

Of course, one may also settle for the weaker statement:

**Conjecture 6.7.** *For $d \geq 3$, the single random current on $\mathbb{Z}^d$ has a unique sharp percolative phase transition, i.e. $\beta_c^{\mathrm{perc}}(\mathbf{P}) = \beta_c^{\mathrm{exp}}(\mathbf{P})$.*

## 6.4 Infinite clusters of the loop $\mathrm{O}(1)$ model on the cut open lattice

The technical reason that we cannot extend our result from infinite expectation of cluster sizes to percolation is that we have difficulties in controlling the variations of where wrap-arounds occur. If $\eta$ is trivial and $\gamma$ is a wrap-around, then $\eta \Delta \gamma$ contains at least one wrap-around by Lemma 4.3, but there is no reason to suspect that this wrap-around intersects $\gamma$ at all. In an artificial setup that we will now sketch, we can overcome this barrier.

Consider the graph $\mathcal{Z}^d$ obtained from $\mathbb{Z}^d$ by removing all edges from a fixed hyperplane with the exception of a single edge, $e$. Call the graph resulting from this procedure $\mathcal{Z}^d$. We also consider a similar cut-up $\mathcal{H}$ version of the hexagonal lattice $\mathbb{H}$. Just as in the non-cut-up case, we may consider a quotient of $\Lambda_n \cap \mathcal{Z}^d$ as a subgraph of the torus and carry out our arguments from before (and similarly for $\Lambda_n^{\mathbb{H}} \cap \mathcal{H}$).

However, by construction, we know that any wrap-around in the cut-open torus must use the edge $e$. Therefore, as the size of the torus grows to infinity, we find that the edge $e$ is part of an infinite cluster with constant probability, and conclude that the corresponding loop $\mathrm{O}(1)$ model percolates.

**Theorem 6.8.** *For $p > p_c(\phi_{\mathbb{Z}^d})$, there exists an infinite cluster of $\ell_{p, \mathcal{Z}^d}$ with positive probability.*

It follows from the stronger result in [11] that the critical parameter for the random-cluster model on the half space is the same as for the full space, so it holds by monotonicity that $p_c(\phi_{\mathbb{Z}^d}) = p_c(\phi_{\mathcal{Z}^d})$ (as well as $p_c(\phi_{\mathbb{H}}) = p_c(\phi_{\mathcal{H}})$).

**Theorem 6.9.** *Even though $\mathcal{H} \subset \mathbb{H}$, there exists a $p > 0$ such that*

$$\ell_{p, \mathcal{H}}[0 \leftrightarrow \infty] > 0 = \ell_{p, \mathbb{H}}[0 \leftrightarrow \infty].$$

One reason that we show the argument here is that we speculate that finer control of the variations of the wrap-arounds arising from the combinatorial argument could help shed light on the remaining questions.

# Acknowledgements

We thank Ioan Manolescu and Peter Wildemann for helpful discussions and Aran Raoufi and Franco Severo for getting us started on the problem in the first place. BK and FRK acknowledge the Villum Foundation for funding through the QMATH center of Excellence (Grant No. 10059) and the Villum Young Investigator (Grant No. 25452) programs. UTH acknowledges funding from Swiss SNF.

# 5. Quantum Walks in Random Magnetic Fields

# Quantum walks in random magnetic fields

Frederik Ravn Klausen, Christopher Cedzich, Albert H. Werner

### Abstract

A model for quantum walks in magnetic fields where the plaquette phases are i.i.d. random is introduced. In a regime analogous to the strong disorder Anderson localziation is expected. We prove an a priori initial scale estimate as well as exponential decay of fractional moments of the Greens function. The proofs generalize the approach to the unitary Anderson model given in [16, 17], thereby overcoming several additional complications stemming from the internal degree of freedom and the Aharonov-Bohm effect.

## 1 Introduction

Quantum walks are mathematical models that describe the dynamics of single and few-particle quantum systems on lattice structures. They constitute a model for quantum simulation in discrete time for single and few-particle quantum systems with an internal degree of freedom. The transport properties have been extensively studied, showcasing ballistic transport for coherent evolution, a transition to classical, i.e. diffusive behaviour in case of decoherence as well as Anderson localization for disordered on-site potentials [21].

Recently, a discrete version of the minimal coupling principle for Hamiltonian systems has been established, allowing for the study of quantum walks in the presence of external gauge fields such as electro-magnetic fields [10]. For electric fields, the spectrum, as well as the propagation behaviour, turns out to sensitively depend on the degree of irrationality of the electric field [11] with generic fields leading to Anderson localization [12]. Similarly, results on the spectral properties of the underlying unitary operators have been obtained in the case of magnetic fields [8] and also for a unitary analogue of the almost Mathieu operator [9].

In this paper, we focus on the case quantum walks subject to randomly fluctuating magnetic fields for two-dimensional quantum walks. In the Hamiltonian case, this model system has been studied exhaustively in the literature before, both in the ideal case of a homogeneous magnetic field [20, 29] as well as in fields that randomly fluctuate [26, 28, 14] albeit without an additional internal degree of freedom. This internal degree adds complexity to the analysis. When we consider the *A*-fields then the internal degree of freedom results in equal and opposite phases acquired by a particle traversing an edge either forwards or backwards. Introducing a rank 2 perturbation instead of rank 1 perturbations studied in previous works [16] and [21]. In fact, showing that parts of the strategy from [16] and [21] adapt to this more complicated setting is one of the main contributions of this paper.

It shall be noted that because of the fluctuations in the field the usual approach cannot be applied anymore. This approach consists of choosing a clever gauge, which trivializes the gauge potential in one of the lattice dimensions. Then, this lattice dimension is admissible for being Fourier transformed to the torus, yielding the so-called "almost-Mathieu" operator. This is an innocently looking system on the one-dimensional lattice which has intricate properties and thus was extensively studied as an emblematic quasi-periodic Schrödinger operator in the mathematical literature.

In the following, we first introduce the model of a quantum walk in a random magnetic field that we suspect localizes. For the model, we try to follow the route of the fractional moment approach to localization pioneered by Aizenman and Molchanov [4] for the (self-adjoint) Anderson model. This was later extended to the unitary Anderson model in [16] and the strategy was used in [21].

Many proofs of (spectral or dynamical) localization entail first proving an a priori estimate and then doing an iteration to prove exponential decay. We prove the initial scale estimate in Theorem 2.4, by generalizing the arguments from [16] (and the thesis of Hamza [17]). Then we continue to the proving exponential decay of fractional moments of the Greens function. Again our strategy is to generalize the strategy from [16] and [21] where we artificially put-in reflecting boundary conditions in a box around a distinguished point. An interesting complication is that the phases are no longer independent (since only the magnetic fluxes are independent) even when the walk is decoupled. However, this dependence is only through the Aharonov-Bohm effect [1], which we will study in detail to overcome the complications.

We face difficulties in transferring the exponential decay of fractional moments to dynamical localization. These difficulties also stem from the internal degree of freedom, which give pairs of phases that are equal and opposite. That is that the (conditional) distrubtion does not have density with respect to the Lebesgue measure.

Finally, in Section 8 we discuss ideas about how to generalize the arguments that spectral and dynamical localization follows from exponential decay of fractional moments of the Greens function.

## Acknowledgements

C.C. was supported in part by the Deutsche Forschungsgemeinschaft (DFG, German Research Foundation) under the grant number 441423094. F.R.K. and A.H.W. acknowledge support from the Villum Foundation through the QMATH center of Excellence(Grant No. 10059) and the Villum Young Investigator (Grant No. 25452) programs. We thank Ulrik Thinggaard Hansen for sharing his ideas on resampling of conditionally independent variables.

# 2 Model and results

## 2.1 Setting

Let us begin by describing the system. The Hilbert space we work on is $\mathcal{H} = \ell^2(\mathbb{Z}^2) \otimes \mathbb{C}^2$ with canonical basis

$$\delta_x^s = \delta_x \otimes e_s, \qquad x = (x_1, x_2) \in \mathbb{Z}^2, \, s \in \{+, -\},$$

where $\{\delta_x : x \in \mathbb{Z}^2\}$ denotes the standard basis of $\ell^2(\mathbb{Z}^2)$ and $e_+ = [1, 0]^\top$ and $e_- = [0, 1]^\top$ span the internal degree of freedom $\mathbb{C}^2$. Sometimes will we write $|x\rangle$ instead of $\delta_x$ and also

$$\delta_x^\pm = |x, \pm\rangle \tag{2.1}$$

to simplify the notation of scalar products. Following this notation we sometimes speak of elements in the lattice $\mathbb{Z}^2 \times \{+, -\}$ as $x^\pm$.

On $\mathcal{H}$ we study the random unitary operator $W_\omega = W_\omega(C_1, C_2)$ defined by

$$W_\omega := D_\omega W_0. \tag{2.2}$$

Here, $W_0 = W_0(C_1, C_2)$ is the deterministic translation-invariant quantum walk

$$W_0 = S_1 C_1 S_2 C_2, \tag{2.3}$$

which is fully specified by the **coin operators** $C_1$ and $C_2$ which locally rotate the internal degree of freedom. We assume them to be independent of position for which we can write

$$C_i = \mathbb{1}_{\ell^2(\mathbb{Z}^2)} \otimes \begin{pmatrix} c_{11}^i & c_{12}^i \\ c_{21}^i & c_{22}^i \end{pmatrix}. \tag{2.4}$$

The state-dependent **shift operators** $S_\alpha$, $\alpha = 1, 2$, relate neighbouring cells. They are defined by

$$S_\alpha |x, \pm\rangle = |x \pm e_\alpha, \pm\rangle. \tag{2.5}$$

To define the random diagonal unitary $D_\omega$, we consider the probability space $(\Omega, \Sigma, \mathbb{P})$ with $\Omega = \mathbb{T}^{\mathbb{Z}^2}$, $\Sigma$ generated by the cylinder sets and $\mathbb{P} = \otimes_{x \in \mathbb{Z}^2} \mu_x$ with each $\mu_x$ a probability measure on $\mathbb{T}$ which we assume to be independent of $x$, i.e. $\mu_x \equiv \mu$. Moreover, we assume each $\mu_x$ to be absolutely continuous with respect to the Lebesgue measure with bounded density, i.e.

$$d\mu(F) = \varphi(F)dF, \qquad \varphi \in L^\infty(\mathbb{T}). \tag{2.6}$$

Below, we shall denote by $\mathbb{E}$ the expectation with respect to $\mu$. In addition to the assumption in (2.6) we also assume that $\varphi(t) \neq 0$ a.s. and that $\frac{1}{\varphi}$ is bounded, that is

$$\frac{1}{\varphi} \in L^\infty(\mathbb{T}). \tag{2.7}$$

On $(\Omega, \Sigma, \mathbb{P})$ we consider independent and identically distributed random variables $F(x) : \Omega \to \mathbb{T}$ defined by

$$F_\omega(x) = \omega_x$$

and we introduce two random functions $\theta^+, \theta^- : \mathbb{Z}^2 \to \mathbb{T}$ by setting

$$\theta^+(x_1, x_2) = -\sum_{k=0}^{x_2-1} F(x_1, k), \qquad \theta^-(x_1, x_2) = \sum_{k=0}^{x_2-1} F(x_1 + 1, k). \tag{2.8}$$

Note that this implies

$$F(x_1, x_2) = \theta^+(x_1 + 1, x_2) - \theta^+(x_1, x_2) = -\theta^-(x_1, x_2) + \theta^-(x_1 - 1, x_2), \qquad (2.9)$$

and, importantly,

$$\theta^+(x_1, x_2) = -\theta^-(x_1 - 1, x_2). \qquad (2.10)$$

Then, the diagonal operator $D_\omega$ on $\mathcal{H}$ is given by

$$D_\omega \delta_x^s = e^{-i\theta_\omega^s(x)} \delta_x^s. \qquad (2.11)$$

We shall call $\theta$ **magnetic phases** and $F$ a **magnetic field**. Consequentially, we shall call $W_\omega$ a (random) magnetic walk. The motivation for this nomenclature is detailed in Section 3 below. In short, if we would take the $\theta$ as given, then (2.9) defines $F$ as a discrete derivative in $x_1$-direction. We shall see below, that $D_\omega$ implements a discrete magnetic field in Landau gauge.

**Remark 2.1.** *At first sight, the setting here looks similar to that in [16, 22, 23] where a translation invariant quantum walk is multiplied by a random diagonal operator just as in (2.2). However:*

1. *The phases $\theta^+$ and $\theta^-$ in the definition of $D_\omega$ are not independent of each other. Indeed, they are correlated as in (2.10).*

2. *We emphasize that the randomly chosen object is the i.i.d. random variable $F$ and not the phases $\theta^+$ and $\theta^-$ in the definition of $D_\omega$. As we shall explain below, there is a hidden "gauge freedom" in the choice of $D_\omega$.*

An important observation is that $W_\omega$ is an ergodic operator: denoting by $\tau_a, a \in \mathbb{Z}^2$ the ergodic shift $(\tau_a \omega)(x) = \omega(x + a)$ on $\Omega$, it is straightforward to see that $F_\omega(x - a) = F_{\tau_a \omega}(x)$ which implies $\theta_\omega^\pm(x - a) = \theta_{\tau_a \omega}^\pm(x)$. Since $W_0$ is translation-invariant, $W_\omega$ is ergodic with respect to the lattice translations. Since the general theory of ergodic operators (see e.g. [7, Section V]) carries over from the self-adjoint to the unitary setting, this observation has profound consequences. In particular, it implies that the spectrum as well as its components are almost surely independent of the field configuration, i.e. deterministic.

Our goal is to show localization in the "strongly disordered" case, i.e. close to the setting where $W_\omega$ completely localizes because it is block-diagonal. In one-dimensional shift-coin walks, the coins with this property are the completely off-diagonal ones. More generally, shift-coin walks in arbitrary dimensions are block-diagonal for coins given by permutations without fixed points [22].

Similarly, for the model in (2.3) there are two possible coin configurations for which $W_0$ decouples into a block-diagonal form, namely either $C_1^r$ is diagonal and $C_2^r$ is off-diagonal, or vice versa. We denote this set of **reflecting coin configurations** by

$$\mathcal{C}_r = \left\{ (C_1, C_2) \in U(2) \times U(2) : c_{11}^i = c_{22}^i = 0, |c_{11}^j| = |c_{22}^j| = 1, \{i, j\} = \{1, 2\} \right\}. \qquad (2.12)$$

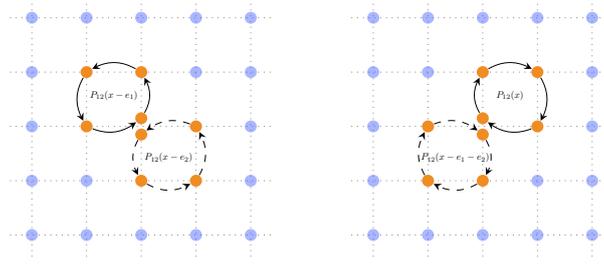

Figure 1: The closed orbits of $W_0^2$ in the reflecting case. Left: $(C_1^r, C_2^r) = (\mathbb{1}, \sigma_x)$. Right: $(C_1^r, C_2^r) = (\sigma_x, \mathbb{1})$

These are the only coin configurations that yield pure point spectrum of the unperturbed walk $W_0$ with closed orbits and in that way they correspond to the infinitely strong disorder limit in the Hamiltonian case. In addition, a coin close to the reflecting coin can be said to correspond to large disorder in the Hamiltonian case.

**Lemma 2.2.** *The spectrum of $W_0(C_1^r, C_2^r)$ is pure point if and only if $(C_1^r, C_2^r) \in \mathcal{C}_r$.*

The "only if" direction, one calculates that

$$W_0(C_1, C_2)^2 |x, e_i\rangle \in \text{span}\{|x, e_i\rangle\}, \tag{2.13}$$

i.e. $W_0(C_1, C_2)$ is block-diagonal in the basis $\{|x, e_i\rangle, W_0(C_1, C_2)|x, e_i\rangle\}_{i=\pm 1}$ with off-diagonal blocks. The "if" direction follows by an argument analogous to the proof of [22, Lemma 1].

The distance of a coin configuration $(C_1, C_2) \in U(2) \times U(2)$ to $\mathcal{C}_r$ is defined as

$$\text{dist}((C_1, C_2), \mathcal{C}_r) = \inf_{(C_1^r, C_2^r) \in \mathcal{C}_r} \left( \|C_1 - C_1^r\|^2 + \|C_2 - C_2^r\|^2 \right)^{1/2}. \tag{2.14}$$

For $(C_1^r, C_2^r) \in \mathcal{C}_r$ note that

$$W_0(C_1, C_2) - W_0(C_1^r, C_2^r) = S_1(C_1 - C_1^r)S_2 C_2 + S_1 C_1^r S_2 (C_2 - C_2^r) \tag{2.15}$$

which by the unitarity of the shift operators and the coins implies

$$\begin{aligned}
\|W_0(C_1, C_2) - W_0(C_1^r, C_2^r)\| &= \|(C_1 - C_1^r)S_2 C_2 + C_1^r S_2 (C_2 - C_2^r)\| \\
&\leq \|C_1 - C_1^r\| + \|C_2 - C_2^r\| \\
&\leq c \max_{i=1,2} \|C_i - C_i^r\|.
\end{aligned}$$

which we will need for the perturbation argument in [22] to work.

## 2.2  The main result

After having introduced the objects of interest, we can now state our main results. The main theorem that we prove here is the exponential decay of fractional moment of the Green's function

$$G(z) := (W - z)^{-1}, \qquad G(k^\pm, l^\pm, z) := \langle k^\pm, G(z)l^\pm \rangle,$$

in expectation for strong disorder, i.e. close to the reflecting case.

**Theorem 2.3.** *Let $W = W_\omega$ be the random quantum walk defined in (2.2). Then there exists $\varepsilon > 0$ such that for all $C_1, C_2$ with $\mathrm{dist}((C_1, C_2), \mathcal{C}_r) < \varepsilon$ such that for all $s \in (0, \frac{1}{3})$, there exist constants $\mu, C > 0$ such that and all $x^\pm, y^\pm \in \mathbb{Z}^2 \times \{+, -\}$ it holds that*

$$\mathbb{E}_\omega \left[ |G(x^\pm, y^\pm, z)|^s \right] \leq C e^{-\mu |x-y|}$$

*for all $z \in \mathbb{C}$ with $\frac{1}{2} < |z| < 2$.*

The first step in order to prove such a bound on the fractional moments is the following initial scale estimate. Here the expectation value $\mathbb{E}_{\{\theta_k, \theta_l\}}$, is over $\{\theta_k, \theta_l\}$ keeping all other variables fixed, a process that is difficult to define for non-independent potentials and we elaborate on it in Section 5.

**Theorem 2.4.** *For any $k^\pm, l^\pm \in \mathbb{Z}^2 \times \{-1, +1\}$, any $0 < s < 1$, and $W_\omega = D_\omega S$ defined above there exists a constant $C(s) > 0$ such that*

$$\mathbb{E}_{\{\theta_k, \theta_l\}} \left[ |G(k^\pm, l^\pm, z)|^s \right] \leq C(s).$$

This estimate we show in Section 6 before Theorem 2.3 is proved in Section 7.

Based on both numerics (a subset of which is shown on Figure 2) and that the obstacles to the proof seem more technical than fundamental we conjecture dynamical localization (see similar statement in [22]):

**Conjecture 2.5.** *Let $W_\omega(C_1, C_2)$ be the walk (2.3) in a random magnetic field. Then there exists $\varepsilon > 0$ such that for all $C_1, C_2$ with $\mathrm{dist}((C_1, C_2), \mathcal{C}_r) < \varepsilon$ there exists $c > 0$ and $\eta > 0$ such that for all $x^\pm, y^\pm \in \mathbb{Z}^2 \times \{+, -\}$*

$$\mathbb{E}_\omega \left[ \sup_{t \in \mathbb{Z}} |\langle x^\pm, W_\omega^t(C_1, C_2) y^\pm \rangle| \right] \leq c e^{-\eta |x-y|}. \tag{2.16}$$

# 3  Motivation: two-dimensional quantum walks in random magnetic fields

Let us describe the physical origin of the model given in (2.2). This is not essential for deriving the results in this paper, yet it explains why we study the particular type of correlations of the phases $\theta_\omega^+$ and $\theta_\omega^-$ described in (2.10) as well as the gauge freedom of the model.

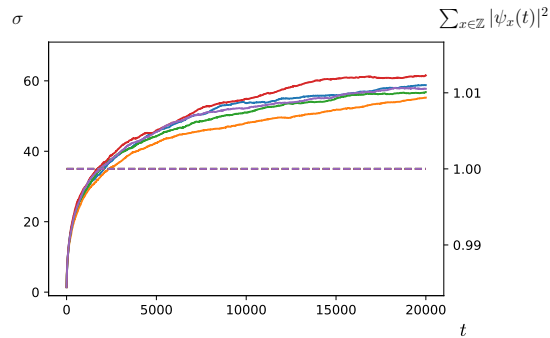

Figure 2: The root mean square $\sigma$ of a walk with $(\theta_1, \theta_2) = (10^{-1}, 1/4 - 10^{-1}) = (.1, .15)$ for 20000 timesteps on a lattice of size $2 * 2000 + 1$ in both lattice directions (5 runs with fluxes uniformly sampled from $[-\pi, \pi)$). The dashed straight line visualizes, that despite the absorbing boundary conditions practically no probability is lost due to finite size effects, i.e. that we chose the lattice large enough. Note that the choices of parameters are pretty close to the balanced setting $(\theta_1, \theta_2) = (1/8, 1/8) = (.125, .125)$, yet one clearly sees localization setting in.

Let us briefly recall the construction from [10] where a **magnetic field** on $\mathbb{Z}^d$ is realized by "magnetic" translations $T_\alpha, \alpha = 1, \ldots, d$. In the current, two-dimensional setting these are unitary operators on $\ell^2(\mathbb{Z}^2)$ that relate neighbouring local Hilbert spaces, i.e. $T_\alpha : \mathcal{H}_x \to \mathcal{H}_{x+\hat{\alpha}}$ for $\alpha = \pm 1, \pm 2$, but commute only up to a $U(1)$-valued multiplication operator, i.e.

$$T_1^* T_2^* T_1 T_2 = P_{12}. \tag{3.1}$$

Evaluating the left side locally corresponds to transporting around a "plaquette", i.e., around an elementary loop in $\mathbb{Z}^2$, see Figure 3. This corresponds to multiplication by an element of the (local) holonomy group which we here take to be (a subgroup of) $U(1)$.

Magnetic translations can always be expressed as lattice translations decorated by a phase $U_\alpha(x)$ which depends on position and direction [10, Lemma III.2], i.e.,

$$T_\alpha = t_\alpha U_\alpha(Q), \tag{3.2}$$

where $Q\delta_x = x\delta_x$ is the position operator, and $U_{-\alpha}(Q) = U_\alpha(Q - \hat{\alpha})^{-1}$ by unitarity of $T_\alpha$. Plugging this into (3.1), the plaquette phases locally evaluate to

$$P_{12}(x) = U_1(x)^{-1} U_2(x + e_1)^{-1} U_1(x + e_2) U_2(x). \tag{3.3}$$

Note that the physical quantity $P$ implementing the magnetic field does not fully fix the $U_\alpha$ [10]: there is a "gauge freedom" that consists in letting $U_\alpha(x) \mapsto V(x + \hat{\alpha}) U_\alpha(x) V^*(x)$ where the **gauge transformation** $V$ is a unitary that multiplies by a phase that only depends on position, but not on the lattice direction $\alpha$. In concrete applications it is often advantageous to fix a particular gauge to carry out concrete calculations. A convenient choice for our purposes is the so-called **Landau gauge**, in which the phase of the translations $T_2$ are trivial, i.e. $U_2(x) \equiv 1$ for all $x \in \mathbb{Z}^2$, whereas $U_1$ is determined recursively via

$$U_1(x_1, 0) = 1, \qquad U_1(x + e_2) = P(x)U_1(x), \tag{3.4}$$

and hence $U_1(x + e_2) = P(x)P(x - e_2) \cdots P(x_1, 0)$. For homogeneous fields where $P(x) \equiv P$ is independent of position, (3.4) reduces to $U_1(x) = P^{x_2}$.

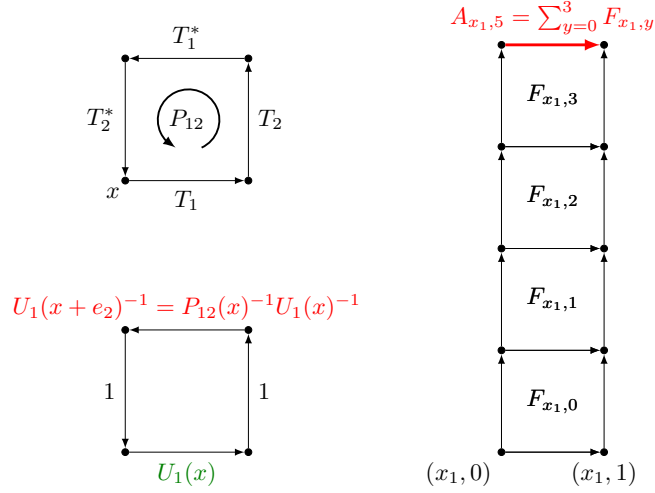

Figure 3: Left: The plaquettes in (3.1) and the "Landau" gauge of (3.4). Right: Figure indicating the flux $F(x_1, 1) + F(x_1, 2) + F(x_1, 3) + F(x_1, 4)$ accumulated when traversing the edge $((x_1, 5), (x_1 + 1, 5))$ shown in red, see (3.7).

An abstract **quantum walk** $W$ is placed in a magnetic field realized by magnetic translations $T_\alpha$ using "discrete minimal coupling" [10]. This amounts to replacing the lattice translations $t_\alpha$ in the definition of $W$ by the corresponding $T_\alpha$, which directly parallels the usual minimal coupling scheme in systems described by a Hamiltonian. Importantly, gauge transforming the $T_\alpha$ amounts to conjugating the walk by the gauge transform, i.e. $W \mapsto VWV^*$ [10, Lemma IV.2]. Thus, $W$ and the gauge transformed $VWV^*$ are isospectral, and statements about Green's functions of either imply statements about Green's functions of the other.

The operators $U_\alpha(Q)$ as well as the plaquette phase $P = P(Q)$ take values in the unit circle $\partial \mathbb{D}$. Writing $\partial \mathbb{D}$ additively, i.e. parameterizing it by phase angles in $\mathbb{T} = \mathbb{R}/(2\pi\mathbb{Z})$, relates the abstract description of discrete magnetic fields to the concrete model given in (2.2). To this end, we write $U_\alpha(Q) = \exp(iA_\alpha(Q))$ with $A_\alpha : \mathbb{Z}^2 \to \mathbb{T}$. Similarly, we write $P(x) = \exp(iF(x))$, where $F(x)$ is the "magnetic flux" through the plaquette at $x$. Then, with the discrete derivative

$(d_\alpha f)(x) = f(x + e_\alpha) - f(x)$, (3.3) takes the familiar form

$$F(x) = d_1 A_2(x) - d_2 A_1(x).\tag{3.5}$$

In this additive form, the Landau gauge defined in (3.4) translates to $A_2(x) \equiv 0$ and $A_1(x_1, x_2) = F(x_1, x_2) + A_1(x_1, x_2 - 1)$ with "initial condition" $A_1(x_1, 0) = 0$, which implies

$$A_1(x_1, x_2) = \sum_{k=1}^{x_2-1} F(x_1, k).\tag{3.6}$$

This is the phase that is picked up when traversing the edge $(x_1, x_2) \to (x_1 + 1, x_2)$, see Figure 3. It corresponds to the total flux through the area enclosed by the loop

$$(x_1, x_2) \to (x_1, 0) \to (0, 0) \to (x_1, 0) \to (x_1 + 1, x_2).\tag{3.7}$$

Comparing (3.6) with the random phases $\theta^+$ and $\theta^-$ defined in (2.8), we can now relate the random quantum walk $W_\omega$ to the walk $W_0$ placed in a random magnetic field, which ultimately explains the correlations given in (2.10):

**Lemma 3.1.** *The random quantum walk $W_\omega$ corresponds to $W_0$ when placed in a random magnetic field $F_\omega$ in Landau gauge.*

*Proof.* Let $F$ be a magnetic field such that $T_1^* T_2^* T_1 T_2 = e^{iF}$ and fix Landau gauge. Then, placing $W_0$ into the field $F$ via discrete minimal coupling amounts to replacing $t_1 \mapsto T_1$ only since in Landau gauge $T_2 \equiv t_2$. Thus,

$$S_1 \mapsto \sum_{s=\pm} T_1^s \otimes P_s = t_1 e^{iA_1(Q)} \otimes P_+ + e^{-iA_1(Q)} t_1^{-1} \otimes P_- = e^{iA_1(Q-\hat{1})} t_1 \otimes P_+ + e^{-iA_1(Q)} t_1^{-1} \otimes P_- = D S_1,$$

where

$$D = e^{iA_1(Q-\hat{1})} \otimes P_+ + e^{-iA_1(Q)} \otimes P_-.$$

Setting $A_1(x_1, x_2) = -\theta^+(x_1 + 1, x_2)$ yields the statement of the lemma. $\qquad\square$

The particular dependence structure of the phases in (2.10) can thus be understood by imagining the phases to be attached to edges between the lattice points in $\mathbb{Z}^2$ rather than to the lattice points themselves as in [22, 16, 17]. Requiring that traversing an edge in opposite directions the picked up phase should vanish yields our model.

# 4 Preliminary observations

## 4.1 The reflecting case or "maximal disorder"

For our model of interest defined in (2.3), there are two possible coin configurations for which $W_0$ decouples into a block-diagonal form, namely either $C_1$ is diagonal and $C_2$ is off-diagonal, or vice versa. We denote this set by

$$\mathcal{C}_r = \{(C_1, C_2) \in U(2) \times U(2) : c_{11}^i = c_{22}^i = 0, |c_{11}^j| = |c_{22}^j| = 1, \ i \neq j = 1, 2\}.\tag{4.1}$$

In the spirit of [22] we write

$$\mathcal{H}_x^i = \text{span}\{|x, e_i\rangle, W_0(C_1, C_2)|x, e_i\rangle\} \tag{4.2}$$

with $\mathcal{H}_x^i = \mathcal{H}_y^j$ if and only if $|y, j\rangle \in \mathcal{H}_x^i$, and write $\mathcal{H}_x = \mathcal{H}_x^1 \oplus \mathcal{H}_x^2$.

**Remark 4.1.** *1. The subspaces $\mathcal{H}_x$ are not orthogonal for different $x$. Indeed, for $C_1$ off-diagonal and $C_2$ diagonal one has $\mathcal{H}_x^1 = \mathcal{H}_{x-e_1+e_2}^2$ whereas for $C_1$ diagonal and $C_2$ off-diagonal $\mathcal{H}_x^1 = \mathcal{H}_{x-e_1-e_2}^2$, see also Figure 1. However, $\mathcal{H}_x^i \perp \mathcal{H}_y^i$ for all $x \neq y$ such that*

$$\mathcal{H} = \bigoplus_{x \in \mathbb{Z}^2} \mathcal{H}_x^1 = \bigoplus_{x \in \mathbb{Z}^2} \mathcal{H}_x^2. \tag{4.3}$$

*2. For $(C_1, C_2) \in \mathcal{C}_r$ the subspaces $\mathcal{H}_x^i$ are invariant also in the setting with magnetic field. This is obvious in Landau gauge where the field is implemented by pre-multiplication by a diagonal unitary in (2.2).*

Since for $(C_1, C_2) \in \mathcal{C}_r$ the walk $W_\omega(C_1, C_2)$ leaves each $\mathcal{H}_x^+$ invariant we obtain the following estimate on its spectrum:

**Lemma 4.2.** *Let $(C_1, C_2) \in \mathcal{C}_r$. Then for any arc $A \subset \mathbb{T}$ with $|A| < \pi$ we have*

$$\mathbb{P}[\sigma(W_\omega(C_1, C_2)\mid_{\mathcal{H}_x^+}) \cap A = \emptyset] \geq 1 - 2\|\varphi\|_\infty \, |A|,$$

*where $\varphi$ is the density of the distribution of $F$.*

*Proof.* Notice how $W_\omega^2 \mid_{\mathcal{H}_x^+} = cF(x) \, \mathbb{1} \mid_{\mathcal{H}_x^+}$ with $c = \det(C_1 C_2)$, and that $W_\omega$ is off-diagonal. Hence the eigenvalues of $W_\omega(C_1, C_2) \mid_{\mathcal{H}_x^+}$ are $\{\pm c^{1/2} F(x)^{1/2}\}$. Since the events that each of the eigenvalues are in $A$ are disjoint it follows that

$$\mathbb{P}(\sigma(W_\omega \mid_{H_x^+}) \cap A = 1 - \mathbb{P}(F(x) \in A) - \mathbb{P}(-F(x) \in A) \geq 1 - 2\|\varphi\|_\infty \, |A|.$$

□

## 4.2 Finite restrictions

Below we do a geometric decoupling argument where we use the reflecting coins. In the argument, we "decouple" the walk inside a finite box from the outside. A box $\Lambda = \Lambda_L$ has dimension $2L + 1$ in both lattice directions, and we achieve the decoupling by changing the coins of the walk $W_\omega$ on its boundary. Here and in the following we let $(C_1^r, C_2^r) \in \mathcal{C}_r$ be any pair of reflecting coins. Explicitly, let

$$C_{i,x}^L = \begin{cases} (C_1^r, C_2^r) & \|x\|_\infty = L - 1, L, L + 1 \\ C_i & \text{else,} \end{cases} \tag{4.4}$$

this means that close to the boundary of $\Lambda^L$, where $\|x\|_\infty = L$, $W_\omega(C_1^L, C_2^L)$ has closed orbits as described in Section 4.1, see Figure 4.

In, other words, we let $C_1^L = \oplus_x C_x$ where $C_x = C_x^r$ whenever $L - 1 \leq |x| \leq L + 1$ and $C_1$ otherwise and similarly for $C_2^L$, see Figure 4. Then, $T^L = W(C_1, C_2) - W(C_1^L, C_2^L)$ has finite rank with non-vanishing matrix elements only in a small annulus of radius $L$. We even show that its norm is bounded.

**Lemma 4.3.** *Suppose that* $\|(C_{i,x} - C_{i,r})\| \leq \varepsilon$ *for some* $\varepsilon > 0$. *Then the operator* $T^L = W(C_1, C_2) - W(C_1^L, C_2^L)$ *satisfies*

$$\|T^L\| \leq 2C\varepsilon.$$

*Proof.* From Schur's criterion [24, p.143] we obtain

$$\|C_i - C_i^L\|^2 = \| \bigoplus_{x \in \partial \Lambda_L} (C_{i,x} - C_{i,r})\|^2$$

$$\leq \Big( \sup_j \sum_k |\langle j|, \bigoplus_{x \in \partial \Lambda_L} (C_{i,x} - C_{i,r})|k\rangle| \Big) \Big( \sup_k \sum_j |\langle j|, \bigoplus_{x \in \partial \Lambda_L} (C_{i,x} - C_{i,r})|k\rangle| \Big).$$

Thus, the statement of the lemma follows from the bound

$$\sup_j \sum_k |\langle j|, \bigoplus_{x \in \partial \Lambda_L} (C_{2,x} - C_{2,r})|k\rangle| \leq \sup_j 2\varepsilon = 2\varepsilon.$$

Concretely,

$$\|T^L\| = \|S_1 C_1 S_2 C_2 - S_1 C_1^L S_2 C_2^L\| \leq \|S_1 C_1 S_2 (C_2 - C_2^L)\| + \|S_1 (C_1 - C_1^L) S_2 C_2^L\|$$

$$\leq C\|C_2 - C_2^L\| + C\|C_1 - C_1^L\|$$

$$\leq 2C\varepsilon.$$

$$\square$$

Now define $\mathcal{H}^{\Lambda_L} = \bigoplus_{x \in \Lambda_L} \mathcal{H}_x^+$ and similarly $\mathcal{H}^{\Lambda_L^c}$. Since the particles get stuck once they get close to the boundary, the operator $W(C_1^L, C_2^L)$ leaves the subspaces $\mathcal{H}^{\Lambda_L}$ and $\mathcal{H}^{\Lambda_L^c}$ invariant, and we can decompose

$$W(C_1, C_2) = W(C_1^L, C_2^L)^L \oplus W(C_1^L, C_2^L)^{L^c} + T^L.$$

Since this construction is independent of the magnetic field, we can decompose $W_0(C_1^L, C_2^L)$ in the same way. Later we will oftentimes denote these operators by $W^L, W^{L^c}$ and $W_0^L, W_0^{L^c}$ respectively. Similarly, we write

$$G^L = \big(W(C_1^L, C_2^L) - z\big)^{-1} \tag{4.5}$$

etc. for the corresponding Greens functions.

If all coins inside a finite box $\Lambda_L$ are totally reflecting, we have that

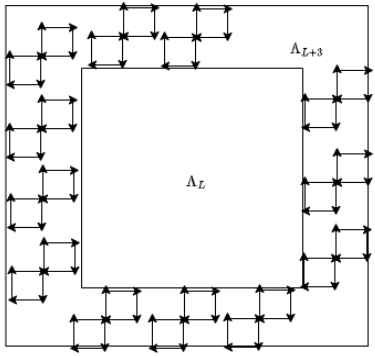

Figure 4: A sketch of the domains $\Lambda_L$ and $\Lambda_{L+3}$ and how the we interchanged the coin in $\Lambda_{L+3} \setminus \Lambda_L$ with the reflecting coin.

**Lemma 4.4.** *Let* $(C_1^r, C_2^r) \in \mathcal{C}_r$. *Then, there is a constant* $c > 0$ *such that for any* $z \notin \partial \mathbb{D}$ *and any* $\eta > 0$ *we have for all integers* $L > 2$ *that*

$$\mathbb{P}\left[\text{dist}\left(z, \sigma(W(C_1^r, C_2^r)^L\right) \leq \eta\right] \leq cL^2\eta.$$

*Proof.* The Hilbert space $\mathcal{H}^{\Lambda_L}$ consists of $(2L+1)^2$ invariant subspaces of $W(C_1^r, C_2^r)^L$, namely $\mathcal{H}_{x,y}$ for $(x,y) \in \Lambda_L$. Hence the eigenvalues of $W^{\Lambda_L}$ are given by the eigenvalues of $W$ restricted to each subspace. Whether they are in an arc $A \subset \partial \mathbb{D}$ is given entirely by the corresponding $F_{x,y}$. Since all of those are independent we have

$$\mathbb{P}[\sigma(W(C_1^r, C_2^r)^L \mid_{H^{\Lambda_L}}) \cap A = \emptyset] \geq (1 - 2\|\varphi\|_\infty |A|))^{(2L+1)^2}.$$

Then since the intersection of a ball of radius $\eta$ with the unit circle is at most $2\pi\eta$ we have for $\eta$ sufficiently small that

$$\mathbb{P}\left[\text{dist}\left(z, \sigma(W^{\Lambda_L}(C_1^r, C_2^r)\right) \leq \eta\right] \leq 1 - (1 - 2\|\varphi\|_\infty(2\pi\eta))^{(2L+1)^2} \leq 2(2L+1)^2\|\varphi\|_\infty 2\pi\eta.$$

$\square$

## 5   The correlations of the system

In the proof of localization of the unitary Anderson model [16] a similar setup to ours was used. The main difference is that there the phases were independent, while in the current setting we have the dependence structure that has its origin in the choice of Landau gauge. In the independent setup, the value of a single phase is independent of the value of all phases around it and therefore the conditional density is just the density. In our setup, we will face two additional technical challenges.

The first technical challenge for our setup is to define the conditional density of a set of phases given a fixed value of all the other phases. As it will turn out, each of the phases has a density with respect to the Lebesgue measure, and therefore conditioning on a single value is the same as conditioning on a measure zero event. That means that the usual notion of conditional probabilities needs to be extended, and the appropriate theoretical framework turns out to be that of Markov kernels, which we introduce next. The second challenge is that we have infinitely many phases. Since there is no infinite-dimensional Lebesgue measure, i.e. no (product) Lebesgue measure on $\mathbb{R}^{\mathbb{N}}$, it is not clear what measure the conditional density should be with respect to. Yet, as we argue in Section 5.2, the locality of the phases in the Landau gauge will provide us with a work-around to that problem.

## 5.1 Conditional measures and their densities

In the following, $\mathcal{J} \subset \mathbb{Z}^2$ will denote a finite set and let $(\Omega, \mathcal{A}, \mathbb{P})$ be a probability space. We need to define conditional distributions of $\Theta_{\mathcal{J}}$ given a fixed configuration of the phases outside $\mathcal{J}$ that is $\Theta_{\mathcal{J}^c}$.

**Definition 5.1** (cf. (10.1.4) in [19]). *Let $(X, \mathcal{A}_X)$ and $(Y, \mathcal{A}_Y)$ be measurable spaces and let $T : \Omega \to Y$ be a measurable function/random variable. A **Markov kernel** is a function $Q(\cdot \mid \cdot) : \mathcal{A}_X \times Y \to [0, 1]$ that satisfies*

1. *$Q(\cdot \mid t) \equiv Q(\cdot \mid T = t)$ is a probability measure on $(X, \mathcal{A}_X)$ for every $t \in Y$, and*

2. *$t \mapsto Q(A \mid t)$ is a measurable function for every $A \in \mathcal{A}_X$.*

One may have several worries before embarking on that. The first worry is that we are conditioning on an event $\{\Theta_{\mathcal{J}^c} = x_{\mathcal{J}^c}\}$ which has measure zero. In order for this conditioning to be well-defined we need the conditional distribution to be regular.

**Definition 5.2** (cf. (10.1.6) in [19]). *In the setting above let $S : \Omega \to X$ be another measurable function/random variable. We say that $P_{S|T}(\cdot \mid \cdot)$ is a **regular conditional distribution** of $S$ given $T$ if $P_{S|T}(\cdot \mid \cdot)$ is a Markov kernel and the following relation holds*

$$\mathbb{P}[S \in A, T \in B] = \int_B P_{S|T}(A \mid t) P_T(dt),$$

*where $P_T$ is the marginal measure of $T$, $A$ and $B$ are events in $\mathcal{A}_X$ and $\mathcal{A}_Y$, respectively, and $\mathbb{P}$ is the probability measure on $\Omega$.*

The second worry is concerned with the densities. If we considered the model on a finite subset $\mathcal{J}$ of the lattice $\mathbb{Z}^2$, then the marginals always have a density with respect to the corresponding finite-dimensional version of the Lebesgue measure. However, for our particular model in question, it will turn out that only finitely many of the phases outside $\mathcal{J}$ carry all the information about the phases in $\mathcal{J}$. Using that finite set $\partial \mathcal{J}^+$ we can construct a density of the corresponding conditional measure.

First, we state the following proposition that exhibits $P_{S|T}(\cdot \mid t)$ as the conditional measure.

**Proposition 5.3** (cf. (10.3.2) in [19]). *In the setup from above, suppose that $\psi : (X \times Y, \mathcal{A}_X \otimes \mathcal{A}_Y) \to \mathbb{R}$ is a random variable. Then*

$$\mathbb{E}[\psi(S,T)] = \int_X P_T(dt) \int_Y \psi(s,t) P_{S|T}(ds \mid t). \tag{5.1}$$

## 5.2 Conditional densities of phases

After having introduced a bit of abstract theory let us connect to our concrete problem with the magnetic phases stemming from the magnetic fluxes. Recall that we assume that the distribution of fluxes $F$ has common density $\varphi$, such that both $\varphi$ and $\frac{1}{\varphi}$ are bounded (cf. (2.6), (2.7)).

Recall from (2.10) that the phases $\theta^{\pm}(x_1, x_2)$ satisfy

$$\theta^+(x_1, x_2) = -\theta^-(x_1 - 1, x_2) \tag{5.2}$$

for all $x = (x_1, x_2) \in \mathbb{Z}^2$, and the distribution of $\theta_k^+$ given all phases other than $\theta_{k-e_1}^-$ is absolutely continuous with respect to the Lebesgue measure. For ease of notation, we shall write $\theta_k = \theta_k^+$ for $k \in \mathbb{Z}^2$, and from now on consider mainly the phases $\{\theta_k\}_{k \in \mathbb{Z}^2}$ (corresponding to the +-phases). For any $K \subset \mathbb{Z}^2$ we write $\Theta_K = \{\theta_k\}_{k \in K}$ and for a vector $x \in \mathbb{R}^{\mathbb{Z}^2}$ we write $x_K$ for the restriction of $x$ to $K$. Later we will show that for any finite set $K$ then $\Theta_K$ has density with respect to the Lebesgue measure and we will denote such densities by $\tau$

Let $\mathcal{J} \subset \mathbb{Z}^2$ be a finite set and $X = \mathbb{T}^{\mathcal{J}^c}$ and $Y = \mathbb{T}^{\mathcal{J}}$ be probability spaces. Let $\Theta_{\mathcal{J}^c} : \Omega \to X$ and $\Theta_{\mathcal{J}} : \Omega \to Y$ be random variables with joint distribution of the $\theta$'s. Now, the regular conditional distribution $P_{\Theta_{\mathcal{J}}|\Theta_{\mathcal{J}^c}}(\cdot \mid \cdot)$ of $\Theta_{\mathcal{J}}$ given $\Theta_{\mathcal{J}^c}$ exists in general [19, Section 10.29] and it is a Markov kernel such that

$$\mathbb{P}(\Theta_{\mathcal{J}} \in A, \Theta_{\mathcal{J}^c} \in B) = \int_B P_{\Theta_{\mathcal{J}}|\Theta_{\mathcal{J}^c}}(A \mid x_{\mathcal{J}^c}) P_{\Theta_{\mathcal{J}^c}}(dx_{\mathcal{J}^c})$$

Now, the central observation about the phases in question is that just taking the phases in the neighborhood of $\mathcal{J}$ into account already carries all outside information. We shall call this neighbourhood the "boundary" of $\mathcal{J}$ and denote it by $\partial \mathcal{J}^+$ and define it by

$$\mathcal{J}^+ = \{(x,y) \mid |y - y_0| \le 1 \text{ for a point } (x, y_0) \in \mathcal{J}\}$$

as well as

$$\partial \mathcal{J}^+ = \mathcal{J}^+ \backslash \mathcal{J}. \tag{5.3}$$

### 5.2.1 Block decomposition

Let us first restrict our attention to one horizontal strip. In this strip we identify blocks of neighbouring plaquettes whose boundary phases are labeled by $\mathcal{J}$. These blocks are all conditionally independent given $\mathcal{J}^c$. In Landau gauge $\theta_n$ and $\theta_m$ are independent whenever the first coordinate of $n$ and $m$ is different. Therefore, we can consider each vertical strip separately. Let us fix such a strip and let $\mathcal{J}_1$ label the phases $\{\theta_n\}_{n \in \mathcal{J}_1}$ in that strip. Denote by $\{\theta_n\}_{n \notin \mathcal{J}_1}$ all the phases in this vertical strip except for the phases labeled by $\mathcal{J}_1$. In other words, we are left with considering the conditional expectation values

$$\mathbb{E}[(\theta_n)_{n \in \mathcal{J}_1} \mid \{\theta_n\}_{n \notin \mathcal{J}_1}].$$

Let us now split $\mathcal{J}_1$ up into blocks. Since all the phases are located in the same vertical strip we can order them and write $\{\theta_n\}_{n \in \mathbb{Z}}$. Since $\mathcal{J}_1$ is a finite set, we can look at the blocks of consecutive phases in $\mathcal{J}$. I.e. there exists a $j \leq |\mathcal{J}_1|$ and $n_i, a_i \in \mathbb{Z}$ for $i = 1, \ldots j$ with $a_i \geq 0$ such that $n_i$ labels the $i$-th block and $a_i$ labels the $a_i - 1$-th phase in the $i$-th block. We naturally have that

$$\bigcup_{i=1}^{j} \{n_i, \ldots, n_i + a_i\} = \mathcal{J}_1$$

and the union on the left side is disjoint. Let $\Theta_i = (\theta_{n_i}, \ldots, \theta_{n_i+a_i})$ be the set of phases of one such block. Then, by conditional independence, it holds that

$$\begin{aligned}
\mathbb{E}[(\theta_n)_{n \in \mathcal{J}_1} \mid \{\theta_n\}_{n \notin \mathcal{J}_1}] &= \mathbb{E}[(\Theta_i)_{i \in [j]} \mid \{\theta_n\}_{n \notin \mathcal{J}_1}] \\
&= (\mathbb{E}[\Theta_i \mid \{\theta_n\}_{n \notin \mathcal{J}_1}])_{i \in [j]} \\
&= (\mathbb{E}[\Theta_i \mid (\theta_{n_i-1}, \theta_{n_i+a_i+1})])_{i \in [j]},
\end{aligned}$$

since each of the blocks $\Theta_i$ depend only on the edge directly before and after as we now prove.

**Lemma 5.4.** *For every finite set $\mathcal{J} \subset \mathbb{Z}^2$ there exists a finite set $\partial \mathcal{J}^+ \subset \mathbb{Z}^2$ such that*

$$\mathbb{E}[\Theta_{\mathcal{J}} \mid \Theta_{\mathcal{J}^c} = x_{\mathcal{J}^c}] = \mathbb{E}[\Theta_{\mathcal{J}} \mid \Theta_{\partial \mathcal{J}^+} = x_{\partial \mathcal{J}^+}].$$

*Proof.* There exist measurable functions $\varphi_1 : (\mathbb{T}^{\mathcal{J}^c}, \mathcal{B}(\mathbb{T}^{\mathcal{J}^c})) \to \mathbb{R}, \varphi_2 : (\mathbb{T}^2, \mathcal{B}(\mathbb{T}^2)) \to \mathbb{R}$ such that

$$\varphi_1(\Theta_{\mathcal{J}_1^c}) = \mathbb{E}[\Theta_i \mid \Theta_{\mathcal{J}_1^c}] = \mathbb{E}[\Theta_i \mid (\theta_{n_i-1}, \theta_{n_i+a_i+1})] = \varphi_2((\theta_{n_i-1}, \theta_{n_i+a_i+1}))$$

and so $\varphi_1$ only depends on $(\theta_{n_i-1}, \theta_{n_i+a_i+1})$ and so it follows that

$$\mathbb{E}[\Theta_i \mid \Theta_{\mathcal{J}_1^c} = x_{\mathcal{J}_1^c}] = \mathbb{E}[\Theta_i \mid (\theta_{n_i-1}, \theta_{n_i+a_i+1}) = (x_{n_i-1}, x_{n_i+a_i+1})].$$

$\square$

Inserting, $\phi(s) = 1_B(s)$ in [19, (10.3.5)] it follows that the conditional measures are equal. That is for every $B \in \mathcal{B}(\mathbb{T}^{\mathcal{J}})$ (the Borel $\sigma$-algebra) it holds that

$$P_{\Theta_{\mathcal{J}}|\Theta_{\mathcal{J}^c}}(B \mid x_{\mathcal{J}^c}) = P_{\Theta_{\mathcal{J}}|\Theta_{\partial\mathcal{J}^+}}(B \mid x_{\partial\mathcal{J}^+}). \tag{5.4}$$

Define $\mathcal{J}^c \backslash \partial\mathcal{J}^+ = R$. Now, for any function $\psi$ of $\Theta_{\mathcal{J}}$ and $\Theta_{\mathcal{J}^c}$ we have by Proposition 5.3

$$\mathbb{E}[\psi(\Theta_{\mathcal{J}}, \Theta_{\mathcal{J}^c})] = \int_{\mathbb{T}^{\mathcal{J}^c}} \mu_{\Theta_{\mathcal{J}^c}}(dx_{\mathcal{J}^c}) \int_{\mathbb{T}^{\mathcal{J}}} \psi(x_{\mathcal{J}}, x_{\mathcal{J}^c}) P_{\Theta_{\mathcal{J}}|\Theta_{\mathcal{J}^c}}(dx_{\mathcal{J}} \mid x_{\mathcal{J}^c})$$

$$= \int_{\mathbb{T}^{\mathcal{J}^c}} \mu_{\Theta_{\mathcal{J}^c}}(dx_{\mathcal{J}^c}) \int_{\mathbb{T}^{\mathcal{J}}} \psi(x_{\mathcal{J}}, x_{\partial\mathcal{J}^+}, x_R) P_{\Theta_{\mathcal{J}}|\Theta_{\partial\mathcal{J}^+}}(dx_{\mathcal{J}} \mid x_{\partial\mathcal{J}^+})$$

Now, since $\mathcal{J}$ and $\partial\mathcal{J}^+$ are finite (disjoint) sets the corresponding densities (with respect to the Lebesgue measure) of $\Theta_{\mathcal{J}}$ and $\Theta_{\partial\mathcal{J}^+}$ exist and so does the joint density. Therefore, we can define the conditional density, that is the density of the conditional measure $P_{\Theta_{\mathcal{J}}|\Theta_{\partial\mathcal{J}^+}}(\,\cdot\mid x_{\partial\mathcal{J}^+})$.

With this machinery, we can also put on firm ground what we mean by integrating out the variables $\theta_k$ and $\theta_l$. For $\mathcal{J} = \{k, l\}$ we denote it by

$$\mathbb{E}_{\{\theta_k, \theta_l\}}[\psi(\Theta_{\mathcal{J}}, \Theta_{\mathcal{J}^c})] = \int_{\mathbb{T}^{\mathcal{J}}} \psi(x_{\mathcal{J}}, x_{\partial\mathcal{J}^+}, x_R) P_{\Theta_{\mathcal{J}}|\Theta_{\partial\mathcal{J}^+}}(dx_{\mathcal{J}} \mid x_{\partial\mathcal{J}^+}).$$

Having cut down our conditioning from the infinite set $\mathcal{J}^c$ to the finite set $\partial\mathcal{J}^+$, the conditional density with respect to the Lebesgue measure becomes possible. Since both $\Theta_{\partial\mathcal{J}^+}$ and $\Theta_{\mathcal{J}}$ have densities $f_{\Theta_{\partial\mathcal{J}^+}}, f_{\Theta_{\mathcal{J}}}$ with respect to the Lebesgue measure on $\mathbb{T}^{\mathcal{J}}$ such that $f_{\Theta_{\partial\mathcal{J}^+}}(x_{\partial\mathcal{J}^+}) > 0$, the density of $P_{\Theta_{\mathcal{J}}|\Theta_{\partial\mathcal{J}^+}}(\,\cdot\mid x_{\partial\mathcal{J}^+})$ is (cf. [19, Section 10.6])

$$f_{\Theta_{\mathcal{J}}|\Theta_{\partial\mathcal{J}^+}}(x_{\mathcal{J}} \mid x_{\partial\mathcal{J}^+}) = \frac{f_{\Theta_{\mathcal{J}}, \Theta_{\partial\mathcal{J}^+}}(x_{\mathcal{J}} \mid x_{\partial\mathcal{J}^+})}{f_{\Theta_{\partial\mathcal{J}^+}}(x_{\partial\mathcal{J}^+})}. \tag{5.5}$$

In Proposition 5.5 below we prove that $\|f_{\Theta_{\mathcal{J}}|\Theta_{\partial\mathcal{J}^+}}(\,\cdot\mid\cdot)\|_\infty < \infty$. We will use this fact to obtain the a priori estimate in Theorem 2.4.

### 5.2.2 Boundedness of conditional density

Let $\mathcal{J} \subset \mathbb{Z}^2$ be a finite set labeling the phases we later want to resample. The conditional distribution of $(\theta_n)_{n \in \mathcal{J}}$ given all the other phases is continuous with respect to the Lebesgue measure.

One important consequence of the proof is that knowing the value of $\theta_{\partial\mathcal{J}^+}$ contains exactly the same information as all of $\theta_{\mathcal{J}^c}$.

**Proposition 5.5.** *Let $\mathcal{J} \subset \mathbb{Z}^2$ be finite. In Landau gauge, the conditional distribution*

$$\tau_{\theta_{\mathcal{J}}|\theta_{\partial\mathcal{J}^+}}(\,\cdot\mid\cdot) \tag{5.6}$$

*exists and is bounded on $[0, 2\pi]^{|\mathcal{J}|} \times [0, 2\pi]^{|\partial\mathcal{J}^+|}$.*

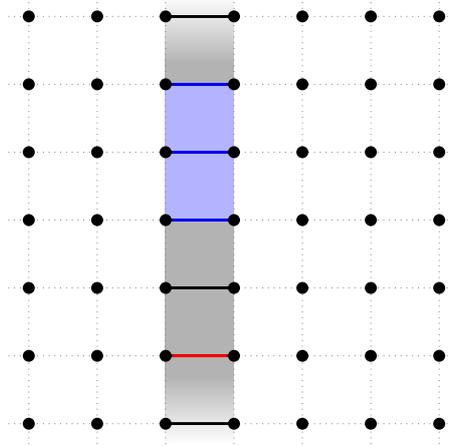

Figure 5: The construction of blocks in $\mathcal{J}$: the red edge is $\Theta_{i_1}$ and the blue edges are $\Theta_{i_2}$

*Proof.* We continue the decomposition into blocks from above. Each block (in a single vertical strip) is labeled by $i \in [j]$ above. To simplify notation we get rid of the $i$ and write $\mathbb{E}[\,\Theta_i \mid (\theta_{n_i-1}, \theta_{n_i+a_i+1})] = \mathbb{E}[\Theta \mid (\theta_{n-1}, \theta_{n+a+1})]$. Now, we write it in terms of the fields $F(x)$. Using the Landau gauge and that we reduced the problem to a fixed horizontal strip in Step 1, we can consider the array of fields also as one-dimensional $\{F_n\}_{n \in \mathbb{Z}}$ such that by (3.4) that $\theta_n = \sum_{i=1}^{n} F_i$. Therefore

$$\Theta = (\theta_n, \dots, \theta_{n+a}) = \left( \sum_{i=1}^{n} F_i, \dots, \sum_{i=1}^{n+a} F_i \right) = \theta_{n-1} + \left( \sum_{i=n}^{n} F_i, \dots, \sum_{i=n}^{n+a} F_i \right).$$

and so since $\theta_{n+a+1} - \theta_{n-1} = \sum_{i=n}^{n+a+1} F_i$,

$$\mathbb{E}[\Theta \mid (\theta_{n-1}, \theta_{n+a+1})] = \theta_{n-1} + \mathbb{E}\left[ \left( \sum_{i=n}^{n} F_i, \dots, \sum_{i=n}^{n+a} F_i \right) \mid (\theta_{n-1}, \theta_{n+a+1}) \right]$$

$$= \theta_{n-1} + \mathbb{E}\left[ \left( \sum_{i=n}^{n} F_i, \dots, \sum_{i=n}^{n+a} F_i \right) \mid \sum_{i=n}^{n+a+1} F_i \right].$$

Our goal is now to calculate the conditional density of $\left( \sum_{i=n}^{n} F_i, \dots, \sum_{i=n}^{n+a} F_i \right)$ given $\sum_{i=n}^{n+a+1} F_i$.

We are left with considering $\mathbb{E}\left[ \left( \sum_{i=n}^{n} F_i, \dots, \sum_{i=n}^{n+a} F_i \right) \mid \sum_{i=n}^{n+a+1} F_i \right]$. As in (5.5) the conditional density is the joint density divided by the marginal density, we set out to calculate the joint density of $\left(\left(\sum_{i=n}^{n} F_i, \ldots, \sum_{i=n}^{n+a} F_i\right), \sum_{i=n}^{n+a+1} F_i\right)$. To this end, notice that

$$\left(\sum_{i=n}^{n} F_i, \ldots, \sum_{i=n}^{n+a} F_i\right) = \begin{pmatrix} 1 & 0 & 0 & \ldots & 0 \\ 1 & 1 & 0 & \ldots & 0 \\ 1 & 1 & 1 & \ldots & 0 \\ \vdots & \vdots & \vdots & \ddots & \vdots \\ 1 & 1 & 1 & \ldots & 1 \end{pmatrix} \begin{pmatrix} F_n \\ F_{n+1} \\ F_{n+2} \\ \ldots \\ F_{n+a} \end{pmatrix} =: M_{a+1} \begin{pmatrix} F_n \\ F_{n+1} \\ F_{n+2} \\ \ldots \\ F_{n+a} \end{pmatrix}$$

Note that $\det(M_{a+1}) = 1$ and that its inverse is given by

$$M_{a+1}^{-1} = \begin{pmatrix} 1 & 0 & 0 & \ldots & 0 \\ -1 & 1 & 0 & \ldots & 0 \\ 0 & -1 & 1 & \ldots & 0 \\ \vdots & \vdots & \vdots & \ddots & \vdots \\ 0 & 0 & 0 & \ldots & 1. \end{pmatrix}$$

Thus, by the density transformation theorem the density $f$ of $\left(\sum_{i=n}^{n} F_i, \ldots, \sum_{i=n}^{n+a} F_i\right)$ is given by

$$f(x_0, \ldots, x_a) = \varphi(x_0)\varphi(x_1 - x_0)\varphi(x_2 - x_1) \cdot \ldots \cdot \varphi(x_a - x_{a-1}).$$

and so the joint density of $\left(\left(\sum_{i=n}^{n} F_i, \ldots, \sum_{i=n}^{n+a} F_i\right), \sum_{i=n}^{n+a+1} F_i\right)$ is

$$j(x_0, \ldots, x_a, y) = \varphi(x_0)\varphi(x_1 - x_0)\varphi(x_2 - x_1) \cdot \ldots \cdot \varphi(x_a - x_{a-1})\varphi(y - x_a)$$

The marginal density of $\sum_{i=n}^{n+a+1} F_i$ is just the $a$-fold convolution

$$m(y) = \phi^{*a}(y).$$

Hence, the conditional density corresponding to $\mathbb{E}\left[\left(\sum_{i=n}^{n} F_i, \ldots, \sum_{i=n}^{n+a} F_i\right) \mid \sum_{i=n}^{n+a+1} F_i\right]$ is

$$\tau_{\left(\sum_{i=n}^{n} F_i, \ldots, \sum_{i=n}^{n+a} F_i\right) \mid \sum_{i=n}^{n+a+1} F_i}\left((x_0, \ldots, x_a) \mid y\right) = \frac{1}{\phi^{*a}(y)} \varphi(x_0)\varphi(x_1 - x_0, x_0) \cdot \ldots \cdot \varphi(x_a - x_{a-1})\varphi(y - x_a).$$

Since this distribution is bounded by what we assumed in (2.6), the conditional distribution (5.6) is bounded following the steps above in reverse order. $\qquad\square$

# 6 A priori estimate on fractional moments

We generalize the argument in [16] to models with an internal degree of freedom and the structure as in our magnetic quantum walk.

We will try to follow Hamza's thesis [17] as close as possible and much of the material is also contained in [30]. However, there will be substantial differences as changing two phases corresponds to a rank-4 pertubation instead of merely a rank-2 pertubation. Our main theorem is the equivalent of [16, Theorem 3.1] namely an a priori estimate on the Greens function $G(k^\pm, l^\pm, z) = \langle k^\pm, (W-z)^{-1} l^\pm \rangle$. In fact, below we rather work with a modified resolvent related to the operator valued Carathéodory function of $W$, i.e. $(W+z)(W-z)^{-1}$. Bounds on this operator imply bounds on the resolvent via the identity

$$(W-z)^{-1} = \frac{1}{2z}[(W+z)(W-z)^{-1} - \mathbb{1}].\tag{6.1}$$

We have the following a priori estimate:

**Theorem 2.4.** *For any $k^\pm, l^\pm \in \mathbb{Z}^2 \times \{-1, +1\}$, any $0 < s < 1$, and $W_\omega = D_\omega S$ defined above there exists a constant $C(s) > 0$ such that*

$$\mathbb{E}_{\{\theta_k, \theta_l\}}\left[|G(k^\pm, l^\pm, z)|^s\right] \leq C(s).$$

The rest of this section is devoted to proving this a priori estimate. Throughout the proof we assume that $k \neq l$. In the case $k = l$ the proof goes through mutatis mutandis.

**Remark 6.1.**

1. *The perturbations we will consider in the proof are of rank 4. If instead of changing the phases $\theta_k$ and $\theta_l$ we changed a single field $F_x$ the corresponding perturbation would be of infinite rank. However, the proof is actually fairly general and also works for perturbations of infinite rank.*

2. *In some sense varying one $\theta_k$ and all of them corresponding to varying one flux is kind of equivalent since the latter just corresponds to varying two adjacent fluxes $F_x \to F_x + \delta$ and $F_{x+1} \to F_x - \delta$. In that sense we do not "leave the model" whenever we vary $(\theta_k^+, \theta_k)$ and leave all other phases invariant. Nevertheless, one might go through the proof of the a priori estimate and see that it also works with the infinite rank pertubation corresponding to varying the flux.*

## 6.1 Basic definitions

Fix two lattice points $k \neq l \in \mathbb{Z}^2$ in the $+$-subspace. Suppose that $D|k^+\rangle = e^{-i\theta_k}|k^+\rangle$, $D|l^+\rangle = e^{-i\theta_l}|l^+\rangle$ and therefore

$$D|k - e_1, -\rangle = e^{-i\theta_{k-e_1}^-}|k - e_1, -\rangle = e^{i\theta_k}|k - e_1, -\rangle.$$

To ease notation, we define a new basis in which the $-$-subspace is shifted by one lattice site to the left, i.e. we set

$$|k^+\rangle := |k, +\rangle, \qquad |k^-\rangle := |k - e_1, -\rangle.$$

In this new basis, we define $A^+ = \{k^+, l^+\}$ and $A^- = \{k^-, l^-\}$ as the $+$- and $-$-subspaces at $k$ and $l$, respectively, and set $A = A^+ \cup A^-$. Define new variables $\alpha = \frac{1}{2}\left(\theta_k + \theta_l\right)$ and $\beta = \frac{1}{2}\left(\theta_k - \theta_l\right)$, and for $j \in \mathbb{Z}^2$

$$\eta_j = \begin{cases} \pm\alpha, & j \in A^{\pm}, \\ 0 & j \notin A \end{cases}$$

$$\xi_j = \begin{cases} \pm\beta, & j \in \{k^{\pm}, l^{\mp}\}, \\ 0 & j \notin A \end{cases}$$

and

$$\hat{\theta}_j = \begin{cases} 0 & j \in A \\ \theta_j & j \notin A \end{cases}$$

with corresponding diagonal operators $D_\alpha$, $D_\beta$ and $\hat{D}$ defined by

$$D_\alpha |j^{\pm}\rangle = e^{-i\eta_j}|j^{\pm}\rangle,$$

and similarly for $D_\beta$ and $\hat{D}$. Acting with $D_\alpha D_\beta$ on the basis states at $k^+, k^-, l^+, l^-$ gives

$$\begin{aligned} \alpha + \beta = \theta_k, &\qquad -\alpha - \beta = -\theta_k, \\ \alpha - \beta = \theta_l, &\qquad -\alpha + \beta = -\theta_l, \end{aligned}$$

respectively, which are the correct phases in the sense that $D_\alpha D_\beta |j\rangle = D_\omega |j\rangle$ for $j \in A$. Define the unitary operator

$$V = D_\beta \hat{D} S \tag{6.2}$$

which does not depend on $\alpha$ and note that $W = D_\alpha V$. Define the subspace $\mathcal{H}_A = \mathrm{span}\{V^{-1}|j\rangle : j \in A\}$. Notice that $V^{-1}|j\rangle$ for $j \in A$ is an orthonormal basis for $\mathcal{H}_A$, and denote by $P$ be the orthgonal projection onto $\mathcal{H}_A$ which has rank 4. Then, for $j \in A$

$$V^{-1}WV^{-1}|j\rangle\ = V^{-1}D_\alpha|j\rangle = e^{i\eta_j}V^{-1}|j\rangle$$

such that on $\mathcal{H}_A$

$$V^{-1}W = \mathrm{diag}(e^{-i\alpha}, e^{i\alpha}, e^{-i\alpha}, e^{i\alpha}) = V^{-1}D_\alpha V =: L_\alpha. \tag{6.3}$$

Note that $L_\alpha$ is only defined on $\mathcal{H}_A$. Hence we can decompose $V^{-1}W$ on the whole Hilbert space as

$$V^{-1}W = \mathbb{1} - P + L_\alpha P$$

with $L_\alpha$ diagonal. Notice that this means (cf. [30, (4.5.10)])

$$W - V = V(L_\alpha - \mathbb{1})P. \tag{6.4}$$

since $(L_\alpha - \mathbb{1})$ maps $\mathcal{H}_A$ to $\mathcal{H}_A$.

Following [30], we define for $z \notin \mathbb{T}$ the operator valued Carathéodory functions $K$ and $\hat{K}$ of $W$ and $V$ on $\mathcal{H}_A$ as

$$K = K_z = P(W + z)(W - z)^{-1}P \tag{6.5}$$

and

$$\hat{K} = \hat{K}_z = P(V + z)(V - z)^{-1}P, \tag{6.6}$$

respectively [30, Prop. 4.5.4]. Since the Carathéodory function satisfies

$$(V + z)(V - z)^{-1} + \left((V + z)(V - z)^{-1}\right)^* = 2(\mathbb{1} - |z|^2)(V - z)^{-1}\left((V - z)^{-1}\right)^*,$$

we conclude that for $|z| > 1$ we have

$$2\Re e(\hat{K}_z) = \hat{K}_z + \hat{K}_z^* < 0. \tag{6.7}$$

Therefore $\langle \Im m(-i\hat{K})x, x \rangle \geq 0$ for all $x \in \mathcal{H}$ which we take as our definition for a "disspative" operator, i.e. $-i\hat{K}$ is dissipative.

## 6.2 A formula for $K$

We now prove a formula for $K$ ensuring that the inverses in (6.5) are well-defined (cf. [30, (4.5.18)])

**Lemma 6.2.** *Let $K$ and $\hat{K}$ be as in (6.5) and (6.6). Let $\mathcal{D} = \mathrm{diag}(-1, 1, -1, 1)$ and assume that $\alpha \notin \{0, \pi\}$ such that $L_\alpha \pm \mathbb{1}$ is invertible. In that case, define $M := (L_\alpha + \mathbb{1})(L_\alpha - \mathbb{1})^{-1}$. Then it holds that*

$$(L_\alpha - \mathbb{1})K(L_\alpha - \mathbb{1})^{-1} = (M + \hat{K})^{-1} + \left(\mathcal{D}\hat{K}^{-1}\mathcal{D} + M^{-1}\right)^{-1}. \tag{6.8}$$

*Proof.* From $(x + z)(x - z)^{-1} = 1 + 2z(x - z)^{-1}$ and the second resolvent identity $(A - z)^{-1} - (B - z)^{-1} = (A - z)^{-1}(B - A)(B - z)^{-1}$ it follows from (6.4) that

$$
\begin{aligned}
K - \hat{K} &= P(\mathbb{1} + 2z(W - z)^{-1} - \mathbb{1} - 2z(V - z)^{-1})P \\
&= -2zP\left((V - z)^{-1} - (W - z)^{-1}\right)P \\
&= -2zP\left((V - z)^{-1}(W - V)(W - z)^{-1}\right)P \\
&= -2zP(V - z)^{-1}VP(L_\alpha - \mathbb{1}_A)P(W - z)^{-1}P \\
&= -\frac{1}{2}P(\mathbb{1} + (V + z)(V - z)^{-1})P(L_\alpha - \mathbb{1}_A)P((W + z)(W - z)^{-1} - \mathbb{1})P \\
&= -\frac{1}{2}(P + \hat{K})P(L_\alpha - \mathbb{1}_A)P(K - P) \\
&= \frac{1}{2}(\mathbb{1}_A + \hat{K})(L_\alpha - \mathbb{1}_A)(\mathbb{1}_A - K)
\end{aligned}
$$

where we used (6.1). In the following, we simplify notation and often write $\mathbb{1}$ instead of $\mathbb{1}_A$. This now means that

$$(2 + (\mathbb{1} + \hat{K})(L_\alpha - \mathbb{1}))K = (\mathbb{1} + \hat{K})(L_\alpha - \mathbb{1}) + 2\hat{K}$$

and hence

$$(L_\alpha + \mathbb{1} + \hat{K}(L_\alpha - \mathbb{1}))K = L_\alpha - \mathbb{1} + \hat{K}(L_\alpha + \mathbb{1}). \tag{6.9}$$

Notice first that $L_\alpha \pm \mathbb{1}$ is invertible for $\alpha \neq 0, \pi$. Let us define the real-valued function

$$m(\alpha) := i\left(\frac{1 + e^{-i\alpha}}{1 - e^{-i\alpha}}\right) = \cot(\alpha/2) \tag{6.10}$$

which clearly is odd, i.e., $m(\alpha) = -m(-\alpha)$. It follows that

$$\begin{aligned} M := (L_\alpha + \mathbb{1})(L_\alpha - \mathbb{1})^{-1} &= \mathrm{diag}\left(\frac{e^{-i\alpha} + 1}{e^{-i\alpha} - 1}, \frac{e^{i\alpha} + 1}{e^{i\alpha} - 1}, \frac{e^{-i\alpha} + 1}{e^{-i\alpha} - 1}, \frac{e^{i\alpha} + 1}{e^{i\alpha} - 1}\right) \\ &= -i\,\mathrm{diag}(m(\alpha), m(-\alpha), m(\alpha), m(-\alpha)) \\ &= i \cdot m(\alpha)\,\mathrm{diag}(-1, 1, -1, 1) \\ &= i \cdot m(\alpha)\mathcal{D} \end{aligned}$$

where $\mathcal{D} = \mathrm{diag}(-1, 1, -1, 1)$ satisfies $\mathcal{D}^2 = \mathbb{1}$ and thus $\mathcal{D} = \mathcal{D}^{-1}$.

We now argue that $L_\alpha + \mathbb{1} + \hat{K}(L_\alpha - \mathbb{1})$ is invertible. Since

$$L_\alpha + \mathbb{1} + \hat{K}(L_\alpha - \mathbb{1}) = (M + \hat{K})(L_\alpha - \mathbb{1}), \tag{6.11}$$

it suffices to prove that $M + \hat{K}$ is invertible. By (6.10), $m(\alpha)$ is real so $M$ is skew-symmetric, i.e., $M^* = -M$. We decompose $M + \hat{K}$ in real and imaginary parts as follows

$$M + \hat{K} = \frac{1}{2}\left(\hat{K} + \hat{K}^*\right) + \frac{1}{2}\left(2M + \hat{K} - \hat{K}^*\right). \tag{6.12}$$

The second term in this expansion is skew-symmetric which implies that its numerical range $W(A) := \{\langle \psi, A\psi \rangle : \|\psi\| = 1\}$ is a subset of the the imaginary axis. The numerical range of the first (self-adjoint) part is the positive reals by (6.7). By the Toeplitz-Hausdorff Theorem [32, 18] the spectrum is a subset of (the closure) of the convex hull of the numerical range, and therefore must be strictly contained in the left half of the complex plane and therefore the operator $M + \hat{K}$ must be invertible.

Thus, we can isolate $K$ in (6.9). Together with (6.11) we obtain

$$\begin{aligned} K &= (L_\alpha - \mathbb{1})^{-1}(M + \hat{K})^{-1}\left(L_\alpha - \mathbb{1} + \hat{K}(L_\alpha + \mathbb{1})\right) \\ &= (L_\alpha - \mathbb{1})^{-1}(M + \hat{K})^{-1}(L_\alpha - \mathbb{1}) + (L_\alpha - \mathbb{1})^{-1}(M + \hat{K})^{-1}\hat{K}(L_\alpha + \mathbb{1}) \end{aligned}$$

which means that

$$
\begin{aligned}
(L_\alpha - \mathbb{1})K(L_\alpha - \mathbb{1})^{-1} &= (M + \hat{K})^{-1} + (M + \hat{K})^{-1}\hat{K}M \\
&= (M + \hat{K})^{-1} + (M^{-1}\hat{K}^{-1}(M + \hat{K}))^{-1} \\
&= (M + \hat{K})^{-1} + (M^{-1}\hat{K}^{-1}M + M^{-1})^{-1} \\
&= (M + \hat{K})^{-1} + (\mathcal{D}\hat{K}^{-1}\mathcal{D} + M^{-1})^{-1}.
\end{aligned}
$$

<div align="right">□</div>

## 6.3 Boundedness of Green's functions

We now prove some preliminary results that will help us prove the boundedness of the Green's function. It is a standard trick, which we slightly extend, that for $i, j \in A$ we have

$$
\begin{aligned}
\langle i, (W + z)(W - z)^{-1} j \rangle &= \langle i, (W + z)(W - z)^{-1} VV^{-1} j \rangle \\
&= \langle VV^{-1}i, (W + z)(W - z)^{-1} WL_\alpha^{-1} V^{-1} j \rangle \\
&= e^{i\alpha(j)} \langle VV^{-1}i, W(W + z)(W - z)^{-1} V^{-1} j \rangle \\
&= e^{i\alpha(j)} \langle W^* VV^{-1}i, (W + z)(W - z)^{-1} V^{-1} j \rangle \\
&= e^{i\alpha(j)} \langle L_\alpha^* V^{-1}i, KV^{-1} j \rangle \\
&= e^{i(\alpha(j) - \alpha(i))} \langle V^{-1}i, KV^{-1} j \rangle
\end{aligned}
$$

where we used the definition of $L_\alpha$ in (6.3) and $\alpha(i) = \pm\alpha$ depending on $i$. Thus

$$
|\langle i, (W + z)(W - z)^{-1} j \rangle| = |\langle V^{-1}i, KV^{-1} j \rangle|
$$

Furthermore, we have that

$$
\begin{aligned}
\langle V^{-1}i, (L_\alpha - \mathbb{1})K(L_\alpha - \mathbb{1})^{-1} V^{-1} j \rangle &= \langle (L_\alpha - \mathbb{1})^{-1} V^{-1}i, K(L_\alpha - \mathbb{1})^{-1} V^{-1} j \rangle \\
&= \frac{e^{+i\alpha(i)} - 1}{e^{-i\alpha(j)} - 1} \langle V^{-1}i, KV^{-1} j \rangle
\end{aligned}
$$

which means that

$$
|\langle V^{-1}i, (L_\alpha - \mathbb{1})K(L_\alpha - \mathbb{1})^{-1} V^{-1} j \rangle| = |\langle V^{-1}i, KV^{-1} j \rangle| = |\langle i, (W + z)(W - z)^{-1} j \rangle|. \quad (6.13)
$$

In the next subsection we will show that

**Lemma 6.3.** *For $0 < s < 1$ and $|z| \neq 1$ we have*

$$
\int_0^{2\pi} d\alpha |\langle V^{-1}i, (L_\alpha - \mathbb{1})K(L_\alpha - \mathbb{1})^{-1} V^{-1} j \rangle|^s \leq C(s)
$$

This lemma allows us to finish the proof of Theorem 2.4. If we let $\hat{\mathbb{E}}$ denote the expectation with respect to the marginal measure of $\{\theta_r\}_{r \neq k,l}$ we obtain by Proposition 5.3.

$$
\begin{aligned}
\mathbb{E}[|\langle i, (W+z)(W-z)^{-1}j\rangle|^s] &= \hat{\mathbb{E}}\left[\int_{[0,2\pi]^2} d\mu(\theta_k, \theta_l \mid \{\theta_r\}_{r \neq k,l})|\langle i, (W+z)(W-z)^{-1}j\rangle|^s\right] \\
&= \hat{\mathbb{E}}\left[\int_{[0,2\pi]^2} d\theta_k d\theta_l \tau(\theta_k, \theta_l \mid \{\theta_r\}_{r \neq k,l})|\langle i, (W+z)(W-z)^{-1}j\rangle|^s\right] \\
&\leq C\hat{\mathbb{E}}\left[\int_{[0,2\pi]^2} d\theta_k d\theta_l|\langle V^{-1}i, (L_\alpha - \mathbb{1})K(L_\alpha - \mathbb{1})^{-1}V^{-1}j\rangle|^s\right] \\
&\leq C\hat{\mathbb{E}}\left[\int_{[0,2\pi]^2} d\beta d\alpha|\langle V^{-1}i, (L_\alpha - \mathbb{1})K(L_\alpha - \mathbb{1})^{-1}V^{-1}j\rangle|^s\right] \\
&\leq 2\pi C(s)
\end{aligned}
$$

where we used Proposition 5.5 and the definition of $\tau$ as well as (6.13). This finishes the proof of Theorem 2.4. $\qquad\square$

## 6.4 Proof of Lemma 6.3

To prove Lemma 6.3 we first obtain the following estimate:

$$
\begin{aligned}
\int_0^{2\pi} d\alpha &|\langle V^{-1}i, (L_\alpha - \mathbb{1})K(L_\alpha - \mathbb{1})^{-1}V^{-1}j\rangle|^s \\
&= \int_0^{2\pi} d\alpha|\langle V^{-1}i, ((M+\hat{K})^{-1} + (\mathcal{D}\hat{K}^{-1}\mathcal{D} + M^{-1})^{-1})V^{-1}j\rangle|^s \\
&= \int_0^{2\pi} d\alpha|\langle V^{-1}i, (M+\hat{K})^{-1}g\rangle|^s + |\langle f, (\mathcal{D}\hat{K}^{-1}\mathcal{D} + M^{-1})^{-1}V^{-1}j\rangle|^s \\
&\leq \int_0^{2\pi} d\alpha \left\|(M+\hat{K})^{-1}\right\|^s + \left\|(\mathcal{D}\hat{K}^{-1}\mathcal{D} + M^{-1})^{-1}\right\|^s,
\end{aligned}
$$

where we used (6.8) and that $|x+y|^s \leq |x|^s + |y|^s$. We continue by handling each term separately.

Then, notice that $M = im(\alpha)\mathcal{D}$ with $m(\alpha) = i\left(\frac{1+e^{-i\alpha}}{1-e^{-i\alpha}}\right) = \cot(\alpha/2)$ as in (6.10). Substituting $x = m(\alpha)$ we have $\left|\frac{dx}{d\alpha}\right| = \frac{x^2+1}{2}$ and so

$$
\begin{aligned}
\int_0^{2\pi} d\alpha \left\|(M+\hat{K})^{-1}\right\|^s &= \int_{-\infty}^{\infty} \frac{2}{x^2+1}\left\|(M+\hat{K})^{-1}\right\|^s dx \\
&\leq 2\sum_{n \in \mathbb{Z}} \frac{1}{(|n|-1)^2 + 1}\int_n^{n+1}\left\|(ix\mathcal{D} + \hat{K})^{-1}\right\|^s dx \leq C(s),
\end{aligned}
$$

where we used Lemma 6.5 below. Similarly, for the second term here we do the substitution $y = m^{-1}(\alpha)$ and we use Lemma 6.4 below, where we saw that $\mathcal{D}\hat{K}^{-1}\mathcal{D}$ was still dissipative. Then we can do the same estimates.

**Lemma 6.4.** *Suppose for matrices $D$ and $K$ that $D^2 = \mathbb{1}$, $D^* = D$ and $K + K^* < 0$. Then $(DKD) + (DKD)^* < 0$.*

*Proof.* For any vector $v$ it holds that

$$\langle v, DKD + (DKD)^* v \rangle = \langle v, D(K + K^*)Dv \rangle = \langle Dv, (K + K^*)Dv \rangle < 0$$

since $D$ is invertible. $\qquad\square$

One of the main differences between the otherwise very similar approaches [16] and [17] is the use of the dissipativity to get a bound of the following type. In [17], the general infinite dimensional version is used and in [16], a $2 \times 2$ version is proved using explicit matrix calculations. Here we give a third proof using the numerical range and pseudospectra.

**Lemma 6.5.** *Let $0 < s < 1$, and $D$ and $K$ such that $D^2 = \mathbb{1}$, $D^* = D$, and $K$ is dissipative, i.e. $K + K^* < 0$, respectively. Then*

$$\int_n^{n+1} \left\| (K + ivD)^{-1} \right\|^s dv \le C(s),$$

*for all $n \in \mathbb{Z}$.*

To prove this lemma, we need the following definition:

**Definition 6.6.** *For any $\varepsilon > 0$, the pseudospectrum of $A$ is defined as*

$$\Lambda_\varepsilon(A) = \{ z \in \mathbb{C} \mid \left\| (zI - A)^{-1} \right\| \ge \frac{1}{\varepsilon} \}$$

Moreover, we need the following general result on the connection between the pseudospectrum and the numerical range [33, (17.9)]:

**Theorem 6.7.** *Let $W(A)$ be the numerical range of $A$ and let $\Delta_\varepsilon$ be the closed disc with radius $\varepsilon$ and with center $0$. Then*

$$\Lambda_\varepsilon(A) \subset W(A) + \Delta_\varepsilon.$$

Now we can go ahead and prove the above lemma:

*Proof of Lemma 6.5.* We decompose $K + ivD$ into real and imaginary part as in (6.12) which yields

$$K + ivD = \frac{1}{2}(K + K^*) + \frac{1}{2}(2ivD + K - K^*)$$

since $K + K^* < 0$ and therefore we must have $\sup\{\lambda \mid \lambda \in \sigma(K + K^*)\} = -r < 0$. Moreover, $\operatorname{dist}(\sigma(K + ivD), 0) \ge r$ since the skew-hermitian part contributes only something purely imaginary to the numerical range, and the spectrum is contained in the closure of the numerical range. Thus, since $\sigma(A^{-1}) = \{\lambda^{-1} \mid \lambda \in \sigma(A)\}$ for any invertible element $A$ in an Banach algebra we have that $\sup\{|\lambda| \mid \lambda \in \sigma((K+ivD)^{-1})\} \leq r^{-1}$. Luckily, some theorems on pseudospectra can help us out here.

Now we know that the real part of the numerical range of $(K+ivD)$ is less than $-r < 0$. That means that the real part of $W(K+ivD) + \Delta_{r/2}$ is less than $-r/2 < 0$. Therefore, by Theorem 6.7 we have that $0 \notin \Lambda_{r/2}(K+ivD)$ which by definition of the pseudospectrum means that $\|(K+ivD)^{-1}\| \leq 2/r$. Hence

$$\int_n^{n+1} \left\| (K+ivD)^{-1} \right\|^s dv \leq C(s) \leq \left( \frac{2}{r} \right)^s = C(s).$$

<div align="right">□</div>

# 7 Proof of exponential decay

In Theorem 2.4 we proved the analog of [16, Theorem 3.1] which has a similar setting as us. We will need this a priori estimate repeatedly in what follows. But before we go into the proof of exponential decay of fractional moments in case of strong disorder we derive some spectral properties in the reflecting case corresponding to maximal disorder.

## 7.1 Finite Volume Restrictions

For what follows, recall from Section 4.2 the notations $W^L = W^{\Lambda_L}(C_1, C_2)$ and $\tilde{W}^L = W^{\Lambda_L}(C_1^r, C_2^r)$. Denote the corresponding resolvents by $G^L = (W^L - z)^{-1}$ and $\tilde{G}^L = (\tilde{W}^L - z)^{-1}$ for some $z \notin \mathbb{T}$. To ease notation we do not write out their dependence on $z$ explicitly.

**Proposition 7.1.** *Let $u, v \in \mathbb{Z}^2 \times \{-1, +1\}$ such that $|u - v| > 2$. Then for any $\eta > 0$, any $0 < s < 1$ and $p > \frac{1}{1-s}$ it holds that*

$$\mathbb{E}[|\langle u, G^L v \rangle|^s] \leq C \left( (L^2 \eta)^{\frac{1}{p}} + \frac{\left\| \tilde{W}^L - W^L \right\|^s}{\eta^{2s}} \right). \tag{7.1}$$

*Moreover, for any $a > 0$ if $\|C_{i,x} - C_{i,r}\| = L^{-2(ap+2) - \frac{a}{s}}$ then there exists a $C > 0$ such that for any $u, v$ with $|u - v| > 2$ then*

$$\mathbb{E}[|\langle u, G^L v \rangle|^s] \leq \frac{C}{L^a}. \tag{7.2}$$

*Proof.* Recall the resolvent identity

$$G^L = \tilde{G}^L + G^L(\tilde{W}^L - W^L)\tilde{G}^L.$$

Before stating estimations we note that due to the invariance of $\mathcal{H}_x^+$ and the power series formula for $\tilde{G}^L$ we have that $\langle x^+, \tilde{G}^L \psi \rangle$ is $0$ except for when $\psi \in \mathcal{H}_x^+$. So if $u, v$ are two states of distance larger than 2 then manifestly $|\langle u, \tilde{G}^L v \rangle| = 0$. That means that when we find matrix elements for $G$ using the resolvent equation and inserting identities we have

$$
\begin{aligned}
|\langle u, G^L v \rangle| &= |\langle u, G^L (\tilde{W}^L - W^L) \tilde{G}^L v \rangle| \\
&= \sum_\alpha \left| \langle u, G^L \alpha \rangle \langle \alpha, (\tilde{W}^L - W^L) \tilde{G}^L v \rangle \right| \\
&\le 2 \sum_\alpha \left| \frac{\langle u, G^L \alpha \rangle}{\operatorname{dist}(\sigma(\tilde{W}^L), z)} \| \tilde{W}^L - W^L \| \right| \\
&\le c \frac{\| \tilde{W}^L - W^L \|}{\operatorname{dist}(\sigma(\tilde{W}^L), z) \operatorname{dist}(\sigma(W^L), z)}
\end{aligned}
\tag{7.3}
$$

Here we also used that $\| \tilde{G}^L \| = \| (\tilde{W}^L - z)^{-1} \| \le 1/\operatorname{dist}(\sigma(\tilde{W}^L), z)$ and equivalently for $W^L$. In the last step, we used that there are only finitely many states with a distance of at most 2 from $|\alpha\rangle$.

Now, we evaluate the expectation in (7.1) which to this end we split into two parts: a small part of $\Omega$ where the spectrum is close to $z$ and a large part where the spectrum is far from $z$ since in this case we can handle the denominators. The way to formalize this is to define

$$
M_\eta(z) = \{ \omega \in \Omega \mid \operatorname{dist}(\sigma(\tilde{W}^L), z) > \eta \}
$$

and then separately bound each of the terms in

$$
\mathbb{E}[\operatorname{id}_{M_\eta(z)} |\langle u, G^L v \rangle|^s] + \mathbb{E}[\operatorname{id}_{M_\eta(z)^c} |\langle u, G^L v \rangle|^s].
\tag{7.4}
$$

We start with the first term. By Lemma 4.4, Hölder's inequality with conjugates $p, q$ such that $qs < 1$, and Theorem 2.4 we obtain for sufficiently small $\eta > 0$

$$
\mathbb{E}[\operatorname{id}_{M_\eta(z)^c} |\langle u, G^L v \rangle|^s] \le \mathbb{P}[M_\eta(z)^c]^{\frac{1}{p}} \mathbb{E}[|\langle u, G^L v \rangle|^{sq}]^{\frac{1}{q}} \le C \left( L^2 \eta \right)^{\frac{1}{p}}.
\tag{7.5}
$$

Turning to the second part we now use that the numerators in (7.3) cannot explode. First note that if $\| W^L - \tilde{W}^L \| \le c\eta$ then their spectra are close. Indeed we use the relation [22, (3.38)]

$$
\operatorname{dist}(\sigma(\tilde{W}^L), z) > \eta \quad \Leftrightarrow \quad \operatorname{dist}(\sigma(W^L), z) > \frac{\eta}{2}
$$

which follows from the normality of unitaries and that it holds for normal operators that $\| (A - z)^{-1} \| = \operatorname{dist}(\sigma(A), z)$. Thus, both terms in the denominator of (7.3) will be large. After using that we can afford to forget the indicator function.

$$
\mathbb{E}[\operatorname{id}_{M_\eta(z)} |\langle u, G^L v \rangle|^s] \le c\eta^{-2s} \mathbb{E}[\operatorname{id}_{M_\eta(z)} \| \tilde{W}^L - W^L \|^s] \le c \frac{\| (\tilde{W}^L - W^L) \|^s}{\eta^{2s}}.
\tag{7.6}
$$

Putting things together we have for any $0 < s < 1$ and $p > \frac{1}{1-s}$ that

$$
\mathbb{E}[|\langle u, G v \rangle|^s] \le C \left( \left( L^2 \eta \right)^{\frac{1}{p}} + \frac{\| (\tilde{W}^L - W^L) \|^s}{\eta^{2s}} \right).
$$

To see the second statement we use Lemma 4.3 to obtain the bound $\|(\tilde{W}^L - W^L)\|^s \leq 2c\|C_{i,x} - C_{i,r}\|$. Again as in [22] for any $a > 0$ we can pick the scale of $\eta$ such that $\eta = \frac{c}{L^{ap+d}}$ which ensures that $(L^2\eta)^{\frac{1}{p}} = \frac{c}{L^a}$. Then our condition on $\|C_{i,x} - C_{i,r}\|$ becomes that $\|C_{i,x} - C_{i,r}\| = L^{-2(ap+2)-\frac{a}{s}}$ which we acheive by assumption. $\qquad\square$

## 7.2 Almost independence and the Aharonov-Bohm effect

Recall, how we proved that the conditional density is bounded in Proposition 5.5. In the following we make substantial modifications to the proof that will ensure that the resampling strategy from [16, Theorem 3.1] can be adapted to our setting, which is more complicated due to the peculiar independence structure of $\{\theta_i\}$ in our case.

We start by considering the gauge invariance of Greens functions.

### 7.2.1 Gauge invariance of Greens functions

Rewriting

$$(W - z)^{-1} = -\frac{1}{z}\left(1 - \left(\frac{W}{z}\right)\right)^{-1} = -\frac{1}{z}\sum_{n\geq 0}\left(\frac{W}{z}\right)^n \qquad (7.7)$$

we can write the Greens function as

$$G(u,v) = \langle u, (W-z)^{-1}v\rangle = -\sum_{n\geq 0} z^{-(n+1)}\langle u, W^n v\rangle. \qquad (7.8)$$

As $W = D_\omega W_0(C_1, C_2) = D_\omega W_0$ it holds that $W^n = (D_\omega W_0)^n$ and we can expand the matrix element $\langle u, W^n v\rangle$ as a sum of paths of length $n$, i.e.,

$$\langle u, W^n v\rangle = \langle u, (D_\omega W_0)^n v\rangle = \sum_{\gamma:\gamma(0)=v,\gamma(n)=u} \prod_{i=0}^{n-1} W_0(\gamma(i),\gamma(i+1))e^{i\theta(\gamma(i),\gamma(i+1))} \qquad (7.9)$$

For any path $\gamma$ of length $n$ from $u$ to $v$ define

$$w_z(\gamma) = -\prod_{i=0}^{n-1} z^{-(n+1)}W_0(\gamma(i),\gamma(i+1))$$

as well as the contribution of the magnetic field accumulated along $\gamma$,

$$\theta(\gamma) = \sum_{i=0}^{n-1}\theta(\gamma(i),\gamma(i+1)). \qquad (7.10)$$

It is a (direct) consequence of the discussion around [10, (20)] that

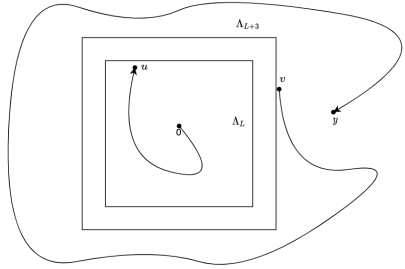

Figure 6: Sketch of the situation with the resolvents in Proposition 7.8. With the arrows we denote the Greens functions $\hat{G}^{(L)}(0, u)$ and $\hat{G}^{(L+3)}(v, y)$. Notice that the Green's functions inside and outside the box are not independent, but the outer Greens function only depends on the inner phases through the total flux through all the plaquettes in $\Lambda_L$. When $L$ becomes very large this dependence becomes smooth with high probability and thus with an extra constant $\|\frac{1}{f_L}\|_\infty$ we can prove all the results as if we had independence.

**Lemma 7.2.** *Suppose that $\gamma, \tilde{\gamma}$ are two paths from $u$ to $v$. Then*

$$\theta(\gamma) - \theta(\tilde{\gamma})$$

*is gauge-independent. In other words, $\theta(\gamma) - \theta(\tilde{\gamma})$ is completely determined by $\{F(x)\}_{x \in \mathbb{Z}^2}$.*

*Proof.* Let us denote by $\gamma^{-1}$ the inverse of a path $\gamma$ which is obtained from inverting the steps and as well as their order. Then $\tilde{\gamma}^{-1}\gamma$ is a loop from $u$ to $u$, so the phase $\theta(\tilde{\gamma}^{-1}\gamma) = \theta(\gamma) - \theta(\tilde{\gamma})$ picked up along it is an element of the holonomy group (at $u$). Since our gauge group is $U(1)$ and therefore abelian, the statement follows directly from [10, (20)]. $\quad\square$

Finally, we are concerned with expectation values, so it is a good sanity check to see that these are indeed gauge invariant.

**Lemma 7.3.** *For any $u, v \in \ell^2(\mathbb{Z}^2) \otimes \mathbb{C}^2$, $|\langle u, Gv\rangle|$ is gauge invariant.*

*Proof.* By [10, Lemma IV.2], $W$ transforms under a gauge transformation $V$ as $W \mapsto VWV^*$. Since $V$ is unitary, we have that

$$|\langle u, (W-z)^{-1}v\rangle| \mapsto |\langle Vu, (VWV^* - z)^{-1}Vv\rangle| = |\langle Vu, V(W-z)^{-1}V^*Vv\rangle| = |\langle u, (W-z)^{-1}v\rangle|.$$

$$\square$$

### 7.2.2 Winding numbers

Let us define the winding number through the Alexander numbering introduced in [6] (see also [27, Lemma 2].For any closed curve $\gamma$ in the plane the winding number of a point $p$ with respect to the curve $\gamma$ we denote by $w_\gamma(p)$. Notice that the curve partitions the plane into several connected regions, one of which is unbounded.

1. The winding numbers of the curve around two points in the same region are equal.

2. The winding number around (any point in) the unbounded region is zero.

3. The winding numbers for any two adjacent regions differ by exactly 1; the region with the larger winding number appears on the left side of the curve (with respect to motion down the curve).

In the particular case where $\gamma$ is a closed curve that only traverses edges of $\mathbb{Z}^2$. All points in a plaquette $P$ are part of the same region and thus the winding number of a plaquette $P$ is well-defined and we write it as $w_\gamma(P)$.

Let us give a more explicit construction of the gauge independence of $\theta(\gamma)$, which is the accumulation of phase along the path $\gamma$ defined in (7.10). We note that in some sense the following is a generalization of Stokes' theorem.

**Lemma 7.4.** *Let $\gamma$ be a closed curve in $\mathbb{R}^2$ that only traverses edges of $\mathbb{Z}^2$. The following formula holds*

$$\theta(\gamma) = \sum_{P \in \mathbb{Z}^2} F(P) w_\gamma(P)$$

*Proof.* Let us work in a Landau gauge where the point 0 is picked so that $\gamma$ is completely above 0. Since the quantity is gauge independent it holds in any gauge. For any plaquette $P$ pick any point $p \in P$. Consider the vertical line $V_p$ through $p$. The plaquette $P$ contributes to the sum

$$\theta(\gamma) = \sum_{i=1}^{n} \theta(\gamma(i), \gamma(i+1))$$

exactly for the $i$'s in the sum where $V_p$ crosses the edge $(\gamma(i), \gamma(i+1))$ above $p$. The contribution is $F(P)$ times the number of times that $V_p$ crosses the edge $(\gamma(i), \gamma(i+1))$ above $p$ counted with signs. But by definition of the winding number this is exactly $F(P) w_\gamma(P)$. Since this holds for any plaquette $P$ the formula follows. $\square$

### 7.2.3 Proving the Aharonov Bohm effect

Let us now continue with providing probabilistic insight on the Aharonov Bohm effect. In the following let $\mathbf{F} = (F_j)_{j \in \mathbb{Z}^2}$.The insight is that what happens outside of $\Lambda_L$ is conditionally independent from what is inside the box $\Lambda_L$ conditioned on

$$F_L = \sum_{x \in \Lambda_L} F(x). \tag{7.11}$$

The sum is understood modulo $2\pi$. Let $\mu_{F_L}$ be the marginal measure of $F_L$ and let $P_{\mathbf{F}|F_L}(\cdot \mid \phi)$ be the conditional measure of $\mathbf{F}$ given $F_L = \phi$. Then, for any measurable function $\psi(\mathbf{F}, F_L)$ it holds by [19, (10.3.2), (10.3.4)] that

$$\mathbb{E}[\psi(\mathbf{F}, F_L)] = \int_{\mathbb{T}} \mu_{F_L}(d\phi) \int_{\mathbb{T}^{2^2}} \psi(s, \phi) P_{\mathbf{F}|F_L}(ds \mid \phi)$$

$$= \int_{\mathbb{T}} \mu_{F_L}(d\phi) \mathbb{E}[\psi(\mathbf{F}, F_L) \mid F_L = \phi].$$

So if we define $\mathbb{E}_\phi$ by $\mathbb{E}_\phi = \int_{\mathbb{T}} \mu_{F_L}(d\phi)$ we have the relation

$$\mathbb{E}[\psi(\mathbf{F}, F_L)] = \mathbb{E}_\phi[\mathbb{E}[\psi(\mathbf{F}, F_L) \mid F_L = \phi]]. \tag{7.12}$$

**Proposition 7.5** (Formalization of the Aharonov Bohm effect). *Let $u \in \Lambda_L$ and $v, y \in \Lambda_{L+3}^c$. Then*

1. $|\langle 0, G^L u\rangle|$ *is independent of* $\{F(x)\}_{x \notin \Lambda_L}$, *and*

2. $|\langle v, G^{(L+3)}y\rangle|$ *depends only on* $\{F(x)\}_{x \in \Lambda_L}$ *through the* $F_L$.

*That means that,* $|\langle 0|G^L|u\rangle|$ *and* $|\langle v|G^{(L+3)}|y\rangle|$ *are conditionally independent given* $F_L$. *In particular, it holds that for all* $u \in \Lambda_L$ *and* $v, y \in \Lambda_{L+3}^c$ *and* $0 < s < 1$ *that*

$$\mathbb{E}\left(|\langle 0|G^L|u\rangle|^s|\langle v|G^{(L+3)}|y\rangle|^s\right) = \mathbb{E}_\phi\left[\mathbb{E}\left(|\langle 0, G^L u\rangle|^s|\langle v, G^{(L+3)}y\rangle|^s \mid F_L = \phi\right)\right]$$

$$= \mathbb{E}_\phi\left[\mathbb{E}\left(|\langle 0, G^L u\rangle|^s \mid F_L = \phi\right) \mathbb{E}\left(|\langle v, G^{(L+3)}y\rangle|^s \mid F_L = \phi\right)\right].$$

*Proof.* (1) follows from the construction of $G^L$. So let us turn our attention to (2).

Using the decomposition of the Greens function and the construction of $G^L$ above, it holds that

$$|\langle 0, G^L u\rangle| = |\sum_{\gamma: 0 \to u, \gamma \subset \Lambda_L} w_z(\gamma) e^{i(\theta(\gamma) - \theta(\gamma_0))}|,$$

where $\gamma_0$ is any path from $0$ to $u$ that completely lies within $\Lambda_L$. Similarly,

$$|\langle v, G^{(L+3)}y\rangle| = |\sum_{\gamma: v \to y, \gamma \subset \Lambda_{L+3}^c} w_z(\gamma) e^{i(\theta(\gamma) - \theta(\tilde\gamma_0))}|,$$

where $\tilde\gamma_0$ is any path from $v$ to $y$ in $\Lambda_{L+3}^c$. Then the concatenation of $\gamma$ and the reversal of $\tilde\gamma_0$ is a closed path from $v$ to $v$ through $u$ in $\Lambda_{L+3}^c$. Denote this path by $\gamma_c$. Notice that $\theta(\gamma_c) = \theta(\gamma) - \theta(\tilde\gamma_0)$. Since $\gamma_c$ is a path in $\Lambda_{L+3}^c$ all plaquettes $P \in \Lambda_L$ have the same winding number with respect to $\gamma_c$. Let that number be $w_{\gamma_c}(\Lambda_L)$. By Lemma 7.4 it holds that

$$\theta(\gamma) - \theta(\tilde\gamma_0) = \theta(\gamma_c) = w_{\gamma_c}(\Lambda_L)F_L + \sum_{P \in \Lambda_L^c} F(P)w_{\gamma_c}(P) \tag{7.13}$$

where we used (7.11). It follows that there exist functions $\varphi, \psi$ such that

$$|\langle v, G^{(L+3)}y\rangle| = \varphi(F_L, \{F_j\}_{j \notin \Lambda_L}) \qquad (7.14)$$

and $|\langle 0, G^L u\rangle| = \psi(\{F_j\}_{j \in \Lambda_L})$. Hence $|\langle v, G^{(L+3)}y\rangle|$ is conditionally independent of $|\langle 0, G^L u\rangle|$ given $F_L$ and the formula follows (cf. [19, (6.15.5)] ) . $\qquad \square$

### 7.2.4 Independence up to the Aharonov Bohm effect

In the following, we will need that things that happen inside and outside of a closed box are independent to continue following the strategy of [16]. However, by the Aharonov-Bohm effect [1] this is not the case. Yet, by conditioning on the total flux through $\Lambda_L$ can we obtain the following approximate result. Recall from Section 4.2 how we constructed the operator $G^L$ (cf. (4.5)).

**Lemma 7.6** (Factorization using Aharonov Bohm effect). *Let $f_L$ be the density of the random variable $F_L = \sum_{x \in \Lambda_L} F(x)$ representing the total flux through $\Lambda_L$. For $|y| \geq L + 2$ and $u \in \Lambda_L, v \in \Lambda_{L+3}^c$ we have that*

$$\mathbb{E}\left[|\langle 0, G^L u\rangle|^s |\langle v, G^{(L+3)}y\rangle|^s\right] \leq \left\|\frac{1}{f_L}\right\|_\infty \mathbb{E}\left[|\langle 0, G^L u\rangle|^s\right] \mathbb{E}\left[|\langle v, G^{(L+3)}y\rangle|^s\right]$$

*for all $0 < s < 1$.*

*Proof.* By Proposition 7.5 we obtain that

$$\mathbb{E}\left[|\langle 0, G^L u\rangle|^s |\langle v, G^{(L+3)}|\rangle\rangle y|^s\right] = \mathbb{E}_\phi\left[\mathbb{E}\left[|\langle 0, G^L u\rangle|^s \mid F_L = \phi\right] \mathbb{E}\left[|\langle v, G^{(L+3)}y\rangle|^s \mid F_L = \phi\right]\right].$$

Notice that $|\Lambda_L| = L^2$. Then consider first the term with $|\langle 0, G^L u\rangle|^s$, which only depends on the finitely many variables $\{F_j\}_{j \in \Lambda_L}$. Therefore the conditional density $f_{\mathbf{F}_{L^2}|F_L}(x \mid \phi)$ is well-defined and it holds that

$$f_{\mathbf{F}_{L^2}|F_L}(x \mid \phi) = \frac{f_{\mathbf{F}_{L^2},F_L}(x,\phi)}{f_L(\phi)}.$$

The joint density is given by

$$f_{\mathbf{F}_{L^2},F_L}(x,\phi) = \prod_{i=1}^{L^2} \varphi(x_i) \mathrm{id}_{\left\{\sum_{i=1}^{L^2} x_i = \phi \pmod{2\pi}\right\}}(x,\phi). \qquad (7.15)$$

So it holds that

$$\mathbb{E}\left[|\langle 0, G^L u\rangle|^s \mid F_L = \phi\right] = \int_{\mathbb{T}^{L^2}} |\langle 0, G^L u\rangle|^s f_{\mathbf{F}_{L^2}|F_L}(x \mid \phi) \lambda^{L^2}(dx)$$

$$\leq \|\frac{1}{f_L}\|_\infty \mathbb{E}[|\langle 0, G^L u\rangle|^s].$$

Now the total expression becomes easier.

$$\mathbb{E}_{\phi}\left[\mathbb{E}\left[|\langle 0, G^L u\rangle|^s \mid F_L = \phi\right]\mathbb{E}\left[|\langle v, G^{(L+3)}y\rangle|^s \mid F_L = \phi\right]\right]$$
$$= \|\frac{1}{f_L}\|_{\infty}\mathbb{E}[|\langle 0, G^L u\rangle|^s]\mathbb{E}_{\phi}\left[\mathbb{E}\left(|\langle v, G^{(L+3)}y\rangle|^s \mid F_L = \phi\right)\right]$$
$$= \|\frac{1}{f_L}\|_{\infty}\mathbb{E}[|\langle 0, G^L u\rangle|^s]\mathbb{E}[|\langle v, G^{(L+3)}y\rangle|^s].$$

$\square$

Now, as $F_L = \sum_{x \in \Lambda_L} F(x)$ and the $F(x)$ are independent and have density $\varphi$ then the density of $F_L$ is the iterated convolution of $\varphi$. Now, for two densities $f, g$ it holds that

$$\inf_{x \in [0,2\pi]}\{(f \star g)(x)\} = \int_0^{2\pi} f(x-y)g(y)dy \geq \inf_{x \in [0,2\pi]}\{f(x)\}\int_0^{2\pi} g(y)dy = \inf_{x \in [0,2\pi]}\{f(x)\}.$$

Using this consideration repeatedly, we conclude that for each $n \in \mathbb{N}$,

$$\inf_{x \in [0,2\pi]}\{\varphi^{\star n}(x)\} \geq \inf_{x \in [0,2\pi]}\{\varphi(x)\} \geq c > 0.$$

and therefore that $\|\frac{1}{f_L}\|_{\infty} \geq c > 0$ for all $L$.

## 7.3 Resampling arguments

In this section, we generalize the resampling arguments from [16] so that they can be applied in our setting where we do not quite have independence. According to [16] the resampling strategy was developed in [3] and [13].

### 7.3.1 Defining resampling for dependent random variables

This section is devoted to defining resampling abstractly. However, we will be guided by our concrete example when we present the theory.

Let $\mathcal{J}$ be a finite set and $X = \mathbb{T}^{\mathcal{J}^c}$ and $Y = \mathbb{T}^{\mathcal{J}}$. Then $X \times Y = \mathbb{T}^{\mathbb{Z}^2}$. Denote by $\mathcal{A}_X$ the sigma algebra generated by cylinder Borel sets and $\mathcal{A}_Y = \mathcal{B}(\mathbb{T}^{\mathcal{J}})$. Then $\mathcal{A}_X \otimes \mathcal{A}_Y$ is the sigma algebra generated by cylinder Borel sets and $(\mathbb{T}^{\mathbb{Z}^2}, \mathcal{B}(\mathbb{T}^{\mathbb{Z}^2})) = (X \times Y, \mathcal{A}_X \otimes \mathcal{A}_Y)$. Let $\mu$ be a probability measure on $X \times Y$. Let $\Theta : X \times Y \to X \times Y = \mathbb{T}^{\mathbb{Z}^2}$ be a corresponding random variable (that is $\mu(A) = \mathbb{P}(\Theta \in A)$). Denote the marginals of $\Theta$ by $\Theta_X$ and $\Theta_Y$. Let $\mathbb{E}(\cdot \mid \Theta_X)$ be the conditional expectation. For every $B \in \mathcal{A}_Y$ and every $x \in X$ then define a measure $\nu_x$ by $\nu_x(B) = \mathbb{E}[\mathbb{1}_B \mid \Theta_X = x]$. Then $\nu_x(\cdot) = P_{\Theta_Y|\Theta_X}(\cdot \mid x)$ is the conditional measure. Let further $\nu$ be the marginal of $\mu$ on $X$ then we have the following decomposition using Proposition 5.3

$$\mu(C) = \int_X \int_Y \mathrm{id}_C(x,y)d\nu_x(y)d\nu(x). \tag{7.16}$$

Consider the measure space $(X \times Y^2, \mathcal{A}_X \otimes \mathcal{A}_Y^{\otimes 2})$, with $\nu$ a measure on $X$ and $\nu_x^{\otimes 2}$ on $Y^2$. Define the measure $\tilde{\mu}$ for any $A \in \mathcal{A}_X \otimes \mathcal{A}_Y^{\otimes 2}$ by

$$\tilde{\mu}(A) = \int_X \int_{Y^2} 1_A(x, y, \hat{y}) d\nu_x^{\otimes 2}(y, \hat{y}) d\nu(x). \tag{7.17}$$

Since $\nu_x$ was a Markov kernel then $\nu_x^{\otimes 2}$ is also a Markov kernel. The corresponding distribution $(\Theta_X, \Theta_Y, \Theta_{\hat{Y}})$ satisfies that $\Theta_Y$ and $\Theta_{\hat{Y}}$ are conditionally independent given $\Theta_X$.

Now, we can define the expectation with respect to the marginal onto $\mathcal{J}^c$

$$\mathbb{E}^{\mathcal{J}^c}[G] = \int_X G \nu(dx). \tag{7.18}$$

Further for each $x \in X$ we can define the conditional measure of the original phases

$$\mathbb{E}^{\mathcal{J}}[G] = \int_Y G \nu_x(dy) \tag{7.19}$$

and the conditional measure of the resampled phases

$$\hat{\mathbb{E}}^{\mathcal{J}}[G] = \int_{\hat{Y}} G \nu_x(d\hat{y}). \tag{7.20}$$

In particular,

$$\hat{\mathbb{E}}^{\mathcal{J}} \mathbb{E}^{\mathcal{J}}[G] = \int_{Y \times \hat{Y}} G(\nu_x)^{\otimes 2}(d(y, \hat{y})) = \mathbb{E}[G \mid \Theta_X = x_X] \tag{7.21}$$

Letting $\mathbb{E}_{\text{all}}$ be the full expectation value corresponding to the measure $\tilde{\mu}$ we can state the following Lemma.

**Lemma 7.7.** *Suppose that $G_1$ only depends on $\theta_{\mathcal{J}^c}$ and $\theta_{\mathcal{J}}$ and $\hat{G}_2$ only depends on $\theta_{\mathcal{J}^c}$ and $\hat{\theta}_{\mathcal{J}}$. Then*

$$\mathbb{E}_{\text{all}}[G_1 \hat{G}_2] = \mathbb{E}^{\mathcal{J}^c}[\hat{\mathbb{E}}^{\mathcal{J}} \mathbb{E}^{\mathcal{J}}[G_1 \hat{G}_2]] = \mathbb{E}^{\mathcal{J}^c}[\mathbb{E}^{\mathcal{J}}[G_1] \hat{\mathbb{E}}^{\mathcal{J}}[\hat{G}_2]] = \mathbb{E}^{\mathcal{J}^c} \mathbb{E}^{\mathcal{J}} \hat{\mathbb{E}}^{\mathcal{J}}[G_1 \hat{G}_2].$$

### 7.3.2 Using resampling to construct iteration

Let us now embark on constructing the iteration using resampling arguments following [16, Proposition 13.1] with some important caveats that we will elaborate on shortly.

Let us first define

$$\hat{D} = \sum_{n \in \mathcal{J}} \left( e^{-i\theta_n} - e^{-i\hat{\theta}_n} \right) |n\rangle\langle n|.$$

and define $\hat{W}^L, \hat{G}^L$ as in [16]. Notice how the only difference between $G^L$ and $\hat{G}^L$ is that $\hat{G}^L$ depends on the resampled phases $\hat{\theta}_n$ whereas $G^L$ depends on the original ones. This thus gives the equation

$$\hat{\mathbb{E}}^{\mathcal{J}} \left[ \hat{G}^L \right] = \mathbb{E}^{\mathcal{J}} \left[ G^L \right]. \tag{7.22}$$

Let $t = \|T^L\|$ which is bounded by Lemma 4.3. Now, we are ready to prove the equivalent of [22, Proposition 3.12], which related the infinite volume resolvents to the finite volume ones.

**Proposition 7.8.** *For every $0 < s < \frac{1}{3}$ we have that*

$$\mathbb{E}[|\langle 0, Gy \rangle|^s] \leq C(s)^2 \| \frac{1}{f_L} \|_\infty ct^{2s} \sum_{u, L-2 \leq |u| \leq L+2} \mathbb{E}[|\langle 0, G^L u \rangle|^s] \sum_{v', L+2 \leq |v'| \leq L+4} \mathbb{E}[|\langle v', G^{(L+3)} y \rangle|^s].$$

*Proof.* We use similar reasoning as in both papers [22, (3.53)] and [16, (13.16)] to see that

$$\mathbb{E}[|\langle 0, Gy \rangle|^s] \leq ct^{2s} \sum_{(u,u') \in \mathcal{H}_L, (v,v') \in \mathcal{H}_{L+3}} \mathbb{E}\left[|\langle 0, G^L u \rangle|^s |\langle u', Gv \rangle|^s |\langle v', G^{(L+3)} y \rangle|^s\right].$$

Now, the proof of resampling (cf. [16, Proposition 13.1]) needs to be reconsidered. In the proof from [16, Proposition 13.1] the terms $|\langle 0, G^L u \rangle|^s$ and $|\langle v', G^{(L+3)} y \rangle|^s$ are independent, but correlated through the factor $|\langle u', Gv \rangle|^s$. For us, they are independent up to the Aharonov-Bohm effect, but still correlated through the factor $|\langle u', Gv \rangle|^s$. Nevertheless, with some care we can still use the same resampling strategy.

We now do the same resolvent equations as in the proof from [16, Proposition 13.1]. Notice that in this process we can do this formally, in sense that these resolvent equations are true for any set of parameters $\theta_n$ and $\hat{\theta}_n$.

We then obtain since the total expression is independent of $\{\hat{\theta}_n\}_{n \in \mathcal{J}}$ and $\hat{\mathbb{E}}^{\mathcal{J}}(1) = 1$ that

$$\mathbb{E}\left[|\langle 0, G^L u \rangle|^s |\langle u', Gv \rangle|^s |\langle v', G^{(L+3)} y \rangle|^s\right]$$
$$= \mathbb{E}_{\text{all}}\left[|\langle 0, G^L u \rangle|^s |\langle u', Gv \rangle|^s |\langle v', G^{(L+3)} y \rangle|^s\right]$$
$$\leq \mathbb{E}^{\mathcal{J}^c} \mathbb{E}^{\mathcal{J}} \hat{\mathbb{E}}^{\mathcal{J}} \Bigg[ \left(|\langle 0, \hat{G}^L u \rangle|^s + |\langle 0, \hat{G}^L \hat{D} W_0^L G^{(L)} u \rangle|^s\right) |\langle u', Gv \rangle|^s$$
$$\times \left(|\langle v', \hat{G}^{(L+3)} y \rangle|^s + |\langle v', G^{(L+3)} \hat{D} W_0^{(L+3)} \hat{G}^{(L+3)} y \rangle|^s\right) \Bigg]$$
$$=: A_1 + A_2 + A_3 + A_4$$

which like in [16] has 4 terms $A_1, A_2, A_3, A_4$. We now consider the modified bound for each term.

**A1:** To bound the term we first use Lemma 7.7. Then we use Theorem 2.4 and Lemma 7.6 as well as (7.22) to obtain

$$A_1 := \mathbb{E}\mathbb{E}^{\mathcal{J}}\left[|\langle 0, \hat{G}^L u \rangle|^s |\langle u', Gv \rangle|^s |\langle v', \hat{G}^{(L+3)} y \rangle|^s\right]$$
$$\leq \mathbb{E}^{\mathcal{J}^c}\left[\hat{\mathbb{E}}^{\mathcal{J}}\left[|\langle 0, \hat{G}^L u \rangle|^s |\langle v', \hat{G}^{(L+3)} y \rangle|^s\right] \mathbb{E}^{\mathcal{J}}\left[|\langle u'|G|v \rangle|^s\right]\right]$$
$$\leq C(s) \mathbb{E}^{\mathcal{J}^c} \hat{\mathbb{E}}^{\mathcal{J}}\left[|\langle 0, \hat{G}^L u \rangle|^s |\langle v', \hat{G}^{(L+3)} y \rangle|^s\right]$$
$$\leq C(s) \| \frac{1}{f_L} \|_\infty \mathbb{E}\left[|\langle 0, G^L u \rangle|^s\right] \mathbb{E}\left[|\langle v', G^{(L+3)} y \rangle|^s\right].$$

**A4:** Then the bound for the $A_4$ term:

$$A_4 := \mathbb{E}_{\text{all}}\left[|\langle 0, \hat{G}^L \hat{D} W_0^{\Lambda_L} G^{(L)} u\rangle|^s |\langle u', Gv\rangle|^s |\langle v', G^{(L+3)} \hat{D} W_0 \hat{G}^{(L+3)} y\rangle|^s\right]$$

$$\leq \mathbb{E}^{\mathcal{J}^c} \hat{\mathbb{E}}^{\mathcal{J}} \mathbb{E}^{\mathcal{J}}\left[|\langle 0, \hat{G}^L \hat{D} W_0^{\Lambda_L} G^{(L)} u\rangle|^s |\langle u', Gv\rangle|^s |\langle v', G^{(L+3)} \hat{D} W_0 \hat{G}^{(L+3)} y\rangle|^s\right]$$

$$\leq \mathbb{E}^{\mathcal{J}^c} \hat{\mathbb{E}}^{\mathcal{J}}\left[\mathbb{E}^{\mathcal{J}}\left[|\langle 0, \hat{G}^L \hat{D} W_0^{\Lambda_L} G^{(L)} u\rangle|^{3s}\right]^{\frac{1}{3}} \mathbb{E}^{\mathcal{J}}\left[|\langle u', Gv\rangle|^{3s}\right]^{\frac{1}{3}} \mathbb{E}^{\mathcal{J}}\left[|\langle v', G^{(L+3)} \hat{D} W_0 \hat{G}^{(L+3)} y\rangle|^{3s}\right]^{\frac{1}{3}}\right]$$

$$\leq \mathbb{E}^{\mathcal{J}^c} \hat{\mathbb{E}}^{\mathcal{J}}\left[\mathbb{E}^{\mathcal{J}}\left[|\langle 0, \hat{G}^L \hat{D} W_0^{\Lambda_L} G^{(L)} u\rangle|^{3s}\right]^{\frac{1}{3}} C(3s)^{\frac{1}{3}} \mathbb{E}^{\mathcal{J}}\left[|\langle v', G^{(L+3)} \hat{D} W_0 \hat{G}^{(L+3)} y\rangle|^{3s}\right]^{\frac{1}{3}}\right]$$

where we used Hölder on the integrals over the old variables first and use that for $s < \frac{1}{3}$ we can bound the middle term by Theorem 2.4. This leaves us with the two factors $|\langle 0, \hat{G}^L \hat{D} W_0^{\Lambda_L} G^{(L)} u\rangle|^{3s}$ as well as $|\langle v', G^{(L+3)} \hat{D} W_0 \hat{G}^{(L+3)} y\rangle|^{3s}$. Those we can estimate as in [16] with an extra factor of $C(s)\|\frac{1}{f_L}\|_\infty$ when we have to split up the expectations. But this term is constant so we have the same estimates.

Let us do the first factor explicitly as in [16, (13.30)].

$$|\langle 0, \hat{G}^L \hat{D} W_0^{\Lambda_L} G^{(L)} u\rangle|^{3s} \leq \sum_{n \in \mathcal{J} \cap \Lambda_L} |\langle 0, \hat{G}^L n\rangle| \, \langle n, W_0^{\Lambda_L} G^{(L)} u\rangle|^{3s}.$$

Now, it holds that

$$|\langle n, W_0^{\Lambda_L} G^{(L)} u\rangle| = |\langle n, DS(DS - z)^{-1} u\rangle| = |\langle n, ((DS - z) + z)(DS - z)^{-1} u\rangle|$$
$$= |\langle n, u\rangle \, + z\langle n, (DS - z)^{-1} u\rangle| \leq 1 + |z| \, |\langle n, G^L u\rangle|.$$

Thus, using that $|z| \leq 2$ by Theorem 2.4 again,

$$\mathbb{E}^{\mathcal{J}}\left[|\langle 0, \hat{G}^L \hat{D} W_0^{\Lambda_L} G^{(L)} u\rangle|^{3s}\right] \leq \sum_{n \in \mathcal{J} \cap \Lambda_L} |\langle 0, \hat{G}^L n\rangle|^{3s} \mathbb{E}^{\mathcal{J}}\left[\, 1 + |z||\langle n, \, G^L u\rangle|^{3s}\right]$$
$$\leq C(s) \sum_{n \in \mathcal{J} \cap \Lambda_L} |\langle 0, \hat{G}^L n\rangle|^{3s}.$$

Similarly, (and similarly as in [16]) we obtain that

$$\mathbb{E}^{\mathcal{J}}\left[|\langle v', G^{(L+3)} \hat{D} W_0 \hat{G}^{(L+3)} y\rangle|^{3s}\right] \leq \sum_{n \in \mathcal{J} \cap \Lambda_{L+3}^c} |\langle n, \hat{G}^{(L+3)} y\rangle|^{3s}.$$

Now, using that $\mathcal{J}$ has a fixed number of elements there is a constant $C > 0$ such that $\left(\sum_{n \in \mathcal{J}} \, x_n^{3s}\right)^{\frac{1}{3}} \leq C \sum_{n \in \mathcal{J}} \, x_n^s$ for positive reals $x_n$. It follows that

$$A_4 \leq C \sum_{n \in \mathcal{J} \cap \Lambda_L, n' \in \mathcal{J} \cap \Lambda_{L+3}^c} \mathbb{E}^{\mathcal{J}^c} \hat{\mathbb{E}}^{\mathcal{J}}\left[|\langle 0, \hat{G}^L n\rangle|^s |\langle n', \hat{G}^{(L+3)} y\rangle|^s\right]$$
$$= C \sum_{n \in \mathcal{J} \cap \Lambda_L, n' \in \mathcal{J} \cap \Lambda_{L+3}^c} \mathbb{E}\left[|\langle 0, G^L n\rangle|^s |\langle n', G^{(L+3)} y\rangle|^s\right]$$
$$\leq C\|\frac{1}{f_L}\|_\infty \sum_{n \in \mathcal{J} \cap \Lambda_L} \mathbb{E}\left[|\langle 0, G^L n\rangle|^s\right] \sum_{n' \in \mathcal{J} \cap \Lambda_{L+3}^c} \mathbb{E}\left[|\langle n', G^{(L+3)} y\rangle|^s\right].$$

Now, for the terms $A_2$ and $A_3$ we can use exactly the same methods as for the $A_4$ term by replacing the Hölder with a Cauchy-Schwarz since there is now only two terms. Then we do the analysis as in the $A_4$-part, but now only for one of the factors each time obtaining the same split up of the expectations with the inner and outer part. In total, we get

$$\mathbb{E}[|\langle 0, Gy\rangle|^s] \leq C(s)^2 \|\frac{1}{f_L}\|_\infty c t^{2s} \sum_{u, L-2\leq |u|\leq L+2} \mathbb{E}\left[|\langle 0, G^L u\rangle|^s\right] \sum_{v', L+2\leq |v'|\leq L+4} \mathbb{E}\left[|\langle v', G^{(L+3)}y\rangle|^s\right].$$

$\square$

Next we bound the resolvent $G^{(L+3)}$ in terms of the full resolvent similarly to [16, Proposition 13.2].

**Proposition 7.9.** *For every $s < \frac{1}{3}$ we have that*

$$\mathbb{E}[|\langle 0, Gy\rangle|^s] \leq C(s)t^{2s}(1 + t^s L) \sum_{u, L-2\leq |u|\leq L+2} \mathbb{E}[|\langle 0, G^L u\rangle|^s] \sum_{x', L+2\leq |x'|\leq L+4} \mathbb{E}[|\langle x', Gy\rangle|^s].$$

*Proof.* We do another resampling argument. First, we use the resolvent equation to write

$$G^{(L+3)} = G + G^{(L+3)}\ T^{L+3}\ G.$$

which in turn implies that

$$\mathbb{E}[|\langle v', G^{(L+3)}y\rangle|^s] \leq \mathbb{E}\ [|\langle v', Gy\rangle|^s] + C \sum_{(w,w')\in\partial\Lambda_{L+3}} \mathbb{E}\ \left[|\langle v', G^{(L+3)}w\rangle|^s|\langle w', Gy\rangle|^s\right]. \quad (7.23)$$

Now, we focus on the term $\mathbb{E}\ \left[|\langle v', G^{(L+3)}w\rangle|^s|\langle w', Gy\rangle|^s\right]$ and where we can resample the fields corresponding to changing the phases at $v', w$ and $w'$ in a similar way as above. Following [16] we denote the resampled phases with $\tilde{\mathbb{E}}$ instead of $\hat{\mathbb{E}}$ and the corresponding expectation with respect to $\tilde{\mu}$ as $\tilde{\mathbb{E}}_{\text{all}}$. Now,

$$\mathbb{E}\left[|\langle v', G^{(L+3)}w\rangle|^s|\langle w', Gy\rangle|^s\right] = \tilde{\mathbb{E}}_{\text{all}}\left[|\langle v', G^{(L+3)}w\rangle|^s|\langle w', \tilde{G}y\rangle|^s\right]$$
$$+ \tilde{\mathbb{E}}_{\text{all}}\left[|\langle v', G^{(L+3)}w\rangle|^s|\langle w', G\tilde{D}S\tilde{G}y\rangle|^s\right],$$

where $\tilde{D}$ is defined analogously as $\hat{D}$ above. Let us start bounding the first term. By Lemma 7.7 and Theorem 2.4 on the first term

$$\tilde{\mathbb{E}}_{\text{all}}\left[|\langle v', G^{(L+3)}w\rangle|^s|\langle w', \tilde{G}y\rangle|^s\right] \leq \mathbb{E}^{\mathcal{J}^c}\left[\mathbb{E}^{\mathcal{J}}\left[|\langle v', G^{(L+3)}w\rangle|^s\right]\tilde{\mathbb{E}}^{\mathcal{J}}\left(|\langle w', \tilde{G}y\rangle|^s\right)\right]$$
$$\leq C(s)\ \mathbb{E}^{\mathcal{J}^c}\tilde{\mathbb{E}}^{\mathcal{J}}\left[|\langle w', \tilde{G}y\rangle|^s\right]$$
$$= C(s)\mathbb{E}[|\langle w', Gy\rangle|^s]$$

$$(7.24)$$

where we also used (7.22). Then, let us turn to the second term. First, we can again use a resolvent equation to write

$$|\langle w', G\tilde{D}S\tilde{G}y\rangle|^{2s} \leq C\sum_{l\in\mathcal{J}} |\langle w', Gl\rangle|^{2s} \, |\langle l, D^{-1}DS\tilde{G}y\rangle|^{2s} \leq C\sum_{l\in\mathcal{J}} |\langle w', Gl\rangle|^{2s} \, |\langle l, 1+z\tilde{G}y\rangle|^{2s}$$

$$\leq C\sum_{l\in\mathcal{J}} |\langle w', Gl\rangle|^{2s} \, |\langle l, \tilde{G}y\rangle|^{2s}.$$

So that we now obtain

$$\tilde{\mathbb{E}}_{\text{all}}\left[|\langle v', G^{(L+3)}w\rangle|^s|\langle w', G\tilde{D}S\tilde{G}y\rangle|^s\right] \leq \mathbb{E}^{\mathcal{J}^c}\tilde{\mathbb{E}}^{\mathcal{J}}\mathbb{E}^{\mathcal{J}}\left[|\langle v', G^{(L+3)}w\rangle|^s|\langle w', G\tilde{D}S\tilde{G}y\rangle|^s\right]$$

$$\leq \mathbb{E}^{\mathcal{J}^c}\tilde{\mathbb{E}}^{\mathcal{J}}\left[\mathbb{E}^{\mathcal{J}}\left[|\langle v', G^{(L+3)}w\rangle|^{2s}\right]^{\frac{1}{2}}\mathbb{E}^{\mathcal{J}}\left[|\langle w', G\tilde{D}S\tilde{G}y\rangle|^{2s}\right]^{\frac{1}{2}}\right]$$

$$\leq C(s)\,\mathbb{E}^{\mathcal{J}^c}\tilde{\mathbb{E}}^{\mathcal{J}}\left[\mathbb{E}^{\mathcal{J}}\left[|\langle w', G\tilde{D}S\tilde{G}y\rangle|^{2s}\right)^{\frac{1}{2}}\right]$$

$$\leq C(s)\sum_{l\in\mathcal{J}}\ \mathbb{E}^{\mathcal{J}^c}\tilde{\mathbb{E}}^{\mathcal{J}}\left[\mathbb{E}^{\mathcal{J}}\left[\ |\langle w', Gl\rangle|^{2s}\ |\langle l, \tilde{G}y\rangle|^{2s}\right]^{\frac{1}{2}}\right]$$

$$\leq C(s)\sum_{l\in\mathcal{J}}\ \mathbb{E}^{\mathcal{J}^c}\tilde{\mathbb{E}}^{\mathcal{J}}\left[\ |\langle l, \tilde{G}y\rangle|^s\mathbb{E}^{\mathcal{J}}\left[\ |\langle w', Gl\rangle|^{2s}\right]^{\frac{1}{2}}\right]$$

$$\leq C(s)\sum_{l\in\mathcal{J}}\ \mathbb{E}^{\mathcal{J}^c}\tilde{\mathbb{E}}^{\mathcal{J}}\left[\ |\langle l, \tilde{G}y\rangle|^s\right] = C(s)\sum_{l\in\mathcal{J}}\mathbb{E}\left[|\langle l, Gy\rangle|^s\right]. \tag{7.25}$$

Now, (7.24) and (7.25) yields that

$$\mathbb{E}\left[|\langle v', G^{(L+3)}w\rangle|^s|\langle w', Gy\rangle|^s\right] \leq C(s)\mathbb{E}\left[|\langle w', Gy\rangle|^s\right] + C(s)\sum_{l\in\mathcal{J}}\mathbb{E}\left[|\langle l, Gy\rangle|^s\right].$$

Combining this result with (7.23) and Proposition 7.8 yields the proof. $\qquad\square$

## 7.4 Iteration and exponential decay of fractional moments Greens function.

In this section we check that the iteration from [22, Section 3.4] also works in our case and we prove Theorem 2.3.

**Theorem 2.3.** *Let $W = W_\omega$ be the random quantum walk defined in (2.2). Then there exists $\varepsilon > 0$ such that for all $C_1, C_2$ with $\text{dist}((C_1, C_2), \mathcal{C}_r) < \varepsilon$ such that for all $s \in (0, \frac{1}{3})$, there exist constants $\mu, C > 0$ such that and all $x^\pm, y^\pm \in \mathbb{Z}^2 \times \{+, -\}$ it holds that*

$$\mathbb{E}_\omega\left[|G(x^\pm, y^\pm, z)|^s\right] \leq Ce^{-\mu|x-y|}$$

*for all $z \in \mathbb{C}$ with $\frac{1}{2} < |z| < 2$.*

| | Self-adjoint | Unitary |
|---|---|---|
| 2nd moment | [15] | [16] |
| EigenCorrelators | [2] | [25] |

Table 1: Overview of the relevant literature for proving that a fractional moment estimate of the Greens function implies localization of i.i.d potentials in the self-adjoint case and i.i.d. phase in the unitary case. The two methods are the 2nd-moment estimate method and the eigenfunction correlator method in the self-adjoint and in the unitary case.

*Proof of Theorem 2.3.* We combine Proposition 7.1 and Proposition 7.9 as in [22, Section 3.4]. That is

$$\mathbb{E}[|\langle 0, Gy \rangle|^s] \leq C(s)t^{2s}(1 + t^s L)\frac{L^2}{L^a} \max_{x', L+2 \leq |x'| \leq L+4} \mathbb{E}[|\langle x', Gy \rangle|^s].$$

where we chose $t \leq L^{-2(ap+2)-\frac{a}{s}}$. Now, we can take $a$ large enough so that $tL \to 0$ as $L \to \infty$ and hence the front factor tends to 0. Now, using translation invariance this gives us an iterative proof of exponential decay (cf. [22, Section 3.4]). $\square$

# 8 Relating Greens functions to dynamical localization.

This section contains some considerations about dynamical localization in the fractional moment method and what barriers we see to obtain Conjecture 2.5.

The fractional moment approach to localization entails proving a priori and exponential decay estimates of the expectations for fractional moments of the Green function. In the case of the standard Anderson model, the unitary Anderson model or when the phases $\theta_i$ are i.i.d. different methods exist for relating the estimates to dynamical localization. For the discussion we follow the spirit of [31]. In the self-adjoint case, there is an approach second moment estimates [15] and one using eigenfunction correlators. In the unitary case, the second moment method was "unitarized" in [16] and also used in [21]. On the other hand, the eigenfunction correlator approach of [2] (see also [5, Theorem 7.7]) was "unitarized" in [25].

We have not succeeded on using any of the approaches for the model of a quantum walk in a random magnetic field that we introduce and study in this paper. Generalizing the unitary approach to eigenfunction correlators in [25] seems intractable so we find it natural to try to follow the steps of [16]. However, as we saw in the proofs above substantial difficulties arise since the phases are no longer i.i.d. These difficulties that are centred around the appropriate generalization of [16, Prop. 5.1] seem at the moment difficult to overcome (although admittedly with natural modifications one can get quite far, but the final estimate corresponding to [16, (5.24)] we were not able to obtain). Having obtained an appropriate generalization of [16, Prop. 5.1] we believe that the rest of the proof of dynamical localization (Conjecture 2.5) is downstream.

# 6. Spectra of generators of Markovian evolution in the thermodynamic limit: From non-Hermitian to full evolution via tridiagonal Laurent matrices

# Spectra of generators of Markovian evolution in the thermodynamic limit: From non-Hermitian to full evolution via tridiagonal Laurent matrices

Frederik Ravn Klausen, Albert H. Werner

### Abstract

We determine spectra of single-particle translation-invariant Lindblad operators on the infinite line. In the case where the Hamiltonian is given by the discrete Laplacian and the Lindblad operators are rank $r$, finite range and translates of each other, we obtain a representation of the Lindbladian as a direct integral of finite range bi-infinite Laurent matrices with rank-$r$-perturbations. By analyzing the direct integral we rigorously determine the spectra in the general case and calculate it explicitly for several types of dissipation e.g. dephasing, and coherent hopping. We further use the detailed information about the spectrum to prove gaplessness, absence of residual spectrum and a condition for convergence of finite volume spectra to their infinite volume counterparts. We finally extend the discussion to the case of the Anderson Hamiltonian, which enables us to study a Lindbladian recently associated with localization in open quantum systems.

## 1 Introduction

Schrödinger operators and their spectra are one of the central objects studied in mathematical physics. Indeed, spectral properties encode many important physical properties such as the speed of propagation as described by the RAGE theorem [2, 3, 4].

Going beyond the closed system paradigm described by Schrödinger operators and unitary dynamics, a natural setting is that of Markovian time-evolution, described by a completely positive dynamical semi-group. These systems have been extensively studied from a quantum information perspective in particular by considering the generator of such evolutions, i.e. the Lindblad generator [5]. Here, gaps in the spectrum around the origin in the complex plane provide information about relaxation times towards the non-equilibrium steady state of the system (as we will discuss in Section 2.2). Understanding the interplay between disorder (e.g. in the form of a random potential) and dissipation (e.g. thermal noise) is emerging as an important problem [6, 7, 8, 9]

In addition, over the past decades, the theory of non-Hermitian Hamiltonians (i.e Non-self-adjoint Schrödinger operators) has developed rapidly [10, 11, 12, 13]. One motivation for these investigations has been that non-Hermitian Hamiltonians model open quantum systems if quantum jumps are neglected [14]. The theory of non-Hermitian Hamiltonians is also closely connected to the study of tridiagonal Laurent matrices (see [15, 16, 17] and references therein). In particular, the stability of the spectra under perturbations has been investigated [18].

In this paper, we extend this connection to the case of Markovian evolution in the single particle regime (illustrated in Figure 1). For Lindbladians with translation-invariant Hamiltonians and Lindblad operators $L_k$ that are rank-one, finite range and translates of each other we prove in Theorem 3.8 that the entire Lindbladian can be rewritten as a direct integral of tridiagonal Laurent matrices $T(q)$ corresponding to the non-Hermitian evolution subject to a rank-one perturbation $F(q)$ which corresponds to the quantum jump terms. The construction of $T(q)$ and $F(q)$ is explicit. This decomposition allows us to explore the spectral effects of the quantum jumps rigorously.

The Lindbladian $\mathcal{L}$ is non-normal and so the notion pseudo-spectrum provides information about the operator in not encoded in the spectrum as discussed in great detail in [19, 10]. To determine the spectrum of the Lindbladian from the direct integral decomposition we extend the use of pseudo-spectra by providing a result of independent interest concerning the spectrum of a direct integral in terms of the pseudo-spectra of its fibers and thereby generalizing the corresponding result for the direct sum [20]. This is the content of Theorem 3.12.

The combination of the direct integral decomposition and Theorem 3.12 enables us to obtain information about the spectrum of the Lindbladian $\mathcal{L}$.

In Section 4 we discuss some abstract consequences of the direct integral decomposition. First, we use the decomposition to prove that Lindbladians in the class we consider only have approximate point spectrum. Second, we discuss convergence of finite volume spectra to their infinite volume counterparts and finally we prove that the Lindbladians in question are always gapless or have an infinite dimensional kernel.

In Section 5, we then use the direct integral decomposition more concretely to completely determine the spectrum of some Lindbladians which have received attention in the physics literature as the one-particle sector of open spin chains [21, 22, 23] thereby complementing the exact results on the spectrum from [24]. We are particularly interested in an example where the dissipators are non-normal and where the system shows signs of localization in an open quantum system [25]. Our rigorous analytic results also complement the large body of very recent work on random Lindblad operators studied from a random matrix theory point of view [26, 27, 28, 29].

Finally, to connect to examples to the open quantum system with disorder, in Theorem 6.1 we prove a Lindbladian analogue of the Kunz-Soulliard theorem from the theory of random operators.

In contrast to many previous investigations, we work directly on the entire lattice $\mathbb{Z}$. That allows us to utilise translation-invariance, Laurent matrices and some tools from random operator theory that do not work for finite systems. Furthermore, working in infinite volume directly allows us to determine closed formulas for the spectra explicitly. From one point of view, one can look at these formulas as approximations to (some) large finite-volume systems. In Theorem 4.5 we give conditions that ensure this convergence.

With some exceptions, [30, 31, 32, 7, 9], the study of Lindbladian evolutions has, in recent decades, focused on finite spin systems. We therefore first present some results with conditions for the Lindbladian $\mathcal{L}$ to be bounded as an operator on respectively the space of bounded, Hilbert-Schmidt and trace-class operators.

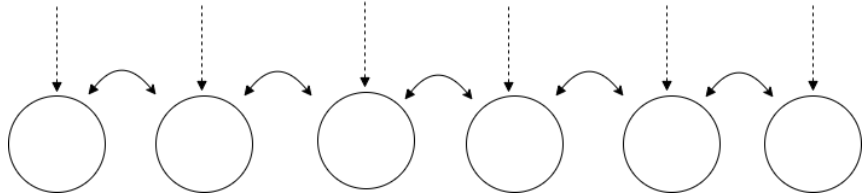

Figure 1: An example of the systems that we discuss in the following. We have a single particle on a lattice with hopping between lattice sites and local dissipation. In this example, the hopping and dissipation have the shortest possible range, but the methods presented apply as long as these properties stay local.

## 2    Lindblad systems on the infinite lattice

We consider the Markovian, open-system dynamics of a single quantum particle on the one-dimensional lattice described by the Hilbert space $\mathcal{H} = \ell^2(\mathbb{Z})$ and a distinguished (position) basis $\{|k\rangle\}_{k \in \mathbb{Z}}$.

The underlying completely positive dynamical semi-group is generated by a *Lindbladian* $\mathcal{L}$ : $\mathcal{B}(\mathcal{H}) \to \mathcal{B}(\mathcal{H})$ of the form [33, 5]

$$\mathcal{L}(\rho) = -i[H, \rho] + G \sum_k L_k \rho L_k^* - \frac{1}{2}(L_k^* L_k \rho + \rho L_k^* L_k), \tag{1}$$

where $H$ is the Hamiltonian of the system (a self-adjoint bounded operator) and $G > 0$ is the coupling constant of the dissipation and we have used $A^*$ to denote the adjoint of an operator $A$. We will refer to the second term as the dissipative part. The operators $L_k \in \mathcal{B}(\mathcal{H})$ implementing the dissipative part are referred to as Lindblad operators. The semi-group determines the time-evolution of a state $\rho(t)$ at time $t \geq 0$, which is given by

$$\rho(t) = e^{t\mathcal{L}}(\rho_0),$$

where $\rho(0) = \rho_0$ is the state at time $t = 0$.

Later, we will often choose our $L_k$ to act locally as for example $L_k = |k\rangle\langle k|$. In the following, we will refer to (1) as the Lindblad form. We will similarly say that an operator $\tilde{\mathcal{L}}$ is in the adjoint Lindblad form if

$$\tilde{\mathcal{L}}(X) = i[H, X] + G \sum_k L_k^* X L_k - \frac{1}{2}(L_k^* L_k X + X L_k^* L_k), \tag{2}$$

which describes the evolution of observables in the Heisenberg picture, whereas the evolution generated on states according to (1) is refered to as the Schrödinger picture. The case of a Markovian evolution on the infinite line is slightly under-represented in the literature as many works, in particular from the quantum information side, where open-system dynamics has been studied extensively, restrict themselves to finite dimensions. The infinite case adds some additional complications that we clarify without using the lattice structure of $\ell^2(\mathbb{Z})$.

Oftentimes, we will use a decomposition of $\mathcal{L}$ in terms of a non-Hermitian evolution part $\mathcal{L}_{\mathrm{NHE}}$ and a quantum jump term $\mathcal{L}_J$ according to

$$\mathcal{L}(\rho) = -i(H_{\mathrm{eff}}\,\rho - \rho H_{\mathrm{eff}}^*) + G\sum_k L_k^* \rho L_k = \mathcal{L}_{\mathrm{NHE}}(\rho) + \mathcal{L}_J(\rho).$$

where $\mathcal{L}_J(\rho) = G\sum_k L_k^* \rho L_k$ and the effective non-Hermitian Hamiltonian is defined by

$$H_{\mathrm{eff}}\ = H - \frac{iG}{2}\ \sum_k L_k^* L_k. \tag{3}$$

The Lindblad form ensures that the spectrum is always contained within the half-plane of the complex plane with non-positive real part (see also [34, 35]).

Furthermore, for finite-dimensional systems $\mathcal{L}(X^*) = \mathcal{L}(X)^*$ which implies that the spectrum is invariant under complex conjugation $\sigma(\mathcal{L}) = \overline{\sigma(\mathcal{L})}$. This we prove also in the infinite-dimensional case in Lemma A.4.

In finite dimensions, it is always the case that there is a steady state $\rho_\infty$ of the dynamics that satisfies $\mathcal{L}(\rho_\infty) = 0$ (see for example [35, Proposition 5]). It was discussed in [36, 37, 38, 39] how the symmetries of $L_k$ are inherited by $\mathcal{L}$ and the steady state. However, translation invariant infinite volume Lindbladians have not been studied extensively although some results exist [40, 7].

## 2.1 Boundedness of $\mathcal{L}$ as an operator on Schatten spaces

In finite dimensions, the set of density matrices is defined to be the set of positive matrices with unit trace. In infinite dimensions, this notion is generalised to positive trace-class operators with unit trace. In the following, we will denote the space of trace-class operators by $\mathrm{TC}(\mathcal{H})$, the Hilbert-Schmidt operators by $\mathrm{HS}(\mathcal{H})$ and the space of bounded operators by $\mathcal{B}(\mathcal{H})$. See for example [10, 41] for more details on these spaces. All three spaces are Banach spaces with regards to their respective norms and $\mathrm{HS}(\mathcal{H})$ is also a Hilbert space with the inner product

$$\langle X, Y\rangle = \mathrm{Tr}(X^*Y).$$

In order to relate the spectra of $\mathcal{L}$ on these different Banach spaces, we will use interpolation methods. These methods rely on the Schatten classes that interpolate between $\mathrm{TC}(\mathcal{H}), \mathrm{HS}(\mathcal{H})$ and $\mathcal{B}(\mathcal{H})$. To define them we consider operators $A \in \mathcal{B}(\mathcal{H})$ with and define for $p \in (1, \infty)$ the Schatten-$p$-norm of $A$ via

$$\|A\|_p = \mathrm{Tr}(|A|^p)^{\frac{1}{p}},$$

where $|A| = (AA^*)^{\frac{1}{2}}$. The Schatten-$p$-class $\mathcal{S}_p$ then consists of all bounded operators with finite $p$-norm. For $p = \infty$ we set $\mathcal{S}_\infty = \mathcal{K}(\mathcal{H})$, the compact operators on $\mathcal{H}$. Furthermore, the Schatten classes interpolate between these spaces in the sense that $\mathcal{S}_1 = \mathrm{TC}(\mathcal{H}), \mathcal{S}_2 = \mathrm{HS}(\mathcal{H})$. For more information see [42] where the Schatten classes are treated extensively.

We first show that under a boundedness assumption on the $L_k$ any Lindblad generator of the form (1) will be a bounded operator on $\mathrm{TC}(\mathcal{H}), \mathrm{HS}(\mathcal{H})$ as well as $\mathcal{B}(\mathcal{H})$. To this end, assume that $H \in \mathcal{B}(\mathcal{H})$ and that

$$(\mathcal{A}_1): \text{ Both } \left\{\sum_{k \in \mathbb{Z}: |k| \leq n} L_k L_k^*\right\}_{n \in \mathbb{N}} \text{ and } \left\{\sum_{k \in \mathbb{Z}: |k| \leq n} L_k^* L_k\right\}_{n \in \mathbb{N}} \text{ converge weakly in } \mathcal{B}(\mathcal{H}),$$

where we recall a sequence of operators $\{A_n\}_{n \in \mathbb{N}} \subset \mathcal{B}(\mathcal{H})$ converges weakly to an operator $A \in \mathcal{B}(\mathcal{H})$ if for all $v \in \mathcal{H}$ it holds that $\|A_n v - A v\| \to 0$.

A similar level of generality was used in [43] and [44]. We will use the Riesz-Thorin interpolation theorem in a non-commutative version, where the operators are defined on the Schatten classes $\mathcal{S}_p$. We state it here for convenience.

We say that a pair of operators $A_p \in \mathcal{B}(\mathcal{S}_p)$ and $A_q \in \mathcal{B}(\mathcal{S}_q)$ are defined *consistently* if for any $\rho \in \mathcal{S}_p \cap \mathcal{S}_q$ it holds that $A_p \rho = A_q \rho$. If this is the case, we abuse notation by writing $A \in \mathcal{B}(\mathcal{S}_p), \mathcal{B}(\mathcal{S}_q)$. Notice that since the Lindbladian is defined through the Lindblad form (1) for some fixed operators $H, L_k$. Then $\mathcal{L}$ is consistently defined and we abuse notation by writing both $\mathcal{L} \in \mathcal{B}(\mathcal{S}_p)$ and $\mathcal{L} \in \mathcal{B}(\mathcal{S}_q)$, whenever it is the case.

**Theorem 2.1** ([45, Section IX.4]). *Let $p, q \geq 1$ and $A \in \mathcal{B}(\mathcal{S}_p), \mathcal{B}(\mathcal{S}_q)$ consistently, then $A \in \mathcal{B}(\mathcal{S}_{r_t})$ where $\frac{1}{r_t} = \frac{t}{p} + \frac{1-t}{q}$ for each $t \in [0, 1]$ with*

$$\|A\|_{r_t \to r_t} \leq \|A\|_{p \to p}^t \|A\|_{q \to q}^{1-t}.$$

The theorem enables us to prove that Assumption $(\mathcal{A}_1)$ is enough to ensure boundedness on all Schatten spaces, but we state the more relevant ones here for clarity.

**Lemma 2.2.** *Suppose that $\mathcal{L}$ is of the Lindblad form (1) and that $(\mathcal{A}_1)$ holds. Then*

$$\mathcal{L} \in \mathcal{B}(\mathcal{B}(\mathcal{H})), \mathcal{B}(\mathrm{HS}(\mathcal{H})), \mathcal{B}(\mathrm{TC}(\mathcal{H})).$$

*This is also true if $\mathcal{L}$ is of the adjoint Lindblad form (2).*

*Proof.* The case $\mathcal{L} \in \mathcal{B}(\mathrm{TC}(\mathcal{H}))$ follows from [44, prop 6.4]. In Appendix A.2 we prove that $\mathcal{L} \in \mathcal{B}(\mathcal{B}(\mathcal{H}))$. Then, to see that $\mathcal{L} \in \mathcal{B}(\mathrm{HS}(\mathcal{H}))$ we use the non-commutative Riesz-Thorin theorem. Note that since the operator $\mathcal{L}$ is bounded on $\mathcal{B}(\mathcal{H})$ it is also bounded as an operator on the compact operators $\mathcal{K}(\mathcal{H}) = \mathcal{S}_\infty$. Thus, we obtain that

$$\|\mathcal{L}\|_{2 \to 2} \leq \|\mathcal{L}\|_{1 \to 1}^{\frac{1}{2}} \|\mathcal{L}\|_{\infty \to \infty}^{\frac{1}{2}}.$$

This shows that $\mathcal{L} \in \mathcal{B}(\mathrm{HS}(\mathcal{H}))$. For further discussions and similar results see also [46, 47]. The result for the adjoint Lindblad form follows in the same way because of the assumption $(\mathcal{A}_1)$. □

We have now proven that the Lindbladian $\mathcal{L}$ is an element of the Banach algebras $\mathcal{B}(\mathrm{TC}(\mathcal{H}))$, $\mathcal{B}(\mathrm{HS}(\mathcal{H}))$ and $\mathcal{B}(\mathcal{B}(\mathcal{H}))$, where $\mathcal{B}(\mathrm{HS}(\mathcal{H}))$ is also a $C^*$-algebra.

In the main text we will only consider the case $\mathcal{B}(\mathrm{HS}(\mathcal{H}))$. In particular, we only determine the spectrum exactly there. But in Appendix A.1 we make some remarks on spectral independence of Lindblad operators.

In the following, we will be concerned with the spectrum of $A$ in each of these algebras. For any Banach algebra $\mathcal{A}$ the spectrum of an operator $A$ with respect to the algebra $\mathcal{A}$ is defined as follows

$$\sigma_{\mathcal{A}}(A) = \{\lambda \in \mathbb{C} \mid A - \lambda \text{ is not invertible in } \mathcal{A}\}.$$

Furthermore, we define the approximate point spectrum of an operator in a Banach algebra $\mathcal{A} = \mathcal{B}(X)$ for some Banach space $(X, \|\cdot\|_X)$, is given by

$$\sigma_{\mathcal{A},\mathrm{appt}}(A) = \{\lambda \in \mathbb{C} \mid \exists (\psi_n)_{n \in \mathbb{N}} \subset X, \|\psi_n\|_X = 1, \lim_{n \to \infty} \| (A - \lambda)\psi_n\|_X = 0\}.$$

A sequence $(\psi_n)_{n \in \mathbb{N}}$ corresponding to a point $\lambda \in \mathbb{C}$ as above we will call a *Weyl sequence* corresponding to $\lambda$. It is always the case that $\sigma_{\mathcal{A},\mathrm{appt}}(A) \subset \sigma_{\mathcal{A}}(A)$ and for normal operators equality holds [41, 12.11]. We will prove in Theorem 4.1 that the equality also holds in many of our cases of interest. The set

$$\sigma_{\mathcal{A},\mathrm{res}}(A) = \sigma_{\mathcal{A}}(A) \backslash \sigma_{\mathcal{A},\mathrm{appt}}(A)$$

is called the residual spectrum of $A$.

In the following, we will be particularly interested in the case where the Banach algebra $\mathcal{A} = \mathcal{B}(\mathcal{S}_p)$. It is a classical result that $\mathcal{S}_p$ (and in turn $\mathcal{A}$) is Banach algebra in itself, and that $\mathcal{S}_p$ is an ideal in $\mathcal{B}(\mathcal{H})$ for all $p \in [1, \infty]$. The Calkin algebra $\mathcal{Q}(\mathcal{H})$ is defined by $\mathcal{Q}(\mathcal{H}) = \mathcal{B}(\mathcal{H}) \backslash \mathcal{K}(\mathcal{H})$ and the spectrum of an operator $A$ in the Calkin algebra we call the essential spectrum $\sigma_{\mathrm{ess}}(A) = \sigma_{\mathcal{Q}(\mathcal{H})}(A)$.

In the following, we are mainly concerned with translation-invariant operators with finite range. We formalize this by the following two assumptions.

($\mathcal{A}_2 a$):    *Finite range $r < \infty$*: $\langle y|, L_k|x\rangle = 0$ whenever $\max\{|x - k|, |y - k|\} > r$

($\mathcal{A}_2 b$):    *Translation-invariance*: $S_1 L_k S_{-1} = L_{k+1}$ for each $k \in \mathbb{Z}$

where the operator $S_n$ on $\mathcal{H}$ is defined by $(S_n \psi)(x) = \psi(x - n)$ with the convention that $S = S_1$. Notice that since $\mathcal{H} = \ell^2(\mathbb{Z})$ the operator $S_n$ is unitary and $S_n^{-1} = S_n^* = S_{-n}$.

If we further assume that the Hamiltonian $H$ is translation-invariant meaning that $H = S_1 H S_1^*$, then Assumption ($\mathcal{A}_2 b$) implies that $\mathcal{L}$ is translation-*covariant*, i.e. it satisfies that

$$\mathcal{L}(S_1 \rho S_1^*) = S_1 \ \mathcal{L}(\rho) S_1^*. \tag{4}$$

It is easy to see that (4) implies that $\mathcal{L}(S_n \rho S_n^*) = S_n \ \mathcal{L}(\rho) S_n^*$ for all $n \in \mathbb{Z}$. Furthermore, it is also true that ($\mathcal{A}_2 a$) and ($\mathcal{A}_2 b$) imply ($\mathcal{A}_1$). To see this, notice that ($\mathcal{A}_2 b$) implies that $\sum_k L_k L_k^*$ is translation-invariant and therefore constant on diagonals. Now, ($\mathcal{A}_2 a$) implies that there are only finitely many non-zero diagonals and the diagonal entries are finite. Thus, $\sum_k L_k L_k^* = \sum_{n=-r}^{r} \alpha_n S_n$ for $\alpha_n \in \mathbb{C}$ for $-r \le n \le r$, which is a bounded operator.

## 2.2 Relation between spectra and dynamics for Lindblad systems

In the rest of the paper, our focus will be on determining the spectra of certain infinite-volume open quantum systems. In the Hamiltonian case, there is a clear dynamical interpretation of the spectra and the different types of spectra. However, due to non-normality of Lindblad operators, the dynamical implications of the spectra are more subtle and the details of the topic are still under discussion in the physics literature [49]. We discuss our knowledge in both finite and infinite dimensional cases.

*Finite dimensions:* In the finite-dimensional, case the relationship between eigenvalues of $\mathcal{L}$ and the time evolution is given through the Jordan normal form. I.e. $\mathcal{L} = S\Lambda S^{-1}$ where $S$ is invertible and $\Lambda$ is of a certain almost-diagonal form. However, even in the cases where $\Lambda$ is diagonal, $\mathcal{L}$ is not necessarily normal. The analysis is counter-intuitive to the person trained in Hamiltonian formalism due to the peculiarities of non-normality. In particular, the mathematical guarantees for convergence of the semigroup are much much weaker in the normal case and they scale much worse with the dimension of the system (see e.g. [50]).

Another peculiarity is the fact that all eigenvectors of $\mathcal{L}$ are traceless, which we describe in the following remark.

**Remark 2.3.** *All eigenvectors of $\mathcal{L}$ with eigenvalues not equal to 0 are traceless. To see that, note that $e^{t\mathcal{L}}$ is trace-preserving for all $t \in [0, \infty)$, so it holds that*

$$\mathrm{Tr}(\rho) = \mathrm{Tr}\left(e^{t\mathcal{L}}(\rho)\right) = e^{\lambda t}\,\mathrm{Tr}(\rho).$$

*Thus, if $\mathrm{Tr}(\rho) \neq 0$ then for all $t \in [0, \infty)$ we get $1 = e^{\lambda t}$ which implies that $\lambda = 0$.*

We can give guarantees about the dynamics in terms of the spectral gap $g$ of $\mathcal{L}$ which we define as follows

$$g = \sup\left\{\mathrm{Re}(\lambda) \mid \lambda \in \sigma(\mathcal{L})\backslash\{0\}\right\}.$$

In the case where we have a unique steady state $\rho_\infty$, we can get a dynamical guarantee for the speed of decay towards the steady state in terms of the gap $g$. Namely that $\left\|e^{t\mathcal{L}}(\rho) - \rho_\infty\right\| \leq Ce^{tg}$ where $C > 0$ is a constant that depends heavily on the size of the Jordan blocks of the systems.

In the literature, the cases where $\mathcal{L}$ is not diagonalizable are called exceptional points, there is evidence that these points can also lead to faster decay towards the steady state [51], although the mathematical guarantee gets worse.

*Infinite dimensions:* In infinite dimensions the relationship between spectra and dynamics can break down due to Jordan blocks of unbounded size (and more generally the breakdown of the Jordan normal form), due to the lack of a trace class steady state (a phenomenon that we will encounter in most examples in Section 5) and due to the lack of a spectral gap (which we prove for our models in Theorem 4.10).

However, we will encounter situations where $\sigma(\mathcal{L})$ has two or more disconnected parts. Suppose for simplicity that we just have two parts $\sigma(\mathcal{L}) = \Sigma_A \dot{\cup} \Sigma_B$, with $\sup\{\mathrm{Re}(z) \mid z \in$ $\Sigma_A\} \ \leq g$ for some gap $g \in (-\infty, 0)$ and such that there exists is a closed continuous curve encircling only $\Sigma_A$. Then we can define the Riesz projections by contour integration to get a decomposition of $\mathcal{H} = \mathcal{H}_A \oplus \mathcal{H}_B$ for two orthogonal subspaces $\mathcal{H}_A, \mathcal{H}_B$ such that $\mathcal{L}$ leave each of the two subspaces invariant. Thus, we can decompose $\mathcal{L} = \mathcal{L}_A \oplus \mathcal{L}_B$. Suppose that $\rho \in \mathcal{H}_A$ then, since $\mathcal{L}_A$ is the generator of a semigroup by [52], it holds that

$$\left\| e^{t\mathcal{L}}(\rho) \right\| = \ \left\| e^{t\mathcal{L}_A}(\rho) \right\| \leq C e^{tg} \ \|\rho\|$$

for some constant $C > 0$. Thus, if $\rho = \rho_A + \rho_B$ with $\rho_A \in \mathcal{H}_A$ and $\rho_B \in \mathcal{H}_B$ we see that the part $\rho_A$ decays quickly.

We leave it to future work to establish a stronger relationship between spectra and dynamics in the infinite-dimensional case. In particular, one cannot use our work to gain many rigorous guarantees about the evolution of infinite open quantum systems, but we consider the results presented as steps towards such rigorous guarantees. Furthermore, due to the apparent convergence of the spectra of some finite-dimensional Lindbladians (see Theorem 4.5 and the discussion in Section 7) one can also view the method presented here as a way to compute large volume approximations to finite systems (which have discrete spectra and where the relation between spectra and dynamics is clearer).

In fact, given the question on spectral independence raised in the previous section one might even ask which Banach algebra (potentially $\mathcal{B}(\mathcal{S}_p)$ for some $p \in [1, \infty]$) enables us to transfer knowledge from spectra to dynamics of states $\rho \in \mathcal{S}_1, \rho \geq 0$ and $\mathrm{Tr}(\rho) = 1$.

# 3   Direct integral decompositions and their spectra

In this section, we specialize to the case where the Hamiltonian $H \in \mathcal{B}(\ell^2(\mathbb{Z}))$ is the discrete Laplacian $-\Delta$ defined in (41). From now on, we will use Dirac notation, to do that we define for a $k \in \mathbb{Z}$ the vector $|k\rangle = e_k$, and we let $\langle k|$ be the corresponding dual vector. On caveat here is that since we will be working with non-selfadjoint operators, then $\langle k|, A|k'\rangle \neq \langle k|A, |k'\rangle$ and therefore the comma in the inner product is important and we write it consistently.

Since $H$ only enters through into the Lindbladian given by (1) the commutator $[H, \cdot]$ does not change when disregarding the term $-2|k\rangle\langle k|$. Thus, we will often work with

$$H = -\tilde{\Delta} = -\sum_{k \in \mathbb{Z}} |k\rangle\langle k+1| \ + \ |k+1\rangle\langle k| = -(S + S^*), \tag{5}$$

where the operator $S_n$ on $\mathcal{H}$ was defined by $(S_n\psi)(x) = \psi(x - n)$ and we used the convention that $S = S_1$ for the shift operator. For the Lindblad operators $L_k$, we made the following assumptions in the previous section.

$(\mathcal{A}_2 a):$   *Finite range* $r < \infty:$ $\langle y|, L_k|x\rangle = 0$ whenever $\max\{|x - k|, |y - k|\} \ > r$.

$(\mathcal{A}_2 b):$   *Translation-invariance:* $S_n L_k S_{-n} = L_{k+n}$.

Sometimes, we will also need the assumption

$(\mathcal{A}_2 c):$   *rank-one*: $\mathrm{Rank}(L_k) = 1$.

Notice that in particular, we do not assume that each of the $L_k$ is normal or self-adjoint and in fact, one of our motivating examples has non-normal $L_k$. Notice that $H$ is translation-invariant in the sense that $H = S_n H S_n^*$. The assumption $(\mathcal{A}_2 b)$ implies that $\mathcal{L}$ is translation-covariant in the sense of (4).

We will use a sequence of isometric isomorphisms of Hilbert spaces to obtain our main result. Therefore we briefly recall some facts about them. We call a map between separable Hilbert spaces $U : \mathcal{H}_1 \to \mathcal{H}_2$ an *isometric isomorphism* if it is linear, bijective and preserves the inner product $\langle U(x), U(y) \rangle_{\mathcal{H}_2} = \langle x, y \rangle_{\mathcal{H}_1}$. We recall that every isometric isomorphism of separable Hilbert spaces is a unitary operator [53, Theorem 5.21] . Now, $U$ gives rise to an isomorphism $C^*$-algebras $U^\uparrow : \mathcal{B}(\mathcal{H}_1) \to \mathcal{B}(\mathcal{H}_2)$ given by conjugation

$$U^\uparrow(A) = U A U^*. \tag{6}$$

Where we have introduced the notation $U^\uparrow$ since we will compose many isomorphisms the reader should notice that

$$(UV)^\uparrow = U^\uparrow V^\uparrow. \tag{7}$$

## 3.1 Review of Fourier transformations and symbol curves for Laurent operators

We will use the Fourier transformation extensively and therefore we review it before continuing. Following the normalization convention in [1, (A.9)] we let $\mathcal{F} : \ell^2(\mathbb{Z}) \to L^2([0, 2\pi])$ be defined by

$$(\mathcal{F}\psi)(q) = \frac{1}{\sqrt{2\pi}} \sum_{x \in \mathbb{Z}} e^{-iqx} \psi(x),$$

for any $\psi \in \ell^2(\mathbb{Z})$. Then $\mathcal{F}$ is unitary and its adjoint, the inverse Fourier transform, $\mathcal{F}^{-1} : L^2([0, 2\pi]) \to \ell^2(\mathbb{Z})$ is given by

$$(\mathcal{F}^{-1}\varphi)(x) = \frac{1}{\sqrt{2\pi}} \int_0^{2\pi} e^{iqx} \varphi(q) dq, \tag{8}$$

for every $\varphi \in L^2([0, 2\pi])$. We summarise these observations as follows.

**Theorem 3.1.** *The operator $\mathcal{F} : \ell^2(\mathbb{Z}) \to L^2([0, 2\pi])$ is a well defined bounded operator and it is an isometric isomorphism. For the operator $\mathcal{F}^{-1} : L^2([0, 2\pi]) \to \ell^2(\mathbb{Z})$ defined by (8) satisfies $\mathcal{F}^{-1} = \mathcal{F}^*$ as well as*

$$\mathcal{F}\mathcal{F}^{-1} = \mathbb{1}_{L^2([0,2\pi])} \quad and \quad \mathcal{F}^{-1}\mathcal{F} = \mathbb{1}_{\ell^2(\mathbb{Z})}.$$

Furthermore, $\mathcal{F}$ maps the standard basis $\{|j\rangle\}_{j \in \mathbb{Z}}$ to the Fourier basis consisting of functions $\{\frac{1}{\sqrt{2\pi}} e^{-ikj}\}_{j \in \mathbb{Z}}$. It is standard that if $T$ is a translation invariant operator on $\ell^2(\mathbb{Z})$ (defined as $STS^* = T$). Then the corresponding operator $\mathcal{F}T\mathcal{F}^{-1} \in \mathcal{B}(L^2([0, 2\pi]))$ will be a multiplication operator.

The Fourier transform transforms Laurent operators into multiplication operators, so we briefly recall the definition of a Laurent operator. A *Laurent operator* $T \in \mathcal{B}(\ell^2(\mathbb{Z}))$ is an operator if for all $i, j, n \in \mathbb{Z}$ it holds $\langle i|, T|j \rangle = \langle i + n|, L|j + n \rangle$, thus, in position basis, we can write $T$ as a bi-infinite matrix that is constant on the diagonals.

$$T = \begin{pmatrix} \dots & \dots & \dots & \dots & \dots \\ \dots & a_0 & a_1 & a_2 & \dots \\ \dots & a_{-1} & a_0 & a_1 & \dots \\ \dots & a_{-2} & a_{-1} & a_0 & \dots \\ \dots & \dots & \dots & \dots & \dots \end{pmatrix}.$$

In other words, $T$ is a banded matrix. If $a_n = 0$, whenever $|n| > r$ we say that $T$ is $r$-*diagonal*. In that case,

$$T = \sum_{i=-r}^{r} a_i S_i.$$

**Lemma 3.2.** *Suppose that* $T \in \mathcal{B}(\ell^2(\mathbb{Z}))$ *is translation invariant with constant entries* $a_i$ *on the* $i$-th *diagonal and* $a_i = 0$ *for* $|i| \geq r$. *Then the Fourier transform of* $T$ *is defined by*

$$\mathcal{F}T\mathcal{F}^* \in \mathcal{B}(L^2([0, 2\pi]))$$

*is a multiplication operator multiplying with* $a_T(q) = \sum_{j=-r}^{r} a_j e^{iqj}$. *In particular, if* $S$ *is the shift operator then*

$$\mathcal{F}^{\uparrow}(S) = \mathcal{F}S\mathcal{F}^* = e^{-iq}$$

*as an operator on* $L^2([0, 2\pi])$.

*Proof.* Consider for any $j \in \mathbb{Z}$ the operator $T = \sum_{n \in \mathbb{Z}} |n\rangle\langle n + j| = S_{-j}$. Then for $q \in [0, 2\pi]$ it holds for any $\phi \in L^2([0, 2\pi])$ that

$$(\mathcal{F}T\mathcal{F}^*\phi)(q) = \frac{1}{\sqrt{2\pi}} \sum_{x \in \mathbb{Z}} e^{-iqx}(T\mathcal{F}^*\phi)(x) = \frac{1}{\sqrt{2\pi}} \sum_{x \in \mathbb{Z}} e^{-iqx}(\mathcal{F}^*\phi)(x + j) \tag{9}$$

$$= \frac{1}{\sqrt{2\pi}} \sum_{x \in \mathbb{Z}} e^{-iqx} \frac{1}{\sqrt{2\pi}} \int_{[0, 2\pi]} e^{iq'(x+j)}\phi(q')dq' \tag{10}$$

$$= e^{iqj}\phi(q), \tag{11}$$

where we used Plancherel's Theorem in the last step. Now, by linearity if

$$T = \sum_{j=-r}^{r} a_j \sum_{n \in \mathbb{Z}} |n\rangle\langle n + j|,$$

it holds that

$$(\mathcal{F}T\mathcal{F}^*\phi)(q) = \left(\sum_{j=-r}^{r} a_j e^{iqj}\right)\phi(q).$$

<div style="text-align: right">□</div>

The curve $a_T(k) = \sum_{j=-r}^{r} a_j e^{iqj}$ can also be viewed as a function from the unit circle defining $z = e^{iq}$, then $a_T(z) = \sum_{j=-r}^{r} a_j z^j$ we denote this curve the *symbol curve*. Since the Fourier transformation is unitary and using that the essential spectrum is the part of the spectrum that is not isolated eigenvalues with finite multiplicity [54, IV.5.33] we obtain the following corollary.

**Corollary 3.3.** *For any Laurent operator $T \in \mathcal{B}(\ell^2(\mathbb{Z}))$ it holds that*

$$\sigma(T) = \{a_T(z) \mid z \in \mathbb{T}\} = \sigma_{\text{ess}}(T).$$

Furthermore, as explained in [55, Theorem 1.2] if $0 \notin \sigma(T)$ then the inverse of $T$ is unitarily equivalent to a multiplication with the inverse symbol

$$a_{T^{-1}}(z) = a_T^{-1}(z) = \frac{1}{a_T(z)}.$$

In concrete applications, we consider tridiagonal matrices $T(q)$ and therefore, we review some results about tridiagonal Laurent operators and their invertibility in Appendix A.7. If $T(q)$ is tridiagonal we let $\alpha, \beta, \gamma : [0, 2\pi] \to \mathbb{C}$ be the entries of $T(q)$. Thus, the symbol curve is given by the image of $\mathbb{T}$ under the map $a$ defined by

$$a(z) = \alpha z^{-1} + \beta + \gamma z.$$

In the tridiagonal case, the symbol curve is a (possibly degenerate) ellipse.

## 3.2 From translation-invariance to a direct integral decomposition

We now show how we can use translation-invariance to get a direct integral decomposition in the case where we consider $\mathcal{L} \in \mathcal{B}(\text{HS}(\ell^2(\mathbb{Z})))$.

To vectorize we will need the tensor product of Hilbert spaces. Thus, we define $\ell^2(\mathbb{Z}) \otimes \ell^2(\mathbb{Z})$ to be the set of all formal symbols $v \otimes w$ where $v, w \in \ell^2(\mathbb{Z})$ with inner product $\langle v_1 \otimes w_1, v_2 \otimes w_2 \rangle = \langle v_1, v_2 \rangle \langle w_1, w_2 \rangle$ and taking closure with respect to that inner product. In that way there is an isomorphism $\ell^2(\mathbb{Z}) \otimes \ell^2(\mathbb{Z}) \to \ell^2(\mathbb{Z}^2)$ given by $|i\rangle \otimes |j\rangle \to |i, j\rangle$ and extending by linearity. While we attempt at making many of the other isomorphisms explicit we will regard $\ell^2(\mathbb{Z}) \otimes \ell^2(\mathbb{Z})$ as the same space as $\ell^2(\mathbb{Z}^2)$ with this isomorphism in mind.

### 3.2.1 Vectorization of the space of Hilbert-Schmidt operators

Vectorization is a formalization of the idea of thinking of matrices as vectors. The space of Hilbert-Schmidt operators is particularly amenable to vectorization. To formalize this we follow ([56, 57, (4.88)]) and define $\mathrm{vec} : \mathrm{HS}(\ell^2(\mathbb{Z})) \to \ell^2(\mathbb{Z}) \otimes \ell^2(\mathbb{Z})$ by extending linearly

$$\mathrm{vec}(|i\rangle\langle j|) = |i\rangle \otimes |j\rangle.$$

**Lemma 3.4.** *The map* $\mathrm{vec} : \mathrm{HS}(\ell^2(\mathbb{Z})) \to \ell^2(\mathbb{Z}^2)$ *is a isometric isomorphism.*

*Proof.* If $A \in \mathrm{HS}(\ell^2(\mathbb{Z})) = \mathcal{S}_2(\ell^2(\mathbb{Z}))$ is an operator then the 2-norm of $A$ is given by $\|A\|_{\mathcal{S}_2}^2 = \sum_{i,j \in \mathbb{Z}} |A_{i,j}|^2$ where $A_{i,j} = \langle i|, A|j\rangle$ is the matrix elements of $A$. Thus, $\|A\|_{\mathcal{S}_2} = \|\{A_{i,j}\}_{i,j \in \mathbb{Z}}\|_2$ where $\|\cdot\|_2$ is the norm on $\ell^2(\mathbb{Z}^2)$. $\qquad\square$

The vectorized form of a Hilbert-Schmidt operator $A = \sum_{i,j \in \mathbb{Z}} A_{i,j}|i\rangle\langle j|$ is given by

$$\mathrm{vec}(A) = \sum_{i,j \in \mathbb{Z}} A_{i,j}|i\rangle \otimes |j\rangle.$$

### 3.2.2 Explicit vectorization of the Lindbladian

The vectorization of a product is given by

$$\mathrm{vec}(ABC) = A \otimes C^T \mathrm{vec}(B) \tag{12}$$

analogous to [57, (4.84)][1], where $C^T$ is the transpose of $C$. It follows that (with some more details, with a slightly different convention, given in [57])

$$\mathrm{vec}(\mathcal{L}(\rho)) = \left( -i(H \otimes \mathbb{1}) + i(\mathbb{1} \otimes H^T) + G \sum_k L_k \otimes (L_k^*)^T - \frac{1}{2}L_k^* L_k \otimes \mathbb{1} - \frac{1}{2}\mathbb{1} \otimes (L_k^* L_k)^T \right) \mathrm{vec}(\rho) \tag{13}$$

Recall the definition of the lifted isomorphism $\mathrm{vec}^{\uparrow} : \mathcal{B}(\mathrm{HS}(\mathcal{H})) \to \mathcal{B}(\ell^2(\mathbb{Z}) \otimes \ell^2(\mathbb{Z}))$ from (6) and the definition of $H_{\mathrm{eff}}$ from (3). The next lemma follows from (13).

**Lemma 3.5.** *The vectorization of the Lindbladian is given by*

$$\mathrm{vec}^{\uparrow}(\mathcal{L}) = \left( -i(H \otimes \mathbb{1}) + i(\mathbb{1} \otimes H^T) + G \sum_k L_k \otimes (L_k^*)^T - \frac{1}{2}L_k^* L_k \otimes \mathbb{1} - \frac{1}{2}\mathbb{1} \otimes (L_k^* L_k)^T \right) \tag{14}$$

$$= -iH_{\mathrm{eff}} \otimes \mathbb{1} + i\mathbb{1} \otimes \overline{H_{\mathrm{eff}}} + G \sum_k L_k \otimes \overline{L_k}. \tag{15}$$

---

[1]We make a with a slight change of notation. The vectorization map in [57, (4.84)] is defined as $|i\rangle\langle j| \to |j\rangle \otimes |i\rangle$.

Notice that the shift $S_1 \rho S_1^*$ from (4) corresponds to the shift $(x, y) \mapsto (x+1, y+1)$ on $l^2(\mathbb{Z}^2)$. That $\mathcal{L}$ is translation-covariant means that it is covariant under these joint translations. Thus, since we only have translation invariance in one coordinate direction we can only hope to utilize it with a Fourier transform in one of the two variables. One may also think of this as relative and absolute position with respect to the diagonal. This change of coordinates was suggested in [58] and it was also used in [9].

Thereby we should be able to decompose the superoperator $\mathcal{L}$ (which if we consider $\mathcal{L}$ as an operator on Hilbert–Schmidt operators can be viewed as an operator on $\ell^2(\mathbb{Z}^2)$) into $q$-dependent operators on $\ell^2(\mathbb{Z})$, where $q$ is the Fourier variable. This is formalized using the direct integral, that we introduce since it is important to us both for the main theorem and its applications. More information, for example on the details of measurability, can be obtained in [59, XII.16].

### 3.2.3 The direct integral of Hilbert spaces

In the following, suppose that $\{\mathcal{H}_q\}_{q \in I}$ is a family of Hilbert spaces indexed by some index set $I$ (in the following $I = [0, 2\pi]$). The direct integral of Hilbert spaces over an index set $I$ is the set of families of vectors in each of the Hilbert spaces, it is written by $\int_I^\oplus \mathcal{H}_q dq$ and it consists of equivalence classes up to sets of measure 0 of vectors $v$ such that $v_q \in \mathcal{H}_q$ for each $q \in I$.

The space $\int_I^\oplus \mathcal{H}_q dq$ is again a Hilbert space with inner product given by

$$\langle v, w \rangle_{\int_I^\oplus \mathcal{H}_q dq} = \int_I \langle v_q, w_q \rangle_{\mathcal{H}_q} \, dq.$$

Now, if $A(q)$ is a bounded operator on $\mathcal{H}_q$ for each $q$ and the family $q \mapsto A(q)$ is measurable then we can define the integral operator $A = \int_I^\oplus A(q) dq$. Naturally, it acts as $\int_I^\oplus A(q) dq v = \int_I^\oplus A(q) v(q) dq$. If for an operator $A$ on $\int_I^\oplus \mathcal{H}_q dq$ the converse is true, i.e. there exists a measurable family of operators $\{A(q)\}_{q \in I}$ such that $A = \int_I^\oplus A(q) dq$ then we say that $A$ is *decomposable*. The norm of a decomposable operator is given as

$$\|A\| = \operatorname*{esssup}_{q \in I} \, \|A(q)\|. \tag{16}$$

With these definitions at hand, we see that an equivalent way of interpreting $\int_I^\oplus \mathcal{H}_q dq$ is as the space of $l^2(\mathbb{Z})$ valued $L^2$-functions from $[0, 2\pi]$, which we write as $L^2\left([0, 2\pi], \ell^2(\mathbb{Z})\right)$.

More formally, we introduce $L^2\left([0, 2\pi], \ell^2(\mathbb{Z})\right)$ as the space of (equivalence classes of) functions $f : [0, 2\pi] \to \ell^2(\mathbb{Z})$ such that $\int_{[0,2\pi]} \|f(q)\|_2^2 dq < \infty$. The inner product on $L^2\left([0, 2\pi], \ell^2(\mathbb{Z})\right)$ given by

$$\langle f, g \rangle_{L^2([0,2\pi], \ell^2(\mathbb{Z}))} = \int_{[0,2\pi]} \langle f(q), g(q) \rangle_{\ell^2(\mathbb{Z})} dq$$

and using this inner product the space becomes a Hilbert space. This space is relevant to us because the map $I : L^2([0, 2\pi]) \otimes \ell^2(\mathbb{Z}) \to L^2\left([0, 2\pi], \ell^2(\mathbb{Z})\right)$ defined by $I(g \otimes \psi) = \psi_g$ where $\psi_g(q) = g(q)\psi$ for any $q \in [0, 2\pi], g \in L^2\left([0, 2\pi]\right), \psi \in \ell^2(\mathbb{Z})$ is an isometric isomorphism.

### 3.2.4 From Hilbert-Schmidt operators to direct integrals

Before continuing we introduce the unitary operator $C : \ell^2(\mathbb{Z}) \otimes \ell^2(\mathbb{Z})$ by

$$C|j,k\rangle \ = |j,k-j\rangle.$$

Since $f : \mathbb{Z}^2 \to \mathbb{Z}^2$ given by $f(j,k) = (j,k-j)$ is a bijection, $C$ maps an orthonormal basis to an orthonormal basis so it is unitary (and hence an isometric isomorphism) and its inverse is given by

$$C^*|j,k\rangle \ = C^{-1}|j,k\rangle \ = |j,k+j\rangle.$$

The following lemma shows that conjugation by the operator $C$ transforms an operator with joint translation invariance to an operator with translation invariance in the first tensor factor and it will be useful for us. We were made aware of this trick in [58].

**Lemma 3.6.** *The unitary operator $C$ satisfies the following relations*

*(i)* $C(\mathbb{1} \otimes S)C^* = \mathbb{1} \otimes S$

*(ii)* $C(S \otimes \mathbb{1})C^* = S \otimes S^*$

*(iii)* $C(S \otimes S)C^* = S \otimes \mathbb{1}$.

*Proof.* We prove this by straightforward computation. For any $k, j \in \mathbb{Z}$ it holds that

$$C(\mathbb{1} \otimes S)C^*|j,k\rangle = C(\mathbb{1} \otimes S)|j,k+j\rangle = C|j,k+j+1\rangle = |j,k+1\rangle = (\mathbb{1} \otimes S)|j,k\rangle$$

and so (i) follows. Similarly,

$$C(S \otimes \mathbb{1})C^*|j,k\rangle = C(S \otimes \mathbb{1})|j,k+j\rangle = C|j+1,k+j\rangle = |j+1,k-1\rangle = S \otimes S^*|j,k\rangle.$$

The third relation follows from multiplying the previous two. $\square$

For ease of notation we define $\mathcal{F}_1 = \mathcal{F} \otimes \mathbb{1}$ the Fourier transform in the first coordinate. Notice that by Lemma 3.2

$$\mathcal{F}_1 \left( \sum_{k \in \mathbb{Z}} S_k |a\rangle \langle b| S_k^* \otimes |a'\rangle \langle b'| \right) \mathcal{F}_1^* = \mathcal{F} S_{a-b} \ \mathcal{F}^* \otimes |a'\rangle \langle b'| = e^{-iq(a-b)} |a'\rangle \langle b'|. \tag{17}$$

The following central lemma sums up the isomorphisms discussed in this subsection and it is the particular decomposition of a Hilbert space as a direct integral that we are going to use.

**Lemma 3.7.** *The maps* $\mathrm{vec}, \mathcal{F}_1, I$ *implement the following isometric isomorphisms of Hilbert spaces*

$$\mathrm{HS}(\ell^2(\mathbb{Z})) \stackrel{\mathrm{vec}}{\cong} \ell^2(\mathbb{Z}) \otimes \ell^2(\mathbb{Z}) \stackrel{\mathcal{F}_1}{\cong} L^2([0,2\pi]) \otimes \ell^2(\mathbb{Z}) \stackrel{I}{\cong} L^2\left([0,2\pi], \ell^2(\mathbb{Z})\right) = \int_{[0,2\pi]}^{\oplus} \ell^2(\mathbb{Z})_q \, dq. \tag{18}$$

As the operator $C : \ell^2(\mathbb{Z}) \otimes \ell^2(\mathbb{Z}) \to \ell^2(\mathbb{Z}) \otimes \ell^2(\mathbb{Z})$ was an isometric isomorphism that means that also the map $\mathcal{J} : \mathrm{HS}(\ell^2(\mathbb{Z})) \to \int_{[0,2\pi]}^{\oplus} \ell^2(\mathbb{Z})_q dq$ defined by

$$\mathcal{J} = I \circ \mathcal{F}_1 \circ C \circ \mathrm{vec} \tag{19}$$

is an isometric isomorphism. As we will see, the lifted $\mathcal{J}^{\uparrow}$, corresponding to conjugation by $\mathcal{J}$ will help us obtain the representation of $\mathcal{L}$ on which the remaining results of this paper are based.

### 3.3   The direct integral decomposition

Let us state and prove our main result.

**Theorem 3.8.** *Suppose that $\mathcal{L}$ is of the form (1) with Lindblad operators $L_k$ satisfying assumption $\mathcal{A}_2 a), \mathcal{A}_2 b)$. Then if we let $\mathcal{J} : \mathrm{HS}(\ell^2(\mathbb{Z})) \to \int_{[0,2\pi]}^{\oplus} \ell^2(\mathbb{Z})_q dq$ be the isometric isomorphism defined in (19) it holds that*

$$\mathcal{J}^{\uparrow}(\mathcal{L}) = \int_{[0,2\pi]}^{\oplus} T(q) + F(q) dq,$$

*with $T(q)$ a bi-infinite $r$-diagonal Laurent operator and $F(q)$ a finite rank operator with finite range for each $q \in [0, 2\pi]$. If $H_{\mathrm{eff}} = H - \frac{iG}{2} \sum_k L_k^* L_k = \sum_{l=-r}^{r} h_l S_l$ then*

$$T(q) = -i \sum_{l=-r}^{r} h_l e^{iql} S_l^* + i \sum_{l=-r}^{r} \overline{h_l} S_l.$$

*Moreover, in the case $L_0 = |\phi\rangle\langle\psi|$ is rank-one with coefficients $|\phi\rangle = \sum_r \alpha_r |r\rangle$ and $|\psi\rangle = \sum_r \beta_r |r\rangle$ then*

$$F(q) = G \left( \sum_{r_1, r_2} \alpha_{r_1} e^{iqr_1} \ \overline{\alpha_{r_2}} \ |r_2 - r_1\rangle \right) \left( \sum_{r_1', r_2'} \beta_{r_1'} e^{-iqr_1'} \ \overline{\beta_{r_2'}} \langle r_2' - r_1'| \right).$$

*so in particular $F(q)$ is rank-one.*

*Proof.* We have to consider

$$\mathcal{J}^{\uparrow}(\mathcal{L}) = (I \circ \mathcal{F}_1 \circ C \circ \mathrm{vec})^{\uparrow}(\mathcal{L}) = I^{\uparrow} \mathcal{F}_1^{\uparrow} C^{\uparrow} \mathrm{vec}^{\uparrow}(\mathcal{L}),$$

where the notation was introduced in (6) and we use (7). Thus, we start by considering the vectorization of the Lindbladian, using Lemma 3.5 we see that

$$\mathrm{vec}^{\uparrow}(\mathcal{L}) = -i H_{\mathrm{eff}} \otimes \mathbb{1} + i \mathbb{1} \otimes \overline{H_{\mathrm{eff}}} + G \sum_{k \in \mathbb{Z}} L_k \otimes \overline{L_k}.$$

We now deal with the three terms separately. For the first term we use Lemma 3.6 ii) and Lemma 3.2 to obtain

$$\mathcal{F}_1^\uparrow\left(C^\uparrow(H_{\text{eff}}\otimes\mathbb{1})\right)=\sum_{l=-r}^r h_l\mathcal{F}_1^\uparrow\left(C^\uparrow(S_l\otimes\mathbb{1})\right)=\sum_{l=-r}^r h_l(\mathcal{F}\otimes\mathbb{1})^\uparrow(S_l\otimes S_l^*)=\sum_{l=-r}^r h_l e^{-iql}\otimes S_l^*$$

where we abused notation slightly in denoting the function $q\mapsto e^{-iql}$ by $e^{-iql}$. Thereby,

$$(I\circ(\mathcal{F}\otimes\mathbb{1})\circ C)^\uparrow(H_{\text{eff}}\otimes\mathbb{1})=\sum_{l=-r}^r h_l e^{-iql}S_l^*.$$

Similarly, for the second term we use Lemma 3.6 i) and Lemma 3.2 to obtain

$$\mathcal{F}_1^\uparrow\left(C^\uparrow(\mathbb{1}\otimes\overline{H_{\text{eff}}})\right)=\sum_{l=-r}^r \overline{h_l}\mathcal{F}_1^\uparrow\left(C^\uparrow(\mathbb{1}\otimes S_l)\right)=\sum_{l=-r}^r \overline{h_l}\mathcal{F}_1^\uparrow(\mathbb{1}\otimes S_l)=\sum_{l=-r}^r \overline{h_l}1\otimes S_l,$$

where $1\in L^2([0,2\pi])$ is the constant function 1. We conclude that

$$I^\uparrow\left(\mathcal{F}_1^\uparrow\left(C^\uparrow(H_{\text{eff}}\otimes\mathbb{1})\right)\ \right)=\sum_{l=-r}^r \overline{h_l}S_l.$$

Finally, for the quantum jump terms we do something slightly different. Notice first that

$$
\begin{aligned}
C^\uparrow\left(\sum_{k\in\mathbb{Z}}L_k\otimes\overline{L_k}\right)&=C^\uparrow\left(\sum_{k\in\mathbb{Z}}S_kL_0S_k^*\otimes S_k\overline{L_0}S_k^*\right)\\
&=C\sum_{k\in\mathbb{Z}}(S_k\otimes S_k)(L_0\otimes\overline{L_0})(S_k^*\otimes S_k^*)C^*\\
&=\sum_{k\in\mathbb{Z}}C(S_k\otimes S_k)C^*C(L_0\otimes\overline{L_0})C^*C(S_k^*\otimes S_k^*)C^*\\
&=\sum_{k\in\mathbb{Z}}(S_k\otimes\mathbb{1})C(L_0\otimes\overline{L_0})C^*(S_k^*\otimes\mathbb{1}),
\end{aligned}
$$

where we used Lemma 3.6 iii).

Since the operator $C(L_0\otimes\overline{L_0})C^*$ is local around $|0\rangle\otimes|0\rangle\langle0|\otimes\langle0|$ then the Fourier transform in the first coordinate is a local operator ($q$-dependent matrix) by the computation in (17). To see the explicit form in the rank-one case, we first write

$$L_0\otimes\overline{L_0}=\left(\sum_{r_1,r_2}\alpha_{r_1}\ \overline{\alpha_{r_2}}|r_1,r_2\rangle\right)\left(\sum_{r_1',r_2'}\beta_{r_1'}\ \overline{\beta_{r_2'}}\langle r_1',r_2'|\right).$$

Thus,

$$C\sum_{r_1,r_2}\alpha_{r_1}\ \overline{\alpha_{r_2}}|r_1,r_2\rangle\sum_{r_1',r_2'}\beta_{r_1'}\ \overline{\beta_{r_2'}}\langle r_1',r_2'|C^*=\sum_{r_1,r_2}\alpha_{r_1}\ \overline{\alpha_{r_2}}|r_1,r_2-r_1\rangle\sum_{r_1',r_2'}\beta_{r_1'}\ \overline{\beta_{r_2'}}\langle r_1',r_2'-r_1'|.$$

Now, using the relation (17) yields that

$$F(q) = \left( \sum_{r_1, r_2} \alpha_{r_1} e^{iqr_1} \ \overline{\alpha_{r_2}} \ |r_2 - r_1\rangle \right) \left( \sum_{r'_1, r'_2} \beta_{r'_1} e^{-iqr'_1} \ \overline{\beta_{r'_2}} \langle r'_2 - r'_1| \right).$$

$\square$

By reading off coefficients we obtain the following Corollary.

**Corollary 3.9.** *Suppose that $\mathcal{L}_\Delta(\rho) = -i[\tilde{\Delta}, \rho]$ corresponding to Hamiltonian evolution with the discrete Laplacian. Then*

$$\mathcal{J}^\uparrow(\mathcal{L}_\Delta) = (1 - e^{-iq})S + (1 - e^{iq})S^*.$$

## 3.4 Spectrum of direct integral of operators

In the case of a self-adjoint, translation-invariant operator $A$ the spectrum of $A$ coincides with the union of the spectra of the operators contained in the direct integral representation of $A$ after the Fourier-transform [59, XIII.85]. However, in our case, due to the non-normality of $\mathcal{L}$, the information about the pointwise spectrum of $T(q) + F(q)$ may not be sufficient to determine the spectrum of $\mathcal{L}$. In fact, already for the case of the direct sum (direct integral with respect to the counting measure), the spectrum is not the union of the spectra of the fibers as may be seen from the following example.

**Example 3.10** ([60, Problem 98]). *Let $\mathcal{H} = \bigoplus_{n \geq 2} \mathbb{C}^n$ and $A = \bigoplus_{n \geq 2} A_n$ with*

$$A_2 = \begin{pmatrix} 0 & 1 \\ 0 & 0 \end{pmatrix}, A_3 = \begin{pmatrix} 0 & 1 & 0 \\ 0 & 0 & 1 \\ 0 & 0 & 0 \end{pmatrix}, \dots.$$

*then $1 \in \sigma(A)$, but $\sigma(A_n) = 0$ for all $n \geq 2$.*

Instead, the correct concept to recover such a connection between the operator $N$ and the operators forming its direct integral decomposition turns out to be the *pseudospectrum*, which where the resolvent has large norm. More precisely, for a bounded operator $B \in \mathcal{B}(X)$ for a Banach space $X$, we define the $\varepsilon$-*pseudospectrum* of $B$ as the set $\sigma_\varepsilon(B) \subset \mathbb{C}$ for which $\|(B - \lambda)^{-1}\| \geq \frac{1}{\varepsilon}$, where we set $\|B^{-1}\| = \infty$ whenever $B$ is not invertible. It is easy to see that $\sigma(B) \subset \sigma_\varepsilon(B)$ as well as

$$\sigma(B) = \bigcap_{\varepsilon > 0} \sigma_\varepsilon(B). \tag{20}$$

It is instructive to see the desired connection in the case of a direct-sum operator before turning to the direct integral. The following is a slight reformulation of [20, Theorem 5].

**Lemma 3.11** ([20, Theorem 5]). *Suppose that $\mathcal{H} = \oplus_{n \in \mathbb{N}} \mathcal{H}_n$ where each $\mathcal{H}_n$ is a separable Hilbert space. Let $A_n \in \mathcal{B}(\mathcal{H}_n)$ for each $n \in \mathbb{N}$ and $A = \oplus_{n \in \mathbb{N}} A_n$ be a bounded operator on $\mathcal{H}$. Then for all $\varepsilon > 0$ it holds that*

$$\sigma_\varepsilon(A) = \bigcup_{n \in \mathbb{N}} \sigma_\varepsilon(A_n) \ \text{and} \ \sigma(A) = \bigcap_{\varepsilon > 0} \bigcup_{n \in \mathbb{N}} \sigma_\varepsilon(A_n).$$

*Proof.* Let first $\lambda \in \cup_{n \in \mathbb{N}} \sigma_\varepsilon(A_n)$, then there exists an $n_0 \in \mathbb{N}$ such that $\|(A_{n_0} - \lambda)^{-1}\| \geq \frac{1}{\varepsilon}$. Thus, $\sup_{n \in \mathbb{N}} \|(A_n - \lambda)^{-1}\| \geq \frac{1}{\varepsilon}$ and hence $\lambda \in \sigma_\varepsilon(A)$.

For the converse inclusion, we use contraposition. So suppose that $\lambda \notin \cup_{n \in \mathbb{N}} \sigma_\varepsilon(A_n)$. Then for all $n \in \mathbb{N}$ it holds that $\|(A_n - \lambda)^{-1}\| \leq \frac{1}{\varepsilon}$. Thus, $\sup_{n \in \mathbb{N}} \|(A_n - \lambda)^{-1}\| \leq \frac{1}{\varepsilon}$. It follows that $\oplus_{n \in \mathbb{N}} (A_n - \lambda)^{-1}$ is a well-defined bounded operator. Since

$$\left( \bigoplus_{n \in \mathbb{N}} (A_n - \lambda)^{-1} \right)(A - \lambda) = \bigoplus_{n \in \mathbb{N}} (A_n - \lambda)^{-1}(A_n - \lambda) = \mathbb{1},$$

we see that $(A - \lambda)$ is invertible. As $\|(A - \lambda)^{-1}\| \leq \frac{1}{\varepsilon}$ then $\lambda \notin \sigma_\varepsilon(A)$. The second relation follows by (20). $\qquad\square$

To state the analogue of Lemma 3.11 for the direct integral, we first need to the define the *essential union* with respect to the Lebesgue measure on some interval (the following also holds for other measures, but we consider the Lebesgue measure for clarity). If $M_q$ is a family of measurable sets we say that $x \in \bigcup_{q \in I}^{\text{ess}} M_q$ if and only if there exists a set $M$ of positive measure such that $M \subset \{q \mid x \in M_q\}$. The following proof is an extension of the proof techniques just employed and reduces to the case of Lemma 3.11 in the case of the counting measure. For more information on direct integrals and their spectral theory see [61, 62]. A related theorem is proven in the self-adjoint case in [59, XIII.85] and spectra and direct integrals were also studied in [63].

**Theorem 3.12.** *Let $I \subset \mathbb{R}$ be an interval and $\mathcal{H} = \int_I^\oplus \mathcal{H}_q \, dq$ for some family of separable Hilbert spaces $\{\mathcal{H}_q\}_{q \in I}$. Suppose that $\{A(q)\}_{q \in I}$ is a measurable family of bounded operators $A(q)$ on $\mathcal{H}_q$ and that $A = \int_I^\oplus A(q) \, dq \in \mathcal{B}(\mathcal{H})$. Then for all $\varepsilon > 0$ it holds that*

$$\sigma(A) \subset \bigcup_{q \in I}^{\text{ess}} \sigma_\varepsilon(A(q)).$$

*Moreover,*

$$\sigma(A) = \bigcap_{\varepsilon > 0} \left( \bigcup_{q \in I}^{\text{ess}} \sigma_\varepsilon(A(q)) \right).$$

*Proof.* We do the proof again by contraposition. So suppose that $\lambda \notin \bigcup_{q \in I}^{\text{ess}} \sigma_\varepsilon(A(q))$. Then

$$\left\| (A(q) - \lambda)^{-1} \right\|_{\mathcal{H}_q} \leq \frac{1}{\varepsilon}$$

for almost all $q \in I$. Thus, $\text{esssup}_{q \in I} \left\| (A(q) - \lambda)^{-1} \right\|_{\mathcal{H}_q} \leq \frac{1}{\varepsilon}$. It follows that $\int_I^{\oplus} (A(q) - \lambda)^{-1} dq$ is a well defined bounded operator. Accordingly, considering

$$\left( \int_I^{\oplus} (A(q) - \lambda)^{-1} \right) dq (A - \lambda) = \int_I^{\oplus} (A(q) - \lambda)^{-1} (A(q) - \lambda) dq = \mathbb{1} \,,$$

we see that $(A - \lambda)$ is invertible. Thus $\lambda \notin \sigma(A)$.

To see the converse inclusion, suppose that $\lambda \in \bigcap_{\varepsilon > 0} \bigcup_{q \in I}^{\text{ess}} \sigma_\varepsilon(A(q))$. Thus, for each $n \in \mathbb{N}$ there exists a set $I_n$ such that $|I_n| > 0$ with $\left\| (A(q) - \lambda)^{-1} \right\|_{\mathcal{H}_q} \geq n$ for all $q \in I_n$. Now, our goal is to construct a vector in the direct integral of the Hilbert spaces out of this family of vectors which has large norm after application of the resolvent. For each $q \in I_n$ and $n \in \mathbb{N}$ there exists a $v_{q,n} \in \mathcal{H}_q$ with $\|v_{q,n}\|_{\mathcal{H}_q} = 1$ and such that $\left\| (A(q) - \lambda)^{-1} v_{q,n} \right\|_{\mathcal{H}_q} \geq \frac{n}{2}$. Now, defining $w_n = \int_{[0,2\pi]}^{\oplus} v_{q,n} \, \mathbb{1}_{q \in I_n} dq$ [2] then

$$\|w_n\|^2 = \int_{[0,2\pi]} \|v_{q,n} \, \mathbb{1}_{q \in I_n}\|_{\mathcal{H}_q}^2 dq = \int_{I_n} 1 dq = |I_n|.$$

Furthermore, it holds that

$$\left\| \int_{[0,2\pi]}^{\oplus} (A(q) - \lambda)^{-1} \, dq \, w_n \right\|^2 = \left\| \int_{[0,2\pi]}^{\oplus} (A(q) - \lambda)^{-1} v_{q,n} \, \mathbb{1}_{q \in I_n} dq \right\|^2$$
$$= \int_{I_n} \left\| (A(q) - \lambda)^{-1} v_{q,n} \right\|_{\mathcal{H}_q}^2 dq \geq \int_{I_n} \left( \frac{n}{2} \right)^2 dq = |I_n| \, \left( \frac{n}{2} \right)^2.$$

So we conclude that $\left\| \int_{[0,2\pi]}^{\oplus} (A(q) - \lambda)^{-1} dq \right\| \geq \frac{n}{2}$. This means that $\int_{[0,2\pi]}^{\oplus} (A(q) - \lambda)^{-1} \, dq$ is not bounded. By [62, Lemma 1.3] we have that if $A - \lambda = \left( \int_I^{\oplus} (A(q) - \lambda) \right) dq$ is invertible then the inverse is given by $(A - \lambda)^{-1} = \left( \int_I^{\oplus} E(q) \right) dq$ where $E(q) = (A(q) - \lambda)^{-1}$ for almost all $q$. Thus, we can conclude that then this inverse would also not be bounded and hence $A - \lambda$ is not invertible and it holds that $\lambda \in \sigma(A)$.

<div align="right">□</div>

## 3.5 Spectrum of non-Hermitian Evolution and of the full Lindbladian

We now gradually move from the abstract operator-theoretic picture to the concrete cases of non-Hermitian and Markovian Evolution. With Theorem 3.12 in mind, we need to determine the essential union of the pseudo-spectra. One way to do that is using some continuity in $q$ in the pseudospectra of $T(q)$. The continuity is reminiscent of a theorem for self-adjoint operators which also finds the spectrum of the direct integral in terms of its fibers [64]. We prove it using resolvent estimates in the Appendix A.4. A very recent related result that, in some sense, is in between the generality of Theorem 3.12 and Theorem 3.13 appeared in [65].

---

[2] One may worry whether there is a measurable choice of $q \mapsto v_{q,n}$. This concern we address in Appendix A.3.

In the following, we say that a family of operators $\{B(q)\}_{q \in I}$ is *norm continuous* if the function $q \mapsto \|B(q)\|$ is continuous. From Theorem 3.8 it is clear that our assumptions imply this continuity since the operators $T(q)$ and $F(q)$ are respectively $2r$-diagonal and finite range with coefficients that polynomials in $\{e^{iq}, e^{-iq}\}$ and hence continuous functions in $q$ (for completeness we write this out in Lemma A.10).

**Theorem 3.13.** *Let $I \subset \mathbb{R}$ be a compact and suppose that $\{A(q)\}_{q \in I}$ is norm continuous. Then*

$$\bigcup_{q \in I} \sigma(A(q)) = \sigma \left( \int_I^\oplus A(q) dq \right).$$

We give the proof in Appendix A.4. We emphasize that the statement of Theorem 3.13 may look innocent, but it in some sense a statement of continuity of the pseudospectrum. Indeed, the spectrum $q \mapsto \sigma(A(q))$ may be very discontinuous even when $q \mapsto \|A(q)\|$ is continuous (see e.g. [66, Example 4.1]), but $q \mapsto \sigma_\varepsilon(A(q))$ will be continuous for every $\varepsilon > 0$. So when the norm of the $(A(q) - \lambda)^{-1}$ blows up for one $q$ the pseudospectrum of $qs$ in the neighbourhood can feel it and $\lambda$ ends up in the essential union. The compactness of $I$ ensures that if $\|(A(q_n) - \lambda)^{-1}\| \to \infty$ for some $\{q_n\}_{n \in \mathbb{N}} \subset I$ then the sequence has a subsequentially limit $q_0$, and we can then prove that $\lambda$ is part of the spectrum of $A_{q_0}$.

As mentioned, we can use the Theorem 3.13 directly on our direct integral decomposition from Theorem 3.8 to obtain the following corollary.

**Corollary 3.14.** *Let $\mathcal{L} \in \mathcal{B}(\mathrm{HS}(\mathcal{H}))$ be a Lindbladian of the form (1) satisfying assumption $\mathcal{A}_2 a)$ and $\mathcal{A}_2 b)$ and let $T(q)$ and $F(q)$ be as in Theorem 3.8. Then*

$$\sigma(\mathcal{L}) = \bigcup_{q \in [0, 2\pi]} \sigma(T(q) + F(q)). \tag{21}$$

*Furthermore, for the non-Hermitian evolution $\mathcal{T} = \int_{[0,2\pi]}^\oplus T(q) dq$ it holds that*

$$\sigma(\mathcal{T}) = \bigcup_{q \in [0, 2\pi]} \sigma(T(q)), \ \text{and} \ \sigma(\mathcal{T}) \subset \sigma(\mathcal{L}).$$

*Proof.* The first two identities follow directly from Theorem 3.8, Theorem 3.13. So we only prove the last inclusion. Now, since by Theorem 3.8 we know that $F(q)$ is finite rank for each $q$ and as the essential spectrum of an operator is invariant under finite rank perturbations [54, IV.5.35]

$$\sigma(T(q)) = \sigma_{\mathrm{ess}}(T(q)) = \sigma_{\mathrm{ess}}(T(q) + F(q)) \subset \sigma(T(q) + F(q)) \subset \sigma(\mathcal{L}),$$

where we also used that $T(q)$ is translation-invariant, which implies that it only has essential spectrum, see Corollary 3.3. Since these inclusions are true for each $q \in [0, 2\pi]$ we obtain the inclusion. □

Now, for our applications in the next section, we assume that each $L_k$ is rank-one, so it follows from Theorem 3.8 that $F(q) = |\Gamma_L(q)\rangle\langle\Gamma_R(q)|$ is also rank one. In that case, we can strengthen Theorem 3.13 even further.

**Corollary 3.15.** *Let $\mathcal{L} \in \mathcal{B}(\mathrm{HS}(\mathcal{H}))$ be a Lindbladian of the form (1) satisfying assumption $\mathcal{A}_2a)$ and $\mathcal{A}_2b)$ and let $T(q)$ and $F(q)$ be as in Theorem 3.8. Assume further that $F(q) = |\Gamma_L(q)\rangle\langle\Gamma_R(q)|$ is rank-one. Then*

$$\sigma(\mathcal{L}) = \bigcup_{q\in[0,2\pi]}\sigma(T(q)) \cup \left\{\lambda \in \mathbb{C} \mid \langle\Gamma_R(q)|(T(q)-\lambda)^{-1}|\Gamma_L(q)\rangle = -1\right\}. \tag{22}$$

*Proof.* By the decomposition in Theorem 3.13 it suffices to prove that

$$\sigma(T(q)+F(q)) = \sigma(T(q)) \cup \left\{\lambda \in \mathbb{C} \mid \langle\Gamma_R(q)|(T(q)-\lambda)^{-1}|\Gamma_L(q)\rangle = -1\right\}$$

for each $q \in [0, 2\pi]$. This is known as rank-one update. For completeness, we give a proof in Appendix A.5. $\square$

## 3.6 Direct sum decomposition for finite systems with periodic boundary conditions

Before we continue, let us remark that the decomposition also works in finite volume with periodic boundary conditions. For simplicity we let $\{1, \dots, n\} = [n]$ and $\mathbb{T}_n = \left\{\frac{2\pi k}{n} \mid k = 1, \dots n\right\}$.

We consider the Hilbert space $\mathcal{H}_n = \ell^2([n]) = \mathrm{span}\{\ |j\rangle, j = 0 \dots n-1 \}$. We consider the shift

$$S^{(n)} |j\rangle = |j+1 \pmod{n}\rangle, \tag{23}$$

which is a unitary on $\mathcal{H}_n$ and satisfies

$$(S^{(n)})^n = \mathbb{1}. \tag{24}$$

We say that an $n \times n$ matrix $C_n^{\mathrm{per}}$ is *circulant* if it is of the form

$$C_n^{\mathrm{per}} = \begin{pmatrix} a_0 & a_{n-1} & \cdots & a_2 & a_1 \\ a_1 & a_0 & a_{n-1} & & a_2 \\ \vdots & a_1 & a_0 & \ddots & \vdots \\ a_{n-2} & & \ddots & \ddots & a_{n-1} \\ a_{n-1} & a_{n-2} & \cdots & a_1 & a_0 \end{pmatrix}.$$

some times we will denote the matrix $C_n^{\mathrm{per}}[a_0, a_1, \ldots, a_n]$. Similarly, to the infinite-dimensional case, we define the *symbol curve* by

$$a_{C_n^{\mathrm{per}}}(z) \; = \sum_{i=0}^{n} a_i z^i \tag{25}$$

for $z \in \mathbb{T}$ (the unit circle).

We can still make sense of vectorization in the sense that the map $\mathrm{vec}^{(n)} : \mathrm{HS}(\ell^2([n])) \to \ell^2([n]) \otimes \ell^2([n])$ and it has similar properties as in Lemma 3.5.

### 3.6.1 Discrete Fourier transform

We let $\omega_n = e^{\frac{2\pi i}{n}}$ be the $n$'th root of unity. Following [66] the discrete Fourier transform on $\mathcal{H}_n$ is the matrix

$$\mathcal{F}^{(n)} \; = \frac{1}{\sqrt{n}} \begin{pmatrix} 1 & 1 & \cdots & 1 & 1 \\ 1 & \omega_n & \omega_n^2 & & \omega_n^{n-1} \\ \vdots & \omega_n^2 & \omega_n^4 & \ddots & \vdots \\ 1 & & \ddots & \ddots & \omega_n^{(n-2)(n-1)} \\ 1 & \omega_n^{n-1} & \cdots & \omega_n^{(n-1)(n-2)} & \omega_n^{(n-1)(n-1)} \end{pmatrix}, \tag{26}$$

which is unitary. It simultaneously diagonalises all circulant matrices.

**Proposition 3.16.** *Let $C_n^{\mathrm{per}}[a_0, a_1, \ldots, a_n]$ be a circulant matrix. Then the operator*

$$D = \mathcal{F}^{(n)} C_n^{\mathrm{per}} \mathcal{F}^{(n)*}$$

*is diagonal with $\{a_{C_n^{\mathrm{per}}}(z), z \in \mathbb{T}_n\}$ on the diagonals.*

### 3.6.2 The map $C^{(n)}$

Similarly as before we define $C^{(n)} : \ell^2([n]) \otimes \ell^2([n]) \to \ell^2([n]) \otimes \ell^2([n])$ by

$$C^{(n)}|j\rangle \, |k\rangle = |j\rangle|k - j \pmod{n}\rangle.$$

Again, since $f : [n] \to [n]$ given by $f(j,k) = (j, k - j \pmod n)$ is a bijection, $C^{(n)}$ maps an orthonormal basis to an orthonormal basis so it is unitary (and hence an isometric isomorphism) and its inverse is given by

$$(C^{(n)})^*|j, k\rangle \; = (C^{(n)})^{-1}|j, k\rangle \; = |j, k + j \pmod{n}\rangle.$$

The finite-dimensional analogue of Lemma 3.6 follows with the same proof.

**Lemma 3.17.** *The operator $(C^{(n)})^\uparrow$ satisfies the following relations*

*(i)* $(C^{(n)})^\uparrow \, (\mathbb{1} \otimes S^{(n)}) = \mathbb{1} \otimes S^{(n)}$

*(ii)* $(C^{(n)})^\uparrow \, (S^{(n)} \otimes \mathbb{1}) = S^{(n)} \otimes (S^{(n)})^*$

*(iii)* $(C^{(n)})^\uparrow \, (S^{(n)} \otimes S^{(n)}) = S^{(n)} \otimes \mathbb{1}.$

### 3.6.3 Periodic boundary conditions

Suppose that $H$ is an infinite (self-adjoint) banded matrix with range $r$, in other words, an $r$-diagonal Laurent operator with $h_{-r}, h_{-r+1}, \ldots, h_{r-1}, h_r$ on the diagonals. Let $n \geq 2r$ then we can define the $n$ dimensional version of $H$ with periodic boundary conditions as

$$H_n^{\mathrm{per}} = C_n^{\mathrm{per}}[h_{-r}, h_{-r+1}, \ldots, h_{r-1}, h_r].$$

For the Lindblad terms suppose that we start with one Lindblad operator $L_0$ with range at most $r$. This operator, even though it strictly speaking is infinite-dimensional, we can view as a $n \times n$ matrix (where we pad with zeros appropriately). Then define for every $k \in \mathbb{Z}$ the operator

$$L_k = (S^{(n)})^k \, L_0 (S^{(n)*})^k$$

and notice that $L_k = L_{k \pmod{n}}$ by (24). Now, given the values $[h_{-r}, h_{-r+1}, \ldots, h_{r-1}, h_r]$ and $L_0$ we define $\mathcal{L}_n^{\mathrm{per}} \colon \mathrm{HS}\left(\ell^2\left([n]\right)\right) \to \mathrm{HS}\left(\ell^2\left([n]\right)\right)$ by

$$\mathcal{L}_n^{\mathrm{per}}(\rho) = -i[H_n^{\mathrm{per}}, \rho] + G \sum_{k \in [n]} L_k \rho L_k^* - \frac{1}{2}\{L_k^* L_k, \rho\}. \tag{27}$$

### 3.6.4 The full isometric isomorphism

Define $\mathcal{F}_1^{(n)} = \mathcal{F}^{(n)} \otimes \mathbb{1}$ to be the discrete Fourier transform in the first coordinate and we also need to define a map $I^{(n)} \colon \mathbb{C}^n \otimes V \to \oplus_{i=0}^{n-1} V^i$ where each $V^i$ is a copy of $V$ and $I^{(n)}$ is defined by

$$I^{(n)}(|i\rangle \otimes |v\rangle) = \underbrace{0 \oplus \cdots \oplus 0}_{i-1} \oplus v \oplus 0 \oplus \cdots \oplus 0.$$

In other words, $I^{(n)}$ maps $|i\rangle \otimes |v\rangle$ to a $v$ in the $i$'th direct summand. If $V$ is a Hilbert space then $I^{(n)}$ is an isometric isomorphism. Then, similarly to the decomposition above we get that

$$\mathrm{HS}\left(\ell^2\left([n]\right)\right) \overset{\mathrm{vec}^{(n)}}{\cong} \ell^2\left([n]\right) \otimes \ell^2\left([n]\right) \overset{\mathcal{F}_1^{(n)}}{\cong} L^2\left(\mathbb{T}_n\right) \otimes \ell^2\left([n]\right) \overset{I^{(n)}}{\cong} \bigoplus_{q \in \mathbb{T}_n} \ell_q^2([n]).$$

We now use the map $\mathcal{J}^{(n)} \colon \mathrm{HS}\left(\ell^2\left([n]\right)\right) \to \bigoplus_{q \in \mathbb{T}_n} \ell_q^2([n])$ defined by

$$\mathcal{J}^{(n)} = I^{(n)} \circ \mathcal{F}_1^{(n)} \circ C^{(n)} \circ \mathrm{vec}^{(n)} \tag{28}$$

Mimicking the proof of Theorem 3.8 we obtain.

**Theorem 3.18.** *Suppose that $n \geq 2r$ and $\mathcal{L}_n^{\mathrm{per}}$ is of the form* (27). *Then with $\mathcal{J}^{(n)}$ defined in* (28)

$$\mathcal{J}^{(n)\uparrow}(\mathcal{L}) = \bigoplus_{q \in \mathbb{T}_n} \left(T_n^{\mathrm{per}}(q) + F_n(q)\right),$$

*with $T_n^{\mathrm{per}}(q)$ an $r$-diagonal circulant $n \times n$ matrix and $F_n(q)$ a finite rank operator with finite range (uniformly in $n$).*

If $H_{\text{eff}} = H_n^{\text{per}} - \frac{iG}{2} \sum_{k\in[n]} L_k^* L_k = \sum_{l=-r}^{r} h_l (S^{(n)})^l$ *then*

$$T_n^{\text{per}}(q) = -i \sum_{l=-r}^{r} h_l e^{iql} (S^{(n)*})^l + i \sum_{l=-r}^{r} \overline{h_l} (S^{(n)})^l.$$

*Moreover, in the case* $L_0 = |\phi\rangle\langle\psi|$ *is rank-one with coefficients* $|\phi\rangle = \sum_r \alpha_r |r\rangle$ *and* $|\psi\rangle = \sum_r \beta_r |r\rangle$ *then*

$$F_n(q) = G \left( \sum_{r_1,r_2} \alpha_{r_1} e^{iqr_1} \ \overline{\alpha_{r_2}} \ |r_2 - r_1\rangle \right) \left( \sum_{r_1',r_2'} \beta_{r_1'} e^{-iqr_1'} \ \overline{\beta_{r_2'}} \langle r_2' - r_1'| \right).$$

*so in particular* $F_n(q)$ *is rank-one.*

### 3.7 Spectral consequences for finite dimensional systems

It follows directly from Theorem 3.18 that

$$\sigma\left(\mathcal{L}_n^{\text{per}}\right) = \sigma\left(\bigoplus_{q\in\mathbb{T}_n} \left(T_n^{\text{per}}(q) + F_n(q)\right)\right) = \bigcup_{q\in\mathbb{T}_n} \sigma\left(T_n^{\text{per}}(q) + F_n(q)\right) \tag{29}$$

This formula motivates our interest in the set $\sigma\left(T_n^{\text{per}}(q) + F_n(q)\right)$, which was the object of study in [66]. We will specialise on the case where $F_n(q) = |\Gamma_L(q)\rangle\langle\Gamma_R(q)|$. In that case, we get the following weaker form of the corresponding infinite-dimensional lemma (Cor. 3.15, see the corresponding statement, which is given in the proof in Appendix A.5). The interpretation of the lemma that curve in the additional spectrum (see Figure 6 for a plot of the curve) gets well approximated by the finite-dimensional periodic system, but we do not get as detailed information as in the infinite volume case in Cor. 3.15.

**Lemma 3.19.** *Suppose that* $F_n(q) = |\Gamma_L\rangle\langle\Gamma_R|$ *has rank one.*
*Let* $\mathcal{S}_n = \{\lambda \in \mathbb{C} \mid \langle\Gamma_R|(T_n^{\text{per}} - \lambda)^{-1}|\Gamma_L\rangle = -1\}$. *Then*

$$\mathcal{S}_n\backslash\sigma(T_n^{\text{per}}) \subset \sigma\left(T_n^{\text{per}} + F_n\right) \subset \sigma(T_n^{\text{per}}) \cup \mathcal{S}_n.$$

*Proof.* Similar what we do in Appendix A.5, but without using essentiality we do not get equality. □

Lemma 3.19 motivates spending efforts finding the matrix elements of tridiagonal circulant matrices that are relevant to our motivating examples. Let the entries of the three main diagonals be $\alpha$, $\beta, \gamma$. To find the specific form of the inverse circulant it turns out that the solutions $\lambda_1, \lambda_2$ to the quadratic equation

$$\alpha + \beta z + \gamma z^2 = 0 \tag{30}$$

and knowledge of whether the two solutions satisfy $|\lambda_1| < 1 < |\lambda_2|$ is important. In Appendix A.7 we return to the equation and the question, which we settle for some of our relevant example of Lindblad operators.

Let us also notice that the inverse of an invertible circulant $C_n$ (where we omit the per) is given by

$$C_n^{-1} = \mathcal{F}_n D^{-1} \mathcal{F}_n^*,$$

where $D$ is the diagonal matrix from Proposition 3.16. So using (26) the matrix elements of $C_n^{-1}$ are given by (cf. [66])

$$\langle j |, C_n^{-1} | k \rangle = \frac{1}{n} \sum_{l=0}^{n-1} \frac{\bar{\omega}_n^{l(j-1)} \omega_n^{l(k-1)}}{a_{C_n}^{\mathrm{per}}(\omega_n^l)}.$$

With a tedious calculation, we obtain the following, which may be a calculation of some independent interest. To state it more concisely, let $[a]_n$ be the representative between 0 and $n-1$ of $a$.

**Lemma 3.20.** *Let $n$ be prime and suppose that $C_n$ is an $n \times n$ circulant with $\alpha, \beta, \gamma$ on the three main diagonals such that $\gamma \neq 0$. Suppose that $\lambda_1$ and $\lambda_2$ are solutions to (30) such that $|\lambda_2| < 1 < |\lambda_1|$. Then*

$$\langle j |, C_n^{-1} | k \rangle = \frac{1}{\gamma(\lambda_1 - \lambda_2)} \left( \frac{1}{1 - \lambda_1^{-n}} \left( \mathbb{1}[j \neq k] \left( \frac{1}{\lambda_1} \right)^{[j-k]_n} + \mathbb{1}[j = k] \frac{1}{\lambda_1^n} \right) + \frac{1}{1 - \lambda_2^n} \lambda_2^{[k-j]_n} \right).$$

*In particular, our matrix element of particular interest can be found by the following formula*

$$\langle 0 |, C_n^{-1} | 0 \rangle = \frac{1}{\gamma(\lambda_1 - \lambda_2)} \left( \frac{1}{\lambda_1^n - 1} + \frac{1}{1 - \lambda_2^n} \right). \tag{31}$$

*Proof.* Consider first the symbol curve with its corresponding polynomial that has roots $\lambda_1$ and $\lambda_2$

$$a_{C_n}(z) = z^{-1} \alpha + \beta + \gamma z = \gamma z^{-1} (\lambda_1 - z)(\lambda_2 - z).$$

Then

$$\frac{1}{a_{C_n}^{\mathrm{per}}(z)} = \frac{z}{\gamma} \frac{1}{\lambda_1 - \lambda_2} \left( \frac{1}{z - \lambda_1} - \frac{1}{z - \lambda_2} \right) = \frac{1}{\gamma(\lambda_1 - \lambda_2)} \left( \sum_{n=1}^{\infty} \left( \frac{z}{\lambda_1} \right)^n + \sum_{n=0}^{\infty} \left( \frac{\lambda_2}{z} \right)^n \right).$$

This means that we have to consider $\sum_{r=1}^{\infty}\left(\frac{\omega_n^l}{\lambda_1}\right)^r$ if $n$ is prime then the set $\{\omega_n^{lr} \mid r \in [n]\}$ has $n$ elements, it is a full cycle, and thus

$$\sum_{r=1}^{\infty}\left(\frac{\omega_n^l}{\lambda_1}\right)^r = \sum_{r=1}^{n}\frac{\omega_n^{lr}}{\lambda_1^r}\sum_{j=0}^{\infty}\left(\frac{1}{\lambda_1^n}\right)^j = \sum_{r=1}^{n}\frac{\omega_n^{lr}}{\lambda_1^r}\frac{1}{1-\lambda_1^{-n}} = \frac{1}{1-\lambda_1^{-n}}\frac{\omega_n^l}{\lambda_1}\sum_{r=0}^{n-1}\left(\frac{\omega_n^l}{\lambda_1}\right)^r.$$

Similarly,

$$\sum_{r=0}^{\infty}\left(\omega_n^{-l}\lambda_2\right)^r = \sum_{r=0}^{n-1}\omega_n^{-lr}\lambda_2^r\sum_{j=0}^{\infty}(\lambda_2^n)^j = \sum_{r=0}^{n}\omega_n^{-lr}\lambda_2^r\frac{1}{1-\lambda_2^n} = \frac{1}{1-\lambda_2^n}\sum_{r=0}^{n-1}\left(\omega_n^{-l}\lambda_2\right)^r.$$

So,

$$\frac{1}{a_{C_n}^{\mathrm{per}}(\omega_n^l)} = \frac{1}{\gamma(\lambda_1-\lambda_2)}\sum_{r=0}^{n-1}\frac{1}{1-\lambda_1^{-n}}\frac{\omega_n^l}{\lambda_1}\left(\frac{\omega_n^l}{\lambda_1}\right)^r + \frac{1}{1-\lambda_2^n}\left(\omega_n^{-l}\lambda_2\right)^r.$$

Now,

$$\langle j|,C_n^{-1}|k\rangle = \frac{1}{n}\sum_{l=0}^{n-1}\frac{\omega_n^{l(k-j)}}{a(\omega_n^l)} = \frac{1}{n}\frac{1}{\gamma(\lambda_1-\lambda_2)}\sum_{r=0}^{n-1}\sum_{l=0}^{n-1}\left(\frac{1}{1-\lambda_1^{-n}}\frac{\omega_n^l}{\lambda_1}\left(\frac{\omega_n^l}{\lambda_1}\right)^r + \frac{1}{1-\lambda_2^n}\left(\omega_n^{-l}\lambda_2\right)^r\right)\omega_n^{l(k-j)}.$$

The first term can be written as

$$\sum_{r=0}^{n-1}\frac{1}{1-\lambda_1^{-n}}\left(\frac{1}{\lambda_1}\right)^r\frac{1}{\lambda_1}\sum_{l=0}^{n-1}\omega_n^l\omega_n^{lr}\omega_n^{l(k-j)} = n\frac{1}{1-\lambda_1^{-n}}\left(\frac{1}{\lambda_1}\right)^{[j-k-1]_n}\frac{1}{\lambda_1}$$

$$= n\frac{1}{1-\lambda_1^{-n}}\left(\mathbb{1}\,[j\neq k]\left(\frac{1}{\lambda_1}\right)^{[j-k]_n} + \mathbb{1}[j=k]\frac{1}{\lambda_1^n}\right),$$

where we used that

$$\sum_{l=0}^{n-1}\omega_n^l\omega_n^{lr}\omega_n^{l(k-j)} = \sum_{l=0}^{n-1}\omega_n^{l(1+r+k-j)} = n\,\mathbb{1}[r \equiv j-k-1 \pmod n].$$

Similarly, for the second term:

$$\sum_{r=0}^{n-1}\frac{1}{1-\lambda_2^n}\lambda_2^r\sum_{l=0}^{n-1}\omega_n^{-lr}\omega_n^{l(k-j)} = n\frac{1}{1-\lambda_2^n}\lambda_2^{[k-j]_n}$$

In total, this means that

$$\langle j|,C_n^{-1}|k\rangle = \frac{1}{\gamma(\lambda_1-\lambda_2)}\left(\frac{1}{1-\lambda_1^{-n}}\left(\mathbb{1}[j\neq k]\left(\frac{1}{\lambda_1}\right)^{[j-k]_n} + \mathbb{1}[j=k]\frac{1}{\lambda_1^n}\right) + \frac{1}{1-\lambda_2^n}\lambda_2^{[k-j]_n}\right).$$

As for the last formula notice that

$$\langle 0|,C_n^{-1}|0\rangle = \frac{1}{\gamma(\lambda_1-\lambda_2)}\left(\frac{1}{1-\lambda_1^{-n}}\left(\frac{1}{\lambda_1^n}\right) + \frac{1}{1-\lambda_2^n}\right) = \frac{1}{\gamma(\lambda_1-\lambda_2)}\left(\frac{1}{\lambda_1^n-1} + \frac{1}{1-\lambda_2^n}\right).\qquad \square$$

### 3.7.1  Catching $n-1$ eigenvalues of $C_n^{\mathrm{per}}$ close to $\sigma(T)$.

For some tridiagonal Laurent operators we consider functions $R_T : \mathbb{C}\backslash\sigma(T) \to \mathbb{C}$ defined by $R_T(z) = \langle 0|, (T-z)^{-1}|0\rangle + 1$. It follows from [54, Theorem III-6.7] that $R_T$ is holomorphic (cf. Lemma 4.7). We can apply Lemma 3.20 to get the following observation.

**Proposition 3.21.** *Let $T$ be a tridiagonal Laurent operator and let $C_n^{\mathrm{per}}$ be the corresponding $n\times n$ circulant. Suppose that there exists exactly one $z_0 \in \mathbb{C}\backslash\sigma(T)$ such that $\langle 0|, (T-z_0)^{-1}|0\rangle = -1$, and where $z_0$ is a simple pole of $R_T$. Suppose that the solutions $\lambda_1, \lambda_2$ to (30) with $\beta$ replaced by $\beta - z_0$ satisfy $|\lambda_2| < 1 < |\lambda_1|$. Let $K$ be any compact subset of $\mathbb{C}\backslash\sigma(T)$ containing $z_0$. Then for sufficiently large prime $n$, it holds that exactly $n-1$ eigenvalues of $C_n^{\mathrm{per}} + |0\rangle\langle 0|$ are outside $K$ and exactly one eigenvalue is inside $K$ and it converges to $z_0$ when $n \to \infty$ along the primes.*

*Proof.* For every $z \in K$ consider the Laurent operator with $\alpha, \beta - z, \gamma$ on the diagonals and consider the solutions $\lambda_1(z), \lambda_2(z)$ to (30) with $\beta$ replaced by $\beta - z$. Notice that any solution $\lambda_i(z)$ of (30) with absolute value 1 corresponds to $z \in \sigma(T)$. Therefore the two solutions (which for a suitable choice of labelling are analytic functions of $z$) must satisfy $|\lambda_2(z)| < 1 < |\lambda_1(z)|$ for every $z \in K$. The functions $|\lambda_1(z)|, |\lambda_2(z)|$ are therefore well defined continous functions on $K$, so the extreme value theorem implies that $|\lambda_1| \geq c_1 > 1$ and $\lambda_2 \leq c_2 < 1$ for some constants $c_1, c_2 > 0$ uniformly on $K$.

Let $\mathbb{P}$ be the set of primes. For each $n \in \mathbb{P}$ define the function $f_n : K \to \mathbb{C}$ by

$$f_n(z) = \langle 0|, (C_n - z)^{-1}|0\rangle + 1$$

and define also $f : K \to \mathbb{C}$ by

$$f(z) = \langle 0|, (T - z)^{-1}|0\rangle + 1.$$

It follows from [54, Theorem III-6.7] that $f_n$ and $f$ are all holomorphic functions. From Lemma 3.20 and Lemma A.13 we get the explicit form of the functions and it follows that $f_n \xrightarrow{\mathrm{lcu}} f$.

Suppose that $\{z_n\}_{n\in\mathbb{P}} \subset K$ such that $f_n(z_n) = 0$. Since $K$ is compact take a convergent subsequence $z_n \to \tilde{z}$. It follows from $|\lambda_1| \geq c_1 > 1$ and $|\lambda_2| \leq c_2 < 1$ and the formula for the inverse (31), that

$$-1 = \frac{1}{\gamma(\lambda_1(\tilde{z}) - \lambda_2(\tilde{z}))}.$$

Since $z_0$ was the unique solution to $-1 = \langle 0|, (T-z)^{-1}|0\rangle$, we get by Lemma A.13 that $\tilde{z} = z_0$.

Furthermore, if (on a subsequence) there exists $z_n^1 \neq z_n^2$ satisfying $f_n(z_n^1) = 0 = f_n(z_n^2)$ for all $n \in \mathbb{P}$ then it contradicts Hurwitz' theorem (see Theorem 4.6), since $z_0$ was a simple pole of $R_T$. So we conclude that for sufficiently large $n$ then there is only one $z_n \in K$ that satisfies $f_n(z_n) = 0$ and (since any convergent subsequence converges to $z_0$) it holds that $z_n \to z_0$. Now, as $C_n^{\mathrm{per}} + F_n$ has $n$ eigenvalues counted with multiplicity. Then at least $n-1$ of them must be inside $k$. □

With the decomposition from Theorem 3.8 in mind, one should view the previous result as a result for fixed $q$. To be able to say something about the gap one would need some uniformity of the estimate in $q_n \to 0$. Probably the exponential decay of the differences in the matrix elements in Lemma 3.20 compared to Lemma A.13 could be key here. In that case, the calculation could probably be made rigorous.

We also notice that these arguments with exponential decay may potentially shed light on whether the numerical observation that many eigenvalues lie on the real axis comes from a symmetry constraint (as argued in [26, Appendix B.9]) or that the eigenvalues in the periodic system tend to the eigenvalues of the full Lindbladian exponentially fast (cf. Lemma 3.20).

# 4 General applications of the direct integral decomposition

Before continuing with the concrete applications in the next section, we discuss some more abstract consequences of the results presented in the previous section.

## 4.1 Approximate point spectrum of Lindblad operators

For normal operators, the residual spectrum is always empty [41, Lemma 12.11], whereas it might not be the case for non-normal operators. In this section, we use the results of the previous sections to prove, under our standing assumptions, that the spectrum of $\mathcal{L}$ is purely approximate point, i.e. the residual spectrum is empty. A result that we believe is of independent interest due to the normality aspect of $\mathcal{L}$ that it indicates. Furthermore, one could hope that the theorem could take part in resolving Question A.2 and Question A.6 concerning spectral independence. For completeness, we note that covariance in the sense of (4) also implies that the entire spectrum is essential.

**Theorem 4.1.** *Let $\mathcal{H} = \ell^2(\mathbb{Z})$ and consider $\mathcal{L} \in \mathcal{B}(\mathrm{HS}(\mathcal{H}))$. Under Assumptions $(\mathcal{A}_2 a)$-$(\mathcal{A}_2 c)$ it holds that*

$$\sigma_{\mathrm{appt}}(\mathcal{L}) = \sigma(\mathcal{L}).$$

*Proof.* We do the split up as in Theorem 3.8 as $\mathcal{L} = \mathcal{T} + \mathcal{F}$, with $\mathcal{T} = \int_{[0,2\pi]}^{\oplus} T(q)dq$ and $\mathcal{F} = \int_{[0,2\pi]}^{\oplus} F(q)dq$. Then we show approximate point spectrum in several steps starting with Laurent operators.

*Step 1: Laurent operators only have approximate point spectrum.* Consider a Laurent operator $T(q)$. The spectrum of $T(q)$ is given by the symbol curve, which we define in Section 3.1, (cf. Theorem 3.3). The symbol curve is a curve given by a polynomial of finite degree (see Appendix A.7) and the trace of the symbol curve must therefore equal its boundary. Hence, since the boundary of the spectrum is approximate point spectrum by [10, Problem 1.18] it holds that

$$\sigma(T(q)) = \partial\sigma(T(q)) \subset \sigma_{\mathrm{appt}}(T(q)).$$

Thus, for fixed $q \in [0, 2\pi]$ and $\lambda \in \sigma(T(q))$ there exists a Weyl sequence $\{v_{q,n}\}_{n \in \mathbb{N}}$ with $\|v_{q,n}\| = 1$ such that $\|(T(q) - \lambda)v_{q,n}\| \to 0$ as $n \to \infty$ (cf. Lemma A.10) .

*Step 2: Direct integrals of Laurent operators only have approximate point spectrum.* Let us prove that $\lambda \in \sigma(\mathcal{T})$ is part of the approximate point spectrum, i.e. we construct a Weyl sequence for $\mathcal{T}$. From Corollary 3.14 it holds that $\sigma(\mathcal{T}) = \bigcup_{q \in [0, 2\pi]} \sigma(T(q))$, so let $\lambda \in \sigma(T(q_0))$. From Step 1 we have a Weyl sequence $v_n$ corresponding to $\lambda$ for the operator $T(q_0)$. Now, define $w_n = \int_{[0, 2\pi]}^{\oplus} v_n \sqrt{n} \mathbb{1}_{[q_0 - \frac{1}{2n}, q_0 + \frac{1}{2n}]} \, dq$. Then

$$\|w_n\|^2 = \int_{[0, 2\pi]} \|v_n\|^2 n \mathbb{1}_{[q_0 - \frac{1}{2n}, q_0 + \frac{1}{2n}]} \, dq = 1.$$

Furthermore,

$$\left\| \int_{[0, 2\pi]}^{\oplus} (T(q) - \lambda) dq w_n \right\|^2 = \int_{q_0 - \frac{1}{2n}}^{q_0 + \frac{1}{2n}} n \|(T(q) - \lambda)v_n\|^2 dq$$

$$\leq \int_{q_0 - \frac{1}{2n}}^{q_0 + \frac{1}{2n}} n \left( \|(T(q) - T(q_0))v_n\| + \|(T(q_0) - \lambda)v_n\| \right)^2 dq \to 0,$$

where we used the continuity bound that $\|(T(q) - T(q_0))\| \to 0$ as $q \to q_0$ (cf. Lemma A.10) as well as $\|(T(q_0) - \lambda)v_n\| \to 0$ as $n \to \infty$.

*Step 3: Direct integrals of Laurent operators with finite range perturbations only have approximate point spectrum.* Assume that $\lambda \in \sigma(\mathcal{L})$. Then we split up into cases according to whether $\lambda \in \sigma(\mathcal{T})$

*3a) Spectrum of the non-Hermitian evolution is approximate point:* Suppose first that $\lambda \in \sigma(\mathcal{T})$. Then translation-invariance of $T(q)$ means that $S_1 T(q) S_{-1} = T(q)$. Consider a Weyl sequence $v_n$ for $T(q)$. Then $\{S_a v_n\}_{n \in \mathbb{N}}$ is also a Weyl sequence for $T(q)$ corresponding to $\lambda$ for any $a \in \mathbb{Z}$ since

$$\|T(q) S_a v_n\| = \|S_a T(q) S_{-a} S_a v_n\| = \|S_a T(q) v_n\| = \|T(q) v_n\| \to 0.$$

Since $F(q)$ is finite range and $v_n$ is finite norm for every $\varepsilon > 0$ there exist an $a \in \mathbb{Z}$ such that $\|F(q) S_a v_n\| \leq \varepsilon$. Thus, there exists a sequence $a_n$ such that $\|F(q) S_{a_n} v_n\| \leq \frac{1}{n}$. Now, $w_n = S_{a_n} v_n$ is a Weyl sequence for $T(q) + F$ since

$$\|(T(q) + F) S_n v_n\| \leq \|T(q) v_n\| + \|F S_n v_n\| \to 0.$$

By an argument as in Step 2, we can use continuity to conclude that this eigenvector in the fiber gives rise to an approximate eigenvector in the direct integral.

*3b) Additional spectrum from quantum jump terms is approximate point :* Suppose that $\lambda \notin \sigma(\mathcal{T})$ then by Corollary 3.15 there is a $q$ such that $\langle \Gamma_R | (T(q) - \lambda)^{-1} | \Gamma_L \rangle = -1$. Now, consider the matrix acting on the vector $v = (T(q) - \lambda)^{-1} | \Gamma_L \rangle$

$$(T(q) - \lambda + |\Gamma_L\rangle\langle\Gamma_R|) \, (T(q) - \lambda)^{-1} \, |\Gamma_L\rangle = |\Gamma_L\rangle + \underbrace{\langle \Gamma_R | (T(q) - \lambda)^{-1} |\Gamma_L\rangle}_{=-1} |\Gamma_L\rangle = 0.$$

So we conclude that $(T(q) + |\Gamma_L\rangle\langle\Gamma_R|)v = \lambda v$ which means that $(T(q) - \lambda)^{-1} |\Gamma_L\rangle$ is an eigenvector of $(T(q) + |\Gamma_L\rangle\langle\Gamma_R|)$ with eigenvalue $\lambda$. Again, by an argument as in Step 2, we can use continuity to conclude that this eigenvector in the fiber gives rise to an approximate eigenvector in the direct integral. □

Theorem 4.1 shows that every point in the spectrum has a corresponding Weyl sequence. Although Weyl sequences are not eigenfunctions they hint towards what an eigenfunction looks like. Now, we prove that the spectrum with origin in the quantum jumps terms are approximately classical, which is a notion we define as follows. We caution the reader that we only consider $\mathcal{L} \in \mathcal{B}(\mathrm{HS}(\mathcal{H}))$ and so the Weyl sequences are also only in the Hilbert-Schmidt sense.

We say that such a Weyl sequence $\{\rho_n\}_{n\in\mathbb{N}} \subset \mathrm{HS}(\mathcal{H})$ is *approximately classical* if (in the position basis $\{|k\rangle\}_{k\in\mathbb{Z}}$) it holds that there exist $C, \mu > 0$ such that for all $n$ sufficiently large and $x, y \in \mathbb{Z}$ it holds that

$$|\langle x|\rho_n|y\rangle| \leq Ce^{-\mu|x-y|}.$$

**Proposition 4.2.** *Let $\mathcal{L}$ satisfy the conditions of Corollary 3.15 and let $\mathcal{T} = \int_{[0,2\pi]}^{\oplus} T(q)dq$ where $T(q)$ is tridiagonal for each $q \in [0, 2\pi]$. Suppose that $\lambda \in \sigma(\mathcal{L}) / \sigma(\mathcal{T})$ then there exists a Weyl sequence for $\lambda$ which is approximately classical.*

*Proof.* By Corollary 3.15 there is a $q_0$ such that $(T(q_0) - \lambda)$ is invertible and it holds that

$$\langle\Gamma_R|(T(q_0) - \lambda)^{-1}|\Gamma_L\rangle = -1.$$

This implies

$$((T(q_0) - \lambda) + |\Gamma_L\rangle\langle\Gamma_R|)((T(q_0) - \lambda)^{-1}|\Gamma_L\rangle) = |\Gamma_L\rangle + \langle\Gamma_R|(T(q_0) - \lambda)^{-1}|\Gamma_L\rangle|\Gamma_L\rangle = 0.$$

So $(T(q_0) - \lambda)^{-1}|\Gamma_L\rangle$ is in the kernel of $((T(q_0) - \lambda) + |\Gamma_L\rangle\langle\Gamma_R|)$ for fixed $q_0$. Let $N = \|(T(q_0) - \lambda)^{-1}|\Gamma_L\rangle\|$. Then, as before, define for each $n \in \mathbb{N}$ the vector $v_n$ by

$$v_n = \frac{\sqrt{n}}{N} \int_{[q_0 - \frac{1}{2n}, q_0 + \frac{1}{2n}]}^{\oplus} (T(q_0) - \lambda)^{-1}|\Gamma_L\rangle \ dq.$$

Then $\|v_n\| = 1$ and $v_n$ is a Weyl sequence for $\int_{[0,2\pi]}^{\oplus}(T(q) + |\Gamma_L\rangle\langle\Gamma_R|)dq$ since by continuity

$$\left\| \int_{[0,2\pi]}^{\oplus} (T(q) + |\Gamma_L\rangle\langle\Gamma_R|)dq \int_{[q_0 - \frac{1}{2n}, q_0 + \frac{1}{2n}]}^{\oplus} \frac{\sqrt{n}}{N} (T(q_0) - \lambda)^{-1}|\Gamma_L\rangle \ dq \right\|^2$$
$$= \frac{n}{N^2} \int_{q_0 - \frac{1}{2n}}^{q_0 + \frac{1}{2n}} \left\| (T(q) + |\Gamma_L\rangle\langle\Gamma_R|)(T(q_0) - \lambda)^{-1}|\Gamma_L\rangle \right\|^2 dq \to 0.$$

In the case where $(T(q_0) - \lambda)$ is tridiagonal, we can explicitly determine the inverse. We now want to unwind the Fourier transform. Back in $l^2(\mathbb{Z}) \otimes \ell^2(\mathbb{Z})$ we get that

$$v_n = \frac{\sqrt{n}}{N} \sum_{x,y \in \mathbb{Z}} \int_{q_0 - \frac{1}{2n}}^{q_0 + \frac{1}{2n}} e^{iqx} \langle y| \ (T(q_0) - \lambda)^{-1} |\Gamma_L\rangle \ dq |x\rangle \ |y\rangle.$$

Now, as $|\Gamma_L\rangle$ is local around 0, we can consider the matrix elements $\langle y| \ (T(q_0) - \lambda)^{-1} |j\rangle$, which by Lemma A.13 are exponentially decaying in $|y|$. Thus,

$$|v_n(x,y)| \ \leq \frac{\sqrt{n}}{N} \int_{q_0 - \frac{1}{2n}}^{q_0 + \frac{1}{2n}} \left| \langle y| \ (T(q_0) - \lambda)^{-1} |\Gamma_L\rangle \right| \ dq \leq e^{-c|y|}.$$

for some $c > 0$ and thus $v_n$ is an approximately classical Weyl sequence corresponding to $\lambda$. $\quad\square$

For example, in case $|\Gamma_L\rangle = |0\rangle$ we get that $\langle y| \ (T(q_0) - \lambda)^{-1} |0\rangle = \frac{\lambda_2(q_0)^y}{\sqrt{(\beta - z)^2 - 4\alpha\gamma}}$ where $|\lambda_2(q_0)| < 1$. Thus, we see that all coefficients are exponentially decaying away from the diagonal. From $|\lambda_2(q_0)|$ we can even calculate the coherence length. In Section 6.2 we prove a weak extension, i.e. there cannot be additional spectrum with Weyl sequences which are not concentrated along the diagonal. In that sense, we can say that the states are classical.

Notice that there is no eigenvector in the direct integral picture and this is intimately related to the absence of steady states and which space the operator is defined on. In particular, notice that for the case of dephasing noise if we instead defined $\mathcal{L}$ on $\mathcal{B}(\mathcal{B}(\mathcal{H}))$ then $\mathcal{L}(\mathbb{1}) = 0$ and in that case, the identity would be a steady state which normalizable. Since $\mathbb{1}$ is neither Hilbert-Schmidt nor trace-class $\mathcal{L}$ does not have a steady state although $0 \in \sigma(\mathcal{L})$.

## 4.2 Sufficient conditions for convergence of finite volume spectra to infinite volume spectra

In this section, we achieve a first result for convergence of the spectra finite volume Lindbladians with periodic boundary conditions to their infinite volume spectra counterparts.

To do that, we study the convergence behaviour of spectra of finite size approximations of Laurent and Toeplitz matrices to their infinite counterpart, a topic of central interest in numerical analysis [55]. For Lindbladians this subject is, to our knowledge, still untouched, but in view of Corollary 3.14 under our assumptions, we can in some sense reduce the question of whether $\sigma(\mathcal{L}_N^{\mathrm{per}}) \to \sigma(\mathcal{L})$ to the better studied questions of convergence in the case of Laurent and Toeplitz matrices. Since the spectrum of an operator in finite volume is the set of eigenvalues, the spectrum is independent of which Schatten class we think of the operator as acting on. This provides further motivation for studying the spectrum of $\mathcal{L}$ as an operator on $\mathrm{HS}(\mathcal{H})$.

To study convergence of subsets of $\mathbb{C}$ we use the Hausdorff metric which is a measure of distance between subsets $X, Y \subset \mathbb{C}$ defined by

$$d_H(X,Y) = \max\left\{\sup_{x \in X} d(x,Y), \sup_{y \in Y} d(X,y)\right\}.$$

where $d$ is the distance in $\mathbb{C}$. However, we will not use the definition, but only its characterization in Theorem 4.3 below.

Following [66] for a sequence of subsets of the complex plane $\{S_n\}_{n\in\mathbb{N}}$ which are all nonempty define $\liminf_{n\to\infty} S_n$ as the set of all $\lambda \in \mathbb{C}$ that are limits of a sequence $\{\lambda_n\}_{n\in\mathbb{N}}$ which satisfies $\lambda_n \in S_n$. Conversely, $\limsup_{n\to\infty} S_n$ is defined as all subsequential limits of such sequences $\{\lambda_n\}_{n\in\mathbb{N}}$ with $\lambda_n \in S_n$. A central characterization of the Hausdorff metric is then the following.

**Theorem 4.3.** *Let $S$ and the members of the sequence $\{S_n\}_{n\in\mathbb{N}}$ be nonempty compact subsets of $\mathbb{C}$ then $S_n \to S$ in the Hausdorff metric if and only if*

$$\limsup_{n\to\infty} S_n = S = \liminf_{n\to\infty} S_n.$$

See [67, Sections 3.1.1 and 3.1.2] or [68, Section 2.8] for a proof. In [66] convergence is proven for periodic boundary conditions for tridiagonal matrices and diagonal perturbations except that it has not been proven that the symbol curve is fully captured by the finite size approximations. We will not use the result, but we state it for comparison since it is of a similar flavour as the condition that we have in Theorem 4.5. Recall the definition of $T_n^{\mathrm{per}}$ from Theorem 3.18.

**Theorem 4.4** ([66, Corollary 1.3]). *Suppose that $T$ is tridiagonal with $\alpha, \beta, \gamma \in \mathbb{C}$ on the diagonal and that $K = \mathrm{diag}\,(K_{11}, \ldots, K_{mm})$ for some $m \in \mathbb{N}$. Then*

$$\lim_{n\to\infty} \left(\sigma(T_n^{\mathrm{per}} + P_n K P_n) \cup \sigma(T)\right) = \sigma(T + K)$$

*where $P_n$ is the projection onto the sites $\{0, \ldots, n\}$.*

It is further conjectured that $\lim_{n\to\infty} \sigma(T_n^{\mathrm{per}} + K) = \sigma(T+K)$ under the same assumptions (see [66, Conjecture 7.3]).

Now, these results are not quite strong enough for our purposes, but we believe that one can show the assumption in the following theorem are satisfied at least for the model of local dephasing that we consider in Section 5.1. As we will see in Figures 5 and 6 in the next section it is important that we use periodic boundary conditions for $\mathcal{L}$.

Recall that $\mathbb{T}_n = \left\{\frac{2\pi k}{n} \mid k = 1, \ldots n\right\}$. Let us call a pair $(\{q_n\}_{n\in\mathbb{N}}, \{a_n\}_{\mathbb{N}})$ of a sequence $\{q_n\}_{n\in\mathbb{N}} \subset [0, 2\pi]$ and a sequence $\{a_n\}_{n\in\mathbb{N}} \subset \mathbb{N}$ *consistent* if $a_n \to \infty$ and $q \in \mathbb{T}_{a_n}$ for every $n \in \mathbb{N}$.

**Theorem 4.5.** *Suppose that $\mathcal{L}$ satisfies the conditions of Theorem 3.8 and let $T(q)$ be the corresponding bi-infinite $r$-diagonal Laurent operator and $F(q)$ the finite rank operator with finite range for each $q \in [0, 2\pi]$. Suppose further that $q \mapsto \sigma(T(q) + F(q))$ is continuous with respect to the Hausdorff metric. Assume for any consistent pair $(\{q_n\}_{n\in\mathbb{N}}, \{a_n\}_{n\in\mathbb{N}})$ where $q_n \to q_0$ that*

$$\sigma\left(T_{a_n}^{\mathrm{per}}(q_n) + F_{a_n}(q_n)\right) \to \sigma\left(T(q_0) + F(q_0)\right).$$

*Then it holds that as $n \to \infty$*

$$\sigma(\mathcal{L}_n^{\mathrm{per}}) \to \sigma(\mathcal{L}).$$

*Proof.* We use the characterization of Haussdorff convergence from Theorem 4.3 and compactness of spectra repeatedly. Let us first prove that $\sigma(\mathcal{L}) \subset \liminf_{N \to \infty} \sigma(\mathcal{L}_N^{\mathrm{per}})$. So let $\lambda \in \sigma(\mathcal{L})$, then by Corollary 3.14 it holds that $\lambda \in \sigma(T(q_0) + F(q_0))$ for some $q_0 \in [0, 2\pi]$. Now find a sequence $q_n \in \bigcup_{N \in \mathbb{N}} \mathbb{T}_N$ such that $q_n \to q_0$. By continuity of $q \mapsto \sigma(T(q) + F(q))$, we can find a sequence $\lambda_n \in \sigma(T(q_n) + F(q_n))$ such that $\lambda_n \to \lambda$.

For each $q_n$ there is a sequence $\{a_m\}_{m \in \mathbb{N}} \subset \mathbb{T}_{a_m}$ such that $a_m \to \infty$ and $q_n \in \mathbb{T}_{a_m}$ for each $m \in \mathbb{N}$. Now, since $\sigma\left(T_{a_m}^{\mathrm{per}}(q_n) + F_{a_m}(q_n)\right) \to \sigma\left(T(q_n) + F(q_n)\right)$ as $m \to \infty$. That is, there exists a sequence $\lambda_n^m \in \sigma\left(T_{a_m}^{\mathrm{per}}(q_n) - F_{a_m}(q_n)\right)$ such that $\lambda_n^m \to \lambda_n$ as $m \to \infty$. Then we can finish with a diagonal argument. I.e. for each $n$ find $k(n) \in \mathbb{N}$ such that $\left| \lambda_n^{k(n)} - \lambda_n \right| \leq \frac{1}{n}$ and then sequence $\lambda_n^{k(n)} \in \sigma\left(T_{a_{k(n)}}^{\mathrm{per}}(q_n) + F_{a_{k(n)}}(q_n)\right)$ satisfies that

$$\left| \lambda_n^{k(n)} - \lambda \right| \leq \left| \lambda_n^{k(n)} - \lambda_n \right| + |\lambda_n - \lambda| \to 0$$

as $n \to \infty$. Thus, we conclude that $\sigma(\mathcal{L}) \subset \liminf_{N \to \infty} \sigma(\mathcal{L}_N^{\mathrm{per}})$.

Let us then prove that $\limsup_{N \to \infty} \sigma(\mathcal{L}_N^{\mathrm{per}}) \subset \sigma(\mathcal{L})$. So let $\lambda_n \in \sigma(\mathcal{L}_{a_n}^{\mathrm{per}})$ and suppose that $\lambda_n \to \lambda$. By Theorem 3.18 it holds that

$$\sigma(\mathcal{L}_{a_n}^{\mathrm{per}}) = \bigcup_{q \in \mathbb{T}_{a_n}} \sigma\left(T_{a_n}^{\mathrm{per}}(q) + F_{a_n}(q)\right).$$

I.e. there is a sequence $q_n \in \mathbb{T}_{a_n}$ such that $\lambda_n \in \sigma\left(T_{a_n}^{\mathrm{per}}(q_n) + F_{a_n}(q_n)\right)$. Now, going to a subsequence $q_n \to q_0$ by compactness of $\mathbb{T}$ and so on that subsequence

$$\sigma\left(T_{a_n}^{\mathrm{per}}(q_n) + F_{a_n}(q_n)\right) \to \sigma\left(T(q_0) + F(q_0)\right).$$

by assumption. Since $\sigma\left(T(q_0) + F(q_0)\right) \subset \sigma(\mathcal{L})$ by Corollary 3.14 the theorem follows.

$\square$

## 4.3 Gaplessness of translation-covariant Lindblad generators

As a last application of the theory developed we prove gaplessness of translation-covariant Lindblad generators. We see in Theorem 4.10 how the Lindbladians that we study are gapless as long as we know that 0 is in the spectrum of $\mathcal{L}$. One might notice that the paper [7], which has a setup which is fairly similar to ours, assumes that there is a spectral gap. In a similar, but different way, this is also the case in [9]. Here, we say that $\mathcal{L}$ has a gap if 0 is an isolated point in the spectrum, where we with *isolated point* mean that there exist a ball $B_r(0)$ of radius $r > 0$ such that $\sigma(\mathcal{L}) \cap B_r(0) = \{0\}$.

To prove gaplessness we first go through some preliminaries. First, we recall Hurwitz's theorem from complex analysis (here in the formulation of [66], see references therein for a proof).

**Theorem 4.6** (Hurwitz). *Let $G \subset \mathbb{C}$ be an open set, let $f$ be a function that is analytic in $G$ and does not vanish identically, and let $\{f_n\}_{n \in \mathbb{N}}$ be a sequence of analytic functions in $G$ that converges to $f$ uniformly on compact subsets of $G$. If $f(\lambda) = 0$ for some $\lambda \in G$, then there is a sequence $\{\lambda_n\}$ of points $\lambda_n \in G$ such that $\lambda_n \to \lambda$ and $f_n(\lambda_n) = 0$ for all sufficiently large $n$.*

For some $r \in \mathbb{N}$ we let $T(q)$ be an $r$-diagonal Laurent operator with diagonals $a_i : S^1 \to \mathbb{C}$ smooth functions for $i = -r, \ldots, r$. The following lemma follows from [54, Theorem III-6.7]. Here and in the following $\varrho(A)$ denotes the resolvent set of $A$, that is the complement of the spectrum of $A$

**Lemma 4.7.** *Let $U$ be an open subset of $\varrho(T(q))$ and $k \in \mathbb{Z}$ then the function $R_q : U \to \mathbb{C}$ defined by*

$$R_q(z) = \langle 0 |, (T(q) - z)^{-1} | k \rangle$$

*is holomorphic.*

The next lemma follows from the resolvent equation, continuity and $r$-diagonality.

**Lemma 4.8.** *Let $q_n \to q_0$ be a convergent sequence in $S^1$ and suppose that $V$ is an open set such that $V \subset \varrho(T(q_n)) \cap \varrho(T(q_0))$ for all $n \in \mathbb{N}$. Assume that $\sup_{z \in V, n \in \mathbb{N}} \| (T(q_n) - z)^{-1} \| \leq C < \infty$ and $\sup_{z \in V} \| (T(q_0) - z)^{-1} \| \leq C < \infty$ for some $C > 0$. Then*

$$R_{q_n}(\cdot) \to R_{q_0}(\cdot)$$

*locally uniformly on $V$.*

*Proof.* Take any compact set $K \subset V$. Then by the resolvent equation, it follows that

$$
\begin{aligned}
|R_{q_n}(z) - R_{q_0}(z)| &= \left| \langle 0 |, (T(q_n) - z)^{-1} - (T(q_0) - z)^{-1} | k \rangle \right| \\
&\leq \left\| (T(q_n) - z)^{-1} - (T(q_0) - z)^{-1} \right\| \\
&\leq \left\| (T(q_n) - z)^{-1} \right\| \| T(q_n) - T(q_0) \| \left\| (T(q_0) - z)^{-1} \right\| \\
&\leq C^2 \| T(q_n) - T(q_0) \|
\end{aligned}
$$

Since $a_i$ are all smooth functions $\| T(q_n) - T(q_0) \| \to 0$ whenever $q_n \to q_0$ (cf. Lemma A.10). Thus, $|R_{q_n}(z) - R_{q_0}(z)| \to 0$ uniformly for $z \in K$. $\qquad \square$

The following lemma follows from the definition of local uniform convergence.

**Lemma 4.9.** *Suppose that $f_n^i : V \to \mathbb{C}$ is holomorphic for each $n \in \mathbb{N}$ $-r \leq i \leq r$ such that $\{f_n^i\}_{n \geq 0} \to f^i$ (lcu). Let further $a_i : S^1 \to \mathbb{C}$ be smooth functions and $q_n \to q$. Then*

$$\left\{ \sum_{i=-r}^{r} a_i(q_n) f_n^i \right\}_{n \in \mathbb{N}} \xrightarrow{\text{(lcu)}} \left\{ \sum_{i=-r}^{r} a_i(q) f^i \right\}_{n \in \mathbb{N}} .$$

We are now ready to prove the gaplessness of infinite-volume Lindbladians.

**Theorem 4.10.** *Suppose that $\mathcal{L}$ has $H = -\Delta$ and $\{L_k\}_{k \in \mathbb{Z}}$ satisfies $\mathcal{A}_2a) - \mathcal{A}_2c)$. Suppose $0 \in \sigma(\mathcal{L})$. Then $\mathcal{L}$ is gapless or has an infinite dimensional kernel.*

Notice, that if we add a random potential to $H$ and $\mathcal{L}$ is gapless then it stays gapless by Theorem 6.1.

*Proof.* If (as is the case in Section 5.2) the non-Hermitian evolution is gapless we are done by Corollary 3.14. So suppose that the non-Hermitian evolution is gapped around 0. By Lemma 4.7 the function $R_q(\cdot)$ is holomorphic on $\varrho(T(q))$ and so on the complement of $\bigcup_{q \in [0,2\pi]} \sigma(T(q)) = \sigma(\mathcal{T})$, $R_q$ are holomorphic for each $q \in [0, 2\pi]$. Thus, $0 \notin \sigma(\mathcal{T})$ then $\varrho(\mathcal{T})$ is an open set containing 0 and thus $B_{2r}(0) \subset \varrho(\mathcal{T})$ for some $r > 0$. Now, let $V = B_r(0)$ be the (open) ball around 0 with radius $r$.

Now, since $0 \in \sigma(\mathcal{L})$ it holds by Corollary 3.15 that there exists a $q_0 \in [0, 2\pi]$ such that

$$\langle \Gamma_R(q_0)|, (T(q_0) - z)^{-1} |\Gamma_L(q_0) \rangle + 1 = 0.$$

We prove that the assumptions of Lemma 4.8 are satisfied in Appendix A.6.

**Lemma 4.11.** *For the functions $R_q$ and $R_{q_0}$ and the set $V$ constructed above, the assumptions of Lemma 4.8 are satisfied.*

We conclude that for any sequence $q_n \to q_0$ if we define $f_n : V \to \mathbb{C}$ given by

$$f_n(z) = \langle \Gamma_R(q_n)|, (T(q_n) - z)^{-1} |\Gamma_L(q_n) \rangle + 1$$

as well as

$$f(z) = \langle \Gamma_R(q_0)|, (T(q_0) - z)^{-1} |\Gamma_L(q_0) \rangle + 1,$$

then by Lemma 4.9 we conclude that $\{f_n\}_{n \in \mathbb{N}} \overset{(\text{lcu})}{\to} f$ on $V$. Furthermore, it holds that $f(0) = 0$. Now, since $f$ is analytic and $f$ is not identically zero, then every zero of the function is isolated and there is a ball of some radius $r > 0$ such that $z = 0$ is the only zero of $f$ on $B_r(0)$. Then since $\{f_n\}_{n \in \mathbb{N}} \overset{(\text{lcu})}{\to} f$ by Hurwitz's theorem (cf. Theorem 4.6) for every $a > 0$ then for sufficiently large $n$ such that $f_n$ has a zero $z_n$ with $|z_n| < a$.

We define $I \subset [0, 2\pi]$ to be the set

$$I = \left\{ q \in [0, 2\pi] \mid \langle \Gamma_R(q)|, T(q)^{-1} |\Gamma_L(q) \rangle + 1 \neq 0 \right\},$$

that is all $q \in [0, 2\pi]$ such that $0 \notin \sigma(T(q) + F(q))$.

Now, either there is a sequence $\{q_n\}_{n \in \mathbb{N}} \subset I$ such that $q_n \downarrow q_0$ or there exists an $\varepsilon > 0$ such that $(q_0 - \varepsilon, q_0 + \varepsilon) \cap [0, 2\pi] \subset I^c$.

In the first case, that $f_n$ has a zero at $z_n$ means that $\langle \Gamma_R(q_n)|, (T(q_n) - z_n)^{-1} |\Gamma_L(q_n) \rangle = -1$, which again means that $z_n \in \sigma(\mathcal{L})$. As $\{q_n\}_{n \in \mathbb{N}} \subset I$ then $z_n$ converges to 0 without being equal to zero. In this case, $\mathcal{L}$ is gapless.

In the second case, for every $q \in (q_0 - \varepsilon, q_0 + \varepsilon) \cap [0, 2\pi] \subset I$ there exists a normalized vector $|v(q)\rangle \in \ell^2(\mathbb{Z})$ such that

$$(T(q) + F(q))|v(q)\rangle = 0.$$

Now, split $(q_0 - \varepsilon, q_0 + \varepsilon) \cap [0, 2\pi] \subset I$ up into $N$ disjoint intervals $I_1, \ldots I_N$ and define

$$w_i = \frac{1}{\sqrt{|I_i|}} \int_{[0, 2\pi]}^{\oplus} v(q) \mathbb{1}_{q \in I_i} \, dq.$$

Then $\langle w_i, w_j \rangle = \delta_{i,j}$ and it holds that

$$\mathcal{L} w_i = \frac{1}{\sqrt{|I_i|}} \int_{[0, 2\pi]}^{\oplus} (T(q) + F(q)) v(q) \mathbb{1}_{q \in I_i} \, dq = 0.$$

Thus, we conclude that the kernel of $\mathcal{L}$ is at least $N$ dimensional. Since this holds for any $N \in \mathbb{N}$ the kernel of $\mathcal{L}$ must be infinite-dimensional. $\qquad\square$

# 5 Examples of spectra of translation-covariant Lindbladians

In this section, we apply the techniques developed so far to several open quantum systems studied in the literature such as local dephasing and decoherent hopping. In all cases, we determine an expression for the spectrum of the Lindblad generator.

In the following, we study Lindbladians $\mathcal{L}$ defined on $\mathrm{HS}(\ell^2(\mathbb{Z}))$ which as in (1) are of the form

$$\mathcal{L}(\rho) = -i[H, \rho] + G \sum_k L_k \rho L_k^* - \frac{1}{2}(L_k^* L_k \rho + \rho L_k^* L_k),$$

where $G > 0$ is the strength of the dissipation and $L_k$ are the Lindblad operators.

We will use Theorem 3.8 to rewrite $\mathcal{L}$ as a direct integral of the operators $T(q) + F(q)$ where $T(q)$ is a banded Laurent operator (recall the definition from Section 3.3) and $F(q)$ is finite range and finite rank. Further by Corollary 3.14 the spectrum of $\mathcal{L}$ is the union of the spectra of $T(q) + F(q)$.

From now on we only consider the case where all of the $L_k$ are rank one. We saw how that implies that $F(q) = |\Gamma_L(q)\rangle\langle\Gamma_R(q)|$ is rank one for each $q \in [0, 2\pi]$ and by Corollary 3.15 can find as the union of $\bigcup_{q \in [0, 2\pi]} \sigma(T(q))$ with all solutions $z \in \mathbb{C}$ to the equation

$$\langle \Gamma_R(q)|, (T(q) - z)^{-1} |\Gamma_L(q)\rangle \ = -1.$$

Recall also the symbol curve introduced in Lemma 3.2. In concrete applications, we consider tridiagonal matrices $T(q)$ and therefore, we review some results about tridiagonal Laurent operators and their invertibility in Appendix A.7. If $T(q)$ is tridiagonal we let $\alpha, \beta, \gamma : [0, 2\pi] \to \mathbb{C}$ be the entries of $T(q)$. Thus, the symbol curve is, by Lemma 3.2, given by

$$a(z) = \alpha z^{-1} + \beta + \gamma z, z \in \mathbb{T}$$

which is a (possibly degenerate) ellipse. In particular, for the dephasing example considered in the next section, we see from (34) that $\alpha(q) = i(1 - e^{-iq}), \beta(q) = -G, \gamma(q) = i(1 - e^{iq})$.

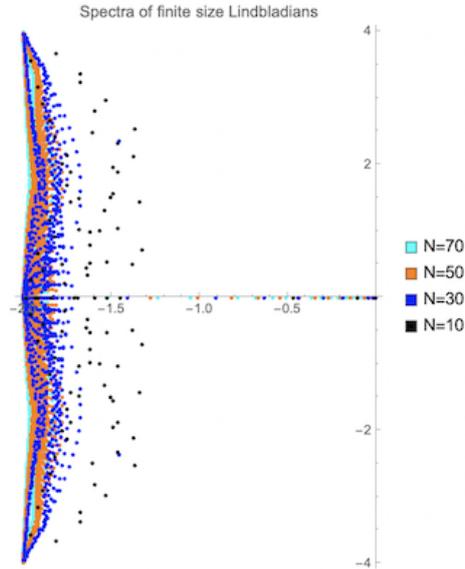

Figure 2: Spectrum of the Lindbladian corresponding to local dephasing (see (33)) restricted to $N = 10, 30, 50, 70$ lattice sites for $G = 2$ where we have larger $N$ plotted first and they therefore appear behind smaller values of $N$. Notice how it seems that most of the spectrum converges to $-2 + i[-4, 4]$ as $N$ gets larger, but that we also see some spectrum in the interval $[-2, 0]$.

## 5.1  Local dephasing

One prominent example, which has been discussed in the physics literature is local dephasing. The spectrum was investigated numerically with free boundary conditions in [22] and analytically many of the same considerations were made in finite volume with periodic boundary conditions [21, 69]. We plot some numerical results in finite volume in Figure 2.

The model is given by choosing the Hamiltonian as nearest neighbour hopping

$$H = -\tilde{\Delta} = -\sum_{k \in \mathbb{Z}} |k\rangle\langle k+1| \ + |k+1\rangle\langle k| \tag{32}$$

and the local Lindblad operators as local dephasing, i.e.  projectors onto single lattice sites $k \in \mathbb{Z}$

$$L_k = |k\rangle\langle k| \,. \tag{33}$$

We notice that each Lindblad generator satisfies $L_k = L_k^* = L_k^* L_k = |k\rangle\langle k|$. Now, define $Q$ by $Q = \sum_k L_k^* L_k = \mathbb{1}$ and therefore the model satisfies the assumptions of Theorem 3.8. Using Theorem 3.8, we can first determine the spectrum of the non-Hermitian part of the evolution $\mathcal{T} = \int_{[0,2\pi]}^{\oplus} T(q) dq$ in Theorem 3.8 without the quantum jumps $\mathcal{F} = \int_{[0,2\pi]}^{\oplus} F(q) dq$.

**Proposition 5.1** (Spectrum of NHE local dephasing)**.** *Let $H = -\Delta$ as in (32) and let the Lindblad operators $L_k$ be given as $|k\rangle\langle k|$, then the generator of the non-Hermitian evolution $\mathcal{T} \in \mathcal{B}(\mathrm{HS}(\ell^2(\mathbb{Z})))$ satisfies*

$$\sigma(\mathcal{T}) = -G + i[-4, 4].$$

*Proof.* Using Theorem 3.8 and $Q = \sum_k L_k^* L_k = \mathbb{1}$, we see that $T(q)$ is given by

$$T(q) = i(1 - e^{-iq})S + i(1 - e^{iq})S^* - G \ . \tag{34}$$

For fixed $q$ this corresponds to a tridiagonal Laurent operator with symbol curve $T(q, \theta) = i(1 - e^{-iq})e^{i\theta} + i(1 - e^{iq})e^{-i\theta} - G$. By Theorem 3.13 and Theorem 3.3, it follows that

$$\sigma(\mathcal{T}) = \bigcup_{q \in [0, 2\pi]} \sigma(T(q)) = \bigcup_{q \in [0, 2\pi]} \{T(q, \theta) \mid \theta \in [0, 2\pi]\} \ .$$

Putting in the explicit form of the parameters from (34), we find

$$T(q, \theta) = -G + 2i(\cos(\theta) - \cos(\theta - q)) \ .$$

Since $\theta$ and $q$ can be varied independently over the interval $[-\pi, \pi]$ we can realize any value in $-G + i[-4, 4]$ as claimed. $\qquad\square$

From (34) we see that $\alpha(q) = i(1 - e^{-iq}), \beta(q) = -G, \gamma(q) = i(1 - e^{iq})$ are the entries of the tridiagonal matrix $T(q)$.

**Remark 5.2** (Transfer matrix picture)**.** *Notice that the proof of Proposition 5.1 is in some sense equivalent to finding out that the polynomial $\alpha x^{-1} + \beta + \gamma x$ has a root of absolute value 1 if and only if $z \in -G + [-4i, 4i]$, where $\alpha, \beta, \gamma : [0, 2\pi] \to \mathbb{C}$ are the diagonals of the tridiagonal matrix $T(q)$. This has an interpretation in terms of transfer matrices: If one considers the Laurent operator with $\alpha, \beta$ and $\gamma$ on the diagonal and solves it iteratively then a root of $\alpha x^{-1} + \beta + \gamma x$ of absolute value less than 1 corresponds to an exponentially decaying and therefore normalizable solution. Conversely, a root of absolute value strictly larger than 1 corresponds to an exponentially increasing and therefore not normalizable solution. In this picture, one can interpret the rank-one perturbation $|\Gamma_R\rangle\langle\Gamma_L|$ as boundary condition at 0 for an iterative solution for positive and negative entries.*

Thus, we have computed the spectrum of the non-Hermitian Hamiltonian. Turning now to the full Lindblad generator $\mathcal{L}$, we can characterize its spectrum depending on the dissipation strength by including the quantum jump part as a perturbation to the non-Hermitian part.

**Proposition 5.3** (Full spectrum for local dephasing).  *Let $H = -\tilde{\Delta}$ be the (modified) Laplacian defined in (32) and let the Lindblad operators $L_k$ be given as $|k\rangle\langle k|$, then the Lindblad generator $\mathcal{L}$ satisfies*

$$\sigma(\mathcal{L}) = (-G + i[-4, 4]) \cup \begin{cases} [-G, 0], & \text{if } G \leq 4 \\ [-G + \sqrt{G^2 - 16}, 0], & \text{otherwise.} \end{cases}.$$

*Proof.* Since $T(q)$ is translation invariant its spectrum is essential. Thus, when $T(q)$ is compactly perturbed the spectrum can only be extended. So, as in the proof of Corollary 3.14, we get from Proposition 5.1 that

$$-G + [-4i, 4i] = \sigma(\mathcal{T}) \subset \sigma(\mathcal{L}).$$

The coefficients of the perturbation also follow from our consideration in Section 3. Based on Theorem 3.8, we find that $F(q) = G\,|0\rangle\langle 0|$, independent of $q$, which means that we may choose $|\Gamma_L\rangle = |\Gamma_R\rangle = \sqrt{G}\,|0\rangle$. Hence, both $|\Gamma_L\rangle$ and $|\Gamma_R\rangle$ are bounded vectors and $q \mapsto \langle 0|(T(q) - z)^{-1}|0\rangle$ is continuous for $z \notin -G + [-4i, 4i]$. Therefore, we may determine $\sigma(\mathcal{L})$ from Corollary 3.15 which leads to the condition

$$G\langle 0|(T(q) - z)^{-1}|0\rangle = -1. \tag{35}$$

Thus, we now have to compute $\langle 0|(T(q) - z)^{-1}|0\rangle$ which we elaborate on in Appendix A.7. Define $\lambda_+$ and $\lambda_-$ by (with a convention on taking square roots of complex numbers elaborated on in Appendix A.7 )

$$\lambda_{\pm} = -\frac{\beta}{2\gamma} \pm \sqrt{\left(\frac{\beta}{2\gamma}\right)^2 - \frac{\alpha}{\gamma}}. \tag{36}$$

Let further, $|\lambda_2| \leq |\lambda_1|$ such that $\{\lambda_1, \lambda_2\} = \{\lambda_+, \lambda_-\}$. Then notice that the conditions of Lemma A.12 are satisfied and we may therefore $|\lambda_2| < 1 < |\lambda_1|$. Thus, we can use Lemma A.13 to find the inverse of the Laurent operator to obtain

$$-1 = (-1)^{\mathbb{1}[|\lambda_-| < 1 < |\lambda_+|]} \frac{G}{\sqrt{(\beta - z)^2 - 4\gamma}}. \tag{37}$$

Our strategy is to square the equation, solve to find a set of possible $z$ and then reinsert into (37) to see which sign is correct. The potential solutions satisfy

$$z = \beta \pm \sqrt{G^2 + 4\alpha\gamma}.$$

In the specific example with dephasing noise, we saw that $\beta = -G$, $\alpha = -ie^{-iq} + i$ and $\gamma = -ie^{iq} + i$. Thus, strictly speaking, the considerations above only apply when $q \neq 0$, in the case $q = 0$ we see that $z = 0$ is the only solution to (35).

As $\alpha\gamma = 2(\cos(q) - 1)$, we consider

$$z = -G \pm \sqrt{G^2 + 8(\cos(q) - 1)}.$$

As $G^2 + 8(\cos(q) - 1) \in [-16 + G^2, G^2]$ it is natural to consider the cases $G < 4$ and $G \geq 4$. In the case $G < 4$ we have $-16 + G^2 < 0$. Thus,

$$\bigcup_{q \in [0, 2\pi]} \{\sqrt{G^2 + 8(\cos(q) - 1)}\} = i[0, \sqrt{16 - G^2}] \cup [0, G].$$

Therefore, the potential values of $z$ are

$$z \in [-G, 0] \cup [-2G, -G] \cup \left( -G + i \left[ -\sqrt{16 - G^2}, \sqrt{16 - G^2} \right] \right).$$

Notice that since $\left[ -\sqrt{16 - G^2}, \sqrt{16 - G^2} \right] \subset [-4i, 4i]$ this does not give additional spectrum.

It now remains to check the sign of the remaining possible solutions. From (37) we see that the square root must yield a (positive) real number and we at the same time need that $|\lambda_+| < 1 < |\lambda_-|$, so we only need to deal with the case where $0 \leq G^2 + 8(\cos(q) - 1) \leq G^2$. Then

$$\beta - z = \mp_z \sqrt{G^2 + 8(\cos(q) - 1)}.$$

Using $\mp_z$ and $\mp_\lambda$ to denote two, on the outset, independent signs we obtain from (36) that

$$2\gamma\lambda_\pm = -(\beta - z) \pm_\lambda \sqrt{(\beta - z)^2 - 4\alpha\gamma} = \pm_z \sqrt{G^2 + 8(\cos(q) - 1)} \pm_\lambda \sqrt{G^2}.$$

If $\pm_z = +$ then $|\lambda_-| < |\lambda_+|$ and $\pm_z = -$ then $|\lambda_+| < |\lambda_-|$. Thus, from (37) we see that $\pm_z = +$ is the only valid solution. Thus, only $z \in [-G, 0]$ are valid solutions.

In the case $G \geq 4$: It holds that $-16 + G^2 \geq 0$ and therefore there is only one segment $[\sqrt{G^2 - 16}, G]$. This observation translates into

$$z \in [-G + \sqrt{G^2 - 16}, 0] \cup [-2G, -G - \sqrt{G^2 - 16}].$$

Using a similar argument as above one finds that only the part $[-G + \sqrt{G^2 - 16}, 0]$ has the correct sign.

$\square$

We finish this subsection with the following remarks.

*Emergence of two timescales:* From Proposition 5.3 we see that if we change $G$ from a value below 4 to a value above 4 the spectrum transitions from being connected to consist of two connected components. Furthermore, increasing the dissipation strength $G$, the connected component of the spectrum containing $\{0\}$ shrinks. This indicates the emergence of two timescales in the dynamics. The first one corresponding to the fast decay at rate $e^{-tG}$ and a second one with a much slower decay. On the infinite lattice it is difficult to discuss the density of states, but it is noticed numerically in [22], that there are of the order of $L$ eigenvalues on the real axis close to 0 and $L^2 - L$ eigenvalues with real part close to $-G$. In general, such a phenomenon can be explained from symmetry as in [26, Appendix B.9].

*Heuristic calculation of the finite-volume spectral gap:* In the dephasing case, that we study in section 5.1, it holds the solutions on the curve are given by

$$z = -G + \sqrt{G^2 + 8(\cos(q) - 1)}.$$

Thus, naively we can let $q = \frac{2\pi}{N}$ and so $\cos\left(\frac{2\pi}{N}\right) - 1 \approx \frac{1}{2}\left(\frac{2\pi}{N}\right)^2 = \frac{2\pi^2}{N^2}$ and thus as $\sqrt{1+x} \approx 1 + \frac{x}{2}$ for small $x$ we obtain that

$$z \approx -G + \sqrt{G^2 + 8\frac{2\pi^2}{N^2}} \approx -G + G\left(1 + \frac{16\pi^2}{G^2N^2}\right) = \frac{16\pi^2}{GN^2}.$$

This formula was also obtained in finite volume with periodic boundary conditions by Znidanic in [22, (8)].

## 5.2 Non-normal dissipators

Next, we turn to a model with non-normal Lindblad operators. The following family of dissipative models was studied in [70, 25, 71, 8]. For the Hamiltonian $H$ we still take the discrete Laplacian given in (32). The Lindblad operators are of the form

$$L_k = \left(|k\rangle + e^{i\delta}|k+l\rangle\right)\left(\langle k| - e^{-i\delta}\langle k+l|\right) \tag{38}$$

for some $\delta \in [0, 2\pi]$ and $l \in \mathbb{N}$. Notice that $L_k$ is not normal. We have that

$$L_k^* L_k = 2\,|k\rangle\langle k| + 2\,|k+l\rangle\langle k+l| - 2e^{i\delta}\,|k\rangle\langle k+l| - 2e^{-i\delta}\,|k+l\rangle\langle k|.$$

Define $Q$ by

$$Q = \sum_{k\in\mathbb{Z}} L_k^* L_k = 4\mathbb{1} - 2D_l,$$

where $D_l$ is the operator which has $e^{i\delta}$ and $e^{-i\delta}$ on the $l$'th sub- and superdiagonal. To compute the form of the rank-one perturbation notice that all $\alpha_r$ from Theorem 3.8 are 0 except for $\alpha_0 = 1, \alpha_l = e^{i\delta}$. Similarly, $\beta_0 = 1, \beta_l = -e^{-i\delta}$ and all other entries are 0.

Thus, from Theorem 3.8 we obtain

$$\frac{|\Gamma_L\rangle}{\sqrt{G}} = \sum_{r_1,r_2} \alpha_{r_1}\,\overline{\alpha_{r_2}}e^{ir_1q}|r_2 - r_1\rangle = (1 + e^{ilq})|0\rangle + e^{i\delta}e^{ilq}|-l\rangle + e^{-i\delta}|l\rangle$$

as well as

$$\frac{\langle \Gamma_R|}{\sqrt{G}} = \sum_{r_1', r_2'} \beta_{r_1'} e^{-ir_1'q} \overline{\beta_{r_2'}} \langle r_2' - r_1'| = (1 + e^{-ilq}) \langle 0| - e^{-i\delta} e^{-ilq} \langle -l| - e^{i\delta} \langle l|.$$

We see that also the operator non-Hermitian evolution $T$ is given by

$$\left(-i\tilde{\Delta} - \frac{G}{2}Q\right) \otimes \mathbb{1} + \mathbb{1} \otimes \left(i\tilde{\Delta} - \frac{G}{2}\overline{Q}\right) = \left(-i\tilde{\Delta} + GD_l\right) \otimes \mathbb{1} + \mathbb{1} \otimes \left(i\tilde{\Delta} + G\overline{D_l}\right) - 4G\mathbb{1} \otimes \mathbb{1}$$

which is tridiagonal. Let us examine how this $Q$ transforms under the shift and Fourier transformation in the first variable that we do in Theorem 3.8. The term $-4G\mathbb{1} \otimes \mathbb{1}$ is left invariant. The term $(iGD_l) \otimes \mathbb{1}$ becomes:

$$Ge^{i\delta} e^{-ilq} \sum_j |j\rangle\langle j - l| + Ge^{-i\delta} e^{iql} \sum_j |j\rangle\langle j + l|.$$

For the term $\mathbb{1} \otimes \left(G\overline{D_l}\right)$ we get

$$Ge^{-i\delta} \sum_j |j\rangle\langle j + l| + Ge^{i\delta} \sum_j |j\rangle\langle j - l|.$$

In total,

$$T(q) = G\sum_j e^{-i\delta} \left(1 + e^{iql}\right) |j\rangle\langle j + l| + e^{i\delta} \left(1 + e^{-ilq}\right) |j\rangle\langle j - l| - 4G\mathbb{1} \otimes \mathbb{1} + i(1 - e^{-iq})S + i(1 - e^{iq})S^*.$$

We can see that the full operator has non-zero entries in 5 diagonals if $l > 1$ so from now on we will assume that $l = 1$ which is also the case mainly studied in [25] and [70]. In that case, we find the infinite band matrix with diagonals:

$$\alpha = i(1 - e^{iq}) + Ge^{-i\delta} \left(1 + e^{iq}\right) \qquad \beta = -4G \qquad \gamma = i(1 - e^{-iq}) + Ge^{i\delta} \left(1 + e^{-iq}\right).$$

Let us first determine the union of the $q$-wise spectra without the perturbation $F(q)$. See also Figure 4 for an illustration.

**Proposition 5.4.** *The spectrum of the non-Hermitian evolution for the Lindbladian from* (38) *in the case $l = 1$ is given by*

$$\sigma(\mathcal{T}) = \bigcup_{q, \theta \in [0, 2\pi]} \{-4G + 2i\cos(\theta) - 2i\cos(q - \theta) + 2G\cos(\delta + \theta) + 2G\cos(q - \delta - \theta)\}.$$

*For $\delta = 0$ and $\delta = \pi$ this set is the convex envelope of the points $-8G, -4i, 0, 4i$.*

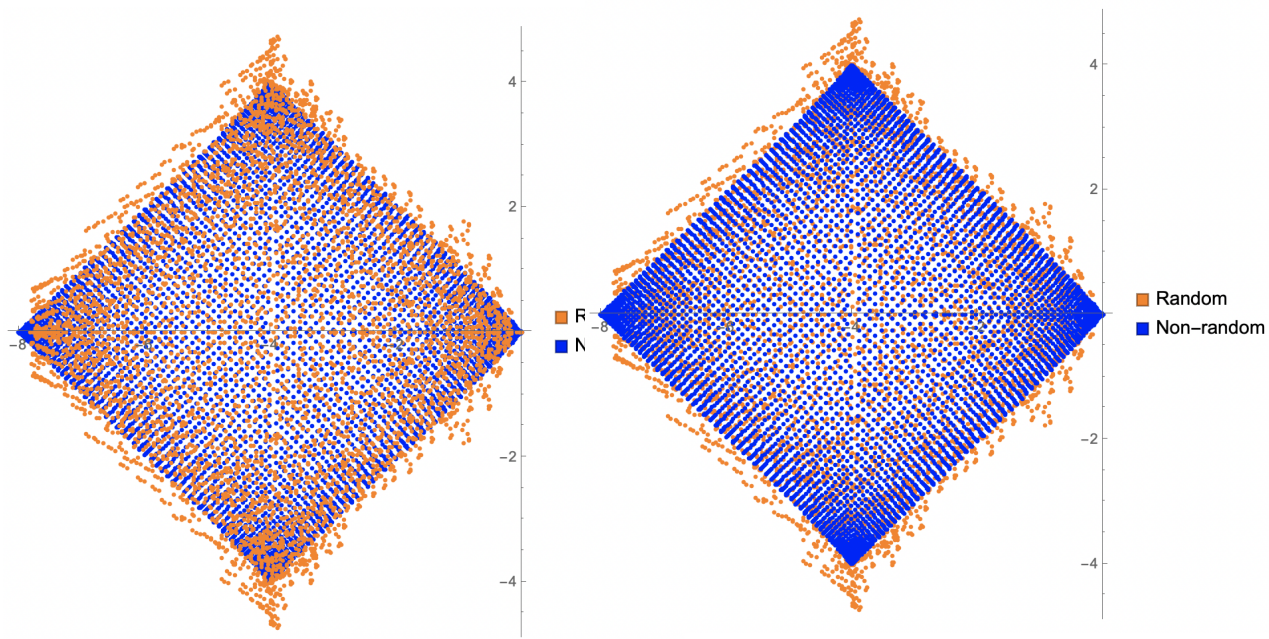

Figure 3: Spectrum $\mathcal{L}$ with non-normal Lindblad operators given by (38) with a random potential in orange and without in blue in the complex plane. In the right picture blue points are plotted on the top and on the left it is the orange points. In this case $G = 1$, the lattice size $n = 70$ and the support of the distribution of the strength of the external potential $V = 2$. Notice how the blue points are ordered very regularly except that there is a vertical hole in the middle and those eigenvalues tend to collapse to the real axis. It seems that the main effect of the external field is to push the eigenvalues vertically.

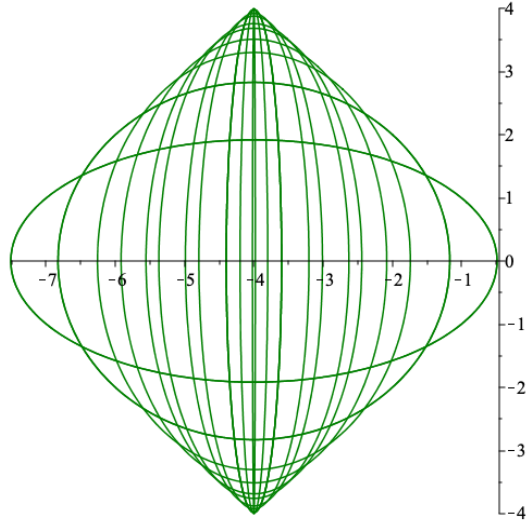

Figure 4: The ellipses corresponding to the spectrum of the Laurent matrices $T(q)$ in the model with non-normal Lindblad operators see for example (39). For the plot, we chose $\delta = 0$ corresponding to Proposition 5.4. Notice how the union of the ellipses make up the quadrilateral described in Proposition 39, compare also to the shape in Figure 3.

*Proof.* As before, it holds by Corollary 3.14 and the fact that the spectrum of a Laurent operator is the image of the symbol curve that

$$\sigma(\mathcal{T}) = \bigcup_{q \in [0,2\pi]} \sigma(T(q)) = \bigcup_{q \in [0,2\pi]} \left\{ \delta(q)e^{-i\theta} + \beta + \gamma(q)e^{i\theta} \mid \theta \in [0,2\pi] \right\}. \tag{39}$$

The result now follows from a direct computation. For $\delta = 0$ and $\delta = \pi$ it reduces to

$$\sigma(\mathcal{T}) = \bigcup_{q,\theta \in [0,2\pi]} \left\{ -4G + \cos(\theta)\left(2i \pm 2G\right) + \cos(q-\theta)\left(-2i \pm 2G\right) \right\}$$

$$= \bigcup_{x,y \in [-2,2]} \left\{ -4G + x\left(i \pm G\right) + y\left(-i \pm G\right) \right\} = \text{conv}(-8G, -4i, 0, 4i).$$

$\square$

Thus, we have proven that $\mathcal{T}$ is gapless even before adding the quantum jump term and therefore also afterwards when considering the full $\mathcal{L}$. We now turn to study the spectral effects of the perturbation.

Notice that for $\delta = 0$ then $\alpha\gamma = 2G^2\cos(q) + 2G^2 + 2\cos(q) - 2 \in \mathbb{R}$. We now ask whether the set

$$\bigcup_{q \in [0,2\pi]} \left\{ z \in \mathbb{C} \mid \langle \Gamma_R | (T(q) - z)^{-1} | \Gamma_L \rangle = -1 \right\}$$

increases the spectrum. Absorbing $z$ into $\beta$ yields that $\beta = -4G - z$ and from Theorem 3.8 we find that $|\Gamma_L\rangle$ and $\langle\Gamma_R|$ have all entries zero except for the $-1, 0, 1$st entries, and that part of each of the vectors are

$$|\Gamma_L\rangle = \sqrt{G} \begin{pmatrix} e^{iq} \\ 1 + e^{iq} \\ 1 \end{pmatrix} \text{ as well as } \langle\Gamma_R| = \sqrt{G} \begin{pmatrix} -1 & 1 + e^{-iq} & -e^{-iq} \end{pmatrix},$$

in the basis $\{\,|-1\rangle, |0\rangle, |1\rangle\,\}$ highlighting that we do allow for $q$-dependent $|\Gamma_L\rangle$ and $\langle\Gamma_R|$.

Now, let $\omega = \sqrt{\beta^2 - 16\alpha\gamma}$ and then from Lemma A.13 the equation reduces to

$$
\begin{aligned}
-\frac{\omega}{G} = \frac{\langle\Gamma_R|\; T^{-1}\;|\;\Gamma_L\rangle}{G} &= \begin{pmatrix} -1 & 1 + e^{-iq} & -e^{-iq} \end{pmatrix} \begin{pmatrix} 1 & \frac{1}{\lambda_1} & \frac{1}{\lambda_1^2} \\ \lambda_2 & 1 & \frac{1}{\lambda_1} \\ \lambda_2^2 & \lambda_2 & 1 \end{pmatrix} \begin{pmatrix} e^{iq} \\ 1 + e^{iq} \\ 1 \end{pmatrix} \\
&= -e^{-iq} + 2 + 2\cos(q) - e^{iq} + \lambda_1^{-1}\left(-1 - e^{-iq} + e^{iq} + 1\right) \\
&\quad - \lambda_1^{-2} + \lambda_2(1 + e^{-iq} - 1 - e^{iq}) - \lambda_2^2 \\
&= 2 + 2\left(\frac{1}{\lambda_1} - \lambda_2\right)\sin(q) - \left(\frac{1}{\lambda_1^2} + \lambda_2^2\right),
\end{aligned}
$$

where $\lambda_1, \lambda_2$ are defined right after (36). This equation in $z$ can be numerically solved for fixed $q$ (and $G$). Numerical evidence suggests that the curve stays inside the quadrilateral $\mathrm{conv}(-8G, -4i, 0, 4i)$ and therefore leads us to the following conjecture.

**Conjecture 5.5.** *For the Lindbladian $\mathcal{L}$ with Lindblad operators given by (38) and non-Hermitian evolution $A$ it holds that*

$$\sigma(\mathcal{L}) = \sigma(\mathcal{T}) = \mathrm{conv}(-8G, -4i, 0, 4i).$$

The reader can further compare with the plot of the spectrum in finite volume in Figure 3. Notice further, how Corollary 3.14 gives us the inclusion $\sigma(\mathcal{L}) \supset \sigma(\mathcal{T})$. In particular, we obtain that $\mathcal{L}$ is gapless (independently of the conjecture). Interestingly, this gaplessness could be related to the dynamical behaviour in a random potential observed in [25]. A topic that we return to in Section 6.

## 5.3 Incoherent hopping

We now turn to an incoherent hopping studied numerically in [22]. Here, the Lindblad dissipators are hopping terms, they are given by $L_k = |k\rangle\langle k + l|$. The numerical finding for finite sections with free boundary conditions of the lattice [22] is that the gap is uniformly positive as the length of the lattice increases. We find the spectrum for periodic boundary conditions.

**Theorem 5.6.** *Let $l \in \mathbb{Z}$. For the Lindbladian with $H = -\Delta$ and $L_k = |k\rangle\langle k+l|$ it holds that*

$$\sigma(\mathcal{L}) = -G + i[-4, 4] \cup \bigcup_{q \in [0, 2\pi]} \left\{ -G \pm_q \sqrt{e^{-2iql}G^2 + 8(\cos(q) - 1)} \right\}$$

*where $\pm_q$ is either $+$ or $-$ for each $q \in [0, 2\pi]$.*

*Proof.* Notice that $L_k^* L_k = |k+l\rangle\langle k+l|$ so still $Q = \sum_k L_k^* L_k = \mathbb{1}$. so $\left(-i\tilde{\Delta} - \frac{G}{2}Q\right) \otimes \mathbb{1} + \mathbb{1} \otimes \left(i\tilde{\Delta} - \frac{G}{2}\overline{Q}\right)$ is tridiagonal with

$$\alpha = i(1 - e^{iq}), \quad \beta = -G, \quad \gamma = i(1 - e^{-iq})$$

on the diagonals. Furthermore,

$$F(q) = G \left( \sum_{r_1, r_2} \alpha_{r_1} \overline{\alpha_{r_2}} e^{ir_1 q} |r_2 - r_1\rangle \right) \left( \sum_{r_1', r_2'} \beta_{r_1'} e^{-ir_1' q} \overline{\beta_{r_2'}} \langle r_2' - r_1'| \right)$$

with $\alpha_r = \delta_{r,0}$ and $\beta_r = \delta_{r,1}$. Thus,

$$|\Gamma_L\rangle = \sqrt{G} \sum_{r_1, r_2} \alpha_{r_1} \overline{\alpha_{r_2}} e^{ir_1 q} |r_2 - r_1\rangle = \sqrt{G}|0\rangle \quad \text{and}$$

$$\langle \Gamma_R| = \sqrt{G} \left( \sum_{r_1', r_2'} \beta_{r_1'} e^{-ir_1' q} \overline{\beta_{r_2'}} \langle r_2' - r_1'| \right) = \sqrt{G} e^{-iql} \langle 0|.$$

That means,

$$\langle \Gamma_R| \frac{1}{T - z} |\Gamma_L\rangle = e^{-iql} G \langle 0| \frac{1}{T - z} |0\rangle = \frac{(-1)^{\mathbb{1}[|\lambda_-| \, < 1 < |\lambda_+|]} e^{-iql} G}{\sqrt{(\beta - z)^2 - 4\alpha\gamma}}.$$

So by Corollary 3.15 we need to solve

$$-1 = (-1)^{\mathbb{1}[|\lambda_-| \, < 1 < |\lambda_+|\,]} \frac{e^{-iql} G}{\sqrt{(\beta - z)^2 - 4\alpha\gamma}}. \tag{40}$$

Squaring yields

$$1 = \frac{e^{-2iql} G^2}{(\beta - z)^2 - 4\alpha\gamma},$$

so solving for $z$ gives

$$z = \beta \pm \sqrt{e^{-2iql}G^2 + 4\alpha\gamma} = -G \pm \sqrt{e^{-2iql}G^2 + 8(\cos(q) - 1)},$$

where we again, as in the proof of Proposition 5.3, have to throw some of the solutions away according to get the correct sign in (40). $\qquad \square$

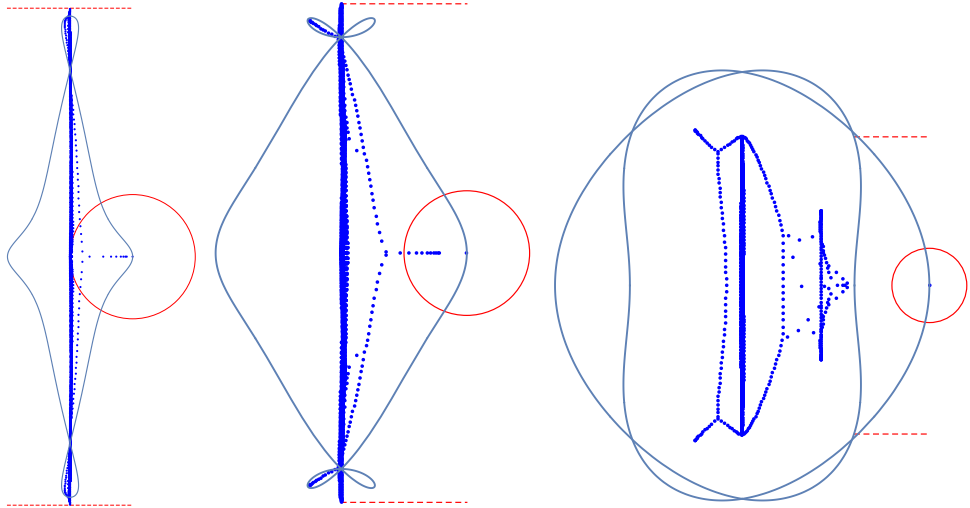

Figure 5: Exact diagonalization of the Lindbladian $\mathcal{L}$ in Section 5.3 for $l = 1$, $G = 1, 2, 5$ and $N = 70$ comparison of the predicted curve with numerics with free boundary conditions. The red circle is the unit circle. Notice how the two curves do not match due to the non-Hermitian Skin effect.

In Figure 5 and 6 we explicitly plot the solutions as a function of $q$ as well as the spectra obtained by exact diagonalization of $\mathcal{L}$ in finite volume. Notice how the predicted spectrum fits well with the numerical spectra for finite size systems only for the system with periodic boundary conditions, as is consistent with Theorem 4.5. This dramatic dependence on boundary conditions is sometimes called the non-Hermitian Skin effect [72, 73, 74]. It states that the spectra of non-Hermitian operators may exhibit dramatic dependence on boundary conditions. One could also view the difference between Toeplitz and Laurent operators this way. Furthermore, it is a feature of the non-Hermitian skin effect that the spectrum is pushed inwards and real eigenvalues start to appear [75]. This effect we also seen in Figure 5 and 6.

## 5.4 Single particle sector of a quantum exclusion process

In a many-body setting a model with $L_k = |k\rangle\langle k+1|$ and $L'_k = |k+1\rangle\langle k|$ was studied analytically in [76]. We briefly discuss how to adapt our methods to the single particle sector of that case and rederive the spectrum of the Lindbladian. Going through the proof of Theorem 3.8, we see that we just get two independent contributions that diagonalize in the same way. Since we still use discrete Laplacian as the Hamiltonian, it holds that $\alpha = i(1 - e^{iq})$ and $\gamma = i(1 - e^{-iq})$ are unchanged. On the other hand, we get the sum $\sum_k L_k^* L_k = \mathbb{1}$ twice, which means that $\beta = -2G$. Therefore we have from Proposition 5.1 the spectrum of the non-Hermitian evolution $\sigma(\mathcal{T}) = -2G + [-4i, 4i]$.

We now turn to the spectrum of the full Lindbladian.

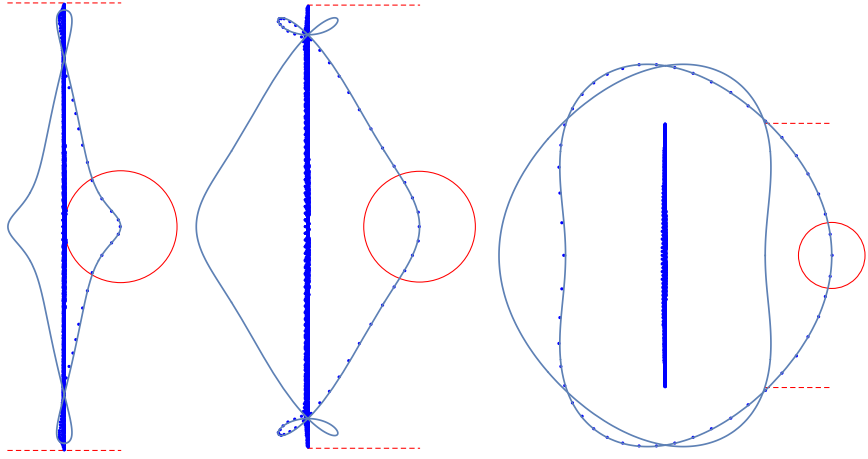

Figure 6: Exact diagonalization of the Lindbladian in Section 5.3 for $l = 1$, $G = 1, 2, 5$ and $N = 50$ comparison of the predicted curve with numerics with periodic boundary conditions. The red circle is the unit circle. Notice how the numerics and the analytical spectrum in the infinite volume fit.

**Theorem 5.7** (Hopping both ways). *If the system has Lindblad operators $L_k = |k\rangle\langle k+1|$ and $L'_k = |k+1\rangle\langle k|$ then the spectrum is given by*

$$\sigma(\mathcal{L}) = [-2G, 0] \cup \{-2G + i[-4, 4]\}.$$

*Proof.* We saw that the spectrum of the non-Hermitian evolution was given by $\sigma(\mathcal{T}) = -2G + i[-4, 4]$. Now, we calculate the spectrum of the quantum jump terms as follows:

$$F_1(q) = G\,|0\rangle\langle 0|\,e^{-iq}$$
$$F_2(q) = G\,|0\rangle\langle 0|\,e^{iq}.$$

Thus, in total the quantum jump contribution is

$$F(q) = F_1(q) + F_2(q) = 2G\cos(q)\,|0\rangle\langle 0|,$$

which is still rank-one and in contrast to the incoherent hopping discussed in Theorem 5.6 this expression is real. We use the same method to compute the spectrum as before

$$-1 = \langle\Gamma_R|\frac{1}{T - z}|\Gamma_L\rangle = \pm\frac{2\cos(q)G}{\sqrt{(\beta - z)^2 - 4\alpha\gamma}},$$

squaring and solving for $z$ and inserting $\alpha, \beta, \gamma$ yields that

$$z = -2G \pm \sqrt{4\cos^2(q)G^2 + 8(\cos(q) - 1)}.$$

Let us analyze the function $f(q) = 4\cos^2(q)G^2 + 8(\cos(q) - 1)$. It has extremal points $q$ satisfying

$$0 = 8\cos(q)\sin(q)G^2 + 8\sin(q).$$

If $q \notin \{0, \pi\}$ then

$$\cos(q) = -\frac{1}{G^2}$$

which has a solution if and only if $G \geq 1$.

This means that if $G < 1$ then the only extremal points are at $q \in \{0, \pi\}$. The values are $f(0) = 4G^2$ and $f(\pi) = 4G^2 - 16$. Thus, by continuity and the intermediate value theorem, it holds that the range of $f$ is $[4G^2 - 16, 4G^2] = [4G^2 - 16, 0] \cup [0, 4G^2]$. The first interval corresponds to the segments $\pm i[0, 2\sqrt{4 - G^2}] \subset i[-4, 4]$. The second one to $[0, 2G]$ analogously to the dephasing case from Section 5.1.

For $G \geq 1$ there is in addition the solution $\cos(q) = -\frac{1}{G^2}$ which has values $-\frac{4}{G^2} - 8$. Since $G \geq 1$, the potential values of $f$ are extended to the interval $\left[-\frac{4}{G^2} - 8, 0\right] \cup [0, 4G^2]$ which corresponds to solutions $i\left[-\sqrt{8 + \frac{4}{G^2}}, +\sqrt{8 + \frac{4}{G^2}}\right]$ and $[0, 2G]$ respectively. Still, since $G \geq 1$, it holds that $\sqrt{8 + \frac{4}{G^2}} \leq 4$ and thus $i\left[-\sqrt{8 + \frac{4}{G^2}}, +\sqrt{8 + \frac{4}{G^2}}\right] \subset i[-4, 4]$. Thus, the spectrum is not extended in that case. $\qquad\square$

# 6 Lindbladians with random potentials

In this section, we study models as in the previous section where we have added a random potential $V$ to the Hamiltonian $H$. The potential $V = \sum_{n \in \mathbb{Z}} V(n) |n\rangle\langle n|$ is a random such that $(V(n))_{n \in \mathbb{Z}}$ is i.i.d. uniformly distributed potential in some range $[-\lambda, \lambda]$ for some $\lambda > 0$. In the closed system case, the study of operators of that type was initiated in the celebrated work by Anderson [77] and has led to the field of random operator theory and the topic of Anderson localization.

Although, some results exist (e.g. [25, 9]), it is not clear what the effects of a random potential in an open quantum system are. In addition, there has recently been interest in random Lindblad systems from the point of view of random matrix theory [78, 27]. Our methods are however more along the lines of random operator theory [1]. Different methods that were also more random operator theoretic were pioneered in [9].

To study the spectrum of our random Lindbladians we use the following Lindbladian version of the Kunz–Souliard Theorem from [79] generalised in [80]. We first need a bit of notation. Recall from Section 2 that we say that a (super-)operator $\mathcal{L} \in \mathcal{B}(\mathrm{HS}(\mathcal{H}))$ is *translationally covariant* if for all $\rho \in \mathrm{HS}(\mathcal{H})$

$$\mathcal{L}(S\rho S^{-1}) = S^{-1}\,\mathcal{L}(\rho)S,$$

where $S$ is the translation with respect to the computational basis in $\mathrm{HS}(\mathcal{H})$. In the examples without a random field, the Lindbladian is translation-invariant. We can separate the action of the potential by $\mathcal{E}_V$, where $\mathcal{E}_V(\rho) = -i[V, \rho]$.

We also need the numerical range of an operator $A$ which we define as follows:

$$W(A) = \{\langle v, Av\rangle \mid \|v\| = 1\}.$$

For example $W(\mathcal{E}_V) = -i[\lambda, \lambda]$. The numerical range will be useful because its closure is an upper bound to the spectrum by the Toeplitz–Hausdorff Theorem [81, 82].

**Theorem 6.1.** *Let $\mathcal{L}_0$ be a translation-covariant operator on $\mathrm{HS}(\mathcal{H})$, which satisfied that $\sigma_{\mathrm{appt}}(\mathcal{L}_0) = \sigma(\mathcal{L}_0)$, (as ensured for our models of interest in Theorem 4.1). Let further $\{V(n)\}_{n\in\mathbb{Z}}$ be the i.i.d. random potential from above. Define $\mathcal{L} = \mathcal{L}_0 + \mathcal{E}_V$ which is also an operator on $\mathrm{HS}(\mathcal{H})$. Then almost surely it holds that*

$$\sigma(\mathcal{L}_0) \subset \sigma(\mathcal{L}) \subset \overline{W(\mathcal{L}_0) + W(\mathcal{E}_V)},$$

*where the second inclusion holds surely.*

*Proof.* Notice first that by the Toeplitz–Hausdorff Theorem [81, 82] it holds that

$$\sigma(\mathcal{L}) \subset \overline{W(\mathcal{L})} \subset \overline{W(\mathcal{L}_0) + W(\mathcal{E}_V)}.$$

For the first inclusion, consider $\lambda \in \sigma_{\mathrm{appt}}(\mathcal{L}_0)$ and let $\{\rho_n\}_{n\in\mathbb{N}}$ be a Weyl sequence corresponding to $\lambda$ for $\mathcal{L}_0$. Without loss of generality assume that each $\rho_n$ is compactly supported in the sense that in the computational basis, we have that $\rho_n(x,y) = 0$ for $(x,y) \notin \Lambda_{R_n} \subset \mathbb{Z}^2$ for some large but finite box $\Lambda_{R_n} \subset \mathbb{Z}^2$. Then define

$$\Omega_n = \left\{ \text{for some } j_n \in \mathbb{Z} : \sup_{(x,y)\in\mathrm{supp}(\rho_n)} |(-i)(V(x+j_n) - V(y+j_n))| \leq \frac{1}{n} \right\},$$

and notice that each $\Omega_n$ is a set of full measure (one way to see this is that for any $k \in \mathbb{N}$ there is almost surely an $a \in \mathbb{Z}$ such that $|V(a+j)| \leq \frac{1}{2n}$ for any $j \in \{1, \ldots, k\}$). Hence $\Omega_0 = \cap_{n\in\mathbb{N}}\Omega_n$ must have probability 1 as well. Thus, we can almost surely for each $n \in \mathbb{N}$ find integers $j_n \in \mathbb{Z}$ such that $|(-i)(V(x+j_n) - V(y+j_n))| \leq \frac{1}{n}$.

Now, define the sequence $\{\gamma_n\}_{n\in\mathbb{N}} \subset \mathrm{HS}(\mathcal{H})$ by $\gamma_n = \rho_n(\cdot - j_n)$, where $\rho_n(\cdot - j_n) \in \mathrm{HS}(\mathcal{H})$ is defined by $\langle x|, \rho_n(\cdot - j_n)|y\rangle = \rho_n(x-j_n, y-j_n)$, e.g. $\rho_n(\cdot - j_n)$ is the operator $\rho_n$ shifted by $j_n$ in both coordinates.

Then $\{\gamma_n\}$ is a Weyl sequence for the eigenvalue $\lambda$ since by translation covariance of $\mathcal{L}_0$ it holds that

$$\|(\mathcal{L} - \lambda)(\gamma_n)\| \leq \|(\mathcal{L}_0 - \lambda)(\rho_n)\| + \|\mathcal{E}_V(\gamma_n)\|.$$

The first term is small since $\rho_n$ is a Weyl sequence for $\mathcal{L}_0$. For the second term, we estimate:

$$||\mathcal{E}_V(\rho_n(\cdot - j_n))|| = \left\|\mathcal{E}_V\left(\sum_{x,y\in\text{supp }\rho_n+j_n}\rho_n(x-j_n,y-j_n)|x\rangle\langle y|\right)\right\|$$

$$= \left\|\sum_{x,y\in\text{supp }\rho_n+j_n}((-i)(V(x)-V(y)))\rho_n(x-j_n,y-j_n)|x\rangle\langle y|\right\|$$

$$= \left\|\sum_{x,y\in\text{supp }\rho_n}((-i)(V(x+j_n)-V(y+j_n)))\rho_n(x,y)|x+j_n\rangle\langle y+j_n|\right\|.$$

Now, if $|(-i)(V(x+j_n)-V(y+j_n))| \leq \frac{1}{n}$ for each $(x,y)\in\text{supp}(\rho_n)$ then by definition of the Hilbert Schmidt inner product it holds that

$$\left\|\sum_{x,y\in\text{supp }\rho_n}((-i)(V(x+j_n)-V(y+j_n)))\rho_n(x,y)|x+j_n\rangle\langle y+j_n|\right\|^2$$

$$= \sum_{x,y\in\text{supp }\rho_n}|((-i)(V(x+j_n)-V(y+j_n)))\rho_n(x,y)|^2$$

$$\leq \frac{1}{n^2}\sum_{x,y\in\text{supp }\rho_n}|\rho_n(x,y)|^2 \leq \frac{1}{n^2}.$$

It follows that $||\mathcal{E}_V(\rho_n(\cdot - j_n))|| \leq \frac{1}{n}$ upon such a choice of $j_n$, which finished the proof.

$\square$

In the following sections, we will see how both of the two inclusions in Theorem 6.1 are in general strict.

Furthermore, one can see in Figure 3 how it seems as if the random field generally pushes eigenvalues vertically as would be the straightforward generalization of the theorem by Kunz and Soulliard [79]. However, the effect is much stronger in the bulk of the spectrum, whereas close to $\{0\}$ it does not seem as if the spectrum changes. A model that describes this phenomenon which is exactly solvable (even without the theory we have developed) is the following.

## 6.1 Exactly solvable model with random potential

Define on the Hilbert space $\ell^2(\mathbb{Z})$ the Lindbladian by $H=0$ and Lindblad operators $L_k=|k\rangle\langle k|$. Then for any $i,j\in\mathbb{Z}$ we have that

$$\mathcal{L}_0(|i\rangle\langle j|) = G\sum_k |k\rangle\langle k||i\rangle\langle j||k\rangle\langle k| - \frac{1}{2}\{|k\rangle\langle k|,|i\rangle\langle j|\} = G\left(|i\rangle\langle i|\ \delta_{i,j}-|i\rangle\langle j|\right).$$

This means that all states of the form $|i\rangle\langle j|$ are eigenvalues with eigenvalue $-G$ if $i\neq j$ and $0$ if $i=j$. In particular, all states of the form $|i\rangle\langle i|$ for $i\in\mathbb{Z}$ are steady states. We conclude that $\sigma\left(\mathcal{L}_0\right)=\{0,-G\}$. Notice that in this case, we do have trace-class steady states in infinite volume.

To use the Lindblad analogue of Kunz–Soulliard we first need to consider the numerical range of $\mathcal{L}_0$. However, $\mathcal{L}_0$ is normal (indeed self-adjoint). That means that the numerical range is the convex hull of the spectrum i.e. $[-G,0]$.

We can find the exact spectrum of the model in a random potential by noticing that

$$\mathcal{L}(|i\rangle\langle i|)=0+\mathcal{E}_V(|i\rangle\langle i|)=(V(i)-V(i))\,|i\rangle\langle i|=0$$

as well as

$$\mathcal{L}(|i\rangle\langle j|)=-G\,|i\rangle\langle j|\ +\mathcal{E}_V(|i\rangle\langle j|)=-G\,|i\rangle\langle j|+i(V(i)-V(j))\,|i\rangle\langle j|\,.$$

Hence the spectrum $\sigma\left(\mathcal{L}\right)=\left\{-G+\lambda\ i[-1,1]\right\}\cup\{0\}$ . Notice, how this is an example where both inclusions in Theorem 6.1 are strict. In the case at hand we can see that this happens because $\mathcal{L}$ leaves both the diagonal and off-diagonal subspaces of $\mathrm{HS}(\ell^2(\mathbb{Z}))\cong\ell^2(\mathbb{Z}^2)$ invariant. The potential $\mathcal{E}_V$ is zero on the diagonal part and therefore only the spectrum of the off-diagonal can be extended by $V$.

In the case of $H=-\Delta$ the two subspaces get mixed, the effects are more complicated and rigorous guarantees are harder to obtain. This is the aim of the next section.

## 6.2   Improving the upper bound on the spectrum

In the following, we prove an upper bound for the Anderson model with local dephasing. The model recently attracted attention in [83, 84]. In the following, we use the model to exemplify how we can use a new method to prove a non-trivial upper bound to the spectrum using the numerical range.

**Theorem 6.2.** *Consider the Anderson model* $H=-\tilde{\Delta}+V$ *with local dephasing defined in* (33). *Then for any* $\rho\in\mathrm{HS}(\mathcal{H})$ *with* $\|\rho\|_2=1$ *such that*

$$\sum_{x\in\mathbb{Z}}|\rho(x,x)|^2=a\in[0,1]$$

*it holds that*

$$\langle\rho,\mathcal{L}\rho\rangle\in Ga+\langle\rho,\mathcal{E}_V\rho\rangle+\langle\rho,\mathcal{T}\rho\rangle\in G(a-1)+i[f(a,\lambda),f(a,\lambda)]$$

*with* $f$ *defined by* $f(a,L)=4\left(1-a+\ 2\sqrt{a}\ \sqrt{1-a}\right)+(1-a)\lambda$. *It follows that*

$$\sigma(\mathcal{L})\subset\bigcup_{a\in[0,1]}\left(G(a-1)+i[f(a,\lambda),f(a,\lambda)]\right).$$

We have plotted the upper bound to the spectrum in Figure 7.

*Proof.*   We first rewrite $\mathcal{L}$ using Theorem 3.8.

$$\mathcal{L} = \mathcal{L}_0 + \mathcal{E}_V = \int_{[0,2\pi]}^{\oplus} T(q)dq + \int_{[0,2\pi]}^{\oplus} F(q)dq + \mathcal{E}_V = \mathcal{T} \; + \mathcal{F} + \mathcal{E}_V.$$

Again, we want to use the upper bound to the spectrum given the numerical range $\sigma(\mathcal{L}) \subset \overline{W(\mathcal{L})}$, so we try to bound the numerical range of each term. So let $\rho \in \mathrm{HS}(\mathcal{H})$ with $\|\rho\|_2 = 1$. Notice first how that identity looks in our two pictures: Define $|\rho(q)\rangle = (\mathcal{F}_1 \circ C \circ \mathrm{vec}(\rho))(q) \in \ell^2(\mathbb{Z})$ Since $(\mathcal{F}_1 \circ C \circ \mathrm{vec}(\rho))$ is an isometric isomorphism it follows that

$$1 = \|\rho\|_2^2 = \sum_{x,y \in \mathbb{Z}} |\rho(x,y)|^2 = \sum_{n \in \mathbb{Z}} \int_0^{2\pi} |\langle \rho(q)|n\rangle|^2 dq.$$

Explicit calculation yields that

$$|\rho(q)\rangle = \frac{1}{\sqrt{2\pi}} \sum_{x,y \in \mathbb{Z}} \rho(x,y)e^{-iqx} \, |y - x\rangle,$$

which again means that

$$\langle 0|\rho(q)\rangle = \sum_{x \, \in \mathbb{Z}} \rho(x,x)e^{-iqx}.$$

So it follows that

$$\int_0^{2\pi} |\langle \rho(q)|0\rangle|^2 dq = \sum_{x \in \mathbb{Z}} |\rho(x,x)|^2 = a,$$

where $a \in [0,1]$ is a real number that indicates how classical $\rho$ is. Notice that then $\sum_{x,y|x \neq y} |\rho(x,y)|^2 = 1 - a$. Now, we go term by term. For the first term, we estimate

$$\langle \rho, \mathcal{E}_V \rho \rangle = \mathrm{Tr}\left( \sum_{x,x',y,y' \in \mathbb{Z}} \overline{\rho(x',y')} |x'\rangle\langle y'| \; i(V(x) - V(y))\rho(x,y)|x\rangle\langle y| \right)$$

$$= i \sum_{x,y|x \neq y} |\rho(x,y)|^2 (V(x) - V(y)) \in i[-(1-a)\lambda, (1-a)\lambda]$$

since $V$ is supported in $[0,\lambda]$. Notice that the more classical a state is the less it is affected by an external field.

For the $T(q)$ term we know from (34) in the proof of Proposition 5.1 that $T(q)$ is tridiagonal with $\alpha, \beta, \gamma$ on the diagonals such that $\overline{\alpha(q)} = -\gamma(q)$ and that $\beta(q) = \beta$ is constant. Then

$$\langle \rho, \mathcal{T} dq \rho \rangle = \langle \rho, \int_{[0,2\pi]}^{\oplus} T(q) dq \rho \rangle = \int_0^{2\pi} \langle \rho(q), T(q)\rho(q) \rangle dq$$

$$= \sum_{n \in \mathbb{Z}} \int_0^{2\pi} \langle \rho(q)|n\rangle (\alpha(q)\langle n+1|\rho(q)\rangle$$
$$+ \ \beta(q)\langle n|\rho(q)\rangle + \gamma(q)\langle n-1|\rho(q)\rangle) dq$$

$$= \beta(q) \sum_{n \in \mathbb{Z}} \int_0^{2\pi} |\langle n|\rho(q)\rangle|^2 dq$$

$$+ \sum_{n \in \mathbb{Z}} 2i \int_0^{2\pi} \mathrm{Im}\left(\gamma(q)\langle \rho(q)|n\rangle\langle n-1|\rho(q)\rangle\right) dq$$

$$= \beta(q) + \sum_{n \in \mathbb{Z}} 2i \int_0^{2\pi} \mathrm{Im}\left(\gamma(q)\langle \rho(q)|n\rangle\langle n-1|\rho(q)\rangle\right) dq.$$

We want to bound the second term. Define the sequence $l_{q,n} = \langle \rho(q)|n\rangle$. And let $l_{q,n}^{\geq j} = l_{q,n}\mathbb{1}_{n \geq j}$ and similarly with $l_{q,n}^{\leq j}$. Using Cauchy–Schwarz in a few different spaces we obtain

$$\left| \sum_{n \in \mathbb{Z}} \int_0^{2\pi} \mathrm{Im}\left(\gamma(q) l_{q,n}\overline{l_{q,n-1}}\right) dq \right| \leq \|\gamma\|_\infty \int_0^{2\pi} \left| \sum_{n \in \mathbb{Z}} l_{q,n}\overline{l_{q,n-1}} \right| dq$$

$$\leq \|\gamma\|_\infty \int_0^{2\pi} \left| \sum_{n \leq -1} l_{q,n}\overline{l_{q,n-1}} \right| + \left| \sum_{n \geq 2} l_{q,n}\overline{l_{q,n-1}} \right|$$
$$+ \ \left| l_{q,0}\overline{l_{q,-1}} \right| + \left| l_{q,1}\overline{l_{q,0}} \right| dq$$

$$\leq \|\gamma\|_\infty \int_0^{2\pi} \|l_q^{\leq -1}\|_2 \|l_q^{\leq -2}\|_2 + \|l_q^{\geq 1}\|_2 \|l_q^{\geq 2}\|_2 + \ \left| l_{q,0}\overline{l_{q,-1}} \right| + \left| l_{q,1}\overline{l_{q,0}} \right| dq$$

$$\leq \|\gamma\|_\infty \left( 1 - a + \sqrt{\int_0^{2\pi} |\langle \rho(q)|0\rangle|^2 dq} \left( \|l_{q,-1}\|_2 + \| l_{q,1}\|_2 \right) \right)$$

$$\leq \|\gamma\|_\infty \left( 1 - a + 2\sqrt{a} \sqrt{1-a} \right).$$

Notice that in particular, this vanishes as $a \to 1$, i.e. for classical $\rho$. Notice also that in our case of interest $\beta(q) = -G$ and $\|\gamma\|_\infty = 2$ yielding the second part of the bound.

For the $F(q) = |\Gamma_L\rangle\langle\Gamma_R|$ we obtain

$$\langle \rho, \int_{[0,2\pi]}^{\oplus} |\Gamma_L\rangle\langle\Gamma_R| dq \rho \rangle = \int_0^{2\pi} \langle \rho(q)|\Gamma_L\rangle\langle\Gamma_R|\rho(q)\rangle dq.$$

Thus, in the case where $|\Gamma_L\rangle\langle\Gamma_R| = G|0\rangle\langle 0|$ in which case the term becomes

$$G \int_0^{2\pi} |\langle \rho(q)|0\rangle|^2 dq = G \sum_{x \in \mathbb{Z}} |\rho(x,x)|^2 = Ga.$$

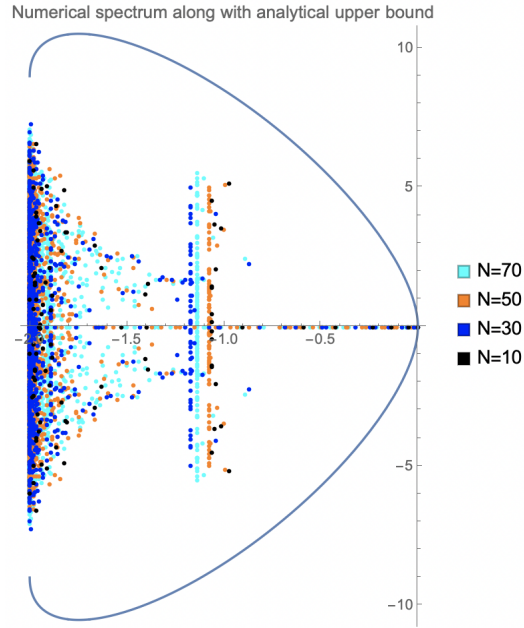

Figure 7: The numerical spectrum of the dephasing Lindbladian $\mathcal{L}$ in a random potential studied in Section 6.2 for $N \in \{10, 30, 50, 70\}, G = 2, V = 5$ along with the analytically calculated upper bound for the numerical range (and therefore the spectrum) sketched. Notice the line with real part close to $-1$ which is consistent with, but not predicted by our analysis. Notice further how it seems that the spectrum does not extend much when the real part is less than $1$.

$\square$

Notice that $\mathrm{Re}(\langle \rho, \mathcal{L}\rho \rangle) = G(a - 1) = \mathrm{Re}(\langle \rho, \mathcal{L}_0\rho \rangle)$. This means that the states as seen from the numerical range get more classical closer to 0. This also implies the absence of a non-classical eigenvalue since one would then be able to find it in the numerical range. Thus, for the dephasing model, we see that the long-lived states are the classical even in the presence of an external (random) potential. In other words, the states that survive longest are very diagonal so we see very explicitly how the dephasing noise suppresses coherences something which was also noted for a simpler model in Chapter 8 of [85].

## 6.3 Further discussion of spectral effects of random potentials

Spectra of random Lindbladians have been studied in random matrix theory approaches in for example [78]. There a lemon-like shape of the spectrum was found. This shape is reminiscent of the spectrum in both Figure 3 and Figure 7 where there seems to be a tendency that the spectrum close to 0 extends less in the direction of the imaginary axis. This is also mimicked in the exactly solvable example in Section 6.1 and the discussion in the previous section. Furthermore, inspecting the non-random spectrum with and without the perturbation one can see how there is a similar phenomenon with eigenvalues jumping from the bulk of the non-Hermitian spectrum and down to the real line [22]. An analytical explanation stemming from the symmetries of the Lindbladian is given in [26]. The one also sees in a corresponding RMT model [28] and the previous example indicates a mechanism for this behaviour.

# 7    Discussion and further questions

We believe that the methods developed here can be applied to many one-particle open quantum systems of Lindblad form. In particular, we believe that many of our results generalize to $\mathbb{Z}^d$ for $d \geq 1$. Further, directions for the study could be many-body systems – in fact the model with non-normal Lindblad operators we studied in Section 5.2 is already a candidate for many-body localization in the many-body setting [86]. A simple example where the number of particles is not fixed could be to have a Lindbladian where the particle number can only decrease. If the system starts with one excitation, such models are used to describe the transport of an excitation during photosynthesis [6]. This would be one way to study the dynamics of similar systems which are not as intimately related to the spectra of $\mathcal{L}$ as discussed in Section 2.2. Another approach could be through the transport in the steady state as is for example done for a non-random system in [87]. A system that is almost within our framework, where transport in the steady state is studied is discussed in [88].

We have also seen how the quantum-jump terms increase the number of real eigenvalues, this effect was also observed in [22]. We can now explain the phenomenon corresponding to our first example in that case. For finite-dimensional systems an explanation was given using symmetries of the Lindbladian in [26, Appendix B.9]. The proof there relies on some rather old results, so it would be of value to obtain a self-contained proof. Along this line of investigation, one could try to obtain results on the density of states. Since relatively many eigenvalues are confined to the real line the density of states will potentially be challenging to define.

Furthermore, we have seen how the spectrum for the finite size numerics in some cases matches the analytic results for all of $\mathbb{Z}$ fairly well. This we have given an explanation of in Theorem 4.5, but it still remains to give a proof of ([66, Conjecture 7.3]), both in general and in our cases of interest. As shown by the examples in Figures 5 and 6 it will be important in that regard to use periodic boundary conditions, as it is the case in the corresponding question for Laurent operators compared to Toeplitz operators in [89].

Another direction of investigation is to expand on the spectral theory of Lindblad operators in a random potential and possibly prove a stronger lower bound to Theorem 6.1. In the self-adjoint case the spectrum and dynamics of the Hamiltonian in infinite volume are related through the RAGE theorem [2, 3, 4] much closer related than what we can prove for $\mathcal{L}$. Such a relation is probably difficult to recover since the notion of continuous spectrum breaks down in the non-normal case. However, we can still define the point spectrum of $\mathcal{L}$ to be spectrum that corresponds to normalizable eigenvectors. In that case, we can ask, in analogy with the case of the Anderson model, whether for large enough strength of the random potential $\lambda$ it is the case that $\sigma(\mathcal{L})$ has only point spectrum, in the sense of exponentially decaying (normalizable) eigenfunctions.

# Acknowledgements

The authors acknowledge support from the Villum Foundation through the QMATH center of Excellence(Grant No. 10059) and the Villum Young Investigator (Grant No. 25452) programs. The authors would like to warmly thank Simone Warzel, Daniel Stilck Franca, Alex Bols, Jacob Fronk, Paul Menczel and Cambyse Rouzé for discussions as well as Nick Weaver for the useful input [48, 58].

# A    Appendix

In the appendix, we provide some of the technicalities that we have postponed from the main part for clarity.

## A.1    Remarks on spectral independence of Lindblad operators

Let us make some further remarks on the spectrum of Lindblad operators as an operator on the different Schatten spaces $\mathcal{S}_p$. If an operator $A$ is bounded on all Schatten spaces consistently, i.e. $A \in \mathcal{B}(\mathcal{S}_p)$ for all $1 \leq p \leq \infty$. Then we say that $A$ satisfies *spectral independence* if for all $1 \leq p, q \leq \infty$ it holds that

$$\sigma_{\mathcal{B}(\mathcal{S}_p)}(A) = \sigma_{\mathcal{B}(\mathcal{S}_q)}(A).$$

In words this means that the spectrum of $A$ is independent of which Banach algebra $\mathcal{B}(\mathcal{S}_p)$. To show how involved the situation is we start with the following remark.

**Remark A.1.** *Consider the Lindbladian with $H = V$ for some onsite operator $V = \sum_i V(i)|i\rangle\langle i|$ and the dissipative part is $0$. In that case $\mathcal{L}(\rho) = -i[V, \rho]$. It holds that $\mathcal{L}(|i\rangle\langle j|) = -i(V(i) - V(j))(|i\rangle\langle j|)$ for each $i, j \in \mathbb{N}$. Thus, from one point of view, $\mathcal{L}$ is a Schur multiplier with the matrix with $(i,j)$'th index $-i(V(i) - V(j))$. If $\sup_i |V(i)| < \infty$ then the Schur multiplier with bounded coefficients. In [48] an example of a Schur multiplier with bounded coefficients, although not of this form, that does not map trace class operators to trace class operators is given.*

However, the operator discussed in [48] is very far from having Lindblad form which is a much stronger condition. Therefore, we can ask under which assumptions, for example, locality of the Lindblad operators, that spectral independence holds.

**Question A.2.** *Suppose that $H = \sum_i h_i$ and such that $h_i, L_k$ have range bounded uniformly by some constant. Is $\sigma_{S_p}(\mathcal{L})$ independent of $p$?*

*In particular, in the case $\mathcal{L}(\rho) = -i[V, \rho]$ where $V|x\rangle = V(x)|x\rangle$ such that $|V(x)| \leq 1$ is it the case that $\sigma_{S_p}(\mathcal{L})$ is independent of $p$?*

In particular, we are concerned with the case $\mathcal{L}(\rho) = -i[H, \rho]$ where $H = -\Delta + V$ for some diagonal non-translation-invariant potential $V$ as an operator from $S_p$ to $S_p$ (i.e. $V$ could be a random potential, that we will study later). Here $\Delta$ is the discrete Laplacian (in Dirac notation elaborated on at the beginning of Section 3) defined by

$$H = -\Delta = -\sum_{k \in \mathbb{Z}} |k\rangle\langle k+1| \ + \ |k+1\rangle\langle k| - 2\,|k\rangle\langle k|\,, \tag{41}$$

where $|k\rangle$ is the position eigenstate at site $k \in \mathbb{Z}$.

It is natural to conjecture that $\sigma_{S_p}(\mathcal{L})$ is independent of $p \in [1, \infty]$, but this is beyond our current methods. However, we do present some partial results.

**Proposition A.3.** *Suppose that $1 \leq p \leq p' \leq \infty$ are Hölder conjugates, $\frac{1}{p} + \frac{1}{p'} = 1$ and that $\mathcal{L} \in \mathcal{B}(S_p)$ and $\mathcal{L} \in \mathcal{B}(S_{p'})$. Then $\mathcal{L} \in \mathcal{B}(S_q)$ for all $p \leq q \leq p'$ and it holds that*

$$\sigma_{S_q}(\mathcal{L}) \subset \sigma_{S_p}(\mathcal{L}) \cup \sigma_{S_{p'}}(\mathcal{L}).$$

*Proof.* The first claim follows directly from the non-commutative Riesz-Thorin theorem (Theorem 2.1). For the second claim notice that if $z \notin \sigma_{S_p}(\mathcal{L}) \cap \sigma_{S_{p'}}(\mathcal{L})$ then $(\mathcal{L} - z)^{-1}$ is bounded both from $S_p$ to $S_p$ and from $S_{p'}$ to $S_{p'}$. Again, by the non-commutative Riesz-Thorin theorem this means that $(\mathcal{L} - z)^{-1}$ is bounded from $S_q$ to $S_q$ for all $q \in [p, p']$ and so $z \notin \sigma_{S_q}(\mathcal{L})$ and we are done by contraposition. $\square$

In fact, we can say a bit more using the Lindblad form. Let us start with the following lemma proving that the spectrum is invariant under complex conjugation in our infinite volume setting.

**Lemma A.4.** *Suppose that $\mathcal{L}$ is of the Lindblad form (1) and that $(\mathcal{A}_1)$ holds. Then for all $p \in [1, \infty]$ it holds that $\sigma_{S_p}(\mathcal{L})$ is closed under complex conjugation*

$$\sigma_{S_p}(\mathcal{L}) = \overline{\sigma_{S_p}(\mathcal{L})}.$$

*Proof.* Suppose first that $\lambda \in \sigma_{\mathrm{appt}, S_p}(\mathcal{L})$ there exists a sequence $\{\rho_n\}_{n \in \mathbb{N}} \subset S_p(\mathcal{H})$ where

$$\|(\mathcal{L} - \lambda)(\rho_n)\|_p \to 0.$$

Then consider the modified Weyl sequence $\{\rho_n^*\}_{n\in\mathbb{N}}$ which satisfies that $\|\rho_n^*\|_p = \|\rho_n\|_p = 1$. This is a Weyl sequence for $\bar{\lambda}$ since

$$\left\|(\mathcal{L} - \bar{\lambda})(\rho_n^*)\right\|_p = \|((\mathcal{L} - \lambda)(\rho_n))^*\|_p = \|(\mathcal{L} - \lambda)(\rho_n)\|_p \to 0.$$

We conclude that $\bar{\lambda} \in \sigma_{\mathrm{appt},\mathcal{S}_p}(\mathcal{L}) \subset \sigma_{\mathcal{S}_p}(\mathcal{L})$.

Now, assume that $\lambda \notin \sigma_{\mathrm{appt},\mathcal{S}_p}(\mathcal{L})$. Let $p'$ be the Hölder conjugate of $p$ and consider $\mathcal{L}^\dagger : \mathcal{S}_{p'} \to \mathcal{S}_{p'}$ be the Banach space adjoint of $\mathcal{L} : \mathcal{S}_p \to \mathcal{S}_p$. It holds that $\lambda \in \sigma_{\mathrm{res},\mathcal{S}_p}(\mathcal{L}) = \sigma_{p,\mathcal{S}_{p'}}(\mathcal{L}^\dagger) = \sigma_{p,\mathcal{S}_{p'}}(\tilde{\mathcal{L}})^3$, where $\sigma_{p,\mathcal{S}_{p'}}$ here denotes the point spectrum in the algebra $\mathcal{S}_{p'}$. Thus, there exists an operator $X \in \mathcal{S}_{p'}$ such that $\tilde{\mathcal{L}}(X) = \lambda X$ which implies that $\tilde{\mathcal{L}}(X^*) = \tilde{\mathcal{L}}(X)^* = \bar{\lambda}X^*$ and thus $\bar{\lambda} \in \sigma_{p,\mathcal{S}_{p'}}(\tilde{\mathcal{L}}) = \sigma_{\mathrm{res},\mathcal{S}_p}(\mathcal{L}) \subset \sigma_{\mathcal{S}_p}(\mathcal{L})$ which means that $\bar{\lambda} \in \sigma_{\mathcal{S}_p}(\mathcal{L})$. □

Now, we collect the results in the following theorem using that $\mathcal{L}$ and $\tilde{\mathcal{L}}$ on $\mathcal{S}_p$ and $\mathcal{S}_{p'}$ are (Banach)-adjoints.

**Theorem A.5.** *Suppose that $1 \leq p \leq p' \leq \infty$ are Hölder conjugates, $\frac{1}{p} + \frac{1}{p'} = 1$ and that $\mathcal{L} \in \mathcal{B}(\mathcal{S}_p)$ and $\mathcal{L} \in \mathcal{B}(\mathcal{S}_{p'})$. Then $\mathcal{L} \in \mathcal{B}(\mathcal{S}_q)$ for all $p \leq q \leq p'$ and it holds that*

$$\sigma_{\mathcal{S}_q}(\mathcal{L}) \subset \sigma_{\mathcal{S}_p}(\mathcal{L}) \cup \sigma_{\mathcal{S}_{p'}}(\mathcal{L}) = \sigma_{\mathcal{S}_p}(\mathcal{L}) \cup \sigma_{\mathcal{S}_p}(\tilde{\mathcal{L}}).$$

In obtaining Theorem A.5 we did not use Lemma A.4 and therefore we did not use the property $\mathcal{L}(\rho^*) = \mathcal{L}(\rho)^*$. A way to improve the Theorem A.5 would be to settle the following question positively.

**Question A.6.** *Suppose that $\mathcal{L}$ is of the Lindblad form (1) and that $(\mathcal{A}_1)$ holds. Is it then the case that*

$$\overline{\sigma_{\mathcal{S}_p}(\mathcal{L})} = \sigma_{\mathcal{S}_p}(\tilde{\mathcal{L}})?$$

In that case, (which we consider fairly natural since it is true in finite dimensions) we would then obtain that for any $1 \leq p \leq q \leq 2$ $\sigma_{\mathcal{S}_q}(\mathcal{L}) \subset \sigma_{\mathcal{S}_p}(\mathcal{L})$ as well as the reverse inclusion in the case $2 \leq q \leq p \leq \infty$. In particular, it would hold that $\sigma_{\mathcal{S}_2}(\mathcal{L}) \subset \sigma_{\mathcal{S}_1}(\mathcal{L})$ meaning that our lower bounds to $\sigma_{\mathcal{S}_2}(\mathcal{L})$ would be lower bounds to $\sigma_{\mathcal{S}_1}(\mathcal{L})$.

## A.2 Proof of Lemma 2.2

We prove that $\mathcal{L} \in \mathcal{B}(\mathcal{B}(\mathcal{H}))$. It is easy to see that by $(\mathcal{A}_1)$ the commutator and anti-commutator terms in the Lindbladian are bounded. Thus, we are left with the operator $\mathcal{J}$ defined by $\mathcal{J}(X) = \sum_k L_k X L_k^*$.

---

[3] One may argue that the Banach space adjoint of $\mathcal{L}$ has the adjoint Lindblad form, that is $\mathcal{L}^\dagger = \tilde{\mathcal{L}}$ as follows. Every $\mathcal{S}_{p'} = \mathcal{S}_p'$ since every $A \in \mathcal{S}_{p'}$ defines a linear functional on $\mathcal{S}_p$ through $A(\rho) = \langle A, \rho \rangle$ for every $\rho \in \mathcal{S}_p$. Then for every $A \in \mathcal{S}_{p'}, \rho \in \mathcal{S}_p$ it holds that $\mathcal{L}^\dagger(A)(\rho) = A(\mathcal{L}(\rho)) = \langle A, \mathcal{L}(\rho) \rangle_{\mathrm{HS}} = \langle \tilde{\mathcal{L}}(A), \rho \rangle_{\mathrm{HS}} = \tilde{\mathcal{L}}(A)(\rho)$.

Let $N$ be the weak limit of $\sum_k L_k L_k^*$. By assumption $\|N\|_\infty < \infty$. Since $\sum_k L_k L_k^*$ is also self-adjoint it follows by the spectral radius theorem for every $|x\rangle \in H$ that

$$\sum_k \| L_k^* |x\rangle \|^2 = \sum_k \langle x| , L_k L_k^* |x\rangle \ = \langle x| , \sum_k L_k L_k^* |x\rangle \ \le \|N\|_\infty \|x\|^2.$$

Then consider $X \in \mathcal{B}(\mathcal{H})$ with $X \ge 0$. Then $X = AA^*$ for some $A \in \mathcal{B}(\mathcal{H})$ and it holds that

$$\sum_k L_k X L_k^* = \sum_k L_k AA^* L_k^*$$

is a sum of positive operators and therefore positive. We bound the norm

$$\langle x|, \sum_k L_k AA^* L_k^* |x\rangle = \sum_k \langle x|, L_k AA^* L_k^* |x\rangle = \sum_k \|A^* L_k^* |x\rangle\|^2 \ \le \sum_k \|A^*\|^2 \|L_k^* |x\rangle\|^2$$
$$\le \|A^*\|^2 \|N\|_\infty \|x\|^2.$$

Again, by self-adjointness of $\sum_k L_k AA^* L_k^*$ the spectral radius theorem holds and therefore if the numerical range is bounded then so is the norm. We conclude that

$$\left\| \sum_k L_k AA^* L_k^* \right\| \ \le \|A\|^2 \|N\|_\infty.$$

Now, for any element $X \in \mathcal{B}(\mathcal{H})$ we write $X = P_1 - P_2 + iP_3 - iP_4$ where $P_1, P_2, P_3, P_4$ are all positive and satisfy $\|P_i\| \le \|X\|$ for each $i = 1, \ldots, 4$ (see [90, Theorem 11.2 and 9.4]). Then

$$\|\mathcal{J}(X)\| \le 4\|X\|\|N\|_\infty,$$

where we also used the $C^*$-identity since we could write $P_i = A_i A_i^*$ and that it holds that

$$\|A_i\|^2 = \|A_i A_i^*\| = \|P_i\| \ \le \|X\|.$$

We conclude that $\mathcal{J}$ as an operator on $\mathcal{B}(\mathcal{H})$ has a norm bounded by $4\|N\|_\infty$.

## A.3  Measurability in the proof of Theorem 3.12

In this appendix, we prove that for each fixed $n \in \mathbb{N}$ there exists a measurable choice of the vectors $q \mapsto v_{q,n}$ in the proof of Theorem 3.12. We do that with inspiration from [91] and let $\{a_m\}_{m \in \mathbb{N}}$ be a countable dense subset of $\mathcal{H}$ which does not contain 0. Then define $b_m = \frac{a_m}{\|a_m\|}$. Recall that $I_n$ is a set such that $|I_n| > 0$ with $\|(A(q) - \lambda)^{-1}\| \ \ge n$. Now, consider the function $N : I_n \to \mathbb{N}$ defined by

$$N(q) = \min \left\{ m \in \mathbb{N} \mid \|(A(q) - \lambda)^{-1} b_m\| \ge \frac{n}{2} \right\}.$$

Notice that $N$ is well-defined since $(A(q) - \lambda)^{-1}$ is bounded and hence continuous and by density of $\{a_m\}_{m \in \mathbb{N}}$. We claim that $N$ is also $\mathcal{B}(I_n) - \mathcal{P}(\mathbb{N})$ measurable, where $\mathcal{B}(I_n)$ is the Borel $\sigma$-algebra on $I_n$. To see that, notice first that since $q \mapsto (A(q) - \lambda)^{-1}$ is measurable then also $q \mapsto \|(A(q) - \lambda)^{-1} b_m\|$ will be measurable for each $m$. Thus, the set

$$\left\{ q \in I_n \mid \|(A(q) - \lambda)^{-1} b_m\| \geq \frac{n}{2} \right\}$$

is measurable. Now, it holds that

$$N^{-1}(\{1, \ldots, k\}) = \bigcup_{m=1}^{k} \left\{ q \in I \mid \|(A(q) - \lambda)^{-1} b_m\| \geq \frac{n}{2} \right\}$$

and this is sufficient to prove measurability of $N$. Now, since any function from $(\mathbb{N}, \mathcal{P}(\mathbb{N}))$ to any measure space is measurable and compositions of measurable functions are measurable it holds that $q \mapsto b_{N(q)}$ is measurable. We then pick $v_{q,n} = b_{N(q)}$.

## A.4 Proof of Theorem 3.13

*Proof.* We use the characterisation from Theorem 3.12. Whenever $\lambda \notin \sigma(A(q))$ define the resolvent $R(q) = (A(q) - \lambda)^{-1}$. Then if both $R(p), R(q)$ exist it follows from the resolvent equation that

$$| \|R(p)\| - \|R(q)\| | \leq \|R(p)\| \, \|A(p) - A(q)\| \, \|R(q)\|. \tag{42}$$

"$\subset$": We prove the following stronger assertion.

**Claim A.7.** *For any $p_0 \in [0, 2\pi]$ and $\varepsilon > 0$ it holds that*

$$\sigma_\varepsilon(A(p_0)) \subset \bigcup_{p \in [0, 2\pi]}^{\mathrm{ess}} \sigma_{2\varepsilon}(A(p)). \tag{43}$$

Let us first prove the claim.

*Proof of Claim.* To prove the claim, let $\lambda \in \sigma_\varepsilon(A(p_0))$.

*Case 1:* Suppose first that $\lambda \notin \sigma(A(p_0))$. This means that $\frac{1}{\varepsilon} \leq \|R(p_0)\| < \infty$. Suppose that $\|R(p_0)\| \leq K < \infty$. Assume for contradiction that there is a sequence $p_n \to p_0$ such that $\|R(p_n)\| \to \infty$. By continuity (cf. Lemma A.10) there exists a $\delta > 0$ such that $\|A(p) - A(q)\| \leq \frac{1}{2K}$ as long as $|p - q| < \delta$. If we pick $N$ such that $|p_n - p_0| < \delta$ for $n \geq N$ this means that

$$| \|R(p_0)\| - \|R(p_n)\| | \leq \|R(p_0)\| \, \|A(p_0) - A(p_n)\| \, \|R(p_n)\| \leq \frac{\|R(p_n)\|}{2K}.$$

Thus, $\frac{\|R(p_n)\|}{2} \leq \|R(p_0)\|$ and hence $\|R(p_0)\| = \infty$ which is a contradiction. We conclude that there exists a $\gamma > 0$ such that $\sup_{p:|p-p_0| \leq \gamma} \|R(p)\| \leq K_2 < \infty$. Thus, the resolvent bound

(42) ensures that there is an $\theta > 0$ such that if $|p - p_0| \leq \theta$ then $\|R(p)\| \geq \frac{1}{2\varepsilon}$. This means that $\lambda \in \sigma_{2\varepsilon}(A(p))$ for all $p$ with $|p - p_0| \leq \theta$. Again this implies that $\lambda \in \bigcup_{p \in [0, 2\pi]}^{\text{ess}} \sigma_{2\varepsilon}(A(p))$, proving the claim in the case $\lambda \notin \sigma(A(p_0))$.

*Case 2:* Now, let $\lambda \in \sigma(A(p_0)) \subset \sigma_{\varepsilon}(A(p_0))$. Absorb $\lambda$ into $A$ for ease of notation. Now, suppose that $(p_n)_{n \in \mathbb{N}}$ is a sequence converging to $p_0$ and assume that $\|R(p_n)\| \leq C$ uniformly in $n$. Then

$$| \, \|R(p_n)\| - \|R(p_m)\| \, | \leq \|R(p_n)\| \, \|A(p_n) - A(p_m)\| \, \|R(p_m)\| \leq C^2 \| \, A(p_n) - A(p_m)\|,$$

so the sequence $(R(p_n))_{n \in \mathbb{N}}$ is Cauchy and hence convergent to some $R_0$. It holds by norm continuity that

$$\begin{aligned} \|R(p_n)A(p_0) - \mathbb{1}\| \; &= \|R(p_n)(A(p_0) - A(p_n))\| \\ &\leq \|R(p_n)\| \; \|A(p_0) - A(p_n)\| \; \leq C \; \|A(p_0) - A(p_n)\| \to 0. \end{aligned}$$

Similarly, $\|A(p_0)R(p_n) - \mathbb{1}\| \; \to 0$. Since further

$$\|R_0 A(p_0) - \mathbb{1}\| \; \leq \|R(p_n)A(p_0) - \mathbb{1}\| \; + \|R_0 - R(p_n)\| \; \|A(p_0)\| \; \to 0,$$

we conclude that $R_0 A(p_0) = \mathbb{1}$ and similarly $A(p_0)R_0 = \mathbb{1}$. This contradicts that $A(p_0)$ is invertible. So we conclude that no such sequence $p_n$ exists. Hence for our given $\varepsilon > 0$ there is a $\delta > 0$ such that $\|R(p)\| \geq \frac{1}{\varepsilon}$ for every $p \in [p_0 - \delta, p_0 + \delta]$. This means that $\lambda \in \bigcup_{p \in [0, 2\pi]}^{\text{ess}} \sigma_{2\varepsilon}(A(p))$. $\qquad \square$

So we conclude that for all $\varepsilon > 0$ it holds that

$$\sigma_{\varepsilon}(A(p_0)) \subset \bigcup_{p \in [0, 2\pi]}^{\text{ess}} \sigma_{2\varepsilon}(A(p)) \tag{44}$$

and therefore by Theorem 3.12

$$\sigma(A(p_0)) \subset \bigcap_{\varepsilon > 0} \sigma_{\varepsilon}(A(p_0)) \subset \bigcap_{\varepsilon > 0} \bigcup_{p \in [0, 2\pi]}^{\text{ess}} \sigma_{2\varepsilon}(A(p)) = \sigma \left( \int_{[0, 2\pi]}^{\oplus} A(p) dp \right)$$

Since this holds for any $p_0 \in [0, 2\pi]$ we have proven one inclusion.

"$\supset$": For the converse let $\lambda \in \bigcap_{\varepsilon > 0} \left( \bigcup_{q \in I}^{\text{ess}} \sigma_{\varepsilon}(A(q)) \right)$. Suppose for contradiction that $\lambda \notin \sigma(A(q))$ for any $q \in I$. It then follows from the resolvent bound (42) that the function $N : I \to \mathbb{R}$ defined by $N(q) = \|(A(q) - \lambda)^{-1}\|$ is continuous. Now,

$$S_n = \left\{ q \in I \mid \lambda \in \sigma_{\frac{1}{n}}(A(q)) \right\} \; = \left\{ q \in I \mid \big\| \, A(q) - \lambda)^{-1} \big\| \geq n \right\} \; = N^{-1}\left( [n, \infty) \right).$$

Since $N$ is continuous and $[n, \infty)$ is closed we must have that $S_n$ is closed. Further, we must have that $S_n \subset S_{n-1}$ and that $S_n$ is non-empty for each $n \in \mathbb{N}$. Thus, by the finite intersection property (since $I$ is compact) it holds that $\bigcap_{n \in \mathbb{N}} S_n$ is non-empty. So let $q_0 \in S_n$ for each $n \in \mathbb{N}$. Then $\|(A(q_0) - \lambda)^{-1}\| \geq n$ for all $n \in \mathbb{N}$ and thus $\lambda \in \sigma(A(q_0))$ which is a contradiction. So we conclude that $\lambda \in \sigma(A(q))$ for at least one $q \in I$ and then $\lambda \in \bigcup_{q \in I} \sigma(A(q))$.

$\square$

## A.5 Spectrum of rank-one perturbation of Laurent operators

In this appendix, we prove the following relation that was used in the proof of Corollary 3.15.

$$\sigma(T(q) + F(q)) = \sigma(T(q)) \cup \left\{ \lambda \in \mathbb{C} \mid \langle \Gamma_R | (T(q) - \lambda)^{-1} | \Gamma_L \rangle = -1 \right\}.$$

" $\supset$: " From the proof of Corollary 3.14 we saw that $\sigma(T(q)) \subset \sigma(T(q) + F(q))$. So let $\lambda \notin \sigma(T(q))$ and $\langle \Gamma_R | (T(q) - \lambda)^{-1} | \Gamma_L \rangle = -1$. Assume for contradiction that $T(q) + F(q) - \lambda$ is invertible. Let for ease of notation $T = T(q) - \lambda$. Then the resolvent equation states that

$$\frac{1}{T + |\Gamma_L\rangle\langle\Gamma_R|} = \frac{1}{T} - \frac{1}{T + |\Gamma_L\rangle\langle\Gamma_R|} |\Gamma_L\rangle\langle\Gamma_R| \frac{1}{T}. \tag{45}$$

Suppose that $\langle \Gamma_R | \frac{1}{T} | \Gamma_L \rangle = -1$ then multiplying with $|\Gamma_L\rangle$ from the right yields that $T^{-1}|\Gamma_L\rangle = 0$, which is a contradiction.

" $\subset$: " Assume that $T$ is invertible and that $\langle \Gamma_R | \frac{1}{T} | \Gamma_L \rangle \neq -1$. It is then straightforward to check that the following operator is well defined and an inverse to $T + |\Gamma_L\rangle\langle\Gamma_R|$

$$\frac{1}{T} - \frac{1}{\langle \Gamma_R | \frac{1}{T} | \Gamma_L \rangle + 1} \frac{1}{T} |\Gamma_L\rangle\langle\Gamma_R| \frac{1}{T},$$

so we conclude that $T + |\Gamma_L\rangle\langle\Gamma_R|$ is invertible.

## A.6 Resolvent norm estimates

*Proof of Lemma 4.11.* First, $\sup_{z \in V} \|(T(q) - z)^{-1}\| \leq C < \infty$ for some $C > 0$ To see that let $z_n \in V$ be a sequence such that $\|(T(q) - z_n)^{-1}\| \to \infty$. Since $\bar{V}$ is compact then $z_n$ has a convergent subsequence and it holds that $z_n \to z_0$ with $\|(T(q) - z_0)^{-1}\| = \infty$, hence $z_0 \in \sigma(T(q))$ which is a contradiction to the construction of $V$.

Second, we show $\sup_{z \in V, n \in \mathbb{N}} \|(T(q_n) - z)^{-1}\| \leq C < \infty$. We first need the following claim.

**Claim A.8.** *For any $z \in V$ then*

$$\sup_{n \in \mathbb{N}} \|(T(q_n) - z)^{-1}\| < \infty$$

*Proof.* Suppose for contradiction that

$$\sup_{n \in \mathbb{N}} \left\| (T(q_n) - z)^{-1} \right\| = \infty$$

Then for every $\varepsilon > 0$, there is a $n$ large enough such that for a subsequence

$$z \in \sigma_\varepsilon(T(q_n)) \subset \bigcup_{q \in [0, 2\pi]}^{\mathrm{ess}} \sigma_{2\varepsilon}(T(q)),$$

where the inclusion follows from (44). Thus,

$$z \in \bigcap_{\varepsilon > 0} \bigcup_{q \in [0, 2\pi]}^{\mathrm{ess}} \sigma_{2\varepsilon}(T(q)) = \bigcup_{q \in [0, 2\pi]} \sigma(T(q)),$$

which is a contradiction. □

**Claim A.9.** $\sup_{z \in V, n \in \mathbb{N}} \left\| (T(q_n) - z)^{-1} \right\| \leq C < \infty$

*Proof.* Assume first for contradiction that $\sup_{z \in V, n \in \mathbb{N}} \left\| (T(q_n) - z)^{-1} \right\| = \infty$. Then find a subsequence $(q_n, z_n)$ such that $\left\| (T(q_n) - z_n)^{-1} \right\| \to \infty$, by compactness of $\bar{V}$ and $[0, 2\pi]$ a suitable subsubsequence converges to some point $(q_0, z_0) \in [0, 2\pi] \times V$.

Assume now for contradiction that $\left\| (T(q_0) - z_0)^{-1} \right\| < \infty$ For any $(q, z) \in [0, 2\pi] \times V$ it follows by the resolvent equations that

$$
\begin{aligned}
\left| \left\| (T(q_0) - z_0)^{-1} \right\| - \left\| (T(q) - z)^{-1} \right\| \right| &\leq \left\| (T(q_0) - z_0)^{-1} - (T(q) - z)^{-1} \right\| \\
&\leq \left\| (T(q_0) - z_0)^{-1} - (T(q) - z_0)^{-1} \right\| \\
&\quad + \left\| (T(q) - z_0)^{-1} - (T(q) - z)^{-1} \right\| \\
&\leq \left\| (T(q_0) - z_0)^{-1} \right\| \| T(q_0) - T(q) \| \left\| (T(q) - z_0)^{-1} \right\| \\
&\quad + \left\| (T(q) - z_0)^{-1} \right\| |z - z_0| \left\| (T(q) - z)^{-1} \right\| \\
&\leq C \| T(q_0) - T(q) \| + C |z - z_0| \left\| (T(q) - z)^{-1} \right\|.
\end{aligned}
$$

Thus, for any $z$ such that $|z - z_0| \leq \frac{1}{2C}$ then we conclude for any $q \in [0, 2\pi]$ that

$$
\begin{aligned}
\left\| (T(q_0) - z_0)^{-1} \right\| &\geq \left\| (T(q) - z)^{-1} \right\| - \left| \left\| (T(q_0) - z_0)^{-1} \right\| - \left\| (T(q) - z)^{-1} \right\| \right| \\
&\geq \left\| (T(q) - z)^{-1} \right\| - C \| T(q_0) - T(q) \| - \frac{1}{2} \left\| (T(q) - z)^{-1} \right\| \\
&= \frac{1}{2} \left\| (T(q) - z)^{-1} \right\| - C \| T(q_0) - T(q) \|.
\end{aligned}
$$

Thus, if $(q, z) \to (q_0, z_0)$ and $\|(T(q) - z)^{-1}\| \to \infty$ then $\|(T(q_0) - z_0)^{-1}\| = \infty$ which is a contradiction to $\|(T(q_0) - z_0)^{-1}\| < \infty$ and so we conclude that $\|(T(q_0) - z_0)^{-1}\| = \infty$. That means that $z_0 \in \sigma(T(q_0))$ since $z_0 \in \bar{V}$ this contradicts the construction of $V$ and therefore $\sup_{z \in V, n \in \mathbb{N}} \|(T(q_n) - z)^{-1}\| = \infty$ and so the claim is proven. $\qquad \square$

$\qquad \square$

For completeness, we write out the norm continuity of $q \mapsto T(q)$ in the case that is relevant to us.

**Lemma A.10.** *Let* $r \in \mathbb{N}$ *and* $T(q)$ *be an* $r$-*diagonal Laurent operator with smooth functions* $a_i : [0, 2\pi] \to \mathbb{C}$ *on the* $i$'*th diagonal for* $-r \leq i \leq r$. *Then if* $q \to q_n$ *it holds that*

$$\|T(q) - T(q_n)\| \to 0.$$

*Proof.* The proof is a simple computation.

$$\|T(q_n) - T(q)\| = \left\| \sum_{i=-r}^{r} (a_i(q) - a_i(q_n)) S^i \right\| \leq \sum_{i=-r}^{r} |a_i(q) - a_i(q_n)| \to 0$$

as $q \to q_n$ by continuity the functions $a_i$. $\qquad \square$

## A.7  Invertibility of bi-infinite tridiagonal Laurent matrices

In this appendix, we discuss the inversion of bi-infinite tridiagonal Laurent matrices.

In Corollary 3.15 we saw that the solutions to the equation

$$\langle \Gamma_R | (T(q) - \lambda)^{-1} | \Gamma_L \rangle = -1$$

were part of the spectrum of $\mathcal{L}$ extending the spectrum of the non-Hermitian evolution.

In our applications, $T(q) - \lambda$ is a tridiagonal Laurent operator and thus to find explicit expressions we need to evaluate matrix elements of the inverse of such a matrix. For a tridiagonal operator with $\alpha, \beta$ and $\gamma$ on the diagonal the symbol curve is given by

$$a(z) = \alpha z^{-1} + \beta + \gamma z, \, z \in \mathbb{T},$$

which is a (possibly degenerate) ellipse. Thus, the spectrum always forms a (possibly degenerate) ellipse in that case. In the cases where our Lindblad operators are supported on at most two lattice sites, by Theorem 3.8 the corresponding Laurent operator will be tridiagonal.

If the matrix has $\alpha(q), \beta(q), \gamma(q)$ on the diagonals it will be useful to study the following equation

$$\alpha + \beta x + \gamma x^2 = 0. \tag{46}$$

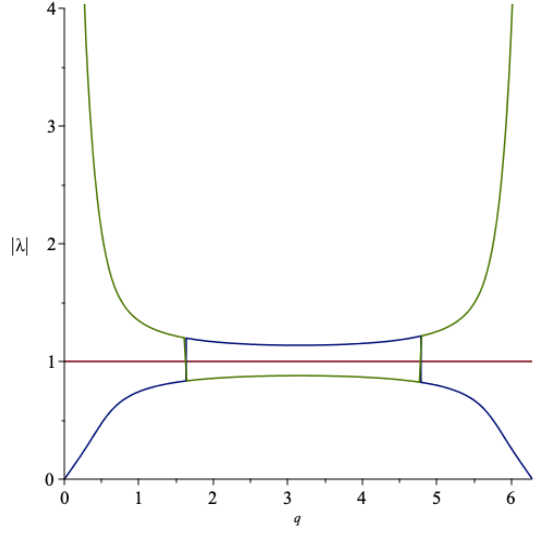

Figure 8: Plot of the absolute value of the two solutions to (46), for relevant parameters in the dephasing model for $G = 1$ and $\lambda = -0.5 + i$.

In particular, we would like to know whether the two solutions $\lambda_1, \lambda_2$ to the equation satisfy that

$$|\lambda_1| < 1 < |\lambda_2|.$$

To define square roots we will use the following convention.

**Convention A.11.** *For any $z \in \mathbb{C}$ the two branches $\pm\sqrt{z}$ are defined such that $\mathrm{Re}(+\sqrt{z}) \geq 0$ and $\mathrm{Re}(-\sqrt{z}) \leq 0$. If $\mathrm{Re}(+\sqrt{z}) = 0$ then we use the convention that $\mathrm{Im}(+\sqrt{z}) \geq 0$.*

The following lemma will be useful. We also sketch the situation in Figure 8.

**Lemma A.12.** *Let $\alpha, \beta, \gamma : S^1 \to \mathbb{C}$ be continuous functions. Suppose that $T(q)$ is a family of tridiagonal Laurent matrices with $\alpha(q), \beta(q), \gamma(q)$ on the diagonals, which is invertible for all $q$. Assume further that $\gamma(q) = 0$ for at most one $q$ and that there exists $q_0, q_1$ such that*

$$|\alpha(q_0)| \geq |\gamma(q_0)| \quad and \quad |\alpha(q_1)| \leq |\gamma(q_1)|.$$

*Let $\lambda_1(q), \lambda_2(q)$ be the two solutions of $\alpha(q) + \beta(q)x + \gamma(q)x^2 = 0$ such that $|\lambda_1| \leq |\lambda_2|$. Then for all $q$*

$$|\lambda_2(q)| < 1 < |\lambda_1(q)|.$$

*Proof.* For all $q$ such that $\gamma(q) \neq 0$ we can find the roots as

$$\lambda_{\pm} = -\frac{\beta}{2\gamma} \pm \sqrt{\left(\frac{\beta}{2\gamma}\right)^2 - \frac{\alpha}{\gamma}}, \tag{47}$$

where we used the convention. From Vieta's formula, we also know that $\lambda_+ \lambda_- = \frac{\alpha}{\gamma}$. Notice that since we have assumed that $T(q)$ is invertible we know that $|\lambda_+|$ and $|\lambda_-|$ are never equal to 1.

Thus, for at least one $q_0$ it holds that $|\lambda_+ \lambda_-| \geq 1$. That means either $|\lambda_1(q_0)| \geq |\lambda_2(q_0)| \geq 1$ or $|\lambda_1(q_0)| < 1 < |\lambda_2(q_0)|$. Similarly since for at least one $q_1$ it holds that $|\lambda_+ \lambda_-| \leq 1$ this means that either $|\lambda_1(q_1)| < 1 < |\lambda_2(q_1)|$ or $|\lambda_1(q_1)| \leq |\lambda_2(q_1)| < 1$.

Now, since the solutions are given by equation (47) it means that $\lambda_-(q), \lambda_+(q)$ are continuous functions of $p$ as long as $\gamma(q) \neq 0$ which happens in at most one point $q_2$. Since $S^1 \backslash \{q_2\}$ is connected the image of the set under the map $p \mapsto (\lambda_-(q), \lambda_+(q))$ is still connected. Since we have the properties for the points $q_0$ and $q_1$ and we at the same time never have eigenvalues with absolute value 0 it must mean that we are in the case

$$|\lambda_1(q)| < 1 < |\lambda_2(q)|$$

for all $q \neq q_2$. For $q = q_2$ we have the equation $\alpha(q_2) + \beta(q_2)x = 0$.

$\square$

Knowledge about the modulus of the solutions can be transferred into explicitly knowing the inverse of the operator. The two solutions $\lambda_\pm$ are defined through (47) using Convention A.11. However, to be able to deal with the solutions with largest and smallest absolute values define $\lambda_1, \lambda_2$ such that $\{\lambda_1, \lambda_2\} = \{\lambda_+, \lambda_-\}$ and $|\lambda_2| \leq |\lambda_1|$. Whether $\lambda_+$ is equal to $\lambda_1$ or $\lambda_2$ will introduce a sign in the following lemma. This we write as $(-1)^{\mathbb{1}[|\lambda_-| < 1 < |\lambda_+|]}$ where $\mathbb{1}$ is the indicator function. Finally, we note that for finite size matrices, the problem of inverting Laurent operators is a lot more intricate and has been studied in [92]. For related results see [93, Chap 3].

**Lemma A.13.** *Suppose that $T$ is an invertible tridiagonal Laurent operator with $\alpha, \beta, \gamma$ on the diagonals and such that $\gamma \neq 0$. Let $\lambda_\pm$ be given by (47) and $\lambda_1, \lambda_2$ as above. Assume that $|\lambda_2| < 1 < |\lambda_1|$ as is for example ensured by Lemma A.12. Then for $k \geq 0$ it holds that*

$$\langle n| \, T^{-1} \, |n+k\rangle \; = \frac{(-1)^{\mathbb{1}[|\lambda_-| < 1 < |\lambda_+|]}}{\lambda_1^k \sqrt{\beta^2 - 4\alpha\gamma}} \; \text{ as well as } \; \langle n| \, T^{-1} \, |n-k\rangle \; = \frac{\lambda_2^k (-1)^{\mathbb{1}[|\lambda_-| < 1 < |\lambda_+|]}}{\sqrt{\beta^2 - 4\alpha\gamma}}.$$

*In particular, for $k = 0$ we have*

$$\langle n| \, T^{-1} \, |n\rangle \; = \frac{(-1)^{\mathbb{1}[|\lambda_-| < 1 < |\lambda_+|]}}{\sqrt{\beta^2 - 4\alpha\gamma}}.$$

The proof is a rather standard computation that we repeat for completeness.

*Proof.* First, $\sigma(T)$ is the image of the symbol curve $a(z) = \alpha z^{-1} + \beta + \gamma z$ for $z \in \mathbb{T}$. Since $T$ is invertible it holds that $a(z) \neq 0$ for all $z \in \mathbb{T}$ and therefore, using Theorem 3.3, the symbol curve of the inverse is given by

$$\frac{1}{a(z)} = \frac{1}{\frac{\alpha}{z} + \beta + \gamma z} = \frac{z}{\alpha + \beta z + \gamma z^2} = \frac{z}{\gamma(\frac{\alpha}{\gamma} + \frac{\beta}{\gamma}z + z^2)}.$$

We can rewrite the denominator $\gamma(z - \lambda_+)(z - \lambda_-)$ with

$$\lambda_\pm = \frac{-\beta}{2\gamma} \pm \sqrt{\left(\frac{\beta}{2\gamma}\right)^2 - \frac{\alpha}{\gamma}}\,.$$

Notice that

$$\lambda_+ \lambda_- = \frac{\alpha}{\gamma}\,,\ \lambda_+ + \lambda_- = -\frac{\beta}{\gamma}\,,\ \text{and}\ \lambda_+ - \lambda_- = 2\sqrt{\left(\frac{\beta}{2\gamma}\right)^2 - \frac{\alpha}{\gamma}}\,.$$

Now, the assumption $|\lambda_2| < 1 < |\lambda_1|$ has implications on how we write this up as a geometric series:

$$
\begin{aligned}
\frac{1}{a(z)} &= \frac{z}{\gamma(z - \lambda_+)(z - \lambda_-)} = \frac{z}{\gamma(\lambda_1 - \lambda_2)}\ \left(\frac{1}{z - \lambda_1} - \frac{1}{z - \lambda_2}\right) \\
&= \frac{z}{\gamma(\lambda_1 - \lambda_2)}\ \left(-\frac{1}{\lambda_1}\frac{1}{1 - \frac{z}{\lambda}} - \frac{1}{z}\frac{1}{1 - \frac{\lambda_2}{z}}\right) \\
&= \frac{z}{\gamma(\lambda_1 - \lambda_2)}\ \left(-\frac{1}{\lambda_1}\sum_{n=0}^{\infty}(\frac{z}{\lambda_1})^n - \frac{1}{z}\sum_{n=0}^{\infty}(\frac{\lambda_2}{z})^n\right) \\
&= \frac{1}{\gamma(\lambda_2 - \lambda_1)}\ \left(\sum_{n=1}^{\infty}(\frac{z}{\lambda_1})^n + \sum_{n=0}^{\infty}(\frac{\lambda_2}{z})^n\right).
\end{aligned}
$$

From this formula, we can read off the coefficients. The computation

$$\gamma(\lambda_2 - \lambda_1) = (-1)^{\mathbb{1}[|\lambda_-| \, < 1 < |\lambda_+| \,]}\ \gamma(\lambda_+ - \lambda_-) = (-1)^{\mathbb{1}[|\lambda_-| \, < 1 < |\lambda_+| \,]}\ \sqrt{\beta^2 - 4\alpha\gamma}$$

proves the last formula. The formulas for $k \geq 1$ then follow from reading off the coefficients. $\qquad\square$

# 7. Exponential Decay of Coherences in Steady States of Open Quantum Systems

# Exponential decay of coherences in open quantum systems with large disorder

FREDERIK RAVN KLAUSEN, SIMONE WARZEL

**Abstract**

We consider the steady state of single-particle open quantum systems described by the Lindblad master equation with local terms. For systems where the non-hermitian evolution is either gapped or strongly disordered, we show exponential decay of off-diagonal matrix elements of any steady state in the position basis, e.g. exponentially decaying coherences. The gap exists whenever any level of local dephasing is present in the system. In the case of sufficiently strong disorder (in the Hamiltonian) we also assume that the inverse gap of the effective non-hermitian evolution increases at most polynomially in the system size, a condition that is satisfied for a type of dissipation recently associated with localization in open quantum systems. We describe how the result in the disordered case can be viewed as an extension of Anderson localization to open quantum systems. Finally, we bootstrap the result to get guarantees on the decoherence for averages of the finite time evolution starting from any initial state.

## 1 Introduction

Open quantum systems describe the effective evolution of a quantum system coupled to an external environment. Under natural assumptions, the evolution of open quantum systems is governed by the Lindblad master equation [39, 29].

In this paper, we focus on local single-particle open quantum systems. These systems have been extensively studied in the position basis, as highlighted in [56, 37, 22, 25, 34], and have found wide applications, including in biology, where they have been employed to understand dephasing-assisted transport [47].

One of the most intriguing quantum effects within the paradigm of single-particle closed quantum systems is Anderson localization, first predicted by Anderson [7]. Rigorous results on localization beyond the paradigm of (single-particle) random Schrödinger operators are limited but exist [23, 10]. In the case of open quantum systems, an important result is a slow-down of the dynamics from ballistic to diffusive [27]. In addition, the effect of disorder and relations to localization open quantum systems has been an object of a recent study [55, 51, 54, 50].

The primary focus of this paper is to establish two key results pertaining to the decoherence of steady states in Lindbladian models comprising local terms (cf. Assumption 2.4). First, if there is any dephasing in the system (cf. Assumption 2.6), we prove that any steady state of the system will exhibit exponentially strong decoherences with constants that are uniform in the system size. Second, if we consider the Hamiltonian to be a random operator and the norm of the resolvent of the non-hermitian evolution is polynomially bounded in system size (cf. Assumption 2.5) then the same result holds for sufficiently large disorder. We also get a bound on the (Abel-) averaged time evolution starting from any initial state.

We note that these implications are standard in the single qubit case (cf. [45, 8.4.1]). The result also resemble results on spectral gaps implying clustering of correlations for Liouvillians that are for example discussed in [15, 33, 43, 30]. The relations between spectral gaps, their closing in infinite volume and decoherence in time were also discussed for example [14, 37]

Our methods rely heavily on the fractional moment approach to random Schrödinger operators developed by Aizenman and Molchanov [2] and the technicalities are particularly inspired by the work of Aizenman and Warzel on Anderson localization for multiparticle systems [3, 4].

It will be natural for us to split up the full Lindblad evolution in the non-hermitian evolution and the quantum jump terms (see e.g. [19]). Surprisingly, we only require knowledge of the non-hermitian evolution to obtain our bounds on the steady state. Therefore, the study of non-hermitian Anderson models is in natural connection to our line of inquiry. One way of viewing our results is that we provide conditions on the decay of fractional moments of resolvent of the non-hermitian evolution that ensure decoherent steady states. Thus, we draw on intuition from non-hermitian Anderson models studied in [31, 26, 6] and one can view our results and methods as a motivation for studying those models.

In a sense, that we describe in Section 3, our result is a direct extension of Anderson localization to open quantum systems. Thereby, the result complements the work of Fröhlich and Schenker [27] who discuss the evolution along the diagonal starting from the state $|0\rangle\langle0|$, whereas we discuss the evolution away from the diagonal in configuration space. Both results are also related to the findings in [16].

Our results may also have implications for collapse theories of quantum mechanics introduced in [28], as we rigorously demonstrate that macroscopic superpositions in the position basis are very fragile. In other words, we demonstrate that single-particle cat states are impossible under weak assumptions. Additionally, our findings contribute to the resource theories of coherence [9, 44].

We emphasize that whenever all $L_k$ are normal (i.e. $L_k^* L_k = L_k L_k^*$) then $\rho_\infty = \frac{1}{|\Lambda|}$ is a steady state, where $|\Lambda|$ is the dimension of the system. Hence, our results are less surprising in those cases, and the implications for the steady states are trivial. With that in mind, our first task, after introducing the setup, is to provide motivating examples, some of which entail non-normal $L_k$. These examples feature Lindblad operators that are non-normal and have been used in dissipative engineering [52, 21]. Since all the examples have coherent steady states and one might view our results as proving the limits of dissipative engineering. While we only study properties of the single particle sector, we note that properties of the steady state sometimes can be lifted from the single particle to a many-body setup (e.g. in [56, 18]). In the future, the limits for dissipative engineering that we prove for single particle systems may be extended to the dissipative preparation of topological states which is discussed in [11, 20, 34].

## 2 Setup and Examples

We consider a single-particle open quantum system. To avoid the mathematical complications of infinite dimensional open quantum systems we will always be looking at finite, but arbitrarily large, subsets $\Lambda \subset \mathbb{Z}^d$ for $d \geq 1$. We emphasize that everything in our setup works on (subsets of) amenable graphs, but we state it on subsets of $\mathbb{Z}^d$.

As in the study of Schrödinger operators, we mainly think of $\Lambda$ as a subset of position space and so we consider the finite-dimensional Hilbert space $\ell^2(\Lambda)$. We then consider evolution of states, that is positive operators $\rho \in \mathcal{B}(\ell^2(\Lambda))$ with $\text{Tr}(\rho) = 1$. The generator of Markovian evolution, henceforth the Lindbladian $\mathcal{L}_\Lambda$, describes the dynamics of states $\rho \in \mathcal{B}(\ell^2(\Lambda))$ and so it is an operator on $\mathcal{B}(\ell^2(\Lambda))$. The Lindbladian has the following form, known as the Lindblad master equation [39, 29],

$$\dot\rho = \mathcal{L}_\Lambda(\rho) = -i[H, \rho] - \frac{1}{2}\sum_k (L_k^* L_k \rho + \rho L_k^* L_k) + \sum_k L_k \rho L_k^*, \tag{1}$$

where the sum is over finitely many $k$. The dynamics of a state in the open quantum system is then described through the semi-group $e^{t\mathcal{L}}$ for any $t > 0$.

Steady states of the dynamics we denote by $\rho_\infty$ and they satisfy that $\dot\rho_\infty = \mathcal{L}(\rho_\infty) = 0$. In the finite-dimensional space we are working on, the existence of steady states is guaranteed [8, Proposition 5], however, they might be non-unique (see e.g. [5] for discussions of these systems). Our results hold for any steady state, regardless of whether we have uniqueness or not, so we refrain from discussions about uniqueness and refer the interested reader to [46] and references therein.

We will assume that $H = \sum_n h_n$ is a sum of finitely many local terms $h_n$ and that $L_k$ are local and elaborate on the locality assumptions in the next section.

Notice that, in finite dimensions, if all the $L_k$ are normal then $\frac{1}{|\Lambda|}$ is a steady state, where $|\Lambda|$ is the number of lattice points in $|\Lambda|$ and therefore the dimension of the system.

## 2.1 Example: Dissipative engineering

Dissipative engineering is the study of using carefully chosen dissipation as a method of preparing quantum states. In the many-body case, examples of coherent states prepared by dissipation are given in [52].

In this section, we give examples, that fit our setup of local single-particle Lindbladians $\mathcal{L}_\Lambda$ with Lindblad operators $L_k$. The examples were investigated from a spectral point of view in one dimension in [37].

Here and in the following, we will also make extensive use of Dirac notation, even though it has to be used with some care in the non-normal case. We single out (what we refer to as) position basis. Notationally, we let $|k\rangle$ denote the vector in $\ell^2(\Lambda)$ with a 1 only at the site $k \in \Lambda$. Conversely, we let $\langle k|$ denote the (conjugate) transpose of the vector.

One of our main motivating examples was introduced in the many-body setting in [21] to describe how dissipation may construct states with desired properties. Its physical realization was discussed [40]. More recently, this dissipation was associated with localization in single-particle open quantum systems in [55] and robustness of the dissipative control was discussed in [51] also in the single-particle case. It was discussed in connection to many-body localization in [50].

**Example 2.1** (Coherence creation). *Suppose that $\Lambda \subset \mathbb{Z}$ and consider the Hilbert space $\ell^2(\Lambda)$. Consider Lindblad operators of the form*

$$L_k = (|k\rangle + |k+l\rangle)(\langle k| - \langle k+l|) \tag{2}$$

*for some $l \in \mathbb{Z}$, such that $k, k+l \in \Lambda$. These $L_k$ are not normal and*

$$L_k^* L_k = 2|k\rangle\langle k| + 2|k+l\rangle\langle k+l| - 2|k\rangle\langle k+l| - 2|k+l\rangle\langle k|.$$

A steady state of a system with consisting solely of this dissipation (and $H = 0$) denoted $\rho_\infty$ is given by $\langle x, \rho_\infty y \rangle = \frac{1}{|\Lambda|}$ for all $x, y \in \Lambda$. Thus, these Lindblad operators actively create coherence in the system. As explained in [21, Sec. II.A] the operator $L_k$ can be thought of as a pumping process where (upon translating the setup to second quantization) the operator $(\langle k| - \langle k+1|)$ annihilates out-of-phase superpositions and then the term $(|k\rangle + |k+1\rangle)$ recreates in-phase superpositions.

We remark that Example 2.1 can be generalized to more general graphs $H = (V, E)$. To do that define for every $e = (vw) \in E$ the operator.

$$L_e = (|v\rangle + |w\rangle)(\langle v| - \langle w|).$$

Then,

$$L_e^* L_e = 2|v\rangle\langle v| + 2|w\rangle\langle w| - 2|v\rangle\langle w| - 2|w\rangle\langle v|,$$

is symmetric under interchange of $v$ and $w$. Furthermore,

$$\sum_{e \in E, e=(v,w)} L_e^* L_e = 2 \sum_{v \in V} \deg(v) |v\rangle\langle v| - 2 \sum_{e \in E, e=(v,w)} |v\rangle\langle w| + |w\rangle\langle v| = 2\Delta_H^G, \tag{3}$$

where $\Delta_H^G$ is the graph Laplacian of the graph $H$.

**Example 2.2** (Local dephasing). *Local dephasing is defined through the Lindblad operators $L_k = |k\rangle\langle k|$ for each $k \in \Lambda$.*

If we only have dephasing then the maximally mixed state $\frac{\mathbb{1}}{|\Lambda|}$ is a steady state for any choice of the Hamiltonian $H$. In the case $H = 0$, all states of the form $|i\rangle\langle i|$ are steady states. The following example, was studied in [37, 56]. Whether it has the maximally mixed state as its steady state depends on the boundary conditions (in the natural one-dimensional setup).

**Example 2.3** (Incoherent hopping). *Let $\Lambda = (V, E)$ be a finite directed graph. Then the incoherent hopping Lindbladian is defined by having a term for every $e = (v \rightarrow w)$ a Lindblad term $L_e = |v\rangle\langle w|$. In particular, for the line graph $[0, n] \subset \mathbb{Z}$ the Lindblad terms are of the form $L_k = |k\rangle\langle k+1|$ for $0 \leq k \leq n-1$.*

The example has similarities with an exclusion process, that, at least in the many-body case, gives examples of open quantum systems that do not have the maximally mixed state as the steady state (see e.g. [48]).

## 2.2 Example Hamiltonians: The Anderson model

For the Hamiltonian $H$ we also assume that it consists of local bounded terms $h_n$. Our standard example of a Hamiltonian is the discrete Laplacian, is given by $H = \Delta$, defined in (3). Central to our enquiry is the notion of a random potential $V$, where $V|x\rangle = V(x)|x\rangle$ are all i.i.d. with bounded, compactly supported density with respect to the Lebesgue measure. For any local (non-random) Hamiltonian $H_0$ and any $\lambda \geq 0$ we say that the system is disordered with strength $\lambda > 0$ if the Hamiltonian is of the form.

$$H = H_0 + \lambda V.$$

In case $H_0$ is the discrete Laplacian we say that $H$ is the Anderson model, introduced by Anderson [7] and since the subject of intense study.

## 2.3 Locality assumptions

Overall, we assume that $\mathcal{L}_\Lambda$ is a sum of local terms with an overall bound $R$ on the radius of the support of local terms. In the present context, we define the support of the operator $A$ on $\ell^2(\Lambda)$ as follows

$$\mathrm{supp}(A) = \{x \in \Lambda \mid \exists y \in \Lambda : \langle y, Ax \rangle \neq 0 \text{ or } \langle x, Ay \rangle \neq 0\}. \tag{4}$$

To formalize the locality of $H$ we consider $H = \sum_n h_n$. In quantum optics and numerical simulation of open quantum systems, the terms in the Lindbladian are often split up into two parts [42]: the quantum jump term and the non-hermitian evolution. The non-hermitian evolution we can write as

$$D_\Lambda = -i \sum_n h_n - \frac{1}{2} \sum_k L_k^* L_k = -i \sum_{Z \subset \Lambda} h_Z - K_Z, \tag{5}$$

where we on the right hand side have grouped the $h_n$ and $L_k^* L_k$ in such a way that

$$h_Z = \sum_{n:\mathrm{supp}(h_n)=Z} h_n, \qquad K_Z = \frac{1}{2} \sum_{k:\mathrm{supp}(L_k)=Z} L_k^* L_k. \tag{6}$$

Notice that $\mathrm{supp}(h_Z) = Z$ and $\mathrm{supp}(K_Z) \subset Z$.

The second term $\sum_{k \in \mathbb{Z}} L_k \rho L_k^*$ is known as the quantum jump term. Generally, the quantum jump term is more complicated to work with. One of the insights in this paper is how properties of the time-evolution of $\rho$ and more specifically the steady state $\rho_\infty$ can be inferred only from knowledge of (the Green function of) the non-hermitian evolution. Throughout, we will have the following locality assumptions on $\mathcal{L}_\Lambda$.

**Assumption 2.4.** *We say that a familiy of Lindbladians $\mathcal{L} = \{\mathcal{L}_\Lambda\}_\Lambda$ is local if there exists an $R$ and and $N$ such that for all $\Lambda$ it holds that $\mathcal{L}_\Lambda$ is a Lindbladian of the form* (1) *with decomposition according to* (5) *satisfying*

- (Finite-Range) *It holds that $h_Z = 0 = K_Z$, whenever $\text{diam}(Z) > R$.*

- (Finite Norm) *There is a $N > 0$ such that $\|h_Z\| \le N$ and $\|K_Z\| \le N$, for all $Z \subset \Lambda$.*

## 2.4 Gap assumptions on the non-hermitian evolution

We will need some extra assumptions to get the technicalities of the argument to work. To introduce them, define for a finite subgraph $H \subset \Lambda \subset \mathbb{Z}^d$ the points that are at most distance $R$ from the boundary $H$, by

$$\partial_R^\Lambda H = \{v \in H \mid \text{dist}(v, \Lambda \setminus H) \le R\}.$$

In particular, we will use the inner vertex boundary defined by $\partial H = \partial_1 H$. Define also the boundary operator

$$B_{\partial_R H}^\Lambda = \sum_{v \in \partial_R^\Lambda H} |v\rangle\langle v| \,. \tag{7}$$

For any set $\Gamma \subset \Lambda$, define the non-hermitian evolution in $\Gamma$ by

$$D_\Gamma = -i \sum_{Z \subset \Gamma} h_Z - K_Z. \tag{8}$$

Hence,

$$\text{Re}(D_\Gamma) = -\sum_{Z \subset \Gamma} K_Z,$$

is a negative semi-definite operator. In the following, we let $|\Gamma|$ denote the number of vertices in a subgraph $\Gamma$ and for two vertices $x, y$ then $|x - y|$ denote the graph distance (in $\Lambda$) between $x$ and $y$. Let further $B_L^\Lambda(v) = \{w \in \Lambda \mid |v - w| \le L\}$ be the ball of radius $L$ around $v$ in $\Lambda$. Now, the following assumption is satisfied for our motivating examples of local dephasing and coherence creation as we show in Section 2.5.

**Assumption 2.5** (Inverse polynomially slow gap closing of non-hermitian evolution). *There exist constants $c, c_0, \alpha > 0$ such that for any finite connected set $\Lambda \subset \mathbb{Z}^d$, any vertex $v \in \Lambda$ and any $L < \frac{1}{2} \text{diam}(\Lambda)$ it holds that*

$$\frac{c}{|L|^\alpha} \, \mathbb{1}_{B_L^\Lambda(v)} \le c_0 B_{\partial_R(B_L^\Lambda(v))}^\Lambda - \text{Re}(D_{B_L^\Lambda(v)}).$$

Recall that the operator $D_{B_L^\Lambda(v)}$ is dissipative and thus has spectrum and numerical range in the left half of the complex plane. However, the non-hermitian evolution can have a gap that closes in the thermodynamic limit (as was proved for Example 2.1 in [37]). However, due to a technical trick that we will elaborate on in Section 4.3 are able to impose Dirichlet boundary conditions on the non-hermitian evolution. With Dirichlet boundary conditions the gap still vanishes in the thermodynamic limit, but it only vanishes polynomially fast (as we explain in Appendix C).

In the case of local dephasing, see Example 2.2, a stronger assumption is satisfied.

**Assumption 2.6** (Constant gap of non-hermitian evolution). *There exists a constant $\gamma > 0$ such that for any finite subset $\Gamma \subset \Lambda \subset \mathbb{Z}^d$ then*

$$\mathrm{Re}(D_\Gamma) \leq -\gamma \mathbb{1}_\Gamma.$$

## 2.5 Motivating examples satisfy gap assumptions

For a graph $H = (V, E)$ recall the graph Laplacian $\Delta_H^G$ is defined by

$$\Delta_H^G = \sum_{v \in V} \deg(v) \, |v\rangle\langle v| - \sum_{e \in E, e = (v,w)} |v\rangle\langle w| + |w\rangle\langle v|.$$

If we let some vertices $\partial_v G \subset V$ be boundary vertices then we can define the Dirichlet Laplacian by

$$\Delta_H^D = \Delta_H^G + \sum_{k \in \partial_v H} |k\rangle\langle k|.$$

The boundary operator $\sum_{k \in \partial_v H} |k\rangle\langle k|$ we will also denote by $B_H$. Note that all three operators $\Delta_H^G, \Delta_H^D$ and $B_H$ are positive.

Let us argue that our motivating example of coherence creation from Example 2.1 satisfies Assumption 2.5. To do it recall from (3) how for any $\Gamma \subset \Lambda$

$$\mathrm{Re}(D_\Gamma) = -\frac{1}{2} \sum_{e \in \Gamma} L_e^* L_e = -\Delta_\Gamma^G,$$

the graph Laplacian in $\Gamma$. Therefore,

$$\Delta_\Gamma^D = B_{\partial_1 \Gamma}^\Lambda - \mathrm{Re}(D_\Gamma) \leq B_{\partial_R \Gamma}^\Lambda - \mathrm{Re}(D_\Gamma).$$

Now, suppose that $\Lambda \subset \mathbb{Z}^d$ is finite and connected and let us consider $\Gamma = B_L^\Lambda(v)$ for any vertex $v \in \Lambda$ and $L < \frac{1}{2} \mathrm{diam}(\Lambda)$. Then $\left| B_L^\Lambda(v) \right| \leq (2L)^d$. Furthermore, since $\Lambda$ is connected then $\left| \partial_v B_L^\Lambda(v) \right| \geq 1$, thereby we obtain from Lemma C.2 that the assumption is satisfied with $\alpha = 4d$.

For local dephasing Example 2.2 it holds that

$$\mathrm{Re}(D_x) = -\frac{1}{2} \sum_{k \in B_D^x} \gamma \, |k\rangle\langle k| = -\frac{\gamma}{2} \mathbb{1}_{\Lambda_x}$$

and so Assumption 2.6 is satisfied.

In the example of incoherent hopping Example 2.3 we get that

$$\mathrm{Re}(D_x) = -\frac{1}{2} \sum_{k \in B_D^x} |k+l\rangle\langle k+l|.$$

While one can do a technical trick as in Section 4.3, this example cannot quite be taken to satisfy Assumption 2.6. However, one can still get an exponential decay estimate with respect to the distance to the nearest point in the graph that has no edge point out of it. We note that the steady state in the one-dimensional case can be explicitly computed.

# 3 Main result

In situations where the Lindbladian $\mathcal{L}_\Lambda$ has multiple steady states (for example in the case of pure Hamiltonian evolution), it may occur that $e^{t\mathcal{L}_\Lambda}(\rho_0)$ does not have a limit. However, we can still time-average $e^{t\mathcal{L}_\Lambda}(\rho_0)$. The Abel average is a particular time average that will be useful since it directly relates to the resolvent.

For any $\varepsilon > 0$ and state $\rho_0$ we define the Abel average $\rho_\varepsilon$ by

$$\rho_\varepsilon = \varepsilon \int_0^\infty e^{-t\varepsilon}\, e^{t\mathcal{L}_\Lambda}(\rho_0)dt = -\varepsilon(\mathcal{L}_\Lambda - \varepsilon)^{-1}(\rho_0). \tag{9}$$

Since $e^{t\mathcal{L}_\Lambda}$ maps states to states it means that $\rho_\varepsilon$ is a convex combination of states and therefore a state. In particular, it holds that $|\rho_\varepsilon(x,y)| \leq 1$ for all $\varepsilon > 0$ and $x, y \in \Lambda$. Since $e^{t\mathcal{L}_\Lambda}(\rho_0)$ is the time evolution of the state $\rho$ until time $t$, we interpret $\rho_\varepsilon$ as the time average up to timescales of $\frac{1}{\varepsilon}$.

In that spirit, it holds that $\varepsilon \to 0$ the state $\rho_0$ gets projected the kernel of $\mathcal{L}_\Lambda$, that is the subspace of steady states,

$$\lim_{\varepsilon \downarrow 0} \rho_\varepsilon = \lim_{\varepsilon \downarrow 0} -\varepsilon(\mathcal{L}_\Lambda - \varepsilon)^{-1}(\rho_0) = P_{\ker(\mathcal{L}_\Lambda)}(\rho_0). \tag{10}$$

Naturally, in the case $\mathcal{L}$ has a unique steady state $\rho_\infty$, then $P_{\ker(\mathcal{L}_\Lambda)}(\rho_0) = \rho_\infty$ for every initial state $\rho_0$.

**Theorem 3.1.** *Let $\mathcal{L}$ be local in the sense of Assumption 2.4 such that the non-hermitian evolution satisifies Assumption 2.5. For sufficiently large disorder $\lambda > 0$, there exist constants $C, \mu > 0$ such that for any connected set $\Lambda \subset \mathbb{Z}^d$ and any $x, y \in \Lambda, \varepsilon \in (0,1)$, initial state $\rho_0$, and any $s \in (0,1)$ there exists a $C_s > 0$ such that*

$$\mathbb{E}|\rho_\varepsilon(x,y)| \leq Ce^{-\mu|x-y|} + \varepsilon^{2s-1}\, C_s. \tag{11}$$

*Furthermore, for any measurable choice of steady state $\omega \mapsto \rho_\infty(\omega)$ of $\mathcal{L}_\Lambda$ it holds that*

$$\mathbb{E}|\rho_\infty(x,y)| \leq Ce^{-\mu|x-y|}. \tag{12}$$

In the case of any degree of local dephasing we have the following deterministic result. We prove both results in Section 5.3.

**Theorem 3.2.** *Let $\mathcal{L}$ be local in the sense of Assumption 2.4 such that the non-hermitian evolution satisifies Assumption 2.6 for some $\gamma > 0$. Then there exist $C, \mu > 0$ such that for any $\Lambda \subset \mathbb{Z}^d$, any $x, y \in \Lambda$, initial state $\rho_0$, $\varepsilon \in (0,1)$, and any $s \in (0,1)$ there exists a $C_s > 0$ such that*

$$|\rho_\varepsilon(x,y)| \leq Ce^{-\mu|x-y|} + \varepsilon^{2s-1}\, C_s. \tag{13}$$

*In particular, for any steady state $\rho_\infty$ of $\mathcal{L}_\Lambda$ then*

$$|\rho_\infty(x,y)| \leq Ce^{-\mu|x-y|}. \tag{14}$$

One application of the result is to Lindbladians of the form $\mathcal{L} = \mathcal{L}_1 + \mathcal{L}_2$, where $\mathcal{L}_1, \mathcal{L}_2$ are two Lindbladian such that one of them satisfies Assumption 2.6. Thus, if the Lindbladian $\mathcal{L}_1$ is the local dephasing Lindbladian from Example 2.2 then no matter how weak the dephasing $\gamma > 0$ is, it will dominate the steady state of any local Lindbladian $\mathcal{L} = \mathcal{L}_1 + \mathcal{L}_2$. Or phrased in another way, that the steady-state $\rho_\infty$ of the Lindbladian in the example Example 2.1 for $H = 0$ is not stable under perturbations, potentially an aspect of the non-normality of $\mathcal{L}$.

Finally, let us describe why our result can be viewed as an extension to open quantum systems. Consider the case where there are no Lindblad terms $L_k$ and the system is only governed by unitary evolution corresponding to the Hamiltonian $H$. The evolution of the system on states $\rho$ is given by $\mathcal{L}(\rho) = -i[H, \rho]$. To see that, for any eigenfunction $\psi$ of $H$ it holds that $|\psi\rangle\langle\psi|$ is a steady state $\rho_\infty$ of the Hamiltonian. Thus, our main result in the disordered case Theorem 3.1 states, if it was applicable, that there exist constants $C, \mu > 0$ such that for every eigenfunction $\psi$ then

$$\mathbb{E}(|\psi(x)\psi(y)|) \leq Ce^{-\mu|x-y|},$$

which shows that the eigenfunctions must be exponentially localized.

# 4 Decomposition of $\mathcal{L}_\Lambda$ given vertices $x$ and $y$

In this section, we discuss a decomposition $\Lambda$ that will allow us to reduce the problem from the full Lindbladian to the effective non-hermitian evolution.

## 4.1 Decomposition of $\Lambda \subset \mathbb{Z}^d$

We consider the following change of notation.

$$\mathcal{L}_\Lambda = \sum_{Z \subset \Lambda} \ell_Z, \tag{15}$$

where $\ell_Z$ consist of all terms $L_k$ and $h_n$ that have support $Z$. That is

$$\ell_Z(\rho) = \sum_{k:\mathrm{supp}(L_k)=Z} L_k \rho L_k^* - \frac{1}{2}(L_k^* L_k \rho + \rho L_k^* L_k) - i \sum_{n:\mathrm{supp}(h_n)=Z} [h_n, \rho], \tag{16}$$

which in the terms of $h_Z$ and $K_Z$ defined in (6) means that

$$\ell_Z(\rho) = i\rho h_Z - ih_Z\rho + K_Z\rho + \rho K_Z + \sum_{k:\mathrm{supp}(L_k)=Z} L_k \rho L_k^*. \tag{17}$$

In all of the following, we suppose that $|x - y| \geq 6R$. Then define the regions

$$\Lambda_x = B_{\frac{|x-y|}{3}}(x) = \left\{ a \in \mathbb{Z}^d \mid |x - a| \leq \frac{|x-y|}{3} \right\}.$$

Figure 1: The points $x$ and $y$ and the construction of $\Lambda_x, \Lambda_y$ and $\Lambda_\perp^{x,y}$. Any term that acts non-trivially in $\Lambda_x$ and $\Lambda_y$ can only act non-trivially a safety distance of $R$ into $\Lambda_\perp^{x,y}$.

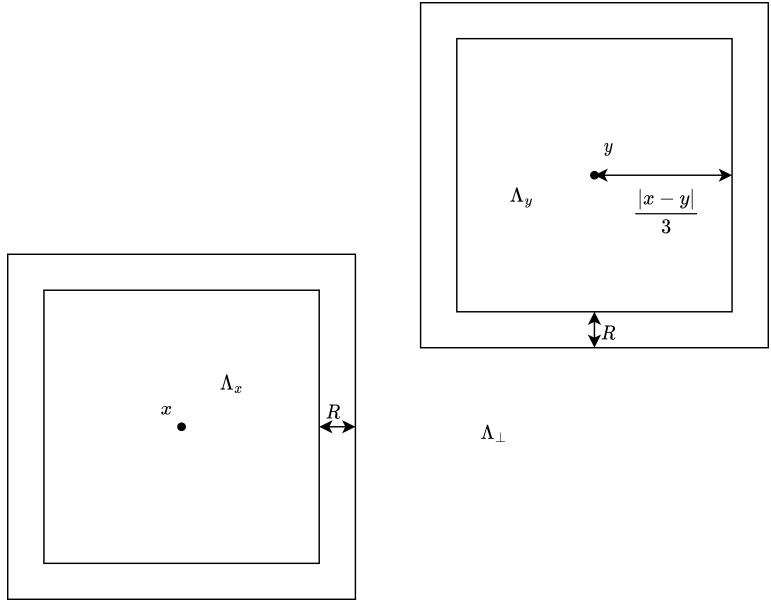

and $\Lambda_y = B_{\frac{|x-y|}{3}}(y)$. Let $\Lambda_\perp^{x,y} = \Lambda \setminus \left( B_{\frac{|x-y|}{3}}(y) \cup B_{\frac{|x-y|}{3}}(y) \right)$. We assume throughout that $\Lambda_x, \Lambda_y \subset \Lambda$. See also the construction in Figure 1.

We now aim to split up the Lindbladian in $\mathcal{L}_\Lambda$ into an operator $\mathcal{L}_\Lambda^0$ that is local on $\Lambda_x, \Lambda_y$ and $\Lambda_\perp^{x,y}$ and an operator $\mathcal{L}_\Lambda^\partial$ that connects the regions. To do that, let

$$\mathcal{L}_\Lambda = \mathcal{L}_\Lambda^0 + \mathcal{L}_\Lambda^\partial, \tag{18}$$

with

$$\mathcal{L}_\Lambda^0 = \sum_{Z \subset \Lambda_x} \ell_Z + \sum_{Z \subset \Lambda_y} \ell_Z + \sum_{Z \subset \Lambda_\perp^{x,y}} \ell_Z.$$

That means that $\mathcal{L}_\Lambda^{x,y}$ consists of all the terms that are supported in $\Lambda_x, \Lambda_y$ or $\Lambda_\perp^{x,y}$ and $\mathcal{L}_\Lambda^{\partial_{x,y}}$ consists of all the terms that connect $\Lambda_x \cup \Lambda_y$ to the complement $\Lambda_\perp^{x,y}$.

Next, we introduce vectorization which one can use to view the operators $\mathcal{L}_\Lambda, \mathcal{L}_\Lambda^0, \mathcal{L}_\Lambda^\partial$ as operators on $\ell^2(\Lambda \times \Lambda)$.

## 4.2 Vectorization

The superoperator $\mathcal{L}$ is complicated to work with, so it is natural to vectorize it. This we can do as follows, for a more detailed discussion see [38, (4.88)]. Here we consider only finite sets $\Lambda$, which means that $\mathcal{B}(\ell^2(\Lambda)) = \mathrm{HS}(\ell^2(\Lambda))$, which will ease the discussion, for a discussion for infinite dimensional spaces see for example [37, Section 3.1]. Let us define a linear map $\mathrm{vec} : \mathcal{B}(\ell^2(\Lambda)) \to \ell^2(\Lambda) \otimes \ell^2(\Lambda)$ by $\mathrm{vec}(|i\rangle \langle j|) = |i\rangle |j\rangle = |i,j\rangle$, then it holds (and is best verified by straightforward computation) that

$$\mathrm{vec}(ABC) = (A \otimes C^T)\mathrm{vec}(B),$$

for $A, B, C \in \mathcal{B}(\ell^2(\Lambda))$ and where $T$ denotes the transpose.

Let us note that when the bounded operators on $\ell^2(\Lambda)$ for a finite set $\Lambda$ are viewed with the Hilbert Schmidt norm $\mathrm{HS}(\ell^2(\Lambda))$, given by $\|A\|_2^2 = \sum_{i,j} |A(i,j)|^2$. Then the map $\mathrm{vec} : \mathrm{HS}(\ell^2(\Lambda)) \to \ell^2(\Lambda \times \Lambda)$ is an isometric isomorphism. We write $\mathcal{B}(\ell^2(\Lambda)) \cong \ell^2(\Lambda \times \Lambda)$ to indicate this isomorphism.

This means for example that $-i[H, \rho]$ in vectorized form becomes

$$\mathrm{vec}(-i[H, \rho]) = \mathrm{vec}(-iH\rho\mathbb{1} + i\mathbb{1}\rho H) = -iH \otimes \mathbb{1}\mathrm{vec}(\rho) + \mathbb{1} \otimes iH^T\mathrm{vec}(\rho)$$

and so the operator $\mathcal{L}_H$ defined by $\mathcal{L}_H(\rho) = -i[H, \rho]$ is $-iH \otimes \mathbb{1} + i(1 \otimes H^T)$ in vectorized form.

In the vectorized picture, we obtain a block diagonal form of the operator $\mathcal{L}_\Lambda^0$. As the finite set $\Lambda$ is the disjoint union of $\Lambda_i$ for $i \in \{x, y, \perp\}$, we can consider the Lindbladian $\mathcal{L}$ which is an operator on $\mathcal{B}(\ell^2(\Lambda)) \cong \ell^2(\Lambda \times \Lambda) = \oplus_{i,j \in \{x,y,\perp\}} \ell^2(\Lambda_i) \otimes \ell^2(\Lambda_j)$. The following lemma states that $\mathcal{L}_\Lambda^0$ is block diagonal with respect to this decomposition.

**Lemma 4.1.** *On the space $\mathcal{B}(\ell^2(\Lambda)) \cong \ell^2(\Lambda \times \Lambda) = \oplus_{i,j \in \{x,y,\perp\}} \ell^2(\Lambda_i) \otimes \ell^2(\Lambda_j)$ the operator $\mathcal{L}_\Lambda^0$ has block diagonal form*

$$\mathcal{L}_\Lambda^0 = \oplus_{i,j \in \{x,y,\perp\}} \mathcal{L}_{i,j}.$$

*All the eigenvalues of each of the operators $\mathcal{L}_{i,j}$ for $i, j \in \{x, y, \perp\}$ have non-positive real part. Moreover, for any $\varepsilon > 0$ and $u, v \in \Lambda$,*

$$\langle (x, y), (\mathcal{L}_\Lambda^0 - \varepsilon)^{-1}(u, v)\rangle = \mathbb{1}_{u \in \Lambda_x} \mathbb{1}_{v \in \Lambda_y} \langle (x, y), (\mathcal{L}_{x,y} - \varepsilon)^{-1}(u, v)\rangle.$$

*Proof.* The decomposition follows if $\mathcal{L}_\Lambda^0(u, v)$ is supported in $\ell^2(\Lambda_i) \otimes \ell^2(\Lambda_j)$ if $u \in \Lambda_i$ and $v \in \Lambda_j$ for $i, j \in \{x, y, \perp\}$. This follows directly from the construction of $\mathcal{L}_\Lambda^0$. Since $\mathcal{L}_\Lambda^0$ has Lindblad form all its eigenvalues have negative real part and since the eigenvalues of a direct sum of operators are the union of the eigenvalues it follows that all eigenvalues of each of the operators $\mathcal{L}_{i,j}$ for $i, j \in \{x, y, \perp\}$ have negative real part. The last statement follows directly. $\qquad\square$

Notice that the operators $\mathcal{L}_{i,i}$ have Lindblad form, but the operator $\mathcal{L}_{x,y}$ does not. With the vectorization at hand, we can state how the finite-range assumption influences the support of $\mathcal{L}_\Lambda^\partial$. To do that, we consider the points that are at most distance $R$ from the boundary

$$\delta_R(\Lambda_x \cup \Lambda_y) = \{v \in \mathbb{Z}^d \mid \mathrm{dist}(v, \partial\Lambda_x) \leq R \text{ or } \mathrm{dist}(v, \partial\Lambda_y) \leq R\}.$$

Let also
$$\mathcal{C}_R(x,y) = \Lambda \times \delta_R(\Lambda_x \cup \Lambda_y) \cup \delta_R(\Lambda_x \cup \Lambda_y) \times \Lambda,$$
which will upper bound the support of $\mathcal{L}_\Lambda^\partial$ as we now show, see also Figure 2. Note that here we define the support of a vectorized superoperator analogously to (4) with position space $\Lambda$ interchanged with configuration space $\Lambda \times \Lambda$.

**Lemma 4.2.** *It holds that*
$$\mathrm{supp}(\mathcal{L}_\Lambda^\partial) \subset \mathcal{C}_R(x,y).$$
*In particular, for any $(u,v) \in \mathrm{supp}(\mathcal{L}_\Lambda^\partial)$ then either $|x-u| \geq \frac{|x-y|}{3} - R$ or $|v-y| \geq \frac{|x-y|}{3} - R$ and thus for any $(u,v) \in \mathcal{C}_R(x,y)$ it holds that*

$$|x-u| + |v-y| \geq \frac{|x-y|}{3} - R.$$

*Proof.* Suppose that $(u,v) \in \mathrm{supp}(\mathcal{L}_\Lambda^\partial)$. Then there is a term $h_n$ or $L_k$ with support not contained in any of the sets $\Lambda_x, \Lambda_y, \Lambda_{x,y}^\perp$, that is nonzero upon acting on $|u\rangle\langle v|$ from either the left or the right. Due to the finite-range it means that either $u \in \delta_R(\Lambda_x \cup \Lambda_y)$ or $v \in \delta_R(\Lambda_x \cup \Lambda_y)$. $\square$

Let us abbreviate $D_{\Lambda_x}$ defined in (8) by $D_x$ and slightly different we define $D_y$ such that

$$D_y^T = i \sum_{Z \subset \Gamma} h_Z - K_Z$$

where $T$ denotes the transpose. Let us look only at the off-diagonal term $\mathcal{L}_{x,y}$ defined in Lemma 4.1. The central insight is that all the cross terms vanish since the $L_k$ terms only act on either $\Lambda_x$ or $\Lambda_y$. Thus, $\mathcal{L}_{x,y}$ depends on the non-hermitian evolution only and it holds that

$$\mathcal{L}_{x,y}\,(|x\rangle\langle y|) = D_x\,|x\rangle\langle y|\ + |x\rangle\langle y|\,D_y^T.$$

Hence, we obtain the following vectorized form

$$\mathcal{L}_{x,y}\ = D_x \otimes \mathbb{1} + \mathbb{1} \otimes D_y,$$

where $D_x$ and $D_y$ are both dissipative, and therefore $\mathcal{L}_{x,y}$ is dissipative as well. We summarize the finding in the following lemma.

**Lemma 4.3.** *The off-diagonal operator $\mathcal{L}_{x,y}$ on $\ell^2(\Lambda_x) \otimes \ell^2(\Lambda_y)$ in the decomposition in Lemma 4.1 has the following form*
$$\mathcal{L}_{x,y}\ = D_x \otimes \mathbb{1} + \mathbb{1} \otimes D_y.$$
*with $D_x$ and $D_y$ being dissipative operators acting on $\ell^2(\Lambda_x)$ and $\ell^2(\Lambda_y)$ respectively. In particular, the operator does not have any contribution to the quantum jumps.*

The decomposition means that for $\mathcal{L}_{x,y}$ there is no interaction between $\Lambda_x$ and $\Lambda_y$ and furthermore for any real $\varepsilon > 0$ then $\mathcal{L}_{x,y} - \varepsilon$.

Figure 2: The configuration space $\Lambda \times \Lambda$ with the points $x$ and $y$ and the support of $\mathcal{L}_\Lambda^\partial$. The coordinate axes are indicated with dotted lines and the support of $\mathcal{L}_\Lambda^\partial$ with full lines. Notice how it has a double cross structure and how the point $(x,y)$ in $\Lambda_x \times \Lambda_y$ is distance $\frac{|x-y|}{3}$ from the support of $\mathcal{L}_\Lambda^\partial$.

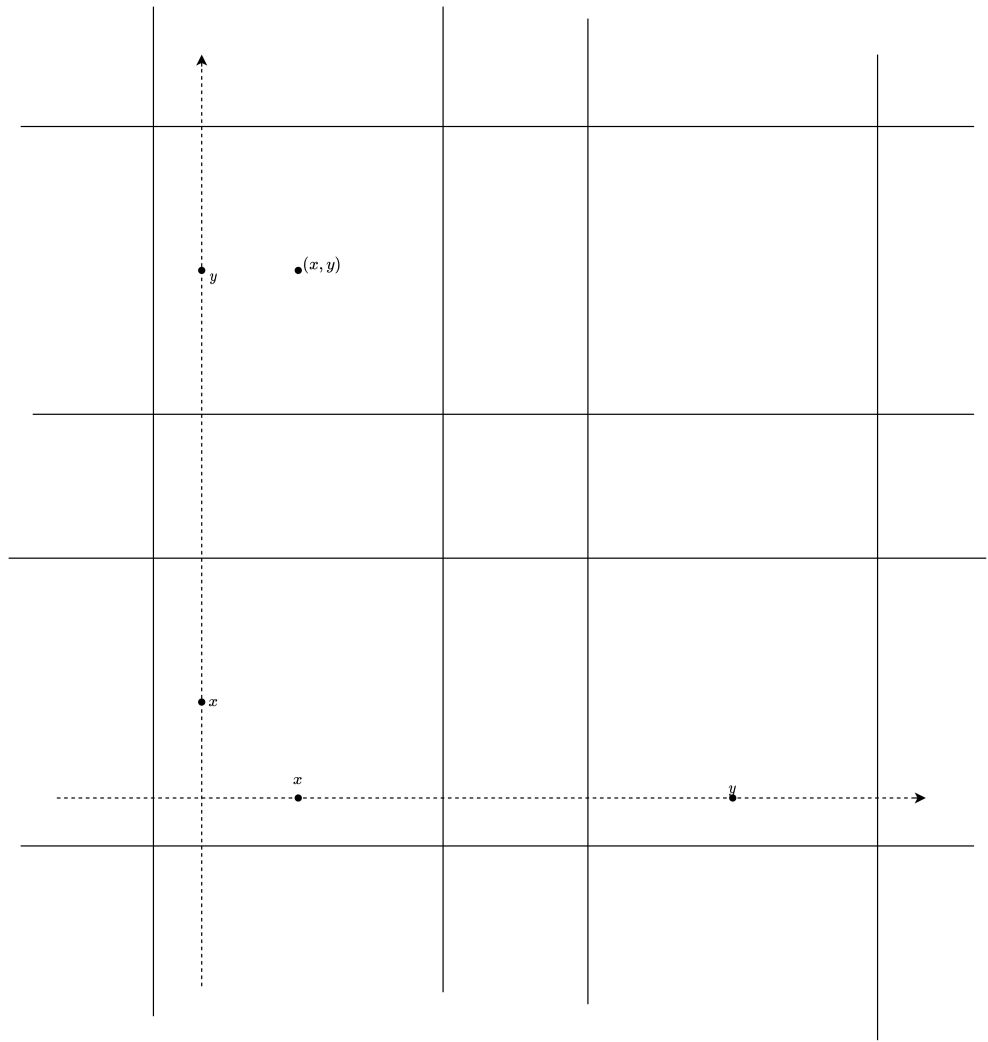

### 4.3 Getting additional boundary terms in $\mathcal{L}_{x,y}$

In the following, we will use a slight extension of the construction of $\mathcal{L}_{x,y}$ above to get additional boundary terms, that were used to relax the assumptions in Assumption 2.5 so that it is satisfied for our motivating example. For the box $\Lambda_x$ and $\Lambda_y$ consider their boundaries $\partial\Lambda_x$ and $\partial\Lambda_y$ that we recall are defined by $\partial\Lambda_i = \{v \in \Lambda \mid \operatorname{dist}(v, \Lambda \setminus \Lambda_i) = 1\}$ for $i \in \{x, y\}$. Both of these are non-empty since $\Lambda$ is not contained in $\Lambda_x \cup \Lambda_y$. Define then for $i \in \{x, y\}$ the non-zero boundary dephasing operators

$$B_i = \sum_{k \in \partial\Lambda_i} |k\rangle\langle k|. \tag{19}$$

Let further $\mathcal{B} = B_x \otimes \mathbb{1} + \mathbb{1} \otimes B_y$. Define the modified operators $\tilde{\mathcal{L}}_\Lambda^0 = \mathcal{L}_\Lambda^0 + \mathcal{B}$ and $\tilde{\mathcal{L}}_\Lambda^\partial = \mathcal{L}_\Lambda^\partial - \mathcal{B}$. Then the split up becomes

$$\mathcal{L}_\Lambda = \mathcal{L}_\Lambda^0 + \mathcal{L}_\Lambda^\partial = \tilde{\mathcal{L}}_\Lambda^0 + \tilde{\mathcal{L}}_\Lambda^\partial. \tag{20}$$

Now, the operator $\tilde{\mathcal{L}}_\Lambda^0$ is still reducing on the subspace $\ell^2(\Lambda_i) \otimes \ell^2(\Lambda_j)$ and the operator $\tilde{\mathcal{L}}_\Lambda^\partial$ still satisfies $\operatorname{supp}(\tilde{\mathcal{L}}_\Lambda^\partial) \subset \mathcal{C}_R(x, y)$, which ensures the structure sketched on Figure 2. For the $(x, y)$ block of $\tilde{\mathcal{L}}_\Lambda^0$ we now, since $B_y^T = B_y$, obtain the operator

$$\tilde{\mathcal{L}}_{x,y} = \mathcal{L}_{x,y} + B_x \otimes \mathbb{1} + \mathbb{1} \otimes B_y = (D_x + B_x) \otimes \mathbb{1} + \mathbb{1} \otimes (D_y + B_y).$$

**Remark 4.4.** *The operator $\tilde{\mathcal{L}}_\Lambda^0$ does not have Lindblad form (and the Lindblad form will not be necessary for us). On the other hand, the operator $B_x \otimes \mathbb{1} + \mathbb{1} \otimes B_y$ could arise as the $(x, y)$ block of the decomposition from a Lindbladian corresponding to Lindblad operators $L_k = |k\rangle\langle k|$ for each $k \in \partial\Lambda_i$.*

### 4.4 Fractional moment estimates on the effective evolution

The fractional moment method, initiated in [2], is a successful approach to localization of random operators. Here we briefly sketch how the arguments for strong disorder carry over to the dissipative non-selfadjoint case. That is,

$$A_\lambda = A_0 + i\lambda V, \tag{21}$$

for a dissipative non-random operator $A_0$. The randomness is still independent and diagonal $V |x\rangle = V(x) |x\rangle$ where are all i.i.d with bounded density $\rho$ with respect to the Lebesgue measure. Furthermore, we assume that $\rho$ has compact support.

In particular, one may go through the proofs of fractional moment estimates in the Hamiltonian case and try to generalize them to the case of a local, dissipative operator. The strategy works in the case of large disorder whenever the non-hermitian evolution $A$ is finite-range with bounded coefficients. This is the case whenever $\mathcal{L}_\Lambda$ is local in the sense of Assumption 2.4.

In the following, let $\mathbb{C}_- = \{z \in \mathbb{C} \mid \mathrm{Re}(z) < 0\}, \overline{\mathbb{C}_-} = \{z \in \mathbb{C} \mid \mathrm{Re}(z) \leq 0\}$ and $\mathbb{C}_+ = \mathbb{C}_-^c$. We also define the numerical range of an operator $A$ by

$$W(A) = \{\langle \psi, W\psi \rangle \mid \|\psi\| = 1\}. \tag{22}$$

For any $S \subset \Lambda$ we also let $\mathbb{E}_S$ denote the expectation with respect to the variables $\{V(x)\}_{x \in S}$. Then for any $z \in \mathbb{C}_+$ it holds that $A_\lambda - z$ is invertible and we define the Green function $G_\Lambda(x, y; z)$ by $G_\Lambda(x, y; z) = \langle x, (A_\lambda - z)^{-1} \, y \rangle$. Then, the following a priori estimate follows from rank one perturbation theory analogous to the self-adjoint case, see [4, (5.4), (6.16)] for details.

$$\mathbb{E}_{\{x\}}[|G_\Lambda(x, x; z)|^s] \leq \frac{C_s}{\lambda^s}. \tag{23}$$

Furthermore, by a similar exercise in rank-2 perturbation theory (cf. [4, Theorem 8.3]), we obtain under the same conditions that

$$\mathbb{E}_{\{x,y\}}[|G_\Lambda(x, y; z)|^s] \leq \frac{C_s}{\lambda^s}. \tag{24}$$

The following theorem, which is central to our proof is a modification of [4, Corollary 10.1] to the non-selfadjoint case and the case of a range 1 operator to a range $R$ operator. Here for a bounded operator on $\ell^2(\Lambda)$ we define

$$\|A_0\|_{\infty,\infty} = \sup_x \sum_y |A_0(x, y)|.$$

**Theorem 4.5.** *For any bounded operator $A_0$ on $\ell^2(\Lambda)$ satisfying $A_0(u, v) = 0$ for $|u - v| > R$ and $\mathrm{Re}(W(A_0)) \leq a$ for some $a \in \mathbb{R}$. Then there exists $\lambda_0 > 0$ such that for all $\lambda \geq \lambda_0$, then for all $s \in (0, 1)$, there exist $C_s, \mu_s > 0$ such that for any $x, y \in \Lambda$, and any $z \in \mathbb{C}$ with $\mathrm{Re}(z) > a$,*

$$\mathbb{E}[|G_\Lambda(x, y; z)|^s] \leq C_s e^{-\mu_s |x-y|}.$$

*Proof.* We do suitable modifications to [4, Corollary 10.1]. As in the one-step bound there, we first use the resolvent equation, which we can use without issue since $\mathrm{Re}(z) > a$. Whenever $x \neq y$ it holds that

$$G_\Lambda(x, y; z) = \sum_{y' \neq x} G_\Lambda(x, x; z) A_0(x, y') \, G_{\Lambda \backslash \{x\}}(y', y; z).$$

Since $A_0(x, y') = 0$ whenever $|x - y'| \geq R$, one can view this as one step in a walk with stepsize $R$. Thus,

$$\mathbb{E}_\Lambda[|G_\Lambda(x, y; z)|^s] \leq \sum_{y' \neq x} \mathbb{E}_\Lambda \left[|G_\Lambda(x, x; z) A_0(x, y') \, G_{\Lambda \backslash \{x\}}(y', y; z)|^s\right].$$

Now since $\left|G_{\Lambda\setminus\{x\}}(y', y; z)\right|^s$ is independent of the value of the potential at $x$, we can integrate the potential of $x$ out first and we get that using (23),

$$\mathbb{E}_\Lambda\left[|G_\Lambda(x, y; z)|^s\right] \leq \frac{\|A_0\|_{\infty,\infty}\, C_s}{\lambda^s} \sum_{y': 0 < |y'-x| < R} \mathbb{E}_{\Lambda\setminus\{x\}}\left[\left|G_{\Lambda\setminus\{x\}}(y', y; z)\right|^s\right].$$

We can iterate this up to $\frac{|x-y|}{R}$ times. In every iteration, we get less than $R^d$ terms. Then we obtain using (24) that

$$\mathbb{E}_\Lambda\left[|G_\Lambda(x, y; z)|^s\right] \leq \left(\frac{C_s\|A_0\|_{\infty,\infty}}{\lambda^s}\right)^{\frac{|x-y|}{R}} (2R)^{d\frac{|x-y|}{R}} \frac{C_s}{\lambda^s}.$$

Thus, picking $\lambda$ sufficiently large the expression yields exponential decay. $\qquad\square$

# 5  Proof of exponential decay of coherences

In this section, we connect the previous results and the assumptions that we have shown that our motivating examples satisfy to prove our main theorem. But before that, we first prove exponential decay of the modified resolvent of $\mathcal{L}_\Lambda$. In all of the following, we assume that the model satisfies Assumption 2.5 and that $x, y$ are fixed vertices that are taken as starting point of construction of $\mathcal{L}_{x,y}$.

## 5.1  Exponential decay of resolvent of $\mathcal{L}_{x,y}$

Let us lift the exponential decay of fractional moments from Section 4.4 to the case of the resolvent of $\mathcal{L}_{x,y}$ using the assumptions in Section 2.5.

**Lemma 5.1.**  *Under the assumptions of Theorem 3.1, then it holds that for any $(u, v) \in \Lambda_x \times \Lambda_y$ that*

$$\mathbb{E}\left[\left|\langle(x, y), (\mathcal{L}_{x,y} - \varepsilon)^{-1}(u, v)\rangle\right|\right] \leq C|x - y|^{\alpha d} e^{-\mu|x-u|} e^{-\mu|v-y|}. \tag{25}$$

*Proof.* By Lemma B.2 for a contour $\Gamma$ that encloses and is disjoint from the pseudospectrum $\sigma_r(D_x - \varepsilon)$ for some $r > 0$, then for any $s \in (0, 1)$,

$$\begin{aligned}
\left|\langle(x, y), (\mathcal{L}_{x,y} - 2\varepsilon)^{-1}(u, v)\rangle\right| &= \left|\left\langle(x, y), \frac{1}{(D_x - \varepsilon) \otimes \mathbb{1} + \mathbb{1} \otimes (D_y - \varepsilon)}(u, v)\right\rangle\right| \\
&\leq \int_\Gamma \left|\left\langle x, \frac{1}{z + D_x - \varepsilon}u\right\rangle\right|\left|\left\langle y, \frac{1}{z - (D_y - \varepsilon)}v\right\rangle\right|\frac{dz}{2\pi} \\
&\leq c^{2-2s}|\Lambda_x|^{2\alpha(1-s)} \int_\Gamma \left|\left\langle x, \frac{1}{z + D_x - \varepsilon}u\right\rangle\right|^s\left|\left\langle y, \frac{1}{z - (D_y - \varepsilon)}v\right\rangle\right|^s\frac{dz}{2\pi},
\end{aligned}$$

where we used Assumption 2.5, as well as the technical trick in Section 4.3 to obtain the boundary term in Assumption 2.5. Now, since $D_x$ and $D_y$ are independent it holds by Theorem 4.5 that

$$\mathbb{E}\big|\langle(x,y),(\mathcal{L}_{x,y} - 2\varepsilon)^{-1}(u,v)\rangle\big| \leq C|\Lambda_x|^{2\alpha(1-s)} \int_\Gamma \mathbb{E}_1\Big|\langle x,\frac{1}{z+D_x-\varepsilon}u\rangle\Big|^s \mathbb{E}_2\Big|\langle y,\frac{1}{z-(D_y-\varepsilon)}v\rangle\Big|^s\frac{dz}{2\pi}$$
$$\leq C|\Lambda_x|^{2\alpha(1-s)} \int_\Gamma e^{-\mu|x-u|}e^{-\mu|v-y|}\,\frac{dz}{2\pi}.$$

The claim then follows by noticing that since the density of the potentials $V(x)$ is compactly supported the curve $\Gamma$ running around the spectrum can always be taken to have finite length and so $|\Gamma| \leq C\|D_x - \varepsilon\| \leq C\,\lambda$ for some constant $C > 0$. The result not follows letting $s = 1/2$ and noting that $|\Lambda_x| = \big|\frac{x-y}{3}\big|^d$. $\qquad\square$

**Lemma 5.2.** *Under the assumptions of Theorem 3.2, then for any $(u,v) \in \Lambda_x \times \Lambda_y$,*

$$\Big|\langle(x,y),\frac{1}{\mathcal{L}_{x,y} - \varepsilon}(u,v)\rangle\Big| \leq Ce^{-\mu|x-u|}e^{-\mu|v-y|}. \tag{26}$$

*Proof.* For any $z \in i\mathbb{R}$ it holds that $\mathrm{Re}(z + D_x - \varepsilon) \geq \gamma\mathbb{1}_{\Lambda_x}$ and thus using that the psuedospectrum is bounded by the inverse distance from the spectrum in (30) with the non-normal Combes-Thomas from Theorem A.2 there exist constants $C,\mu > 0$ such that for all $\varepsilon > 0$ then

$$\Big|\langle x,\frac{1}{z+D_x-\varepsilon}u\rangle\Big| \leq Ce^{-\mu|x-u|}\ .$$

Then using Lemma B.2 as in the proof of Lemma 5.1 we get that

$$\Big|\langle(x,y),\frac{1}{(\mathcal{L}_{x,y} - 2\varepsilon)}(u,v)\rangle\Big| = \Big|\langle(x,y),\frac{1}{(D_x-\varepsilon)\otimes\mathbb{1}+\mathbb{1}\otimes(D_y-\varepsilon)}(u,v)\rangle\Big|$$
$$\leq C|\Gamma|e^{-\mu|x-u|}e^{-\mu|v-y|}.$$

$$\square$$

## 5.2 Abel averaged states and their properties

Recall the definition of the Abel average from (9). Here we also use the Abel average of the reduced resolvent

$$\rho_\varepsilon^0 = -\varepsilon(\mathcal{L}_\Lambda^0 - \varepsilon)^{-1}(\rho_0). \tag{27}$$

**Proposition 5.3.** *Under the assumptions of Theorem 3.1,*

$$\mathbb{E}\big|\rho_\varepsilon^0(x,y)\big| \leq \varepsilon^{2s-1}\sum_{u\in\Lambda_x,v\in\Lambda_y}|\rho_0(u,v)|C_s e^{-\mu_s(|x-u|+|v-y|)}.$$

*In particular, $\mathbb{E}|\rho_\varepsilon^0(x,y)| \leq \varepsilon^{2s-1}C_s$. In the non-random case, the assumptions of Theorem 3.2, the same bounds hold without the expectation.*

Figure 3: Sketch of the contour $\Gamma$ (dashed) and the numerical range of $D_y$ and $-D_x$. The reader may notice the similarity with the contour in [4, Theorem 7.7].

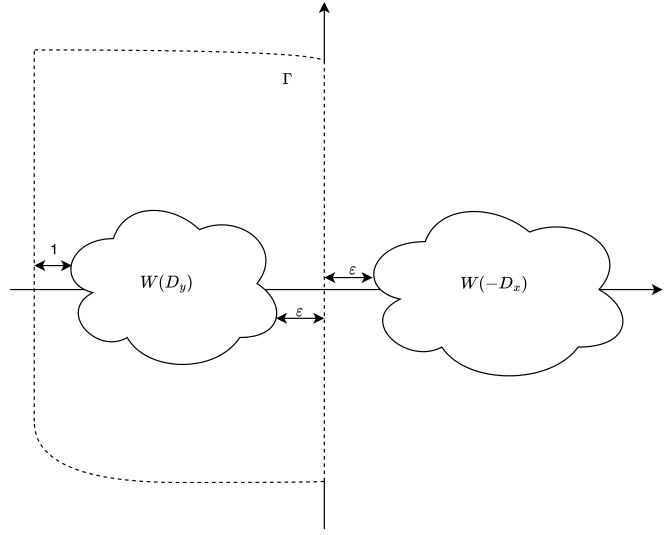

*Proof.* Using again Lemma B.2 we first see that for a contour $\gamma$ in $\overline{\mathbb{C}^-}$ enclosing $\sigma(D_y - \varepsilon)$ then

$$\left| \left\langle (x,y), \frac{1}{(\mathcal{L}_{x,y} - 2\varepsilon)}(u,v) \right\rangle \right| = \left| \left\langle (x,y), \frac{1}{(D_x - \varepsilon \otimes \mathbb{1} + \mathbb{1} \otimes D_y - \varepsilon}(u,v) \right\rangle \right|$$
$$\leq \int_\Gamma \left| \left\langle x, \frac{1}{z + D_x - \varepsilon} u \right\rangle \right| \left| \left\langle y, \frac{1}{z - (D_y - \varepsilon)} v \right\rangle \right| \frac{dz}{2\pi}.$$

Let us now explicitly take $\Gamma$ to be along the imaginary axis and closed up around $\sigma(D_y - \varepsilon)$ with a distance of at least 1 away from the numerical range of $D_y - \varepsilon$. That is, in the complement of the bounded set $W(D_y - \varepsilon) + B_1(0)$, see Figure 3. Thus, for any $z$ on the contour $\Gamma$ then $\left\| \frac{1}{z + D_x - \varepsilon} \right\| \leq \frac{1}{\varepsilon}$ since $\mathrm{Re}(z) \leq 0$ and $\left\| \left\langle y, \frac{1}{z - (D_y - \varepsilon)} v \right\rangle \right\| \leq \frac{1}{\varepsilon}$ on the segment on the imaginary axis and on the remaining part of the contour we have that $\left\| \left\langle y, \frac{1}{z - (D_y - \varepsilon)} v \right\rangle \right\| \leq 1 \leq \frac{1}{\varepsilon}$, since the norm of the resolvent of an operator is at most the distance to the numerical range (a fact that is proven in for example [35, V. 3.2]). Therefore we obtain,

$$\left| \left\langle (x,y), \frac{1}{(\mathcal{L}_{x,y} - 2\varepsilon)}(u,v) \right\rangle \right| \leq \varepsilon^{2s-2} \int_\Gamma \left| \left\langle x, \frac{1}{z + D_x - \varepsilon} u \right\rangle \right|^s \left| \left\langle y, \frac{1}{z - (D_y - \varepsilon)} v \right\rangle \right|^s \frac{dz}{2\pi}.$$

Now, combined with (30) in Appendix A and using that $D_x$ and $D_y$ are independent by Theorem 4.5 we get that

$$\mathbb{E} \left| \left\langle (x,y), \frac{1}{(\mathcal{L}_{x,y} - 2\varepsilon)}(u,v) \right\rangle \right| \leq \varepsilon^{2s-2} |\Gamma| C_s e^{-\mu_s(|x-u|+|v-y|)}$$

and hence that

$$\mathbb{E}\left|\rho_\varepsilon^0(x,y)\right| \leq \varepsilon \sum_{u \in \Lambda_x, v \in \Lambda_y} |\rho_0(u,v)| \varepsilon^{2s-2} |\Gamma| C_s e^{-\mu_s(|x-u|+|v-y|)}.$$

Using that the density of the potentials $V(x)$ is compactly supported as before then proves the first part of the proposition. Now, since $|\rho_0(u,v)| \leq 1$ for each $u, v$ and the exponential the terms are summable the second part follows. The non-random case follows by replacing the use of Theorem 4.5 by an application of Theorem A.2. $\qquad\square$

To prove the main theorem we also need the following technical lemma.

**Lemma 5.4.** *For any $u, v \in \Lambda$ it holds that $\left\|\mathcal{L}_\Lambda^\partial(u,v)\right\| \leq C$.*

*Proof.* Let us first focus on $\|\ell_Z(u,v)\|$. Then from (17) is follows that

$$\|\ell_Z(u,v)\| \leq \sum_{k:\mathrm{supp}(L_k)=Z} \|L_k \, |u\rangle\langle v| \, L_k^\star\| + \|K_Z\| \, \||u\rangle\langle v|\| + \, \|h_Z\|\||u\rangle\langle v|\|.$$

Furthermore, $\|L_k \, |u\rangle\langle v| \, L_k^\star\| \leq \|L_k\| \, \|L_k^\star\|\||u\rangle\langle v|\| = \|L_k L_k^\star\|$ and it holds that

$$\sum_{k:\mathrm{supp}(L_k)=Z} \|L_k L_k^\star\| \ \leq \ \sum_{k:\mathrm{supp}(L_k)=Z} \mathrm{Tr}(L_k L_k^\star) = \mathrm{Tr}(K_Z) \leq |Z|\|K_Z\|$$

since each of the operators $L_k L_k^\star$ are positive operators on the same $|Z|$-dimensional Hilbert space. Since $\ell_Z = 0$ whenever, $\mathrm{diam}(Z) \geq R$ we can assume that $|Z| \leq R^d$. In conclusion, by the finite norm assumption in Assumption 2.4

$$\|\ell_Z(u,v)\| \leq C,$$

where the constant depends only on $N, R$ and $d$. Now, as there are finitely many $Z \subset \mathbb{Z}^d$ with $\mathrm{diam}(Z) < R$ and $\{u,v\} \cap Z \neq \emptyset$ we have that

$$\left\|\mathcal{L}_\Lambda^\partial(u,v)\right\| \ \leq \ \sum_{Z \subset \mathrm{supp}(\mathcal{L}_\Lambda^\partial)} \|\ell_Z(u,v)\|\mathbb{1}\{\{u,v\} \cap Z \neq \emptyset\} \leq C.$$

$\qquad\square$

## 5.3 Proof of main theorems: Exponential decay of coherences

We are now ready to prove the main theorems.

*Proof of Theorem 3.1.* We note that to define $\mathcal{L}_\Lambda^0$ in Equation (18) we needed to assume that $|x-y| \geq 6R$. However, since $\rho_\varepsilon$ is a state then $|\rho_\varepsilon(x,y)| \leq 1$ and so we prove the result for large $|x-y|$ yields the result for all $x, y$.

Let $\rho_\varepsilon^0$ be the Abel average under the modified Lindbladian from (27) and let us prove under the assumptions in the theorem that

$$\mathbb{E}|\rho_\varepsilon(x,y)| \le \mathbb{E}|\rho_\varepsilon^0(x,y)| + Ce^{-\mu|x-y|}. \tag{28}$$

Once (28) is established, the theorem follows from Proposition 5.3.

To prove (28) we consider the split up $\mathcal{L}_\Lambda = \mathcal{L}_\Lambda^0 + \mathcal{L}_\Lambda^\partial$ from Equation (18). Using a resolvent equation yields

$$(\mathcal{L}_\Lambda - \varepsilon)^{-1} = (\mathcal{L}_\Lambda^0 - \varepsilon)^{-1} + (\mathcal{L}_\Lambda^0 - \varepsilon)^{-1}\, \mathcal{L}_\Lambda^\partial (\mathcal{L}_\Lambda - \varepsilon)^{-1}$$

and thus

$$\begin{aligned}
\rho_\varepsilon = -\varepsilon(\mathcal{L}_\Lambda - \varepsilon)^{-1}(\rho_0) &= -\varepsilon(\mathcal{L}_\Lambda^0 - \varepsilon)^{-1}(\rho_0) + (\mathcal{L}_\Lambda^0 - \varepsilon)^{-1}\,\mathcal{L}_\Lambda^\partial\left((-\varepsilon)(\mathcal{L}_\Lambda - \varepsilon)^{-1}(\rho_0)\right) \\
&= \rho_\varepsilon^0 + (\mathcal{L}_\Lambda^0 - \varepsilon)^{-1}(\mathcal{L}_\Lambda^\partial(\rho_\varepsilon)).
\end{aligned}$$

Now, using the finite-range and finite norm assumption in the form of Lemma 5.4, that $\mathcal{L}_\Lambda^\partial$ is supported in $\mathcal{C}_R(x,y)$ by Lemma 4.2, the explicit form of $\mathcal{L}_\Lambda^\partial$ as well as $|\rho_\varepsilon(u',v')| \le 1$ there exists a constant $C > 0$ such that

$$\begin{aligned}
\left|\langle x, (\mathcal{L}_\Lambda^0 - \varepsilon)^{-1}\,(\mathcal{L}_\Lambda^\partial(\rho_\varepsilon))y\rangle\right| &\le \sum_{(u,v)\in\mathcal{C}_R(x,y)} \left|\langle(x,y),(\mathcal{L}_\Lambda^0 - \varepsilon)^{-1}\mathcal{L}_\Lambda^\partial(u,v)\rangle\right||\rho_\varepsilon(u,v)| \\
&\le C\sum_{(u,v)\in\mathcal{C}_R(x,y)} \left|\langle(x,y),(\mathcal{L}_\Lambda^0 - \varepsilon)^{-1}(u,v)\rangle\right|.
\end{aligned}$$

In particular, from the block diagonal form of $\mathcal{L}_\Lambda^0$ as stated in Lemma 4.1 we get that

$$\left|\rho_\varepsilon(x,y) - \rho_\varepsilon^0(x,y)\right| \le C\sum_{(u,v)\in\mathcal{C}_R(x,y)} \left|\langle(x,y),(\mathcal{L}_{x,y} - \varepsilon)^{-1}(u,v)\rangle\right|\mathbb{1}_{u\in\Lambda_x}\mathbb{1}_{v\in\Lambda_y}.$$

Now, taking expectations and using Lemma 5.1 yields that

$$\begin{aligned}
\mathbb{E}\left|\rho_\varepsilon(x,y) - \rho_\varepsilon^0(x,y)\right| &\le C\sum_{(u,v)\in\mathcal{C}_R(x,y)} \mathbb{E}\left[\left|\langle(x,y),(\mathcal{L}_{x,y} - \varepsilon)^{-1}(u,v)\rangle\right|\right] \\
&\le C|\Lambda_x|^\alpha \sum_{(u,v)\in\mathcal{C}_R(x,y)} e^{-\mu|x-u|}e^{-\mu|v-y|}\mathbb{1}_{u\in\Lambda_x}\mathbb{1}_{v\in\Lambda_y} \\
&\le Ce^{-\mu\frac{|x-y|}{3}}|\Lambda_x|^\alpha|\Lambda_x|\ |\Lambda_y| \le Ce^{-\mu|x-y|}.
\end{aligned}$$

for some adjusted $C,\mu > 0$. An application of the triangle inequality now establishes (28). $\qquad\square$

The main theorem in the deterministic case follows using similar considerations.

*Proof of Theorem 3.2.* Using the same strategy as above replacing Lemma 5.1 with Lemma 5.2 one proves the inequality

$$|\rho_\varepsilon(x,y)| \le \left|\rho_\varepsilon^0(x,y)\right| + Ce^{-\mu|x-y|}. \tag{29}$$

The deterministic part of Proposition 5.3 then finishes the proof. $\qquad\square$

## 6   Concluding remarks

Let us first remark that we made extensive use of the Lindblad form, but we did not use that $\mathcal{L}$ is dissipative with respect to an inner product. To caution the reader we note that this is not necessarily the case. It is for example discussed in [53, Section 2.2] that whenever $\mathcal{L}$ is not unital (that is $\mathcal{L}(\mathbb{1}) = 0$) then $\mathrm{Re}(\mathcal{L})$ (defined with respect to the Hilbert-Schmidt inner product) has positive eigenvalues [1]. Instead, in this paper, we relied on the dissipativity of the non-Hermitian evolution.

Second, in our investigations, we had to use the gap assumptions introduced in Section 2.4 to obtain the proof. We would like to provide some insights into the relationship between the gap on the imaginary axis and dark states. A dark state $|\psi\rangle \in \ell^2(\Lambda)$ is defined by $L_k |\psi\rangle = 0$ for all $k$ as well as $H |\psi\rangle = E |\psi\rangle$ for some $E \in \mathbb{R}$. Then, $\mathcal{L}(|\psi\rangle \langle\psi|) = 0$, so all dark states are steady states. Now, let us discuss the connection between dark states and the gap assumption.

- The state $\rho_0$ such that $\rho_0(x,y) = \frac{1}{|\Lambda|}$ for all $x, y \in \Lambda$ is dark for the example of coherence creation in Example 2.1 without disorder.

- Only dark states can have spectrum on the imaginary axis. This can be demonstrated from the following computation. Let $|v\rangle$ is an eigenvector of the non-hermitian evolution with eigenvalue $z$ on the imaginary axis. Then,

$$0 = \mathrm{Re}(z) = \mathrm{Re}(i\langle v, Hv\rangle \ - \frac{1}{2}\sum_k \ \langle v, L_k^* L_k v\rangle) = -\frac{1}{2}\sum_k \ \| L_k |v\rangle \|^2.$$

  In particular, $|v\rangle$ must also be an eigenstate of $H$. These considerations are also true in an approximative sense.

- For operators with random potentials it is unlikely that the eigenstates are approximately dark. This phenomenon could give rise to a Lifshitz tail behaviour for the spectrum close to the imaginary axis.

- We could replace the use of the gap assumption in the proof of Lemma 5.1 with a probabilistic estimate. Thereby, establishing the Lifshitz tails may be way to extend our method beyond Assumption 2.5.

## A   Resolvent estimates for non-normal operators

In this Appendix, we consider resolvent estimates for non-normal operators. First, we discuss the pseudospectrum in the right half plane for dissipative operators and then we turn to the non-normal Combes-Thomas estimate.

---

[1] Nevertheless, all the eigenvalues of $\mathcal{L}$ has a non-positive real part as was discussed in [8, 32].

## A.1 Psuedospectrum in the right half plane for dissipative operators.

The $\varepsilon$-pseudospectrum of an operator $A$ which we denote $\sigma_\varepsilon(A)$ is also characterised as follows

$$\sigma_\varepsilon(A) = \{\lambda \in \mathbb{C} \mid \|(A - \lambda)v\| < \varepsilon, \|v\| = 1\}.$$

At the same time, we note that our definition of dissipativity (which is that $\mathrm{Re}(\langle x, Ax \rangle) \leq 0$ for all vectors $x$) is equivalent to

$$\|(\lambda - A)x\| \geq \lambda \|x\|$$

for all $\lambda > 0$ (see e.g. [24, Proposition 3.23]). From those two observations, we extract the following observations for dissipative operators $A$ on a Hilbert space then for any $r > 0$ it holds that $r \in \sigma_{\frac{1}{r}}(A)$. Suppose that $\mathrm{Re}(\langle x, Ax \rangle) \leq -a < \mathrm{Re}(b)$, then

$$\left\| (A - b\mathbb{1})^{-1} \right\| \leq \frac{1}{|\mathrm{Re}(b) + a|}. \tag{30}$$

## A.2 Combes-Thomas for non-normal operators

In the following, we present the Combes-Thomas estimate from [4, Theorem 10.5] with the appropriate changes that have to be made because of non-normality. Most importantly, the distance $\mathrm{dist}(\sigma(A), z)$ has to be replaced by the $\|(A - z)^{-1}\|$. Originally, the estimate was proven in the self-adjoint case by Combes and Thomas [17] and the discussion in [4, Sec. 10.3] followed the presentation in [1, App II]. Let us start by stating the following lemma. The proof follows by estimating the Neumann series.

**Lemma A.1.** *Suppose $z \notin \sigma_\varepsilon(A)$ and $\|B\| < \varepsilon$. Then $\|(A - z)^{-1}\| \leq \frac{1}{\varepsilon}$ and*

$$\left\| (A + B - z)^{-1} \right\| \leq \frac{1}{\|(A - z)^{-1}\|^{-1} - \|B\|} \leq \frac{1}{\varepsilon - \|B\|}.$$

We now prove a non-normal Combes-Thomas estimate. To do that we define for any $\alpha > 0$

$$S_\alpha = \sqrt{\left( \sup_x \sum_y |A(x,y)| \ e^{\alpha|x-y|} \right) \left( \sup_y \sum_x |A(x,y)| \ e^{\alpha|x-y|} \right)}. \tag{31}$$

**Theorem A.2** (Non-normal Combes-Thomas). *Suppose that $A$ is such that for some $\alpha > 0$ that $S_\alpha < \infty$. Let $z \notin \sigma_\varepsilon(A)$. If $\mu < \alpha$ and $S_\mu < \varepsilon$ then*

$$\left| \langle x, \ (A - z)^{-1} y \rangle \right| \leq \frac{1}{\varepsilon - S_\mu} \exp(-\mu|x - y|). \tag{32}$$

*Further, if $\varepsilon < 2S_\alpha$ then*

$$\left| \langle x, \ (A - z)^{-1} y \rangle \right| \leq \frac{2}{\varepsilon} \exp\left( -\frac{\alpha \varepsilon}{2 S_\alpha} |x - y| \right). \tag{33}$$

*Proof.* Take fixed $y$ and let $R > 0$ be arbitrary. The operator $M$ by

$$M \ket{x} = \exp(\mu \min\{ \; d(x,y), R \} \;)$$

is bounded, invertible and self-adjoint for any $\mu \in (0, \infty)$. Next, notice that for any $x$ such that $|x - y| \leq R$ then

$$\langle x, \; (A - z)^{-1} y \rangle e^{\mu d(x,y)} \; = \langle x, M(H - z)^{-1} \, M^{-1} \, y \rangle = \langle x, (H + B - z)^{-1} \, y \; \rangle$$

as in [4, Theorem 10.5]. Define also similarly again $B = MHM^{-1} \, - H$.

It holds that $\|B\| \;\; \leq \sqrt{\|B\|_{1,1} \, \|B\|_{\infty,\infty}} \;\; = S_\mu \leq S_\alpha$ by interpolation (cf. [4, Proposition 10.6]). Now, if $z \notin \sigma_\varepsilon(A)$ and $S_\mu < \varepsilon$ then

$$\left\| (A + B - z)^{-1} \right\| \; \leq \frac{1}{\|(A - z)^{-1}\|^{-1} \; - \|B\|} \leq \frac{1}{\varepsilon \; - \|B\|} \leq \frac{1}{\varepsilon - S_\mu}.$$

For a proof of (33) we use the inequality $\frac{e^{\mu d} - 1}{e^{\alpha d} - 1} \leq \frac{\mu}{\alpha}$ in both terms in $S_\alpha$ to see that we still have that $S_\mu \leq \frac{\mu}{\alpha} \, S_\alpha$ and the same bound follows. $\qquad\square$

# B  Integral representation of the $\mathcal{L}_{x,y}$ resolvent

Before we start we breifly reconsider when the formula $\frac{1}{A} \; = \int_0^\infty e^{tA} dt$ makes sense for matrices $A$. The following follows from [49, Theorem 15.1].

**Lemma B.1.** *Suppose that* $\mathrm{Re}(A) \leq a < 0$. *Then*

$$-\frac{1}{A} \; = \int_0^\infty e^{tA} dt$$

*where* $\int_0^\infty e^{tA} dt$ *is the matrix defined by* $\langle x|, \int_0^\infty e^{tA} dt \; \ket{y} = \int_0^\infty \langle x|, \; e^{tA} \ket{y} \, dt \;$.

In the self-adjoint case, non-interacting systems were studied using a similar decomposition in the approach to the multiparticle systems in [3, (5.2)].

**Lemma B.2.** *For any pair of dissipative operators* $A_1, A_2$ *such that* $\sup \mathrm{Re}(W(A_i)) \leq -\varepsilon$ *for* $i \in \{1, 2\}$ *for some* $\varepsilon > 0$. *Suppose that* $\Gamma$ *is a closed continuous curve in* $\{z \mid \mathrm{Re}(z) \leq 0\}$ *that encloses and is disjoint from* $\sigma_r(A_2)$, *for some* $r > 0$. *Then*

$$-\frac{1}{A_1 \otimes \mathbb{1} + \mathbb{1} \otimes A_2} \; = \int_\Gamma \frac{dz}{2\pi i} \;\; \frac{1}{z + A_1} \otimes \; \frac{1}{z - A_2} \;.$$

*Thus, it holds that*

$$\langle (x,y), \; \frac{1}{A_1 \otimes \mathbb{1} + \mathbb{1} \otimes A_2} \; (u,v) \rangle = \int_\Gamma \frac{dz}{2\pi i} \;\; \langle x, \; \frac{1}{z + A_1} \; u \rangle \langle y, \; \frac{1}{z - A_2} v \rangle.$$

*Proof.* By assumption $A_1 \otimes \mathbb{1} + \mathbb{1} \otimes A_2$ is dissipative and invertible. Therefore

$$-\frac{1}{A_1 \otimes \mathbb{1} + \mathbb{1} \otimes A_2} = \int_0^\infty dt e^{t(A_1 \otimes \mathbb{1} + \mathbb{1} \otimes A_2)} = \int_0^\infty dt e^{t(A_1 \otimes \mathbb{1})} e^{-t(\mathbb{1} \otimes A_2)} = \int_0^\infty dt e^{tA_1} \otimes e^{tA_2}.$$

Now, by the Dunford-Taylor formula [49, Theorem 15.1] if $\Gamma$ is a contour that encloses $\sigma(A_2)$ then it holds that

$$e^{tA_2} = \int_\Gamma \frac{dz}{2\pi i} \; e^{tz} \frac{1}{z - A_2}.$$

Therefore,

$$-\frac{1}{A_1 \otimes \mathbb{1} + \mathbb{1} \otimes A_2} = \int_0^\infty dt \int_\Gamma \frac{dz}{2\pi i} e^{t(A_1 + z)} \otimes \frac{1}{z - A_2}.$$

Since $\Gamma$ is disjoint from $\sigma_r(A_2)$ it holds that $\left\| \frac{1}{z - A_2} \right\| \le \frac{1}{r}$ uniformly in $z \in \Gamma$. Further, since $\sup \mathrm{Re}(W(A_1)) \le -\varepsilon$ and $\Gamma$ is in the left half of the complex plane it holds that $\left\| e^{t(A_1 + z)} \right\| \le e^{-t\varepsilon}$. Thus,

$$\left\| e^{t(A_1 + z)} \otimes \frac{1}{z - A_2} \right\| \le \frac{e^{-t\varepsilon}}{r} < \infty,$$

which means that we can swap the integrals using Fubini. Thus, since by assumption $A_1 + z$ is always invertible we get that

$$-\frac{1}{A_1 \otimes \mathbb{1} + \mathbb{1} \otimes A_2} = \int_\Gamma \frac{dz}{2\pi i} \int_0^\infty dt e^{t(A_1 + z)} \otimes \frac{1}{z - A_2} = \int_\Gamma \frac{dz}{2\pi i} \frac{1}{z + A_1} \otimes \frac{1}{z - A_2}.$$

$\square$

# C   Lower bounding the Dirichlet Laplacian

Recall the definition of the graph and Dirichlet Laplacians from Section 2.5. The second largest eigenvalue of the graph Laplacian carries much information about the graph and has been intensively studied in the field of spectral graph theory, where it is known as the algebraic connectivity of the graph. For us, it will mainly play a technical role through the following theorem.

**Theorem C.1** ([41, Theorem 4.2]). *Let $H$ be a finite graph with $n$ vertices. Let $\lambda_2$ be the second smallest eigenvalue of $\Delta_H^G$ the graph Laplacian on $H$. Then*

$$\lambda_2 \ge \frac{4}{n \, \mathrm{diam}(H)}.$$

The next lemma extends the bound above to derive a lower bound on the Dirichlet Laplacian, that we give for completeness. Although we believe that a sufficient result can be obtained from the general theory on the discrete Dirichlet problem (see e.g. [12, 13, 36] and references therein).

**Lemma C.2.** *Let $H$ be a connected graph with $n$ vertices and (vertex) boundary $\partial_v H$ and let $\rho = \frac{|\partial_v H|}{n}$ be proportion of boundary vertices. Then it holds that*

$$\Delta_H^D \geq \frac{\rho - \rho^2}{16} \lambda_2 \mathbb{1}_H \geq \frac{\rho - \rho^2}{24} \frac{1}{n \operatorname{diam}(H)} \mathbb{1}_H. \tag{34}$$

*where $\lambda_2$ is the second smallest eigenvalue of $\Delta^H$. In particular, if there is at least one boundary vertex, i.e $|\partial_v H| \geq 1$, it holds that*

$$\Delta_H^D \geq \frac{1}{24n^4} \mathbb{1}_H. \tag{35}$$

*Proof.* Since $H$ is connected $\Delta_H^G$ has a unique eigenvalue 0 with eigenvector $|1\rangle = \frac{1}{\sqrt{n}}(1, \ldots 1)$.

Now, for any normalized vector $|\psi\rangle$ we can write as $|\psi\rangle = a |1\rangle + b |\varphi\rangle$ with $\langle 1|\varphi\rangle = 0$ and $|a|^2 + |b|^2 = 1$. Thus,

$$\langle \psi| \Delta_H^D |\psi\rangle = \langle \psi| \Delta_H^G |\psi\rangle + \langle \psi| B_H |\psi\rangle \geq |b|^2 \langle \varphi| \Delta_H^G |\varphi\rangle \geq |b|^2 \lambda_2. \tag{36}$$

This means our goal is to prove that $|b|^2$ is not too small for any eigenvector of $\Delta_H^D$.

So suppose that $|\psi\rangle = a |1\rangle + b |\varphi\rangle$ is an eigenvector of $\Delta_H^D$ corresponding to the (smallest) eigenvalue $\lambda_0$. Then,

$$\lambda_0 a |1\rangle + \lambda_0 b |\varphi\rangle = \Delta^D |\psi\rangle = a B_H |1\rangle + b(\Delta_H^D) |\varphi\rangle.$$

And thus, defining $|B_H\rangle = B_H |1\rangle$ and its normalization $\left| \hat{B}_H \right\rangle$ and norm $B$ then notice that

$$B_H |1\rangle = \frac{1}{\sqrt{n}} \sum_{k \in \partial G} |k\rangle$$

meaning that $B = |\partial H| n^{-1}$ as well as $\langle 1|B_H\rangle = |\partial H| n^{-1} = \rho$. In particular $B^2 \left| \left\langle 1 \middle| \hat{B}_H \right\rangle \right|^2 = |\langle 1|B_H\rangle|^2 = \rho^2$.

Now, we continue with

$$B^2 |a|^2 \left\| \left( \frac{\lambda_0}{B} |1\rangle - \left| \hat{B}_H \right\rangle \right) \right\|^2 = |b|^2 \left\| (\Delta^D - \lambda_0) |\varphi\rangle \right\|^2.$$

Using Lemma C.3 below and that $\lambda_0 \leq \left\| \Delta^D \right\|$ yields

$$(1 - |b|^2)(\rho - \rho^2) = B^2 |a|^2 \left( 1 - \left| \left\langle 1 \middle| \hat{B}_H \right\rangle \right|^2 \right) \leq |b|^2 4 \left\| \Delta^D \right\|^2.$$

so that if $J = \frac{(\rho - \rho^2)}{4 \left\| \Delta^D \right\|^2} \leq 1$ then it holds that

$$\frac{(\rho - \rho^2)}{8 \left\| \Delta^D \right\|^2} = \frac{J}{2} \leq \frac{J}{J + 1} \leq |b|^2.$$

Combining this result with (36) yields the main statement. For the second statement. Notice that $\operatorname{diam}(H) \leq n$ and that $\rho - \rho^2 = (1-\rho)\rho \geq \min(\rho^2, (1-\rho)^2)$. If all vertices are boundary vertices, then the operator $B_H = \mathbb{1}$ and the statement is true. So, if neither none nor all vertices are boundary vertices then $\min(\rho^2, (1-\rho)^2) \geq \frac{1}{n^2}$ and the last statement follows. $\qquad \square$

**Lemma C.3.** *Suppose that $|v\rangle$ and $|w\rangle$ are unit vectors then for any complex number $z$ it holds that*

$$\||v\rangle + z\,|w\rangle\|^2 \geq 1 - |\langle v|w\rangle|^2. \tag{37}$$

*Proof.* It is a calculation: Set first $z = e^{i\theta}x$ for real number $x$ and let $|\tilde{w}\rangle = e^{i\theta}\,|w\rangle$. Since

$$\||v\rangle + x\,|\tilde{w}\rangle\|^2 = 1 + 2x\operatorname{Re}(\langle v|\tilde{w}\rangle) + x^2$$

the minimal value is attained at $x = -\operatorname{Re}(\langle v|\tilde{w}\rangle)$. So the minimum is

$$\||v\rangle + x\,|\tilde{w}\rangle\|^2 \geq 1 - \operatorname{Re}(\langle v|\tilde{w}\rangle)^2 \geq 1 - |\langle v|\tilde{w}\rangle|^2 = 1 - |\langle v|w\rangle|^2.$$

$\qquad \square$

# Acknowledgements

FRK thanks the Villum Foundation for support through the QMATH center of Excellence (Grant No.10059) and the Villum Young Investigator (Grant No.25452) programs. SW was partially supported by the DFG under grants EXC-2111-390814868 and CRC-TRR 352.